\documentclass[12pt]{report}
\usepackage{graphicx}
\usepackage{amsmath,amssymb,amsthm}

\begin{document}
\begin{center}{\large\bf Finite Mathematics, Finite Quantum Theory and Applications to Gravity and Particle Theory}\end{center}

\vskip 1em \begin{center} {\large Felix M. Lev} \end{center}
\vskip 1em \begin{center} {\it Artwork Conversion Software Inc.,
509 N. Sepulveda Blvd, Manhattan Beach, CA 90266, USA
(Email:  felixlev314@gmail.com)} \end{center}

\begin{flushleft}{\it Abstract:}\end{flushleft} 
We argue that the main reason of crisis in quantum theory is that nature, which is fundamentally discrete and even finite, is described by
classical mathematics involving the notions of infinitely small, continuity etc. Moreover, since classical mathematics has its own foundational problems which cannot be resolved (as follows, in particular, from G\"{o}del's incompleteness theorems), the ultimate physical theory cannot be based on that
mathematics.  
In the first part of the work we discuss inconsistencies in standard quantum theory and reformulate the theory such that it can be naturally generalized to a formulation based on finite mathematics. 
It is shown that: a) as a consequence of inconsistent definition of standard position operator, predictions of the theory contradict
the data on observations of stars; b) the cosmological acceleration and gravity can be treated
simply as {\it kinematical} manifestations of quantum de Sitter symmetry, {\it i.e. the cosmological constant problem does not exist, and for describing those
phenomena the notions of dark energy, space-time background and gravitational interaction are not needed}.
In the second part we first prove that classical mathematics  is a special degenerate case of finite mathematics in the formal limit
when the characteristic $p$ of the field or ring in the latter goes to infinity. {\bf This implies that mathematics describing nature at the most fundamental level involves only a finite number of numbers while the notions of limit and 
infinitely small/large and the notions constructed from them (e.g. continuity, derivative and integral) are needed only in calculations describing nature approximately}.
In a quantum theory based on finite mathematics,  the de Sitter gravitational constant depends on $p$ and disappears in the formal limit $p\to\infty$, i.e. gravity is a consequence of finiteness of nature. The application to particle theory
gives that the notion of a particle and its antiparticle is only approximate and, as a consequence: a) the electric charge and the baryon and lepton quantum numbers can be only approximately conserved; b) particles which in standard theory are treated as neutral (i.e. coinciding with their antiparticles) cannot be elementary. We argue that only Dirac singletons can be true elementary particles and discuss a conjecture that
classical time $t$ manifests itself as a consequence of the fact that $p$ changes, i.e. $p$ and not $t$ is
the true evolution parameter.

\begin{flushleft} PACS: 02.10.Hh, 11.30.Fs, 11.30.Ly, 12.90.+b\end{flushleft}

\begin{flushleft} Keywords: quantum theory, finite fields and rings, de Sitter invariance, gravity\end{flushleft}

\vfill\eject

\tableofcontents

\begin{center} {\bf  List of abbreviations} \end{center}
\begin{flushleft} AdS: anti-de Sitter\end{flushleft}
\begin{flushleft} CC: cosmological constant\end{flushleft}
\begin{flushleft} dS: de Sitter \end{flushleft}
\begin{flushleft}  FQT: finite quantum theory \end{flushleft}
\begin{flushleft} GR: General Relativity\end{flushleft}
\begin{flushleft} GWs: gravitational waves\end{flushleft}
\begin{flushleft} IR: irreducible representation\end{flushleft}
\begin{flushleft} LQG: Loop Quantum Gravity\end{flushleft}
\begin{flushleft}  NT: nonrelativistic theory\end{flushleft}
\begin{flushleft}  QFT: quantum field theory\end{flushleft}
\begin{flushleft}  RT: relativistic theory\end{flushleft}
\begin{flushleft} WF: wave function \end{flushleft}
\begin{flushleft} WPS: wave packet spreading \end{flushleft}

\chapter{Introduction}
\label{Ch1}

The main problem in accepting new theories is probably the following. Our experience is based on generally acknowledged
theories and everything not in the spirit of this experience is treated as contradicting common sense.  A known example is that from the 
point of view of classical mechanics it seems meaningless that the velocity $v=0.999c$ is possible while the velocity $v=1.001c$ is not.
The reason of this judgement is that the experience based on classical mechanics works only for 
velocities $v\ll c$ and
extrapolation of this experience to cases where $v$ is comparable to $c$ is not correct. 

The discovery of quantum theory was a revolutionary breakthrough in physics. One of its lessons is that quantum phenomena cannot be explained in terms of common sense based on everyday experience. Quantum theory has achieved impressive successes in describing some experimental data with an unprecedented accuracy. Nevertheless, the current situation in quantum theory can be characterized as crisis because there are no strong indications that existing modern theories (e.g. string theory, loop quantum gravity, noncommutative geometry etc.) will solve known fundamental problems (e.g. constructing quantum theory of gravity, constructing S-matrix beyond perturbation theory etc.).

It is important to note that, although the philosophies of quantum and classical theories considerably differ each other, quantum theory inherited many its notions from classical one. For example, quantum theory is based on classical mathematics involving the notions of infinitely small, continuity, differentiability etc. As discussed below, those notions are natural from the point of view of macroscopic experience but are not natural on quantum level. Another unnatural feature of fundamental quantum theories (e.g. Quantum Electrodynamics, Electroweak Theory and Quantum Chromodynamics) is that their construction involves space-time which is a pure classical notion but the final results are formulated exclusively in terms of the S-matrix in momentum space without mentioning space-time at all.

In this chapter we discuss whether classical mathematics and standard notions of space-time, symmetry and interaction should be used for constructing ultimate quantum theory. 

\section{What is the main reason of crisis in quantum theory?}
\label{crisis}

The notions of infinitely small, continuity etc. were proposed by Newton and Leibniz more than 300 years ago
and later were substantiated by Cauchy, Weierstrass, and Riemann. At that times people did not know about atoms and elementary particles. On the basis of everyday experience they believed that any macroscopic object can be divided into arbitrarily large number of arbitrarily small parts. However, from the point of view of the present knowledge those notions are problematic. For example, a glass of water contains approximately $10^{25}$ molecules.
We can divide this water by ten, million, etc. but when we reach the level of atoms and elementary particles
the division operation loses its usual meaning and we cannot obtain arbitrarily small parts. 

The discovery of atoms and elementary particles indicates that at the very fundamental level nature is discrete.
As a consequence, any description of macroscopic phenomena using continuity and differentiability can be only approximate.
For example, in macroscopic physics it is assumed that spatial coordinates and time are continuous measurable variables.
However, this is obviously an approximation because coordinates cannot be {\it directly} measured with the accuracy better than atomic
sizes and time cannot be measured with the accuracy better than $10^{-18}s$,
which is of the order of atomic size over $c$. 

As a consequence, distances less than atomic ones do not have a physical meaning  As an example, water in the ocean can be described 
by differential equations of 
hydrodynamics but this is only an approximation since matter is discrete.  Another example is that if we draw a line on a sheet of paper 
and look at this line by a microscope then we will see that the line is strongly discontinuous because it consists of atoms. In nature there are no continuous lines
and surfaces; those geometrical notions can describe reality only when sizes of atoms and elementary particles are neglected. 
{\bf In general, geometry and topology can describe nature only in the approximation when sizes of atoms and elementary particles are neglected}.

Note that even the name "quantum theory" reflects a belief that nature is quantized, i.e.
discrete. Nevertheless, when quantum theory was created it was based on classical mathematics developed mainly in the 19th century. One of the greatest successes of the early quantum 
theory was the discovery that energy levels of the
hydrogen atom can be described in the framework of classical mathematics because the Schr\"{o}dinger 
differential operator has a discrete spectrum. This and many other successes of quantum theory were treated 
as indications that all   
problems of the theory can be solved by using classical mathematics. 

As a consequence, even after 90+ years of the existence of quantum theory it is still based on 
classical mathematics. Although the theory contains divergences and other inconsistencies, physicists persistently try to resolve them in the framework of classical mathematics. 

The mathematical formalism of Quantum Field Theory (QFT) is based on continuous space-time and it is assumed
that this formalism works at distances much smaller than atomic ones. The following problem arises: should we pose a question whether such distances have any physical meaning? One might say that this question does not arise because if a theory
correctly describes experiment then, by definition, mathematics used in this theory does have a physical meaning.
In other words, such an approach can be justified {\it a posteriori}. 

However, even if we forget for a moment that
QFT has divergences and other inconsistencies (see Sec. \ref{ST}), the following question arises. On macroscopic level space-time coordinates
are not only mathematical notions but physical quantities which can be measured. Even in the Copenhagen formulation
of quantum theory measurement is an interaction with a classical object. If we know from our macroscopic experience that
space-time coordinates are continuous only with the accuracy of atomic sizes then why do we use continuous space-time at 
much smaller distances and here we treat space-time coordinates only as mathematical objects?

In particle physics distances are never
measured directly and the phrase that the physics of some process is defined by characteristic distances $l$ means
only that if $q$ is a characteristic momentum transfer in this process then $l=\hbar/q$. This conclusion is based
on the assumption that coordinate and momentum representations in quantum theory are related to each other by the
Fourier transform. However, as shown in Chap. \ref{WPS}, this assumption is based neither on strong theoretical 
arguments nor on experimental data.

Many physicists believe that M theory or string theory will become
"the theory of everything". In those theories physics depends on topology of continuous and differentiable
manifolds at Planck distances $l_P\approx 10^{-35}m$. The corresponding value of $q$ is $q\approx 10^{19}Gev/c$, i.e. much greater than the momenta which can be achieved at
modern accelerators.  Nevertheless, the above 
theories are initially formulated in coordinate representation and it is assumed that at Planck distances 
physics still can be described by classical mathematics. Meanwhile 
lessons of quantum theory indicate that it is highly unlikely that at such distances (and even much greater ones) any
continuous topology or geometry can describe physics.

Another example is the discussion of the results \cite{BICEP2} of the BICEP2 collaboration on the B-mode
polarization in CMB. In the literature those results are discussed in view of the problem whether or
not those data can be treated as a manifestation of gravitational waves in the inflationary period of the Universe. 
Different pros and cons are made on the basis of inflationary models combining 
QFT or string theory with General Relativity (GR). The numerical results are essentially model dependent
but it is commonly believed that the inflationary period lasted in the range $(10^{-36}s,10^{-32}s)$ after the Big Bang.
For example, according to Ref. \cite{Guth}, the inflationary period lasted within $10^{-35}s$ during which the
size of the Universe has grown from a patch as small as $10^{-26}m$ to macroscopic scales of the order of a meter. However, the very notions of $(kg,m,s)$ are purely classical and they were first proposed in 1791 by using the conditions on the Earth. So it is very problematic to use those notions for describing the
inflationary stage of the Universe.

For example, now the official definition of the second is "the duration of 9192631770 periods of the radiation corresponding to the transition between the two hyperfine levels of the ground state of the caesium 133 atom." 
While in the modern system of units, $c$ and $\hbar$ are treated as exact quantities which do not change over time and it is postulated that from now on $c=299792458m/s$ and $\hbar=1.054571800\cdot 10^{-34}kg\cdot m^2/s$, the second is treated only as
an approximate quantity. Since the problem of time is one of the most fundamental problems of of quantum theory and it is not clear whether or not the time operator exists (see e.g. the discussion in Sec. \ref{problemoftime} and Chap. \ref{time}), it is not even legitimate to say whether time should be discrete or continuous. The
physical quantity describing the transition is the transition energy $\Delta E$, and the frequency of the radiation {\it is defined} as $\Delta E/\hbar$. The transition energy cannot be the exact quantity because the width of the
transition energies cannot be zero. In addition, the transition energy depends on 
 gravity, electromagnetic fields and other phenomena. In view of all those phenomena the accuracy of one second given in the literature
is in the range $(10^{-18}s,10^{-16}s)$, and the better accuracy cannot be obtained in principle. 

Additional reservations regarding the definition of the second follow. The quantity $\Delta E$ is
of the order of $10^{-26}ev$ and the problem arises whether $ev$ changes over time.
The measurement of energy in $ev$ reflects the fact that we assume Poincare symmetry. However,
as shown in Sec. \ref{symmetry}, dS or AdS symmetries are more general than Poincare
symmetry and for those symmetries the corresponding quantity is dimensionless and equals 
$R\cdot \Delta E$. At present it is adopted that $R$ is of the order of $10^{26}m$ and there is no guaranty that $R$ does not change with time. Also, the measurement of time in seconds assumes that we can have
a set of many almost independent caesium 133 atoms which are not in a strong gravitational field etc. 
In summary, "continuous time" is a part of classical notion of space-time continuum and makes no sense beyond this notion. 

In addition to the fact that times $(10^{-36}s,10^{-32}s)$ cannot be measured in principle, 
at the inflationary stage of the Universe
there were no nuclei and atoms and so it is unclear whether at such conditions time can be defined at all. The philosophy of classical physics is that any physical quantity can be measured with any desired accuracy. However the state of the Universe at that time
could not be classical, and in quantum theory the definition of any physical quantity is a description how this
quantity can be measured, at least in principle. In quantum theory it is not acceptable to say that "in fact" some 
quantity exists but cannot be measured. So in our opinion, description of the inflationary period by times
$(10^{-36}s,10^{-32}s)$ has no physical meaning. Analogously, 
the distances of the order of $10^{-26}m$ have no physical meaning
because there are no phenomena where such distances can be directly measured. So discussing such distances and times is again an example where our classical experience is extrapolated to areas where it is not
meaningful. In addition,  GR is a pure classical theory and its applicability at such time and spatial intervals is highly questionable (see Sec. \ref{ST}). 

It is usually stated that, during inflation, quantum effects were very important. However, since quantum
theory of gravity has not been constructed yet, a problem arises how to take such effects into account.
Many authors propose models where such effects are taken into account by modifying the action of
standard GR. However, in this case the theory describing distances of the order of $10^{-26}m$ and
times of the order of $(10^{-36}s,10^{-32}s)$ in fact remains classical because the theory is still formulated
in terms of classical action on continuous space-time.

Inflationary models are based on the hypothesis that there exists an inflaton field and its characteristics are
fitted for obtaining observable cosmological quantities. This is a trend in modern physics that for describing a set of experimental data many fitting parameters are involved. Other examples of this trend will be discussed below.

In view of these remarks, inflationary models and statements that the BICEP2 results indicate to the existence of primordial gravitational waves are not based on strong theoretical arguments. In addition,  the conclusion of the BICEP2 collaboration has been questioned by several authors (see e.g. Ref. \cite{BICEP2critics}).

Discussions about the role of space-time in quantum theory were rather popular till
the beginning of the 1970s (see Sec. \ref{ST} for a more detailed discussion). As stated in Ref. 
\cite{BLP}, local quantum fields and
Lagrangians are rudimentary notions which will disappear in the
ultimate quantum theory. My observation is that now physicists usually cannot believe that such words
could be written in such a known textbook. The reason is that in view of successes of QCD and electroweak theory
those ideas have become almost forgotten. However, although the successes are rather impressive, 
they do not contribute to resolving inconsistencies in QFT.

It is also very important to note that even classical mathematics itself has its own foundational problems. 
Indeed, as follows from G\"{o}del's incompleteness theorems, no system of axioms can ensure that all facts about natural 
numbers can be proved. Moreover, the system of axioms in classical mathematics cannot demonstrate its own consistency.
Therefore one might expect that the ultimate quantum theory will be based on mathematics which is not only discrete
but even finite. 

The reason why modern quantum physics is based on continuity, differentiability etc. 
is probably historical: although the founders of quantum theory and many physicists who contributed to it were
highly educated scientists, finite mathematics was not (and still is not) a part of standard
physics education. It is usually believed that classical mathematics is
fundamental while finite mathematics is something inferior which is used only in special applications.
However, as we prove in Sec. \ref{finitemath}, the situation is the opposite: classical mathematics is only a degenerate
case of finite one in the formal limit when the characteristic of the ring or field in finite mathematics
goes to infinity. 

In view of efforts to describe discrete nature by continuous mathematics, one could recall the following joke. A group of monkeys is ordered to reach the Moon. For solving this problem each monkey climbs a tree. The monkey who has reached the highest point believes that he has made the greatest progress and is closer to the goal than the other monkeys.


The main problem is the choice of strategy for constructing a new quantum theory. Since 
no one knows for sure what strategy is the best one, different approaches should 
be investigated. Dirac's advice given in Ref. \cite{DirMath} is: {\it "I learned to distrust all
physical concepts as a basis for a theory. Instead one should
put one's trust in a mathematical scheme, even if the scheme
does not appear at first sight to be connected with physics.
One should concentrate on getting an interesting mathematics."} 

I understand this advice such that our macroscopic experience and physical intuition do not
work on quantum level and hence here we can rely only on solid mathematics. However, many physicists do not
think so and believe that Dirac was "The Strangest Man" (this is the title of the book by Graham Farmelo
about Dirac).

In view of the above remarks and Dirac's advice it seems natural that fundamental quantum physics should be
based on finite mathematics rather than the field of complex numbers. Beginning from Chap. \ref{Ch4} we consider 
such an approach. At the same time, one of the key
principles of physics is the correspondence principle. It means that at some conditions any new theory should
reproduce results of the old well tested theory with a high accuracy. Usually the correspondence principle is
applied such that the new theory contains a parameter and reproduces results of the old theory in a formal limit
when the parameter goes to zero or infinity. Known examples are that nonrelativistic
theory is a special degenerate case of relativistic one in the formal limit $c\to\infty$ and classical (i.e. non-quantum)
theory is a special degenerate case of quantum one in the formal limit $\hbar\to 0$ (see however a discussion in Sec.
\ref{CC}).

Hence one should find a formulation
of standard continuous physics which can be naturally generalized to a formulation based on finite
mathematics. This problem is discussed in the first part of this work. Beginning from Chap. \ref{Ch4}
we consider a quantum theory based either on a finite field or even on a finite ring with characteristic $p$. 
This theory  does not contain infinitely small and infinitely large quantities and here
divergences cannot exist in principle. Standard theory can be treated as a special degenerate case of finite one in a formal limit $p\to\infty$.  

\section{Does quantum theory need space-time background?}
\label{ST}

As noted in the preceding section, using continuous space-time coordinates in quantum theory is
highly questionable. In this section we consider this problem in greater details.

The phenomenon of QFT has no analogs in the history of science. There is no branch of 
science where so impressive agreements between theory and experiment have been
achieved. At the same time, the level of 
mathematical rigor in QFT is very poor and, as a result, QFT has several known 
difficulties and inconsistencies. 

At the end of the 40th it was shown that QED correctly reproduces seven digits for the anomalous magnetic moments of the electron and muon and five digits for the Lamb shift. Although those results have been obtained by subtracting one infinity from the other,
 the results were so impressive that now the dominant philosophy of the majority of physicists is  
that agreement with experiment is much more important than the lack of mathematical rigor. However, not all of them think so.
For example, Dirac wrote in Ref. \cite{DirMath}: 
{\it "The agreement
with observation is presumably by coincidence, just like the
original calculation of the hydrogen spectrum with Bohr orbits.
Such coincidences are no reason for turning a blind eye to the
faults of the theory. Quantum electrodynamics is rather like
Klein-Gordon equation. It was built up from physical ideas
that were not correctly incorporated into the theory and it
has no sound mathematical foundation."}  
In addition, QFT fails in quantizing gravity since in units $c=\hbar=1$ the gravitational constant has the 
dimension $length^2$, and, as a consequence, standard
quantum gravity is not renormalizable.

Usually there is no need to require that the level of mathematical rigor in physics
should be the same as in mathematics. However, physicists should
have a feeling that, at least in principle, mathematical
statements used in the theory can be substantiated. The absence
of a well-substantiated QFT by no means can be treated as a
pure academic problem. This becomes immediately clear when one
wants to work beyond perturbation theory. The problem arises to 
what extent the difficulties of QFT can be resolved in the framework of QFT itself or 
QFT can only be a special case of a more general theory based on essentially new 
ideas. The majority of physicists believe that QFT should
be treated \cite{Weinberg} {\it "in the way it is"}, but at the same
time it is \cite{Weinberg} a {\it "low energy approximation to a
deeper theory that may not even be a field theory, but something
different like a string theory"}. 

One of the key ingredients of QFT is the notion of space-time background. We will discuss this notion
in view of the measurability principle, i.e. that a definition of a physical quantity is a description of 
how this quantity should be measured. In particular, the Copenhagen interpretation is based on this 
principle. In this interpretation the process of measurement necessarily implies interaction with a classical object.
This interpretation cannot be universal since it does not consider situations when the Universe does not have
classical objects at all. Meanwhile in cosmological theories there were no classical objects at the
early stages of the Universe. The problem of interpretation of quantum theory is still open but it is commonly
accepted that at least at the present stage of the world the measurability principle is valid.

Since physics is based on mathematics, intermediate stages of physical theories 
can involve abstract mathematical notions but any physical theory should formulate its final results 
only in terms of physical (i.e. measurable) quantities. Typically the theory does not say explicitly how
physical quantities in question should be measured (well-known exclusions are special and general
theories of relativity where the distances should be measured by using light signals) but it is assumed
that in principle the measurements can be performed. In classical
(i.e. non-quantum) theory it is assumed that any physical quantity in the theory can be measured with
any desired accuracy. In quantum theory the measurability principle is implemented
by requiring that any physical quantity can be discussed only in conjunction with an operator defining 
this quantity. However, quantum theory does not specify how the operator of a physical quantity is
related to the measurement of this quantity. 

\subsection{Space-time background in classical theory}

In standard classical mechanics, the space-time background is the four-dimensional Galilei space, the coordinates
$(t,x,y,z)$ of which are in the range $(-\infty,\infty)$. Then an important observation is that, from the point of view of the measurability principle, Galilei space has a physical meaning only as a {\it space of events for real particles} while if particles are absent, the notion 
of empty Galilei space has no physical meaning. Indeed, there is no way to measure coordinates of a space which exists only in our imagination. 
In mathematics one can use different spaces regardless of whether they have a physical meaning or not. 
However, in physics spaces which have no
physical meaning can be used only at intermediate stages. Since in classical mechanics the final results are formulated in terms of Galilei space, this space should be physical.

In classical relativistic mechanics, the space-time background is the four-dimensional Minkowski space and the
above remarks can be applied to this space as well. The distances in Minkowski space are defined by the 
diagonal metric
tensor $\eta_{\mu\nu}$ such that $\mu,\nu=0,1,2,3$ and $\eta_{00}=-\eta_{11}=-\eta_{22}=-\eta_{33}=1$. Minkowski 
space is also the space-time background in
classical electrodynamics. Here the Maxwell equations make it possible to calculate the electric
and magnetic fields, ${\bf E}(t,x,y,z)$ and ${\bf B}(t,x,y,z)$, at each point of Minkowski space. These fields can
be measured by using test bodies at different moments of time and different positions. Hence in
classical electrodynamics, Minkowski space can be physical only in the presence of test bodies but not as an
empty space. 

In GR the range of the coordinates
$(t,x,y,z)$ and the geometry of space-time are dynamical. 
They are defined by the Einstein equations
\begin{equation}
R_{\mu\nu}+\frac{1}{2}g_{\mu\nu}R_c+\Lambda g_{\mu\nu}=(8\pi G/c^4)T_{\mu\nu}
\label{Einstein}
\end{equation}
where $R_{\mu\nu}$ is the Ricci tensor, $R_c$ is the scalar curvature, $T_{\mu\nu}$ is the stress-energy tensor of matter, 
$g_{\mu\nu}$ is the metric tensor, $G$ is the gravitational constant and $\Lambda$ is the 
cosmological constant (CC). In modern quantum theory space-time in GR is treated as a description of 
quantum gravitational field in classical limit. On quantum level each field is
a collection of particles; in particular it is believed that the gravitational field is a collection of gravitons. 
From this point of view the following question arises. Why does $T_{\mu\nu}$ describe the contribution of
electrons, protons, photons and other particles but gravitons are not included into $T_{\mu\nu}$ and are
described separately by a quantized version of $R_{\mu\nu}$? In any case, quantum theory of gravity has not
been constructed yet and gravity is known only at macroscopic level. 

Here the coordinates and the curvature of  space-time are not pure mathematical notions but physical quantities used
for describing the motion of macroscopic bodies. Therefore in the formal limit when matter disappears, those
notions do not have a physical meaning. Meanwhile, in this limit the solutions of
Eq. (\ref{Einstein}) are Minkowski space when $\Lambda=0$, dS space when $\Lambda>0$ 
and AdS space when $\Lambda < 0$. Hence Minkowski, dS or AdS spaces can be only empty spaces, 
i.e. they are not physical. This shows that the formal limit of GR when matter disappears is nonphysical since in this limit the space-time 
background survives and has a curvature - zero curvature in the case of Minkowski space and a nonzero curvature in the 
case of dS or AdS spaces. 

To avoid this problem one might try to treat the space-time background as a reference frame. 
Moreover, in textbooks (see e.g., Ref. \cite{LLII}) the reference frame in GR is defined as a
collection of weightless bodies, each of which is characterized by three numbers (coordinates) and is supplied by a clock. 
However, the approximation of weightless bodies can be valid
only if matter can be divided by any number of parts. In real situations, since the coordinates refer to macroscopic bodies, they can have a physical meaning 
only with the accuracy discussed in Sec. \ref{crisis}. 

In some approaches (see e.g. Ref. \cite{Hikin1}), when matter
disappears, the metric tensor becomes not the Minkowskian one but zero, i.e. the space-time background 
disappears too. Also, as argued in Ref. \cite{Hikin2}, the metric tensor should be dimensionful since 
$g_{\mu\nu}dx^{\mu}dx^{\nu}$ should be scale independent. In this approach the absolute value of the metric tensor 
is proportional to the number of particles in the Universe.

In approaches based on holographic principle it is stated that the space-time background is not fundamental 
but emergent. For example, as noted in Ref. \cite{Verlinde},
"{\it Space is in the first place a device introduced to describe the positions and movements of particles.
Space is therefore literally just a storage space for information...}". This implies that the
emergent space-time background is meaningful only if matter is present. The author of Ref. \cite{Verlinde}
states that in his approach one can recover Einstein equations where the coordinates and curvature refer
to the emergent space-time. However, it is not clear how to treat the fact that the formal limit
when matter disappears is possible and the space-time background formally remains although, if it is emergent, 
it cannot exist without matter.

\subsection{Problem of time in classical and quantum theories}
\label{problemoftime}

As noted above, from the point of view of quantum 
theory, any physical quantity
can be discussed only in conjunction with an operator defining this quantity. As noted by Pauli (see p. 63 of Ref. \cite{Pauli}), 
at early stages of quantum theory some
authors treated time $t$ as an operator commuting with the Hamiltonian as $[H,t] = i\hbar$. However, 
there are several reasons why such a treatment is not correct. For example, one cannot construct the eigenstate
of the time operator with the eigenvalue 5000 BC or 3000 AD.
Also, as it has been pointed out by several authors (see e.g. Ref.
\cite{Leon}), the conjugated operators should necessarily have the
same spectrum, time has the continuous spectrum in the range $(-\infty,\infty)$ while the
Hamiltonian is usually bounded below and a part of its spectrum may be discrete.

It is usually assumed that in quantum theory the quantity
$t$ can be only a classical parameter describing evolution of a quantum system by the time dependent 
Schr\"{o}dinger equation (see e.g. Refs. \cite{Pauli,time}). The usual justification of this equation is that in the formal limit $\hbar\to 0$ it
becomes the Hamilton-Jacobi equation. Moreover, the justification of standard choice for
different operators (e.g. coordinate, momentum, angular momentum operators and others) is that such a choice has a correct classical limit. However,  the correct classical limit does not guarantee the correct behavior on quantum level.
For example, if $A$ and $B$ are two operators such that $B$ becomes zero in classical limit then the operators
$A$ and $A+B$ have the same classical limit but on quantum level they may have considerably different 
properties.

A problem arises why the principle of
quantum theory that every physical quantity is defined by an operator does not apply
to time. In the literature the problem of time is also often formulated such that "the time of GR and of ordinary Quantum Theory are mutually incompatible notions" (see e.g. Ref. \cite{Anderson}).
As noted by several authors, (see e.g. Refs. \cite{RovelliBook,RovelliTime,Keaton}), $t$ cannot be treated as a fundamental
physical quantity. The reason is that all fundamental physical
laws do not require time and the quantity $t$ is obsolete on fundamental level. A
hypothesis that time is an independently flowing fundamental continuous quantity has
been first proposed by Newton. However, a problem arises whether this hypothesis is
compatible with the principle that the definition of a physical quantity is a description
of how this quantity can be measured.

Let us note that even in classical mechanics particle coordinates and time can be treated in different ways.
A standard treatment of this theory is that its goal is to solve equations of motion and get
classical trajectories where coordinates and momenta are functions of $t$. In 
Hamiltonian mechanics the action can be written as $S = S_0-\int Hdt$ where $S_0$ does not 
depend on $t$ and is called the abbreviated action. Then, as
explained in textbooks, the dependence of coordinates and momenta on
$t$ can be obtained from a variational principle with the action $S$. Suppose now that
one wishes to consider a problem which is usually treated as less general: to find
not the dependence of the coordinates and momenta on $t$ but only possible forms of
trajectories in the phase space without mentioning time at all. If the energy is
a conserved physical quantity then, as described in textbooks, this problem
can be solved by using the Maupertuis principle involving only $S_0$.

However, the latter problem {\it is not} less general than the former one. For
illustration we consider the one-body case. Suppose that by using the Maupertuis
principle one has solved the problem with some initial values of coordinates
and momenta. Let $s$ be a parameter characterizing the particle trajectory, i.e. the particle radius-vector ${\bf r}$,
the momentum ${\bf p}$ and the energy $E$ are functions of $s$.  The particle velocity ${\bf v}$ in units $c=1$ 
{\it is defined} as ${\bf v}(s)={\bf p}(s)/E(s)$. At this stage the problem does not contain $t$ yet. One can {\it define}
$t$ by the condition that $dt=|d{\bf r}|/|{\bf v}|$ and
hence the value of $t$ at any point of the trajectory can be obtained by integration. Hence 
the general problem of classical mechanics can be formulated without mentioning $t$ while if
for some reasons one prefers to work with $t$ then its value can flow only in the positive direction since $dt>0$.

Another point of view is that, at least on classical level, time is a primary quantity while the coordinates ${\bf r}$ 
of each free particle should be {\it defined} in terms of momentum and time as 
\begin{equation}
d{\bf r}={\bf v}dt=\frac{{\bf p}}{E}dt
\label{coordmom}
\end{equation}
where $E=(m^2+{\bf p}^2)^{1/2}$ and $m$ is the particle mass.
In this work we will consider only the case of free particles. Then, as shown in Sec. \ref{classeq}, classical
equations of motions follow from Eq. (\ref{coordmom}) without using Hamilton equations,
Lagrange equations or Hamilton-Jacobi equation. Such a definition of coordinates
is similar to that in GR where distances are defined in terms of time needed for light to travel from one point to
another.

Consider now the problem of time in quantum theory. In the case of
one strongly quantum system (i.e. the system which cannot be described in classical
theory) a problem arises whether there exists a quantum analog of the Maupertuis
principle and whether time can be defined by using this analog. This is a difficult
unsolved problem. A possible approach for solving the problem "{\it how to forget time}" has been proposed
in Refs. \cite{RovelliBook,RovelliTime}. 

One can also consider a situation when a quantum system under
consideration is a small subsystem of a big system where the other subsystem - the environment, is
strongly classical. Then one can define $t$ for the environment as described
above. The author of Ref. \cite{Keaton} considers a scenario when the system as a whole is
described by the stationary Schr\"{o}dinger equation $H\Psi	 = E\Psi$	 but the small quantum
subsystem is described by the time dependent Schr\"{o}dinger equation where $t$ is
defined for the environment as $t=\partial S_0/\partial E$.

One might think that this scenario gives a natural solution of the problem of time in quantum theory.
Indeed, in this scenario it is clear why a quantum system is described by the  
Schr\"{o}dinger equation depending on the classical parameter $t$ which is not an operator: because 
$t$ is the physical quantity
characterizing not the quantum system but the environment. This scenario seems also natural because
it is in the spirit of the Copenhagen interpretation of quantum theory: the evolution of a quantum system
can be characterized only in terms of measurements which in the Copenhagen interpretation are treated as
interactions with classical objects. However, as noted in Ref. \cite{Keaton}, this scenario encounters several problems. For example, the environment can be a classical
object only in some approximation and hence $t$ can be only an approximately continuous parameter. 
In addition, as noted above, the Copenhagen interpretation cannot be universal in all situations. 

As noted in Ref. \cite{Keaton}, the above scenario also does not solve the problem of quantum jumps. 
For illustration, consider a photon emitted in the famous
21cm transition line between the hyperfine energy levels of the hydrogen atom. The phrase that the lifetime of 
this transition is of the order of $\tau=10^7$ years is understood such that the width of the level 
is of the order of $\hbar/\tau$ i.e. the uncertainty of the photon energy is $\hbar/\tau$. 
In this situation a description of the system (atom + electric
field) by the wave function (e.g. in the Fock space) depending on a continuous parameter $t$ has no physical meaning 
(since roughly speaking the quantum of time in this process is of the order of $10^7$ years). 
If we accept this explanation then we should acknowledge that in some situations a description of
evolution by a continuous classical parameter $t$ is not physical. 

One of the arguments in favor of the time dependent Schr\"{o}dinger equation is that in classical
approximation it becomes the Hamilton-Jacobi equation. However, there are no experimental
confirmations that it works on quantum level. Fundamental quantum theories (QED, QCD and
Electroweak theories) proceed from the Heisenberg S-matrix
program according to which in quantum theory one can describe only transitions of states from the infinite past when $t\to -\infty$ to the distant future when $t\to +\infty$.  

The authors of Ref. \cite{Leon} state that the Pauli objection can be circumvented if 
 one uses an external system  to track time, so that "time arises as correlations between the system
and the clock". In this case, the time operator can be defined. It is not
conjugate to the system Hamiltonian, but its eigenvalues still satisfy the Schr\"{o}dinger equation for
arbitrary Hamiltonians. Such an approach is to some extent in the spirit of Ref. \cite{Keaton}. The authors of Ref. \cite{Leon} refer to the extensive literature where the time operator has been discussed. In any case,
the problem to deal or not with the time operator depends on the 
physical situation and there is no universal choice of the time operator which follows from first principles of quantum theory. The authors of Ref. \cite{BryanMedved} consider the “timelessness” of the Universe 
"by subdividing a faux Universe into two entangled parts, “the clock” and “the remainder of the Universe”
and this approach is also to some extent in the spirit of Ref. \cite{Keaton}.

\subsection{Do we need local field operators in quantum theory?}
\label{localfields}

While no operator can be associated with time, a problem arises whether it is possible to consistently define the
position operator. This problem is discussed in detail in Chap. \ref{WPS}. However, QFT operates not with position
operators for each particle but with local quantum fields. A non-quantized quantum field 
$\psi(x)=\psi(t,{\bf x})$ 
combines together two irreducible representations (IRs) with positive
and negative energies. The IR with the positive energy is associated with a particle and the
IR with the negative energy is associated with the corresponding antiparticle. From mathematical
point of view, a local quantum field is described by a reducible representation induced not
from the little algebra IRs are induced from but from the Lorenz algebra. The local
fields depend on $x$ because the factor space of the Poincare group over the Lorentz group is Minkowski space.
In that case there is no physical operator corresponding to $x$, i.e. $x$ is not measurable.
Since the fields describe nonunitary representations, their probabilistic interpretation is problematic.
As shown by Pauli \cite{Pauli2} (see also textbooks on QFT, e.g. Chap. 2 in
Ref. \cite{AB}), in the case of fields with an integer spin there is no subspace where the spectrum of the charge operator has a definite sign
while in the case of fields with a half-integer spin there is no subspace where the spectrum of the energy operator has a definite sign. 
It is also known that
the description of the electron in the external field by the Dirac spinor is not
accurate (e.g. it does not take into account the Lamb shift).

A secondly quantized field $\psi(x)$ is an operator in the Fock space and therefore the contribution of each particle is explicitly taken into account. Each particle in the field can be described by its own coordinates (in the approximation when the position operator exists - see Chap. \ref{WPS}).
In view of this fact the following natural question arises: why do we need an extra coordinate $x$ which does not
belong to any particle? This coordinate does not have a physical meaning and is simply a parameter arising from the second quantization of the non-quantized field $\psi(x)$. 

In QED, QCD and Electroweak theory the Lagrangian density 
depends on local quantized fields and the four-vector $x$ in them is associated with a point in Minkowski space. However, $x$ does not have a physical meaning and is
only the the formal integration parameter which is used in the intermediate stage. The goal of the theory is to construct the
$S$-matrix and when the theory is already constructed
one can forget about Minkowski space because no physical quantity depends on $x$. This is in the
spirit of the Heisenberg $S$-matrix program according to which in relativistic quantum theory it is possible to describe only transitions of states from the infinite past when $t\to -\infty$ to the distant future when $t\to +\infty$. Note that the fact that the $S$-matrix is the operator in momentum space does not exclude a possibility that
in some situations it is possible to have a space-time description with some accuracy but not with absolute accuracy. This problem is discussed in detail in Chap. \ref{WPS}.

Hence a problem
arises why we need local fields at all. They are not needed if we consider only systems
of noninteracting particles. Indeed, such systems are described by tensor products of IRs
and all the operators of such tensor products are well defined. Local fields are used for constructing
interacting Lagrangians which in turn, after quantization, define the representation operators
of the Poincare algebra for a system of interacting particles under consideration. 
Hence local fields do not have a direct physical meaning but are only auxiliary notions.

It is known (see e.g. the textbook \cite{Bogolubov}) that quantum interacting local fields
can be treated only as operatorial distributions. A known fact from the theory of 
distributions is that the product of distributions at the same point is not a correct mathematical
operation. Hence if $\psi_1(x)$ and $\psi_2(x)$
are two local operatorial fields then the product $\psi_1(x)\psi_2(x)$ is not well defined. This is known as the
problem of constructing composite operators. A typical approach discussed in the literature is that
the arguments of the field operators $\psi_1$ and $\psi_2$ should be slightly separated and the limit when the
separation goes to zero should be taken only at the final stage of calculations. However, no universal way
of separating the arguments is known and it is not clear whether any separation can resolve the problems
of QFT. Physicists often ignore this problem and
use such products to preserve locality (although the operator of the quantity $x$ does not exist). As a consequence, the representation operators of interacting systems 
constructed in QFT are not well defined and the theory contains anomalies and infinities. Also, one of the known results
in QFT is the Haag theorem and its generalizations (see e.g. Ref. \cite{Haag}) that the interaction picture
in QFT does not exist. We believe it is rather unethical that even in almost all textbooks on QFT this theorem
is not mentioned at all. A detailed discussion of other problems of QFT can be found, for example, in Ref. \cite{QFT}.

While in renormalizable
theories the problem of infinities can be somehow circumvented at the level of perturbation
theory, in quantum gravity infinities cannot be excluded even in lowest orders of
perturbation theory. One of the ideas of the string theory is that if products of fields
at the same points (zero-dimensional objects) are replaced by products where the arguments of
the fields belong to strings (one-dimensional objects) then there is 
hope that infinities will be less singular. However, a similar mathematical inconsistency exists in string theory
as well and here the problem of infinities has not been solved yet. As noted above, in spite of such mathematical problems, QFT is very popular since it has achieved great successes in describing many experimental data.

In quantum theory, if we have
a system of particles, its wave function (represented as a Fock state or in other forms) gives the 
maximum possible information about this system and
there is no other way of obtaining any information about the system except from its wave function.
So if one works with the emergent space, the information encoded in this space
should be somehow extracted from the system wave function.
However, to the best of our knowledge, there is no theory relating the emergent space with the system wave
function. Typically the emergent space is described in the same way as the "fundamental" space, i.e. as
a manifold and it is not clear how the points of this manifold are related to the wave function. 
The above arguments showing that the "fundamental" space is not physical can be applied to the
emergent space as well. In particular, the coordinates of the emergent space are not measurable and
it is not clear what is the meaning of those coordinates where there are no particles at all. 

In Loop Quantum Gravity (LQG), space-time is treated on quantum level as a special state 
of quantum gravitational field (see e.g. Ref. \cite{RovelliLQG}).
This construction is rather complicated and one of its main goals is to have a quantum generalization of 
space-time such that GR should be recovered as a classical limit of quantum theory. However, so far LQG 
has not succeeded in proving that GR is a special case of LQG in classical limit.

In view of this discussion, it is
unrealistic to expect that successful quantum theory of gravity
will be based on quantization of GR or on emergent space-time. The results of GR might
follow from quantum theory of gravity only in situations when
space-time coordinates of {\it real bodies} is a good
approximation while in general the formulation of quantum theory should not 
involve the space-time background at all. One might take objection that coordinates of 
space-time
background in GR can be treated only as parameters defining
possible gauge transformations while final physical results do
not depend on these coordinates. Analogously, although the
quantity $x$ in the Lagrangian density $L(x)$ is not
measurable, it is only an auxiliary tool for deriving equations
of motion in classical theory and constructing Hilbert spaces
and operators in quantum theory. After this construction has
been done, one can safely forget about background coordinates
and Lagrangian. In other words, a problem is whether
nonphysical quantities can be present at intermediate stages of
physical theories. This problem has a long history discussed in
a vast literature. Probably Newton was the first
who introduced the notion of space-time background but, as noted
in a paper in Wikipedia, "Leibniz thought instead that space
was a collection of relations between objects, given by their
distance and direction from one another". As noted above, the assumption that
space-rime exists and has a curvature even when matter is absent is not physical.
We believe that at the fundamental level unphysical notions should not be present
even at intermediate stages. So Lagrangian can be at best
treated as a hint for constructing a fundamental theory. As already noted, the authors of 
Ref. \cite{BLP} state that local quantum fields and
Lagrangians are rudimentary notions which will disappear in the
ultimate quantum theory. 

\subsection{Summary}
\label{subsecsumm}

The goal of the present quantum theory is to construct operators and the S-matrix for the system under consideration. After the construction has been accomplished,
the final formulation of the theory does not contain space-time background and operators depending on $x$. Therefore {\it those notions do not have a physical meaning on quantum level}.
 On the other hand, since quantum theory is treated as 
more general than the classical one, quantum theory should be able to describe space-time coordinates
of real bodies in semiclassical approximation. In particular, quantum theory should explain how photons from distant stars travel to Earth and even how one can recover the motion of macroscopic bodies along classical trajectories (see Chaps. \ref{WPS} and \ref{time} for a detailed discussion).

Let us make a few remarks about the terminology of quantum theory.
The terms "wave function" (WF) and "particle-wave duality" have arisen at the beginning of quantum era in efforts to explain quantum behavior in terms of classical waves but now it is clear that no such explanation exists. The notion of wave
is purely classical; it has a physical meaning only as a way of describing systems of many particles by
their mean characteristics. 

Such notions as frequency and wave length can be applied
only to classical waves, i.e. to systems consisting of many particles. If a particle state vector contains $exp[i(px-Et)/\hbar]$ then by analogy with the theory of
classical waves one might say that the particle is a wave with the frequency $\omega=E/\hbar$ and the 
(de Broglie) wave
length $\lambda=2\pi\hbar/p$. However, such defined quantities $\omega$ and $\lambda$ are not real frequencies
and wave lengths measured e.g. in spectroscopic experiments. 
A striking example showing that on quantum level $\lambda$ does not have the usual meaning is that from the point of view of classical theory an electron having the size of the order of the
Bohr radius cannot emit a wave with $\lambda=21cm$ (this observation has been pointed out to me by Volodya Netchitailo).

In quantum theory the photon 
and other particles are characterized by their energies, momenta and 
other quantities for which there exist well defined operators while the notion of coordinates on quantum
level is a problem which is investigated in this work. Several results of this work (see e.g.
Sec. \ref{experiment}) are good illustrations
that the term "wave function" might be misleading
since in quantum theory it defines not amplitudes of waves but only amplitudes of probabilities. 

For example, the electron has an electric charge $e$ which is
indivisible. So  for the electron the notion of the charge density is meaningless.
Roughly speaking the electron is a point and its coordinate WF $\psi({\bf r})$ 
(if it exists) describes only probabilities to find the electron at different points ${\bf r}$. 
The quantity $e|\psi({\bf r})|^2$ does not have a meaning of charge density. If a decomposition $\psi=\psi_1+\psi_2$ is possible according to the superposition principle, this does not mean splitting the
electron into two parts with the charges $e_1$ and
$e_2$ such that $e_1+e_2=e$.  Therefore classical description of elementary particles is not
adequate. Those remarks will be important in Sec. \ref{experiment}.

So, although in our opinion the term "state vector" is more pertinent than "wave function" we will use the
latter in accordance with the usual terminology, and the phrase that a photon has a frequency $\omega$ and the wave length $\lambda$ will be understood only such that $\omega=E/\hbar$ and $\lambda=2\pi\hbar/|{\bf p}|$.

In classical theory the notion of field, as well as that of wave, is used for describing systems of many particles by their  mean characteristics. For example, the electromagnetic field consists of many photons. In classical theory each
photon is not described individually but the field as a whole is described by the quantities ${\bf E}(x)$ and 
${\bf B}(x)$ which, as noted above, can be measured (in principle) by using macroscopic test bodies. However, as noted above, in quantum theory there is no well defined operator of the four-vector $x$. In particular,
the notions of electric and magnetic fields of an elementary particle have no physical meaning. In view of these
observations and the above remarks about quantum fields we believe that the term "quantum field", as well as the term
"wave function" might be misleading.

\section{Comparison of different physical theories}
\label{symmetry}

One of the known examples is the comparison 
of nonrelativistic theory (NT) with relativistic one (RT). One of the reasons why RT can be treated as
more general is that it contains a finite parameter $c$ and NT can be treated as a special degenerate
case of RT in the formal limit $c\to\infty$. Therefore, by choosing a large value of $c$,  RT can reproduce any result of NT with a high accuracy. On the contrary, when the limit is already taken one cannot return back from NT to RT and NT cannot reproduce all results of RT. It can reproduce only results obtained when $v\ll c$. Other known examples
are that classical theory is a special degenerated case of quantum one in the formal limit $\hbar\to\infty$ and
RT is a special degenerate case of dS and AdS invariant theories in the formal
limit $R\to\infty$ where $R$ is the parameter of contraction from the dS or AdS algebras to the Poincare algebra
(see below). A question arises whether it is possible to give a general definition when theory A is more
general than theory B. In view of the above examples, we propose the following

{\bf Definition 1.3:} {\it Let theory A contain a finite parameter and theory B be obtained from theory A in the formal limit when the parameter goes to zero or infinity. Suppose that with any desired accuracy theory A can reproduce any result of theory B by choosing a value of the parameter. On the contrary, when the limit is already
taken then one cannot return back to theory A and theory B cannot reproduce all results of theory A. Then theory A is more general than theory B and theory B is a special degenerate case of theory A}. A problem arises how to justify 
{\bf Definition 1.3} not only from physical but also from mathematical considerations.

In relativistic quantum theory the usual approach to symmetry on quantum level follows. 
Since the Poincare group is the group of motions of Minkowski space, quantum states should be described by representations of this group. 
This implies that the representation generators commute according to the commutation relations of the Poincare group Lie algebra:
\begin{eqnarray}
&&[P^{\mu},P^{\nu}]=0,\quad [P^{\mu},M^{\nu\rho}]=-i(\eta^{\mu\rho}P^{\nu}-
\eta^{\mu\nu}P^{\rho}),\nonumber\\
&&[M^{\mu\nu},M^{\rho\sigma}]=-i (\eta^{\mu\rho}M^{\nu\sigma}+\eta^{\nu\sigma}M^{\mu\rho}-
\eta^{\mu\sigma}M^{\nu\rho}-\eta^{\nu\rho}M^{\mu\sigma})
\label{PCR}
\end{eqnarray}
where $\mu,\nu=0,1,2,3$, $P^{\mu}$ are the operators of the four-momentum and  $M^{\mu\nu}$ are the operators of Lorentz angular momenta. This approach is in the spirit of Klein's Erlangen program in mathematics.

However, as noted in Sec. \ref{ST}, in quantum theory the notions of space-time background and operators
depending on $x$ do not have a physical meaning, and, as argued in Refs. \cite{symm1401,DS}, the approach should be the opposite. Each system is described by a set of linearly  independent operators.
By definition, the rules how they commute with each other define the symmetry algebra. 
In particular, {\it by definition}, Poincare symmetry on quantum level means that the operators commute
according to Eq. (\ref{PCR}). This definition does not involve Minkowski space at all.

Such a definition of symmetry on quantum level has been proposed in Ref. \cite{BKT} and in
subsequent publications of those authors. I am very grateful to Leonid Avksent'evich Kondratyuk for explaining me 
this definition during our collaboration. I believe that this replacement of standard paradigm is fundamental for understanding 
quantum theory, and I did not succeed in finding a similar idea in the literature. 
This idea is to some extent in the spirit of Ref. \cite{Dir}. Here Dirac proposed different forms of relativistic dynamics  which are defined by choosing which operators in Eq. (\ref{PCR}) are free and which of them are interaction dependent. 

For understanding this definition the following example might be
useful. If we define how the energy should be measured (e.g.,
the energy of bound states, kinetic energy {\it etc.}), we have
a full knowledge about the Hamiltonian of our system. In
particular, we know how the Hamiltonian commutes with
other operators. In standard theory the Hamiltonian is also
interpreted as the operator responsible for evolution in time,
which is considered as a classical macroscopic parameter (see the preceding section). In
situations when this parameter is a good approximate parameter,
macroscopic transformations from the symmetry group
corresponding to the evolution in time have a meaning of
evolution transformations. However, there is no guaranty that
such an interpretation is always valid (e.g. at the very early
stage of the Universe or in the example with the 21cm transition line discussed in the preceding section). 
In general, according to principles of quantum theory, self-adjoint operators in Hilbert spaces
represent observables but there is no requirement that
parameters defining a family of unitary transformations
generated by a self-adjoint operator are eigenvalues of another
self-adjoint operator. A known example from standard
quantum mechanics is that if $P_x$ is the $x$ component of the
momentum operator then the family of unitary transformations
generated by $P_x$ is $exp(iP_xx/\hbar)$ where $x\in
(-\infty,\infty)$ and in some cases such parameters can be identified with
the spectrum of the position operator. At the same time, the
family of unitary transformations generated by the Hamiltonian
$H$ is $exp(-iHt/\hbar)$ where $t\in (-\infty,\infty)$ and
those parameters cannot be identified with a spectrum of a
self-adjoint operator on the Hilbert space of our system. In
the relativistic case the parameters $x$ can be formally
identified with the spectrum of the Newton-Wigner position
operator \cite{NW} but, as noted in the preceding section and shown in Chap. \ref{WPS}, this operator 
does not have all the required properties for the position operator.
So, although the operators $exp(iP_xx/\hbar)$ and
$exp(-iHt/\hbar)$ are formally well defined, their physical interpretation 
as translations in space and time is questionable.

Following Refs. \cite{ECHAYA,symm}, we will compare four theories: classical (i.e. non-quantum) theory, nonrelativistic quantum theory, relativistic quantum theory and dS or AdS quantum theory. All those theories are described by representations of the symmetry 
algebra containing ten linearly independent operators $A_{\alpha}\,\, (\alpha=1,2,...10)$: four energy-momentum operators, three angular momentum operators and three Galilei or Lorentz boost operators. For definiteness we assume that the 
operators $A_{\alpha}$ where $\alpha=1,2,3,4$ refer to energy-momentum operators, the operators $A_{\alpha}$ where $\alpha=5,6,7$ refer to angular momentum operators and the operators $A_{\alpha}$ where $\alpha=8,9,10$ refer to Galilei or 
Lorentz boost operators. Let $[A_{\alpha},A_{\beta}]=ic_{\alpha\beta\gamma}A_{\gamma}$ where summation over repeated indices is assumed. In the theory of Lie algebras the quantities $c_{\alpha\beta\gamma}$ are called the structure constants. 

Let $S_0$ be a set 
of $(\alpha,\beta)$ pairs such that $c_{\alpha\beta\gamma}=0$ for all values of $\gamma$ and $S_1$ be a set of $(\alpha,\beta)$ pairs such that $c_{\alpha\beta\gamma}\neq 0$ at least for some values of $\gamma$. Since
$c_{\alpha\beta\gamma}=-c_{\beta\alpha\gamma}$ it suffices to consider only such $(\alpha,\beta)$ pairs
where $\alpha<\beta$. If $(\alpha,\beta)\in S_0$ then the operators $A_{\alpha}$ and $A_{\beta}$ commute
while if $(\alpha,\beta)\in S_1$ then they do not commute.

Let $(S_0^A,S_1^A)$ be the sets $(S_0,S_1)$ for theory A and $(S_0^B,S_1^B)$ be the sets $(S_0,S_1)$ for theory B. As noted above, we will consider only theories where $\alpha,\beta=1,2,...10$. Then one can prove the following 

{\bf Statement:} {\it Let theory A contain a finite parameter and theory B be obtained from theory A in the formal limit when the parameter goes to zero or infinity. If the sets $S_0^A$ and $S_0^B$ are different and $S_0^A 
\subset S_0^B$ (what equivalent to $S_1^B \subset S_1^A$)
then theory A is more general than theory B and theory B is a special degenerate case of theory A.}

Proof: Let ${\tilde S}$ be the set of $(\alpha,\beta)$ pairs such that $(\alpha,\beta)\in S_1^A$ and
$(\alpha,\beta)\in S_0^B$. Then, in theory B, $c_{\alpha\beta\gamma}=0$ for any $\gamma$. One can choose
the parameter such that in theory A all the quantities $c_{\alpha\beta\gamma}$ are arbitrarily small.
Therefore, by choosing a value of the parameter, theory A can reproduce any result of theory B with any
desired accuracy. When the limit is already taken then, in theory B, $[A_{\alpha},A_{\beta}]=0$ for all
$(\alpha,\beta)\in {\tilde S}$. This means that the operators $A_{\alpha}$ and $A_{\beta}$ become
fully independent and therefore there is no way to return to the situation when they do not commute.
Therefore for theories A and B the conditions of {\bf Definition 1.3} are satisfied.

It is sometimes stated that the expressions in Eq. (\ref{PCR}) are not general enough because they are
written in the system of units $c=\hbar =1$. Let us consider this problem in more details.
The operators $M^{\mu\nu}$ in Eq. (\ref{PCR}) are dimensionless. In particular, standard angular momentum
operators $(J_x,J_y,J_z)=(M^{12},M^{31},M^{23})$ are dimensionless and satisfy the commutation relations
\begin{equation}
[J_x,J_y]=iJ_z,\quad [J_z,J_x]=iJ_y,\quad [J_y,J_z]=iJ_x
\label{J}
\end{equation}
If one requires that the operators $M^{\mu\nu}$
should have the dimension $kg\cdot m^2/sec$ then they should be replaced by $M^{\mu\nu}/\hbar$, respectively.
In that case the new commutation relations will have the same form as in Eqs. (\ref{PCR}) and (\ref{J}) but
the right-hand-sides will contain the additional factor $\hbar$.

The result for the components of angular momentum depends on the system of units. As shown in quantum
theory, in units $\hbar=1$ the result is given by a half-integer
$0, \pm 1/2, \pm 1,...$. We can reverse the order of units
and say that in units where the angular momentum is a half-integer $l$, its
value in $kg\cdot m^2/sec$ is $1.05457162\cdot 10^{-34}\cdot
l\cdot kg\cdot  m^2/s$. Which of those two values has more
physical significance? In units where the angular momentum
components are half-integers, the commutation relations (\ref{J})
do not depend on any parameters. Then the meaning of
$l$ is clear: it shows how big the angular momentum is in
comparison with the minimum nonzero value 1/2. At the same time,
the measurement of the angular momentum in units $kg\cdot
m^2/s$ reflects only a historic fact that at macroscopic
conditions on the Earth in the period between the 18th and 21st
centuries people measured the angular momentum in such units.

We conclude that for quantum theory itself the quantity $\hbar$ is not needed. However, it
is needed for the transition from quantum theory to classical one: we introduce $\hbar$, then 
the operators $M^{\mu\nu}$ have the dimension $kg\cdot m^2/sec$, and since the right-hand-sides
of Eqs. (\ref{PCR}) and (\ref{J}) in this case contain an additional factor $\hbar$, all the 
commutation relations disappear in the formal limit $\hbar\to 0$. Therefore in classical theory
the set $S_1$ is empty and all the $(\alpha,\beta)$ pairs belong to $S_0$. Since in quantum theory
there exist $(\alpha,\beta)$ pairs such that the operators $A_{\alpha}$ and $A_{\beta}$ do not commute
then in quantum theory the set $S_1$ is not empty and, as follows from {\bf Statement},
classical theory is the special degenerate case of quantum one in the formal limit $\hbar\to 0$.
Since in classical theory all operators commute with each other then in this theory operators are not needed
and one can work only with physical quantities. A question why $\hbar$ is as is does not arise
since the answer is: because people want to measure angular momenta in $kg\cdot m^2/sec$.

Consider now the relation between RT and NT. If we introduce the Lorentz boost operators 
$L^j=M^{0j}\,\, (j=1,2,3)$ then Eqs. (\ref{PCR}) can be written as
\begin{eqnarray}
&&[P^0,P^j]=0,\quad [P^j,P^k]=0, \quad [J^j,P^0]=0,\quad [J^j,P^k]=
i\epsilon_{jkl}P^l,\nonumber\\
&&[J^j,J^k]=i\epsilon_{jkl}J^l,\quad [J^j,L^k]=i\epsilon_{jkl}L^l,\quad [L^j,P^0]=iP^j
\label{RT1}
\end{eqnarray}
\begin{equation}
[L^j,P^k]=i\delta_{jk}P^0,\quad [L^j,L^k]=-i\epsilon_{jkl}J^l
\label{RT2}
\end{equation}
where $j,k,l=1,2,3$, $\epsilon_{jkl}$ is the fully asymmetric tensor such that $\epsilon_{123}=1$, 
$\delta_{jk}$ is the Kronecker symbol and a summation over repeated indices is assumed.
If we now define the energy and Galilei boost operators as $E=P^0c$ and $G^j=L^j/c\,\, (j=1,2,3)$,
respectively then the new expressions in Eqs. (\ref{RT1}) will have the same form while instead of
Eq. (\ref{RT2}) we will have
\begin{equation}
[G^j,P^k]=i\delta_{jk}E/c^2,\quad [G^j,G^k]=-i\epsilon_{jkl}J^l/c^2
\label{NT2}
\end{equation}

Note that in relativistic theory itself the quantity $c$ is not needed. One can choose $c=1$ and treat
velocities $v$ as dimensionless quantities such that $v\leq 1$ if tachyons are not
taken into account. One needs $c$ only for transition from RT to NT: when we introduce $c$ then the dimension of velocities becomes $m/s$ and instead of the operators $P^0$ and $L^j$ we work with the operators $E$
and $G^j$, respectively. If $M$ is the Casimir operator for the Poincare algebra defined such that
$M^2c^4=E^2-{\bf P}^2c^2$ then in the formal limit $c\to\infty$ the first expression in Eq. (\ref{NT2})
becomes $[G^j,P^k]=i\delta_{jk}M$ while the commutators in the second expression become zero.
Therefore in NT the $(\alpha,\beta)$ pairs with $\alpha,\beta=8,9,10$ belong to $S_0$ while in RT
they belong to $S_1$. Therefore, as follows from {\bf Statement}, NT is a special degenerate case of RT in 
the formal limit $c\to\infty$. The question why $c\approx 3\cdot 10^8 m/s$ and not, say
$c=7\cdot 10^9 m/s$ does not arise since the answer is: because people want to
measure $c$ in $m/s$. From the mathematical point of view, $c$ is the parameter of contraction from the Poincare algebra to the Galilei one. This
parameter must be finite: the formal case $c=\infty$ corresponds to the situation when the Poincare algebra does not exist because it becomes the Galilei algebra.

In his famous paper "Missed Opportunities" \cite{Dyson} Dyson notes that RT is more general than NT and dS and AdS theories
are more general than RT not only from physical but also from pure mathematical considerations. Poincare group is more symmetric than Galilei one and the transition from the former to the latter at $c\to\infty$ is called contraction. For the first time this notion has been discussed by Inonu and Wigner \cite{IW}.
Analogously dS and AdS groups are more symmetric than Poincare one and the transition from the former to the latter at $R\to\infty$ (described below) also is called contraction. At the same time, since dS and AdS groups are semisimple they have a maximum possible symmetry and cannot be obtained from more symmetric groups by contraction. However, since we treat symmetry not from the point of view of a group of motion for the corresponding background space but from the point of view of commutation relations in the symmetry algebra, we will discuss the relations between the dS and AdS algebra on one hand and the Poincare algebra on the other.

By analogy with the definition of Poincare symmetry on quantum level, the definition of dS symmetry on quantum level should not
involve the fact that the dS group is the group of motions of dS space.
Instead, {\it the definition} is that the operators $M^{ab}$ ($a,b=0,1,2,3,4$, $M^{ab}=-M^{ba}$)
describing the system under consideration satisfy the
commutation relations {\it of the dS Lie algebra} so(1,4), {\it i.e.},
\begin{equation}
[M^{ab},M^{cd}]=-i (\eta^{ac}M^{bd}+\eta^{bd}M^{ac}-
\eta^{ad}M^{bc}-\eta^{bc}M^{ad})
\label{CR}
\end{equation}
where $\eta^{ab}$ is the diagonal metric tensor such that
$\eta^{00}=-\eta^{11}=-\eta^{22}=-\eta^{33}=-\eta^{44}=1$.
The {\it definition} of AdS symmetry on quantum level is given by the same equations
but $\eta^{44}=1$.

With such a definition of symmetry on quantum level, dS and AdS
symmetries are more natural than Poincare symmetry. In the
dS and AdS cases all the ten representation operators of the symmetry
algebra are angular momenta while in the Poincare case only six
of them are angular momenta and the remaining four operators
represent standard energy and momentum. If we {\it define} the
operators $P^{\mu}$ as $P^{\mu}=M^{4\mu}/R$ where $R$ is a parameter with the dimension
$length$ then in the formal
limit when $R\to\infty$, $M^{4\mu}\to\infty$ but the quantities
$P^{\mu}$ are finite, Eqs. (\ref{CR}) become Eqs. (\ref{PCR}). This procedure is called contraction and 
in the given case it is the same for the dS or AdS symmetry. As follows from Eqs. (\ref{PCR}) and
(\ref{CR}), if $\alpha,\beta=1,2,3,4$ then the $(\alpha,\beta)$ pairs belong to $S_0$ in
RT and to $S_1$ in dS and AdS theories. Therefore, as follows from {\bf Statement}, 
RT is indeed
a special degenerate case of dS and AdS theories in the formal limit when $R\to\infty$. 

By analogy with the transitions from quantum theory to classical one and from RT to NT,
one can say that for dS/AdS theories themselves the quantity $R$ is not needed. It is used
because instead of working with dimensionless operators $M^{4\mu}$ people prefer to work
with Poincare momenta $P^{\mu}$ which in system of units $c=\hbar=1$ have the dimension
$1/m$. The quantity $R$ is
fundamental to the same extent as $c$ and $\hbar$, and the question why $R$ is as is has a simple answer:
because people want to measure distances in meters. 

One might pose a question whether or not the values of $\hbar$, $c$ and $R$ may change with time. 
As far as $\hbar$ is concerned,
this is a question that if the angular momentum equals 1/2 then
its value in $kg\cdot  m^2/s$ will always be $1.054571800\cdot
10^{-34}/2$ or not. It is obvious that this is not a problem of
fundamental physics but a problem of definition of the units $(kg,m,s)$. 
In other words, this is a problem of metrology and
cosmology. Moreover, since $\hbar$ is the parameter of contraction from quantum to classical
theory, the very notion of $\hbar$ can have a physical meaning only in situations when
classical physics works with a high accuracy. In particular, there is no reason to believe that
this was the case at the early stage of the Universe. Analogous remarks can be made about
the quantity $c$. However, as noted in Sec. \ref{crisis}, in the modern system of units
it is postulated that the values $\hbar$ and $c$ do not change over time.
Analogously, there is no guaranty that the value of $R$ in meters will be always 
the same and, since $R$ is the parameter of contraction from dS and AdS symmetries to Poincare
symmetry, the very notion of $R$ is meaningful only when Poincare symmetry works with a high
accuracy. 

We have proved that all the three discussed comparisons satisfy the conditions formulated in 
{\bf Definition 1.3}.
Namely, the more general theory contains a finite
parameter and the less general theory can be treated as a special degenerate case of the former in the
formal limit when the parameter goes to zero or infinity. The more general theory can reproduce all results of
the less general one by choosing some value of the parameter. On the contrary, when the limit is already taken
one cannot return back from the less general theory to the more general one. 

In Refs. \cite{lev1a,jpa1,symm1401} we considered properties of dS quantum theory and argued that
dS symmetry is more natural than Poincare one. However, the above discussion proves that dS
and AdS symmetries are not only more natural than Poincare symmetry but more general.
In particular, $R$ is fundamental to the same extent as $\hbar$ and $c$. 
By analogy with the abovementioned fact that $c$ must be finite, $R$ must be finite too: the formal case $R=\infty$ corresponds to the 
situation when the dS and AdS algebras do not exist because they become the Poincare algebra.
{\bf On quantum level $R$ is only the parameter of contraction from dS or AdS algebras to the Poincare one
and has nothing to do with the radius of the dS or AdS space}. We will see in Sec. \ref{antigravity}
that the result for the cosmological acceleration obtained without any geometry in semiclassical approximation to dS quantum theory  is the same as in GR when the radius of the dS space equals $R$.
 
In the literature the notion of the $c\hbar G$ cube of physical theories is sometimes used. The meaning is
that any relativistic theory should contain $c$, any quantum theory should contain $\hbar$ and any gravitation 
theory should contain $G$.
The more general a theory is the greater number of those parameters it contains. In particular, relativistic quantum theory of
gravity is the most general because it contains all the three parameters $c$, $\hbar$ and $G$ while nonrelativistic
classical theory without gravitation is the least general because it contains none of those parameters.

However, this notion is problematic for the following reasons. The Lagrangian of GR is linear in Riemannian curvature $R_c$, but from the point of view 
of symmetry requirements there exist infinitely many
Lagrangians satisfying such requirements. For example, $f(R_c)$ theories of gravity are widely discussed, where there can be many possibilities 
for choosing the function $f$. Then the effective gravitational
constant $G_{eff}$ can considerably differ from standard gravitational constant $G$. It is also argued that GR is a low energy approximation of more general theories involving higher order derivatives. The nature of gravity on quantum level is a problem, and standard canonical quantum gravity is not renormalizable. For those reasons
the quantity $G$ can be treated only as a phenomenological parameter but not fundamental one.

A known problem is: how many independent
dimensionful constants are needed for a complete description of
nature? A paper \cite{Okun} represents a trialogue between
three known scientists: M.J. Duff, L.B. Okun and G.
Veneziano (see also Ref. \cite{Okun2} and references therein). 
The results of their discussions are summarized as
follows: {\it LBO develops the traditional approach with three
constants, GV argues in favor of at most two (within
superstring theory), while MJD advocates zero.} According to
Ref. \cite{W-units}, a possible definition of a
fundamental constant might be such that it cannot be calculated
in the existing theory. 

As follows from the above discussion, the
three fundamental parameters are $(c,\hbar,R)$, and, in contrast to usual statements,
the situation is the opposite: relativistic theory should not contain $c$, quantum theory should not contain $\hbar$ and dS or AdS theories should not contain $R$.
Those three parameters are needed only for transitions from more general
theories to less general ones. The most general dS and AdS quantum theories
do not contain dimensionful quantities at all while the least general nonrelativistic
classical theory contains three dimensional quantities $(kg,m,s)$. Therefore our results support the opinion of the third author in Ref. \cite{Okun}.

In view of the above remarks, one might think that the dS analog of the
energy operator is $M^{40}$. However, in dS theory all the operators $M^{a0}$
$(a=1,2,3,4)$ are on equal footing. This poses a problem whether a parameter
describing the evolution defined by the Hamiltonian is a
fundamental quantity even on classical level.

All the three discussed comparisons of general theories have the following common feature:
the more general theory contains a finite parameter and the less general theory is a special 
degenerate case of the former in the formal limit when the parameter goes to zero or
infinity. As shown in Subsec. \ref{finmath}, the transition from finite mathematics to classical
one also can be described in the framework of such a scheme.

In the existing quantum theory, problems with nonphysical
notions and infinities arise as a result of describing interactions in terms of local quantum
fields. In the present work local quantum fields are not used at all and we apply the notion of symmetry 
on quantum level only to systems of free particles. One might think that such a consideration can be only 
of academic interest. Nevertheless, we will
see below that there is a class of problems where such a consideration gives a new perspective on
fundamental notions of quantum theory. We will consider applications of our approach to the
cosmological constant problem, gravity and particle theory.

Finally, let us define the notion of elementary particle. 
Although theory of elementary particles exists for a rather long period of time, there is no commonly accepted definition of elementary particle in this theory. In the spirit of the above definition of symmetry on quantum level and 
Wigner's approach to Poincare symmetry
\cite{Wigner}, a general definition, not depending on the choice of the classical background 
and on whether we consider a local or nonlocal theory, is  that a particle is elementary if 
the set of its WFs is the space of an IR of the symmetry algebra in the given theory. 
In particular, in Poincare invariant theory an elementary particle is described by an IR of the Poincare algebra,
in dS or AdS theory it is described by an IR of the dS or AdS algebra, respectively, etc.

\section{Remarks on the cosmological constant problem}
\label{CC}

The discovery of the cosmological repulsion (see e.g. Refs.
\cite{Perlmutter,Melchiorri}) has ignited a vast discussion on how this
phenomenon should be interpreted. The majority of authors treat
this phenomenon as an indication that $\Lambda$ is positive and therefore
the space-time background has a positive curvature. According to
Refs. \cite{Perlmutter,Melchiorri,Spergel,Nakamura}, the observational data on the value
of $\Lambda$ indicate that it is non-zero and positive with a confidence of 99\%.
Therefore the possibilities that $\Lambda=0$ or $\Lambda<0$ are
practically excluded. In the approach discussed in Ref. \cite{Hawking}, the "fundamental" quantity
$\Lambda$ is negative while effectively $\Lambda>0$ only on classical level. In our approach the
notion of "fundamental" $\Lambda$ does not exist since we proceed from the commutation relations
(\ref{CR}) which do not contain space-time characteristics. 

The majority of works dealing with the CC problem 
proceed from the assumption that $G$ is the fundamental physical quantity, the goal of the theory is to 
express $\Lambda$ in terms of $G$ and to explain why $\Lambda$ is so small. The usual formulation of the CC problem follows.
In standard QFT one starts from the choice of the space-time background. By analogy with the philosophy of GR,
it is believed that the choice of the Minkowski background is more physical than the choice of the dS or AdS one. Here the quantity $G$ is treated as fundamental and 
the value of $\Lambda$ should be extracted from the vacuum expectation value of
the energy-momentum tensor. The theory contains strong
divergencies and a reasonable cutoff gives for $\Lambda$ a value exceeding
the experimental one by 120 orders of magnitude. This result is expected because in
units $c=\hbar=1$ the dimension of $G$ is $m^2$, the dimension of $\Lambda$ is
$m^{-2}$ and therefore one might think than $\Lambda$ is of the order of $1/ G$
what exceeds the experimental value by 120 orders of magnitude. However, one of the consequences of the results of the preceding section is that the 
CC problem does not exist
because its formulation is based on the incorrect assumption that RT is more general than
dS and AdS theories. 

The discussion in the preceding section indicates that three quantities describing
transitions from more general theories to less general ones are not $(\hbar, c, G)$ but $(\hbar, c, R)$,
and these quantities have nothing to do with gravity. In particular,
RT {\it should not} contain $c$, quantum
theory {\it should not contain} $\hbar$ and dS and AdS theories {\it should not} contain $R$. The problem of
treating $G$ is one of key problems of this work and will be discussed below.

As noted in the preceding section, a standard phrase that RT becomes NT when $c\to\infty$ should be understood
such that if RT is rewritten in conventional units then $c$ will appear and one can take
the limit $c\to\infty$. A more physical description of the
transition is that all velocities in question are much less
than unity. We will see in this section and Sec. \ref{antigravity} that those
definitions are not equivalent. Analogously, a more physical
description of the transition from quantum to classical theory
should be that all angular momenta in question are very large
rather than $\hbar\to 0$.

Consider now what happens if one assumes that dS symmetry is the most general. 
As explained in the preceding section, in our approach dS symmetry has nothing to do with 
dS space but now we consider
standard notion of this symmetry. The dS space is a four-dimensional
manifold in the five-dimensional space defined by
\begin{equation}
x_1^2+x_2^2+x_3^2+x_4^2-x_0^2=R^{'2}
\label{dSspace}
\end{equation}
Here $R'$ is the radius of dS space while in the preceding section the notation $R$
was used 
for the parameter defining contraction from dS and AdS theories to RT. In the formal limit $R'\to\infty$ the action of the dS group in
a vicinity of the point $(0,0,0,0,x_4= R')$ becomes the action
of the Poincare group on Minkowski space. In the literature,
instead of $R'$, the CC $\Lambda=3/R^{'2}$ is often used. The dS space 
can be parameterized without using
the quantity $R'$ at all if instead of $x_a$ ($a=0,1,2,3,4$) we
define dimensionless variables $\xi_a=x_a/R'$. It is also clear
that  the elements of the SO(1,4) group do not depend on $R'$
since they are products of conventional and hyperbolic
rotations. So the dimensionful value of $R'$ appears only if one
wishes to measure coordinates on the dS space in terms of
coordinates of the flat five-dimensional space where the dS
space is embedded in. This requirement does not have a
fundamental physical meaning. Therefore the value of $R'$
defines only a scale factor for measuring coordinates in the dS
space. It is also obvious that if dS symmetry is assumed
from the beginning then the value of $\Lambda$ has no relation
to the value of $G$.

If one assumes that space-time background is fundamental regardless of whether
matter is present or not, then in
the spirit of GR it is natural to think that empty
space-time background is flat, i.e. that $\Lambda=0$ and this was one of the 
subjects of the well-known debates between Einstein and de
Sitter. However, as noted above, it is now
accepted that $\Lambda\neq 0$ and, although it is very small,
it is positive rather than negative. If we accept
parameterization of the dS space as in Eq. (\ref{dSspace}) then
the metric tensor on the dS space is
\begin{equation}
g_{\mu\nu}=\eta_{\mu\nu}-x_{\mu}x_{\nu}/(R^{'2}+x_{\rho}x^{\rho})
\label{metric}
\end{equation}
where $\mu,\nu,\rho = 0,1,2,3$, $\eta_{\mu\nu}$ is the Minkowski metric tensor,  
and a summation
over repeated indices is assumed. It is easy to calculate the
Christoffel symbols in the approximation where all the
components of the vector $x$ are much less than $R'$:
$\Gamma_{\mu,\nu\rho}=-x_{\mu}\eta_{\nu\rho}/R^{'2}$. Then a
direct calculation shows that in the nonrelativistic
approximation the equation of motion for a single particle is
\begin{equation}
{\bf a}={\bf r}c^2/R^{'2}
\label{accel}
\end{equation}
where ${\bf a}$ and ${\bf r}$ are the acceleration and the
radius vector of the particle, respectively.

The acceleration given by Eq. (\ref{accel}) depends on
$c$ although
it is usually believed that $c$ can be present only in
relativistic theory. This illustrates the fact mentioned above
that the transition to nonrelativistic theory
understood as $|{\bf v}|\ll 1$ is more physical than that
understood as $c\to\infty$. The presence of $c$ in Eq.
(\ref{accel}) is a consequence of the fact that this expression is
written in standard units. In nonrelativistic theory $c$ is
usually treated as a very large quantity. Nevertheless, the
last term in Eq. (\ref{accel}) is not large since we assume
that $R'$ is very large.

Suppose now that we have a system of two noninteracting particles and $({\bf r}_i,{\bf a}_i)$
$(i=1,2)$ are their radius vectors and accelerations, respectively. Then Eq. (\ref{accel}) is
valid for each particle if $({\bf r},{\bf a})$ is replaced by $({\bf r}_i,{\bf a}_i)$, respectively.
Now if we define the relative radius vector ${\bf r}={\bf r}_1-{\bf r}_2$ and the 
relative acceleration ${\bf a}={\bf a}_1-{\bf a}_2$ then they will satisfy the same Eq. (\ref{accel})
which shows that the dS antigravity is repulsive. It terms of $\Lambda$ it reads
${\bf a}=\Lambda{\bf r}c^2/3$ and therefore in the AdS case we have attraction rather than repulsion. 

The fact that even a single particle in the Universe has a
nonzero acceleration might be treated as contradicting the law
of inertia but, as already noted, this law has been postulated only for Galilean
or Poincare symmetries and we have ${\bf a}=0$ in the limit
$R'\to\infty$. A more serious problem is that, according to
standard experience, any particle moving with acceleration
necessarily emits gravitational waves, any charged particle
emits electromagnetic waves etc. Does this experience
work in the dS Universe? This problem is intensively discussed in
the literature (see e.g. Ref. \cite{Akhmedov} and
references therein). Suppose we accept that, according to GR,
the loss of energy in gravitational emission is proportional to
the gravitational constant. Then in the given case it might be problematic to apply GR since the constant
$G$ characterizes interaction between different particles and it is not clear
whether it can be used if only one particle exists in the Universe.

In textbooks written before 1998 (when the cosmological
acceleration was discovered) it is often claimed that $\Lambda$ is not needed since its presence
contradicts the philosophy of GR: matter creates curvature of space-time, so  
in the absence of matter space-time should be flat (i.e. Minkowski) while empty dS space is not flat.
For example, the authors of Ref. \cite{LLII}
write that "...there are no convincing reasons, observational and theoretical, for introducing
a nonzero value of $\Lambda$" and that "... introducing to the density of the Lagrange function a constant term which does not depend on the field state would mean attributing to space-time a principally ineradicable curvature which is related neither to matter nor to gravitational waves".

As noted above, such a philosophy has no physical meaning since the goal of curvature is to describe
the motion of real bodies and therefore the curvature does not have a physical
meaning for empty space-time. However, in view of this philosophy,
the discovery of the fact that $\Lambda \neq 0$ has ignited many discussions.
The most popular approach is as follows. One can move the term with $\Lambda$ in Eq. (\ref{Einstein}) from the
left-hand side to the right-hand one. 
Then the term with $\Lambda$ is treated as the stress-energy tensor of a hidden matter which is called
dark energy: $(8\pi G/c^4)T_{\mu\nu}^{DE}=-\Lambda g_{\mu\nu}$. With such an approach one implicitly
returns to Einstein's point of view that a curved space-time cannot
be empty. In other words, this is an
assumption that the Poincare symmetry is fundamental while the dS one
is emergent. With the observed value of $\Lambda$ this
dark energy contains approximately 68\% of the energy of the Universe. In this approach $G$ is treated as a
fundamental constant and one might try to express $\Lambda$ in terms of $G$. As noted above,
in such an approach we have the CC problem which sometimes is called the dark energy problem.

Several authors criticized this approach from the following considerations. GR without the contribution of
$\Lambda$ has been confirmed with a good accuracy in experiments in the Solar System. If $\Lambda$ is as
small as it has been observed then it can have a significant effect only at cosmological distances while
for experiments in the Solar System the role of such a small value is negligible. The authors of Ref.
\cite{Bianchi} titled "Why All These Prejudices Against a Constant?", note that
it is not clear why we should think that only a special case $\Lambda=0$ is allowed. If we accept the 
theory containing a constant $G$ which cannot be calculated and
is taken from the outside then why can't we accept a theory containing two independent 
constants?

In Secs. \ref{antigravity} and \ref{LambdaDiscrete} we show by different methods that, as a consequence of
dS symmetry on quantum level defined in the preceding section, the 
CC problem does not exist and the cosmological acceleration can be easily and naturally explained 
without using any geometry (i.e. dS space, its metric and connection) but simply from semiclassical 
approximation in dS quantum mechanics. Our result for the cosmological acceleration 
coincides with Eq. (\ref{accel}) if $R'=R$.

Concluding this section we note the following. As follows from Eq. (\ref{accel}), the quantity $R'$ can be extracted
from measurements of the relative acceleration in the dS Universe. However, as follows from this equation, the acceleration
is not negligible only if distances between particles are comparable to $R'$. Hence at present a direct measurement
of $R'$ is impossible and conclusions about its value are made indirectly from the data on high-redshift supernovae by using 
different cosmological models. Probably the most often used model is the $\Lambda$CDM one which is based on six parameters.
It assumes that GR is  the correct theory of gravity on cosmological scales and uses the FLRW metric (see e.g. Ref.
\cite{LambdaCDM}). Then the result of Refs. \cite{Spergel,Nakamura} is that with the accuracy of 5\% $\Lambda$ is such
that $R$ is of the order of $10^{26}m$ and in subsequent experiments the accuracy has been improved to 
1\%. This value of $\Lambda$ is also obtained in other cosmological models.
On the other hand, in the literature several alternative models are
discussed where $R'$ considerably differs from $10^{26}m$. It is also important to note that if $\Lambda$ is treated
only as an effective cosmological constant (arising e.g. due to dark energy) then the radius of the Universe does not define the curvature of the dS space. 

In summary, we treat the fact that $\Lambda>0$ as an indication in favor of dS symmetry on quantum level, and this
has nothing to do with gravity, with existence or nonexistence of dark energy and with the problem whether
empty space-time can or cannot be curved. On the other hand, the numerical value of $R'$ is still an open problem.

\section{Is the notion of interaction physical?}
\label{inter}

The fact that problems of quantum theory arise as a result of describing interactions 
in terms of local quantum fields poses the following dilemma. One can either modify the
description of interactions (e.g. by analogy with the string theory where interactions at points 
are replaced by interactions at strings) or investigate whether the notion of interaction is needed at 
all. A reader might
immediately conclude that the second option fully contradicts the existing knowledge and
should be rejected right away. In the present section we discuss whether gravity
might be not an interaction but simply a kinematical manifestation of dS symmetry on quantum level.

Let us consider an isolated system of two particles and pose a question of whether they interact or not.
In theoretical physics there is no unambiguous criterion for answering this question.
For example, in classical nonrelativistic
and relativistic mechanics the criterion is clear and simple: if the relative acceleration of the
particles is zero they do not interact, otherwise they interact. However, those theories are based
on Galilei and Poincare symmetries, respectively and there is no reason to believe that such
symmetries are exact symmetries of nature.

In quantum mechanics the criterion can be as follows. If $E$ is the energy operator of the two-particle system 
and $E_i$ ($i=1,2$) is the energy operator of particle $i$ then one can
formally define the interaction operator $U$ such that 
\begin{equation}
E=E_1+E_2+U
\label{E12}
\end{equation}
Therefore the criterion can be such that the particles do not interact if $U=0$, i.e. $E=E_1+E_2$. 

In QFT the criterion is also clear: the particles interact
if they can exchange by virtual quanta of some fields. For example, the electromagnetic interaction
between the particles means that they can exchange by virtual photons, the gravitational interaction -
that they can exchange by virtual gravitons etc. In that case $U$ in Eq. (\ref{E12}) is an effective
operator obtained in the approximation when all degrees of freedom except those corresponding to the
given particles can be integrated out.

A problem with approaches based on Eq. (\ref{E12}) is that the answer should be given in terms of
invariant quantities while energies are reference frame dependent. Therefore one should consider the
two-particle mass operator. In standard Poincare invariant theory the free mass operator is given
by $M=M_0({\bf q})=(m_1^2+{\bf q}^2)^{1/2}+(m_2^2+{\bf q}^2)^{1/2}$ where the $m_i$ are the particle masses and
${\bf q}$ is the relative momentum operator. In classical approximation ${\bf q}$ becomes the relative
momentum and $M_0$ becomes a function of ${\bf q}$ not depending on the relative distance $r$ between
the particles. Therefore the relative acceleration is zero and this case can be treated as noninteracting.

Consider now a two-particle system in dS invariant theory. As explained in Sec. \ref{symmetry}, 
on quantum level the only consistent
definition of dS invariance is that the operators describing the system satisfy the commutation relations of the
dS algebra. This definition does not involve GR, QFT, dS space and its geometry. A definition of an elementary particle given in that section is that 
the particle is described
by an IR of the dS algebra (see also Secs. \ref{IRsdS} and \ref{elementary}). Therefore a possible definition 
of the free two-particle
system can be such that the system is described by a representation where not only the energy but all other 
operators are given by sums of the corresponding single-particle operators. In representation theory such a 
representation is called the tensor products of IRs.

In other words, we consider only quantum mechanics
of two free particles in dS invariant theory. In that case, as shown in Refs. \cite{lev3,lev1a,jpa1} 
(see also Sec. \ref{antigravity} of the present work), the two-particle mass 
operator can be explicitly calculated. It can be written as
$M=M_0({\bf q})+V$ where $V$ is an operator depending not only on ${\bf q}$. 
In classical approximation $V$
becomes a function depending on $r$. As a consequence, the relative acceleration is not
zero and the result for the relative acceleration describes a known cosmological repulsion 
(sometimes called dS antigravity). 

One might argue that the above situation contradicts the law of inertia according to which if particles
do not interact then their relative acceleration must be zero. However, this law has been postulated
in Galilei and Poincare invariant theories and there is no reason to believe that it will be valid
for other symmetries. Another argument might be such that dS invariance implicitly implies existence of
other particles which interact with the two particles under consideration. Therefore the above situation
resembles a case when two particles not interacting with each other are moving with different accelerations 
in a nonhomogeneous
field and therefore their relative acceleration is not zero. This argument has much in common with the
discussion of whether the empty space-time background can have a curvature and whether a nonzero curvature
implies the existence of dark energy or other fields (see the preceding section). However, as 
argued in the preceding sections, fundamental quantum theory should not involve the empty space-time
background at all. Therefore our result demonstrates that 
the cosmological constant problem does not exist and the cosmological acceleration can be easily (and naturally) 
explained without involving dark energy or other fields.  

In QFT interactions can be only local and there are no 
interactions at a distance (sometimes called direct interactions), when particles interact without an 
intermediate field. In particular, a potential interaction (when the force of the interaction depends only 
on the distance between the particles) can be only a good approximation in situations when the particle 
velocities are much less than $c$. The explanation is such that if the force of the interaction
depends only on the distance between the particles and the distance slightly changes then
the particles will feel the change immediately, but this contradicts the statement that no interaction
can be transmitted with the speed greater than $c$. Although standard QFT is based on
Poincare symmetry, physicists typically believe that the notion of interaction adopted in QFT is valid for 
any symmetry. However, the above discussion shows that the dS antigravity is not caused by exchange of any 
virtual particles. In particular a question about the speed of propagation of dS antigravity in not
physical. In other words, the dS antigravity is an example of a true direct interaction. It is also
possible to say that the dS antigravity is not an interaction at all but simply an inherent property of 
dS invariance. 

In quantum theory, dS and AdS symmetries are widely used for investigating QFT
in curved space-time background. However, it seems rather paradoxical that such a simple case as a free two-body 
system in dS invariant theory has not been widely discussed. According to our observations, such a
situation is a manifestation of the fact that even physicists working on dS QFT are not
familiar with basic facts about IRs of the dS algebra. It is difficult to imagine how standard Poincare
invariant quantum theory can be constructed without involving known results on IRs of the Poincare
algebra. Therefore it is reasonable to think that when Poincare invariance is replaced by dS one,
IRs of the Poincare algebra should be replaced by IRs of the dS algebra. However, physicists working
on QFT in curved space-time argue that fields are more fundamental than particles and therefore there
is no need to involve commutation relations (\ref{CR}) and IRs. In other words, they treat dS symmetry
on quantum level not such that the relations (\ref{CR}) should be valid but such that quantum fields
are constructed on dS space (see e.g. Refs. \cite{Birrel,Wald}).

Our discussion shows that the notion of interaction depends on symmetry. For example, when we consider
a system of two particles which from the point of view of dS symmetry are free (since they are described
by a tensor product of IRs), from the point of view of our experience based on Galilei or Poincare 
symmetries they are not free since their relative acceleration is not zero. This poses a question whether 
not only dS antigravity but other interactions are in fact not interactions but effective
interactions emerging when a higher symmetry is treated in terms of a lower one. 

In particular, is it possible
that quantum symmetry is such that on classical level the relative acceleration of two free particles
is described by the same expression as that given by the Newton gravitational law and corrections
to it? This possibility has been first discussed in Ref. \cite{lev3}. It is clear that this possibility
is not in mainstream according to which gravity is a manifestation of the graviton exchange.  
We believe that until the nature of gravity has been unambiguously understood, different possibilities should be investigated. A very strong argument in favor of our approach is as follows. In contrast to theories based on
Poincare and AdS symmetries, in the dS case the spectrum of the free mass operator is not bounded below
by $(m_1+m_2)$. As a consequence, it is not a problem to indicate states where the mean value of the mass operator
has an additional contribution $-Gm_1m_2/r$ with possible corrections. 
A problem is to understand reasons why macroscopic bodies have such WFs. 

If we accept dS symmetry then the first step is to investigate the structure of dS invariant theory 
from the point of view of IRs of the dS algebra. This problem is discussed in Refs. \cite{lev1a,jpa1,symm1401}.
In Ref. \cite{lev3} we
discussed a possibility that gravity is simply a manifestation of the fact that fundamental quantum theory
should be based not on complex numbers but on a Galois field with a large characteristic $p$ which is a
parameter defining the laws of physics in our Universe. This approach has been discussed in Refs. \cite{lev4,tmf,complex,symm1810} and other publications. In Refs. 
\cite{levgr,physessay} we discussed additional arguments in favor of our hypothesis about gravity.
We believe that the results of the present work give strong indications that our hypothesis is correct.

Another arguments that gravity is not an interaction follow. 
The quantity $G$ defines the gravitational force in the Newton law of gravity. 
Numerous experimental data show that this law works with a very high accuracy. However, this only 
means that $G$ is a good {\it phenomenological} parameter. At the level of the Newton law one cannot prove 
that $G$ is the exact constant which does not change with time, does not depend on masses, distances etc.

General Relativity is a classical (i.e. non-quantum) theory based on the minimum action principle.
Here we have two different quantities which have different dimensions: the 
stress energy tensor of matter and the Ricci tensor describing the curvature of the space-time background. 
Then the Einstein equations (\ref{Einstein}) derived from the minimum action principle show that
$G$ is the coefficient of proportionality between the left-hand and right-hand sides of Eq. (\ref{Einstein}). 
General Relativity cannot calculate it or give a {\it theoretical} explanation why this value should be as is. 
By analogy with the treatment of the quantities $c$ and $\hbar$ in
the preceding section, one might think that $G$ can be treated analogously and its value is as is simply
because we wish to measure
masses in kilograms and distances in meters (in the spirit of Planck units). However, arguments given in the preceding section
indicate that there are no solid reasons to treat $G$ as a fundamental constant.

From the point of view of dS symmetry on quantum level, G cannot be a fundamental constant from the following
considerations. The commutation relations (\ref{CR}) do not depend on any free parameters. 
One might say that this is a consequence of the choice of units where
$\hbar=c=1$. However, as noted in the preceding sections, any fundamental theory
should not involve the quantities $\hbar$ and $c$. A theory based on the above definition of the dS 
symmetry on quantum level cannot involve 
quantities which are dimensionful in units $\hbar=c=1$. In particular, we inevitably come to
conclusion that the gravitational and cosmological constants cannot be fundamental. 

By analogy with the above discussion about gravity, one can pose a question of whether the notions of
other interactions are fundamental or not. In QFT all interactions (e.g. in QED, electroweak theory and QCD)
are introduced according to the
same scheme. One writes the Lagrangian as a sum of free and interaction Lagrangians. The latter 
are proportional to interaction constants which cannot be calculated from the theory and hence can be 
treated only as phenomenological parameters. It is reasonable to believe that the future fundamental
theory will not involve such parameters. For example, one of the ideas of the string theory is that
the existing interactions are only manifestations of how higher dimensions are compactified.

\section{The content of this work}

In Chap. \ref{WPS} we show that in standard nonrelativistic and relativistic quantum theory the position operator 
is defined inconsistently. As a consequence, in standard quantum theory there exist several paradoxes discussed 
in Sec. \ref{experiment}. We propose a definition of the position operator which resolves the paradoxes 
and gives a new look at the construction of quantum theory.

In Chap. \ref{Ch2} we construct IRs of the dS algebra following the book by Mensky \cite{Mensky}. This construction
makes it possible to show that the known cosmological repulsion is simply a kinematical effect in dS quantum
mechanics. The derivation involves only standard quantum mechanical notions. It does not require dealing with
dS space, metric tensor, connection and other notions of Riemannian geometry. As argued in the preceding sections,
fundamental quantum theory should not involve space-time at all. In our approach the cosmological
constant problem does not exist and there is no need to involve dark energy or other fields for explaining this problem.

In Chap. \ref{Ch3} we construct IRs of the dS algebra in the basis where all quantum numbers are discrete. 
In particular, the results of Chap. \ref{WPS} on the position operator and wave packet spreading are generalized to the dS case. This makes it possible to investigate in Chap. \ref{twobody}
for which two-body WFs one can get standard Newton's law of gravity and the results which are
treated as three classical tests of GR.

In Chap. \ref{Ch4} we argue that fundamental quantum theory should be based on finite mathematics
rather than complex
numbers. In our approach, standard theory is a special case of a finite quantum theory (FQT)
in a formal limit when the characteristic $p$ of the field or ring used in FQT becomes infinitely large.
We try to make the presentation 
as self-contained as possible without assuming that the reader is familiar with finite fields or finite rings.
In Chap. \ref{Ch5} we construct semiclassical states in FQT and discuss the problem of calculating 
the gravitational constant. 

In Chap. \ref{AdS} a finite analog of the AdS symmetry is applied to particle theory.
It is shown that in this approach the particles which in standard theory are treated as neutral (i.e. they coincide
with their antiparticles) cannot be elementary. In particular, even the photon cannot be elementary. 
The notion of a
particle and its antiparticle can be only approximate and such additive quantum numbers
as the electric charge and the baryon and lepton quantum numbers can be only approximately
conserved. 

In Chap. \ref{DiracSingletons} we discuss Dirac singletons in FQT and argue  that only Dirac singletons are true elementary particles. 

In Chap. \ref{time} we discuss a conjecture that classical time $t$ is a manifestation of the fact 
that $p$ changes, i.e. $p$ and not $t$ is the true evolution parameter.

Finally, Chap. \ref{Discussion} is discussion.

\chapter{A new look at the position operator in quantum theory}
\label{WPS}

\section{Status of the position operator in quantum theory}
\label{intropos}
\subsection{Historical reasons for choosing standard form of position operator}
\label{historical}

It has been postulated from the beginning of quantum theory that the coordinate and momentum representations of WFs are related to each other by the Fourier transform. One of the historical reasons was that in classical electrodynamics the coordinate and wave vector ${\bf k}$ representations are related analogously and we postulate that 
${\bf p}=\hbar {\bf k}$ where ${\bf p}$ is the particle momentum. Then, although the interpretations of classical fields on one hand and WFs on the other are fully different, from mathematical point of view classical electrodynamics and quantum mechanics have much in common (and such a situation does not seem to be natural). As noted in Subsec. \ref{subsecsumm}, a similarity of classical 
electrodynamics and quantum theory is reflected even in the terminology of the latter. 

One of the examples of the above similarity follows. Consider a WF of the form
$\psi({\bf r},t)=a({\bf r},t)exp[iS({\bf r},t)/\hbar]$, where $S({\bf r},t)$ is the classical action as a function of coordinates and time. Then
\begin{equation}
\frac{\partial \psi({\bf r},t)}{\partial {\bf r}}=[\frac{i}{\hbar}\frac{\partial 
S({\bf r},t)}{\partial {\bf r}}+\frac{1}{a({\bf r},t)}\frac{\partial 
a({\bf r},t)}{\partial {\bf r}}]\psi({\bf r},t)
\label{quasi}
\end{equation}
and analogously for $\partial \psi({\bf r},t)/\partial t$. In the formal limit
$\hbar\to 0$ the second term in the r.h.s. can be neglected and, as explained in textbooks on quantum mechanics (see e.g. Ref. \cite{LLIII})  the 
Schr\"{o}dinger equation becomes the Hamilton-Jacoby equation. This situation is 
analogous to the approximation of geometrical optics in classical electrodynamics 
(see e.g. Ref. \cite{LLII}) when fields contain a rapidly oscillating factor $exp[i\varphi ({\bf r},t)]$ where the function $\varphi({\bf r},t)$ is called eikonal. It satisfies the eikonal equation which coincides with the relativistic Hamilton-Jacobi equation for a particle with zero mass. This is reasonable
in view of the fact that electromagnetic waves consist of photons. 

Another example follows. In classical electrodynamics, a wave packet moving even in empty space inevitably spreads out and this fact has been known
for a long time. For example, as pointed out by Schr\"{o}dinger (see pp. 41-44 in Ref. \cite{Schr}), in standard quantum mechanics a packet does not spread out if a particle is moving in a harmonic oscillator potential in contrast to 
"a wave packet in classical optics, which is dissipated in the course of time". 
However, as a consequence of the
similarity, a free quantum mechanical wave packet inevitably spreads out too. This effect is called wave packet spreading (WPS) and it is described in textbooks and many papers (see e.g. Refs. \cite{Dirac,QM} and references therein). In this chapter the effect is discussed in detail and we argue that
it plays a crucial role in drawing a conclusion on whether standard position operator is consistently
defined.

The requirement that the momentum and position operators are related to each other
by the Fourier transform is equivalent to standard commutation relations between these
operators and to the Heisenberg uncertainty principle (see Sec. \ref{classical}).

A reason for choosing standard form of the position operator is described, for example, in
the Dirac textbook \cite{Dirac}. Here Dirac argues that the momentum and position operators
should be such that their commutator should be proportional to the corresponding classical
Poisson bracket with the coefficient $i\hbar$. However, this argument is not convincing 
because only in very special
cases the commutator of two physical operators is a $c$-number. One can check, for example,
a case of momentum and position operators squared.

In Ref. \cite{Heisenberg} Heisenberg argues in favor of his principle by considering 
{\it Gedankenexperiment} with Heisenberg's microscope. Since that time the problem has been investigated 
in many publications. A discussion of the current status of the
problem can be found e.g. in Ref. \cite{Lahti} and references therein. A general opinion based on those 
investigations is   that Heisenberg's arguments are problematic but the uncertainty principle is valid, although several
authors argue whether standard mathematical notion of uncertainty (see Sec. \ref{classical})
is relevant for describing a real process of
measurement. However, a common assumption in those investigations
is that one can consider uncertainty relations for all the components of the position and momentum operators independently.
Below we argue that this assumption is not based on solid physical arguments. 

\subsection{Problem of consistency of standard position operator}
\label{consistency} 

Usual arguments in favor of choosing standard position and  momentum operators are that these 
operators have correct properties in semiclassical approximation (see e.g. Ref. \cite{LLIII}).
However, this requirement does not define the operator unambiguously. Indeed, if the operator
$B$ becomes zero in semiclassical limit then the operators $A$ and $A+B$ have the same
semiclassical limit. 

 As noted above,
in the main approximation in $1/\hbar$ the Schr\"{o}dinger equation becomes the Hamilton-Jacoby equation if the coordinate WF $\psi({\bf r},t)$ contains a factor $exp[iS({\bf r},t)/\hbar]$.
In textbooks this is usually treated as the correspondence principle between quantum 
and classical theories. However, the following question arises.

As follows from Eq. (\ref{quasi}),  the Hamilton-Jacoby equation is a good approximation for
the Schr\"{o}dinger equation if the index of the exponent changes much faster than the
amplitude $a({\bf r},t)$. Is this correct to define semiclassical approximation by this condition?
Quantum theory fully reproduces the results of classical one when not only this condition is
satisfied but, in addition, the amplitude has a sharp maximum along the classical
trajectory. If the latter is true at some moment of time then, in view of the WPS effect, one
cannot guarantee that this will be true always.

At the beginning of quantum theory the WPS effect has been investigated by de Broglie,
Darwin and Schr\"{o}dinger. The fact that WPS is inevitable has been treated by several authors as unacceptable
and as an indication that standard quantum theory should be modified. For example, de Broglie has proposed to
describe a free particle not by the Schr\"{o}dinger equation but by a wavelet which satisfies a nonlinear equation
and does not spread out (a detailed description of de Broglie's wavelets can be found e.g. in Ref. \cite{Barut}).
Sapogin writes (see Ref. \cite{Sapogin} and references therein) that "Darwin showed 
that such packet quickly and steadily dissipates
and disappears" and proposes an alternative to standard theory which he calls unitary unified
quantum field theory. 

At the same time, it has not been explicitly shown that numerical results on WPS are 
incompatible with experimental data. For example, it is known (see Sec. \ref{NRWPS})
that for macroscopic bodies the effect of WPS is extremely small. Probably it is also believed that in 
experiments on the Earth with atoms 
and elementary particles spreading does not have enough time to manifest itself although we have not 
found an explicit statement on this problem in the literature. 
According to our observations, different physicists have different opinions on the role of
WPS in different phenomena but in any case the absolute majority of physicists do not treat WPS as a drawback of the theory. 

A natural problem arises what happens to photons which can travel from distant objects to Earth even for 
billions of years. As shown in Sec. \ref{experiment}, standard theory predicts that, as a consequence
of WPS, WFs of such photons will have the size of the order of light years or more. Does this contradict observations? We argue that it does and the reason of 
the paradox is that 
standard position operator is not consistently defined. Hence the inconsistent definition of the position 
operator is not only an academic problem but leads to the above paradox.
 
In view of the fact that the coordinate and momentum representations are related to each other by the
Fourier transform, one might think that the position and momentum operators are on equal footing. However,
this is not the case for the following reasons. In quantum theory each
elementary particle is described by an irreducible representation (IR) of the symmetry algebra. 
For example, in Poincare invariant theory the set of momentum operators represents three of ten 
linearly independent representation
operators of the Poincare algebra and hence those operators are consistently defined.
On the other hand, among the representation operators there is no position operator. 
So the assumption that the position operator in momentum representation is $i\hbar\partial /\partial {\bf p}$ should be substantiated.
{\it A simple observation showing that standard definition of the position operator is not consistent follows}.

Consider first a one-dimensional case. As argued in textbooks (see e.g. Ref. \cite{LLIII}),
if the mean value of the $x$ component of the momentum $p_x$ is rather large, the definition of the coordinate operator 
$i\hbar \partial/\partial p_x$ can be justified  but this definition does not have a physical meaning in situations when $p_x$ is small. 
This is clear even from the fact that if $p_x$ is small then $exp(ip_xx/\hbar)$ is not a rapidly oscillating function of $x$.

Consider now the three-dimensional case.
If all the components $p_j$ ($j=1,2,3$) are rather large then all the operators 
$i\hbar \partial/\partial p_j$ can have a physical meaning. A semiclassical WF $\chi({\bf p})$ in
momentum space should describe a narrow distribution around the mean value ${\bf p}_0$. Suppose now that coordinate 
axes are chosen such ${\bf p}_0$ is directed along the $z$ axis. Then the mean values of the $x$ and $y$
components of the momentum operator equal zero and
the operators $i\hbar \partial/\partial p_j$ cannot be physical for $j=1,2$, i.e. in directions perpendicular
to the particle momentum. 
The situation when a definition of an operator is physical or not depending on the choice of coordinate axes is not acceptable. 

\subsection{Remarks on the Schr\"{o}dinger and Dirac equations}
\label{history}

For the subsequent discussion of the position operator, it is useful to recall some known facts about
the Schr\"{o}dinger and Dirac equations. Historically these equations have been first written in coordinate 
space and in textbooks they are still discussed in this form. The equations have played a great role for constructing
quantum theory. However, a problem arises whether the equations are so fundamental as usually believed.

In textbooks on quantum mechanics the Schr\"{o}dinger equation is discussed for different model potentials.
However, the only case when this equation has been unambiguously confirmed by experimental data is
the case of light atoms and especially the case of energy levels of the hydrogen atom. The successful
description of those levels has been immediately treated as a great success of quantum theory.

This equation is nonrelativistic and describes the energy
levels with a high accuracy because the electron in the hydrogen atom is nonrelativistic. The typical velocities
of the electron in the hydrogen atom are of the order of $\alpha c$ where $\alpha \approx 1/137$ is the
fine structure constant.

The Dirac equation for the electron in the hydrogen atom describes the fine structure of the energy levels:
each Schr\"{o}dinger energy level (which depends on $\alpha$ as $\alpha^2$) splits such that the
differences of fine structure energy levels for the given Schr\"{o}dinger energy level are proportional
to $\alpha^4$.

However, even the Dirac equation for the hydrogen energy levels is not exact. For example, the Lamb shift
results in the additional splitting of fine structure energy levels such that the differences between energy
levels within one fine structure energy level are proportional to $\alpha^5$. The Lamb shift cannot be calculated
in the single-particle approximation and can be calculated only in the framework of quantum electrodynamics (QED)
which is treated as a fundamental theory describing electromagnetic interactions on quantum level. 

This theory proceeds from quantizing classical Lagrangian which is only an auxiliary tool for constructing S-matrix. 
As already noted in the preceding chapter, the argument ${\bf x}$ in the Lagrangian density $L(t, {\bf x})$ cannot be treated as a position operator because
$L(t, {\bf x})$ is constructed from field functions which do not have a probabilistic interpretation. When
quantization is accomplished, the results of QED are formulated exclusively in momentum space and the theory does not contain space-time at all. 

From the point of view of the present knowledge, 
 the Schr\"{o}dinger and Dirac equations should be treated as follows.
 As follows from Feynman diagrams for the one-photon exchange, in the approximation 
up to  $(v/c)^2$  the electron in the hydrogen atom can be described in the potential formalism where the potential acts on the WF in momentum space.  So for calculating energy levels one should solve the eigenvalue problem for the Hamiltonian with this potential. This is an integral equation which can be solved by different methods. One of the convenient methods is to apply the Fourier transform and get standard Schr\"{o}dinger or Dirac equation in coordinate representation with the Coulomb potential. Hence the fact that the results for energy levels are in good agreement with experiment shows only that QED defines the potential correctly and {\it standard coordinate Schr\"{o}dinger and Dirac equations are only convenient mathematical ways of solving the eigenvalue problem in the approximation
up to $(v/c)^2$}. For this problem 
the physical meaning of the position operator is not important at all. One can consider other transformations of the original integral equation and define other position operators. The fact that for non-standard choices one might obtain something different from the Coulomb potential is not important on quantum level. On classical level the interaction between two charges can be described by the Coulomb potential but this does not imply that on quantum level the potential in coordinate representation should be necessarily Coulomb.

Let us now consider a hypothetical situation: consider a Universe in which the value of $\alpha$ is of the
order of unity or greater. Then the energy levels cannot be calculated in perturbation theory, and it is not known 
(even if $\alpha$ is small) whether the perturbation series of QED converges or not.
However, the logical structure of QED remains the same. At the same time, the single-particle approximation
is not valid anymore and the Schr\"{o}dinger and 
Dirac equations do not define the hydrogen energy levels even approximately. In other words, in this
situation the application of those equations for calculating the hydrogen energy level does not have a
physical meaning.  

The fact that in our world  the Schr\"{o}dinger and Dirac equations describe the hydrogen energy
level with a high accuracy, is usually treated as a strong argument that the coordinate and momentum representations
should be related to each other by the Fourier transform.  However, as follows from the above considerations,
this fact takes place only because we are lucky that the value of $\alpha$ in our
Universe is small. Therefore this argument is not physical and cannot be used.

\subsection{When do we need position operator in quantum theory?}
\label{when}

\begin{sloppypar}
As follows from the discussion in the preceding subsection, the fact that the Schr\"{o}dinger and Dirac equations describe the hydrogen energy level with a high accuracy cannot be treated as an argument that the
coordinate and momentum representations are related to each other by the Fourier transform.
\end{sloppypar}

Let us also note the following. In the literature the statement that the Coulomb law works with a high accuracy is often 
substantiated from the point of view that predictions of QED have been experimentally confirmed with a high accuracy. 
However, as follows from the above remarks, the meaning of distance on quantum level is not
clear and in QED the law $1/r^2$ can be tested only if we
assume additionally that the coordinate and momentum representations are related to each other by the Fourier transform. 
So a conclusion about the validity of the law can be made only on the basis of macroscopic experiments.
A conclusion made from the results of classical Cavendish and Maxwell experiments is that if the exponent in Coulomb's
law is not 2 but $2\pm q$ then $q<1/21600$. The accuracy of those experiments have been considerably improved in
the experiment \cite{Plimpton} the result of which is $q<2\cdot 10^{-9}$. However, the
Cavendish-Maxwell experiments and the experiment \cite{Plimpton} do not involve pointlike electric charges. 
Cavendish and Maxwell used a 
spherical air condenser consisting of two insulated spherical shells while the authors of Ref. \cite{Plimpton}
developed a technique where the difficulties due to spontaneous ionization and contact potentials were avoided.
Therefore the conclusion that $q<2\cdot 10^{-9}$ for pointlike electric charges requires additional assumptions.  

Another example follows. It is said that the spatial distribution of the electric charge inside a system
can be extracted from measurements of form-factors in the electron scattering on this system. 
However, as noted in Subsec. \ref{subsecsumm}, for elementary particles the notion of
charge distribution is meaningless. In addition, even in the case of a composite system the information about 
the experiment is again given only in terms of
momenta. So conclusions about the spatial distribution can be drawn only if we assume additionally how
the position operator is expressed in terms of momentum variables. 

In view of the above discussion, since the {\it results} of existing fundamental quantum theories 
describing interactions on quantum level (QED, electroweak theory and QCD) are formulated exclusively
in terms of the S-matrix in momentum space without mentioning space-time, {\it for investigating such
stationary quantum problems as calculating energy levels, form-factors etc., the notion of the
position operator is not needed}.

However, the choice of the position operator is important in nonstationary problems when evolution is
described by the time dependent Schr\"{o}dinger equation (with the nonrelativistic or relativistic Hamiltonian). As follows 
from the correspondence principle, 
 quantum theory should reproduce the motion of a particle along
the classical trajectory defined by classical equations of motion. Hence the position operator is needed only
in semiclassical approximation and it should be {\it defined} from additional considerations. 

In standard approaches to quantum theory the existence of space-time background is assumed from the beginning.
Then the position operator for a particle in this background is the operator of multiplication by the
particle radius-vector ${\bf r}$.
As explained in textbooks on quantum mechanics (see e.g. Ref. \cite{LLIII}), the result 
$-i\hbar \partial/\partial{\bf r}$
for the momentum operator can be justified from the requirement that quantum theory should correctly reproduce classical
results in semiclassical approximation. However, as noted above, this requirement does not define the operator 
unambiguously.

As noted in Sec. \ref{symmetry}, an elementary particle in quantum theory is described by an IR of the symmetry algebra. In Poincare invariant theory the IRs can be
implemented in a space of functions $\chi({\bf p})$ such that $\int |\chi({\bf p})|^2d^3{\bf p}<\infty$ 
(see Sec. \ref{momentum}).
In this representation the momentum operator ${\bf P}$ is defined {\it unambiguously} and is simply the operator of multiplication by ${\bf p}$.
A standard {\it assumption} is that the position operator in this representation is $i\hbar \partial/\partial {\bf p}$. However, as argued above, this assumption is not consistent.

The above discussion shows that there is no position operator which unambiguously follows from first
principles of quantum theory. History of physics tells us that it is always desirable to involve the least
possible number of notions and so a problem arises whether the position operator is needed at all.
Also, as argued in Subsec. \ref{problemoftime}, in quantum theory classical time $t$ cannot be a fundamental quantity. At the same time, since quantum theory is treated as more fundamental than classical one, in some approximations the theory should reproduce the results which follow from classical equations of motion.

In Chap. \ref{time} we consider a conjecture that classical time manifests itself as a consequence of the
fact that the parameter $p$ in FQT changes. However, at present this conjecture is in its infancy and only further
investigations can show whether it can be substantiated. For this reason, in the main part of the
work we assume that time $t$ is a classical parameter defining evolution in semiclassical approximation.
In this chapter we also argue that classical coordinates can be defined from pure quantum notion even without
semiclassical approximation. However, since this problem is not well understood yet, in the main part of the
work we assume that for derivation of classical motion from quantum theory one needs a position operator.
As noted above, standard position operator is not consistent. In this chapter we propose a position operator
which for sure is more consistent than standard one. As a consequence, in our approach WPS in directions perpendicular to the particle
momentum is absent regardless of whether the particle is nonrelativistic or relativistic. 
Moreover, for an ultrarelativistic particle the effect of 
WPS is absent at all. Different components of the new position operator do not commute with each other and,
as a consequence, there is no WF in coordinate representation. 

The chapter is organized as follows. In Secs. \ref{classical} and \ref{momentum} we discuss standard 
approach to the position operator in nonrelativistic and relativistic quantum theory, respectively. 
An inevitable consequence 
of this approach is the effect of WPS of the coordinate
WF which is discussed in Secs. \ref{NRWPS} and \ref{RelWPS} for the nonrelativistic and
relativistic cases, respectively. In Sec. \ref{WPW} we discuss a relation between the WPS effects for a classical
wave packet and for photons comprising this packet. In Sec. \ref{coherent} the problem of WPS in coherent states is
discussed. In Sec. \ref{experiment} we show that the WPS effect leads to several paradoxes and  
in standard theory it is not possible to avoid those paradoxes.
Our approach to a consistent definition of the position operator and its application to WPS are discussed 
in Secs. \ref{consistent}-\ref{newWPS}. 

\section{Position operator in nonrelativistic quantum mechanics}
\label{classical}

In quantum theory, states of a system are represented by elements of a projective Hilbert space. The fact that a
Hilbert space $H$ is projective means that if $\psi\in H$ is a state then $const\cdot\psi$ is the same state.
The matter is that not the probability itself but only relative probabilities of different measurement outcomes 
have a physical meaning. In this chapter we will work with states $\psi$ normalized to one, i.e. such that 
$||\psi||=1$ where $||...||$ is a norm. It is defined such that if 
 $(...,...)$ is a scalar product in $H$ then $||\psi||=(\psi,\psi)^{1/2}$.  

In quantum theory every physical quantity is described by a selfadjoint operator. 
Each selfadjoint operator is
Hermitian i.e. satisfies the property $(\psi_2,A\psi_1)=(A\psi_2,\psi_1)$ for any states belonging to the
domain of $A$. If $A$ is an operator of some
quantity then the mean value of the quantity and its uncertainty in state $\psi$ are given by ${\bar A}=(\psi,A\psi)$
and $\Delta A=||(A-{\bar A})\psi||$, respectively. The condition that a quantity corresponding to the operator $A$
is semiclassical in state $\psi$ can be defined such that $\Delta A\ll |{\bar A}|$. This implies that the
quantity can be semiclassical only if $|{\bar A}|$ is rather large. In particular, if ${\bar A}=0$ then the
quantity cannot be semiclassical.

Let $B$ be an operator corresponding to another physical quantity and ${\bar B}$ and $\Delta B$ be the
mean value and the uncertainty of this quantity, respectively. We can write $AB=\{A,B\}/2 + [A,B]/2$
where the commutator $[A,B]=AB-BA$ is anti-Hermitian and the anticommutator $\{A,B\}=AB+BA$ is Hermitian. 
Let $[A,B]=-iC$ and ${\bar C}$ be the mean value of the operator $C$.

A question arises whether two physical
quantities corresponding to the operators $A$ and $B$ can be simultaneously semiclassical in state $\psi$. Since
$||\psi_1||||\psi_2||\geq |(\psi_1,\psi_2)|$, 
\begin{equation}
\Delta A \Delta B\geq \frac{1}{2}|(\psi, (\{A-{\bar A},B-{\bar B}\}+[A,B])\psi)|
\end{equation}
Since $(\psi,\{A-{\bar A},B-{\bar B}\}\psi)$ is real and $(\psi,[A,B]\psi)$ is imaginary, 
\begin{equation}
\Delta A \Delta B \geq \frac{1}{2}|{\bar C}|
\label{uncert}
\end{equation}
This condition is known as a general uncertainty relation between two quantities. A well-known special case is
that if $P$ is the $x$ component of the momentum operator and $X$ is the operator of multiplication by $x$
then $[P,X]=-i\hbar$ and $\Delta p \Delta x\geq \hbar/2$.
The states where $\Delta p \Delta x= \hbar/2$ are called coherent ones. They are treated such that the momentum
and the coordinate are simultaneously semiclassical in a maximal possible extent. A known example is that if
$$\psi(x)=\frac{1}{a^{1/2}\pi^{1/4}}exp[\frac{i}{\hbar}p_0x-\frac{1}{2a^2}(x-x_0)^2]$$
then ${\bar X}=x_0$, ${\bar P}=p_0$, $\Delta x=a/\sqrt{2}$ and $\Delta p=\hbar /(a\sqrt{2})$.

Consider first a one dimensional motion. In standard textbooks on quantum mechanics, the presentation
starts with a WF $\psi(x)$ in coordinate space since it is implicitly assumed that the meaning of
space coordinates is known. Then a question arises why $P=-i\hbar d/dx$ should be treated as the momentum operator.
The explanation follows.

Consider WFs having the form $\psi(x)=exp(ip_0x/\hbar)a(x)$ where the amplitude $a(x)$ has a sharp maximum 
near $x=x_0\in [x_1,x_2]$ such that $a(x)$ is not small only when $x\in [x_1,x_2]$. Then $\Delta x$ is of the
order $x_2-x_1$ and the condition that the coordinate is semiclassical is $\Delta x\ll |x_0|$. 
Since $-i\hbar d\psi(x)/dx=p_0\psi(x)-i\hbar exp(ip_0x/\hbar)da(x)/dx$, $\psi(x)$ will be approximately
the eigenfunction of $-i\hbar d/dx$ with the eigenvalue $p_0$ if $|p_0a(x)|\gg \hbar|da(x)/dx|$. Since 
$|da(x)/dx|$ is of the order of $|a(x)/\Delta x|$, we have a condition
$|p_0\Delta x| \gg \hbar$. Therefore if the momentum operator is $-i\hbar d/dx$, the uncertainty of momentum $\Delta p$
is of the order of $\hbar/\Delta x$, $|p_0|\gg \Delta p$ and this implies that the momentum is also semiclassical. 
At the same time, $|p_0\Delta x|/2\pi\hbar$ is approximately the number of oscillations which the exponent 
makes on the segment $[x_1,x_2]$. Therefore the number of oscillations should be much greater than unity. In particular, 
semiclassical approximation cannot be valid if $\Delta x$ is very small, but on the other hand, $\Delta x$ cannot be
very large since it should be much less than $x_0$. Another justification of the fact that $-i\hbar d/dx$ is the
momentum operator is that in the formal limit $\hbar\to 0$ the Schr\"{o}dinger equation becomes the Hamilton-Jacobi
equation. 

We conclude that the choice of $-i\hbar d/dx$ as the momentum operator is justified from the requirement that
in semiclassical approximation this operator becomes the classical momentum. However, it is obvious that this
requirement does not define the operator uniquely: any operator ${\tilde P}$ such that ${\tilde P}-P$ disappears
in semiclassical limit, also can be called the momentum operator.

One might say that the choice $P=-i\hbar d/dx$ can also be justified from the following considerations. In nonrelativistic
quantum mechanics we assume that the theory should be invariant under the action of the Galilei group, which is a group
of transformations of Galilei space-time. The $x$ component of the momentum operator should be the generator corresponding
to spatial translations along the $x$ axis and $-i\hbar d/dx$ is precisely the required operator. In this consideration
one assumes that the space-time background has a physical meaning while, as discussed in Secs. \ref{ST} and \ref{symmetry}, this is not the case.

As noted in Secs. \ref{ST} and \ref{symmetry}, one should start not from space-time but from a 
symmetry algebra. Therefore in nonrelativistic
quantum mechanics we should start from the Galilei algebra and consider its IRs. 
For simplicity we again consider a
one dimensional case. Let $P_x=P$ be one of representation operators in an IR of the Galilei algebra. We can implement
this IR in a Hilbert space of functions $\chi(p)$ such that 
$\int_{-\infty}^{\infty}|\chi(p)|^2dp < \infty$ and $P$ is
the operator of multiplication by $p$, i.e. $P\chi(p)=p\chi(p)$. 
Then a question arises how the operator of the $x$
coordinate should be defined. In contrast to the momentum operator, the 
coordinate one is not defined by the representation and so it should be defined 
from additional assumptions. Probably a future quantum theory of
measurements will make it possible to construct operators of physical quantities 
from the rules how these quantities should be measured. However, at present 
we can construct necessary operators only from rather intuitive considerations. 

By analogy with the above discussion, one can say that semiclassical WFs should
be of the form $\chi(p)=exp(-ix_0p/\hbar)a(p)$ where the amplitude $a(p)$ has a sharp maximum 
near $p=p_0\in [p_1,p_2]$ such that $a(p)$ is not small only when $p\in [p_1,p_2]$. Then $\Delta p$ is of the
order of $p_2-p_1$ and the condition that the momentum is semiclassical is $\Delta p\ll |p_0|$. 
Since $i\hbar d\chi(p)/dp=x_0\chi(p)+i\hbar exp(-ix_0p/\hbar)da(p)/dp$, $\chi(p)$ will be approximately
the eigenfunction of $i\hbar d/dp$ with the eigenvalue $x_0$ if $|x_0a(p)|\gg \hbar|da(p)/dp|$. Since 
$|da(p)/dp|$ is of the order of $|a(p)/\Delta p|$, we have a condition
$|x_0\Delta p| \gg \hbar$. Therefore if the coordinate operator is $X=i\hbar d/dp$, the uncertainty of coordinate $\Delta x$
is of the order of $\hbar/\Delta p$, $|x_0|\gg \Delta x$ and this implies that the coordinate defined in such a way
is also semiclassical. We can also note that $|x_0\Delta p|/2\pi\hbar$ is approximately the number of oscillations which the exponent 
makes on the segment $[p_1,p_2]$ and therefore the number of oscillations should be much greater than unity. 
It is also clear that semiclassical approximation cannot be valid if $\Delta p$ is very small, but on the other 
hand, $\Delta p$ cannot be very large since it should be much less than $p_0$. 
By analogy with the above discussion,
the requirement that the operator $i\hbar d/dp$ becomes the coordinate in classical limit does not define the
operator uniquely. In nonrelativistic quantum mechanics it is assumed that the coordinate is a well defined
physical quantity even on quantum level and that $i\hbar d/dp$ is the most pertinent choice. 

The above results can be formally generalized to the three-dimensional case. For example, if the coordinate wave
function is chosen in the form
\begin{equation}
\psi({\bf r})=\frac{1}{\pi^{3/4}a^{3/2}}exp[-\frac{({\bf r}-{\bf r}_0)^2}{2a^2}+\frac{i}{\hbar}{\bf p}_0{\bf r}]
\label{psir}
\end{equation}
then the momentum WF is
\begin{equation}
\chi({\bf p})=\int exp(-\frac{i}{\hbar}{\bf p}{\bf r})\psi({\bf r})\frac{d^3{\bf r}}{(2\pi\hbar)^{3/2}}=
\frac{a^{3/2}}{\pi^{3/4}\hbar^{3/2}}exp[-\frac{({\bf p}-{\bf p}_0)^2a^2}{2\hbar^2}-
\frac{i}{\hbar}({\bf p}-{\bf p}_0){\bf r}_0]
\label{chip}
\end{equation}
It is easy to verify that
\begin{equation}
||\psi||^2=\int |\psi({\bf r})|^2d^3{\bf r}=1,\quad ||\chi||^2=\int |\chi({\bf p})|^2d^3{\bf p}=1,
\label{norm}
\end{equation}
the uncertainty of each component of the coordinate operator is $a/\sqrt{2}$ and the uncertainty of each
component of the momentum operator is $\hbar /(a\sqrt{2})$. Hence one might think that Eqs. (\ref{psir}) 
and (\ref{chip}) describe a state which is semiclassical in a maximal possible extent.

Let us make the following remark about semiclassical vector quantities. We defined a quantity as semiclassical if
its uncertainty is much less than its mean value. In particular, as noted above, a quantity cannot be
semiclassical if its mean value is small. In the case of vector quantities we have sets of three physical quantities.
Some of them can be small and for them it is meaningless to discuss whether they are semiclassical or not. We say that
a vector quantity is semiclassical if all its components which are not small are semiclassical and there should be
at least one semiclassical component. 

For example, if the mean value of the momentum ${\bf p}_0$ is directed along the $z$ axes then the $xy$ components of the
momentum are not semiclassical but the three-dimensional vector quantity ${\bf p}$ can be semiclassical
if ${\bf p}_0$ is rather large. However, in that case the definitions of the $x$ and $y$ components of the position
operator as $x=i\hbar \partial/\partial p_x$ and $y=i\hbar \partial/\partial p_y$ become inconsistent.
The situation when the validity of an operator depends on the choice of directions of the coordinate axes
is not acceptable and this fact has been already mentioned in Subsec. \ref{consistency}.   

Let us note that semiclassical states can be constructed not only in momentum or coordinate representations.
For example, instead of momentum WFs $\chi({\bf p})$ one can work in the representation where 
the quantum numbers $(p,l,\mu)$ in WFs $\chi(p,l,\mu)$ mean the magnitude of the momentum $p$, the
orbital quantum number $l$ (such that a state is the eigenstate of the orbital momentum squared ${\bf L}^2$
with the eigenvalue $l(l+1)$) and the magnetic quantum number $\mu$ (such that a state is the eigenvector or
$L_z$ with the eigenvalue $\mu$). A state described by $\chi(p,l,\mu)$ will be semiclassical with respect to those quantum
numbers if $\chi(p,l,\mu)$ has a sharp maximum at $p=p_0$, $l=l_0$, $\mu=\mu_0$ and the widths of the maxima in
$p$, $l$ and $\mu$ are much less than $p_0$, $l_0$ and $\mu_0$, respectively. However, by analogy with the above
discussion, those widths cannot be arbitrarily small if one wishes to have other semiclassical variables
(e.g. the coordinates). Examples of such situations will be discussed in Sec. \ref{newsemicl}.

\section{Wave packet spreading in nonrelativistic quantum mechanics}
\label{NRWPS}

As noted in Subsec. \ref{when}, we treat time in a standard way, 
i.e. that time is a classical parameter such
that the dependence of the WF on time is defined by the Hamiltonian
according to the Schr\"{o}dinger equation.

In nonrelativistic quantum mechanics the Hamiltonian of a free particle with the mass $m$ is $H={\bf p}^2/2m$ and hence, 
as follows from Eq. (\ref{chip}), in the model discussed above the dependence of the momentum WF on $t$ is 
\begin{equation}
\chi({\bf p}, t)=\frac{a^{3/2}}{\pi^{3/4}\hbar^{3/2}}exp[-\frac{({\bf p}-{\bf p}_0)^2a^2}{2\hbar^2}-
\frac{i}{\hbar}({\bf p}-{\bf p}_0){\bf r}_0-\frac{i{\bf p}^2t}{2m\hbar}]
\label{chipt}
\end{equation}
It is easy to verify that for this state the mean value of the operator ${\bf p}$ and the uncertainty of each 
momentum component are the same as for the state $\chi({\bf p})$, i.e. those quantities do not change with time.

Consider now the dependence of the coordinate WF on $t$. This dependence can be calculated by using Eq. (\ref{chipt}) and the fact that
\begin{equation}
\psi({\bf r},t)=\int exp(\frac{i}{\hbar}{\bf p}{\bf r})\chi({\bf p},t)\frac{d^3{\bf p}}{(2\pi\hbar)^{3/2}}
\label{Fourier}
\end{equation}
The result of a direct calculation is
\begin{equation}
\psi({\bf r},t)=\frac{1}{\pi^{3/4}a^{3/2}}(1+\frac{i\hbar t}{ma^2})^{-3/2}exp[-\frac{({\bf r}-{\bf r}_0-{\bf v}_0t)^2}{2a^2(1+\frac{\hbar^2t^2}{m^2a^4})}(1-\frac{i\hbar t}{ma^2})+\frac{i}{\hbar}{\bf p}_0{\bf r}-\frac{i{\bf p}_0^2t}{2m\hbar}]
\label{psirt}
\end{equation}
where ${\bf v}_0={\bf p}_0/m$ is the classical velocity. This result shows that the semiclassical wave packet is
moving along the classical trajectory ${\bf r}(t)={\bf r}_0+{\bf v}_0t$. At the same time, it is now obvious that
the uncertainty of each coordinate depends on time as 
\begin{equation}
\Delta x_j(t)=\Delta x_j(0)(1+\hbar^2t^2/m^2a^4)^{1/2}, \quad (j=1,2,3)
\label{deltax}
\end{equation}
where $\Delta x_j(0)=a/\sqrt{2}$, i.e. the width of the
wave packet in coordinate representation is increasing. This fact, known as the wave-packet spreading (WPS), is 
described in many textbooks and papers (see e.g. the textbooks \cite{Dirac,QM} and references therein).
It shows that
if a state was semiclassical in the maximal extent at $t=0$, it will not have this property at $t>0$ and the
accuracy of semiclassical approximation will decrease with the increase of $t$. The characteristic time of spreading
can be defined as $t_*=ma^2/\hbar$. For macroscopic bodies this is an extremely large quantity
and hence in macroscopic physics the WPS effect can be neglected. In the formal limit
$\hbar\to 0$, $t_*$ becomes infinite, i.e. spreading does not take place. This shows that WPS is a pure
quantum phenomenon. For the first time the result (\ref{psirt}) has been obtained by Darwin in Ref. \cite{Darwin}.

One might pose a problem whether the WPS effect is specific only for Gaussian WFs. One might
expect that this effect will take place in general situations since each component of the standard position
operator $i\hbar \partial/\partial {\bf p}$ does not commute with the Hamiltonian and so the distribution of the 
corresponding physical quantity will be time dependent. A good example showing inevitability of WPS follows.
If at $t=0$ the coordinate WF is $\psi_0({\bf r})$ then, as follows from Eqs. (\ref{chip}) and (\ref{Fourier}),
\begin{equation}
\psi({\bf r},t)=\int exp\{\frac{i}{\hbar}[{\bf p}({\bf r}-{\bf r}')-\frac{{\bf p}^2t}{2m}]\}\psi_0({\bf r}')
\frac{d^3{\bf r}'d^3{\bf p}}{(2\pi\hbar)^3}
\label{tnonrel}
\end{equation}
As follows from this expression, if $\psi_0({\bf r})\neq 0$ only if ${\bf r}$ belongs to a finite vicinity of some 
vector ${\bf r}_0$ then at any $t>0$ the support of $\psi({\bf r},t)$ belongs to the whole three-dimensional space, i.e. the
WF spreads out with an infinite speed. One might think that in nonrelativistic theory this is not
unacceptable since this theory can be treated as a formal limit $c\to\infty$ of relativistic theory. In the next sections 
we will discuss an analogous situation in relativistic theory.

As shown in Ref. \cite{Berry} titled "Nonspreading wave packets", for a one-dimensional WF in the form of an Airy function, spreading does not take place and the maximum of the quantity $|\psi(x)|^2$ propagates with constant acceleration even in the absence of external forces. Those properties of Airy packets have been observed in optical experiments \cite{Siviloglou}. However, since such a WF is not normalizable, the term "wave packet" in the given situation might be misleading since the mean values and uncertainties of the coordinate and momentum cannot be calculated in a standard way. Such a WF can be constructed only in a limited region of space. As explained in Ref. \cite{Berry}, this WF describes not a particle but rather families of particle orbits. As shown in Ref. \cite{Berry}, one can construct a normalized state which is a superposition of Airy functions with Gaussian coefficients and "eventually the spreading due to the Gaussian cutoff takes over". This is an additional argument that the effect of WPS is an
inevitable consequence of standard quantum theory.

Since quantum theory is invariant under time reversal, one might ask the following question: is it possible that the width of the wave packet in coordinate representation decreases over time? From the formal point of view,
the answer is "yes". Indeed, the solution given by Eq. (\ref{psirt}) is valid not only when $t\geq 0$ but when
$t < 0$ as well. Then, as follows from Eq. (\ref{deltax}), the uncertainty of each coordinate is decreasing when $t$
changes from some negative value to zero. However, eventually the value of $t$ will become positive and the quantities
$\Delta x_j(t)$ will grow to infinity. In this chapter we consider situations when a photon is created on
atomic level and hence one might expect that its initial coordinate uncertainties are not large. However, when the
photon travels a long distance to Earth, those uncertainties become much greater, i.e. the term WPS reflects the
physics adequately. 

\section{Mott-Heisenberg problem and its generalization}
\label{Mott}

In 1929 Mott and Heisenberg considered the following problem. Let an alpha-particle 
be emitted by a nucleus in a radioactive decay. Suppose, for simplicity, that the particle
has been emitted in a state with zero angular momentum. Then the momentum WF is spherically symmetric and all directions of the momentum have equal probabilities. However, when the
particle is detected in Wilson's cloud chamber, the registered trajectory is always linear  as if
the particle moved along a classical trajectory. The explanation of the paradox has been 
given in Ref. \cite{Mott}. In this section we consider a general case when it is not assumed 
that the partical WF is spherically symmetric. 


Consider the state (\ref{tnonrel}) after a long period of time such that $D\gg a$ where
$D=\hbar t/(ma)$. As follows from Eq. (\ref{tnonrel}), at this condition the width of the
coordinate WF is of the order $D$. Suppose that the particle is emitted at the
origin such that ${\bf r}_0=0$. Suppose that a measuring device is at the point ${\bf r}_1$
and the size of the device is of the order of $d$. Although the device is macroscopic, we assume
that $D$ is already so large that $D\gg d$. A problem arises at which momentum range the
particle will be detected.

For solving this problem we first project the coordinate WF onto the region of space
belonging to the device. Assume that the projected WF is
\begin{equation}
{\tilde\psi({\bf r},t)}=exp[-\frac{({\bf r}-{\bf r}_1)^2}{2d^2}]\psi({\bf r},t)
\label{project}
\end{equation}
A direct calculation shows that the norm of this state is
\begin{equation}
||{\tilde\psi}||^2=(\frac{d}{D})^3exp[-\frac{({\bf r}_1-{\bf v}_0t)^2}{D^2}]
\label{projnorm}
\end{equation}
This result is obvious because the WF of the packet is not negligible only in the region
having the volume of the order of $D^3$ and so if ${\bf r}_1$ is inside this region then
the probability to detect the particle is of the order of $(d/D)^3$.

If the particle is detected by the device then the measured momentum range is defined
by the Fourier transform of ${\tilde\psi}({\bf r},t)$. A direct calculation gives
\begin{eqnarray}
&&{\tilde\chi({\bf p},t)}=\frac{1}{(2\pi\hbar)^{3/2}}\int exp(-\frac{i}{\hbar}{\bf p}{\bf r})
{\tilde\psi({\bf r},t)}d^3{\bf r}=\nonumber\\
&&f({\bf p},t)exp[-\frac{d^2D^2a^2({\bf p}-m{\bf r}_1/t)^2}
{2\hbar^2(D^2a^2+d^4)}]
\label{star}
\end{eqnarray} 
where $f({\bf p},t)$ contains the dependence on ${\bf p}$ only in the exponent with the
imaginary index. Therefore the probabilities of different momenta are defined by the last
exponent which shows that the distribution of momenta has a sharp peak around the
vector $m{\bf r}_1/t$ pointing to the device. While the width of the momentum 
distribution in the initial packet is of order of
${\hbar}/a$ (see Eq. (\ref{chipt})), the width given by Eq. (\ref{star}) is much narrower.
If for example $D^2a^2\gg d^4$ then the width is of the order of ${\hbar}/d$
and in the opposite case  the width is of the order of ${\hbar}d/(Da)$.

As discussed in Sec. \ref{classical}, in semiclassical approximation the value of the 
momentum can be found by applying the
operation $-i\hbar \partial/\partial{\bf r}$ to the rapidly oscillating exponent. In general the
momentum distribution can be rather wide. However, if the particle is detected in a
vicinity of the point ${\bf r}$ then, as follows from Eq. (\ref{star}),
it will be detected with the momentum close to
$m{\bf r}/t$. This result has the following qualitative explanation. The operation
$-i\hbar \partial/\partial{\bf r}$ applied to the imaginary index of the exponent in Eq. 
(\ref{psirt}) gives exactly $m{\bf r}/t$.

The above results gives the solution of the Mott-Heisenberg problem when
the particle is in the state (\ref{psirt}). However, in this case the WF can be
spherically symmetric only if ${\bf p}_0=0$. This case is of no interest because typically
a particle created in the spherically symmetric state has a nonzero kinetic energy.
We now consider a model where, instead of Eq. (\ref{chip}), the initial particle
momentum WF is
\begin{equation}
\chi({\bf p})=\frac{f({\bf p}/p)}{p}exp[-\frac{1}{2\hbar^2}a^2(p-p_0)^2]
\label{chipn}
\end{equation}
where $p=|{\bf p}|$ and the quantaties $p_0$ and $a$ are 
positive. We assume that $p_0a\gg \hbar$. Then, with a good accuracy, integrals
over $p$ from $0$ to $\infty$ containing the exponent can be replaced by
integrals from $-\infty$ to $\infty$.  By analogy with the calculation in Sec. \ref{classical}, 
one can easily show that ${\bar p}\approx p_0$ and $\Delta p\approx \hbar/(a\sqrt{2})$
and therefore  the $p$-distribution is semiclassical. The dependence of
the momentum WF on $t$ is the same as in Eq. (\ref{chipt}).

The coordinate WF is again given by Eq. (\ref{Fourier}). For calculating this
function in the case when the initial momentum WF is given by Eq. (\ref{chipn})
we need the following auxiliary results:  
\begin{equation} 
\int_0^{\infty} exp[-\frac{1}{2\hbar^2}a^2(p-p_0)^2+\frac{i}{\hbar}pr \xi]dp\approx 
\frac{\hbar}{a}(\frac{2\pi}{1+iD/a})^{1/2}exp[-\frac{(r\xi-p_0t/m)^2}{2a^2(1+iD/a)}]
\label{rxi}
\end{equation}
where $r=|{\bf r}|$ and
\begin{equation}
exp(\frac{i}{\hbar}{\bf pr})=4\pi\sum_{l\mu}i^lj_l(pr/\hbar )Y_{l\mu}^*({\bf p}/p)Y_{l\mu}({\bf r}/r)
\label{flat}
\end{equation}
The last expression is the known decomposition of the flat wave. Here $Y_{l\mu}$ is the
spherical function corresponding to the orbital angular momentum $l$ and its $z$-projection $\mu$
and $j_l$ is the spherical Bessel function. Its asymptotic expression when the argument is large
is $j_l(x)\approx sin(x-\pi l/2)/x$.

Let $f({\bf p}/p)=\sum_{l\mu}c_{lm}Y_{l\mu}({\bf p}/p)$ be the decomposition of the
function $f$ in Eq. (\ref{chipn}) over spherical functions. Then it follows from orthogonality
of spherical functions, Eqs. (\ref{chipn}-\ref{flat}) and the above remarks that
if $(pr/\hbar)\gg 1$ then
\begin{eqnarray}
&&\psi({\bf r},t)=-\frac{i}{ar}(\frac{\hbar}{1+iD/a})^{1/2}exp(-\frac{p_0^2t}{2m\hbar})
\sum_{l\mu}c_{l\mu}Y_{l\mu}({\bf r}/r)\nonumber\\
&&\{exp[-\frac{(r-p_0t/m)^2}{2a^2(1+iD/a)}]-(-1)^lexp[-\frac{(r+p_0t/m)^2}{2a^2(1+iD/a)}]\}
\label{psirn}
\end{eqnarray}
At large distances and times the second term in the figure brackets is negligible and the final
result is
\begin{eqnarray}
\psi({\bf r},t)=-\frac{i}{ar}(\frac{\hbar}{1+iD/a})^{1/2}exp(-\frac{p_0^2t}{2m\hbar})
f({\bf r}/r)exp[-\frac{(r-p_0t/m)^2(1-iD/a)}{2(a^2+D^2)}]
\label{psif}
\end{eqnarray}

{\it Therefore for the initial momentum WF (\ref{chipn}) the coordinate WF at large distances and 
times has the same angular dependence as the momentum WF, and the radial WF spreads out by analogy with Eq. (\ref{psirt}).}

The result (\ref{psif}) gives an obvious solution of the Mott-Heisenberg problem in the case
when the angular dependence of the WF is arbitrary. Indeed, suppose that a particle
is created at the origin and a measuring device is seen from the origin in the narrow angular range
defined by the function ${\tilde f}({\bf r}/r)$. Suppose that the support of ${\tilde f}({\bf r}/r)$
is within the range defined by $f({\bf r}/r)$. Then the projection of the WF (\ref{psif})
onto the device is given by the same expression where $f({\bf r}/r)$ is replaced by 
${\tilde f}({\bf r}/r)$. Since the angular WFs in coordinate and momentum 
representations are the same, the momenta  measured by the device will be in the angular range
defined by the function ${\tilde f}({\bf p}/p)$.

\section{Position operator in relativistic quantum mechanics}
\label{momentum}

The problem of position operator in relativistic quantum theory has been discussed in a wide literature
and different authors have different opinions on this problem. In particular, some authors state that in
relativistic quantum theory no position operator exists. As already noted, the results of fundamental quantum theories
are formulated only in terms of the S-matrix in momentum space without mentioning space-time. This is in
the spirit of the Heisenberg S-matrix program that in relativistic quantum theory it is possible to describe only
transitions of states from the infinite past when
$t\to -\infty$ to the distant future when $t\to +\infty$. On the other hand, since quantum theory is treated as a
theory more general than classical one, it is not possible to fully avoid space and time in quantum theory.
For example, quantum theory should explain how photons from distant objects travel to Earth and even  
how macroscopic bodies are moving along classical trajectories. Hence we can conclude that: 
a) in quantum theory (nonrelativistic and relativistic) we must have a position operator and 
b) this operator has a physical meaning only in semiclassical approximation.

As noted in Sec. \ref{symmetry},  in relativistic quantum theory elementary particles are
described by IRs of the Poincare algebra. 
There exists a wide literature on constructing such IRs.  
In particular, an IR for a spinless particle can be implemented in a space of functions $\xi({\bf p})$ 
satisfying the condition
\begin{equation}
\int |\xi({\bf p})|^2d\rho({\bf p})<\infty, \quad d\rho({\bf p})=\frac{d^3{\bf p}}{\epsilon({\bf p})}
\label{invnorm}
\end{equation}
where $\epsilon({\bf p})=(m^2+{\bf p}^2)^{1/2}$ is the energy of the particle with the mass $m$. The convenience
of the above requirement is that the volume element $d\rho({\bf p})$ is Lorentz invariant. In that case 
it can be easily shown by direct calculations (see e.g. Ref. \cite{current}) that the representation operators 
have the form
\begin{eqnarray}
{\bf L}=-i{\bf p}\times\frac{\partial}{\partial {\bf p}},\quad 
{\bf N}=-i\epsilon({\bf p})\frac{\partial}{\partial {\bf p}}, \quad {\bf P}={\bf p},\quad E=\epsilon({\bf p})
\label{IRoperators}
\end{eqnarray}
where ${\bf L}$ is the orbital angular momentum operator, ${\bf N}$ is the Lorentz
boost operator, ${\bf P}$ is the momentum operator, $E$ is the energy operator and these operators are expressed
in terms of the operators in Eq. (\ref{PCR}) as
$${\bf L}=(M^{23},M^{31},M^{12}),\,\, {\bf N}=(M^{10},M^{20},M^{30}),\,\,
{\bf P}=(P^1,P^2,P^3),\,\,E=P^0$$

For particles with spin these results are modified as follows. For a massive particle with spin $s$ the 
functions $\xi({\bf p})$ also depend on spin projections which can take $2s+1$ values $-s,-s+1,...s$. If ${\bf s}$ is the spin operator then the total angular momentum has an additional term ${\bf s}$ and the Lorentz boost operator has 
an additional term $({\bf s}\times{\bf p})/(\epsilon({\bf p})+m)$ (see e.g. Eq. (2.5) in Ref. \cite{current}). Hence corrections of the spin terms to the quantum 
numbers describing the angular momentum and the Lorentz boost do not exceed $s$. We assume as usual that in semiclassical approximation the quantum numbers characterizing the angular momentum and the Lorentz boost are much greater than
unity and hence in this approximation spin effects can be neglected. For a massless particle with the spin $s$ the
spin projections can take only values $-s$ and $s$ and those quantum numbers have the meaning of helicity. In this
case the results for the representation operators can be obtained by taking the limit $m\to 0$ if the operators
are written in the light front variables (see e.g. Eq. (25) in Ref. \cite{symm1401}). As a consequence,
in semiclassical approximation the spin corrections in the massless case can be neglected as well. Hence for investigating
the position operator we will neglect spin effects and will not explicitly write the dependence of WFs 
on spin projections. 

In the above IRs the representation operators are Hermitian as it should be for operators corresponding to physical
quantities. In standard theory (over complex numbers) such IRs of the Lie algebra can be extended to unitary IRs
of the Poincare group. In particular, in the spinless case the unitary operator $U(\Lambda)$ corresponding to the
Lorentz transformation $\Lambda$ acts in $H$ as (see e.g. Ref. \cite{current})
\begin{equation}
U(\Lambda)\xi(p)=\xi(\Lambda^{-1}p)
\label{U(Lambda)}
\end{equation}

In the literature
the problem of position operator is mainly discussed in the approach when elementary particles are described
by local fields rather than unitary IRs. Below we discuss the both approaches 
but first we consider the case of unitary IRs.

As follows from Eq. (\ref{PCR}), the operator $I_2=E^2-{\bf P}^2$ is the Casimir operator 
of the second order, 
i.e. it is a bilinear combination of representation operators commuting with
all the operators of the algebra. As follows from the known Schur lemma, all states belonging to an IR are
the eigenvectors of $I_2$ with the same eigenvalue $m^2$. Note that Eq. (\ref{IRoperators}) contains only $m^2$
but not $m$. The choice of the energy sign is only a matter of convention
but not a matter of principle. Indeed, the energy can be measured only if the momentum ${\bf p}$ is measured
and then it is only a matter of convention what sign of the square root should be chosen. However, it is
important that the sign should be the same for all particles. For example, if we consider a system of two
particles with the same values of $m^2$ and the opposite momenta ${\bf p}_1$ and ${\bf p}_2$ such that
${\bf p}_1+{\bf p}_2=0$, we cannot define the energies of the particles as $\epsilon({\bf p}_1)$
and $-\epsilon({\bf p}_2)$, respectively, since in that case the total four-momentum of the two-particle
system will be zero what contradicts experiment.

The notation $I_2=m^2$ is justified by the fact that for all known particles $I_2\geq 0$. Then the mass 
$m$ is {\it defined} as the square root of $m^2$ and the sign of $m$ is only a
matter of convention. The usual convention is that $m\geq 0$. However, from mathematical point of view, 
IRs with $I_2<0$ are not prohibited. If the velocity
operator ${\bf v}$ is {\it defined} as ${\bf v}={\bf P}/E$ then for known particles $|{\bf v}|\leq 1$, i.e.
$|{\bf v}|\leq c$ in standard units. However, for IRs with $I_2 < 0$, $|{\bf v}|> c$ and,
at least from the point of view of mathematical construction of IRs, this case is not prohibited. The hypothetical
particles with such properties are called tachyons and their possible existence is widely discussed in
the literature. If the tachyon mass $m$ is also defined as the square root of $m^2$ then this quantity will
be imaginary. However, this does not mean than the corresponding IRs are unphysical since all the operators
of the Poincare group Lie algebra depend only on $m^2$.

As follows from Eqs. (\ref{invnorm}) and (\ref{IRoperators}), in the nonrelativistic approximation 
$d\rho({\bf p})=d^3{\bf p}/m$ and ${\bf N}=-im\partial /\partial{\bf p}$. Therefore in this approximation ${\bf N}$
is proportional to {\it standard} position operator and one can say that the position operator is
in fact present in the description of the IR.

The following remarks are in order. The choice of the volume element in the Lorentz invariant 
form $d\rho({\bf p})$ (see Eq. (\ref{invnorm})) might be convenient from the point of view that then
the Hilbert space can be treated as a space of functions $\xi(p)$ depending on four-vectors $p$ such that
$p^0=\epsilon({\bf p})$ and the norm can be written in the covariant form (i.e. in the form depending only on
Lorentz invariant quantities):
$||\xi||^2=\int |\xi(p)|^2\delta (p^2-m^2)\theta (p^0)d^4p$. However, the requirement of covariance does not have
a fundamental physical meaning. In relativistic theory a necessary requirement is that symmetry is defined by
operators satisfying the commutation relations (\ref{PCR}) and this requirement can be implemented in different 
forms, not necessarily in covariant ones. 

As an illustration, consider the following problem. Suppose that we wish to construct a single-particle coordinate
WF. Such a WF cannot be defined on the whole Minkowski space. This is clear even from the
fact that there is no time operator. The WF can be defined only on a space-like hyperplane of the
Minkowski space. For example, on the hyperplane $t=const$ the WF depends only on ${\bf x}$. Hence
for defining the WF one has to choose the form of the position operator. By analogy with the
nonrelativistic case, one might try to define the position operator as $i\partial/\partial{\bf p}$. However,
if the Hilbert space is implemented as in Eq. (\ref{invnorm}) then this operator is not selfadjoint since $d\rho({\bf p})$
is not proportional to $d^3{\bf p}$. One can perform a unitary transformation 
$\xi({\bf p})\to \chi({\bf p})=\xi({\bf p})/\epsilon({\bf p})^{1/2}$ such that the Hilbert space becomes
the space of functions $\chi({\bf p})$ satisfying the condition $\int|\chi({\bf p})|^2d^3{\bf p}<\infty$. 
It is easy to verify that in this implementation of the IR the operators $({\bf L},{\bf P},E)$ will have
the same form as in Eq. (\ref{IRoperators}) but the expression for ${\bf N}$ will be
\begin{equation}
{\bf N}=-i\epsilon({\bf p})^{1/2}\frac{\partial}{\partial {\bf p}}\epsilon({\bf p})^{1/2}
\label{newN}
\end{equation}
In this case one can {\it define} $i\hbar \partial/\partial{\bf p}$ as a position operator
but now we do not have a situation when the position operator is present among the other representation
operators. 

A problem of the definition of the position operator in relativistic quantum theory has been discussed since
the beginning of the 1930s and it has been noted that when quantum theory is combined with relativity the
existence of the position operator with correct physical properties becomes a problem. The above definition
has been proposed by Newton and Wigner in Ref. \cite{NW}. They worked in the approach when elementary particles
are described by local fields $\Psi(x)$ defined on the whole Minkowski space rather than unitary IRs. As
noted above, such fields cannot be treated as single-particle WFs. The spacial Fourier transform 
of such fields at $t=const$ describes states
where the energy can be positive and negative and this is interpreted such that local quantum fields describe
a particle and its antiparticle simultaneously. Newton and Wigner first discuss the spinless case and consider
only states on the upper Lorentz hyperboloid where the energy is positive. For such states the representation
operators act in the same way as in the case of spinless unitary IRs. With this definition 
the coordinate WF $\psi({\bf r})$ can be again defined by Eq. (\ref{psir}) and a question arises 
whether such a position operator has all the required properties.

For example, in the introductory section of the
textbook \cite{BLP} the following arguments are given in favor of the statement that in relativistic quantum theory 
it is not possible to define a physical position operator.
Suppose that we measure coordinates of an electron with the mass $m$. When the uncertainty of  
coordinates is of the order of $\hbar/mc$, the uncertainty of momenta is of the order of $mc$, the uncertainty
of energy is of the order of $mc^2$ and hence creation of electron-positron pairs is allowed. As a consequence,
it is not possible to localize the electron with the accuracy better than its Compton wave length
${\hbar}/mc$. Hence, for a particle with a nonzero mass exact measurement is possible only
either in the nonrelativistic limit (when $c\to\infty$) or
classical limit (when ${\hbar}\to 0)$. In the case of the photon, as noted by Pauli (see p. 191 of Ref.
\cite{Pauli}), the coordinate cannot be measured with the accuracy better than $\hbar/p$ where $p$ is the magnitude
of the photon momentum. The quantity $\lambda=2\pi\hbar/p$ is called the photon wave length although, as noted
in Subsec. \ref{subsecsumm}, the meaning of this quantity in quantum case might be fully different than in classical one. Since
$\lambda\to 0$ in the formal limit $\hbar\to 0$, Pauli concludes that "Only within the confines of the classical ray
concept does the position of the photon have a physical significance".

Another argument that the Newton-Wigner position operator does not have all the required properties follows.
Since the energy operator acts on the function $\chi({\bf p})$ as $E\chi({\bf p})=\epsilon({\bf p})\chi({\bf p})$
(see Eq. (\ref{IRoperators})) and the energy is an operator corresponding to infinitesimal time translations,
the dependence of the WF $\chi({\bf p})$ on $t$ is given by 
\begin{equation}
\chi({\bf p},t)=exp(-\frac{i}{\hbar}Et)\chi({\bf p})=exp(-\frac{i}{\hbar}\epsilon({\bf p})t)\chi({\bf p})
\label{chiptA}
\end{equation}
Then a relativistic analog of Eq. (\ref{tnonrel}) is 
\begin{equation}
\psi({\bf r},t)=\int exp\{\frac{i}{\hbar}[{\bf p}({\bf r}-{\bf r}')-\epsilon({\bf p})t]\}\psi_0({\bf r}')
\frac{d^3{\bf r}'d^3{\bf p}}{(2\pi\hbar)^3}
\label{trel}
\end{equation}
As a consequence, the Newton-Wigner position operator has the "tail property": if $\psi_0({\bf r})\neq 0$ only if ${\bf r}$ belongs to a finite vicinity of some vector ${\bf r}_0$ then at any $t>0$ the function $\psi({\bf r},t)$ 
has a tail belonging to the whole three-dimensional space, i.e. the
WF spreads out with an infinite speed. Hence at any $t>0$ the
particle can be detected at any point of the space and this contradicts the requirement that no information
should be transmitted with the speed greater than $c$. 

The tail property of the Newton-Wigner position operator has been known for a long time (see e.g. Ref. \cite{Hegerfeldt}
and references therein). It is characterized as nonlocality leading to the action at a distance. Hegerfeldt argues \cite{Hegerfeldt} that this property is rather general because it can be proved assuming that energy is positive and without assuming a specific choice of the position operator. 
The Hegerfeldt theorem \cite{Hegerfeldt} is based on the assumption that there exists an operator $N(V)$
whose expectation defines the probability to find a particle inside the volume $V$. However, 
the meaning of time on quantum level is not clear and for the position operator
proposed in this chapter such a probability does not exist because there is no WF in coordinate
representation (see Sec. \ref{consistent} and the discussion in Sec. \ref{conclusion}). 

One might say that the requirement that no signal can be transmitted with the speed greater than $c$ has been obtained in
Special Relativity which is a classical (i.e. non-quantum) theory operating only with classical space-time
coordinates. For example, in classical theory the velocity of a particle is defined as ${\bf v}=d{\bf r}/dt$
but, as noted above, the velocity {\it should be defined} as ${\bf v}={\bf p}/E$ (i.e. without mentioning 
space-time) and then on classical level it can be shown that ${\bf v}=d{\bf r}/dt$. 
In QFT local quantum fields separated by space-like intervals commute or anticommute
(depending on whether the spin is integer or half-integer) and this is treated as a requirement of causality
and that no signal can be transmitted with the speed greater than $c$. However, as noted above, the physical
meaning of space-time coordinates on quantum level is not clear. Hence from the point of view of quantum theory 
the existence of tachyons is not prohibited. Note also that when two electrically charged particles exchange by a virtual photon, a typical situation is that the four-momentum of the photon is space-like, i.e. the photon is the tachyon. 
We conclude that although in relativistic theory such a behavior might seem undesirable, there is no proof that it must 
be excluded. Also, as argued by Griffiths (see Ref. \cite{Griffiths} and references therein), with a consistent 
interpretation of quantum theory there are no nonlocality and superluminal interactions. In Sec. \ref{conclusion} we argue that the position operator proposed in the present paper sheds a new light on this problem.

Another striking example is a photon emitted in the famous
21cm transition line between the hyperfine energy levels of the hydrogen atom. The phrase that the lifetime of 
this transition is of the order of $\tau=10^7$ years implies that the width of the level is of the order of $\hbar/\tau$, 
i.e. experimentally the uncertainty of the photon energy is $\hbar/\tau$. Hence the uncertainty of the
photon momentum is $\hbar/(c\tau)$ and with the above definition of the coordinate operators the uncertainty of the longitudinal coordinate is $c\tau$, i.e. of the order of $10^7$ light years. Then there is a nonzero probability that 
immediately after its creation at point A the photon can be detected at point B such that the distance between A 
and B is $10^7$ light years. 

A problem arises how this phenomenon should be interpreted. On one hand, one might say that in view of the above
discussion it is not clear whether or not the requirement that no information should be transmitted with the speed greater than $c$ should be a must in relativistic quantum theory. On the other hand (as pointed out to me by Alik Makarov), 
we can know about the photon creation only if the photon is detected and when it 
was detected at point B at the moment of time $t=t_0$, this does not mean that the photon 
traveled from A to B with the speed greater than $c$ since the time of creation has an uncertainty of the order 
of $10^7$ years. Note also that in this situation a description of the system (atom + electric
field) by the WF (e.g. in the Fock space) depending on a continuous parameter $t$ has no physical meaning 
(since roughly speaking the quantum of time in this process is of the order of $10^7$ years). 
If we accept this explanation then we should acknowledge that in some situations a description of
evolution by a continuous classical parameter $t$ is not physical and this is in the spirit of the Heisenberg S-matrix
program. However, this example describes a pure quantum phenomenon while, as noted above, a position operator is
needed only in semiclassical approximation.

For particles with nonzero spin, the number of states in local fields is typically by a factor of two greater
than in the case of unitary IRs (since local fields describe a particle and its antiparticle simultaneously) but those
components are not independent since local fields satisfy a covariant equation (Klein-Gordon, Dirac etc.). In Ref.
\cite{NW} Newton and Wigner construct a position operator in the massive case but say that in the massless one they
have succeeded in constructing such an operator only for Klein-Gordon and Dirac particles while in the case of the
photon the position operator does not exist. On the other hand, as noted above, in the case of unitary IRs different
spin components are independent and in semiclassical approximation spin effects are not important. So in this
approach one might adopt the Newton-Wigner position operator for particles with any spin and any mass. 

We now consider the following problem. Since the Newton-Wigner position operator formally has the same form as in
nonrelativistic quantum mechanics, the coordinate and momentum WFs also are related to each other by the same 
Fourier transform as in nonrelativistic quantum mechanics (see Eq. (\ref{Fourier})). One might think that
this relation is not Lorentz covariant  and pose a question whether in relativistic theory this is acceptable. As noted above, for constructing 
the momentum WF covariance does not have a fundamental physical meaning and is not necessary. A question
arises whether the same is true for constructing the coordinate WF.

Let us note first that if the four-vector $x$ is such that $x=(t,{\bf x})$ then the WF 
$\psi(x)=\psi({\bf x},t)$ can have a physical meaning only if we accept that (at least in some approximations) a
position operator is well defined. Then the function $\psi({\bf x},t)$ describes amplitudes of probabilities for
different values of ${\bf x}$ at a fixed value of $t$. This function cannot describe amplitudes of probabilities for
different values of $t$ because there is no time operator. 

For discussing Lorentz covariance of the coordinate WF it is important to note that, in view of the above
remarks, this function can be defined not in the whole Minkowski space but only on space-like hyperplanes of that
space (by analogy with the fact that in QFT the operators $(P^{\mu},M^{\mu\nu})$ are defined by integrals
over such hyperplanes). They are defined by a time-like unit vector $n$ and the evolution parameter $\tau$ such that 
the corresponding hyperplane is a set of points with the coordinates $x$ satisfying the condition $nx=\tau$.
Wave functions $\psi(x)$ on this hyperplane satisfy the requirement $\int |\psi(x)|^2\delta(nx-\tau)d^4x<\infty$. In a
special case when $n^0=1$, ${\bf n}=0$ the hyperplane is a set of points $(t=\tau,{\bf x})$ and the wave
functions satisfy the usual requirement $\int |\psi({\bf x},t)|^2d^3{\bf x}<\infty$. In the literature coordinate WFs are usually considered without discussions of the position operator
and without mentioning the fact that those functions are defined on space-like hyperplanes (see e.g. Refs.
\cite{SR,Bradler}).

By analogy with the construction of the coordinate WF in Refs.  \cite{SR,Naumov}, it can be defined as
follows. Let ${\tilde x}_0$ be a four-vector and $p$ and $p_0$ be four-vectors
$(\epsilon({\bf p}),{\bf p})$ and $(\epsilon({\bf p}_0),{\bf p}_0)$, respectively. We will see below that momentum WFs describing wave packets can be chosen in the form
\begin{equation}
\xi(p,p_0,{\tilde x}_0)=f(p,p_0)exp(\frac{i}{\hbar}p{\tilde x}_0)
\label{covxi}
\end{equation}
where $f(p,p_0)$ as a function of $p$ has a sharp maximum in the vicinity of $p=p_0$, ${\tilde x}_0=x_0-(nx_0)n$ and
the four-vector $x_0$ has the coordinates $(t,{\bf r}_0)$. Then the coordinate WF can be defined as
\begin{equation}
\psi(x,p_0,{\tilde x}_0)=\frac{1}{(2\pi\hbar)^{3/2}}\int \xi(p,p_0,{\tilde x}_0)exp(-\frac{i}{\hbar}px)d\rho({\bf p})
\label{covpsi}
\end{equation}
Suppose that $f(p,p_0)$ is a covariant function of its arguments, i.e. it can depend only on $p^2$, $p_0^2$ and
$pp_0$. Then, as follows from Eq. (\ref{U(Lambda)}), the function $\psi(x,p_0,{\tilde x}_0)$ is covariant because 
its Lorentz 
transformation is $\psi(x,p_0,{\tilde x}_0)\to \psi(\Lambda^{-1}x, p_0, {\tilde x}_0)$.

The choice of $f(p,p_0)$ in the covariant form might encounter the following problem. For example, the authors of
Ref. \cite{Naumov} propose to consider $f(p,p_0)$ in the form
\begin{equation}
f(p,p_0)=const\, exp[\frac{(p-p_0)^2}{4\sigma^2}]
\label{Naumov}
\end{equation}
The exponent in this expression has the maximum at ${\bf p}={\bf p}_0$ and in the vicinity of the maximum
\begin{equation}
(p-p_0)^2=-({\bf p}-{\bf p}_0)^2+[\frac{({\bf p}_0,{\bf p}-{\bf p}_0)}{\epsilon({\bf p}_0)}]^2+o(|{\bf p}-{\bf p}_0|^2)
\label{p-p0}
\end{equation}
If ${\bf p}_0$ is directed along the $z$ axis and the subscript ${\bot}$ is used to denote the projection of
the vector onto the $xy$ plane then
\begin{equation}
(p-p_0)^2=-({\bf p}_{\bot}-{\bf p}_{0\bot})^2-[\frac{m}{\epsilon({\bf p}_0)}]^2(p_z-p_{0z})^2+o(|{\bf p}-{\bf p}_0|^2)
\label{p-p0B}
\end{equation}
It follows from this expression that if the particle is ultrarelativistic then the width of the momentum distribution
in the longitudinal direction is much greater that in transverse ones and for massless particles the former becomes
infinite. We conclude that for massless particles the covariant parametrization of $f(p,p_0)$ is problematic.

As noted above, the only fundamental requirement on quantum level is that the representation operators should satisfy
the commutation relations (\ref{PCR}) while covariance is not fundamental. Nevertheless, the above discussion shows
that covariance of coordinate WFs can be preserved if one takes into account the fact that they are 
defined on space-like hyperplanes. In particular, covariance of functions $f$ can be preserved if one assumes
that they depend not only on $p$ and $p_0$ but also on $n$. In what follows we consider only the case when the vector 
$n$ is such that $n^0=1$ and ${\bf n}=0$. Let us replace $f(p,p_0)$ by $f({\tilde p},{\tilde p}_0)$ where ${\tilde p}=p-(pn)n$ and ${\tilde p}_0=p_0-(p_0n)n$. Then the four-vectors ${\tilde p}$ and ${\tilde p}_0$ have only nonzero
spatial components equal ${\bf p}$ and ${\bf p}_0$, respectively. As a consequence, any rotationally invariant 
combination of 
${\bf p}$ and ${\bf p}_0$ can be treated as a Lorentz covariant combination of ${\tilde p}$ and ${\tilde p}_0$.

We conclude that with the above choice of the vector $n$ one can work with momentum and coordinate WFs
in full analogy with nonrelativistic quantum mechanics and in that case Lorentz covariance is satisfied. In
particular in that case Eq. (\ref{covpsi}) can be written in the form of Eq. (\ref{Fourier}).

We now consider the photon case in greater details. The coordinate photon WF has been discussed by
many authors. A question arises in what situations this function is needed. As already noted, since the
fundamental theory of electromagnetic interactions is QED, and this theory does not 
contain space-time at all,
for solving quantum problems in the framework of QED the coordinate photon WF is not needed.
However, this function is used in some special problems, for example for describing single-photon interference
and diffraction by analogy with classical theory.

In this chapter we consider only the case of free photons. If we consider a motion of a free particle, 
it is not important in what interactions this particle
participates and, as explained above, if the particle is described by its IR in semiclassical approximation then the particle spin is not important. Hence the effect of WPS for an ultrarelativistic particle does not depend on
the nature of the particle, i.e. on whether the particle is the photon, the proton, the electron etc. 
For this reason we are interested in papers on the photon coordinate WF mainly from the point
of view how the position operator for the free ultrarelativistic particle is defined.

For the first time the coordinate photon WF has been discussed by Landau and Peierls in Ref. \cite{LP}.
However, in the literature it has been stated (see e.g. Refs. \cite{AB} and \cite{SR}) that in QED there is no way 
to define a coordinate photon WF. A section in the textbook \cite{AB} is titled "Impossibility of introducing the
photon WF in coordinate representation". The arguments follow. The electric
and magnetic fields of the photon in coordinate representation are proportional to the Fourier transforms
of $|{\bf p}|^{1/2}\chi({\bf p})$, rather than $\chi({\bf p})$. As a consequence, the quantities ${\bf E}({\bf r})$
and ${\bf B}({\bf r})$ are defined not by $\psi({\bf r})$ but by integrals of $\psi({\bf r})$ over a 
region of the order of the wave length. However, this argument also does not exclude the possibility that 
$\psi({\bf r})$ can have
a physical meaning in semiclassical approximation since, as noted above, the notions of the electric and
magnetic fields of a single photon are problematic. In addition, since $\lambda\to 0$ in the
formal limit $\hbar\to 0$, one should not expect that any position operator in semiclassical approximation
can describe coordinates with the accuracy better than the wave length. Another arguments in favor
of the existence of the coordinate photon WF have been given by Bialynicki-Birula \cite{Bialy}. 

A detailed discussion of the photon position operator can be found in papers by Margaret Hawton and
references therein (see e.g. Ref. \cite{Hawton}). In this approach the photon is described by a local field
and the momentum and coordinate representations are related to each other by standard Fourier transform.
The author of Ref. \cite{Hawton} discusses generalizations of the photon position operator proposed by
Pryce \cite{Pryce}. However, the Pryce operator and its generalizations discussed in Refs. \cite{Bialy,Hawton}
differ from the Newton-Wigner operator only by terms of the order of the wave length. Hence in semiclassical approximation all those operators are equivalent. 

The above discussion shows that on quantum level the physical meaning of the coordinate is a difficult problem but
in view of a) and b) (see the beginning of this section) one can conclude that in semiclassical
approximation all the existing proposals for the position operator are equivalent to the  
Newton-Wigner operator $i\hbar \partial/\partial{\bf p}$. An additional argument in favor of this operator
is that the relativistic nature of the photon might be somehow manifested in the longitudinal direction while
in transverse directions the behavior of the WF should be similar to that in standard
nonrelativistic quantum mechanics. Another argument is that the photon WF in coordinate 
representation constructed by using this
operator satisfies the wave equation in agreement with classical electrodynamics (see Sec. \ref{geom}).

For all the reasons described above, in the next section we consider what happens if the 
space-time evolution of relativistic wave packets is described by using the Newton-Wigner position operator.

\section{Wave packet spreading in relativistic quantum mechanics}
\label{RelWPS}

Consider first a construction of the wave packet for a particle with nonzero mass. A possible way of the
construction follows. We first consider the particle in its rest system, i.e. in the reference frame where
the mean value of the particle momentum is zero. The WF $\chi_0({\bf p})$ in this case can be taken
as in Eq. (\ref{chip}) with ${\bf p}_0=0$. As noted in Sec. \ref{classical}, such a state cannot be 
semiclassical. However, it is possible to obtain a semiclassical state by applying a Lorentz transformation
to $\chi_0({\bf p})$. As a consequence of Eq. (\ref{U(Lambda)}) and the relation between the functions $\xi$ and
$\chi$ 
\begin{equation}
U(\Lambda)\chi_0({\bf p})=[\frac{\epsilon({\bf p}')}{\epsilon({\bf p})}]^{1/2}\chi_0({\bf p}')
\label{Lorentz}
\end{equation} 
where ${\bf p}'$ is the momentum obtained from ${\bf p}$ by the Lorentz transformation $\Lambda^{-1}$. If $\Lambda$ is
the Lorentz boost along the $z$ axis with the velocity $v$ then
\begin{equation}
{\bf p}_{\bot}'={\bf p}_{\bot},\quad p_z'=\frac{p_z-v\epsilon({\bf p})}{(1-v^2)^{1/2}}
\label{p'}
\end{equation}

As follows from this expression, $exp(-{\bf p}^{'2}a^2/2\hbar^2)$ as a function of ${\bf p}$ has the maximum
at ${\bf p}_{\bot}=0$, $p_z=p_{z0}=v[(m^2+{\bf p}_{\bot}^2)/(1-v^2)]^{1/2}$ and near the maximum
$$exp(-\frac{a^2{\bf p}^{'2}}{2\hbar^2})\approx exp\{-\frac{1}{2\hbar^2}[a^2{\bf p}_{\bot}^2+b^2(p_z-p_{z0})^2]\}$$
where $b=a(1-v^2)^{1/2}$ what represents the effect of the Lorentz contraction. If $mv\gg \hbar/a$ (in units where $c=1$) 
then $m\gg |{\bf p}_{\bot}|$  and $p_{z0}\approx mv/(1-v^2)^{1/2}$. In this case the transformed state is semiclassical
and the mean value of the momentum is exactly the classical (i.e. non-quantum) value of the momentum of a particle
with mass $m$ moving along the $z$ axis with the velocity $v$. However, in the opposite case when 
$m\ll \hbar/a$ the transformed state is not semiclassical since the uncertainty of $p_z$ is of the same order
as the mean value of $p_z$.

If the photon mass is exactly zero then the photon cannot have the rest state. However, even if the photon mass
is not exactly zero, it is so small that the condition $m\ll \hbar/a$ is certainly satisfied for any realistic
value of $a$. Hence a semiclassical state for the photon or a particle with a very small mass cannot be obtained
by applying the Lorentz transformation to $\chi_0({\bf p})$ and considering the case when $v$ is very close to unity.
An analogous problem with the covariant description of the massless WF has been discussed in
the preceding section (see Eq. (\ref{p-p0B})). 

The above discussion shows that in the relativistic case the momentum distribution in transverse directions is the
same as in the nonrelativistic case (see also Eq. (\ref{p-p0B})) and the difference arises only for 
the momentum distribution in the longitudinal direction. Let us consider the ultrarelativistic case when 
$|{\bf p}_0|=p_0\gg m$ and suppose that ${\bf p}_0$ is directed along the $z$ axis. As noted in the preceding section,
the formal requirement of Lorentz covariance will be satisfied if one works with rotationally invariant 
combinations of ${\bf p}$ and ${\bf p}_0$. The quantities ${\bf p}_{\bot}^2$ and $(p_z-p_0)^2$ satisfy this condition
because
$${\bf p}_{\bot}^2=[{\bf p}-{\bf p}_0\frac{({\bf p}{\bf p}_0)}{p_0^2}]^2, \quad (p_z-p_0)^2=\frac{1}{p_0^2}[({\bf p}{\bf p}_0)-p_0^2]^2$$

We will describe an ultrarelativistic semiclassical state by a WF 
which is a generalization of the function (\ref{chip}) (see also Eq. (\ref{covxi})):
\begin{equation}
\chi({\bf p},0)=\frac{ab^{1/2}}{\pi^{3/4}\hbar^{3/2}}exp[-\frac{{\bf p}_{\bot}^2a^2}{2\hbar^2}
-\frac{(p_z-p_0)^2b^2}{2\hbar^2}-\frac{i}{\hbar}{\bf p}_{\bot}{\bf r}_{0\bot}-\frac{i}{\hbar}(p_z-p_0)z_0]
\label{chiprel}
\end{equation}
In the general case the parameters $a$ and $b$ defining the momentum distributions in the transverse and 
longitudinal  directions, respectively, can be different. In that case the uncertainty of each transverse component of momentum is 
$\hbar /(a\sqrt{2})$ while the uncertainty of the $z$ component of momentum is $\hbar /(b\sqrt{2})$.
In view of the above discussion one might think that, as a consequence of the Lorentz contraction, the parameter $b$
should be very small. However, the notion of the Lorentz contraction has a physical
meaning only if $m\gg \hbar/a$ while for the photon the opposite relation takes place. We will see below
that in typical situations the quantity $b$ is large and much greater than $a$.

In relativistic quantum theory the situation with time is analogous to that in the nonrelativistic case (see Sec.
\ref{NRWPS}) and time can be treated only as a good approximate parameter describing the evolution according to the Schr\"{o}dinger equation with the
relativistic Hamiltonian. Then, as a consequence of Eq. (\ref{chiptA}), we have that in the ultrarelativistic
case (i.e. when $p=|{\bf p}|\gg m$) 
\begin{equation}
\chi({\bf p}, t)=exp(-\frac{i}{\hbar}pct)\chi({\bf p},0)
\label{chiptphoton}
\end{equation}
Since at different moments
of time the WFs in momentum space differ each other only by a phase factor, the mean value and uncertainty
of each momentum component do not depend on time. In other words, there is no WPS for the WF in momentum
space. As noted in Sec. \ref{NRWPS}, the same is true in the nonrelativistic case.

As noted in the preceding section, in the relativistic case the function $\psi({\bf r},t)$ can be again defined by Eq. (\ref{Fourier})
where now $\chi({\bf p}, t)$ is defined by Eq. (\ref{chiptphoton}). If the variable $p_z$ in the 
integrand is replaced by $p_0+p_z$ then as follows from Eqs. (\ref{Fourier},\ref{chiprel},\ref{chiptphoton})
\begin{eqnarray}
&&\psi({\bf r},t)=\frac{ab^{1/2}exp(i{\bf p}_0{\bf r}/\hbar)}{\pi^{3/4}\hbar^{3/2}(2\pi\hbar)^{3/2}}
\int exp\{-\frac{{\bf p}_{\bot}^2a^2}{2\hbar^2}-\frac{p_z^2b^2}{2\hbar^2}+\frac{i}{\hbar}{\bf p}({\bf r}-{\bf r}_0)\nonumber\\
&&-\frac{ict}{\hbar}[(p_z+p_0)^2+{\bf p}_{\bot}^2]^{1/2}\} d^3{\bf p}
\label{psirtphoton}
\end{eqnarray}

In contrast to the nonrelativistic case where the energy is the quadratic function of momenta and the integration in
 Eq. (\ref{psirt}) can be performed analytically, here the analytical integration is a problem in
view of the presence of square root in Eq. (\ref{psirtphoton}). We will perform the integration
by analogy with the Fresnel approximation in optics and with Ref. \cite{Dillon} where a
similar approximation has been used for discussing the WPS effect in classical electrodynamics.
Considering this effect from quantum point of view
is even simpler since the photon WF satisfies the relativistic Schr\"{o}dinger equation which is linear
in $\partial /\partial t$. As noted in Sec. \ref{geom}, this function also satisfies the wave equation but
it is simpler to consider an equation linear in $\partial /\partial t$ than that quadratic in
$\partial /\partial t$. However, in classical theory there is no such an object as the photon WF and
hence one has to solve either a system of Maxwell equations or the wave equation. 
The Fresnel approximation describes some important features of the relativistic WPS
effect but, as noted below, in this approximation some important features of this effect
are lost.   

The approximation is based on the fact that in semiclassical approximation the quantity $p_0$ should be much greater
than uncertainties of the momentum in the longitudinal and transversal directions, i.e. $p_0\gg p_z$ and
$p_0\gg |{\bf p}_{\bot}|$. Hence with a good accuracy one can expand the square 
root in the integrand in
powers of $|{\bf p}|/p_0$. Taking into account the linear and quadratic terms in the square root we get
\begin{equation}
[(p_z+p_0)^2+{\bf p}_{\bot}^2]^{1/2}\approx p_0+p_z+{\bf p}_{\bot}^2/2p_0
\label{pperp}
\end{equation}
This is analogous to the approximation 
$(m^2+{\bf p}^2)^{1/2}\approx m+{\bf p}^2/2m$ in nonrelativistic case.
Then the integral over $d^3{\bf p}$ can be calculated as a product of integrals over $d^2{\bf p}_{\bot}$
and $dp_z$ and the calculation is analogous to that in Eq. (\ref{psirt}). The result of the calculation is
\begin{eqnarray}
&&\psi({\bf r},t)=[\pi^{3/4}ab^{1/2}(1+\frac{i\hbar ct}{p_0a^2})]^{-1}
exp[\frac{i}{\hbar}({\bf p}_0{\bf r}-p_0ct)]\nonumber\\
&&exp[-\frac{ ({\bf r}_{\bot}-{\bf r}_{0\bot})^2(1-\frac{i\hbar ct}{p_0a^2})}{2a^2(1+\frac{\hbar^2c^2t^2}{p_0^2a^4})}
-\frac{(z-z_0-ct)^2}{2b^2}]
\label{final}
\end{eqnarray}

This result shows that the wave packet describing an ultrarelativistic particle (including a photon) is moving
along the classical trajectory $z(t)=z_0+ct$, in the longitudinal direction there is no spreading while in
transverse directions spreading is characterized by the function 
\begin{equation}
a(t)=a(1+\frac{\hbar^2c^2t^2}{p_0^2a^4})^{1/2}
\label{at}
\end{equation}
The characteristic time of spreading can be defined as $t_*=p_0a^2/\hbar c$. 
The fact that $t_*\to \infty$ in the
formal limit $\hbar\to 0$ shows that in relativistic case WPS also is a pure quantum phenomenon (see the end of
Sec. \ref{NRWPS}). From the formal point of view the result for $t_*$ is the same as in nonrelativistic
theory but $m$ should be replaced by $E/c^2$ where $E$ is the energy of the ultrarelativistic particle. 
This fact could be expected since, as noted above, it is reasonable to think that spreading in directions perpendicular 
to the particle momentum is similar to that in standard nonrelativistic
quantum mechanics. However,
in the ultrarelativistic case spreading takes place only in these directions.
If $t\gg t_*$ the transverse width
of the packet is $a(t)=\hbar ct/(p_0a)$. 

Hence the speed of spreading in perpendicular directions is $v_*=\hbar c/p_0a$.
In the nonrelativistic case different points of the packet are moving with different
velocities and this is not a problem but in the case of the photon one expects
that each point is moving with the speed $c$. However, the Fresnel approximation
creates a problem because different points are moving with different velocities
such that their magnitudes are in the range $[c,(c^2+v_*^2)^{1/2}]$. 

We now consider a model where 
\begin{equation}
\chi({\bf p})=f({\bf p}/p)F(p)/p
\label{relchipn}
\end{equation}
and assume that $f({\bf p}/p)=\sum_{l\mu}c_{l\mu}Y_{l\mu}({\bf p}/p)$ is the
decomposition of the function $f$ over spherical functions.  The dependence
of the momentum WF on $t$ is now defined by Eq. (\ref{chiptphoton}).
In full analogy with the derivation of Eq. (\ref{psirn}) we now get that
\begin{eqnarray}
&&\psi({\bf r},t)=\frac{-i}{(2\pi\hbar)^{1/2}r}\sum_{l\mu}c_{l\mu}Y_{l\mu}({\bf r}/r)
[G(ct-r)-(-1)^lG(ct+r)]
\label{relpsirn}
\end{eqnarray}
where
\begin{equation}
G(\xi)=\int_0^{\infty} F(p)exp(\frac{-i}{\hbar}\xi p )dp
\label{Gxi}
\end{equation}

For reasonable choices of $F(p)$ we will have that at large distances and times $G(ct-r)\gg G(ct+r)$. 
Indeed if, for example, the quantities $p_0$ and $b$ are such that $p_0b\gg \hbar$ then
possible $(F,G)$ choices are: 
\begin{eqnarray}
&&F(p)=exp(-\frac{|p-p_0|b}{\hbar}),\quad G(\xi)=\frac{exp(-ip_0\xi/\hbar)}{b^2+\xi^2};\nonumber\\
&&F(p)=exp(-\frac{(|p-p_0|b)^2}{2\hbar^2}),\quad 
G(\xi)=(2\pi)^{1/2}\frac{\hbar}{b}exp(-\frac{ip_0\xi}{\hbar}-\frac{\xi^2}{2b^2})
\label{FG}
\end{eqnarray}
As follows from Eq. (\ref{relpsirn}), in those cases
\begin{eqnarray}
\psi({\bf r},t)=\frac{-i}{(2\pi\hbar)^{1/2}r}f({\bf r}/r)G(ct-r)
\label{relpsif}
\end{eqnarray}
Therefore at each moment of time $t$ the coordinate WF is not negligible only
inside a thin sphere with the radius $ct$ and the width of the order of $b$.

The conclusion is that, in contrast to the nonrelativistic case, in the ultrarelativistic one 
there is no WPS in the radial direction (by analogy with the Fresnel approximation) and,
by analogy with the result (\ref{psif}), at large distances and times the angular distributions
in momentum and coordinate WFs are the same. Therefore, in full analogy with
the Mott-Heisenberg problem (see Sec. \ref{Mott}), the momenta
of particles detected by a measuring device will be in the angular range defined not by
the function $f({\bf r}/r)$ but by the function ${\tilde f}({\bf r}/r)$ characterizing the
angles at which the device is seen from the origin. In addition, the angular distribution of
momenta characterized by the function $f$ does not depend on time, as well as in the
nonrelativistic case.

If the function $f$ is essentially different from zero only in the range where 
angles between momenta and the $z$-axis are small then the model (\ref{relchipn}) gives
the same qualitative predictions as the Fresnel approximation. Indeed, suppose that
this function is essentially different from zero for angles which are of the order of
$\alpha$ or less, and $\alpha\ll 1$. Then the parameter $b$ in Eq. (\ref{FG}) is
similar to the parameter $b$  in Eq. (\ref{chiprel}). The characteristic magnitude of the transverse
momentum is of the order of $p_{\bot}\approx \alpha p_0$. Let $a$ be defined such that 
$p_{\bot}=\hbar/a$. When the time is greater than a characteristic time
for which the transition from Eq. (\ref{relpsirn}) to Eq. (\ref{relpsif}) is legitimate
(this time can differ from $t_*$ for the Fresnel model) then, since the angular
distributions in the momentum and coordinate WF are the same, the
transversal width of the packet is of the order of $\alpha ct\approx ct\hbar/(p_0a)$ in
agreement with the Fresnel approximation. Therefore {\it if $t$ is greater than some
characteristic time then the width $a(t)$ of the packet is inversely proportional to
the initial width $a(0)=a$.} It is also possible to define $v_*$ by the same expression as
in the Fresnel approximation. If $v_*\ll c$ the only difference between the two models
is that in the Fresnel approximation different points of the packet are moving with different
speeds while in the model (\ref{relchipn}) they are moving with the same speed $c$.
In fact the Fresnel approximation is such that a small arc representing the front of the WF
 in the model (\ref{relchipn}) is replaced by a segment.

\section{Geometrical optics}
\label{geom}

The relation between quantum and classical electrodynamics is known 
and is described in textbooks (see e.g. Ref. \cite{Dirac,AB}). As already 
noted, classical electromagnetic field consists of many photons and in 
classical electrodynamics the photons are not described individually. 
Instead, classical electromagnetic field is described by field strengths
which represent mean characteristics of a large set of photons. For 
constructing the field strengths one can use the photon WFs 
$\chi({\bf p},t)$ or
$\psi({\bf r},t)$ where $E$ is replaced by $\hbar\omega$ and ${\bf p}$ 
is replaced by $\hbar {\bf k}$. In this connection it is interesting to note that
since $\omega$ is a classical quantity used for describing a classical electromagnetic field, 
the photon is a pure quantum
particle since its energy disappears in the formal limit $\hbar\to 0$. Even this
fact shows that the photon cannot be treated as a classical particle and the effect of WPS 
for the photon cannot be neglected.

With the above replacements the functions $\chi$ and $\psi$ do not contain any 
dependence on $\hbar$ (note that the normalization factor $\hbar^{-3/2}$ 
in $\chi({\bf k},t)$ disappears since the normalization integral for 
$\chi({\bf k},t)$ is now over $d^3{\bf k}$, not $d^3{\bf p}$). The 
quantities $\omega$ and ${\bf k}$ are now treated, respectively, as the 
frequency and the wave vector of the classical electromagnetic field, and 
the functions $\chi({\bf k},t)$ and $\psi({\bf r},t)$ are interpreted 
not such that they describe probabilities for a single photon but such that
they describe classical electromagnetic field ${\bf E}({\bf r},t)$ 
and ${\bf B}({\bf r},t)$ which can be constructed from these functions as
described in textbooks on QED (see e.g. Ref. \cite{AB}).

An additional argument in favor of the choice of $\psi({\bf r},t)$ as the coordinate photon WF
is that in classical electrodynamics the 
quantities ${\bf E}({\bf r},t)$ and ${\bf B}({\bf r},t)$ for the free 
field should satisfy the wave equation $\partial^2{\bf E}/c^2 \partial t^2=\Delta {\bf E}$ and 
analogously for ${\bf B}({\bf r},t)$. Hence if 
${\bf E}({\bf r},t)$ and ${\bf B}({\bf r},t)$ are constructed from $\psi({\bf r},t)$ as described
in textbooks (see e.g. Ref. \cite{AB}), they will satisfy the wave equation since, as follows from
Eqs. (\ref{Fourier},\ref{chiprel},\ref{chiptphoton}), $\psi({\bf r},t)$ also satisfies this equation.

The geometrical optics approximation
implies that if ${\bf k}_0$ and ${\bf r}_0$ are the mean values
of the wave vector and the spatial radius vector for a wave packet describing the electromagnetic wave then
the uncertainties $\Delta k$ and $\Delta r$, which are the mean values of $|{\bf k}-{\bf k}_0|$ and
$|{\bf r}-{\bf r}_0|$, respectively, should satisfy the requirements $\Delta k\ll |{\bf k}_0|$ and
$\Delta r\ll |{\bf r}_0|$. In full analogy with the derivation of Eq. (\ref{uncert}), one
can show that for each $j=1,2,3$ the uncertainties of the corresponding projections of the vectors
${\bf k}$ and ${\bf r}$ satisfy the requirement $\Delta k_j\Delta r_j\geq 1/2$
(see e.g. Ref. \cite{LLII}). In particular, an electromagnetic wave satisfies the approximation of geometrical optics
in the greatest possible extent if $\Delta k\Delta r$ is of the order of unity.

The above discussion confirms what has been mentioned in Sec. \ref{intropos} that {\it the effect of WPS in transverse directions takes place not only in quantum theory
but even in classical electrodynamics}. Indeed, since the function $\psi({\bf r},t)$ satisfies the classical wave equation,
the above consideration can be also treated as an example showing that {\it even for a free wave packet in classical
electrodynamics the WPS effect is inevitable}. In the language of classical waves the parameters of spreading can be
characterized by the function $a(t)$ (see Eq. (\ref{at})) and the quantities $t_*$ and $v_*$ such that in terms of the wave length $\lambda=2\pi c/\omega_0$ 
\begin{equation}
a(t)=a(1+\frac{\lambda^2c^2t^2}{4\pi^2a^4})^{1/2},\quad t_*=\frac{2\pi a^2}{\lambda c}, \quad 
v_*=\frac{\lambda c}{2\pi a}
\label{em}
\end{equation}
The last expression can be treated such that if $\lambda\ll a$ then the momentum has the angular
uncertainty of the order of $\alpha=\lambda/(2\pi a)$. This result is natural from the following consideration.
Let the mean value of the momentum be directed along the $z$-axis and the uncertainty of the transverse 
component of the
momentum be $\Delta p_{\bot}$. Then $\Delta p_{\bot}$ is of the order of $\hbar/a$, $\lambda=2\pi\hbar/p_0$ and 
hence $\alpha$ is of the order of 
$\Delta p_{\bot}/p_0\approx \lambda/(2\pi a)$. This is analogous to the known result in classical optics
that the best angular resolution of a telescope with the dimension $d$ is of the order of $\lambda/d$.
Another known result of classical optics is that if a wave encounters an obstacle having the dimension $d$
then the direction of the wave diverges by the angle of the order of $\lambda/d$. 

The inevitability of WPS for a free wave packet in classical electrodynamics is obvious from the following
consideration. Suppose that a classical wave packet does not have a definite value of the momentum. 
Then if $a$ is the initial width of the packet
in directions perpendicular to the mean momentum, one might expect that the width will grow as $a(t)=a+\alpha ct$
and for large values of $t$, $a(t)\approx \alpha ct$. As follows from Eq. (\ref{em}), if $t\gg t_*$ then indeed
$a(t)\approx \alpha ct$. In standard quantum theory we have the same result because the coordinate and momentum wave 
functions are related to each other by the same Fourier transform as the coordinate and ${\bf k}$ distributions
in classical electrodynamics. 

The quantity $N_{||}=b/\lambda$ shows how many oscillations the oscillating exponent in Eq. (\ref{final})
makes in the region where the WF or the amplitude of the classical wave is significantly different
from zero. As noted in Sec. \ref{classical}, for the validity of semiclassical approximation this quantity should be 
very large. In nonrelativistic quantum mechanics $a$ and $b$ are of the same order and hence the same can be said
about the quantity $N_{\bot}=a/\lambda$. As noted above, in the case of the photon we do not know the relation between $a$
and $b$. In terms of the quantity $N_{\bot}$ we can rewrite the expressions for $t_*$ and $v_*$ in Eq. (\ref{em}) as
\begin{equation}
t_*=2\pi N_{\bot}^2 T,\quad v_*=\frac{c}{2\pi N_{\bot}}
\label{tv}
\end{equation}
where $T$ is the period of the classical wave. Hence the accuracy of semiclassical approximation 
(or the geometrical optics approximation in classical
electrodynamics) increases with the increase of $N_{\bot}$.

In Ref. \cite{Dillon} the problem of WPS for classical electromagnetic waves has been discussed 
in the Fresnel approximation for a two-dimensional wave
packet. Equation (25) of Ref. \cite{Dillon} is a special case of Eq. (\ref{pperp}) and the author of 
Ref. \cite{Dillon} shows
that, in his model the wave packet spreads out in the direction perpendicular to the group velocity of the
packet. As noted in the preceding section,
in the ultrarelativistic case the function $a(t)$ is given by the same expression as in the nonrelativistic case
but $m$ is replaced by $E/c^2$. Hence if the results of the preceding section are reformulated in terms of
classical waves then $m$ should be replaced by $\hbar\omega_0/c^2$ and this fact has been pointed out in Ref. \cite{Dillon}.

\section{Wave packet width paradox}
\label{WPW}

We now consider the following important question. We assume that a classical wave packet is a collection of photons.
Let $a_{cl}$ be the quantity $a$ for the classical packet and $a_{ph}$ be a typical value of $a$ for the photons.
What is the relation between $a_{cl}$ and $a_{ph}$? 

My observation is that physicists answer this question in different ways. Quantum physicists usually say  
that in typical situations $a_{ph}\ll a_{cl}$ because $a_{cl}$ is of macroscopic size while in semiclassical
approximation the quantity $a_{ph}$ for each photon can be treated as the size of the region where the photon has 
been created. On the other hand, classical physicists usually say that
$a_{ph}\gg a_{cl}$ and the motivation follows.

Consider a decomposition of some component of classical electromagnetic field into the Fourier series: 
\begin{equation}
A(x)=\sum_{\sigma} \int [a({\bf p},\sigma)u({\bf p},\sigma)exp(-ipx)+
a({\bf p},\sigma)^*u({\bf p},\sigma)^*exp(ipx)]d^3{\bf p}
\label{Acont}
\end{equation}
where $\sigma$ is the polarization, $x$ and $p$ are the four-vectors such that $x=(ct,{\bf x})$ and $p=(|{\bf p}|c,{\bf p})$,
the functions $a({\bf p},\sigma)$ are the same for all the components, the functions $u({\bf p},\sigma)$
depend on the component and $^*$ is used to denote the complex conjugation. Then photons arise as a result of quantization 
when $a({\bf p},\sigma)$ and $a({\bf p},\sigma)^*$ are understood
not as usual function but as operators of annihilation and creation of the photon with the quantum numbers
$({\bf p},\sigma)$ and $^*$ is now understood as Hermitian conjugation. Hence the photon is described by a plane wave 
which has the same magnitude in all points of the
space. In other words, $a_{ph}$ is infinitely large and a finite width of the classical wave packet arises as a result
of interference of different plane waves. 

The above definition of the photon has at least the following inconsistency. If the photon is treated as a particle
then its WF should be normalizable while the plane wave is not normalizable. In textbooks this problem
is often circumvented by saying that we consider our system in a finite box. Then the spectrum of momenta becomes
finite and instead of Eq. (\ref{Acont}) one can write
\begin{equation}
A(x)=\sum_{\sigma} \sum_j [a({\bf p}_j,\sigma)u({\bf p}_j,\sigma)exp(-ip_jx)+a({\bf p}_j,\sigma)^*
u({\bf p}_j,\sigma)^*exp(ip_jx)]
\label{Adiscr}
\end{equation}
where $j$ enumerates the points of the momentum spectrum.

One can now describe quantum electromagnetic field by states in the Fock space where the vacuum vector $\Phi_0$ 
satisfies the condition $a({\bf p}_j,\sigma)\Phi_0=0$, $||\Phi_0||=1$ and the operators commute as
\begin{equation}
[a({\bf p}_i,\sigma_k),a({\bf p}_j,\sigma_l)]=[a({\bf p}_i,\sigma_k)^*,a({\bf p}_j,\sigma_l)^*]=0,
\quad [a({\bf p}_i,\sigma_k),a({\bf p}_j,\sigma_l)^*]=\delta_{ij}\delta_{kl}
\label{adiscr}
\end{equation}
Then any state can be written as
\begin{equation}
\Psi=\sum_{n=0}^{\infty}\sum_{\sigma_1...\sigma_n}\sum_{{\bf p}_1,...{\bf p}_n} 
\chi({\bf p}_1,\sigma_1,...{\bf p}_n,\sigma_n)
a({\bf p}_1,\sigma_1)^*\cdots a({\bf p}_n,\sigma_n)^* \Phi_0
\label{Psidiscrt}
\end{equation}

Classical states are understood such that although the number of photons is large, it is much less than the number
of possible momenta and in Eq. (\ref{Psidiscrt}) all the photons have different momenta (this is analogous to
the situation in classical statistics where mean occupation numbers are much less than unity).
Then it is not important whether the operators $(a,a^*)$ commute or anticommute. However, according to the Pauli theorem on
spin-statistics connection \cite{Pauli2},
they should commute and this allows the existence of coherent states where many photons 
have the same quantum numbers. Such states can be created in lasers and they are not described by classical
electrodynamics. In the next section we consider position operator for coherent states while in this section we
consider only quantum description of states close to classical.

Note that even in some textbooks on quantum optics (see e.g. Ref. \cite{Mandel}) classical and quantum states
are characterized in the opposite way: it is stated that classical states are characterized by large occupation numbers
while quantum states - by small ones. The question what states should be called classical or quantum is not a matter of
convention since in quantum theory there are rigorous criteria for that purpose. In particular, as explained
in textbooks on quantum theory, the exchange interaction is a pure quantum phenomenon which does not have classical
analogs. That's why the Boltzmann statistics (which works when mean occupation numbers are much less than unity
and the exchange interaction is negligible) is classical while the Fermi-Dirac and Bose-Einstein statistics (which
work when mean occupation numbers are of the order of unity or greater and the exchange interaction is important)
are quantum.  

The next problem is that one should take into account that in standard theory the photon momentum spectrum is
continuous. Then the above construction can be generalized as follows. The vacuum state $\Phi_0$ satisfies the
same conditions $||\Phi_0||=1$ and $a({\bf p},\sigma)\Phi_0=0$ while the operators $(a,a^*)$ satisfy the following
commutation relations
\begin{equation}
[a({\bf p},\sigma),a({\bf p}',\sigma')]=[a({\bf p},\sigma)^*,a({\bf p}',\sigma')^*]=0,\quad  
[a({\bf p},\sigma),a({\bf p}',\sigma')^*]=\delta^{(3)}({\bf p}-{\bf p}')\delta_{\sigma\sigma'}
\label{photonaa*}
\end{equation}
Then a general quantum state can be written as
\begin{equation}
\Psi=\sum_{n=0}^{\infty}\sum_{\sigma_1...\sigma_n}\int ...\int \chi({\bf p}_1,\sigma_1,...{\bf p}_n,\sigma_n)
a({\bf p}_1,\sigma_1)^*\cdots a({\bf p}_n,\sigma_n)^* d^3{\bf p}_1\cdots d^3{\bf p}_n \Phi_0
\label{manyphoton}
\end{equation}

In the approximation when a classical wave packet is understood as a collection of independent photons
(see the discussion in Sec. \ref{Discussion}), the state of this packet has the form 
\begin{equation}
\Psi=\sum_{n=0}^{\infty}c_n \prod_{j=1}^n \{\sum_{\sigma_j}\int\chi_j({\bf p}_j,\sigma_j)
a({\bf p}_j,\sigma_j)^*d^3{\bf p}_j\}
\Phi_0
\label{indepphotons}
\end{equation}
where $\chi_j$ is the WF of the $j$th photon and intersections of supports of WFs of different photons
can be neglected. This is an analog of the above situation with the discrete case where it is assumed that 
different photons in a classical wave packet have different momenta. In other words, while the WF of
each photon can be treated as an interference of plane waves, different photons can interfere only
in coherent states but not in classical wave packets. 

We now describe a known generalization of the results on IRs of the Poincare algebra to the description in the Fock space. 
If $A$ is an operator in the space of the photon IR then a generalization of this operator to the case of the
Fock space can be constructed as follows. Any operator in the space of IR can be represented as an integral
operator acting on the WF as
\begin{equation}
A\chi({\bf p},\sigma)=\sum_{\sigma'}\int A({\bf p},\sigma,{\bf p}',\sigma')\chi({\bf p}',\sigma')d^3{\bf p}'
\label{integraloper}
\end{equation} 
For example, if ${\bf A}\chi({\bf p},\sigma)=\partial \chi({\bf p},\sigma)/\partial{\bf p}$ then ${\bf A}$ is
the integral operator with the kernel 
$${\bf A}({\bf p},\sigma,{\bf p}',\sigma')=\frac{\partial 
\delta^{(3)}({\bf p}-{\bf p}')}{\partial{\bf p}}\delta_{\sigma\sigma'}$$
We now require that if the action of the operator $A$ in the space of IR is defined by Eq. (\ref{integraloper})
then in the case of the Fock space this action is defined as
\begin{equation}
A=\sum_{\sigma\sigma'}\int A({\bf p},\sigma,{\bf p}',\sigma')a({\bf p},\sigma)^*a({\bf p}',\sigma')
d^3{\bf p}d^3{\bf p}'
\label{Fockoper}
\end{equation}
Then it is easy to verify that if $A$, $B$ and $C$ are operators in the space of IR satisfying the commutation
relation $[A,B]=C$ then the generalizations of these operators in the Fock space satisfy the same commutation relation.
It is also easy to verify that the operators generalized to the action in the Fock space in such a way are additive,
i.e. for a system of $n$ photons they are sums of the corresponding single-particle operators. In particular, the
energy of the $n$-photon system is a sum of the energies of the photons in the system and analogously for the other
representation operators of the Poincare algebra - momenta, angular momenta and Lorentz boosts.

We are interested in calculating mean values of different combinations of the momentum operator. Since this operator
does not act over spin variables, we will drop such variables in the $(a,a^*)$ operators and in the functions $\chi_j$.
Then the explicit form of the momentum operator is 
${\bf P}=\int {\bf p} a({\bf p})^*a({\bf p})d^3{\bf p}$. Since this operator does not change the number of photons,
the mean values can be independently calculated in each subspace where the number of photons is $N$.

Suppose that the momentum of each photon is
approximately directed along the $z$-axis and the quantity $p_0$ for each photon approximately equals
$2\pi\hbar/\lambda$. If $\Delta p_{\bot}$ is a typical uncertainty of the transversal component of the
momentum for the photons then a typical value of the angular uncertainty for the photons is 
$\alpha_{ph}=\Delta p_{\bot}/p_0\approx \lambda/(2\pi a_{ph})$. The total momentum of the classical wave packet
consisting of $N$ photons is a sum of the photon momenta: ${\bf P}=\sum_{i=1}^N {\bf p}^{(i)}$. 
Suppose that the mean value of ${\bf P}$
is directed along the $z$-axis and its magnitude $P_0$ is such that $P_0\approx Np_0$. The uncertainty of the
$x$ component of ${\bf P}$ is $\Delta P_x={\overline{P_x^2}}^{1/2}$ where
$${\overline{P_x^2}}=\sum_{i=1}^N \overline{(p_x^{(i)})^2}+\sum_{i\neq j;i,j=1}^N\overline{p_x^{(i)}p_x^{(j)}}$$
Then in the approximation of independent photons (see the remarks after Eq. (\ref{indepphotons}))
$${\overline{P_x^2}}=\sum_{i=1}^N \overline{(p_x^{(i)})^2}+\sum_{i\neq j;i,j=1}^N\overline{p_x^{(i)}}\cdot
\overline{p_x^{(j)}}=\sum_{i=1}^N [\overline{(p_x^{(i)})^2}-\overline{p_x^{(i)}}^2]=
\sum_{i=1}^N (\Delta p_x^{(i)})^2$$
where we have taken into account that $\overline{P_x}=\sum_{i=1}^N\overline{p_x^{(i)}}=0$.

As a consequence, if typical values of $\Delta p_{\bot}^{(i)}$ have the the same order of magnitude equal 
to $\Delta p_{\bot}$
then $\Delta P_{\bot}\approx N^{1/2} \Delta p_{\bot}$ and the angular divergence of 
the classical vave packet is 
\begin{equation}
\alpha_{cl}=\Delta P_{\bot}/P_0\approx \Delta p_{\bot}/(p_0N^{1/2})=\alpha_{ph}/N^{1/2}
\label{acl}
\end{equation}
Since the classical wave packet is described by the same wave equation as the photon WF, its angular
divergence can be expressed in terms of the parameters $\lambda$ and $a_{cl}$ such that 
$\alpha_{cl}=\lambda/(2\pi a_{cl})$. Hence $a_{cl}\approx N^{1/2}a_{ph}$ and we conclude that $a_{ph}\ll a_{cl}$.

Note that in this derivation no position operator has been used. Although the quantities $\lambda$ and $a_{ph}$ have
the dimension of length, they are defined only from considering the photon in momentum space because, as noted in Sec.
\ref{momentum}, for individual photons $\lambda$ is understood only as $2\pi\hbar/p_0$, $a_{ph}$ defines the
width of the photon momentum WF (see Eq. (\ref{chiprel})) and is of the order of $\hbar/\Delta p_{\bot}$.
As noted in Secs. \ref{NRWPS} and \ref{RelWPS}, the momentum distribution does not depend on time and hence the
result $a_{ph}\ll a_{cl}$ does not depend on time too. If photons in a classical wave packet could be treated as 
(almost) pointlike particles then photons do not experience WPS while the WPS effect for a classical wave packet 
is a consequence of the fact that different photons in the packet have different momenta. 

However, in standard quantum theory this scenario does not take place for the following reason. 
Let $a_{cl}(t)$ be the quantity $a(t)$ for the classical wave packet and $a_{ph}(t)$ be a typical value of the
quantity $a(t)$ for individual photons. With standard position operator the quantity $a_{ph}(t)$ is interpreted
as the spatial width of the photon coordinate WF in directions perpendicular to the photon
momentum and this quantity is time dependent. As shown in Secs. \ref{RelWPS} and \ref{geom}, $a(0)=a$ but if 
$t\gg t_*$ then $a(t)$ is {\it inversely proportional} to $a$ and the coefficient of
proportionality is the same for the classical wave packet and individual photons (see Eq. (\ref{em})). 
Hence {\it in standard quantum theory we have a paradox that after some period of time} $a_{ph}(t)\gg a_{cl}(t)$
i.e. individual photons in a classical wave packet spread out in a much greater extent than the wave packet
as a whole. We call this situation the wave packet width (WPW) paradox 
(as noted above, different photons in
a classical wave packet do not interfere with each other). The reason of the paradox is obvious:
if the law that the angular divergence of a wave packet is of the order of $\lambda/a$ is applied to both,
a classical wave packet and photons comprising it then the paradox follows from the fact that the quantities $a$ for
the photons are much less than the quantity $a$ for the classical wave packet. Note that in classical case the 
quantity $a_{cl}$ does not have the meaning of $\hbar/\Delta P_{\bot}$ and $\lambda$ is not equal to $2\pi\hbar/P_0$.

\section{Wave packet spreading in coherent states}
\label{coherent}

In textbooks on quantum optics the laser emission is described by the following model (see e.g. 
Refs. \cite{Mandel,Scully}). 
Consider a set of photons having the same momentum ${\bf p}$ and polarization $\sigma$ and, by analogy with the discussion
in the preceding section, suppose that the momentum spectrum is discrete. 
Consider a quantum superposition $\Psi=\sum_{n=0}^{\infty} c_n [a({\bf p},\sigma)^*]^n\Phi_0$ where the coefficients $c_n$ satisfy 
the condition that $\Psi$ is an eigenstate of the annihilation operator $a({\bf p},\sigma)$. Then 
the product of the coordinate and momentum uncertainties has the minimum possible value $\hbar/2$ and, as 
noted in Sec. \ref{classical}, such a state is called coherent. However, the term coherent is sometimes used meaning
that the state is a quantum superposition of many-photon states $[a({\bf p},\sigma)^*]^n\Phi_0$.

In the above model it is not taken into account that (in standard theory) photons emitted by a laser can have only a
continuous spectrum of momenta. Meanwhile for the WPS effect the width of the momentum
distribution is important. In this section we consider a generalization of the above model where the fact that photons 
have a continuous spectrum of momenta is taken into account. This will make it possible to consider the WPS effect
in coherent states.

In the above formalism coherent states can be defined as follows. We assume that all the photons in the state  
Eq. (\ref{manyphoton}) have the same polarization. Hence for describing such states we can drop the quantum number
$\sigma$ in WFs and $a$-operators. We also assume that all photons in coherent states have the same momentum
distribution. These conditions can be satisfied by requiring that coherent states have the form 
\begin{equation}
\Psi=\sum_{n=0}^{\infty}c_n[\int\chi({\bf p})a({\bf p})^*d^3{\bf p}]^n \Phi_0
\label{cohstate}
\end{equation} 
where $c_n$ are some coefficients. Finally, by analogy with the description of coherent states in standard textbooks
on quantum optics one can require that they are eigenstates of the operator $\int a({\bf p})d^3{\bf p}$. 

The dependence of the state $\Psi$ in Eq. (\ref{cohstate}) on $t$ is $\Psi(t)=exp(-iEt/\hbar)\Psi$ where, as follows
from Eqs. (\ref{IRoperators}) and (\ref{Fockoper}), the action of the energy operator in the Fock space is
$E=\int pc a({\bf p})^*a({\bf p})d^3{\bf p}$. Since $exp(iEt/\hbar)\Phi_0=\Phi_0$, it readily follows from
Eq. (\ref{photonaa*}) that 
\begin{equation}
\Psi(t)=\sum_{n=0}^{\infty}c_n[\int\chi({\bf p},t)a({\bf p})^*d^3{\bf p}]^n \Phi_0
\label{cohstatet}
\end{equation}
where the relation between $\chi({\bf p},t)$ and $\chi({\bf p})=\chi({\bf p},0)$ is given by Eq. (\ref{chiptphoton}). 

A problem arises how to define the position operator in the Fock space. If this operator is defined by analogy with
the above construction then we get an unphysical result that each coordinate of the $n$-photon system as a whole is
a sum of the corresponding coordinates of the photons in the system. This is an additional argument that the position
operator is less fundamental than the representation operators of the Poincare algebra and its action should be 
defined from additional considerations. In textbooks on quantum optics the position operator for coherent states 
is usually defined by analogy with the position operator in nonrelativistic quantum mechanics for the harmonic
oscillator problem. The motivation follows. If the energy levels $\hbar\omega (n+1/2)$ of the harmonic oscillator
are treated as states of $n$ quanta with the energies $\hbar\omega$ then the harmonic oscillator problem can be described
by the operators $a$ and $a^*$ which are expressed in terms of the one-dimensional position and momentum operators
$q$ and $p$ as $a=(\omega q+ip)/(2\hbar\omega)^{1/2}$ and $a^*=(\omega q-ip)/(2\hbar\omega)^{1/2}$, respectively. However, as noted
above, the model description of coherent states in those textbooks is one-dimensional because the continuous nature
of the momentum spectrum is not taken into account. In addition, the above results on WPS give indications that 
the position operator in standard theory is not consistently defined. For all these reasons a problem arises whether
the requirement that the state $\Psi$ in Eq. (\ref{cohstate}) is an eigenvector of the operator 
$\int a({\bf p})d^3{\bf p}$
has a physical meaning. In what follows this requirement is not used.

In nonrelativistic classical mechanics the radius vector of a system of $n$ particles as a whole (the radius vector of
the center of mass) is defined as ${\bf R}=(m_1{\bf r}_1+...+m_n{\bf r}_n)/(m_1+...+m_n)$ and in works on 
relativistic classical mechanics it is usually defined as ${\bf R}=(\epsilon_1({\bf p}_1){\bf r}_1+...
+\epsilon_n({\bf p}_n){\bf r}_n)/(\epsilon_1({\bf p}_1)+...+\epsilon_n({\bf p}_n))$ where 
$\epsilon_i({\bf p}_i)=(m_i^2+{\bf p}_i^2)^{1/2}$. Hence if all the particles have the same masses and momenta,
${\bf R}=({\bf r}_1+...+{\bf r}_n)/n$.

These remarks make it reasonable to define the position operator for coherent states as follows. Let $x_j$ be the
$j$th component of the position operator in the space of IR and $A_j({\bf p},{\bf p}')$ be the kernel of this
operator. Then in view of Eq. (\ref{Fockoper}) the action of the operator $X_j$ on the state $\Psi(t)$ in 
Eq. (\ref{cohstate}) can be defined as
\begin{equation}
X_j\Psi(t)=\sum_{n=1}^{\infty}\frac{c_n}{n}\int\int A_j({\bf p}",{\bf p}')a({\bf p}")^*a({\bf p}')
d^3{\bf p}"d^3{\bf p}'[\int\chi({\bf p},t)a({\bf p})^*d^3{\bf p}]^n \Phi_0
\label{Xj}
\end{equation} 

If $\overline{x_j}(t)$ and $\overline{x_j^2}(t)$ are the mean values of the operators $x_j$ and $x_j^2$, 
respectively then as follows from the definition of the kernel of the operator $x_j$
\begin{eqnarray}
&&\overline{x_j}(t)=\int \int \chi({\bf p},t)^*A_j({\bf p},{\bf p}')\chi({\bf p}',t)d^3{\bf p}d^3{\bf p}'\nonumber\\
&&\overline{x_j^2}(t)=\int\int \int \chi({\bf p}",t)^*A_j({\bf p},{\bf p}")^*A_j({\bf p},{\bf p}')
\chi({\bf p}',t)d^3{\bf p}d^3{\bf p}"d^3{\bf p}'
\label{meanxj}
\end{eqnarray} 
and in the case of IR the uncertainty of the quantity $x_j$ is $\Delta x_j(t)=[\overline{x_j^2}(t)-\overline{x_j}(t)^2]^{1/2}$. At the same time, if $\overline{X_j}(t)$ and 
$\overline{X_j^2}(t)$ are the mean values of the operators $X_j$ and $X_j^2$, 
respectively then
\begin{equation}
\overline{X_j}(t)=(\Psi(t),X_j\Psi(t)),\quad \overline{X_j^2}(t)=(\Psi(t),X_j^2\Psi(t))
\label{meanXj}
\end{equation}
and the uncertainty of the quantity $X_j$ is $\Delta X_j(t)=[\overline{X_j^2}(t)-\overline{X_j}(t)^2]^{1/2}$.
Our goal is to express $\Delta X_j(t)$ in terms of $\overline{x_j}(t)$, $\overline{x_j^2}(t)$ and
$\Delta x_j(t)$. 

If the function $\chi({\bf p},t)$ is normalized to one (see Eq. (\ref{norm})) then, as follows from
Eq. (\ref{photonaa*}), $||\Psi(t)||=1$ if
\begin{equation}
\sum_{n=0}^{\infty}n! |c_n|^2=1
\label{normPsi}
\end{equation}
A direct calculation using Eqs. (\ref{photonaa*}), (\ref{Xj}), (\ref{meanxj}) and (\ref{meanXj}) gives
\begin{eqnarray}
&&\overline{X_j}(t)=\overline{x_j}(t)\sum_{n=1}^{\infty}n! |c_n|^2\nonumber\\
&&\overline{X_j^2}(t)=\sum_{n=1}^{\infty}(n-1)!|c_n|^2[\overline{x_j^2}(t)+(n-1)\overline{x_j}(t)^2]
\label{meanxjXj}
\end{eqnarray}
It now follows from Eq. (\ref{normPsi}) and the definitions of the quantities $\Delta x_j(t)$ and 
$\Delta X_j(t)$ that 
\begin{equation}
\Delta X_j(t)^2=(1-|c_0|^2)|c_0|^2\overline{x_j}(t)^2+\sum_{n=1}^{\infty}(n-1)!|c_n|^2\Delta x_j(t)^2
\label{DeltaXj}
\end{equation}
  
Equation (\ref{DeltaXj}) is the key result of this section. It has been derived without using
a specific choice of the single photon position operator. The consequence of this result follows. If the
main contribution to the state $\Psi(t)$ in Eq. (\ref{cohstatet}) is given by
very large values of $n$ then $|c_0|$ is very small and the first term in this expression can be neglected. 
Suppose that the main contribution is given by terms where $n$ is of the order of ${\bar n}$. Then, as follows
from Eqs. (\ref{normPsi}) and (\ref{DeltaXj}), $\Delta X_j(t)$ is of the order of $\Delta x_j(t)/{\bar n}^{1/2}$.
This means that for coherent states where the main contribution is given by very large numbers of photons the effect
of WPS is pronounced in a much less extent than for single photons. 

It is interesting to note that the relation between $\Delta X_j(t)$ and $\Delta x_j(t)$ is analogous to
(\ref{acl}) although those relations describe fully difference situations. In both of them
relative uncertainties for a system of many particles are much less than for a single particle.
Since the WPS effect for photons in laser beams is very small, divergence of the laser beam is
only a consequence of the fact that different photons have different momenta.

\section{Experimental consequences of WPS in standard theory}
\label{experiment}

\subsection{Does light from stars consist of free photons?}
\label{freephotons}

The answer to this question depends on: 1) whether the interaction between the photons is
important; 2) whether their interaction with the interstellar medium is important and 
3) whether coherent photon states play an important role in the star radiation. 

As explained in standard textbooks on QED (see e.g. Ref. \cite{AB}), the photon-photon
interaction can go only via intermediate creation of virtual electron-positron or quark-antiquark pairs. If $\omega$ is
the photon frequency, $m$ is the mass of the charged particle in the intermediate state and $e$ is the electric charge
of this particle then in the case
when $\hbar\omega\ll mc^2$ the total cross section of the photon-photon interaction is \cite{AB}
\begin{equation}
\sigma=\frac{56}{5\pi m^2}\frac{139}{90^2}(\frac{e^2}{\hbar c})^4 (\frac{\hbar\omega}{mc^2})^6
\label{photonphoton}
\end{equation} 
For photons of visible light the quantities $\hbar\omega/(mc^2)$ and $\sigma$ are very small and for radio waves
they are even smaller by several orders of magnitude. At present the effect of the direct
photon-photon interaction has not been detected, and experiments with strong laser fields were only able to 
determine the upper limit of the cross section \cite{gammagamma}. Therefore in the star radiation the interaction between the photons is negligible.

As far as item 2) is concerned, one can note the following.  
The problem of explaining the redshift phenomenon has a long history. Different competing approaches can be 
divided into two big sets which we call Theory A and Theory B. In Theory A the redshift has been originally explained 
as a manifestation of the Doppler effect but in recent years  the cosmological and gravitational redshifts 
have been added to the consideration. In this theory the interaction of photons with the interstellar medium is treated as practically not important. On the contrary, in Theory B, which is often called the tired-light theory, the interaction of photons with the interstellar medium is treated as the main reason for the redshift. The majority of physicists 
believe that Theory A explains the astronomical data better than Theory B because  
any sort of scattering of light would predict more blurring than is seen (see e.g.
the article "Tired Light" in Wikipedia). 

On quantum level a process of propagation of photons in the medium is rather complicated because several
mechanisms of propagation should be taken into account. For example, a possible process is such that a
photon can be absorbed by an atom and reemitted. This process makes it
clear why the speed of light in the medium is less than $c$: because the atom which absorbed the photon is
in excited state for some time before reemitting the photon. However, this process is also important from the
following point of view: even if the coordinate photon WF had a large width before absorption,
as a consequence of the collapse of the WF, the WF of the emitted
photon will have in general much smaller size since after detection the width is
defined only by parameters of the corresponding detector. If the photon encounters many atoms on its way,
this process does not allow the photon WF to spread out significantly. Analogous remarks can
be made about other processes, for example about rescattering of photons on large groups of atoms, rescattering
on elementary particles if they are present in the medium etc. However, such processes have been discussed in
Theory B and, as noted above, they probably result in more blurring than is seen. 

The interaction of photons with the interstellar or interplanetary medium might also be important in view of
hypotheses that the density of the medium is much greater than usually believed. Among the most popular
scenarios are dark energy, dark matter etc. As shown in Ref. \cite{dark} and Chaps. \ref{Ch2} and \ref{twobody}, the phenomenon of the cosmological acceleration can be easily and naturally explained from first
principles of quantum theory without involving dark energy, empty space-background and other artificial notions. 
However, the other scenarios seem to be more realistic and one might expect that they will be intensively investigated. A rather hypothetical
possibility is that the propagation of photons in the medium has something in common with
the induced emission when a photon induces emission of other photons in practically the same direction.
In other words, the interstellar medium amplifies the emission as a laser. This possibility seems to be not 
realistic since it is not clear why the energy levels in the medium might be inverted.  
In view of these remarks, we accept Theory A where it is assumed that with a good accuracy we can treat photons 
as propagating in empty space. 

We now consider item 3). As noted in Sec. \ref{WPW}, a general form of the state vector of 
the electromagnetic field is given by Eq. (\ref{manyphoton}). It has been also noted that 
the two extreme cases of the state vector are as follows. The first case, which can be
called strongly incoherent, is such that the functions 
$\chi_n({\bf p}_1,\sigma_1,...{\bf p}_n,\sigma_n)$ have sharp maxima at
${\bf p}_i={\bf p}_i^0\,\, (i=1,...n)$ and all the values ${\bf p}_i^0$ are considerably
different. The second case, which can be called strongly coherent, is defined by Eq.
(\ref{cohstate}).

The density of radiation coming to us from distant stars is very small. Therefore the
assumption  that this radiation can be described in the framework of classical
electrodynamics is problematic and one might think that this radiation can be treated simply as a
collection of independent photons. However, one of the arguments in favor of this assumption is the
Hanbury Brown and Twiss experiment. Here two photomultiplier tubes separated by about 6 meters, were aimed at Sirius and excellent angular resolution has been achieved. The theoretical explanation of the
experiment (see e.g. Ref. \cite{Scully}) can be given both, in terms of classical optics and
in terms of interference of independent photons (which are not in coherent states). 
In the latter case it is also assumed that radiation from Sirius has the blackbody type.

However, Sirius is the brightest star on the sky and its distance to Earth is "only" 8.6 light years.
The angular resolution of Sirius is at the limit of modern telescope arrays and this resolution is
insufficient for determining radii of other stars.  Conclusions about them are made from the data on luminosity and temperature assuming that the major part of the radiation can be described in
the blackbody model. Therefore even if a conclusion about radiation from Sirius is valid,
this does not mean that the same conclusion is valid for radiation from other stars.

Nevertheless we now consider what conclusions about the 
structure of the states $\Phi(t)$ can be made if we accept that
radiation from distant stars is classical. As noted in Sec. \ref{WPW}, the operator of the vector potential 
has the form (\ref{Acont}).  Therefore the tensor of the electromagnetic field can be written as
\begin{eqnarray}
&&F_{\mu\nu}(x)=\frac{\partial A_{\nu}(x)}{\partial x^{\mu}}-
\frac{\partial A_{\mu}(x)}{\partial x^{\nu}}=\nonumber\\
&&-i\sum_{\sigma}\int [exp(-\frac{i}{\hbar}px)
(p_{\nu}e({\bf p},\sigma)_{\mu}-p_{\mu}e({\bf p},\sigma)_{\nu})a({\bf p},\sigma)-\nonumber\\
&&exp(\frac{i}{\hbar}px)(p_{\nu}e({\bf p},\sigma)_{\mu}-p_{\mu}e({\bf p},\sigma)_{\nu})^*
a({\bf p},\sigma)^*]\frac{d^3{\bf p}}{[(2\pi)^3|{\bf p}|]^{1/2}}
\end{eqnarray}
Since in classical case the mean values of the operators $F_{\mu\nu}(x)$ should not be zero,
it follows from this expression and Eq. (\ref{photonaa*}) that classical states of the electromagnetic
field cannot be states with a fixed number of photons (this fact has been pointed out by the
referee of Ref. \cite{Paradox}). Such states are complex superpositions of states with different numbers of photons.

However, this observation is not sufficient for making a conclusion on whether classical
states are coherent or noncoherent. In Sec. \ref{WPW} we noted that  the Boltzmann statistics is
classical while the Fermi-Dirac and Bose-Einstein statistics are quantum.  
 For pedagogical purposes it is often assumed that the electromagnetic field
is confined within a large finite volume and then the spectrum of momenta becomes discrete.
Then one can work with occupation numbers ${\bar n}({\bf p})$ characterizing mean values of photons
with the momentum ${\bf p}$. For the Boltzmann statistics those numbers are much less than 1
while for the Fermi-Dirac and Bose-Einstein statistics those numbers are of the order of 1 or greater.

Let us note that the Boltzmann statistics is only a necessary but not sufficient condition for the 
electromagnetic field to be classical. In QED, ${\bf E}(x)$ and ${\bf B}(x)$ are the operators in the Fock
space and those operators and not the states satisfy the Maxwell equations. If those operators are
semiclassical then the electromagnetic field is classical because in that case the mean values of the operators
satisfy the Maxwell equations and the uncertainties of the operators are small. The exact requirements on
the numbers of photons and their WFs to guarantee that the field is classical are not known but
intuitively one might think that the density of photons should not be small. That's why, as mentioned
above, it is not clear whether or not the radiation of stars is classical. However, for our purposes it is
important only that, as argued below, photons in the radiation are independent.

In laser emission the number of photons is
much greater than the number of possible states. From the point of view of analogy with
the quantum mechanical oscillator problem the laser states are called coherent if the coefficients
$c_n$ in Eq. (\ref{cohstate}) are such that 
\begin{equation}
a({\bf p},\lambda)\sum_{n=0}^{\infty}c_n[a({\bf p},\lambda)^*]^n\Phi_0=0
\end{equation}
Laser emission can be created only at very special conditions when energy
levels are inverted, the emission is amplified in the laser cavity etc. There are no reasons
to think that such conditions exist on stars. A part of the star
radiation consists of photons emitted from different atomic energy levels and this radiation
is fully spontaneous rather than induced. At the same
time, the main part of the radiation is understood such that it can
be approximately described in the blackbody model. 
Several authors (see e.g. Ref. \cite{Letokhov} and references therein)
discussed a possibility that at some conditions the inverted population and amplification of radiation in stellar
atmospheres might occur and so a part of the radiation can be induced. However, at present this possibility
is not widely accepted. 

A theoretical model describing blackbody radiation (see e.g. Ref. \cite{LLV}) is such that photons are treated
as an ideal Bose gas weakly interacting with matter and such that typical photon energies are not close to  
energies of absorption lines for that matter (hence the energy spectrum of photons is almost continuous). 
It is also assumed that the photons are distributed over states with definite values of momenta. With these 
assumptions one can derive the famous
Planck formula for the spectral distribution of the blackbody radiation (this formula is treated as
marking the beginning of quantum theory). 

As shown in textbooks, the occupation numbers for the blackbody radiation are given by
${\bar n}({\bf p})=1/[exp(E({\bf p})/kT)-1]$ where $E({\bf p})=|{\bf p}|c$, $k$ is the Boltzmann
constant and $T$ is the temperature of the blackbody radiation. If $E\gg kT$ then we get the
result for the Boltzmann statistics with zero chemical potential: ${\bar n}({\bf p})=exp(-kT/E)$.
The energy spectrum of the blackbody radiation has the maximum at $E/kT \approx 2.822$ \cite{LLV}.
Hence in the region of maximum ${\bar n}({\bf p})\approx 0.063$ and the result is close to that given by
the Boltzmann statistics.

When photons emitted by a star leave the area of the black body, their distribution differs from
the blackbody one. As argued, for example, in Ref. \cite{LLV}, this distribution can be described by the
Liouville theorem; in particular this implies that the photons leaving stars are moving
along classical trajectories. However, in any case, since the occupation numbers are small then
$\Phi(t)$ is a superposition of states where photons are independent of each other.
Therefore the effect of WPS for them can be considered for each photon independently
and  the formalism considered in Sec. \ref{RelWPS} applies.

\subsection{Typical properties of photons in the star radiation}

Since wave lengths of such photons are typically much less than all characteristic dimensions
in question one
might think that the radiation of stars can be described in the geometrical optics approximation. As
discussed in Sec. \ref{geom}, this approximation is similar to semiclassical approximation in quantum
theory. This poses a question whether this radiation can be approximately treated as a collection
of photons moving along classical trajectories. 

Consider, for example, the Lyman transition $2P\to 1S$ in the hydrogen atom, which plays
an important role in the star radiation. We first consider the case when the atom is at rest.
Then the mean energy of the photon is $E_0=10.2eV$, its wave length is $\lambda=121.6nm$ and
the lifetime is $\tau=1.6\cdot 10^{-9}s$. The phrase that the 
lifetime is $\tau$ is interpreted such that the uncertainty of the energy is
$\hbar/\tau$. This implies that the uncertainty of the momentum magnitude is $\hbar/c\tau$ 
and $b$ is of the order of $c\tau\approx 0.48m$.
In this case the photon has a very narrow energy distribution since the mean value of the momentum
$p_0=E_0/c$ satisfies the condition $p_0b\gg \hbar$. At the same time, since the orbital angular momentum of the photon
is a small quantity, the function $f(\theta)=f({\bf p}/p)$ in Eq. (\ref{FG}) has the same order of magnitude at all angles and the direction of the photon momentum cannot be semiclassical. If the atom is not
at rest those conclusions remain valid because typically the speed of the atom is much less than $c$.

As pointed out in Sec. \ref{RelWPS}, it follows from Eq. (\ref{relpsif}) that even if 
the function $f(\theta)$ describes a broad 
angular distribution, the star will be visible only in the angular range of the order of
$R/L$ where $R$ is the radius of the star and $L$ is the distance to the star.
The experimental verification of this prediction is problematic since the quantities $R/L$ 
are very small and, as noted above, at present star radii cannot be measured directly. 
In that case we cannot estimate the quantity $b$ as above and it is 
not clear what criteria can be used for estimating the quantity $a$. 
The estimation $a\approx b\approx 0.48m$ seems to be
extremely favorable since one might expect that the value of $a$ is of the order of atomic size, i.e. 
much less than $0.48m$. With this
estimation for yellow light (with $\lambda=580nm$) $N_{\bot}=a/\lambda\approx 8\cdot 10^5$. So the value of $N_{\bot}$ is rather large 
and in view of Eq. (\ref{tv}) one might think that the effect of spreading is not important. 

However, this
is not the case because, as follows from Eq. (\ref{tv}), $t_*\approx 0.008s$. Even in the case of
the Sun the distance to the Earth is approximately $t=8$ light minutes,
 and this time is much greater than $t_*$. Then the value of $a(t)$
(which can be called the half-width of the wave packet) when the packet arrives to the Earth is $v_*t\approx 28km$. 
In this case standard geometrical interpretation does not apply. In addition, if we assume that the initial value of $a$ is of the order of several wave lengths then the value of $N_{\bot}$ is much less and the width of the wave packet 
coming to the Earth even from the Sun is much greater. An analogous estimation shows that even in the favorable scenario the half-width 
of the wave packet coming to the Earth 
from Sirius will be approximately equal to $15\cdot 10^6km$ but in less favorable situations the half-width will
be much greater. Hence we come to the conclusion that even in favorable scenarios the assumption that 
photons are moving along classical trajectories does not
apply and a problem arises whether or not this situation is in agreement with experiment. 

\subsection{Fundamental quantal paradox: standard choice of position operator contradicts observations of stars}
\label{paradox}

As already noted, even if the function $f(\theta)$ describes a broad 
angular distribution, a star will be visible only in the angular range of the order of
$R/L$. Hence one might think
that the absence of classical trajectories does not contradict observations. We now consider this
problem in greater details. For simplicity we first assume that the photon WF is spherically symmetric, i.e. $f({\bf r}/r)=const$.

As follows from Eqs. (\ref{FG}) and (\ref{relpsif}), the WF of the photon coming to Earth
from a distant star is not negligible only within a thin sphere with the radius $ct$ and the width
of the order of $b$. 
On its way to Earth the sphere passes {\it all} stars, planets and other objects
the distance from which to the star is less than $L$ (in particular, even those objects which are from
the star in directions opposite to the direction to Earth). 
A problem arises how to explain the fact that the photon was detected on Earth and
escaped detection by those stars, planets etc. 

One might think that the event when the photon was detected on Earth is
purely probabilistic. The fact that the photon was not detected by the objects on its way
to Earth can be explained such that since the photon WF has a huge size (of the
order of light years or more) the probability of detection even by stars is extremely small
and so it was only a favorable accident that the photon was detected on Earth.

However, if the photon passed stars, planets
and other objects on its way to Earth then with approximately the same probability it can
pass Earth and can be detected on the opposite side of the Earth. In that case we could 
see stars even through the Earth.

Moreover, consider the following experiment. Suppose that we first look at a star and then place 
a small screen between the eye and the star. Then the experiment shows that the star will not
be visible. However, since the photon WF passed many big objects without
interacting with them then with approximately the same probability it can pass the screen.
In that case we could see the star through the screen. 

Those phenomena are not unusual 
in view of our understanding of neutrino physics.
It is known that neutrinos not only can pass the Earth practically without problems but even neutrinos created in the
center of the Sun can easily reach the Earth. The major neutrino detectors are
under the Earth surface and, for example, in the OPERA and ICARUS experiments
neutrinos created at CERN reached Gran Sasso (Italy) after traveling 730km under
the Earth surface. The explanation is that the probability of the neutrino interaction 
with particles comprising the Sun and the Earth is very small.

At low energies the electromagnetic interaction is much stronger than the weak one but,
as follows from the discussion in Secs. \ref{Mott} and \ref{RelWPS}, the probability of interaction 
for photons having cosmic sizes contains the factor $|{\tilde f}/f|^2=(d/D)^2$. 
Therefore it is reasonable to expect that for such photons the 
probability of interaction with particles comprising an object is even much less
than in the above experiments with neutrinos.

In my discussions with physicists some of them proposed to avoid the above paradoxes by using an 
analogy with classical diffraction theory.  Here it is assumed that in optical phenomena
a wave falling on an object cannot penetrate inside the object. Then  
 the wave far from the object does not change, right after the object the wave has
a hole but when its length is much greater than the Rayleigh one the hole disappears and
the wave is practically the same as without diffraction. Those results 
are natural from the point of view
that classical waves consist of many almost pointlike particles. 

Let us now consider an experiment where a photon 
encounters a classical object and the transversal width of the photon coordinate WF
is much greater than the size of the object. By analogy with
diffraction theory one might represent
the photon WF as $\psi=\psi'+\psi''$ where the support of $\psi'$ is outside the
object and the support of $\psi''$ is inside the object. In contrast to diffraction theory, 
this decomposition is ambiguous because coordinates of the object have uncertainties. 
However, one might assume that the decomposition is valid with some accuracy.
Then one might expect that 
after interaction with the object $\psi'$ will not change and  $\psi''$ will be 
absorbed by the object. This statement can be formalized as follows.

Let the object be initially in the ground state $\Psi_g$. Then the initial WF of the system
photon+object is $\psi \Psi_g$. The S-matrix acts on this state as 
\begin{equation}
S(\psi \Psi_g)=S(\psi' \Psi_g)+S(\psi'' \Psi_g)=\psi' \Psi_g+(...)
\label{Greg}
\end{equation}
where (...) consists of states emerging after interaction. 
This expression describes the situation when the photon always interacts with the object but only a small 
part $\psi''$ of the initial WF interacts while the major part $\psi'$ remains intact.  
As a result of interaction, the photon
will be either absorbed by the object or will pass the object. In the latter case the photon
WF $\psi'$ will have a hole by analogy with the behavior of waves after diffraction.
Therefore in any case the photon cannot be detected in the geometrical shadow of the object.

Understanding whether or not Eq. (\ref{Greg}) is acceptable is crucial 
for drawing a conclusion on the above paradoxes. This expression can be justified if evolution is described by a Hamiltonian where interaction of the photon with the object is local. 
As noted in Sec. \ref{WPW}, in fundamental quantum theories elementary particles are described by states in the Fock space, the annihilation and creation operators for a
photon satisfy Eq. (\ref{photonaa*}) and, as a consequence of Eq. (\ref{manyphoton}),
the one-photon state in the Fock space is $\Phi_1=\int \chi({\bf p})a({\bf p})^*d^3{\bf p} \Phi_0$
where spin indices are suppressed.

If $\psi$ is the
coordinate WF defined by Eq. (\ref{Fourier}) then the decomposition $\psi=\psi'+\psi''$ corresponds
to the decomposition $\chi=\chi'+\chi''$ where 
\begin{equation}
\chi'({\bf p})=\int exp(-\frac{i}{\hbar}{\bf p}{\bf r})\psi'({\bf r})\frac{d^3{\bf r}}{(2\pi\hbar)^{3/2}},\quad
\chi''({\bf p})=\int exp(-\frac{i}{\hbar}{\bf p}{\bf r})\psi''({\bf r})\frac{d^3{\bf r}}{(2\pi\hbar)^{3/2}}
\label{chidecomp}
\end{equation}
Therefore the photon state in the Fock space can be represented as
\begin{equation}
\Phi_1=\int [\chi'({\bf p})+\chi''({\bf p})]a({\bf p})^*d^3{\bf p} \Phi_0
\label{Chi}
\end{equation}

In quantum mechanics particles exist during the whole time 
interval $t\in (-\infty,\infty)$ and evolution is defined by the interaction operator acting on 
the particle WF. This operator is local in 
coordinate space if in 
momentum space it acts on the particle WF $\chi$ as the convolution operator. However, if annihilation and creation of
particles are possible then evolution is described by operators acting not on particle WFs but on the operators
$a({\bf p})$ and $a({\bf p})^*$. In approximations when annihilation and creation is not important, evolution can be
reformulated in terms of $\chi$ only. For example, as noted in Subsec. \ref{history}, in the approximation $(v/c)^2$
the electron in the hydrogen atom can be described by the Dirac or Schr\"{o}dinger equation. 
 
However, Feynman diagrams contain only vertices with one photon. Hence in any interaction the photon is first absorbed as a whole, in the intermediate state there is no photon, and in the case when the photon is reemitted this is a new photon. So in the case of interactions the evolution of the photon cannot be described
by an equation where the photon WF exists during the whole time interval $t\in (-\infty,\infty)$,
and the action of the evolution operator on photon states can be defined only in terms of $a({\bf p})$ and $a({\bf p})^*$. As follows from Eq. (\ref{photonaa*}), 
those operators are not local in coordinate space. So it is not possible that they act only on $\psi''$ and do not act on $\psi'$.

In general, if $\Psi$ is the WF of a system, and $\Psi=\Psi_1+\Psi_2$ is a decomposition of this function
then evolutions of $\Psi_1$ and $\Psi_2$ will be independent of each other if the states $\Psi_1$ and $\Psi_2$ have
at least one different conserved quantum number (e.g. angular momentum). However, in the decomposition 
$\psi \Psi_g=\psi' \Psi_g+\psi'' \Psi_g$ the
states are not characterized by a different conserved quantum number and therefore evolutions of the different parts of the decomposition will not be independent.

The crucial difference between diffraction theory and the given case follows. In diffraction theory
it is always known where different parts of the wave are. However, the photon does not
have parts (roughly speaking, it is a point) and its WF describes only probabilities to find
the photon at different points. Hence  the fact that $\psi'' \neq 0$ does not mean that the part 
$\psi''$ of the photon is inside
the object but means only that the probability
to find the photon inside the object is not zero because this probability equals $||\psi''||^2$.
Since this quantity is very small then with the probability very close to unity the photon will
not interact with the object. 

This expectation is also in the spirit of QED. Since in any interaction the initial photon will be first
absorbed as a whole and there will be no photon in the intermediate state, the sizes of reemitted photons 
(if they are created) will be defined by the absorber, and after any interaction the WF of the object will 
not be $\Psi_g$. So there is no part of the photon which does not participate in the interaction, and 
after any interaction WFs of final photons will not have large transverse sizes anymore. 

This is an illustration of the WF collapse: if the photon WF has a large size 
before interaction then, as a result of the WF collapse, after any interaction the WF cannot have a large size.
The WF collapse is a pure quantum phenomenon and there is no analog of the WF collapse in
diffraction theory. 

A possible reason why Eq. (\ref{Greg}) might seem to be acceptable is that the
decomposition $\psi=\psi'+\psi''$ is implicitly (and erroneously) understood as breaking the photon
into two photons with the WFs $\psi'$ and $\psi''$. However, such a decomposition does not
mean that a particle is broken into two parts. Mathematically this is clear from the fact that 
the two-photon state
\begin{equation}
\Phi_{12}=const\int\int \chi'({\bf p}')\chi''({\bf p}'')a({\bf p}')^*a({\bf p}'')^*d^3{\bf p}'d^3{\bf p}'' \Phi_0
\label{Chi2}
\end{equation} 
fully differs from the state (\ref{Chi}).  

We conclude that the photon WF after interaction cannot be $\psi'$ and, instead of Eq. (\ref{Greg}), the result is
\begin{equation}
S(\psi \Psi_g)=c\psi \Psi_g+(...)
\label{Greg2}
\end{equation}
where $1-|c|^2$ is the probability of interaction. Since the probability is small, the quantity
$c$ is very close to unity and the photon will probably pass the objects without any
interaction. In rare cases when interaction happens, the WF of any final photon 
will not have a cosmic size anymore.  Such a photon can reach Earth only if its momentum 
considerably differs from the original one but this contradicts Theory A. So the assumption that the above 
paradoxes can be explained by analogy with diffraction theory is not justified. 

If $f({\bf r}/r)\neq const$ then, as follows from Eqs. (\ref{FG}) and (\ref{relpsif}),
the radial part of the WF is the same as in the spherically symmetric case and, 
as follows from the above discussion, the coordinate WF of the initial photon still has a cosmic size.
Therefore on its way to Earth the photon WF will also encounter stars, planets and
other objects (even if they are far from the line connecting the star and Earth) 
and the same inconsistencies arise.

{\it In summary, since according to standard theory photons emitted by stars have coordinate WFs with cosmic sizes,
the above arguments indicate that the theory contradicts observational data.}

\subsection{Other WPS paradoxes}

In the infrared and radio astronomy wave lengths are much greater than in the optical region but typical values
of $a_{ph}$ are expected to be much greater. As a consequence, here standard quantum theory
encounters the same problems that in the optical region.

In the case of gamma-ray bursts (GRBs) wave lengths are much less than in the optical region but this is 
outweighed by the facts that, according
to the present understanding of the GRB phenomenon (see e.g. Ref. \cite{bursts}), gamma quanta created in GRBs 
typically travel to Earth for billions of years and typical values
of $a_{ph}$ are expected to be much less than in the optical region. The location of sources of GBRs 
are determined with a good
accuracy and the data can be explained only assuming that the gamma quanta are focused into  
narrow jets 
which are observable when Earth lies along the path of those jets. However, in view of the above
discussion, the results on WPS predicted by standard quantum theory are incompatible with the
data on GRBs because, as a consequence of WPS, the probability to detect photons from
GRBs would be negligible. 

Consider now WPS effects for radio wave photons. In radiolocation it is important
that a beam from a directional antenna has a narrow angular distribution and a narrow distribution of wave lengths.
 This makes it possible to communicate even with very distant 
space probes. For this purpose a set of radio telescopes can be used but for simplicity we consider
a model where signals from a space probe are received by one radio telescope having the diameter $D$ of the dish.

The Cassini spacecraft can transmit to Earth at three radio wavelengths: 14cm, 4cm and 1cm \cite{Cassini}.
A radio telescope on Earth can determine the position of Cassini with a good accuracy if it detects photons having
momenta in the angular range of the order of $D/L$ where $L$ is the
distance to Cassini. The main idea of using a system of radio telescopes is to increase the effective value of $D$. 
As a consequence of the fact that the radio signal sent from Cassini has an angular divergence
which is much greater than $D/L$, only a small part of photons in the signal can be detected. 
We consider a case when Cassini was 7AU away from the Earth.

Consider first the problem on classical level. For the quantity $a=a_{cl}$ we take the value of $1m$
which is of the order of the radius of the Cassini antenna. If $\alpha=\lambda/(2\pi a)$ and $L(t)$
is the length of the classical path then, as follows from Eq. (\ref{em}), $a_{cl}(t)\approx L(t)\alpha$. As a 
result, even for $\lambda=1cm$ we have $a_{cl}(t)\approx 1.6\cdot 10^6km$. Hence one 
might expect that only
a $[D/a_{cl}(t)]^2$ part of the photons can be detected. 

Consider now the problem on quantum level. The condition
$t\gg t_*$ is satisfied for both, the classical and quantum problems. Then, as follows from Eq. (\ref{em}), $a_{ph}(t)=a_{cl}(t)a_{cl}/a_{ph}$,
i.e. the quantity $a_{ph}(t)$ is typically greater than $a_{cl}(t)$ and in Sec. \ref{WPW} this effect is called
the WPW paradox. The fact that only photons in the angular range $D/L$ can be detected can be described by
projecting the states $\chi=\chi({\bf p},t)$ (see Eqs. (\ref{chiprel}), and (\ref{chiptphoton})) onto the states
$\chi_1={\cal P}\chi$ where $\chi_1({\bf p},t)=\rho({\bf p})\chi({\bf p},t)$ and the form factor $\rho({\bf p})$ is
significant only if ${\bf p}$ is in the needed angular range. We choose 
$\rho({\bf p})=exp(-{\bf p}_{\bot}^2a_1^2/2\hbar^2)$ where $a_1$ is of the order of $\hbar L/(p_0 D)$. 
Since $a_1\gg a_{ph}$,
it follows from Eqs. (\ref{chiprel}), and (\ref{chiptphoton}) that $||{\cal P}\chi||^2=(a_{ph}/a_1)^2$.  Then, as follows from Eq. (\ref{em}),
$(a_{ph}/a_1)^2$ is of the order of $[D/a_{ph}(t)]^2$ as expected and this quantity is typically much less than 
$[D/a_{cl}(t)]^2$. Hence the WPW paradox would make communications with space probes 
much more difficult. 

We now consider the following problem. The parameter $\gamma$ in General Relativity (GR) 
is extracted from experiments on deflection of light from distant stars by the Sun and from the effect
called Shapiro time delay. The meaning of the effect follows. An antenna
on Earth sends a signal to Mercury, Venus or an interplanetary space probe and receives the reflected signal.
If the path of the signal nearly grazes the Sun then the gravitational influence of the Sun deflects the path
from a straight line. As a result, the path becomes longer by 
$S\approx 75km$ and the signals arrive with a
delay $S/c\approx 250\mu s$. This effect is treated as the fourth test of GR.

The consideration of the both effects in GR is based on the assumption that the photon 
is a pointlike classical particle moving along classical trajectory. In the first case the
photon WF has a cosmic size. In the second case the available experimental data are treated
such that the best test of $\gamma$ has been performed in measuring the Shapiro delay when signals from the DSS-25 antenna \cite{DSS25} were sent to the Cassini spacecraft when 
it was 7AU away from the Earth. As noted above, in that case, even in the most favorable 
scenario $a_{cl}(t)\approx 1.6\cdot 10^6km$ and the quantity $a_{ph}(t)$ is expected to be
much greater. Therefore a problem arises whether the classical consideration in GR is
compatible with the fact that the photon coordinate WFs have very large sizes.

One might think that the compatibility is not a problem because when we detect a photon with the momentum 
pointing to the area near the Sun
we know that this photon moved to us on the trajectory bending near the Sun.
The results of Sec. \ref{RelWPS} indeed show that even if the photon momentum WF
has a broad distribution, the photon detected by a measuring device can be detected only
at the moment of time close to $L/c$ and momentum of the detected photon will point to the star which
emitted this photon. However, quantum formalism does not contain any information about the
photon trajectory from the moment of emission to the moment of detection. One might 
guess that the required trajectory will give the main contribution in the Feynman path integral 
formulation but the proof of this guess is rather complicated.

In summary, by analogy with the consideration in Subsec. \ref{when}, one can conclude 
that quantum theory does not contain any information about trajectories. The notion of
trajectories in quantum theory is a reasonable approximation only in semiclassical 
approximation when a choice of the position operator has been made. However, in
the case of packets with broad coordinate distributions the notion of trajectories
does not have a physical meaning and one cannot avoid quantum consideration of
the problem. In particular, the results of GR on the deflection of light and on the Shapiro
delay are meaningful only
if there is no considerable WPS in quantum theory. In addition, in view of the WPW
paradox, the probability to detect reflected photons in the Shapiro delay experiments can be very small.

One might think that the  WPS effect is
important only if a particle travels a rather long distance. Hence one might expect that in experiments 
on the Earth this effect is negligible.
Indeed, one might expect that in typical experiments on the Earth the time $t$ is so small that $a(t)$ 
is much less than the size of any macroscopic source of light. However, a conclusion that 
the effect of WPS is
negligible for any experiment on the Earth might be premature.

As an example, consider the case of protons in the LHC accelerator. According to Ref. \cite{protons},
protons in the LHC ring injected at the energy $E=450 GeV$ should be accelerated to the energy $E=7 TeV$ within
one minute during which the protons will turn around the $27km$ ring approximately 674729 times. Hence the length
of the proton path is of the order of $18\cdot 10^6km$. The protons cannot be treated as free particles since
they are accelerated by strong magnets. A problem of how the width of the proton WF behaves in the
presence of strong electromagnetic field is very complicated and the solution of the problem is not known.
It is always assumed that the WPS effect for the protons can be neglected. 

We first consider a model problem of
the WPS for a free proton which moves for $t_1=1min$ with the energy in the range $[0.45,7]\, TeV$.
In nuclear physics the size of the proton is usually assumed to be a quantity of the order of $10^{-13}cm$. 
Therefore for estimations we take $a=10^{-13}cm$. Then the quantity $t_*$ defined after Eq. (\ref{at}) is not
greater than $10^{-19}s$, i.e. $t_*\ll t_1$. Hence, as follows from Eq. (\ref{at}), the quantity 
$a(t_1)$ is of the order of $500km$ if $E=7\, TeV$ and  by a factor of
$7/0.45\approx 15.6$ greater if $E=450\, GeV$. 

This fully unrealistic result cannot be treated as a paradox since, as noted
above, the protons in the LHC ring are not free. In the real situation the protons interact
with many real and virtual photons emitted by magnets. For example, this might lead to
the collapse of the proton WF each time when the proton interacts with the
real or virtual photon. This phenomenon is not well studied yet and so  
a problem of what standard theory 
predicts on the width of proton WFs in the LHC ring is far from being obvious.

The last example follows. The astronomical objects called pulsars are treated such that they are neutron stars
with radii much less than radii of ordinary stars. Therefore if mechanisms of pulsar electromagnetic radiation
were the same as for ordinary stars then the pulsars would not be visible. The fact that pulsars are visible is
explained as a consequence of the fact that they emit beams of
light which can only be seen when the light is pointed in the direction of the observer with some periods which 
are treated
as periods of rotation of the neutron stars. In popular literature this is compared with the light of a lighthouse.
However, by analogy with the case of a signal sent from Cassini, only a small part of photons in the beam can reach the Earth.
At present the pulsars have been observed in different regions of the electromagnetic spectrum but the first
pulsar called PSR B1919+21 was discovered in 1967 as a radio wave radiation with $\lambda\approx 3.7m$ \cite{pulsar}.
This pulsar is treated as the neutron star with the radius $R=0.97km$ and the distance from the pulsar to
the Earth is 2283 light years. If for estimating $a_{cl}(t)$ we assume that   
$a_{cl}=R$ then we get $\alpha\approx 6\cdot 10^{-4}$ and $a_{cl}(t)\approx 1.3 ly\approx 12\cdot 10^{12}km$.
Such an extremely large value of spreading poses a problem whether even predictions of classical electrodynamics are
compatible with the fact that pulsars are observable. However, in view of the WPW paradox, 
the value of $a_{ph}(t)$ will be even much greater and no observation of pulsars would be possible.

In view of the above discussion, standard treatment of WPS leads to several fundamental paradoxes.
To the best of our knowledge, those paradoxes have never been discussed in the literature. 
The above discussion shows that at present in standard theory there are no realistic scenarios which can explain
the WPS paradoxes. In the remaining part of the chapter we propose a solution of the problem proceeding from a consistent definition of the position operator. 

\section{Consistent construction of position operator}
\label{consistent}

The above results give grounds to think that the reason of the paradoxes which follow from the behavior of the 
coordinate photon WF in perpendicular directions is that standard definition of the position operator in 
those directions does not correspond to realistic measurements of coordinates. 
Before discussing a consistent construction, let us make the following
remark. On elementary level students treat the mass $m$ and the velocity ${\bf v}$ as primary quantities such
that the momentum is $m{\bf v}$ and the kinetic energy is $m{\bf v}^2/2$. However, from the point of view of
Special Relativity, the primary quantities are the momentum ${\bf p}$ and the total energy $E$ and then the mass
and velocity are defined as $m^2c^4=E^2-{\bf p}^2c^2$ and ${\bf v}={\bf p}c^2/E$, respectively. This example has
the following analogy. In standard quantum theory the primary operators are the position and momentum operators
and the orbital angular momentum operator is defined as their cross product. However,  
the operators ${\bf P}$ and ${\bf L}$ are consistently defined as representation operators of the Poincare algebra
while the definition of the position operator is a problem. Hence a question arises whether the position
operator can be defined in terms of ${\bf P}$ and ${\bf L}$.

One might seek the position operator such that on classical level
the relation ${\bf r}\times{\bf p}={\bf L}$ will take place. Note that on quantum level this relation is not
necessary. Indeed, the very fact that some elementary particles have a half-integer spin
shows that the total angular momentum for those particles does not have the orbital nature but on classical 
level the angular momentum can be always represented as a cross
product of the radius-vector and standard momentum. However, if the values 
of ${\bf p}$ and ${\bf L}$ are
known and ${\bf p}\neq 0$ then the requirement that ${\bf r}\times{\bf p}={\bf L}$ does not define ${\bf r}$ 
uniquely. One can define parallel and perpendicular components of ${\bf r}$ as 
${\bf r}=r_{||}{\bf p}/p+{\bf r}_{\bot}$ where $p=|{\bf p}|$. Then the relation ${\bf r}\times{\bf p}={\bf L}$ 
defines uniquely only ${\bf r}_{\bot}$. Namely, as follows from this relation, 
${\bf r}_{\bot}=({\bf p}\times{\bf L})/p^2$. In view of the fact that on quantum level the operators
${\bf p}$ and ${\bf L}$ do not commute, on this level  
${\bf r}_{\bot}$ should be replaced by a selfadjoint operator 
${\bf {\cal R}}_{\bot}=({\bf p}\times{\bf L}-{\bf L}\times{\bf p})/(2p^2)$. Therefore 
\begin{eqnarray}
&&{\cal R}_{\bot j}=\frac{\hbar}{2p^2}e_{jkl}(p_kL_l+L_lp_k)=\frac{\hbar}{p^2}e_{jkl}p_kL_l-
\frac{i\hbar}{p^2}p_j\nonumber\\
&&=i\hbar\frac{\partial}{\partial p_j}-i\frac{\hbar}{p^2}p_jp_k\frac{\partial}{\partial p_k}-\frac{i\hbar}{p^2}p_j
\label{rbot}
\end{eqnarray}
where $e_{jkl}$ is the absolutely antisymmetric tensor, $e_{123}=1$, a sum over repeated indices is assumed and 
we assume that if ${\bf L}$ is given by Eq. (\ref{IRoperators}) then the orbital momentum is $\hbar{\bf L}$.

We define the operators ${\bf F}$ and ${\bf G}$ such that ${\bf {\cal R}}_{\bot}=\hbar{\bf F}/p$ and
${\bf G}$ is the operator of multiplication by the unit vector ${\bf n}={\bf p}/p$.
A direct calculation shows that these operators satisfy the following relations:
\begin{eqnarray}
&&[L_j,F_k]=ie_{jkl}F_l,\quad [L_j,G_k]=ie_{jkl}F_l,\quad {\bf G}^2=1,\quad {\bf F}^2={\bf L}^2+1 \nonumber\\
&&[G_j,G_k]=0,\quad [F_j,F_k]=-ie_{jkl}L_l\quad e_{jkl}\{F_k,G_l\}=2L_j\nonumber\\
&&{\bf L}{\bf G}={\bf G}{\bf L}={\bf L}{\bf F}={\bf F}{\bf L}=0, \quad {\bf F}{\bf G}=-{\bf G}{\bf F}=i
\label{vectorFG1}
\end{eqnarray}
The first two relations show that ${\bf F}$ 
and ${\bf G}$ are the vector operators as expected. The result for the anticommutator shows
that on classical level ${\bf F}\times {\bf G}={\bf L}$ and the last two relations show that on classical level
the operators in the triplet $({\bf F},{\bf G},{\bf L})$ are mutually orthogonal. 

Note that if the momentum distribution is narrow and such that the mean
value of the momentum is directed along the $z$ axis then it does not mean that on the operator level the $z$
component of the operator ${\bf {\cal R}}_{\bot}$ should be zero. The matter is that the direction of the momentum
does not have a definite value. One might expect that only the mean value of the operator ${\bf {\cal R}}_{\bot}$
will be zero or very small.

In addition, an immediate consequence of the definition (\ref{rbot}) follows: {\it Since the momentum and
angular momentum operators commute with the Hamiltonian, the distribution of all the components of ${\bf r}_{\bot}$
does not depend on time. In particular, there is no WPS in directions defined by ${\bf {\cal R}}_{\bot}$.}
This is also clear from the fact that ${\bf {\cal R}}_{\bot}=\hbar{\bf F}/p$ where the operator ${\bf F}$ acts only
over angular variables and the Hamiltonian depends only on $p$.
On classical level the conservation of ${\bf {\cal R}}_{\bot}$ is obvious since it is defined by the conserving
quantities ${\bf p}$ and ${\bf L}$. It is also obvious that since a free particle is moving along a 
straight line, a vector from the origin perpendicular to this line does not change with time.

The above definition of the perpendicular component of the position operator is well substantiated
since on classical level the relation ${\bf r}\times{\bf p}={\bf L}$ has been verified in numerous experiments.
However, this relation does not make it possible to define the parallel component of the position operator
and a problem arises what physical arguments should be used for that purpose. 

A direct calculation shows that if $\partial/\partial {\bf p}$ is written in terms of $p$ and angular variables then
\begin{equation}
i\hbar\frac{\partial}{\partial {\bf p}}={\bf G}{\cal R}_{||}+{\bf {\cal R}}_{\bot}
\label{decomp1}
\end{equation}
where the operator ${\cal R}_{||}$ acts only over the variable $p$:
\begin{equation}
{\cal R}_{||}=i\hbar (\frac{\partial}{\partial p}+\frac{1}{p})
\label{rparall}
\end{equation}
The correction $1/p$ is related to the fact that the operator ${\cal R}_{||}$ is Hermitian since in variables 
$(p,{\bf n})$ the scalar product is given by
\begin{equation}
(\chi_2,\chi_1)=\int \chi_2(p,{\bf n})^*\chi_1(p,{\bf n})p^2dp do
\end{equation}
where $do$ is the element of the solid angle.

While the components of standard position operator commute
with each other, the operators ${\cal R}_{||}$ and ${\bf {\cal R}}_{\bot}$ satisfy the following commutation relations:
\begin{equation}
[{\cal R}_{||},{\bf {\cal R}}_{\bot}]=-\frac{i\hbar}{p}{\bf {\cal R}}_{\bot},\quad 
[{\cal R}_{\bot j},{\cal R}_{\bot k}]
=-\frac{i\hbar^2}{p^2}e_{jkl}L_l
\label{rparallrbot}
\end{equation}
An immediate consequence of these relations follows: {\it Since the operator ${\cal R}_{||}$ and different
components of ${\bf {\cal R}}_{\bot}$ do not commute with each other, the corresponding quantities cannot be
simultaneously measured and hence there is no WF $\psi(r_{||},{\bf r}_{\bot})$ in 
coordinate representation.}

In standard theory $-\hbar^2 (\partial/\partial{\bf p})^2$ is the operator of the quantity ${\bf r}^2$. As follows
from Eq. (\ref{vectorFG1}), the two terms in Eq. (\ref{decomp1}) are not strictly orthogonal and on the operator level
$-\hbar^2 (\partial/\partial{\bf p})^2\neq {\cal R}_{||}^2+{\bf {\cal R}}_{\bot}^2$.
A direct calculation using Eqs. (\ref{vectorFG1}) and (\ref{decomp1}) gives
\begin{equation}
\frac{\partial^2}{\partial{\bf p}^2}=\frac{\partial^2}{\partial p^2}+\frac{2}{p}\frac{\partial}{\partial p}-
\frac{{\bf L}^2}{p^2},\quad -\hbar^2\frac{\partial^2}{\partial{\bf p}^2}={\cal R}_{||}^2+{\bf {\cal R}}_{\bot}^2-
\frac{\hbar^2}{p^2}
\label{r2}
\end{equation}
in agreement with the expression for the Laplacian in spherical coordinates. In semiclassical approximation, 
$(\hbar^2/p^2)\ll {\bf {\cal R}}_{\bot}^2$ since the eigenvalues of ${\bf L}^2$ are $l(l+1)$, 
in semiclassical states $l\gg 1$ and, as follows from Eq. (\ref{vectorFG1}),
${\bf {\cal R}}_{\bot}^2=[\hbar^2(l^2+l+1)/p^2]$.

As follows from Eq. (\ref{rparallrbot}), $[{\cal R}_{||},p]=-i\hbar$, i.e. in the longitudinal direction the
commutation relation between the coordinate and momentum is the same as in standard theory. One can also calculate
the commutators between the different components of ${\bf {\cal R}}_{\bot}$ and ${\bf p}$. Those commutators are
not given by such simple expressions as in standard theory but it is easy to see that all of them are of the
order of $\hbar$ as it should be.

Equation (\ref{decomp1}) can be treated as an implementation of the relation 
${\bf r}=r_{||}{\bf p}/|{\bf p}|+{\bf r}_{\bot}$ on quantum level. As argued in Secs. \ref{intropos}
and \ref{classical}, standard position operator $i\hbar\partial/\partial p_j$ in the direction $j$ is not consistently
defined if $p_j$ is not sufficiently large. One might think however that since the operator ${\cal R}_{||}$ contains
$i\hbar\partial/\partial p$, it is defined consistently if the magnitude of the momentum is sufficiently large. 

In summary, we propose to define the position operator not by the set $(i\hbar\partial/\partial p_x,
i\hbar\partial/\partial p_y,i\hbar\partial/\partial p_z)$ but by the operators ${\cal R}_{||}$ and ${\bf {\cal R}}_{\bot}$.
Those operators are defined from different considerations. As noted above, the definition of ${\bf {\cal R}}_{\bot}$ is 
based on solid physical facts while the definition of ${\cal R}_{||}$ is expected to be more consistent than the
definition of standard position operator. However, this does not guarantee that the operator ${\cal R}_{||}$
is consistently defined in all situations. As argued in Sec. \ref{validity}, in a quantum theory over a Galois field
an analogous definition is not consistent {\it for macroscopic bodies} (even if $p$ is large) since in that case 
semiclassical approximation is not valid. In the remaining part of this section we assume that for elementary particles
the above definition of ${\cal R}_{||}$ is consistent in situations when semiclassical approximation applies.

One might pose the following question. What is the reason to work with the parallel and perpendicular components
of the position operator separately if, according to Eq. (\ref{decomp1}), their sum is the standard position operator?
The explanation follows.

In quantum theory every physical quantity corresponds to a selfadjoint operator but the theory does not define
explicitly how a quantity corresponding to a specific operator should be measured. There is no guaranty that for
each selfadjoint operator there exists a physical quantity which can be measured in real experiments.

Suppose that there are three physical quantities corresponding to the selfadjoint operators $A$, $B$ and $C$ such that
$A+B=C$. Then in each state the mean values of the operators are related as ${\bar A}+{\bar B}={\bar C}$ but in situations 
when the operators $A$ and $B$ do not commute with each other there is no direct relation between the distributions
of the  physical quantities corresponding to the operators $A$, $B$ and $C$. For example, in situations when the physical 
quantities corresponding to the
operators $A$ and $B$ are semiclassical and can be measured with a good accuracy, there is no guaranty that the physical 
quantity corresponding to the operator $C$ can be measured in real measurements. As an example, the physical meaning
of the quantity corresponding to the operator $L_x+L_y$ is problematic. Another example is the situation with WPS in 
directions perpendicular to the particle momentum. Indeed, as noted above, the physical
quantity corresponding to the operator ${\bf {\cal R}}_{\bot}$ does not experience WPS and, as shown in Sec. \ref{newWPS},
in the case of ultrarelativistic particles there is no WPS in the parallel direction as well. However,  
standard position operator is a sum of noncommuting operators corresponding to well defined physical
quantities and, as a consequence, there are situations when standard position operator defines a quantity which
cannot be measured in real experiments.

\begin{sloppypar}
\section{New position operator and semiclassical states}
\label{newsemicl}
\end{sloppypar}

As noted in Sec. \ref{classical}, in standard theory states are treated as semiclassical in greatest possible extent
if $\Delta r_j \Delta p_j =\hbar/2$ for each $j$ and such states are called coherent. The existence of coherent
states in standard theory is a consequence of commutation relations $[p_j,r_k]=-i\hbar \delta_{jk}$. Since in our
approach there are no such relations, a problem arises how to construct states in which all physical quantities
$p$, $r_{||}$, ${\bf n}$ and ${\bf r}_{\bot}$ are semiclassical.

One can calculate the mean values and uncertainties of the operator ${\cal R}_{||}$ and
all the components of the operator 
${\bf {\cal R}}_{\bot}$ in the state defined by Eq. (\ref{chiprel}). The calculation is not simple since it involves
three-dimensional integrals with Gaussian functions divided by $p^2$. The result is that these operators
are semiclassical in the state (\ref{chiprel}) if $p_0\gg \hbar/b$, $p_0\gg \hbar/a$ and 
$r_{0z}$ has the same order of magnitude as $r_{0x}$ and $r_{0y}$. 

However, a more natural approach follows. Since ${\bf {\cal R}}_{\bot}=\hbar{\bf F}/p$, the operator
${\bf F}$ acts only over the angular
variable ${\bf n}$ and ${\cal R}_{||}$ acts only over the variable $p$, it is convenient to work
in the representation where the Hilbert space is the space of functions $\chi(p,l,\mu)$ such that 
the scalar product is
\begin{equation}
(\chi_2,\chi_1)=\sum_{l\mu}\int_0^{\infty} \chi_2(p,l,\mu)^*\chi_1(p,l,\mu)dp
\label{newscalar}
\end{equation} 
and $l$ and $\mu$ are the orbital and magnetic quantum numbers, respectively, i.e.
\begin{equation}
{\bf L}^2\chi(p,l,\mu)=l(l+1)\chi(p,l,\mu),\quad L_z\chi(p,l,\mu)=\mu\chi(p,l,\mu)
\label{orbmagn}
\end{equation}

The operator ${\bf L}$ in this space does not act over the variable $p$ and the action of the remaining
components is
given by
\begin{equation}
L_+\chi(l,\mu)=[(l+\mu)(l+1-\mu)]^{1/2}\chi(l,\mu-1),\quad
L_-\chi(l,\mu)=[(l-\mu)(l+1+\mu)]^{1/2}\chi(l,\mu+1)
\label{L}
\end{equation}
where the $\pm$ components of vectors are defined such that $L_x=L_++L_-$, $L_y=-i(L_+-L_-)$.

A direct calculation shows that, as a consequence of Eq. (\ref{rbot})
\begin{eqnarray}
&&F_+\chi(l,\mu)=-\frac{i}{2}[\frac{(l+\mu)(l+\mu-1)}{(2l-1)(2l+1)}]^{1/2}l\chi(l-1,\mu-1)\nonumber\\
&&-\frac{i}{2}[\frac{(l+2-\mu)(l+1-\mu)}{(2l+1)(2l+3)}]^{1/2}(l+1)\chi(l+1,\mu-1)\nonumber\\
&&F_-\chi(l,\mu)=\frac{i}{2}[\frac{(l-\mu)(l-\mu-1)}{(2l-1)(2l+1)}]^{1/2}l\chi(l-1,\mu+1)\nonumber\\
&&+\frac{i}{2}[\frac{(l+2+\mu)(l+1+\mu)}{(2l+1)(2l+3)}]^{1/2}(l+1)\chi(l+1,\mu+1)\nonumber\\
&&F_z\chi(l,\mu)=i[\frac{(l-\mu)(l+\mu)}{(2l-1)(2l+1)}]^{1/2}l\chi(l-1,\mu)\nonumber\\
&&-i[\frac{(l+1-\mu)(l+1+\mu)}{(2l+1)(2l+3)}]^{1/2}(l+1)\chi(l+1,\mu)
\label{F1}
\end{eqnarray}
The operator ${\bf G}$ acts on such states as follows
\begin{eqnarray}
&&G_+\chi(l,\mu)=\frac{1}{2}[\frac{(l+\mu)(l+\mu-1)}{(2l-1)(2l+1)}]^{1/2}\chi(l-1,\mu-1)\nonumber\\
&&-\frac{1}{2}[\frac{(l+2-\mu)(l+1-\mu)}{(2l+1)(2l+3)}]^{1/2}\chi(l+1,\mu-1)\nonumber\\
&&G_-\chi(l,\mu)=-\frac{1}{2}[\frac{(l-\mu)(l-\mu-1)}{(2l-1)(2l+1)}]^{1/2}\chi(l-1,\mu+1)\nonumber\\
&&+\frac{1}{2}[\frac{(l+2+\mu)(l+1+\mu)}{(2l+1)(2l+3)}]^{1/2}\chi(l+1,\mu+1)\nonumber\\
&&G_z\chi(l,\mu)=-[\frac{(l-\mu)(l+\mu)}{(2l-1)(2l+1)}]^{1/2}\chi(l-1,\mu)\nonumber\\
&&-[\frac{(l+1-\mu)(l+1+\mu)}{(2l+1)(2l+3)}]^{1/2}\chi(l+1,\mu)
\label{G1}
\end{eqnarray}
and now the operator ${\cal R}_{||}$ has a familiar form ${\cal R}_{||}=i\hbar \partial/\partial p$. 

Therefore by analogy with Secs. \ref{classical} and \ref{NRWPS} one can construct states which are coherent 
with respect to $(r_{||},p)$, i.e. such that 
$\Delta r_{||}\Delta p=\hbar/2$. Indeed (see Eq. (\ref{chip})), the WF
\begin{equation}
\chi(p)=\frac{b^{1/2}}{\pi^{1/4}\hbar^{1/2}}exp[-\frac{(p-p_0)^2b^2}{2\hbar^2}-
\frac{i}{\hbar}(p-p_0)r_0]
\label{pr||}
\end{equation}
describes a state where the mean values of $p$ and $r_{||}$ are $p_0$ and $r_0$, respectively and their uncertainties
are $\hbar /(b\sqrt{2})$ and $b/\sqrt{2}$, respectively. Strictly speaking, the analogy between the given case and
that discussed in Secs. \ref{classical} and \ref{NRWPS} is not full since in the given case the quantity $p$ can be
in the range $[0,\infty)$, not in $(-\infty,\infty)$ as momentum variables used in those sections. However, if 
$p_0b/\hbar \gg 1$ then the formal expression for $\chi(p)$ at $p<0$ is extremely small and so the normalization
integral for $\chi(p)$ can be formally taken from $-\infty$ to $\infty$.

In such an approximation one can define WFs $\psi(r)$ in the $r_{||}$ representation.
By analogy with the consideration in Secs. \ref{classical} and \ref{NRWPS} we define 
\begin{equation}
\psi(r)=\int exp(\frac{i}{\hbar}pr)\chi(p)\frac{dp}{(2\pi\hbar)^{1/2}}
\label{psir||in}
\end{equation}
where the integral is formally taken from $-\infty$ to $\infty$. Then 
\begin{equation}
\psi(r)=\frac{1}{\pi^{1/4}b^{1/2}}exp[-\frac{(r-r_0)^2}{2b^2}+\frac{i}{\hbar}p_0r]
\label{psir||B}
\end{equation}
Note that here the quantities $r$ and $r_0$ have the meaning of coordinates in the direction parallel to the
particle momentum, i.e. they can be positive or negative.

Consider now states where the quantities ${\bf F}$ and ${\bf G}$ are semiclassical.
One might expect that in semiclassical states the quantities $l$ and $\mu$ are very
large. In this approximation, as follows from Eqs. (\ref{F1}) and (\ref{G1}), the action of 
the operators ${\bf F}$ and ${\bf G}$ can be written as
\begin{eqnarray}
&&F_+\chi(l,\mu)=-\frac{i}{4}(l+\mu)\chi(l-1,\mu-1)-\frac{i}{4}(l-\mu)\chi(l+1,\mu-1)\nonumber\\
&&F_-\chi(l,\mu)=\frac{i}{4}(l-\mu)\chi(l-1,\mu+1)+\frac{i}{4}(l+\mu)\chi(l+1,\mu+1)\nonumber\\
&&F_z\chi(l,\mu)=-\frac{i}{2l}(l^2-\mu^2)^{1/2}[\chi(l+1,\mu)+\chi(l-1,\mu)]\nonumber\\
&&G_+\chi(l,\mu)=\frac{l+\mu}{4l}\chi(l-1,\mu-1)-\frac{l-\mu}{4l}\chi(l+1,\mu-1)\nonumber\\
&&G_-\chi(l,\mu)=-\frac{l-\mu}{4l}\chi(l-1,\mu+1)+\frac{l+\mu}{4l}\chi(l+1,\mu+1)\nonumber\\
&&G_z\chi(l,\mu)=-\frac{1}{2l}(l^2-\mu^2)^{1/2}[\chi(l+1,\mu)+\chi(l-1,\mu)]
\label{FGsemicl}
\end{eqnarray}

In view of the remark in Sec. \ref{classical} about semiclassical vector quantities, 
consider a state $\chi(l,\mu)$ such that $\chi(l,\mu)\neq 0$ only if $l\in [l_1,l_2]$, $\mu \in [\mu_1,\mu_2]$
where $l_1,\mu_1>0$, $\delta_1=l_2+1-l_1$, $\delta_2=\mu_2+1-\mu_1$, $\delta_1\ll l_1$, $\delta_2\ll \mu_1$
$\mu_2<l_1$ and $\mu_1\gg (l_1-\mu_1)$. This is the state where the quantity $\mu$ is close to its maximum value $l$.
As follows from Eqs. (\ref{orbmagn}) and (\ref{L}), in this state the quantity $L_z$ is much 
greater than $L_x$ and $L_y$ and, as 
follows from Eq. (\ref{FGsemicl}), the quantities $F_z$ and $G_z$ are small. So on classical level this state
describes a motion of the particle in the $xy$ plane. The quantity $L_z$ in this state is obviously semiclassical 
since $\chi(l,\mu)$ is the eigenvector of the operator $L_z$ with the eigenvalue $\mu$. As follows from 
Eq. (\ref{FGsemicl}), the action of the operators $(F_+,F_-,G_+,G_-)$ on this state can be described by the
following approximate formulas:
\begin{eqnarray}
&&F_+\chi(l,\mu)=-\frac{il_0}{2}\chi(l-1,\mu-1),\quad F_-\chi(l,\mu)=\frac{il_0}{2}\chi(l+1,\mu+1)\nonumber\\
&&G_+\chi(l,\mu)=\frac{1}{2}\chi(l-1,\mu-1),\quad G_-\chi(l,\mu)=\frac{1}{2}\chi(l+1,\mu+1)
\label{FGsimple}
\end{eqnarray}
where $l_0$ is a value from the interval $[l_1,l_2]$.

Consider a simple model when $\chi(l,\mu)=exp[i(l\alpha -\mu\beta)]/(\delta_1\delta_2)^{1/2}$, $l\in [l_1,l_2]$
and $\mu\in [\mu_1,\mu_2]$. Then a simple direct calculation using Eq. (\ref{FGsimple}) gives
\begin{eqnarray}
&&{\bar G}_x=cos\gamma,\quad {\bar G}_y=-sin\gamma\quad {\bar F}_x=-l_0sin\gamma\quad {\bar F}_y=-l_0cos\gamma\nonumber\\
&& \Delta G_x=\Delta G_y=(\frac{1}{\delta_1}+\frac{1}{\delta_2})^{1/2},\quad 
\Delta F_x=\Delta F_y=l_0(\frac{1}{\delta_1}+\frac{1}{\delta_2})^{1/2}
\label{meanGF}
\end{eqnarray}
where $\gamma=\alpha-\beta$. Hence the vector quantities ${\bf F}$ and ${\bf G}$ are semiclassical since either 
$|cos\gamma|$ or $|sin\gamma|$ or both are much greater than $(\delta_1+\delta_2)/(\delta_1\delta_2)$.

\begin{sloppypar}
\section{New position operator and wave packet spreading}
\label{newWPS}
\end{sloppypar}

If the space of states is implemented according to the scalar product (\ref{newscalar}) then the dependence 
of the WF on $t$ is
\begin{equation}
\chi(p, k,\mu,t)=exp[-\frac{i}{\hbar}(m^2c^2+p^2)^{1/2}ct]\chi(p, k,\mu,t=0)
\label{chitnew}
\end{equation}
As noted in Secs. \ref{NRWPS} and \ref{RelWPS}, there is no WPS in momentum space and this is natural in view of momentum conservation.
Then, as already noted, the distribution of the quantity ${\bf r}_{\bot}$ does not depend on time and
this is natural from the considerations described in Sec. \ref{consistent}.

At the same time, the dependence of the $r_{||}$ distribution on time can be calculated in full analogy with
Sec. \ref{NRWPS}. Indeed, consider, for example a function $\chi(p,l, \mu, t=0)$ having the form
\begin{equation}
\chi(p,l, \mu, t=0)=\chi(p,t=0)\chi(l,\mu)
\label{factoriz}
\end{equation}
Then, as follows from Eqs. (\ref{psir||in}) and (\ref{chitnew}),
\begin{equation}
\psi(r,t)=\int exp[-\frac{i}{\hbar}(m^2c^2+p^2)^{1/2}ct+\frac{i}{\hbar}pr]\chi(p, t=0)\frac{dp}{(2\pi\hbar)^{1/2}}
\label{psir||int}
\end{equation}

Suppose that the function $\chi(p, t=0)$  is given by Eq. (\ref{pr||}). Then in full analogy with the calculations
in Sec. \ref{NRWPS} we get that in the nonrelativistic case the $r_{||}$ distribution is defined by the wave
function
\begin{equation}
\psi(r,t)=\frac{1}{\pi^{1/4}b^{1/2}}(1+\frac{i\hbar t}{mb^2})^{-1/2}exp[-\frac{(r-r_0-v_0t)^2}{2b^2(1+\frac{\hbar^2t^2}{m^2b^4})}(1-\frac{i\hbar t}{mb^2})+
\frac{i}{\hbar}p_0r-\frac{ip_0^2t}{2m\hbar}]
\label{psirtnew}
\end{equation}
where $v_0=p_0/m$ is the classical speed of the particle in the direction of the particle momentum. 
Hence the WPS effect in this direction is similar to that given by Eq. (\ref{psirt}) in standard theory.

In the opposite case when the particle is ultrarelativistic, Eq. (\ref{psir||int}) can be written as
\begin{equation}
\psi(r,t)=\int exp[\frac{i}{\hbar}p(r-ct)]\chi(p, t=0)\frac{dp}{(2\pi\hbar)^{1/2}}
\label{ultra}
\end{equation}
Hence, as follows from Eq. (\ref{psir||B}):
\begin{equation}
\psi(r,t)=\frac{1}{\pi^{1/4}b^{1/2}}exp[-\frac{(r-r_0-ct)^2}{2b^2}+\frac{i}{\hbar}p_0(r-ct)]
\label{psir||C}
\end{equation}
In particular, for an ultrarelativistic particle there is no WPS in the direction of particle momentum and
this is in agreement with the results of Sec. \ref{RelWPS}.

We conclude that in our approach an ultrarelativistic particle (e.g. the photon) experiences WPS neither in the
direction of its momentum nor in perpendicular directions, i.e. the WPS effect for an ultrarelativistic particle
is absent at all.

Let us note that the absence of WPS in perpendicular directions is simply a consequence of the fact that 
a consistently defined operator ${\bf {\cal R}}_{\bot}$ commutes with the Hamiltonian.
In quantum theory a physical quantity is called conserved if its operator commutes with the Hamiltonian.
Therefore ${\bf r}_{\bot}$ is a conserved physical quantity. In contrast to classical theory, this 
does not mean that ${\bf r}_{\bot}$ should necessarily have only one value but means
that the ${\bf r}_{\bot}$ distribution does not depend on time.  
 On the other hand, the longitudinal coordinate is not a conserved physical quantity
since a particle
is moving along the direction of its momentum. However, in a special case of ultrarelativistic particle
the absence of WPS is simply a consequence of the fact that the WF given by Eq. (\ref{ultra})
depends on $r$ and $t$ only via a combination of $r-ct$.

\section{Discussion and conclusion}
\label{conclusion}

In this chapter we consider a problem of constructing position operator in quantum theory. As noted in Sec.
\ref{intropos}, this operator is needed in situations where semiclassical approximation works with a high
accuracy. Standard choice of the position operator in momentum space is 
$i\hbar\partial/\partial{\bf p}$. A motivation for this choice is discussed in Sec. \ref{classical}.
We note that this choice is not consistent since $i\hbar\partial/\partial p_j$
cannot be a physical position operator in directions where the momentum is small. 
Physicists did not pay attention
to the inconsistency probably for the following reason: as explained in textbooks, 
transition from quantum to classical theory can be performed such that if the 
coordinate WF contains a rapidly oscillating exponent $exp(iS/\hbar)$, where $S$ is the classical action, then in the
formal limit $\hbar\to 0$ the Schr\"{o}dinger equation becomes the Hamilton-Jacobi equation. 

However, an inevitable
consequence of standard quantum theory is the effect of wave packet spreading (WPS). 
As shown in Sec. \ref{experiment}, if the WPS effect for photons traveling to Earth from distant objects
is as given by standard theory then we have several fundamental paradoxes. The most striking of
them is that standard theory contradicts our experience on observations of stars. 

We propose a new definition of the position operator which we treat as consistent for the following reasons.
Our position operator is defined by two components - in the direction along the momentum
and in perpendicular directions. The first part has a familiar form $i\hbar \partial/\partial p$ and is
treated as the operator of the longitudinal coordinate if the magnitude of
$p$ is rather large. At the same condition the position operator in the perpendicular directions
is defined as a quantum generalization of the relation ${\bf r}_{\bot}\times {\bf p}={\bf L}$.
So in contrast to standard definition of the position operator, the new operator is expected
to be physical only if the {\it magnitude} of the momentum is rather large. 

As a consequence of our construction, WPS in directions perpendicular to the particle
momentum is absent regardless of whether the particle is nonrelativistic or relativistic. 
Moreover, for an ultrarelativistic particle the effect of WPS is absent at all. 

Different components of the new position operator commute with each other only in the formal limit $\hbar\to 0$. 
As a consequence, there is no WF in coordinate representation. In particular, there is no
quantum analog of the coordinate Coulomb potential (see the discussion in Sec. \ref{intropos}). A possibility 
that coordinates can be noncommutative has been first
discussed by Snyder \cite{Snyder} and it is implemented in several modern theories. In those theories the
measure of noncommutativity is defined by a parameter $l$ called the fundamental length  (the role of which can be
played e.g. by the Planck length or the Schwarzschild radius). In the formal limit $l\to 0$ the coordinates become 
standard ones related to momenta by a Fourier transform. 
As shown above, this is unacceptable in view of the WPS paradoxes. One of ideas of those theories
is that with a nonzero $l$ it might be possible to resolve difficulties of standard theory where $l=0$ (see
e.g. Ref. \cite{Smolin} and references therein). At the same time, in our approach 
there can be no notion of fundamental length since commutativity of coordinates takes place only in the
formal limit $\hbar\to 0$.

The absence of the coordinate WF is not unusual. For example, there is no WF
in the angular momentum representation because different components 
of the angular momentum operator commute only in the formal limit $\hbar\to 0$. 
However, on classical level all the commutators can be neglected and different components
of the position vector and angular momentum can be treated independently.

In our approach the uncertainties of each component of the photon momentum and each component of the photon coordinate do not change with time. If in some problem those quantities can be treated as small then the photon can be treated as a pointlike particle moving along classical trajectory. 
So in our approach the coordinate photon WF never has a cosmic size and there
can be no paradoxes discussed in Sec. \ref{experiment}.

In view of the absence of the coordinate WF, such quantum problems as the single-photon
diffraction and interference 
 should be considered only in momentum representation. In particular, if boundary conditions are needed
they should be formulated in that representation. When a problem is solved and characteristic spatial dimensions in
the problem are greater than uncertainties of all the coordinates one can discuss spatial features of the process.

As an example, consider the double-slit experiment which is treated as a strong
confirmation of standard quantum theory. The explanation is that parts of the wave
function of an elementary particle projected on the slits pass the screen and interfere, and
the remaining part is absorbed by the screen. However, in view of the discussion of the paradoxes in
Sec. \ref{experiment}, it is not consistent to
treat the elementary particle by analogy with the classical wave different parts of which interact with the screen differently.
The problem of understanding the experiment is very difficult because here the WF of the elementary particle
does not have an anomalously large size and the particle strongly interacts with the screen.

As noted in Sec. \ref{WPW}, in  standard quantum theory photons
comprising a classical electromagnetic wave packet cannot be (approximately) treated as pointlike particles 
in view of the WPW paradox. However, in our approach, in view of the absence of WPS for massless particles, 
the usual intuition is restored
and photons comprising a divergent classical wave packet can be (approximately) treated as pointlike particles.
Moreover, the phenomenon of divergence of a classical wave packet can now be naturally explained simply as
a consequence of the fact that different photons in the packet have different momenta.

Our consideration also poses
a problem whether the results of classical electrodynamics can be applied for wave packets moving for a long period
of time. For example, as noted in Sec. \ref{experiment}, even classical theory predicts that when a
wave packet emitted in a gamma-ray burst or by a pulsar reaches the Earth, the width of the packet 
is extremely large (while the value predicted
by standard quantum theory is even much greater) and this poses a problem whether such a packet can be detected. 
A natural explanation of why classical theory does not apply in this case follows. 
As noted in Sec. \ref{momentum}, classical electromagnetic fields should be understood as a result of taking mean
characteristics for many photons. Then the fields will be (approximately) continuous if the density of the photons
is high. However, for a divergent beam of photons their density decreases with time. Hence after a long period of
time the mean characteristics of the photons in the beam cannot represent continuous fields. In other words, in this
situation the set of photons cannot be effectively described by classical electromagnetic fields.

The new position operator might also have applications in the problem of neutrino oscillations. As pointed out by several authors 
(see e.g. Ref. \cite{Naumov}) this problem should be considered from the point of
view that for describing observable neutrinos one should treat them as quantum superpositions of wave packets with
different neutrino flavors. Then the choice of the position operator might play an important role.
 
The position operator proposed in this section is also important in view of the following. There exists a wide 
literature discussing the Einstein-Podolsky-Rosen paradox, locality in quantum theory, quantum entanglement, Bell's 
theorem and similar problems (see e.g. Ref. \cite{Griffiths} and references therein). Consider, for example, the
following problem in standard theory. Let at $t=0$ particles 1 and 2 be localized inside finite volumes $V_1$ and $V_2$,
respectively, such that the volumes are very far from each other. Hence the particles don't interact with each other.
However, as follows from Eq. (\ref{trel}), their WFs will overlap at any $t>0$ and hence the interaction 
can be transmitted even with an infinite speed. This is often characterized as quantum nonlocality, entanglement and/or
action at a distance. 

Consider now this problem in the framework of our approach. Since in this approach there is no WF in
coordinate representation, there is no notion of a particle localized inside a finite volume.
Hence a problem arises whether on quantum level the notions of locality or nonlocality have a physical meaning. 
In addition, spreading does not take place in directions perpendicular to the particle momenta and for ultrarelativistic
particles spreading does not occur at all. Hence, at least in the case of ultrarelativistic particles, this kind of 
interaction does not occur in agreement with classical intuition that no interaction can be
transmitted with the speed greater than $c$. This example poses a problem whether the position operator should be
modified not only in directions perpendicular to particle momenta but also in longitudinal directions such that the effect of WPS should be excluded at all.

Our result for ultrarelativistic particles can be treated as ideal: quantum
theory reproduces the motion along a classical trajectory without any spreading. However, this is only a
special case of one free elementary particle. If quantum theory is treated as more general than the classical one
then it should describe not only elementary particles and atoms but even the motion of macroscopic bodies in the Solar
System and in the Universe. We believe that the 
assumption that the evolution of macroscopic bodies can be described by the Schr\"{o}dinger
equation is unphysical. For example, if the motion of the Earth is described by the 
evolution operator $exp[-iH(t_2-t_1)/\hbar]$ where $H$ is the Hamiltonian of the Earth then the quantity 
$H(t_2-t_1)/\hbar$ becomes of the order of
unity when $t_2-t_1$ is a quantity of the order of $10^{-68}s$ if the Hamiltonian is written in 
nonrelativistic form and $10^{-76}s$ if it is written in relativistic form. 
Such time intervals seem to be unphysical and so in the given case
the approximation when $t$ is a continuous parameter seems to be unphysical too. In modern 
theories (e.g. in the Big Bang hypothesis) it is often stated that the Planck time $t_P\approx 10^{-43}s$ 
is a physical minimum time interval. However, at present
there are no experiments confirming that time intervals of the order of $10^{-43}s$ can be measured.

The time dependent Schr\"{o}dinger equation has not been experimentally verified and the major theoretical
arguments in favor of this equation are as follows: a) the Hamiltonian is the generator of the time translation
in the Minkowski space; b) this equation becomes the Hamilton-Jacobi one in the formal limit $\hbar\to 0$.
However, as noted in Sec. \ref{intropos}, quantum theory should not be based on the space-time background
and the conclusion b) is made without taking into account the WPS effect. Hence the problem of describing evolution in quantum theory remains open.

Let us now return to the problem of the position operator.
As noted above, in directions perpendicular to the particle momentum the choice of the position
operator is based only on the requirement that semiclassical approximation should reproduce the standard 
relation ${\bf r}_{\bot}\times{\bf p}={\bf L}$. This requirement 
seems to be beyond any doubts since {\it on classical level} this relation is confirmed in numerous experiments. 
At the same time, the choice $i\hbar \partial/\partial p$
of the coordinate operator in the longitudinal direction is analogous to that in standard theory and hence
one might expect that this operator is physical if the magnitude of $p$ is rather large (see, however, the above
remark about the entanglement caused by WPS).

It will be shown in the next sections that the construction of the position operator described in this
chapter for the case of Poincare invariant theory can be generalized to the case of de Sitter (dS) invariant theory.
In this case the interpretation of the position operator is even
more important than in Poincare invariant theory. The reason is that even the free two-body mass operator
in the dS theory depends not only on the relative two-body momentum but also on the distance between the
particles. 

As argued in Sec. \ref{macroscopicdist}, in dS theory over a Galois field the assumption that the dS analog of the operator
$i\hbar \partial/\partial p$ is the operator of the longitudinal coordinate is not valid 
{\it for macroscopic bodies} 
(even if $p$ is large) since in that case semiclassical approximation is not valid. We have proposed a 
modification of the
position operator such that quantum theory reproduces for the two-body mass operator the mean value  
compatible with the Newton law of gravity. Then a problem arises
how quantum theory can reproduce classical evolution for macroscopic bodies. 

The above examples show that at macroscopic level a consistent definition 
of the transition from quantum to 
classical theory is the fundamental open problem.

\chapter{Basic properties of dS quantum theories}
\label{Ch2}

\section{dS invariance vs. AdS and Poincare invariance}
\label{dSvsAdS}
 
As already mentioned, one of the motivations for this work is to investigate whether standard gravity 
can be obtained in the framework of a free theory. In standard nonrelativistic approximation, 
gravity is characterized by the term $-Gm_1m_2/r$ in the mean value of the mass operator.
Here $m_1$ and $m_2$
are the particle masses and $r$ is the distance between
the particles. Since the kinetic energy is always positive,
the free nonrelativistic mass operator is positive definite
and therefore there is no way to obtain gravity in the
framework of the free theory. Analogously, in Poincare
invariant theory the spectrum of the free two-body mass
operator belongs to the interval $[m_1+m_2,\infty )$ while the existence of gravity necessarily requires
that the spectrum should contain values less than $m_1+m_2$.

In theories where the symmetry algebra is the AdS algebra so(2,3), the structure of IRs
is known (see e.g. Ref. \cite{Evans} and Chap. \ref{AdS}). In particular, for positive
energy IRs the AdS Hamiltonian has the spectrum in
the interval $[m,\infty )$ and $m$ has the meaning 
of the mass. Therefore the situation is pretty much
analogous to that in Poincare invariant theories.
In particular, the free two-body mass operator again
has the spectrum in the interval $[m_1+m_2,\infty )$
and therefore there is no way to reproduce gravitational
effects in the free AdS invariant theory.

As noted in Sec. \ref{CC}, the existing experimental data practically exclude the possibility 
that $\Lambda \leq 0$ and this is a strong argument in favor of dS symmetry vs. Poincare
and AdS ones. As argued in Sect. \ref{symmetry}, quantum theory should start not from space-time but
from a symmetry algebra. Therefore the choice of dS symmetry is natural and the cosmological
constant problem does not exist. However, as noted in Secs. \ref{CC} and \ref{inter}, 
the majority of physicists prefer to start from a flat
space-time and treat Poincare symmetry as fundamental while dS one as emergent. 

In contrast to the situation in Poincare and AdS invariant theories, the free mass operator in dS theory
is not bounded below by the value of $m_1+m_2$. The discussion in Sect. \ref{antigravity} shows that
this property by no means implies that the theory is unphysical. Therefore if one has a choice between Poincare, AdS
and dS symmetries then the only chance to describe gravity in a free theory is to choose
dS symmetry.

\section{IRs of the dS algebra}
\label{IRsdS}

In view of the definition of elementary particle discussed in Secs. \ref{symmetry} and \ref{momentum},
we accept that, {\it by definition}, elementary
particles in dS invariant theory are described by IRs of
the dS algebra by Hermitian operators. For different reasons,
there exists a vast literature not on such IRs but on unitary IRs (UIRs) of
the dS group. References to this literature can be found e.g.,
in our papers \cite{lev1a,lev3,jpa1} where we used the
results on UIRs of the dS group for constructing IRs of the dS
algebra by Hermitian operators. In this section we will
describe the construction proceeding from an excellent
description of UIRs of the dS group in the book by Mensky
\cite{Mensky}. The final result is given by explicit
expressions for the operators $M^{ab}$ in Eqs. (\ref{II20})
and (\ref{II21}). The readers who are not interested in technical
details can skip the derivation.

The elements of the SO(1,4) group will be described in the block form
\begin{equation}
g=\left\|\begin{array}{ccc}
g_0^0&{\bf a}^T&g_4^0\\
{\bf b}&r&{\bf c}\\
g_0^4&{\bf d}^T&g_4^4
\end{array}\right\|\
\label{II5}
\end{equation}
where
\begin{equation}
\label{II6}
{\bf a}=\left\|\begin{array}{c}a^1\\a^2\\a^3\end{array}\right\|, \quad
{\bf b}^T=\left\|\begin{array}{ccc}b_1&b_2&b_3\end{array}\right\|,
\quad r\in SO(3)
\end{equation}
and the subscript $^T$ means a transposed vector.

UIRs of the SO(1,4) group belonging to the principle series of
UIRs are induced from UIRs of the subgroup $H$ (sometimes
called ``little group'') defined as follows \cite{Mensky}. Each
element of $H$ can be uniquely represented as a product of
elements of the subgroups SO(3), $A$ and ${\bf T}$:
$h=r\tau_A{\bf a}_{\bf T}$ where
\begin{equation}
\tau_A=\left\|\begin{array}{ccc}
cosh(\tau)&0&sinh(\tau)\\
0&1&0\\
sinh(\tau)&0&cosh(\tau)
\end{array}\right\|\, \quad
{\bf a}_{\bf T}=\left\|\begin{array}{ccc}
1+{\bf a}^2/2&-{\bf a}^T&{\bf a}^2/2\\
-{\bf a}&1&-{\bf a}\\
-{\bf a}^2/2&{\bf a}^T&1-{\bf a}^2/2
\end{array}\right\|\
\label{II7}
\end{equation}
The subgroup $A$ is one-dimensional and the three-dimensional
group ${\bf T}$ is the dS analog of the conventional
translation group (see e.g., Ref. \cite{Mensky,Mielke}).
We believe it should not cause misunderstandings when 1 is used
in its usual meaning and when to denote the unit element of the
SO(3) group. It should also be clear when $r$ is a true element
of the SO(3) group or belongs to the SO(3) subgroup of the
SO(1,4) group. Note that standard UIRs of the Poincare group
are induced from the little group, which is a semidirect
product of SO(3) and four-dimensional translations and so the
analogy between UIRs of the Poincare and dS groups is clear.

Let $r\rightarrow \Delta(r;{\bf s})$ be an UIR of the group
SO(3) with the spin ${\bf s}$ and $\tau_A\rightarrow
exp(im_{dS}\tau)$ be a one-dimensional UIR of the group $A$,
where $m_{dS}$ is a real parameter. Then UIRs of the group $H$
used for inducing to the SO(1,4) group, have the form
\begin{equation}
\Delta(r\tau_A{\bf a}_{\bf T};m_{dS},{\bf s})=
exp(im_{dS}\tau)\Delta(r;{\bf s})
\label{II8}
\end{equation}
We will see below that $m_{dS}$ has the meaning of the dS mass
and therefore UIRs of the SO(1,4) group are defined by the mass
and spin, by analogy with UIRs in Poincare invariant theory.

Let $G$=SO(1,4) and $X=G/H$ be the factor space (or coset
space) of $G$ over $H$. The notion of the factor space is 
known (see e.g., Refs. \cite{Mackey,Mensky}). Each
element $x\in X$ is a class containing the elements $x_Gh$
where $h\in H$, and $x_G\in G$ is a representative of the class
$x$. The choice of representatives is not unique since if $x_G$
is a representative of the class $x\in G/H$ then $x_Gh_0$,
where $h_0$ is an arbitrary element from $H$, also is a
representative of the same class. It is known that $X$ can
be treated as a left $G$ space. This means that if $x\in X$
then the action of the group $G$ on $X$ can be defined as
follows: if $g\in G$ then $gx$ is a class containing $gx_G$ (it
is easy to verify that such an action is correctly defined).
Suppose that the choice of representatives is somehow fixed.
Then $gx_G=(gx)_G(g,x)_H$ where $(g,x)_H$ is an element of $H$.
This element is called a factor.

The explicit form of the operators $M^{ab}$ depends on the
choice of representatives in the space $G/H$. As explained in
works on UIRs of the SO(1,4) group (see e.g., Ref.
\cite{Mensky}), to obtain the  possible closest analogy between
UIRs of the SO(1,4) and Poincare groups, one should proceed as
follows. Let ${\bf v}_L$ be a representative of the Lorentz
group in the factor space SO(1,3)/SO(3) (strictly speaking, we
should consider $SL(2,C)/SU(2)$). This space can be represented
as the velocity hyperboloid with the Lorentz
invariant measure
\begin{equation}
d\rho({\bf v})=d^3{\bf v}/v_0
\label{II9}
\end{equation}
where $v_0=(1+{\bf v}^2)^{1/2}$. Let $I\in SO(1,4)$ be a matrix
which formally has the same form as the metric tensor $\eta$.
One can show (see e.g., Refs. \cite{Mensky} for details)
that $X=G/H$ can be represented as a union of three spaces,
$X_+$, $X_-$ and $X_0$ such that $X_+$ contains classes ${\bf
v}_Lh$, $X_-$ contains classes ${\bf v}_LIh$ and $X_0$ has
measure zero relative to the spaces $X_+$ and $X_-$ (see also
Sec. \ref{Other}).

As a consequence, the space of UIR of the SO(1,4) group can be
implemented as follows. If $s$ is the spin of the particle
under consideration, then we use $||...||$ to denote the norm
in the space of UIR of the group SU(2) with the spin $s$. Then
the space of UIR is the space of functions $\{f_1({\bf
v}),f_2({\bf v})\}$ on two Lorentz hyperboloids with the range
in the space of UIR of the group SU(2) with the spin $s$ and
such that
\begin{equation}
\int\nolimits [||f_1({\bf v})||^2+
||f_2({\bf v})||^2]d\rho({\bf v}) <\infty
\label{II10}
\end{equation}

It is known that positive energy UIRs of the Poincare and
AdS groups (associated with elementary particles) are
implemented on an analog of $X_+$ while negative energy UIRs
(associated with antiparticles) are implemented on an analog of
$X_-$. Since the Poincare and AdS groups do not contain
elements transforming these spaces to one another, the positive
and negative energy UIRs are fully independent. At the same
time, the dS group contains such elements (e.g. $I$
\cite{Mensky,Mielke}) and for this reason its UIRs can be
implemented only on the union of $X_+$ and $X_-$. Even this
fact is a strong indication that UIRs of the dS group cannot be
interpreted in the same way as UIRs of the Poincare and AdS
groups.

A general construction of the operators $M^{ab}$ follows.
We first define right invariant measures on $G=SO(1,4)$ and
$H$. It is known (see e.g. Ref. \cite{Mackey}) that
for semisimple Lie groups (which is the case for the dS group),
the right invariant measure is simultaneously the left
invariant one. At the same time, the right invariant measure
$d_R(h)$ on $H$ is not the left invariant one, but has the
property $d_R(h_0h) = \Delta(h_0)d_R(h)$, where the number
function $h\rightarrow \Delta(h)$ on $H$ is called the module
of the group $H$. It is easy to show \cite{Mensky} that
\begin{equation}
\Delta(r\tau_A{\bf a}_{\bf T})=exp(-3\tau)
\label{II11}
\end{equation}
Let $d\rho(x)$ be a measure on $X=G/H$ compatible with the
measures on $G$ and $H$. This implies that the measure on $G$
can be represented as $d\rho(x)d_R(h)$. Then one can show
\cite{Mensky} that if $X$ is a union of $X_+$ and $X_-$ then
the measure $d\rho(x)$ on each Lorentz hyperboloid coincides
with that given by Eq. (\ref{II9}). Let the representation
space be implemented as the space of functions $\varphi(x)$ on
$X$ with the range in the space of UIR of the SU(2) group such
that
\begin{equation}
\int\nolimits ||\varphi(x)||^2d\rho(x) <\infty
\label{II12}
\end{equation}
Then the action of the representation operator $U(g)$
corresponding to $g\in G$ is 
\begin{eqnarray}
U(g)\varphi(x)=[\Delta((g^{-1},x)_H)]^{-1/2}
\Delta((g^{-1},x)_H;m_{dS},{\bf s})^{-1}\varphi(g^{-1}x)
\label{II13}
\end{eqnarray}
One can directly verify that this expression defines a unitary
representation. Its irreducibility can be proved in several
ways (see e.g. Ref. \cite{Mensky}).

As noted above, if $X$ is the union of $X_+$ and $X_-$, then
the representation space can be implemented as in Eq.
(\ref{II8}). Since we are interested in calculating only the
explicit form of the operators $M^{ab}$, it suffices to
consider only elements of $g\in G$ in an infinitely small
vicinity of the unit element of the dS group. In that case one
can calculate the action of representation operators on
functions having the support in $X_+$ and $X_-$ separately.
Namely, as follows from Eq. (\ref{II11}), for such $g\in G$,
one has to find the decompositions
\begin{equation}
g^{-1}{\bf v}_L={\bf v}'_Lr'(\tau')_A({\bf a}')_{\bf T}
\label{II14}
\end{equation}
and
\begin{equation}
g^{-1}{\bf v}_LI={\bf v}"_LIr"(\tau")_A({\bf a}")_{\bf T}
\label{II15}
\end{equation}
where $r',r"\in SO(3)$. In this expressions it suffices to
consider only elements of $H$ belonging to an infinitely small
vicinity of the unit element.

The problem of choosing representatives in the spaces
SO(1,3)/SO(3) or SL(2.C)/SU(2) is known in standard
theory. The most usual choice is such that ${\bf v}_L$ as an
element of SL(2,C) is given by
\begin{equation}
{\bf v}_L=\frac{v_0+1+{\bf v}{\bf\sigma}}{\sqrt{2(1+v_0)}}
\label{II16}
\end{equation}
Then by using a known relation between elements of SL(2,C)
and SO(1,3) we obtain that ${\bf v}_L\in SO(1,4)$ is
represented by the matrix
\begin{equation}
{\bf v}_L=\left\|\begin{array}{ccc}
v_0&{\bf v}^T&0\\
{\bf v}&1+{\bf v}{\bf v}^T/(v_0+1)&0\\
0&0&1
\end{array}\right\|\
\label{II17}
\end{equation}

As follows from Eqs. (\ref{II8}) and (\ref{II13}), there is no
need to know the expressions for $({\bf a}')_{\bf T}$ and
$({\bf a}")_{\bf T}$ in Eqs. (\ref{II14}) and (\ref{II15}). We
can use the fact \cite{Mensky} that if $e$ is the
five-dimensional vector with the components
$(e^0=1,0,0,0,e^4=-1)$ and $h=r\tau_A{\bf a}_{\bf T}$, then
$he=exp(-\tau)e$ regardless of the elements $r\in SO(3)$ and
${\bf a}_{\bf T}$. This makes it possible to easily calculate
$({\bf v}'_L,{\bf v}"_L,(\tau')_A,(\tau")_A)$ in Eqs.
(\ref{II14}) and (\ref{II15}). Then one can calculate $(r',r")$ in
these expressions by using the fact that the SO(3) parts of the
matrices $({\bf v}'_L)^{-1}g^{-1}{\bf v}_L$ and $({\bf
v}"_L)^{-1}g^{-1}{\bf v}_L$ are equal to $r'$ and $r"$,
respectively.

The relation between the operators $U(g)$ and $M^{ab}$ 
follows. Let $L_{ab}$ be the basis elements of the Lie algebra
of the dS group. These are the matrices with the elements
\begin{equation}
(L_{ab})_d^c=\delta_d^c\eta_{bd}-\delta_b^c\eta_{ad}
\label{II18}
\end{equation}
They satisfy the commutation relations
\begin{equation}
[L_{ab},L_{cd}]=\eta_{ac}L_{bd}-\eta_{bc}L_{ad}-
\eta_{ad}L_{bc}+\eta_{bd}L_{ac}
\label{II19}
\end{equation}
Comparing Eqs. (\ref{CR}) and (\ref{II19}) it is easy to
conclude that the $M^{ab}$ should be the representation
operators of $-iL^{ab}$. Therefore if $g=1+\omega_{ab}L^{ab}$,
where a sum over repeated indices is assumed and the
$\omega_{ab}$ are such infinitely small parameters that
$\omega_{ab}=-\omega_{ba}$ then $U(g)=1+i\omega_{ab}M^{ab}$.

We are now in position to write down the final expressions for
the operators $M^{ab}$. Their action on functions with the
support in $X_+$ has the form
\begin{eqnarray}
&&{\bf J}^{(+)}=l({\bf v})+{\bf s},\quad {\bf N}^{(+)}=-i v_0
\frac{\partial}{\partial {\bf v}}+\frac{{\bf s}\times {\bf v}}
{v_0+1} \nonumber\\
&& {\bf B}^{(+)}=m_{dS} {\bf v}+i [\frac{\partial}{\partial {\bf v}}+
{\bf v}({\bf v}\frac{\partial}{\partial {\bf v}})+\frac{3}{2}{\bf v}]+
\frac{{\bf s}\times {\bf v}}{v_0+1}\nonumber\\
&& {\cal E}^{(+)}=m_{dS} v_0+i v_0({\bf v}
\frac{\partial}{\partial {\bf v}}+\frac{3}{2})
\label{II20}
\end{eqnarray}
where ${\bf J}=\{M^{23},M^{31},M^{12}\}$, ${\bf
N}=\{M^{01},M^{02},M^{03}\}$, ${\bf
B}=\{M^{41},M^{42},M^{43}\}$, ${\bf s}$ is the spin operator,
${\bf l}({\bf v})=-i{\bf v}\times \partial/\partial {\bf
v}$ and ${\cal E}=M^{40}$. At the same time, the action on functions with the support
in $X_-$ is given by
\begin{eqnarray}
&&{\bf J}^{(-)}=l({\bf v})+{\bf s},\quad {\bf N}^{(-)}=-i v_0
\frac{\partial}{\partial {\bf v}}+\frac{{\bf s}\times {\bf v}}
{v_0+1} \nonumber\\
&& {\bf B}^{(-)}=-m_{dS} {\bf v}-i [\frac{\partial}{\partial {\bf v}}+
{\bf v}({\bf v}\frac{\partial}{\partial {\bf v}})+\frac{3}{2}{\bf v}]-
\frac{{\bf s}\times {\bf v}}{v_0+1}\nonumber\\
&& {\cal E}^{(-)}=-m_{dS} v_0-i v_0({\bf v}
\frac{\partial}{\partial {\bf v}}+\frac{3}{2})
\label{II21}
\end{eqnarray}
Note that the expressions for the action of the Lorentz algebra
operators on $X_+$ and $X_-$ are the same and they coincide
with the corresponding expressions for IRs of the Poincare
algebra. At the same time, the expressions for the action of
the operators $M^{4\mu}$ on $X_+$ and $X_-$ differ by sign.

In deriving Eqs. (\ref{II20}) and (\ref{II21}) we have used only the commutation 
relations (\ref{CR}), no approximations have
been made and the results are exact. In particular, the dS
space, the cosmological constant and the Riemannian geometry
have not been involved at all. Nevertheless, the expressions
for the representation operators is all we need to have the
maximum possible information in quantum theory. 
As shown in the literature (see e.g. Ref. \cite{Mensky}), the above construction of IRs applies to IRs
of the principle series where $m_{dS}$ is a nonzero real parameter. Therefore such IRs
are called massive. 

A problem arises how $m_{dS}$ is related to standard particle mass $m$ in Poincare invariant theory.
In view of the contraction procedure described in Sec. \ref{symmetry}, one can 
assume that $m_{dS} > 0$ and define $m=m_{dS}/R$, ${\bf P}={\bf B}/R$ and $E={\cal E}/R$.
The set of operators $(E,{\bf P})$ is the Lorentz vector 
since its components can be written as $M^{4\nu}/R$ ($\nu = 0,1,2,3$). 
Then, as follows from  Eqs. (\ref{CR}), in the limit
when $R\rightarrow \infty$, $m_{dS}\rightarrow \infty$ but $m_{ds} /R$ is finite,   
one obtains from Eq. (\ref{II20}) a standard positive energy representation of the
Poincare algebra for a particle with the mass $m$ such 
that ${\bf P}=m{\bf v}$ is the particle momentum and $E=mv_0$ is the particle energy. 
Analogously one obtains a negative  energy representation from Eq. (\ref{II21}). Therefore $m$ is 
standard mass in Poincare invariant
theory and the operators of the Lorentz algebra $({\bf N},{\bf J})$ have the same form
for the Poincare and dS algebras. 

In Sect. \ref{CC} we have argued that fundamental physical
theory should not contain dimensionful parameters at all.
In this connection it is interesting to note that the
de Sitter mass $m_{dS}$ is a ratio of the radius of the Universe $R$ to the 
Compton wave length of the particle under consideration.
Therefore even for elementary particles the de Sitter
masses are very large. For example, if $R$ is of the order of 
$10^{26}m$ then the de Sitter masses of the electron,
the Earth and the Sun are of the order of $10^{39}$, $10^{93}$ and $10^{99}$, respectively. 
The fact that even the dS mass of the electron is so large might be an
indication that the electron is not a true elementary particle.

In addition, the present upper level for the photon mass is
$10^{-16}ev$ or less but this conclusion is based on the assumption that coordinate and momentum
representations are related to each other by the Fourier transform. This value
seems to be an extremely tiny quantity. However, the corresponding dS mass is of the order of $10^{17}$ and so even the mass which is treated as extremely small in Poincare
invariant theory might be very large in dS invariant~theory.

The operator ${\bf N}$ contains $i\partial/\partial {\bf v}$ which is
proportional to the standard coordinate operator $i\partial/\partial {\bf p}$. The
factor $v_0$ in ${\bf N}$ is needed for Hermiticity since the volume element is given
by Eq. (\ref{II9}). Such a construction can be treated as a relativistic generalization
of standard coordinate operator and then the orbital part of ${\bf N}$ is proportional to the Newton-Wigner
position operator \cite{NW}. However, as shown in Chap. \ref{WPS}, 
this operator does not satisfy all the requirements for the coordinate operator. 

In Poincare invariant theory the operator
$I_{2P}=E^2-{\bf P}^2$ is the Casimir operator, {\it i.e.}, it
commutes with all the representation operators. According to
the known Schur lemma in representation theory, all
elements in the space of IR are eigenvectors of the Casimir
operators with the same eigenvalue. In particular, they are the
eigenvectors of the operator $I_{2P}$ with the eigenvalue
$m^2$. As follows from Eq. (\ref{CR}), in the dS case the
Casimir operator of the second order is
\begin{eqnarray}
&&I_2 =-\frac{1}{2}\sum_{ab} M_{ab}M^{ab}={\cal E}^2+{\bf N}^2-{\bf B}^2-{\bf J}^2
\label{casimir}
\end{eqnarray}
and a direct calculation shows that for the operators (\ref{II20}) and (\ref{II21})
the numerical value of $I_2$ is $m_{dS}^2-s(s+1)+9/4$.
In Poincare invariant theory the value of the spin is related to the Casimir operator of the fourth
order which can be constructed from the Pauli-Lubanski vector. An analogous construction exists
in dS invariant theory but we will not dwell on this.

\section{Absence of Weyl particles in dS invariant theory}
\label{Weyl}

According to the present theoretical concepts, all 12 fundamental fermions of Standard Model (which is Poincare invariant)
are massless Weyl particles which acquired their masses as a result of the Higgs mechanism. Therefore a problem arises whether there exist analogs
of Weyl particles in dS invariant theory. In Poincare invariant theory, 
Weyl particles are characterized
not only by the condition that their mass is zero but also that they have a definite helicity. 
Typically the term "Weyl particles" refers only to the spin 1/2 particles which are described by the Weyl equation.
However, massless IRs characterized by definite helicity exist for any spin. Since we describe elementary particles only
by IRs, we will use the term "Weyl particles" for massless IRs of any spin characterized by only one value of helicity.  
Note that Weyl particles cannot exist in the theory where, in addition to Poincare invariance, invariance under spatial reflections
is required. The reason is that helicity changes its sign under spatial reflections. For example, IRs describing the photon
contain superpositions of states with the helicity +1 and -1. 

Several
authors investigated dS and AdS analogs of Weyl particles
proceeding from covariant equations on the dS and AdS spaces,
respectively. For example, the authors of Ref.
\cite{Ikeda} show that Weyl particles arise only when dS or
AdS symmetries are broken to Lorentz symmetry. In this work we investigate this problem
from the point of view of IRs.

At the level
of IRs, the existence of analogs of Weyl particles is known in
the AdS case. In Ref. \cite{tmf} we investigated such
analogs by using the results of Refs. \cite{Evans} for
standard IRs of the AdS algebra (i.e. IRs over the field
of complex numbers) and the results of Ref. \cite{Braden}
for IRs of the AdS algebra over a Galois field. Those results are described in Sec. \ref{Singletons}. Here it is
explained that in standard
case the minimum value of the AdS energy for massless IRs with
positive energy is $E_{min}=1+s$. In contrast to the
situation in Poincare invariant theory, where massless
particles cannot be in the rest state, massless particles
in the AdS theory do have rest states and the value of the $z$
projection of the spin in such states can be $-s,-s+1, ..., s$ as
usual. However, for any value of the energy greater than $E_{min}$,
the spin state is characterized only by helicity, which can
take the values either $s$ or $-s$, {\it i.e.}, we have the
same result as in Poincare invariant theory. 

In contrast to
IRs of the Poincare and dS algebra, IRs describing particles in
AdS theory belong to the discrete series of IRs and the energy
spectrum is discrete: $E=E_{min}, E_{min}+1,
..., \infty$. Therefore, strictly speaking, rest states do not
have measure zero although the probability that the
energy is exactly $E_{min}$ is extremely small. Nevertheless, as a consequence
of existence of rest states, the states with helicities  $s$ or $-s$ now belong
to the same IR. Therefore, strictly speaking, in the AdS case Weyl particles with definite helicities
cannot exist even if invariance under spatial reflections is not required.

In Poincare invariant theory, IRs describing Weyl particles can
be constructed by analogy with massive IRs but the little group
is now E(2) instead of SO(3) (see e.g.
Sec. 2.5 in the textbook \cite{Weinberg}). The matter is
that the representation operators of the SO(3) group transform
rest states into themselves but for massless particles there
are no rest states. However, there exists another way of
getting massless IRs: one can choose the variables for massive
IRs in such a way that the operators of massless IRs can be
directly obtained from the operators of massive IRs in the
limit $m\to 0$. This construction has been described by several
authors (see e.g. Refs. \cite{Kondratyuk,Fuda,current} and
references therein) and the main stages follow. First,
instead of the $(0,1,2,3)$ components of vectors, we work with
the so called light front components $(+,-,1,2)$ where
$v^{\pm}=(v^0\pm v^3)/\sqrt{2}$ and analogously for other
vectors. We choose $(v^+,{\bf v}_{\bot})$ as three independent
components of the 4-velocity vector, where ${\bf
v}_{\bot}=(v_x,v_y)$. In these variables the measure (\ref{II9})
on the Lorentz hyperboloid becomes $d\rho(v^+,{\bf
v}_{\bot})=dv^+d{\bf v}_{\bot}/v^+$. Instead of Eq.
(\ref{II16}) we now choose representatives of the SL(2,C)/SU(2)
classes as
\begin{equation}
v_L=\frac{1}{(v_0+v_z)^{1/2}}\left\|\begin{array}{cc}
v_0+v_z&0\\
v_x+iv_y&1
\end{array}\right\|\
\label{II22}
\end{equation}
and by using the relation between the groups SL(2,C) and SO(1,3) we obtain that the form of this representative in the Lorentz group is
\begin{equation}
v_L=\left\|\begin{array}{cccc}
\sqrt{2}v^+&0&0&0\\
\frac{{\bf v}_{\bot}^2}{\sqrt{2}v^+}&\frac{1}{\sqrt{2}v^+}&\frac{v_x}{v^+}&\frac{v_y}{v^+}\\
\sqrt{2}v_x&0&1&0\\
\sqrt{2}v_y&0&0&1
\end{array}\right\|\
\label{II23}
\end{equation}
where the raws and columns are in the order $(+,-,x,y)$.

By using the scheme described in the preceding section, we can
now calculate the explicit form of the representation operators
of the Lorentz algebra. In this scheme the form of those
operators in the IRs of the Poincare and dS algebras is the
same and in the case of the dS algebra the action is the same
for states with the support in $X_+$ and $X_-$. The results of
calculations are:
\begin{eqnarray}
&&M^{+-}=iv^+\frac{\partial}{\partial v^+},\quad M^{+j}=iv^+\frac{\partial}{\partial v^j},\quad M^{12}=l_z({\bf v}_{\bot})+s_z\nonumber\\
&&M^{-j}=-i(v^j\frac{\partial}{\partial v^+}+v^-\frac{\partial}{\partial v^j})-\frac{\epsilon_{jl}}{v^+}(s^l+v^ls_z)
\label{II24}
\end{eqnarray}
where a sum over $j,l=1,2$ is assumed and $\epsilon_{jl}$ has
the components $\epsilon_{12}=-\epsilon_{21}=1$,
$\epsilon_{11}=\epsilon_{22}=0$. In Poincare invariant theories
one can define standard four-momentum $p=mv$ and choose
$(p^+,{\bf p}_{\bot})$ as independent variables. Then the
expressions in Eq. (\ref{II24}) can be rewritten as
\begin{eqnarray}
&&M^{+-}=ip^+\frac{\partial}{\partial p^+},\quad M^{+j}=ip^+\frac{\partial}{\partial p^j},\quad M^{12}=l_z({\bf p}_{\bot})+s_z\nonumber\\
&&M^{-j}=-i(p^j\frac{\partial}{\partial p^+}+p^-\frac{\partial}{\partial p^j})-\frac{\epsilon_{jl}}{p^+}(ms^l+p^ls_z)
\label{II25}
\end{eqnarray}
In dS invariant theory we can work with the same variables if
$m$ is defined as $m_{dS}/R$.

As seen from Eqs. (\ref{II25}), only the operators $M^{-j}$
contain a dependence on the operators $s_x$ and $s_y$ but this
dependence disappears in the limit $m\to 0$. In this limit the
operator $s_z$ can be replaced by its eigenvalue $\lambda$
which now has the meaning of helicity. In Poincare invariant
theory the four-momentum operators $P^{\mu}$ are simply the
operators of multiplication by $p^{\mu}$ and therefore massless
particles are characterized only by one constant---helicity.

In dS invariant theory one can calculate the action of the
operators $M^{4\mu}$ by analogy with the calculation in the
preceding section. The actions of these operators on states
with the support in $X_+$ and $X_-$ differ only by sign and the
result for the actions on states with the support in $X_+$ is
\begin{eqnarray}
\label{II26}
&&M^{4-}=m_{dS}v^-+i[v^-(v^+\frac{\partial}{\partial v^+}+v^j\frac{\partial}{\partial v^j}+\frac{3}{2})-\frac{\partial}{\partial v^+}]+
\frac{1}{v^+}\epsilon_{jl}v^js^l\nonumber\\
&&M^{4j}=m_{dS}v^j+i[v^j(v^+\frac{\partial}{\partial v^+}+v^l\frac{\partial}{\partial v^l}+\frac{3}{2})+\frac{\partial}{\partial v^j}]
-\epsilon_{jl}s^l\nonumber\\
&&M^{4+}=m_{dS}v^++iv^+(v^+\frac{\partial}{\partial v^+}+v^j\frac{\partial}{\partial v^j}+\frac{3}{2})
\end{eqnarray}
If we define $m=m_{dS}/R$ and $p^{\mu}=mv^{\mu}$ then for the
operators $P^{\mu}$ we have
\begin{eqnarray}
\label{II27}
&&P^-=p^-+\frac{ip^-}{mR}(p^+\frac{\partial}{\partial p^+}+p^j\frac{\partial}{\partial p^j}+\frac{3}{2})-
\frac{im}{R}\frac{\partial}{\partial p^+}+\frac{1}{Rp^+}\epsilon_{jl}p^js^l\nonumber\\
&&P^j=p^j+\frac{ip^j}{mR}(p^+\frac{\partial}{\partial p^+}+p^l\frac{\partial}{\partial p^l}+\frac{3}{2})+
\frac{im}{R}\frac{\partial}{\partial p^j}-\frac{1}{R}\epsilon_{jl}s^l\nonumber\\
&&P^+=p^++\frac{ip^+}{mR}(p^+\frac{\partial}{\partial p^+}+p^j\frac{\partial}{\partial p^j}+\frac{3}{2})
\end{eqnarray}
Then it is clear that in the formal limit $R\to\infty$ we
obtain the standard Poincare result. However, when $R$ is
finite, the dependence of the operators $P^{\mu}$ on $s_x$ and
$s_y$ does not disappear. Moreover, in this case we cannot take
the limit $m\to 0$. 

Therefore, in contrast to the situation in Poincare invariant theory, where Weyl particles can be obtained from
massive IRs in the limit $m\to 0$, in dS theory it is not possible to obtain Weyl particles analogously, at least in the case when
elementary particles are described by IRs of the principle
series. Mensky conjectured \cite{Mensky} that massless
particles in dS invariant theory might correspond to IRs of
the discrete series with $-im_{dS}=1/2$ but this possibility
has not been investigated. In any case, in contrast to the
situation in Poincare invariant theory, the limit of massive
IRs when $m\to 0$ does not give Weyl particles and moreover,
this limit does not exist.

\section{Other implementations of IRs}
\label{Other}

In this section we briefly describe two more
implementations of IRs of the dS algebra. The first one is
based on the fact that since SO(1,4)=SO(4)$A{\bf T}$ and
$H$=SO(3)$A{\bf T}$ \cite{Mensky}, there also exists a choice
of representatives which is probably even more natural than
those described above. Namely, we can choose as representatives
the elements from the coset space SO(4)/SO(3). Since the
universal covering group for SO(4) is SU(2)$\times$SU(2) and
for SO(3) --- SU(2), we can choose as representatives the
elements of the first multiplier in the product
SU(2)$\times$SU(2). Elements of SU(2) can be represented by the
points $u=({\bf u},u_4)$ of the three-dimensional sphere $S^3$
in the four-dimensional space as $u_4+i{\bf \sigma}{\bf u}$
where ${\bf \sigma}$ are the Pauli matrices and $u_4=\pm
(1-{\bf u}^2)^{1/2}$ for the upper and lower hemispheres,
respectively. Then the calculation of the operators is similar
to that described above and the results follow. The
Hilbert space is now the space of functions $\varphi (u)$ on
$S^3$ with the range in the space of the IR of the su(2)
algebra with the spin $s$ and such that
\begin{equation}
\int\nolimits ||\varphi(u)||^2du <\infty
\label{II28}
\end{equation}
where $du$ is the SO(4) invariant volume element on $S^3$. The
explicit calculation  shows  that in this case the operators
have the form
\begin{eqnarray}
&&{\bf J}=l({\bf u})+{\bf s},\quad {\bf B}=i u_4
\frac{\partial}{\partial {\bf u}}-{\bf s}, \quad {\cal E}=(m_{dS} +3i/2)u_4+i u_4{\bf u}
\frac{\partial}{\partial {\bf u}} \nonumber\\
&& {\bf N}=-i [\frac{\partial}{\partial {\bf u}}-
{\bf u}({\bf u}\frac{\partial}{\partial {\bf u}})]
+(m_{dS} +3i/2){\bf u}-{\bf u}\times {\bf s}+u_4{\bf s}
\label{II29}
\end{eqnarray}
Since Eqs. (\ref{II10}), (\ref{II20}) and (\ref{II21}) on one
hand and Eqs. (\ref{II28}) and (\ref{II29}) on  the other  are
the  different implementations of  the   same
representation, there exists a unitary operator transforming
functions $f(v)$ into $\varphi (u)$ and operators
(\ref{II20},\ref{II21}) into operators (\ref{II29}). For example in
the spinless case the operators (\ref{II20}) and (\ref{II29}) are
related to each other by a unitary transformation
\begin{equation}
\varphi (u)=exp(-im_{dS}lnv_0)v_0^{3/2}f(v)
\label{II30}
\end{equation}
where the relation between the points of the upper hemisphere
and $X_+$ is ${\bf u}={\bf v}/v_0$ and $u_4=(1-{\bf
u}^2)^{1/2}$. The relation between the points of the lower
hemisphere and $X_-$ is ${\bf u}=-{\bf v}/v_0$ and
$u_4=-(1-{\bf u}^2)^{1/2}$.

The equator of $S^3$ where $u_4=0$ corresponds to $X_0$ and has
measure zero with respect to the upper and lower hemispheres.
For this reason one might think that it is of no interest for
describing particles in dS theory. Nevertheless, while none of the components of $u$ has the
magnitude greater than unity, the set $X_0$ in terms of
velocities is characterized by the condition that $|{\bf v}|$
is infinitely large and therefore standard Poincare
momentum ${\bf p}=m{\bf v}$ is infinitely large too. This poses
a question whether ${\bf p}$ always has a physical meaning.
From mathematical point of view Eq. (\ref{II29}) might seem
more convenient than Eqs. (\ref{II20}) and (\ref{II21}) since
$S^3$ is compact and there is no need to break it into the
upper and lower hemispheres. In addition, Eq. (\ref{II29})
is an explicit implementation of the idea that since in dS
invariant theory all the variables $(x^1,x^2,x^3,x^4)$ are on
equal footing and so(4) is the maximal compact kinematical
algebra, the operators ${\bf M}$ and ${\bf B}$ do not depend on
$m_{dS}$. However, those expressions are not convenient for
investigating Poincare approximation since the Lorentz boost
operators ${\bf N}$ depend on $m_{dS}$.

Finally, we describe an implementation of IRs based on the
explicit construction of the basis in the representation space.
This construction is based on the method of su(2)$\times$su(2)
shift operators, developed by Hughes \cite{Hug} for
constructing UIRs of the group SO(5). It will be convenient for
us to  deal with the set of operators $({\bf J}',{\bf
J}'',R_{ij})$ ($i,j=1,2$) instead  of $M^{ab}$. Here ${\bf J}'$
and ${\bf J}"$ are two independent su(2) algebras ({\it i.e.},
$[{\bf J}',{\bf J}'']=0$). In each of them one chooses as the
basis the operators $(J_+,J_-,J_3)$ such that $J_1=J_++J_-$,
$J_2=-\imath (J_+-J_-)$ and the commutation relations have the
form
\begin{equation}
[J_3,J_+]=2J_+,\quad [J_3,J_-]=-2J_-,\quad [J_+,J_-]=J_3
\label{II31}
\end{equation}
The commutation relations of the operators ${\bf J}'$ and
${\bf J}"$  with $R_{ij}$ have the form
\begin{eqnarray}
&&[J_3',R_{1j}]=R_{1j},\quad [J_3',R_{2j}]=-R_{2j},\quad
[J_3'',R_{i1}]=R_{i1},\nonumber\\
&& [J_3'',R_{i2}]=-R_{i2},\quad
[J_+',R_{2j}]=R_{1j},\quad [J_+'',R_{i2}]=R_{i1},\nonumber\\
&&[J_-',R_{1j}]=R_{2j},\quad [J_-'',R_{i1}]=R_{i2},\quad
[J_+',R_{1j}]=\nonumber\\
&&[J_+'',R_{i1}]=[J_-',R_{2j}]=[J_-'',R_{i2}]=0
\label{II32}
\end{eqnarray}
and the commutation relations of the operators $R_{ij}$
with each other have the form
\begin{eqnarray}
&&[R_{11},R_{12}]=2J_+',\quad
[R_{11},R_{21}]=2J_+'',\nonumber\\
&& [R_{11},R_{22}]=-(J_3'+J_3''),\quad
[R_{12},R_{21}]=J_3'-J_3''\nonumber\\
&& [R_{12},R_{22}]=-2J_-'',\quad [R_{21},R_{22}]=-2J_-'
\label{II33}
\end{eqnarray}
The relation between the sets $({\bf J}',{\bf J}",R_{ij})$ and $M^{ab}$  is given by
\begin{eqnarray}
&&{\bf J}=({\bf J}'+{\bf J}'')/2, \quad {\bf B}=({\bf J}'-{\bf J}'')/2,
\quad M_{01}=i(R_{11}-R_{22})/2, \nonumber\\
&& M_{02}=(R_{11}+R_{22})/2, \quad
M_{03}=-i(R_{12}+R_{21})/2,\nonumber\\
&&M_{04}=(R_{12}-R_{21})/2
\label{II34}
\end{eqnarray}
Then it is easy to see that Eq. (\ref{CR}) follows from
Eqs. (\ref{II32}--\ref{II34}) and {\it vice versa}.

Consider the space of maximal  $su(2)\times su(2)$  vectors,
{\it i.e.},  such vectors $x$ that $J_+'x=J_+''x=0$. Then from
Eqs. (\ref{II32}) and (\ref{II33}) it follows that the operators
\begin{eqnarray}
&&A^{++}=R_{11},\quad  A^{+-}=R_{12}(J_3''+1)-
J_-''R_{11},\quad A^{-+}=R_{21}(J_3'+1)-J_-'R_{11},\nonumber\\
&&A^{--}=-R_{22}(J_3'+1)(J_3''+1)+J_-''R_{21}(J_3'+1)+\nonumber\\
&&J_-'R_{12}(J_3''+1)-J_-'J_-''R_{11}
\label{II35}
\end{eqnarray}
act invariantly on this space. The notations are related to the
property  that if $x^{kl}$  ($k,l>0$) is the maximal
su(2)$\times$su(2) vector and simultaneously the eigenvector of
operators $J_3'$ and $J_3"$ with the eigenvalues $k$ and $l$,
respectively, then  $A^{++}x^{kl}$ is  the  eigenvector  of
the  same operators with the values $k+1$ and $l+1$,
$A^{+-}x^{kl}$ - the eigenvector  with the values $k+1$ and
$l-1$, $A^{-+}x^{kl}$ - the eigenvector with the values  $k-1$
and $l+1$ and $A^{--}x^{kl}$ - the eigenvector with the values
$k-1$ and $l-1$.

The basis in the representation space can be explicitly
constructed assuming that there exists a vector $e^0$ which is
the maximal su(2)$\times$su(2) vector such that
\begin{equation}
J_3'e_0=0,\quad J_3''e_0=se_0,\quad A^{--}e_0=A^{-+}e_0=0,\quad I_2e^0 =[m_{dS}^2-s(s+1)+9/4] e^0
\label{II36}
\end{equation}
Then, as shown in Ref. \cite{lev3}, the full basis of the
representation space consists of vectors
\begin{equation}
e_{ij}^{nr}=(J_-')^i (J_-'')^j(A^{++})^n(A^{+-})^re^0
\label{II37}
\end{equation}
where $n=0,1,2,..., r$ can take only the values $0,1,...,2s$
and for the given $n$ and $s$, $i$ can take the values
$0,1,...,n+r$ and $j$ can take the values $0,1,...,n+2s-r$.

These results show that IRs of the dS algebra can be
constructed purely algebraically without involving analytical
methods of the theory of UIRs of the dS group. As shown in
Ref. \cite{lev3}, this implementation is convenient for
generalizing standard quantum theory to FQT. In Chap. \ref{Ch3} we consider in detail the algebraic
construction of IRs in the spinless case and the results are applied to gravity.

\section{Physical interpretation of IRs of the dS algebra}
\label{InterpretationOfIRs}

In Secs. \ref{IRsdS}--\ref{Other} we discussed mathematical
properties of IRs of the dS algebra. In particular it has been
noted that they are implemented on two Lorentz hyperboloids,
not one as IRs of the Poincare algebra. Therefore the number of
states in IRs of the dS algebra is twice as big as in IRs of
the Poincare algebra. A problem arises whether this is
compatible with a requirement that any dS invariant theory
should become a Poincare invariant one in the formal limit
$R\to\infty$. Although there exists a wide literature on IRs of
the dS group and algebra, their physical interpretation has not
been widely discussed. Probably one of the reasons is that
physicists working on dS QFT treat fields as more fundamental
objects than particles (although the latter are observables
while the former are not).

In his book \cite{Mensky} Mensky notes that, in contrast to
IRs of the Poincare and AdS groups, IRs of the dS group
characterized by $m_{dS}$ and $-m_{dS}$ are unitarily
equivalent and therefore the energy sign cannot be used for
distinguishing particles and antiparticles. He proposes an
interpretation where a particle and its antiparticle are
described by the same IRs but have different space-time
descriptions (defined by operators intertwining IRs with
representations induced from the Lorentz group). Mensky shows
that in the general case his two solutions still cannot be
interpreted as a particle and its antiparticle, respectively,
since they are nontrivial linear combinations of functions with
different energy signs. However, such an interpretation is
recovered in Poincare approximation.

In view of the above discussion, it is desirable to give an
interpretation of IRs which does not involve space-time. In
Ref. \cite{jpa1} we have proposed an interpretation such
that one IR describes a particle and its antiparticle
simultaneously. In this section this analysis is extended.

\subsection{Problems with physical interpretation of IRs}

Consider first the case when the quantity $m_{dS}$ is very
large. Then, as follows from Eqs. (\ref{II20}) and
(\ref{II21}), the action of the operators $M^{4\mu}$ on states
localized on $X_+$ or $X_-$ can be approximately written as
$\pm m_{dS}v^{\mu}$, respectively. Therefore a question arises
whether standard Poincare energy $E$ can be defined as
$E=M_{04}/R$. Indeed, with such a definition, states localized
on $X_+$ will have a positive energy while states localized on
$X_-$ will have a negative energy. Then a question arises
whether this is compatible with the standard interpretation of
IRs, according to which the following requirements should be
satisfied:

{\it Standard-Interpretation Requirements:} Each element of the
full representation space represents a possible physical state
for the given elementary particle. The representation
describing a system of $N$ free elementary particles is the
tensor product of the corresponding single-particle
representations.

Recall that the operators of the tensor product are given by
sums of the corresponding single-particle operators. For
example, if ${\cal E}^{(1)}$ is the operator ${\cal E}$ for
particle 1 and ${\cal E}^{(2)}$ is the operator ${\cal E}$ for
particle 2 then the operator ${\cal E}$ for the free system
$\{12\}$ is given by ${\cal E}^{(12)}={\cal E}^{(1)}+{\cal E}^{(2)}$.
Here it is assumed that the action of the operator
${\cal E}^{(j)}$ ($j=1,2$) in the two-particle space is defined
as follows. It acts according to Eq. (\ref{II20}) or
(\ref{II21}) over its respective variables while over the
variables of the other particle it acts as the identity
operator.

One could try to satisfy the standard interpretation as follows.

A) Assume that in Poincare approximation standard energy
should be defined as $E = \pm {\cal E}/R$ where the plus sign
should be taken for the states with the support in $X_+$ and 
the minus sign---for the states with the support in $X_-$. Then
the energy will always be positive definite.

B) One might say that the choice of the energy sign is only a
matter of convention. Indeed, to measure the energy of a
particle with the mass $m$ one has to measure its momentum
${\bf p}$ and then the energy can be defined not only as
$(m^2+{\bf p}^2)^{1/2}$ but also as $-(m^2+{\bf p}^2)^{1/2}$.
In that case standard energy in the Poincare approximation
could be defined as $E = {\cal E}/R$ regardless of whether the
support of the given state is in $X_+$ or $X_-$.

It is easy to see that either of the above possibilities is
incompatible with Standard-Interpretation Requirements.
Consider, for example, a system of two free particles in the
case when $m_{dS}$ is very large. Then with a high accuracy the
operators ${\cal E}/R$ and ${\bf B}/R$ can be chosen diagonal
simultaneously.

Let us first assume that the energy should be treated according
to B). Then a system of two free particles with equal
masses can have the same quantum numbers as the vacuum (for
example, if the first particle has the energy $E_0=(m^2+{\bf
p}^2)^{1/2}$ and momentum ${\bf p}$ while the second one has
the energy $-E_0$ and the momentum $-{\bf p}$) what obviously
contradicts experiment. For this and other reasons it is known 
that in Poincare invariant theory the particles should
have the same energy sign. Analogously, if the single-particle
energy is treated according to A) then the result for the
two-body energy of a particle-antiparticle system will
contradict experiment.

We conclude that IRs of the dS algebra cannot be interpreted in
the standard way since such an interpretation is physically
meaningless even in Poincare approximation. The above
discussion indicates that the problem is similar to
that with the interpretation of the fact that the Dirac
equation has solutions with both, positive and negative
energies.

As already noted, in Poincare and AdS theories there exist
positive energy IRs implemented on the upper hyperboloid and
negative energy IRs implemented on the lower hyperboloid. In
the latter case Standard-Interpretation Requirements are not
satisfied for the reasons discussed above. However, we cannot
declare such IRs unphysical and throw them away. In QFT quantum
fields necessarily contain both types of IRs such that positive
energy IRs are associated with particles while negative energy
IRs are associated with antiparticles. Then the energy of
antiparticles can be made positive after proper second
quantization. In view of this observation, we will investigate
whether IRs of the dS algebra can be interpreted in such a way
that one IR describes a particle and its antiparticle
simultaneously such that states localized on $X_+$ are
associated with a particle while states localized on $X_-$ are
associated with its antiparticle.

By using Eq. (\ref{II10}), one can directly verify that the
operators (\ref{II20}) and (\ref{II21}) are Hermitian if the scalar
product in the space of IR is defined as follows. Since the
functions $f_1({\bf v})$ and $f_2({\bf v})$ in
Eq. (\ref{II10}) have the range in the space of IR of the
su(2) algebra with the spin $s$, we can replace them by the
sets of functions $f_1({\bf v},j)$ and $f_2({\bf v},j)$,
respectively, where $j=-s,-s+1,...,s$. Moreover, we can combine
these functions into one function $f({\bf v},j,\epsilon)$ where
the variable $\epsilon$ can take only two values, say +1 or -1,
for the components having the support in $X_+$ or $X_-$,
respectively. If now $\varphi({\bf v},j,\epsilon)$ and
$\psi({\bf v},j,\epsilon)$ are two elements of our Hilbert
space, their scalar product is defined as
\begin{equation}
(\varphi,\psi)=\sum_{j,\epsilon}\int\nolimits
\varphi({\bf v},j,\epsilon)^*\psi({\bf v},j,\epsilon)
d\rho({\bf v}
\label{II38})
\end{equation}
where the subscript $^*$ applied to scalar functions
means the usual complex conjugation.

At the same time, we use $^*$ to denote the operator adjoint to
a given one. Namely, if $A$ is the operator in our Hilbert
space then $A^*$ means the operator such that
\begin{equation}
(\varphi,A\psi)=(A^*\varphi,\psi)
\label{II39}
\end{equation}
for all such elements $\varphi$ and $\psi$ that the left hand
side of this expression is defined.

Even in the case of the operators (\ref{II20}) and (\ref{II21}) we
can formally treat them as integral operators with some
kernels. Namely, if $A\varphi=\psi$, we can treat this relation
as
\begin{equation}
\sum_{j',\epsilon'}\int\nolimits
A({\bf v},j,\epsilon;{\bf v}',j',\epsilon')
\varphi({\bf v}',j',\epsilon')d\rho({\bf v}')=
\psi({\bf v},j,\epsilon)
\label{II40}
\end{equation}
where in the general case the kernel $A({\bf v},j,\epsilon;{\bf
v}',j',\epsilon')$ of the operator $A$ is a distribution.

As follows from Eqs. (\ref{II38}--\ref{II40}), if $B=A^*$ then
the relation between the kernels of these operators is
\begin{equation}
B({\bf v},j,\epsilon;{\bf v}',j',\epsilon')=
A({\bf v}',j',\epsilon';{\bf v},j,\epsilon)^*
\label{II41}
\end{equation}
In particular, if the operator $A$ is Hermitian then
\begin{equation}
A({\bf v},j,\epsilon;{\bf v}',j',\epsilon')^*=
A({\bf v}',j',\epsilon';{\bf v},j,\epsilon)
\label{II42}
\end{equation}
and if, in addition, its kernel is real then the kernel is
symmetric, {\it i.e.},
\begin{equation}
A({\bf v},j,\epsilon;{\bf v}',j',\epsilon')=
A({\bf v}',j',\epsilon';{\bf v},j,\epsilon)
\label{II43}
\end{equation}
In particular, this property is satisfied for the operators
$m_{dS} v_0$ and $m_{dS} {\bf v}$ in Eqs. (\ref{II20}) and
(\ref{II21}). At the same time, the operators
\begin{equation}
l({\bf v})\quad -i v_0\frac{\partial}{\partial {\bf v}}
\quad -i [\frac{\partial}{\partial {\bf v}}+{\bf v}({\bf v}\frac{\partial}{\partial
{\bf v}})+\frac{3}{2}{\bf v}]\quad -i v_0({\bf v}\frac{\partial}{\partial {\bf v}}+\frac{3}{2})
\label{II44}
\end{equation}
which are present in Eqs. (\ref{II20}) and (\ref{II21}), are
Hermitian but have imaginary kernels. Therefore, as follows
from Eq. (\ref{II42}), their kernels are antisymmetric:
\begin{equation}
A({\bf v},j,\epsilon;{\bf v}',j',\epsilon')=-
A({\bf v}',j',\epsilon';{\bf v},j,\epsilon)
\label{II45}
\end{equation}

In standard approach to quantum theory, the operators of
physical quantities act in the Fock space of the given system.
Suppose that the system consists of free particles and their
antiparticles. Strictly speaking, in our approach it is not
clear yet what should be treated as a particle or antiparticle.
The considered IRs of the dS algebra describe objects such that
$({\bf v}, j, \epsilon)$ is the full set of their quantum
numbers. Therefore we can define the annihilation and creation
operators $(a({\bf v},j,\epsilon),a({\bf v},j,\epsilon)^*)$ for
these objects. If the operators satisfy the anticommutation
relations then we require that
\begin{equation}
\{a({\bf v},j,\epsilon),a({\bf v}',j',\epsilon')^*\}=
\delta_{jj'}\delta_{\epsilon\epsilon'}v_0
\delta^{(3)}({\bf v}-{\bf v}')
\label{II46}
\end{equation}
while in the case of commutation relations
\begin{equation}
[a({\bf v},j,\epsilon),a({\bf v}',j',\epsilon')^*]=
\delta_{jj'}\delta_{\epsilon\epsilon'}v_0
\delta^{(3)}({\bf v}-{\bf v}')
\label{II47}
\end{equation}
In the first case, any two $a$-operators or any two $a^*$
operators anticommute with each other while in the second case
they commute with each other.

The problem of second quantization can now be formulated such
that IRs should be implemented as Fock spaces, i.e.
states and operators should be expressed in terms of the
$(a,a^*)$ operators. A possible implementation follows.
We define the vacuum state $\Phi_0$ such that it has a unit
norm and satisfies the requirement
\begin{equation}
a({\bf v},j,\epsilon)\Phi_0=0\quad \forall\,\, {\bf v},j,\epsilon
\label{vacuum}
\end{equation}
The image of the state $\varphi({\bf v},j,\epsilon)$ in the
Fock space is defined as
\begin{equation}
\varphi_F=\sum_{j,\epsilon}\int\nolimits \varphi({\bf v},j,\epsilon)a({\bf v},j,\epsilon)^*d\rho({\bf v})\Phi_0
\label{oneparticle}
\end{equation}
and the image of the operator with the kernel $A({\bf v},j,\epsilon;{\bf v}',j',\epsilon')$
in the Fock space is defined as
\begin{equation}
A_F=\sum_{j,\epsilon,j',\epsilon'}\int\nolimits\int\nolimits
A({\bf v},j,\epsilon;{\bf v}',j',\epsilon')
a({\bf v},j,\epsilon)^*a({\bf v}',j',\epsilon')
d\rho({\bf v})d\rho({\bf v}')
\label{II48}
\end{equation}
One can directly verify that this is an implementation of IR in
the Fock space. In particular, the commutation relations in the
Fock space will be preserved regardless of whether the
$(a,a^*)$ operators satisfy commutation or anticommutation
relations and, if any two operators are adjoint in the
implementation of IR described above, they will be adjoint in
the Fock space as well. In other words, we have a $^*$
homomorphism of Lie algebras of operators acting in the space
of IR and in the Fock space.

We now require that in Poincare approximation the energy should
be positive definite. Recall that the operators (\ref{II20}) and
(\ref{II21}) act on their respective subspaces or in other words,
they are diagonal in the quantum number $\epsilon$.

Suppose that $m_{dS} > 0$ and consider the quantized operator
corresponding to the dS energy ${\cal E}$ in Eq. (\ref{II20}).
In Poincare approximation, ${\cal E}^{(+)}=m_{dS} v_0$ is fully
analogous to the standard free energy and therefore, as follows
from Eq. (\ref{II48}), its quantized form is
\begin{equation}
({\cal E}^{(+)})_F=m_{dS}\sum_{j}\int\nolimits v_0
a({\bf v},j,1)^*a({\bf v},j,1)d\rho({\bf v})
\label{II49}
\end{equation}
This expression is fully analogous to the quantized Hamiltonian
in standard theory and it is known that the operator
defined in such a way is positive definite.

Consider now the operator $M_{04}^{(-)}$. In Poincare
approximation its quantized form is
\begin{equation}
({\cal E}^{(-)})_F==-m_{dS}\sum_{j}\int\nolimits v_0
a({\bf v},j,-1)^*a({\bf v},j,-1)d\rho({\bf v})
\label{II51}
\end{equation}
and this operator is negative definite, what is unacceptable.

One might say that the operators $a({\bf v},j,-1)$ and $a({\bf
v},j,-1)^*$ are ``nonphysical'': $a({\bf v},j,-1)$ is the
operator of object's annihilation with the negative energy, and
$a({\bf v},j,-1)^*$ is the operator of object's creation with
the negative energy.

We will interpret the operator $({\cal E}^{(-)})_F$ as that
related to antiparticles. In QFT the annihilation and creation
operators for antiparticles are present in quantized fields
with the coefficients describing negative energy solutions of
the corresponding covariant equation. This is an implicit
implementation of the idea that the creation or annihilation of
an antiparticle can be treated, respectively as the
annihilation or creation of the corresponding particle with the
negative energy. In our case this idea can be implemented
explicitly.

Instead of the operators $a({\bf v},j,-1)$ and $a({\bf
v},j,-1)^*$, we define new operators $b({\bf v},j)$ and $b({\bf
v},j)^*$. If $b({\bf v},j)$ is treated as the ``physical"
operator of antiparticle annihilation then, according to the
above idea, it should be proportional to $a({\bf v},-j,-1)^*$.
Analogously, if $b({\bf v},j)^*$ is the ``physical" operator of
antiparticle creation, it should be proportional to $a({\bf
v},-j,-1)$. Therefore
\begin{equation}
b({\bf v},j)=\eta (j)a({\bf v},-j,-1)^*\quad b({\bf v},j)^*= \eta (j)^*a({\bf v},-j,-1)
\label{II53}
\end{equation}
where $\eta (j)$ is a phase factor such that
\begin{equation}
\eta (j)\eta(j)^*=1
\label{II54}
\end{equation}
As follows from this relations, if a particle is characterized
by additive quantum numbers (e.g., electric, baryon or lepton
charges) then its antiparticle is characterized by the same
quantum numbers but with the minus sign. The transformation
described by Eqs. ({\ref{II53}) and (\ref{II54}) can also be
treated as a special case of the Bogolubov transformation
discussed in a wide literature on many-body theory (see, e.g.,
Chap. 10 in Ref. \cite{Walecka} and references therein).

Since we treat $b({\bf v},j)$ as the annihilation operator and
$b({\bf v},j)^*$ as the creation one, instead of
Eq. (\ref{vacuum}) we should define a new vacuum state
${\tilde \Phi}_0$ such that
\begin{equation}
a({\bf v},j,1){\tilde \Phi}_0=b({\bf v},j){\tilde \Phi}_0=0\quad \forall\,\, {\bf v},j,
\label{II55}
\end{equation}
and the images of states localized in $X_-$ should be defined as
\begin{equation}
\varphi_F^{(-)}=\sum_{j,\epsilon}\int\nolimits \varphi({\bf v},j,-1)b({\bf v},j)^*d\rho({\bf v})
{\tilde \Phi}_0
\label{newoneparticle}
\end{equation}
In that case the $(b,b^*)$ operators should be such that in the
case of anticommutation relations
\begin{equation}
\{b({\bf v},j),b({\bf v}',j')^*\}=
\delta_{jj'}v_0 \delta^{(3)}({\bf v}-{\bf v}'),
\label{II56}
\end{equation}
and in the case of commutation relations
\begin{equation}
[b({\bf v},j),b({\bf v}',j')^*]=
\delta_{jj'}v_0 \delta^{(3)}({\bf v}-{\bf v}')
\label{II57}
\end{equation}
We have to verify whether the new definition of the vacuum and
one-particle states is a correct implementation of IR in the
Fock space. A necessary condition is that the new operators
should satisfy the commutation relations of the dS algebra.
Since we replaced the $(a,a^*)$ operators by the $(b,b^*)$
operators only if $\epsilon=-1$, it is obvious from Eq.
(\ref{II48}) that the images of the operators (\ref{II20}) in the
Fock space satisfy Eq. (\ref{CR}). Therefore we have to
verify that the images of the operators (\ref{II21}) in the Fock
space also satisfy Eq. (\ref{CR}).

Consider first the case when the operators $a({\bf
v},j,\epsilon)$ satisfy the anticommutation relations. By using
Eq. (\ref{II53}) one can express the operators $a({\bf
v},j,-1)$ in terms of the operators $b({\bf v},j)$. Then it
follows from the condition (\ref{II53}) that the operators
$b({\bf v},j)$ indeed satisfy Eq. (\ref{II57}). If the
operator $A_F$ is defined by Eq. (\ref{II48}) and is
expressed only in terms of the $(a,a^*)$ operators at
$\epsilon=-1$, then in terms of the $(b,b^*)$-operators it acts
on states localized in $X_-$ as
\begin{equation}
A_F=\sum_{j,j'}\int\nolimits\int\nolimits
A({\bf v},j,-1;{\bf v}',j',-1)\eta(j')\eta(j)^*b({\bf v},-j)b({\bf v}',-j')^*
d\rho({\bf v})d\rho({\bf v}')
\label{II58}
\end{equation}
As follows from Eq. (\ref{II57}), this operator can be written as
\begin{equation}
A_F=C-\sum_{j,j'}\int\nolimits\int\nolimits
A({\bf v}',-j',-1;{\bf v},-j,-1)\eta(j)\eta(j')^*b({\bf v},j)^*b({\bf v}',j')
d\rho({\bf v})d\rho({\bf v}')
\label{II59}
\end{equation}
where $C$ is the trace of the operator $A_F$
\begin{equation}
C=\sum_j\int\nolimits A({\bf v},j,-1;{\bf v},j,-1)d\rho({\bf v})
\label{II60}
\end{equation}
and in general it is some an indefinite constant. The existence of infinities 
in standard approach is the well-known problem. Usually the infinite
constant is eliminated by requiring that all quantized operators should be
written in the normal form or by using another prescriptions.
However, in dS theory this constant cannot be eliminated since IRs
are defined on the space which is a direct some of $X_+$ and $X_-$, and
the constant inevitably arise when one wishes to have an interpretation of
IRs in terms of particles and antiparticles. In Sec. \ref{VS6} we consider
an example when a constant, which is infinite in standard theory, becomes
zero in FQT.

In this chapter we assume that neglecting the constant C can be somehow
justified. In that case if the operator $A_F$ is defined by
Eq. (\ref{II48}) then in the case of anticommutation
relations its action on states localized in $X_-$ can be
written as in Eq. (\ref{II59}) with $C=0$. Then, taking into
account the properties of the kernels discussed above, we
conclude that in terms of the $(b,b^*)$-operators the kernels
of the operators $(m_{dS}v)_F$ change their sign while the
kernels of the operators in Eq. (\ref{II44}) remain the
same. In particular, the operator $(-m_{dS}v_0)_F$ acting on
states localized on $X_-$ has the same kernel as the operator
$(m_{ds}v_0)_F$ acting on states localized in $X_+$ has in
terms of the $a$-operators. This implies that in Poincare
approximation the energy of the states localized in $X_-$ is
positive definite, as well as the energy of the states
localized in $X_+$.

Consider now how the spin operator changes when the
$a$-operators are replaced by the $b$-operators. Since the spin
operator is diagonal in the variable ${\bf v}$, it follows from
Eq. (\ref{II59}) that the transformed spin operator will
have the same kernel if
\begin{equation}
s_i(j, j')=-\eta(j)\eta(j')^*s_i(-j',-j)
\label{II61}
\end{equation}
where $s_i(j, j')$ is the kernel of the operator $s_i$. For the
$z$ component of the spin operator this relation is obvious
since $s_z$ is diagonal in $(j,j')$ and its kernel is
$s_z(j,j')=j\delta_{jj'}$.If we choose $\eta(j)=(-1)^{(s-j)}$
then the validity of Eq. (\ref{II61}) for $s=1/2$ can be
verified directly while in the general case it can be verified
by using properties of $3j$ symbols.

The above results for the case of anticommutation relations can
be summarized as follows. If we replace $m_{dS}$ by $-m_{dS}$
in Eq. (\ref{II21}) then the new set of operators
\begin{eqnarray}
&&{\bf J}'=l({\bf v})+{\bf s},\quad {\bf N}'=-i v_0
\frac{\partial}{\partial {\bf v}}+\frac{{\bf s}\times {\bf v}}
{v_0+1}, \nonumber\\
&& {\bf B}'=m_{dS} {\bf v}-i [\frac{\partial}{\partial {\bf v}}+
{\bf v}({\bf v}\frac{\partial}{\partial {\bf v}})+\frac{3}{2}{\bf v}]-
\frac{{\bf s}\times {\bf v}}{v_0+1},\nonumber\\
&& {\cal E}'=m_{dS} v_0-i v_0({\bf v}
\frac{\partial}{\partial {\bf v}}+\frac{3}{2})
\label{II62}
\end{eqnarray}
obviously satisfies the commutation relations (\ref{CR}). The
kernels of these operators define quantized operators in terms
of the $(b,b^*)$-operators in the same way as the kernels of
the operators (\ref{II20}) define quantized operators in terms of
the $(a,a^*)$-operators. In particular, in Poincare
approximation the energy operator acting on states localized in
$X_-$ can be defined as $E'={\cal E}'/R$ and in this
approximation it is positive definite.

At the same time, in the case of commutation relation the
replacement of the $(a,a^*)$-operators by the
$(b,b^*)$-operators is unacceptable for several reasons. First
of all, if the operators $a({\bf v},j,\epsilon)$ satisfy the
commutation relations (\ref{II47}), the operators defined by
Eq. (\ref{II53}) will not satisfy Eq. (\ref{II57}). Also,
the r.h.s. of Eq. (\ref{II59}) will now have the opposite
sign. As a result, the transformed operator ${\cal E}$ will
remain negative definite in Poincare approximation and the
operators (\ref{II44}) will change their sign. In particular, the
angular momentum operators will no longer satisfy correct
commutation relations.

We have shown that if the definitions (\ref{vacuum}) and
(\ref{oneparticle}) are replaced by (\ref{II55}) and
(\ref{newoneparticle}), respectively, then the images of both
sets of operators in Eq. (\ref{II20}) and Eq.
(\ref{II21}) satisfy the correct commutation relations in the
case of anticommutators. A question arises whether the new
implementation in the Fock space is equivalent to the IR
described in Sec. \ref{IRsdS}. For understanding the essence of
the problem, the following very simple pedagogical example
might be useful.

Consider a representation of the SO(2) group in the space of
functions $f(\varphi)$ on the circumference $\varphi \in
[0,2\pi]$ where $\varphi$ is the polar angle and the points
$\varphi =0$ and $\varphi =2\pi$ are identified. The generator
of counterclockwise rotations  is $A=-id/d\varphi$ while the
generator of clockwise rotations is $B=id/d\varphi$. The equator
of the circumference contains two points, $\varphi=0$ and
$\varphi=\pi$ and has measure zero. Therefore we can represent
each $f(\varphi)$ as a superposition of functions with the
supports in the upper and lower semi circumferences, $S_+$ and
$S_-$. The operators $A$ and $B$ are defined only on
differentiable functions. The Hilbert space $H$ contains not
only such functions but a set of differentiable functions is
dense in $H$. If a function $f(\varphi)$ is differentiable and
has the support in $S_+$ then $Af(\varphi)$ and $Bf(\varphi)$
also have the support in $S_+$ and analogously for functions
with the support in $S_-$. However, we cannot define a
representation of the SO(2) group such that its generator is
$A$ on functions with the support in $S_+$ and $B$ on functions
with the support in $S_-$ because a counterclockwise rotation
on $S_+$ should be counterclockwise on $S_-$ and analogously
for clockwise rotations. In other words, the actions of the
generator on functions with the supports in $S_+$ and $S_-$
cannot be independent.

In the case of finite dimensional representations, any IR of a
Lie algebra by Hermitian operators can be always extended to an
UIR of the corresponding Lie group. In that case the UIR has a
property that any state is its cyclic vector i.e. the
whole representation space can be obtained by acting by
representation operators on this vector and taking all possible
linear combinations. For infinite dimensional IRs this is not
always the case and there should exist conditions for IRs of
Lie algebras by Hermitian operators to be extended to
corresponding UIRs. This problem has been extensively discussed
in the mathematical literature (see e.g. Ref.
\cite{Mackey}). By analogy with finite dimensional IRs, one
might think that in the case of infinite dimensional IRs there
should exist an analog of the cyclic vector. In Sec.
\ref{Other} we have shown that for infinite dimensional IRs of the
dS algebra this idea can be explicitly implemented by choosing
a cyclic vector and acting on this vector by operators of the
enveloping algebra of the dS algebra. This construction shows
that the action of representation operators on states with the
support in $X_+$ should define its action on states with the
support in $X_-$, i.e. the action of representation operators on states 
with the supports in $X_+$ and $X_-$ are not independent.

\subsection{Example of transformation mixing particles and antiparticles}

We treated states localized in $X_+$ as particles and states
localized in $X_-$ as corresponding antiparticles. However, the
space of IR contains not only such states. There is no rule
prohibiting states with the support having a nonempty
intersection with both, $X_+$ and $X_-$. Suppose that there
exists a unitary transformation belonging to the UIR of the dS
group such that it transforms a state with the support in $X_+$
to a state with the support in $X_-$. If the Fock space is
implemented according to Eqs. (\ref{vacuum}) and
(\ref{oneparticle}) then the transformed state will have the
form
\begin{equation}
\varphi_F^{(-)}=\sum_j\int\nolimits \varphi({\bf v},j)a({\bf v},j,-1)^*d\rho({\bf v})\Phi_0
\label{transform}
\end{equation}
while with the implementation in terms of the $(b,b^*)$
operators it should have the form (\ref{newoneparticle}). Since
the both states are obtained from the same state with the
carrier in $X_+$, they should be the same. However, they cannot
be the same. This is clear even from the fact that in Poincare
approximation the former has a negative energy while the latter
has a positive energy.

Our construction shows that the interpretation of states as
particles and antiparticles is not always consistent. It can be
only approximately consistent when we
consider only states localized either in $X_+$ or in $X_-$ and
only transformations which do not mix such states. In quantum
theory there is a superselection rule (SSR) prohibiting states
which are superpositions of states with different electric,
baryon or lepton charges. In general, if states $\psi_1$ and
$\psi_2$ are such that there are no physical operators $A$ such
that $(\psi_2,A\psi_1)\neq 0$ then the SSR says that the state
$\psi=\psi_1+\psi_2$ is prohibited. The meaning of the SSR is
now widely discussed (see e.g., Ref. \cite{Giulini} and
references therein). Since the SSR implies that the
superposition principle, which is a key principle of quantum
theory, is not universal, several authors argue that the SSR
should not be present in quantum theory. Other authors argue
that the SSR is only a dynamical principle since, as a result
of decoherence, the state $\psi$ will quickly disappear and so
it cannot be observable.

We now give an example of a transformation, which transforms
states localized in $X_+$ to ones localized in $X_-$ and {\it
vice versa}. Let $I\in SO(1,4)$ be a matrix which formally
coincides with the metric tensor $\eta$. If this matrix is
treated as a transformation of the dS space, it transforms the
North pole $(0,0,0,0,x^4=R)$ to the South pole
$(0,0,0,0,x^4=-R)$ and {\it vice versa}. As already explained,
in our approach the dS space is not involved and in Secs.
\ref{IRsdS}--\ref{Other} the results for UIRs of the dS group have
been used only for constructing IRs of the dS algebra. This
means that the unitary operator $U(I)$ corresponding to $I$ is
well defined and we can consider its action without relating
$I$ to a transformation of the dS space.

If ${\bf v}_L$ is a representative defined by Eq.
(\ref{II17}) then it is easy to verify that $I{\bf v}_L=({-\bf
v})_L I$ and, as follows from Eq. (\ref{II13}), if $\psi_1$
is localized in $X_+$ then $\psi_2=U(I)\psi_1$ will be
localized in $X_-$. Therefore $U(I)$ transforms particles into
antiparticles and {\it vice versa}. In Secs. \ref{ST} and \ref{symmetry} we
argued that the notion of empty space-time background is
unphysical and that unitary transformations generated by
self-adjoint operators may not have a usual interpretation. The
example with $U(I)$ gives a good illustration of this point.
Indeed, if we work with dS space, we might expect
that all unitary transformations corresponding to the elements
of the group SO(1,4) act in the space of IR only kinematically,
in particular they transform particles to particles and
antiparticles to antiparticles. However, in QFT in curved space-time 
this is not the case. Nevertheless, this is not treated as an indication that 
standard notion of the dS space is not physical.
Although fields are not observable, in QFT in curved space-time they 
are treated as fundamental and single-particle interpretations of field equations
are not tenable (moreover, some QFT theorists state that particles do not
exist). For example, as shown in Ref. \cite{Akhmedov2}, solutions of fields equations are
superpositions of states which usually are interpreted as a
particle and its antiparticle, and in dS space neither
coefficient in the superposition can be zero. This result is
compatible with the Mensky's one \cite{Mensky} described in the
beginning of this section. One might say that our result is in
agreement with those in dS QFT since UIRs of the dS group
describe not a particle or antiparticle but an object such that
a particle and its antiparticle are different states of this
object (at least in Poincare approximation). However, as noted above,
in dS QFT this is not treated as the fact that dS space is
unphysical.

The matrix $I$ belongs to the component of unity of the group
SO(1,4). For example, the transformation $I$ can be obtained as
a product of rotations by 180 degrees in planes $(1,2)$ and
$(3,4)$. Therefore, $U(I)$ can be obtained as a result of
continuous transformations
$exp[i(M_{12}\varphi_1+M_{34}\varphi_2)]$ when the values of
$\varphi_1$ and $\varphi_2$ change from zero to $\pi$. Any
continuous transformation transforming a state with the carrier
in $X_+$ to the state with the support in $X_-$ is such that
the support should cross $X_0$ at some values of the
transformation parameters. As noted in the preceding section,
the set $X_0$ is characterized by the condition that the
standard Poincare momentum is infinite and therefore, from the
point of view of intuition based on Poincare invariant theory,
one might think that no transformation when the support crosses
$X_0$ is possible. However, as we have seen in the preceding
section, in variables $(u_1,u_2,u_3,u_4)$ the condition $u_4=0$
defines the equator of $S^3$ corresponding to $X_0$ and this
condition is not singular. So from the point of view of dS
theory, nothing special happens when the support crosses $X_0$.
We observe only either particles or antiparticles but not their
linear combinations because Poincare approximation works with a
very high accuracy and it is very difficult to perform
transformations mixing states localized in $X_+$ and $X_-$.

\subsection{Summary}

As follows from the above discussion, {\it objects belonging to
IRs of the dS algebra can be treated as particles or
antiparticles only if Poincare approximation works with a high
accuracy}. As a consequence, {\it the conservation of electric,
baryon and lepton charges can be only approximate}.

At the same time, our discussion shows that the approximation
when one IR of the dS algebra splits into independent IRs for a
particle and its antiparticle can be valid only in the case of
anticommutation relations. Since it is a reasonable requirement
that dS theory should become the Poincare one at certain
conditions, the above results show that {\it in dS invariant
theory only fermions can be elementary}.

Let us now consider whether there exist neutral particles in dS
invariant theory. In AdS and Poincare invariant theories,
neutral particles are described as follows. One first constructs
a covariant field containing both IRs, with positive and
negative energies. Therefore the number of states is doubled in
comparison with the IR. However, to satisfy the requirement
that neutral particles should be described by real (not
complex) fields, one has to impose a relation between the
creation and annihilation operators for states with positive
and negative energies. Then the number of states describing a
neutral field again becomes equal to the number of states in
the IR. In contrast to those theories, IRs of the dS algebra
are implemented on both, upper and lower Lorentz hyperboloids
and therefore the number of states in IRs is twice as big as
for IRs of the Poincare and AdS algebras. Even this fact shows
that in dS invariant theory there can be no neutral particles
since it is not possible to reduce the number of states in an IR.
Another argument is that, as follows from the above
construction, dS invariant theory is not $C$ invariant. Indeed,
$C$ invariance in standard theory means that representation
operators are invariant under the interchange of $a$-operators
and $b$-operators. However, in our case when $a$-operators are
replaced by $b$-operators, the operators (\ref{II20}) become the
operators (\ref{II62}). Those sets of operators coincide only in
Poincare approximation while in general the operators
$M^{4\mu}$ in Eqs. (\ref{II20}) and (\ref{II62}) are
different. Therefore a particle and its antiparticle are
described by different sets of operators. We conclude that {\it
in dS invariant theory neutral particles cannot be elementary}.

\section{dS quantum mechanics and cosmological repulsion}
\label{antigravity}

The results on IRs can be applied not only to elementary
particles but even to macroscopic bodies when it suffices to
consider their motion as a whole. This is the case when the
distances between the bodies are much greater that their sizes.
In this section we consider the operators $M^{4\mu}$ not
only in Poincare approximation but taking into account dS
corrections. If those corrections are small, one can neglect
transformations mixing states on the upper and lower Lorentz
hyperboloids (see the discussion in the preceding section) and
describe the representation operators for a particle and its
antiparticle by Eqs. (\ref{II20}) and (\ref{II62}),
respectively.

We define $E={\cal E}/R$, ${\bf P}={\bf B}/R$ and $m=m_{dS}/R$.
Consider the non-relativistic approximation when $|{\bf v}|\ll
1$. If we wish to work with units where the dimension of
velocity is $m/s$, we should replace ${\bf v}$ by ${\bf
v}/c$. If ${\bf p}=m{\bf v}$ then it is clear from the
expressions for ${\bf B}$ in Eqs. (\ref{II20}) and
(\ref{II62}) that ${\bf p}$ becomes the real momentum ${\bf P}$
only in the limit $R\to\infty$. At this
stage we do not have any coordinate space yet. However, if we
assume that semiclassical approximation is valid, then, by analogy with
standard quantum mechanics, we can {\it define} the position
operator ${\bf r}$ as $i\partial/\partial {\bf p}$. As discussed in Chap. \ref{WPS}, such a definition
encounters problems in view of the WPS effect. However, as noted in this chapter, this effect is a pure
quantum phenomenon and for macroscopic bodies it is negligible. The problem of the cosmological acceleration 
is meaningful only for macroscopic bodies when classical approximation applies.

Since the commutators of ${\cal R}_{||}$ and 
${\bf {\cal R}}_{\bot}$ with different components of ${\bf p}$ are proportional to $\hbar$ and the operator
${\bf r}$ is a sum of the parallel and perpendicular components (see Eq. (\ref{decomp})), in
classical approximation we can neglect those commutators and treat
${\bf p}$ and ${\bf r}$ as usual vectors. Then as follows from Eq. (\ref{II20})
\begin{equation}
{\bf P}= {\bf p}+mc{\bf r}/R, \quad H = {\bf p}^2/2m +c{\bf p}{\bf r}/R,\quad {\bf N}=-m{\bf r}
\label{II64}
\end{equation}
where $H=E-mc^2$ is the classical nonrelativistic Hamiltonian
and, as follows from Eqs. (\ref{II62})
\begin{equation}
{\bf P}= {\bf p}-mc{\bf r}/R,\quad H = {\bf p}^2/2m -c{\bf p}{\bf r}/R,\quad {\bf N}=-m{\bf r}
\label{II65}
\end{equation}
As follows from these expressions, in both cases
\begin{equation}
H({\bf P},{\bf r})=\frac{{\bf P}^2}{2m}-\frac{mc^2{\bf r}^2}{2R^2}
\label{II66}
\end{equation}

The last term in Eq. (\ref{II66}) is the dS correction to
the non-relativistic Hamiltonian. The fact that it depends on $c$ is analogous to
the dependence of the r.h.s. of Eq. (\ref{accel}) on $c$. As noted in Sec. \ref{CC},
this illustrates the fact that the transition to nonrelativistic theory
understood as $|{\bf v}|\ll 1$ is more physical than that
understood as $c\to\infty$. 

Now it follows from the Hamilton equations for the Hamiltonian (\ref{II66}) that the
acceleration is given by Eq. (\ref{accel}) if $R'=R$.  As noted in Sec. \ref{symmetry}, the quantity $R$  has nothing to do with the
radius of the dS space, and the result (\ref{II66}) has been obtained
without using dS space and Riemannian geometry. 

We believe that our result is more fundamental than the result of GR because any classical result
should be a consequence of quantum theory in semiclassical approximation. In GR, $\Lambda=3/R^{'2}$ is
the curvature of the dS space with the radius $R'$ and there is no restriction
on the choice if $R'$; in particular the choice $\Lambda=0$ is possible. However, as noted in
Sec. \ref{symmetry}, the quantity $R$ {\it must be finite}. In our approach $\Lambda=3/R^2$ is
only a formal parameter which has nothing to do with the curvature of the dS space. Therefore
the fact that in the framework of GR the data can be described with $\Lambda>0$ 
should be treated not such that the
space-time background has a curvature (since the notion of the
space-time background is meaningless) but as an indication that
the symmetry algebra is the dS algebra rather than the Poincare
one. {\it Therefore for explaining the fact that $\Lambda>0$ there is no need to involve dark energy or any other quantum fields.}

Another way to show that our results are compatible with GR 
follows. The known result of GR is that if the metric
is stationary and differs slightly from the Minkowskian one
then in the nonrelativistic approximation the curved space-time
can be effectively described by a gravitational potential
$\varphi({\bf r})=(g_{00}({\bf r})-1)/2c^2$. We now express
$x_0$ in Eq. (\ref{dSspace}) in terms of a new variable $t$ as
$x_0=t+t^3/6R^{'2}-t{\bf x}^2/2R^{'2}$. Then the expression for the
interval becomes
\begin{equation}
ds^2=dt^2(1-{\bf r}^2/R^{'2})-d{\bf r}^2-
({\bf r}d{\bf r}/R')^2
\label{II67}
\end{equation}
Therefore, the metric becomes stationary and $\varphi({\bf
r})=-{\bf r}^2/2R^{'2}$ in agreement with Eq. (\ref{II66}) if $R'=R$.

Consider now a system of two free particles described by the
variables ${\bf P}_j$ and ${\bf r}_j$ ($j=1,2$). Define the
standard nonrelativistic variables
\begin{eqnarray}
&&{\bf P}_{12}={\bf P}_1+{\bf P}_2,
\quad {\bf q}_{12}=(m_2{\bf P}_1-m_1{\bf P}_2)/(m_1+m_2)\nonumber\\
&&{\bf R}_{12}=(m_1{\bf r}_1+m_2{\bf r}_2)/(m_1+m_2),\quad
{\bf r}_{12}={\bf r}_1-{\bf r}_2
\label{II68}
\end{eqnarray}
Then, as follows from Eqs. (\ref{II64}) and (\ref{II65}), in the
nonrelativistic approximation the two-particle quantities ${\bf P}$, ${\bf E}$ and ${\bf N}$ are given by
\begin{equation}
{\bf P}= {\bf P}_{12},\quad E = M+\frac{{\bf P}_{12}^2}{2M} -\frac{Mc^2{\bf R}_{12}^2}{2R^2},\quad {\bf N}=-M{\bf R}_{12}
\label{II69}
\end{equation}
where
\begin{equation}
M = M({\bf q}_{12},{\bf r}_{12})=
m_1+m_2 +H_{nr}({\bf r}_{12},{\bf q}_{12}),\quad 
H_{nr}({\bf r},{\bf q})=\frac{{\bf q}^2}{2m_{12}}-\frac{m_{12}c^2{\bf r}^2}{2R^2}
\label{II70}
\end{equation}
and $m_{12}$ is the reduced two-particle mass. 

It now follows from Eqs. (\ref{casimir}) and (\ref{II69})
that $M$ has the meaning of the two-body mass since in the nonrelativistic approximation $M^2=I_2/R^2$ where
now $I_2$ is the Casimir operator of the second order for the two-body system. Therefore $M({\bf
q}_{12},{\bf r}_{12})$ is the internal two-body Hamiltonian. Then, as a consequence of the Hamilton
equations, in semiclassical approximation the relative
acceleration is again given by Eq. (\ref{accel}) with $R'=R$ but now
${\bf a}$ is the relative acceleration and ${\bf r}$ is the
relative radius vector. As noted in Sec. \ref{ST}, equations of motions for systems of free particles
can be obtained even without the Hamilton equations but assuming that the coordinates and momenta
are related to each other by Eq. (\ref{coordmom}). This question is discussed in Sec. \ref{classeq}.

The fact that two free particles have a relative acceleration
is known for cosmologists who consider dS symmetry on
classical level. This effect is called the dS antigravity. The
term antigravity in this context means that the particles
repulse rather than attract each other. In the case of the dS
antigravity the relative acceleration of two free particles is
proportional (not inversely proportional!) to the distance
between them. This classical result (which in our approach has
been obtained without involving dS space and Riemannian
geometry) is a special case of dS symmetry on quantum level
when semiclassical approximation works with a good accuracy.

As follows from Eq. (\ref{II70}), the dS antigravity is not important for local
physics when $r\ll R$. At the same time, at cosmological distances the dS antigravity is much stronger than any other
interaction (gravitational, electromagnetic {\it etc.}). One can consider the quantum two-body problem with the 
Hamiltonian given by Eq. (\ref{II70}). Then it is obvious that the spectrum of the operator $H_{nr}$ is purely
continuous and belongs to the interval $(-\infty,\infty)$ (see also Refs. \cite{lev1a,jpa1} for details).
This does not mean that the theory is unphysical since stationary bound states in standard theory
become quasistationary with a very large lifetime if $R$ is large.

Our final remarks follow. The consideration in this chapter involves only standard quantum-mechanical 
notions and in semiclassical approximation the results on the cosmological acceleration
are compatible with GR. As argued in Sect. \ref{classical}, the standard coordinate operator has
some properties which do not correspond to what is expected from physical intuition; however, at least
from mathematical point of view, at cosmological distances semiclassical approximation is valid with
a very high accuracy. At the same time, as discussed in the next chapters, when distances are much less
than cosmological ones, this operator should be modified. Then, as a consequence of the fact that 
in dS invariant theory the spectrum of the mass operator for a free two-body system is not bounded below by $(m_1+m_2)$
it is possible to obtain gravity as a pure kinematical consequence of dS symmetry on quantum level.

In the literature it is often stated that quantum theory of gravity should become GR in classical approximation.
In Subsec.\ref{localfields} we argue that this is probably not the case because on quantum level
the notion of space-time background does not have a physical meaning. The results of this section are
the arguments in favor of this statement. Indeed, {\it our results for the cosmological acceleration obtained
from semiclassical approximation to quantum theory are compatible with GR} but, since in our approach space-time background is absent from the very beginning, it is not possible to recover theories with space-time
background.

\chapter{Algebraic description of irreducible representations}
\label{Ch3}

\section{Construction of IRs in discrete basis}
\label{S6}

In Sec. \ref{Other} we mentioned a possibility that IRs of the so(1,4) algebra
can be constructed in a pure algebraic approach such that
the basis is characterized only by discrete quantum numbers. In this chapter
a detailed consideration of this approach is given for the spinless case and in the next chapter the
results are applied to gravity. First of all, to make relations between standard
theory and FQT more straightforward, we will modify the commutation relations (\ref{CR}) 
by writing them in the form 
\begin{equation}
[M^{ab},M^{cd}]=-2i (\eta^{ac}M^{bd}+\eta^{bd}M^{ac}-
\eta^{ad}M^{bc}-\eta^{bc}M^{ad})
\label{newCR}
\end{equation}
One might say that these relations are written in units $\hbar/2=c=1$. However, as noted in Sect.
\ref{CC}, fundamental quantum theory should not involve quantities $\hbar$ and $c$ at all, and Eq. (\ref{newCR})
indeed does not contain those quantities. The reason for writing the commutation relations in the form (\ref{newCR})
rather than (\ref{CR}) is that in this case the minimum nonzero value of the angular momentum is 1 instead
of 1/2. Therefore the spin of fermions is odd and the spin of bosons is even. This will be convenient in FQT where 1/2 is a very large number (see Chap. \ref{Ch4}). 

As already noted, the results on IRs can be applied not only to elementary
particles but even to macroscopic bodies when it suffices to consider their motion as a whole. This is the case when the
distances between the bodies are much greater that their sizes. 
In Poincare invariant theory, IRs describing
massless Weyl particles can be obtained as a limit of massive IRs when $m\to 0$ with a special choice of
representatives in the factor space $SL(2,C)/SU(2)$. However, as shown in Sec. \ref{Weyl}, in dS theory such a limit does 
not exist and therefore
there are no Weyl particles in dS theory. In standard theory it is believed that the photon is a true 
massless particle but, as noted in Sec. \ref{IRsdS}, if, for example, $R$ is of the order of $10^{26}m$ 
then the commonly accepted upper limit for the photon dS mass is of the order of $10^{17}$ or less. 
In this and the next chapters we 
assume that the photon can be described by IRs of the principle series discussed above. The case of massless
particles is discussed in Sec. \ref{Singletons}.

In all macroscopic experiments the orbital angular momenta of macroscopic bodies and even photons are
very large. As an example, consider a photon moving in approximately radial direction away from the Earth surface.
Suppose that the photon energy equals the bound energy of the ground state of the hydrogen atom $27.2ev$. Then in units
$c=\hbar=1$ this energy is of the order of $10^7/cm$. Hence even if the level arm of the photon trajectory is of the
order of $1cm$, the value of the orbital angular momentum is of the order of $10^7$. In other experiments with photons
and macroscopic bodies this value is greater by many orders of magnitude. Therefore, as explained in Sec. \ref{momentum}, in semiclassical approximation the spin terms can be neglected. 
Hence our goal is to construct massive spinless IRs in a discrete basis. By analogy with the method
of little group in standard theory, one can first choose states which can be treated as rest ones and then
obtain the whole representation space by acting on such states by certain linear
combinations of representation operators. 

Since ${\bf B}$ is a possible choice of the dS analog of the momentum operator, one
might think that rest states $e_0$ can be defined by the condition ${\bf B}e_0=0$. However, in the
general case this is not consistent since, as follows from Eq. (\ref{newCR}), different components of ${\bf B}$
do not commute with each other: as follows from Eq. (\ref{newCR}) and the definitions of the operators 
${\bf J}$  and ${\bf B}$ in Sect. \ref{IRsdS}, 
\begin{equation}
[J^j,J^k]=[B^j,B^k]=2ie_{jkl}J^l,\quad [J^j,B^k]=2ie_{jkl}B^l 
\label{BJ}
\end{equation} 
where a sum over repeated
indices is assumed. Therefore a subspace of elements $e_0$ such that $B^je_0=0$ ($j=1,2,3$) is not closed
under the action of the operators $B^j$.

Let us define the operators ${\bf J}'=({\bf J}+{\bf B})/2$ and ${\bf J}"=({\bf J}-{\bf B})/2$. As follows
from Eq. (\ref{BJ}), they satisfy the commutation relations
\begin{equation}
[J^{'j},J^{"k}]=0,\quad [J^{'j},J^{'k}]=2ie_{jkl}J^{'l},\quad [J^{"j},J^{"k}]=2ie_{jkl}J^{"l} 
\label{J1J2}
\end{equation}
Since in Poincare limit ${\bf B}$ is much greater than ${\bf J}$, as an analog of the momentum operator 
one can treat
${\bf J}'$ instead of ${\bf B}$. Then one can define rest states $e_0$ by the
condition that ${\bf J}'e_0=0$. In this case the subspace of rest states is defined consistently since
it is invariant under the action of the operators ${\bf J}'$. Since the operators ${\bf J}'$ and ${\bf J}"$
commute with each other, one can define the internal angular momentum of the system as a reduction  
of ${\bf J}"$ on the subspace of rest states. In particular, in Ref. \cite{lev3} we used such a construction 
for constructing IRs of the dS algebra in the method of $SU(2)\times SU(2)$ shift operators proposed by Hughes 
for constructing IRs of the SO(5) group \cite{Hug}. In the spinless case the situation is simpler since
for constructing IRs it suffices to choose only one vector
$e_0$ such that 
\begin{equation}
{\bf J}'e_0={\bf J}"e_0=0, \quad I_2e_0=(w+9)e_0
\label{16}
\end{equation} 
The last requirement reflects the fact that all elements from the representation space
are eigenvectors of the Casimir operator $I_2$ with the same eigenvalue. When the 
representation operators satisfy Eq. (\ref{newCR}), the numerical value of the operator
$I_2$ is not as indicated at the end of Sec. (\ref{IRsdS}) but 
\begin{equation}
I_2=w-s(s+2)+9
\label{I2}
\end{equation}
where $w=m_{dS}^2$. Therefore for spinless particles the numerical value equals $w+9$.

As follows from Eq. (\ref{newCR}) and the definitions of the operators $({\bf J},{\bf N},{\bf B},{\cal E})$
in Secs. \ref{IRsdS} and (\ref{Other}), in addition to Eqs. (\ref{BJ}), the following relations are satisfied:
\begin{eqnarray}
[{\cal E},{\bf N}]= 2i{\bf B},\,\, [{\cal E},{\bf B}]=2i{\bf N}, \,\, [{\bf J},{\cal E}]=0,\,\, 
[B^j,N^k]=2i\delta_{jk}{\cal E},\,\, [J^j,N^k]=2ie_{jkl}N^l  
\label{17}
\end{eqnarray} 
We define $e_1=2{\cal E}e_0$ and 
\begin{equation}
e_{n+1}=2{\cal E}e_n-[w+(2n+1)^2]e_{n-1}
\label{18}
\end{equation}
These definitions make it possible to find $e_n$ for any $n=0,1,2...$. As follows from Eqs. (\ref{BJ}), (\ref{17})
and (\ref{18}), ${\bf J}e_n=0$. 

Our next goal is to prove that ${\bf B}^2e_n=C(n)e_n$ where $C(n)=4n(n+2)$. The proof is by induction. The relation
is obviously satisfied for $n=0$. As follows from Eqs. (\ref{BJ}), (\ref{17}) and (\ref{18}) 
$${\bf B}^2e_{n+1}=2{\cal E}C(n)e_n-[w+(2n+1)^2]C(n-1)e_{n-1}-4iAe_n$$
where $A=\{{\bf B},{\bf N}\}$. Therefore the statement will be proved if
$$-4iAe_n=[C(n+1)-C(n-1)]e_{n+1}-2[C(n)-C(n-1)]{\cal E}e_n$$ This relation also can be proved by induction 
taking into account that it is satisfied for $n=0$ and, as follows from Eqs. (\ref{BJ}) and (\ref{17}), 
$[A,{\cal E}]=-4i({\bf B}^2+{\bf N}^2)$.

Since different elements $e_n$ are the eigenvectors of the selfadjoint operator ${\bf B}^2$ with different
eigenvalues, they are mutually orthogonal. Then, if we assume that $(e_0,e_0)=1$, it follows from Eq. (\ref{18})
that
\begin{equation}
||e_n||^2=(e_n,e_n)=\prod_{j=1}^n [w+(2j+1)^2]
\label{normen}
\end{equation}

We use the notation $J_x=J^1$, $J_y=J^2$,
$J_z=J^3$ and analogously for the operators ${\bf N}$ and ${\bf B}$. Instead of the $(xy)$ components
of the vectors it is sometimes convenient to use the $\pm$ components such that $J_x=J_++J_-$,
$J_y=-i(J_+-J_-)$ and analogously for the operators ${\bf N}$ and ${\bf B}$. We now define the elements
$e_{nkl}$ as
\begin{equation}
e_{nkl}=\frac{(2k+1)!!}{k!l!}(J_-)^l(B_+)^ke_n
\label{19}
\end{equation}
As follows from Eqs. (\ref{BJ}) and (\ref{17}), $e_{nkl}$ is the eigenvector of the operator ${\bf B}^2$
with the eigenvalue $4n(n+2)-4k(k+1)$, the eigenvector of the operator ${\bf J}^2$
with the eigenvalue $4k(k+1)$ and the eigenvector of the operator $J_z$ with the eigenvalue $2(k-l)$.
Therefore different vectors $e_{nkl}$ are mutually orthogonal. As follows from Eqs. (\ref{16}-\ref{19}), 
\begin{equation}
(e_{nkl},e_{nkl})=(2k+1)!C_{2k}^lC_n^kC_{n+k+1}^k\prod_{j=1}^n [w+(2j+1)^2]
\label{21}
\end{equation}
where $C_n^k=n!/[(n-k)!k!]$ is the binomial coefficient. 
At this point we do not normalize basis vectors to one since, as will be discussed below, the
normalization (\ref{21}) has its own advantages.

A direct calculation using Eqs. (\ref{BJ}-\ref{19}) gives
\begin{eqnarray}
&&{\cal E}e_{nkl}=\frac{n+1-k}{2(n+1)}e_{n+1,kl}+\frac{n+1+k}{2(n+1)}[w+(2n+1)^2]e_{n-1,kl}\nonumber\\
&&N_+e_{nkl}=\frac{i(2k+1-l)(2k+2-l)}{8(n+1)(2k+1)(2k+3)}\{e_{n+1,k+1,l}-\nonumber\\
&&[w+(2n+1)^2]e_{n-1,k+1,l}\}-\nonumber\\
&&\frac{i}{2(n+1)}\{(n+1-k)(n+2-k)e_{n+1,k-1,l-2}-\nonumber\\
&&(n+k)(n+1+k)[w+(2n+1)^2]e_{n-1,k-1,l-2}\}\nonumber\\
&&N_-e_{nkl}=\frac{-i(l+1)(l+2)}{8(n+1)(2k+1)(2k+3)}\{e_{n+1,k+1,l+2}-\nonumber\\
&&[w+(2n+1)^2]e_{n-1,k+1,l+2}\}+\nonumber\\
&&\frac{i}{2(n+1)}\{(n+1-k)(n+2-k)e_{n+1,k-1,l}-\nonumber\\
&&(n+k)(n+1+k)[w+(2n+1)^2]e_{n-1,k-1,l}\}\nonumber\\
&&N_ze_{nkl}=\frac{-i(l+1)(2k+1-l)}{4(n+1)(2k+1)(2k+3)}\{e_{n+1,k+1,l+1}-\nonumber\\
&&[w+(2n+1)^2]e_{n-1,k+1,l+1}\}-\nonumber\\
&&\frac{i}{n+1}\{(n+1-k)(n+2-k)e_{n+1,k-1,l-1}-\nonumber\\
&&(n+k)(n+1+k)[w+(2n+1)^2]e_{n-1,k-1,l-1}\}
\label{20A}
\end{eqnarray}
\begin{eqnarray}
&&B_+e_{nkl}=\frac{(2k+1-l)(2k+2-l)}{2(2k+1)(2k+3)}e_{n,k+1,l}-\nonumber\\
&&2(n+1-k)(n+1+k)e_{n,k-1,l-2}\nonumber\\
&&B_-e_{nkl}=\frac{(l+1)(l+2)}{2(2k+1)(2k+3)}e_{n,k+1,l+2}+\nonumber\\
&&2(n+1-k)(n+1+k)e_{n,k-1,l}\nonumber\\
&&B_ze_{nkl}=\frac{(l+1)(2k+1-l)}{2(2k+1)(2k+3)}e_{n,k+1,l+1}-\nonumber\\
&&4(n+1-k)(n+1+k)e_{n,k-1,l-1}\nonumber\\
&&J_+e_{nkl}=(2k+1-l)e_{nk,l-1}\quad J_-e_{nkl}=(l+1)e_{nk,l+1}\nonumber\\
&&J_ze_{nkl}=2(k-l)e_{nkl}
\label{20B}
\end{eqnarray}
where at a fixed value of $n$, $k=0,1,...n$, $l=0,1,...2k$ and if $l$ and $k$ are not in
this range then $e_{nkl}=0$. Therefore, the elements $e_{nkl}$ form a basis of the spinless
IR with a given $w$.

Instead of $l$ we define a new quantum number $\mu =k-l$ which can take values $-k,-k+1,...k$. 
Each element of the representation space can be written as 
$x=\sum_{nk\mu}c(n,k,\mu)e_{nk\mu}$ 
where the set of
the coefficients $c(n,k,\mu)$ can be called the WF in the $(nk\mu)$ representation. As follows from Eqs.
(\ref{20A}) and (\ref{20B}), the action of the representation operators on the WF can be written as

\begin{eqnarray}
&&{\cal E}c(n,k,\mu)=\frac{n-k}{2n}c(n-1,k,\mu)+\frac{n+2+k}{2(n+2)}[w+(2n+3)^2]\nonumber\\
&&c(n+1,k,\mu)\nonumber\\
&&N_+c(n,k,\mu)=\frac{i(k+\mu)(k+\mu-1)}{8(2k-1)(2k+1)}\{\frac{1}{n}c(n-1,k-1,\mu-1)-\nonumber\\
&&\frac{1}{n+2}[w+(2n+3)^2]c(n+1,k-1,\mu-1)\}-\nonumber\\
&&\frac{i(n-1-k)(n-k)}{2n}c(n-1,k+1,\mu-1)+\nonumber\\
&&\frac{i(n+k+2)(n+k+3)}{2(n+2)}[w+(2n+3)^2]c(n+1,k+1,\mu-1)\nonumber\\
&&N_-c(n,k,\mu)=\frac{-i(k-\mu)(k-\mu-1)}{8(2k-1)(2k+1)}\{\frac{1}{n}c(n-1,k-1,\mu+1)-\nonumber\\
&&\frac{1}{n+2}[w+(2n+3)^2]c(n+1,k-1,\mu+1)\}+\nonumber\\
&&\frac{i(n-1-k)(n-k)}{2n}c(n-1,k+1,\mu+1)-\nonumber\\
&&\frac{i(n+k+2)(n+k+3)}{2(n+2)}[w+(2n+3)^2]c(n+1,k+1,\mu+1)\nonumber\\
&&N_zc(n,k,\mu)=\frac{-i(k-\mu)(k+\mu)}{4(2k-1)(2k+1)}\{\frac{1}{n}c(n-1,k-1,\mu)-\nonumber\\
&&\frac{1}{n+2}[w+(2n+3)^2]c(n+1,k-1,\mu)\}-\nonumber\\
&&\frac{i(n-1-k)(n-k)}{n}c(n-1,k+1,\mu)+\nonumber\\
&&\frac{i(n+k+2)(n+k+3)}{n+2}[w+(2n+3)^2]c(n+1,k+1,\mu)
\label{22A}
\end{eqnarray}
\begin{eqnarray}
&&B_+c(n,k,\mu)=\frac{(k+\mu)(k+\mu-1)}{2(2k-1)(2k+1)}c(n,k-1,\mu-1)-\nonumber\\
&&2(n-k)(n+2+k)c(n,k+1,\mu-1)\nonumber\\
&&B_-c(n,k,\mu)=-\frac{(k-\mu)(k-\mu-1)}{2(2k-1)(2k+1)}c(n,k-1,\mu+1)+\nonumber\\
&&2(n-k)(n+2+k)c(n,k+1,\mu+1)\nonumber\\
&&B_zc(n,k,\mu)=-\frac{(k-\mu)(k+\mu)}{(2k-1)(2k+1)}c(n,k-1,\mu)-\nonumber\\
&&4(n-k)(n+2+k)c(n,k+1,\mu)\nonumber\\
&&J_+c(n,k,\mu)=(k+\mu)c(n,k,\mu-1)\,\, J_-c(n,k,\mu)=(k-\mu)c(n,k,\mu+1)\nonumber\\
&&J_zc(n,k,\mu)=2\mu c(n,k,\mu)
\label{22B}
\end{eqnarray}
It is seen from the last expression that the meaning of the quantum number $\mu$ is such that $c(n,k,\mu)$
is the eigenfunction of the operator $J_z$ with the eigenvalue $2\mu$, i.e. $\mu$ is the standard 
magnetic quantum number.

We use ${\tilde e}_{nk\mu}$ to denote basis vectors normalized to one and ${\tilde c}(n,k,\mu)$ 
to denote the WF in the normalized basis. As follows from Eq. (\ref{21}),
the vectors ${\tilde e}_{nk\mu}$ can be defined as 
\begin{equation} 
{\tilde e_{nk\mu}}=\{(2k+1)!C_{2k}^{k-\mu}C_n^kC_{n+k+1}^k\prod_{j=1}^n [w+(2j+1)^2]\}^{-1/2}e_{nk\mu} 
\label{23}
\end{equation}
A direct calculation using 
Eqs. (\ref{21}-\ref{23}) shows that the action of the representation operators on the WF
in the normalized basis is given by

\begin{eqnarray}
&&{\cal E}{\tilde c}(n,k,\mu)=\frac{1}{2}[\frac{(n-k)(n+k+1)}{n(n+1)}(w+(2n+1)^2)]^{1/2}{\tilde c}(n-1,k,\mu)+\nonumber\\
&&\frac{1}{2}[\frac{(n+1-k)(n+k+2)}{(n+1)(n+2)}(w+(2n+3)^2)]^{1/2}{\tilde c}(n+1,k,\mu)]\nonumber\\
&&N_+{\tilde c}(n,k,\mu)=\frac{i}{4}[\frac{(k+\mu)(k+\mu-1)}{(2k-1)(2k+1)(n+1)}]^{1/2}\nonumber\\
&&\{[\frac{(n+k)(n+k+1)}{n}(w+(2n+1)^2)]^{1/2}{\tilde c}(n-1,k-1,\mu-1)-\nonumber\\
&&[\frac{(n+2-k)(n+1-k)}{n+2}(w+(2n+3)^2)]^{1/2}{\tilde c}(n+1,k-1,\mu-1)\}-\nonumber\\
&&\frac{i}{4}[\frac{(k+2-\mu)(k+1-\mu)}{(2k+1)(2k+3)(n+1)}]^{1/2}\nonumber\\
&&\{[\frac{(n-k)(n-k-1)}{n}(w+(2n+1)^2)]^{1/2}{\tilde c}(n-1,k+1,\mu-1)-\nonumber\\
&&[\frac{(n+k+2)(n+k+3)}{n+2}(w+(2n+3)^2)]^{1/2}{\tilde c}(n+1,k+1,\mu-1)]\}\nonumber\\
&&N_-{\tilde c}(n,k,\mu)=-\frac{i}{4}[\frac{(k-\mu)(k-\mu-1)}{(2k-1)(2k+1)(n+1)}]^{1/2}\nonumber\\
&&\{[\frac{(n+k)(n+k+1)}{n}(w+(2n+1)^2)]^{1/2}{\tilde c}(n-1,k-1,\mu+1)-\nonumber\\
&&[\frac{(n+2-k)(n+1-k)}{n+2}(w+(2n+3)^2)]^{1/2}{\tilde c}(n+1,k-1,\mu+1)\}+\nonumber\\
&&\frac{i}{4}[\frac{(k+2+\mu)(k+1+\mu)}{(2k+1)(2k+3)(n+1)}]^{1/2}\nonumber\\
&&\{[\frac{(n-k)(n-k-1)}{n}(w+(2n+1)^2)]^{1/2}{\tilde c}(n-1,k+1,\mu+1)-\nonumber\\
&&[\frac{(n+k+2)(n+k+3)}{n+2}(w+(2n+3)^2)]^{1/2}{\tilde c}(n+1,k+1,\mu+1)]\}\nonumber\\
&&N_z{\tilde c}(n,k,\mu)=-\frac{i}{2}[\frac{(k-\mu)(k+\mu)}{(2k-1)(2k+1)(n+1)}]^{1/2}\nonumber\\
&&\{[\frac{(n+k)(n+k+1)}{n}(w+(2n+1)^2)]^{1/2}{\tilde c}(n-1,k-1,\mu)-\nonumber\\
&&[\frac{(n+2-k)(n+1-k)}{n+2}(w+(2n+3)^2)]^{1/2}{\tilde c}(n+1,k-1,\mu)\}-\nonumber\\
&&\frac{i}{2}[\frac{(k+1-\mu)(k+1+\mu)}{(2k+1)(2k+3)(n+1)}]^{1/2}\nonumber\\
&&\{[\frac{(n-k)(n-k-1)}{n}(w+(2n+1)^2)]^{1/2}{\tilde c}(n-1,k+1,\mu)-\nonumber\\
&&[\frac{(n+k+2)(n+k+3)}{n+2}(w+(2n+3)^2)]^{1/2}{\tilde c}(n+1,k+1,\mu)]\}
\label{24A}
\end{eqnarray}
\begin{eqnarray}
&&B_+{\tilde c}(n,k,\mu)=[\frac{(k+\mu)(k+\mu-1)(n+1-k)(n+1+k)}{(2k-1)(2k+1)}]^{1/2}{\tilde c}(n,k-1,\mu-1)\nonumber\\
&&-[\frac{(k+2-\mu)(k+1-\mu)(n-k)(n+k+2)}{(2k+1)(2k+3)}]^{1/2}{\tilde c}(n,k+1,\mu-1)\nonumber\\
&&B_-{\tilde c}(n,k,\mu)=-[\frac{(k-\mu)(k-\mu-1)(n+1-k)(n+1+k)}{(2k-1)(2k+1)}]^{1/2}{\tilde c}(n,k-1,\mu+1)\nonumber\\
&&+[\frac{(k+2+\mu)(k+1+\mu)(n-k)(n+k+2)}{(2k+1)(2k+3)}]^{1/2}{\tilde c}(n,k+1,\mu+1)\nonumber\\
&&B_z{\tilde c}(n,k,\mu)=-2[\frac{(k-\mu)(k+\mu)(n+1-k)(n+1+k)}{(2k-1)(2k+1)}]^{1/2}{\tilde c}(n,k-1,\mu)\nonumber\\
&&-2[\frac{(k+1-\mu)(k+1+\mu)(n-k)(n+k+2)}{(2k+1)(2k+3)}]^{1/2}{\tilde c}(n,k+1,\mu)\nonumber\\
&&J_+{\tilde c}(n,k,\mu)=[(k+\mu)(k+1-\mu)]^{1/2}{\tilde c}(n,k,\mu-1)\nonumber\\
&&J_-{\tilde c}(n,k,\mu)=[(k-\mu)(k+1+\mu)]^{1/2}{\tilde c}(n,k,\mu+1)\nonumber\\
&& J_z{\tilde c}(n,k,\mu)=2\mu{\tilde c}(n,k,\mu)
\label{24B}
\end{eqnarray}    

\section{Semiclassical approximation}
\label{S7}

Consider now the semiclassical approximation in the ${\tilde e}_{nkl}$ basis. 
As noted in Secs. \ref{IRsdS} and \ref{antigravity}, the operator ${\bf B}$ is the dS 
analog of the usual momentum ${\bf P}$
such that in Poincare limit ${\bf B}=2R{\bf P}$. The operator ${\bf J}$ has the same meaning as
in Poincare invariant theory. Then it is clear from Eqs. (\ref{22A}) and (\ref{22B}) that 
a necessary condition for the semiclassical approximation is that
the quantum numbers $(nk\mu)$ are much greater than 1 (in agreement with the remarks in the preceding section). 
By analogy with the discussion of the semiclassical approximation in Secs. \ref{classical}
and \ref{antigravity},
we assume that a state is semiclassical if its WF has the form
\begin{equation}
{\tilde c}(n,k,\mu)=a(n,k,\mu)exp[i(-n\varphi+k\alpha -\mu\beta)]
\label{qclwf}
\end{equation}
where $a(n,k,\mu)$ is an amplitude, which is not small only in some vicinities of $n=n_0$, $k=k_0$ and 
$\mu=\mu_0$.
We also assume that when the quantum numbers $(nk\mu)$ change by one, the main contribution
comes from the rapidly oscillating exponent. Then, as follows from the first expression 
in Eq. (\ref{24A}), the action of the dS energy operator can be written as
\begin{eqnarray}
{\cal E}{\tilde c}(n,k,\mu)\approx\frac{1}{n_0}[(n_0-k_0)(n_0+k_0)(w+4n_0^2)]^{1/2}cos(\varphi)
{\tilde c}(n,k,\mu)
\label{25}
\end{eqnarray}
Therefore the semiclassical WF is approximately the eigenfunction of the dS
energy operator with the eigenvalue
$$\frac{1}{n_0}[(n_0-k_0)(n_0+k_0)(w+4n_0^2)]^{1/2}cos\varphi.$$

We will use the following notations. When we consider not the action of an operator on the
WF but its approximate eigenvalue in the semiclassical state, we will use for
the eigenvalue the same notation as for the operator and this should not lead to misunderstanding.
Analogously, in eigenvalues we will write $n$, $k$ and $\mu$ instead of $n_0$, $k_0$ and $\mu_0$,
respectively. By analogy with Eq. (\ref{25}) we can consider eigenvalues of the other operators
and the results can be represented as
\begin{eqnarray}
&&{\cal E}=\frac{1}{n}[(n-k)(n+k)(w+4n^2)]^{1/2}cos\varphi\nonumber\\
&&N_x=(w+4n^2)^{1/2}\{-\frac{sin\varphi}{k}[\mu cos\alpha cos\beta +ksin\alpha sin\beta]+\nonumber\\
&&\frac{cos\varphi}{n}[\mu sin\alpha cos\beta -kcos\alpha sin\beta]\}\nonumber\\
&&N_y=(w+4n^2)^{1/2}\{-\frac{sin\varphi}{k}[\mu cos\alpha sin\beta -ksin\alpha cos\beta]+\nonumber\\
&&\frac{cos\varphi}{n}[\mu sin\alpha sin\beta +kcos\alpha cos\beta]\}\nonumber\\
&&N_z=[(k-\mu)(k+\mu)(w+4n^2)]^{1/2}(\frac{1}{k}sin\varphi cos\alpha-
\frac{1}{n}cos\varphi sin\alpha)\nonumber\\
&&B_x=\frac{2}{k}[(n-k)(n+k)]^{1/2}[\mu cos\alpha cos\beta + ksin\alpha sin\beta]\nonumber\\
&&B_y=\frac{2}{k}[(n-k)(n+k)]^{1/2}[\mu cos\alpha sin\beta -ksin\alpha cos\beta]\nonumber\\
&&B_z=-\frac{2}{k}[(k-\mu)(k+\mu)(n-k)(n+k)]^{1/2}cos\alpha\nonumber\\
&&J_x=2[(k-\mu)(k+\mu)]^{1/2}cos\beta\quad J_y=2[(k-\mu)(k+\mu)]^{1/2}sin\beta\nonumber\\
&&J_z=2\mu
\label{26}
\end{eqnarray}
Since ${\bf B}$ is the dS analog of ${\bf p}$ and in classical theory ${\bf J}={\bf r}\times {\bf p}$,
one might expect that ${\bf B}{\bf J}=0$ and, as follows from the above expressions, this is the case.
It also follows that ${\bf B}^2=4(n^2-k^2)$ and ${\bf J}^2=4k^2$.  

In Sec. \ref{antigravity} we described semiclassical WFs by six parameters 
$({\bf r},{\bf p})$ while in the basis ${\tilde e}_{nkl}$ the six parameters are $(n,k,\mu,\varphi,\alpha,\beta)$.
Since in dS theory the ten representation operators are on equal footing,
it is also possible to describe a semiclassical state by semiclassical eigenvalues of these operators.
However, we should have four constraints for them. As follows from Eqs. (\ref{casimir}) and (\ref{II26}),
the constraints can be written as
\begin{equation}
{\cal E}^2+{\bf N}^2-{\bf B}^2-{\bf J}^2=w\quad {\bf N}\times{\bf B}=-{\cal E} {\bf J}
\label{27}
\end{equation}  
As noted in Sec. \ref{antigravity}, in Poincare limit ${\cal E}=2RE$, ${\bf B}=2R{\bf p}$ 
(since we have replaced Eq. (\ref{CR}) by Eq. (\ref{newCR})) and the values
of ${\bf N}$ and ${\bf J}$ are much less than ${\cal E}$ and ${\bf B}$. Therefore the first relation
in Eq. (\ref{27}) is the Poincare analog of the well-known relation $E^2-{\bf p}^2=m^2$. 

The quantities $(nk\mu\varphi\alpha\beta)$ can be expressed in terms of semiclassical eigenvalues 
$({\cal E},{\bf N},{\bf B},{\bf J})$ as follows. The quantities $(nk\mu)$ can be found from the relations
\begin{equation}
{\bf B}^2+{\bf J}^2 = 4n^2,\quad {\bf J}^2 = 4k^2,\quad J_z=2\mu
\label{28}
\end{equation}
and then the angles $(\varphi\alpha\beta)$ can be found from the relations
\begin{eqnarray}
&&cos\varphi =\frac{2{\cal E}n}{B(w+4n^2)^{1/2}},\quad sin\varphi =
-\frac{{\bf B}{\bf N}}{B(w+4n^2)^{1/2}}\nonumber\\
&&cos\alpha=-JB_z/(BJ_{\bot}),\quad sin\alpha=({\bf B}\times{\bf J})_z/(BJ_{\bot})\nonumber\\
&&cos\beta=J_x/J_{\bot},\quad sin\beta=J_y/J_{\bot}
\label{29}
\end{eqnarray}
where $B=|{\bf B}|$, $J=|{\bf J}|$ and $J_{\bot}=(J_x^2+J_y^2)^{1/2}$.
In semiclassical approximation, uncertainties of the quantities $(nk\mu)$ should be such that
$\Delta n\ll n$, $\Delta k\ll k$ and $\Delta \mu\ll \mu$. On the other hand, those uncertainties
cannot be very small since the distribution in $(nk\mu)$ should be such that all the ten
approximate eigenvalues $({\cal E},{\bf N},{\bf B},{\bf J})$ should be much greater than
their corresponding uncertainties. The assumption is that for macroscopic bodies all these
conditions can be satisfied.

In applications it is often considered a case when a classical trajectory is in the $xy$ plane. Then the
classical value of $J_{\bot}$ is zero and Eq. (\ref{29}) does not apply. In that case the classical value 
of $\mu$ is $\pm k$ for the counterclockwise and clockwise motion, respectively. For definiteness we consider
the former case. Then by analogy with the above derivation we have that
\begin{equation}
{\tilde c}(n,k)=a(n,k)exp[-i(n\varphi+k\gamma)]
\label{qclwfxy}
\end{equation}
\begin{eqnarray}
&&{\cal E}=\frac{1}{n}[(n-k)(n+k)(w+4n^2)]^{1/2}cos\varphi\nonumber\\
&&N_x=-(w+4n^2)^{1/2}(sin\varphi cos\gamma +\frac{k}{n}cos\varphi sin\gamma )\nonumber\\
&&N_y=-(w+4n^2)^{1/2}(sin\varphi sin\gamma -\frac{k}{n}cos\varphi cos\gamma )\nonumber\\
&&N_z=B_z=J_x=J_y=0,\quad J_z=2k \nonumber\\
&&B_x=2[(n-k)(n+k)]^{1/2}cos\gamma, \quad B_y=2[(n-k)(n+k)]^{1/2}sin\gamma
\label{26xy}
\end{eqnarray}
where $\gamma=\beta - \alpha$. 

In Sec. \ref{antigravity} we discussed operators in Poincare limit and corrections to them,
which lead to the dS antigravity. A problem arises how the Poincare limit should be defined in the basis
defined in the present chapter.
In contrast to Sec. \ref{antigravity},
we can now work not with the unphysical quantities ${\bf v}$ or ${\bf p}=m{\bf v}$ defined on the
Lorentz hyperboloid but directly with semiclassical eigenvalues of the 
representation operators. In contrast to Sec. \ref{antigravity}, we now {\it define} ${\bf p}={\bf B}/(2R)$,
$m=w^{1/2}/(2R)$ and $E=(m^2+{\bf p}^2)^{1/2}$. Then 
Poincare limit can be defined by the requirement that when $R$ is large, the quantities 
${\cal E}$ and ${\bf B}$ are proportional to $R$ while ${\bf N}$ and ${\bf J}$ do not depend on
$R$. In this case, as follows from Eq. (\ref{27}), in Poincare limit ${\cal E}=2RE$ and 
${\bf B}=2R{\bf p}$.

\section{Position operator in dS theory}
\label{position}

By analogy with
constructing a physical position operator in Sec. \ref{consistent}, the position operator in dS theory can be
found from the following considerations. Since the operators 
${\bf B}$ and ${\bf J}$ are consistently defined as representation operators of the dS algebra and we
have defined ${\bf p}$ as ${\bf B}/2R$, one might seek the position operator such that on classical level
the relation ${\bf r}\times{\bf p}={\bf J}/2$ will take place (the factor 1/2 is a consequence of Eq. (\ref{newCR})). 
On classical level one can define parallel and perpendicular components of ${\bf r}$ as 
${\bf r}=r_{||}{\bf B}/|{\bf B}|+{\bf r}_{\bot}$ and analogously
${\bf N}=N_{||}{\bf B}/|{\bf B}|+{\bf N}_{\bot}$. Then the relation ${\bf r}\times{\bf p}={\bf J}/2$ 
defines uniquely only ${\bf r}_{\bot}$ and it follows from the second relation in Eq. (\ref{27}) that
${\bf N}_{\bot}=-2E{\bf r}_{\bot}$. However, it is not clear yet how $r_{||}$ should be defined and whether
the last relation is also valid for the parallel components of ${\bf N}$ and ${\bf r}$. As follows from the
second relation in Eq. (\ref{29}), it will be valid if $|sin\varphi|=r_{||}/R$, i.e. $\varphi$ is the 
angular coordinate. As noted in Sec. \ref{classical}, semiclassical
approximation for a physical quantity can be valid only in states where this quantity is rather large.
Therefore if $R$ is very large then $\varphi$ is very small if the distances are not cosmological
(i.e. they are much less than $R$). Hence the problem arises whether this approximation 
is valid. This is a very important problem since in standard approach it is assumed that nevertheless $\varphi$
can be considered semiclassically. Suppose first that this is the case and consider corrections to Poincare limit 
in classical limit. 

Since ${\bf B}=2R{\bf p}$ and 
${\bf J}/2={\bf r}_{\bot}\times {\bf p}$ then it follows from Eq. (\ref{28}) that in first order in $1/R^2$ we have
$k^2/n^2={\bf r}_{\bot}^2/R^2$. Therefore as follows from the first expression in Eq. (\ref{26}), in first 
order in $1/R^2$ the results on ${\cal E}$ and ${\bf N}$ can be represented as
\begin{equation}
{\cal E}=2ER(1-\frac{{\bf r}^2}{2R^2}),\quad {\bf N}=-2E{\bf r}
\label{EN}
\end{equation}
Hence the result for the energy is in agreement with Eq. (\ref{II66}) while the result for ${\bf N}$ is
in agreement with Eq. (\ref{II20}).

Consider now constructing the position operator on quantum level. In view of the remarks in Sec. \ref{S6}, we assume 
the approximation $n,k,|\mu|\gg 1$. Let us define Hermitian operators ${\cal A}$ and ${\cal B}$ which act as
\begin{eqnarray}
&&{\cal A}{\tilde c}(n,k,\mu)=\frac{i}{2}[{\tilde c}(n+1,k,\mu)-{\tilde c}(n-1,k,\mu)]\nonumber\\
&&{\cal B}{\tilde c}(n,k,\mu)=\frac{1}{2}[{\tilde c}(n+1,k,\mu)+{\tilde c}(n-1,k,\mu)]
\label{A1A2}
\end{eqnarray}
and the operators ${\bf F}$ and ${\bf G}$ which act as (compare with Eqs. (\ref{F1}) and (\ref{G1}))
\begin{eqnarray}
&&F_+{\tilde c}(n,k,\mu)=-\frac{i}{4}[(k+\mu){\tilde c}(n,k-1,\mu-1)+(k-\mu){\tilde c}(n,k+1,\mu-1)]\nonumber\\
&&F_-{\tilde c}(n,k,\mu)=\frac{i}{4}[(k-\mu){\tilde c}(n,k-1,\mu+1)+(k+\mu){\tilde c}(n,k+1,\mu+1)]\nonumber\\
&&F_z{\tilde c}(n,k,\mu)=\frac{i}{2}\sqrt{k^2-\mu^2}[{\tilde c}(n,k-1,\mu)-{\tilde c}(n,k+1,\mu)]
\label{F}
\end{eqnarray} 
\begin{eqnarray}
&&G_+{\tilde c}(n,k,\mu)=\frac{1}{4k}[(k+\mu){\tilde c}(n,k-1,\mu-1)-(k-\mu){\tilde c}(n,k+1,\mu-1)]\nonumber\\
&&G_-{\tilde c}(n,k,\mu)=-\frac{1}{4k}[(k-\mu){\tilde c}(n,k-1,\mu+1)-(k+\mu){\tilde c}(n,k+1,\mu+1)]\nonumber\\
&&G_z{\tilde c}(n,k,\mu)=-\frac{\sqrt{k^2-\mu^2}}{2k}[{\tilde c}(n,k-1,\mu)+{\tilde c}(n,k+1,\mu)]
\label{B/B}
\end{eqnarray}
Then, as follows from Eqs. (\ref{24A}) and (\ref{24B}), the representation operators can be written as
\begin{eqnarray}
&&{\cal E}{\tilde c}(n,k,\mu)=\frac{\sqrt{n^2-k^2}}{n}(w+4n^2)^{1/2}B,\quad 
{\bf N}=-(w+4n^2)^{1/2}({\cal A}{\bf G}+\frac{1}{n}{\cal B}{\bf F})\nonumber\\
&&{\bf B}=2\sqrt{n^2-k^2}{\bf G},\quad J_{\pm}{\tilde c}(n,k,\mu)=\sqrt{k^2-\mu^2}{\tilde c}(n,k,\mu\mp 1)\nonumber\\
&&J_z{\tilde c}(n,k,\mu)=2\mu{\tilde c}(n,k,\mu)
\label{generators}
\end{eqnarray}
and, as follows from Eqs. (\ref{F},\ref{G},\ref{generators})
\begin{eqnarray}
&&[J_j,F_k]=2ie_{jkl}F_l,\quad [J_j,G_k]=2ie_{jkl}G_l,\quad {\bf G}^2=1,\quad {\bf F}^2=k^2 \nonumber\\
&&[G_j,G_k]=0,\quad [F_j,F_k]=-\frac{i}{2}e_{jkl}J_l,\quad e_{jkl}\{F_k,G_l\}=J_j\nonumber\\
&&{\bf J}{\bf G}={\bf G}{\bf J}={\bf J}{\bf F}={\bf F}{\bf J}=0, \quad {\bf F}{\bf G}=-{\bf G}{\bf F}=i
\label{vectorFG}
\end{eqnarray}
The first two relations show that ${\bf F}$ 
and ${\bf G}$ are the vector operators as expected. The third relation shows that ${\bf G}$ can be treated as 
an operator of the unit vector along the direction of the momentum. The result for the anticommutator shows
that on classical level ${\bf F}\times {\bf G}={\bf J}/2$ and the last two relations show that on classical level
the operators in the triplet $({\bf F},{\bf G},{\bf J})$ are mutually orthogonal. Hence we have a full
analogy with the corresponding results in Poincare invariant theory (see Sec. \ref{consistent}).

Let us define the operators} ${\cal R}_{||}$ and ${\bf {\cal R}}_{\bot}$ as
\begin{equation}
{\cal R}_{||}=R{\cal A},\quad {\bf {\cal R}}_{\bot}=\frac{R}{n}{\bf F}
\label{roper}
\end{equation}
Then taking into account that $(w+4n^2)^{1/2}=2RE$, the expression for ${\bf N}$ in Eq. (\ref{generators}) 
can be written as
\begin{equation}
{\bf N}=-2E{\cal R}_{||}{\bf G}-2E{\cal B}{\bf {\cal R}}_{\bot}
\label{NA}
\end{equation}
If the function ${\tilde c}(n,k,\mu)$ depends on $\varphi$ as in Eq. (\ref{qclwf}) and $\varphi$ is of the order of $r/R$
then, as follows from Eq. (\ref{A1A2}), in the approximation when the terms of the order of $(r/R)^2$ in ${\bf N}$ can be 
neglected, ${\cal B}\approx 1$. In the approximation when $n$ can be replaced by a continuous variable $Rp$ 
\begin{equation}
{\cal R}_{||}=i\hbar \frac{\partial}{\partial p},\quad {\bf {\cal R}}_{\bot}=\frac{\hbar}{p}{\bf F}
\label{roperB}
\end{equation}
where the dependence on $\hbar$ is restored. Hence in this approximation
\begin{equation}
{\bf N}=-2E{\cal R}_{||}{\bf G}-2E{\bf {\cal R}}_{\bot}
\label{NB}
\end{equation}
and this result can be treated 
as an implementation of the decomposition ${\bf N}=N_{||}{\bf B}/|{\bf B}|+{\bf N}_{\bot}$ on the
operator level. The semiclassical result ${\bf N}=-2E{\bf r}$ will take place if in semiclassical
approximation ${\cal R}_{||}$ can be replaced by $r_{||}$ and ${\bf {\cal R}}_{\bot}$ can be replaced by
${\bf r}_{\bot}$.

In the approximation when $n$ can be replaced by the continuous variable $Rp$, the commutation relations between
${\cal R}_{||}$, different components of ${\bf {\cal R}}_{\bot}$ and different components of ${\bf p}=p{\bf G}$
are the same as in Sec. \ref{consistent}. Hence the operators ${\cal R}_{||}$ and ${\bf {\cal R}}_{\bot}$ can be treated as the parallel and transverse components of the position operator in dS theory. In particular, 
by analogy with the consideration in Chap. \ref{WPS} we can conclude that
in dS theory there is no WPS in directions transverse to ${\bf B}$ and there is no WF in coordinate
representation. 

We now investigate the properties of the operators ${\cal A}$ and ${\cal B}$ since, as shown in the next chapter,
these operators are present in the two-body mass and distance operators. 
The relations between the operators ${\cal A}$, ${\cal B}$ and $n$ are
\begin{equation}
[{\cal A},n]=i{\cal B},\quad [{\cal B},n]=-i{\cal A},\quad [{\cal A},{\cal B}]=0,\quad {\cal A}^2+{\cal B}^2=1
\label{[]}
\end{equation}

As noted in Sec. \ref{classical}, in standard quantum theory the semiclassical 
WF in momentum space contains a factor $exp(-ipx)$. Since $n$ is now the dS analog
of $pR$, we assume that ${\tilde c}(n,k,\mu)$ contains a factor $exp(-in\varphi)$, i.e. the angle $\varphi$ is the dS
analof of $r_{||}/R$. It is reasonable to expect that since all the ten representation operators of the dS algebra
are angular momenta, in dS theory one should deal only with angular coordinates which are dimensionless.
If we assume that in semiclassical approximation the main contributions
in ${\cal A}{\tilde c}(n,k,\mu)$ and ${\cal B}{\tilde c}(n,k,\mu)$ come from the rapidly oscillating exponent then 
\begin{equation}
{\cal A}{\tilde c}(n,k,\mu)\approx sin\varphi {\tilde c}(n,k,\mu), \quad {\cal B}{\tilde c}(n,k,\mu)\approx 
cos\varphi{\tilde c}(n,k,\mu)
\label{varphi}
\end{equation}
in agreement with the first two expressions in Eq. (\ref{29}). Therefore $\varphi$ is indeed the dS analog of $r_{||}/R$
and if $r_{||}\ll R$ we recover the result that $N_{||}\approx -2Er_{||}$. Eq. (\ref{varphi}) can be treated in such
a way that ${\cal A}$ is the operator of the quantity $sin\varphi$ and ${\cal B}$ is the operator of the quantity 
$cos\varphi$. However, the following question arises.
As noted in Sect. \ref{classical}, semiclassical approximation for a quantity can be correct only if this
quantity is rather large. At the same time, we assume that ${\cal A}$ is the operator of the quantity which
is very small if $R$ is large.

If $\varphi$ is small, we have $sin\varphi\approx \varphi$ and in this approximation ${\cal A}$ can be treated
as the operator of the angular variable $\varphi$. This seems natural since, as shown in Sec. \ref{consistent},
in Poincare invariant theory the operator
of the longitudinal coordinate is $id/dp$ and ${\cal A}$ is the finite difference
analog of derivative over $n$. When $\varphi$ is not small, the argument that ${\cal A}$ is the operator of the 
quantity $sin\varphi$ follows. Since 
$$arcsin\varphi = \sum_{l=0}^{\infty}\frac{(2l)!\varphi^{2l+1}}{4^l(l!)^2(2l+1)}$$ then
$$\Phi  = \sum_{l=0}^{\infty}\frac{(2l)!{\cal A}^{2l+1}}{4^l(l!)^2(2l+1)}$$
can be treated as the operator of the quantity $\varphi$. Indeed, as follows from this expression and Eq. (\ref{[]}),
$[\Phi,n]=i$ what is the dS analog of the relation $[{\cal R}_{||},p]=i\hbar$ (see Sec. \ref{consistent}).

\chapter{Two-body systems in discrete basis}
\label{twobody}

\section{Two-body mass operator and the cosmological acceleration in discrete basis}
\label{LambdaDiscrete}

Consider now a system of two free particles in dS theory. As follows from Eq. (\ref{casimir}), in this case the Casimir operator of the second order is 
\begin{eqnarray}
&&I_2 =-\frac{1}{2}\sum_{ab} (M_{ab}^{(1)}+M_{ab}^{(2)})(M^{ab(1)}+M^{ab(2)})
\label{twobodycasimir}
\end{eqnarray}
As explained in the preceding chapter, for our purposes spins of the particles can be neglected. Then, as follows
from Eq. (\ref{I2})
\begin{equation}
I_2=w_1+w_2+2{\cal E}_1{\cal E}_2+2{\bf N}_1{\bf N}_2-2{\bf B}_1{\bf B}_2-2{\bf J}_1{\bf J}_2+18
\label{2bcasimir}
\end{equation}
where the subscripts 1 and 2 are used to denote operators for particle 1 and 2, respectively. By analogy with Eq. 
(\ref{I2}), one can define the two-body operator $W$, which is an analog of the quantity $w$:
\begin{equation}
I_2=W-{\bf S}^2+9
\label{W}
\end{equation}
where ${\bf S}$ is the two-body spin operator which is the total angular momentum in the rest frame
of the two-body system. Then, as follows from Eqs. (\ref{2bcasimir}) and (\ref{W}), 
\begin{equation}
W=w_1+w_2+2(w_1+4n_1^2)^{1/2}(w_2+4n_2^2)^{1/2}-2F-2{\bf B}_1{\bf B}_2-2{\bf J}_1{\bf J}_2+{\bf S}^2+9
\label{W2}
\end{equation}
where in this chapter we use $F$ to denote the operator
\begin{equation}
F=(w_1+4n_1^2)^{1/2}(w_2+4n_2^2)^{1/2}-{\cal E}_1{\cal E}_2+2{\bf N}_1{\bf N}_2
\label{newF}
\end{equation}

Let $I_{2P}$ be the Casimir operator of the second order in Poincare invariant theory.  If $E$ is the two-body
energy operator in Poincare invariant theory and ${\bf P}$ is the two-body Poincare momentum then 
$I_{2P}=E^2-{\bf P}^2$. This operator is sometimes called the mass operator squared although in general  
$I_{2P}$ is not positive definite (e.g. for tachyons). However, for macroscopic bodies it is positive definite,
i.e. can be represented as $M_0^2$, the classical value of which is $M_0^2=m_1^2+m_2^2+2E_1E_2-2{\bf p}_1{\bf p}_2$.
As follows from Eq. (\ref{W2})
\begin{equation}
W=W_0-2F-2{\bf J}_1{\bf J}_2+{\bf S}^2+9
\label{W3}
\end{equation}
where 
\begin{equation}
W_0=w_1+w_2+2(w_1+4n_1^2)^{1/2}(w_2+4n_2^2)^{1/2}-2{\bf B}_1{\bf B}_2=4R^2M_0^2
\label{W0}
\end{equation}

Consider first the case when semiclassical approximation is valid. 
In Sec. \ref{antigravity} we discussed operators in Poincare limit and corrections to them,
which lead to the dS antigravity. A problem arises how the dS antigravity can be recovered in the discrete basis 
defined in the preceding chapter. Let us assume that the longitudinal part of the position operator is such that Eq. (\ref{EN}) is valid. Then as follows from Eq. (\ref{EN}), $F=2E_1E_2{\bf r}^2$
where ${\bf r}={\bf r}_1-{\bf r}_2$. Let $M^2=W/4R^2$ be the mass squared in
Poincare invariant theory with dS corrections. In the nonrelativistic approximation the last three terms 
in the r.h.s. of Eq. (\ref{W3}) can be neglected. Then if $M=m_1+m_2+H_{nr}$ where
$H_{nr}$ is the nonrelativistic Hamiltonian in the c.m. frame, it follows from Eq. (\ref{W3}) and the expression for 
$F$ that in first order in $1/R^2$
\begin{equation}
H_{nr}({\bf r}, {\bf q}) =\frac{{\bf q}^2}{2m_{12}} -
\frac{m_{12}{\bf r}^2}{2R^2}
\label{33}
\end{equation}
i.e. the same result as that given by Eq. (\ref{II70}). As a consequence, the result for the cosmological
acceleration obtained in the discrete basis is the same as in the basis discussed in Chap. \ref{Ch2}. Note that the correction to the Hamiltonian is always negative and proportional
to $m_{12}$ in the nonrelativistic approximation.

In deriving Eq. (\ref{33}), as well as in deriving Eq. (\ref{II70}), the notions of dS space, metric and connection have not been used.  
This is an independent argument that the cosmological acceleration is simply a
kinematical effect in dS theory and can be explained without dark energy, empty space-time and other
artificial notions. 

Consider now a general case, i.e. we will not assume that Eq. (\ref{EN}) is necessarily valid. Then, as follows from
Eq. (\ref{generators})
\begin{eqnarray}
&&F=(w_1+4n_1^2)^{1/2}(w_2+4n_2^2)^{1/2}G\nonumber\\
&&G=1-\{\frac{1}{n_1n_2}[\sqrt{(n_1^2-k_1^2)(n_2^2-k_2^2)}+{\bf F}_1{\bf F}_2]{\cal B}_1{\cal B}_2+
{\cal A}_1{\cal A}_2{\bf G}_1{\bf G_2}+\nonumber\\
&&\frac{1}{n_1}{\bf F}_1{\bf G}_2{\cal B}_1{\cal A}_2+\frac{1}{n_2}{\bf G}_1{\bf F}_2{\cal A}_1{\cal B}_2\}
\label{twobodyF}
\end{eqnarray}
where the single-particle operators $({\cal A}_j,{\cal B}_j,{\bf F}_j,{\bf G}_j)$ ($j=1,2$) are defined in Sec. \ref{position}.

\section{Two-body relative distance operator}
\label{semiclass}

In Sec. \ref{position} we discussed semiclassical approximation for the single-particle position operator 
in dS theory. In this section we investigate how the relative distance operator can be defined in this
theory. As already noted, among the operators of the dS algebra there are no
operators which can be identified with the distance operator but there are reasons to think that in semiclassical
approximation the values of $E$ and ${\bf N}$ are given by Eq. (\ref{EN}). From the point of view of our
experience in Poincare invariant theory, the dependence of $E$ on ${\bf r}$ might seem to be unphysical
since the energy depends on the choice of the origin. However, only invariant quantities have
a physical meaning; in particular the two-body mass can depend only on relative distances which do not
depend on the choice of the origin. 

In view of Eq. (\ref{EN}) one might think that the operator ${\tilde{\bf D}}={\cal E}_2{\bf N}_1-{\cal E}_1{\bf N}_2$ might
be a good operator which in semiclassical approximation is proportional to $E_1E_2{\bf r}$ at least in the
main order in $1/R^2$. However, the operator ${\bf D}$ defining the relative distance should satisfy the following
conditions. First of all, it should not depend on the motion of the two-body system as a whole; in particular it
should commute with the operator which is treated as the total momentum in dS theory. As noted in Sec. \ref{S6}, the single-particle operator ${\bf J}'$ is a better candidate for the total momentum operator than ${\bf B}$.
Now we use ${\bf J}'$ to denote the total two-particle operator ${\bf J}_1'+{\bf J}_2'$.
Analogously, we use ${\bf J}"$ to denote the total two-particle operator ${\bf J}_1"+{\bf J}_2"$. As noted
in Sec. \ref{S6}, ${\bf J}"$ can be treated as the internal angular momentum operator. Therefore, since
${\bf D}$ should be a vector operator with respect to internal rotations, it should properly commute with ${\bf J}"$.
In summary, the operator ${\bf D}$ should satisfy the relations
\begin{equation}
[J^{'j},D^k]=0,\quad [J^{"j},D^k]=2ie_{jkl}D^l
\label{DJ}
\end{equation}
By using Eqs. (\ref{BJ}) and (\ref{17}) one can explicitly verify that the operator 
\begin{equation}
{\bf D}={\cal E}_2{\bf N}_1-{\cal E}_1{\bf N}_2-{\bf N}_1\times {\bf N}_2
\end{equation}
indeed satisfies Eq. (\ref{DJ}). If Poincare approximation is satisfied with a high accuracy then obviously
${\bf D}\approx{\tilde{\bf D}}$.

In contrast to the situation in standard quantum mechanics, different components of ${\bf D}$ do not commute
with each other and therefore are not simultaneously measurable. As shown in Chap. \ref{WPS}, if in Poincare 
invariant theory the position operator is defined in a consistent way, its different components also
do not commute with each other (see Sec. \ref{consistent}). However, since $[{\bf D}^2,{\bf J}"]=0$, by
analogy with quantum mechanics one can choose $({\bf D}^2,{\bf J}^{"2},J_z^{"})$ as a set of diagonal operators.
The result of explicit calculations is
\begin{equation}
{\bf D}^2=({\cal E}_1^2+{\bf N}_1^2)({\cal E}_2^2+{\bf N}_2^2)-({\cal E}_1{\cal E}_2+{\bf N}_1{\bf N}_2)^2-
4({\bf J}_1{\bf B}_2+{\bf J}_2{\bf B}_1)-4{\bf J}_1{\bf J}_2
\label{D2}
\end{equation} 
It is obvious that in typical situations the last two terms in this expression are much less than the first two
terms and for this reason we accept an approximation
\begin{equation}
{\bf D}^2\approx ({\cal E}_1^2+{\bf N}_1^2)({\cal E}_2^2+{\bf N}_2^2)-({\cal E}_1{\cal E}_2+{\bf N}_1{\bf N}_2)^2
\label{D2B}
\end{equation} 
Then, as follows from Eqs. (\ref{generators}), (\ref{newF}) and (\ref{twobodyF}), if $n_1,n_2\gg 1$ then
\begin{equation}
{\bf D}^2\approx (w_1+4n_1^2)(w_2+4n_2^2)(2-G)G
\label{D2C}
\end{equation}
Hence the knowledge of the operator $G$ is needed for calculating both, the two-body mass and distance operators. 

At this point no assumption that semiclassical approximation is valid has been made. If Eq. (\ref{EN}) is valid
then, as follows from Eq. (\ref{D2B}), in first order in $1/R^2$
${\bf D}^2=16E_1^2E_2^2R^2r^2$ where $r=|{\bf r}|$. In particular, in the nonrelativistic approximation 
${\bf D}^2=16m_1^2m_2^2R^2r^2$, i.e.
${\bf D}^2$ is proportional to $r^2$ what justifies treating ${\bf D}$ as a dS analog of the relative distance
operator. 
 
By analogy with standard theory, we can consider the two-body system in its c.m. frame.
Since we choose ${\bf B}+{\bf J}$ as the dS analog of momentum, the c.m. frame can be defined by the
condition ${\bf B}_1+{\bf J}_2+{\bf B}_2+{\bf J}_2=0$. Therefore, as follows from Eq. (\ref{28}), $n_1=n_2$.
This is an analog of the condition that the magnitudes of particle momenta in the c.m. frame are the same.
Another simplification can be achieved if the position of particle 2 is chosen as the origin. Then ${\bf J}_2=0$,
${\bf J}_1=({\bf r}_{\bot}\times {\bf B}_1)/2R$, $B_2=2n_2$. In quantum theory these relations can be only
approximate if semiclassical approximation is valid. Then, as follows from Eqs. (\ref{vectorFG}) and (\ref{roper}),
the expression for $G$ in Eq. (\ref{twobodyF}) has a much simpler form:
\begin{equation}
G=1-\frac{\sqrt{n_1^2-k_1^2}}{n_1}({\cal B}_1{\cal B}_2-{\cal A}_1{\cal A}_2)
\label{GC}
\end{equation}
In the approximation when ${\cal B}_i$ can be replaced by $cos\varphi_i$ and ${\cal A}_i$ - by $sin\varphi_i$
($i=1,2$), we can again recover the above result ${\bf D}^2=16E_1^2E_2^2R^2r^2$ if $|\varphi_1+\varphi_2|=r_{||}/R$
since $|\varphi_i|=|r_{||i}|/R$, $k_1^2/n_1^2=r_{\bot}^2/R^2$ and the particle momenta are approximately antiparallel.

We conclude that if standard semiclassical approximation is valid then dS corrections to the two-body mass
operator are of the order of $(r/R)^2$. This result is in agreement with standard intuition that dS corrections can
be important only at cosmological distances while in the Solar System the corrections are negligible. On the
other hand, as already noted, those conclusions are based on belief that the angular distance 
$\varphi$, which is of the order of $r/R$, can be considered semiclassically in spite of the fact that it is very
small. In the next section we investigate whether this is the case. Since from now on we are interested only
in distances which are much less than cosmological ones, we will investigate what happens if all corrections of 
the order of $r/R$ and greater are neglected. In particular, we accept the approximation that 
$|{\bf B}_1|=2n_1$, $|{\bf B}_2|=2n_2$ and the c.m. frame is defined by the condition ${\bf B}_1+{\bf B}_2=0$.

By analogy with standard theory, it is convenient to consider the two-body mass operator if individual particle
momenta $n_1$ and $n_2$ are expressed in terms of the total and relative momenta $N$ and $n$. In the c.m. frame
we can assume that ${\bf B}_1$ is directed along the positive direction of the $z$ axis and then ${\bf B}_2$ is
directed along the negative direction of the $z$ axis. Therefore the quantum number $N$ characterizing the total
dS momentum can be defined as $N=n_1-n_2$. In nonrelativistic theory the relative
momentum is defined as ${\bf q}=(m_2{\bf p}_1-m_1{\bf p}_2)/(m_1+m_2)$ and in relativistic theory as 
${\bf q}=(E_2{\bf p}_1-E_1{\bf p}_2)/(E_1+E_2)$.
Therefore, taking into account the fact that in the c.m. frame the particle momenta are directed in opposite
directions, one might define $n$ as $n=(m_2n_1+m_1n_2)/(m_1+m_2)$ or $n=(E_2n_1+E_1n_2)/(E_1+E_2)$. These
definitions involve Poincare masses and energies. Another possibility is $n=(n_1+n_2)/2$. In all these
cases we have that $n\to (n+1)$ when $n_1\to (n_1+1),\,\,n_2\to (n_2+1)$ and 
$n\to (n-1)$ when $n_1\to (n_1-1),\,\,n_2\to (n_2-1)$. In what follows, only this feature is important. 

Although so far we are working in standard dS quantum theory
over complex numbers, we will argue in the next chapters that fundamental quantum theory should be based on
finite mathematics.
We will consider a version of quantum theory where complex numbers are replaced by a finite field field or
finite ring. 
Let $\psi_1(n_1)$ and $\psi_2(n_2)$ be the functions describing the dependence of single-particle WFs
on $n$. Then in our approach
only those functions $\psi_1(n_1)$ and $\psi_2(n_2)$ are physical which have a finite support in 
$n_1$ and $n_2$, respectively. Therefore we assume that $\psi_1(n_1)$ 
can be different from zero only if $n_1\in [n_{1min},n_{1max}]$ and analogously for $\psi_2(n_2)$. 
If $n_{1max}=n_{1min}+\delta_1-1$ then a necessary condition that $n_1$ is 
semiclassical is $\delta_1 \ll n_1$. At the same time, since $\delta_1$ is the dS analog of $\Delta p_1R$ and $R$ is very
large, we expect that $\delta_1\gg 1$. We use $\nu_1$ to denote $n_1-n_{1min}$. Then if 
$\psi_1(\nu_1)=a_1(\nu_1)exp(-i\varphi_1 \nu_1)$,
we can expect by analogy with the consideration in Sect. \ref{classical} that the state $\psi_1(\nu_1)$ will be
semiclassical if $|\varphi_1\delta_1|\gg 1$ since in this case the exponent makes many oscillations on 
$[0,\delta_1]$. Even this condition indicates that $\varphi_1$ cannot be extremely small. Analogously we
can consider the WF of particle 2, define $\delta_2$ as the width of its dS momentum distribution and
$\nu_2=n_2-n_{2min}$. 
\begin{figure}[!ht]
\centerline{\scalebox{1.1}{\includegraphics[scale=0.5]{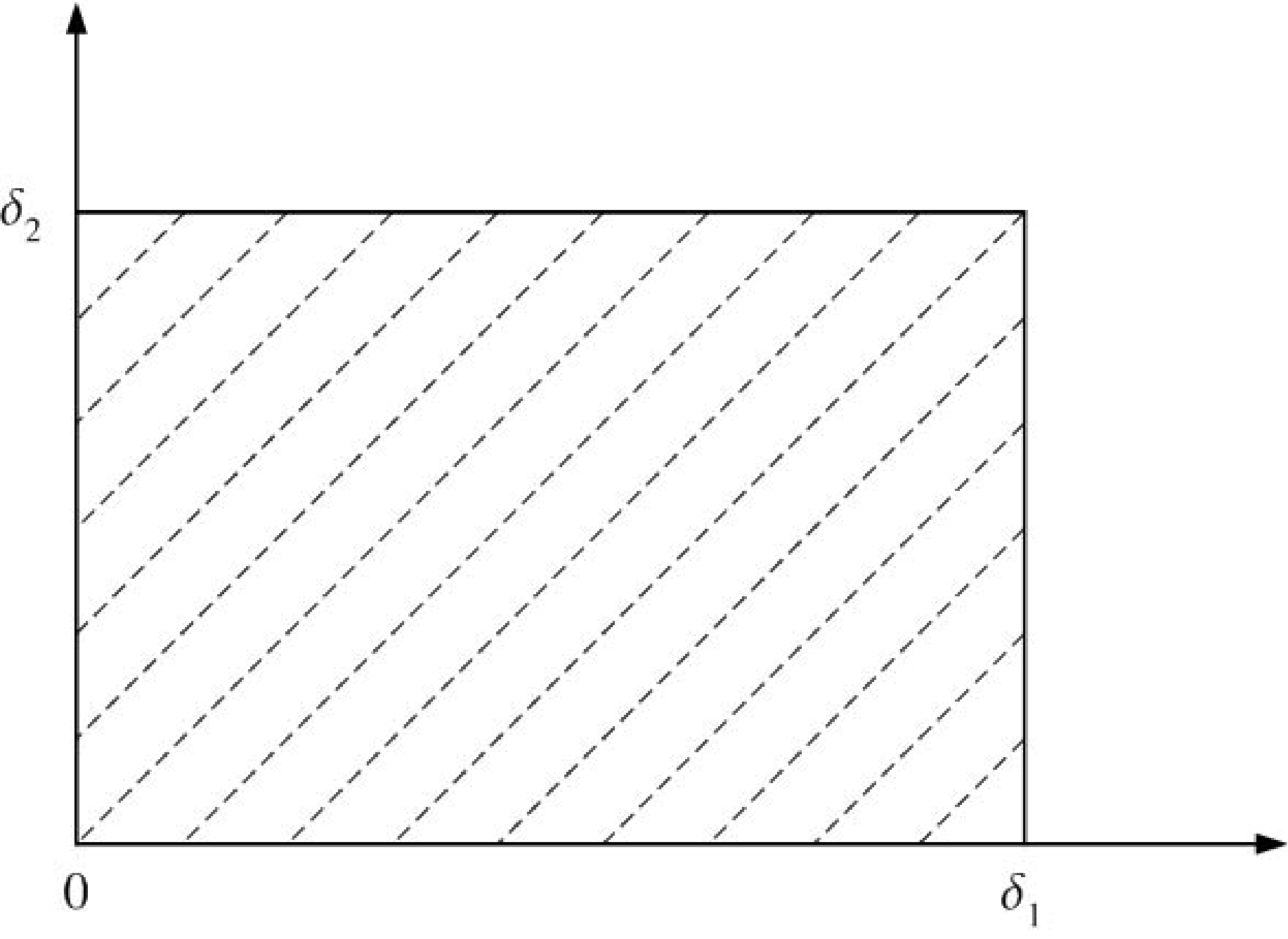}}}
\caption{
  Range of possible values of $N$ and $n$.
}
\label{Fig.1}
\end{figure}
The range of possible values of $N$ and $n$ is shown in Fig. \ref{Fig.1} where it is assumed that 
$\delta_1 \geq \delta_2$. The minimum and maximum values of $N$ are $N_{min}=n_{1min}-n_{2max}$ and 
$N_{max}=n_{1max}-n_{2min}$, respectively. Therefore $N$ can take $\delta_1+\delta_2$ values.
Each incident dashed line represents a set of states with the same value of $N$ and different values of $n$.
We now use $n_{min}$ and $n_{max}$ to define the minimum and maximum values of the relative dS momentum
$n$. For each fixed value of $N$ those values are different, i.e. they are functions of $N$. 
Let $\delta(N)=n_{max}-n_{min}$ for a given value of $N$. It is easy to see that $\delta(N)=0$ when
$N=N_{min}$ and $N=N_{max}$ while for other values of $N$, $\delta(N)$ is a natural number in the range
$(0,\delta_{max}]$ where $\delta_{max}=min(\delta_1,\delta_2)$. The total number of values of $(N,n)$ is
obviously $\delta_1\delta_2$, i.e.
\begin{equation}
\sum_{N=Nmin}^{Nmax}\delta(N)=\delta_1\delta_2
\label{states}
\end{equation}

As follows from Eq. (\ref{A1A2})
\begin{equation}
({\cal B}_1{\cal B}_2-{\cal A}_1{\cal A}_2)\psi_1(n_1)\psi_2(n_2)=
\frac{1}{2}[\psi_1(n_1+1)\psi_2(n_2+1)+\psi_1(n_1-1)\psi_2(n_2-1)]
\label{actionB}
\end{equation}
Therefore in terms of the variables $N$ and $n$ 
\begin{equation}
({\cal B}_1{\cal B}_2-{\cal A}_1{\cal A}_2)\psi(N,n)=
\frac{1}{2}[\psi(N,n+1)+\psi(N,n-1)]
\label{actionC}
\end{equation}
Hence the operator $({\cal B}_1{\cal B}_2-{\cal A}_1{\cal A}_2)$ does not act on the variable $N$ while its
action on the variable $n$ is described by the same expressions as the actions of the operators ${\cal B}_i$
($i=1,2$) on the corresponding WFs. Therefore, considering the two-body system, we will use
the notation ${\cal B}={\cal B}_1{\cal B}_2-{\cal A}_1{\cal A}_2$ and formally the action of this operator 
on the internal WF is the same as in the second expression in Eq. (\ref{A1A2}).
With this notation and with neglecting terms of the order of $r/R$ and higher, Eqs. (\ref{W3}) and (\ref{GC}) 
can be written as
\begin{equation}
G=1-{\cal B},\quad W=W_0-2(w_1+4n_1^2)^{1/2}(w_2+4n_2^2)^{1/2}G
\label{GW}
\end{equation} 

Since both, the operator ${\bf D}^2$ and the dS correction to the operator $W$ are defined by the same operator 
$G$, physical quantities corresponding to ${\bf D}^2$ and $W$ will be semiclassical or not depending on whether
the quantity corresponding to $G$ is semiclassical or not. As follows from Eq. (\ref{[]}), the spectrum of the 
operator ${\cal B}$ can be only in the range [0,1] and therefore, as follows from Eq. (\ref{GW}), the same is
true for the spectrum of the operator $G$. Hence, as follows from Eq. (\ref{GW}), any dS correction to the operator $W$ is negative and in the nonrelativistic approximation is proportional to particle masses.

\section{Validity of semiclassical approximation}
\label{validity}

Since classical mechanics works with a very high accuracy at macroscopic level, one might think that
the validity of semiclassical approximation at this level is beyond any doubts. However, to the best
of our knowledge, this question has not been investigated quantitatively. As discussed in Sect. \ref{classical},
such quantities as coordinates and momenta are semiclassicall if their uncertainties are much less than
the corresponding mean values. Consider WFs describing the motion of macroscopic bodies as
a whole (say the WFs of the Sun, the Earth, the Moon etc.). It is obvious that uncertainties of
coordinates in these WFs are much less than the corresponding macroscopic dimensions. 
What are those uncertainties for the Sun, the Earth, the 
Moon, etc.? What are the uncertainties of their momenta? In standard quantum mechanics, the validity of  
semiclassical approximation is defined by the product $\Delta r \Delta p$ while each uncertainty by
itself can be rather large. On the other hand, as shown in Chap. \ref{WPS}, standard position operator
should be reconsidered. Do we know what scenario for the distribution of momenta and coordinates 
takes place for macroscopic bodies?

In this section we consider several models of the function $\psi(n)$ where it is be possible to explicitly calculate
${\bar G}$ and $\Delta G$ and check whether the condition $\Delta G\ll |{\bar G}|$
(showing that the quantity $G$ in the state $\psi$ is semiclassical) is satisfied.
As follows from Eq. (\ref{[]}), $[G,n]=i{\cal A}$ where formally the action of this operator 
on the internal WF is the same as in the first expression in Eq. (\ref{A1A2}). Therefore, as
follows from Eq. (\ref{uncert}), $\Delta G \Delta n \geq {\bar{\cal A}}/2$.

As noted in Sect. \ref{classical}, one might think that a necessary condition for the validity of
semiclassical approximation is that the exponent in the semiclassical WF makes many oscillations
in the region where the WF is not small. We will consider WFs $\psi(n)$ 
containing $exp(-i\varphi n)$ such that $\psi(n)$ can be different from zero only if $n\in [n_{min},n_{max}]$.
Then, if $\delta=n_{max}-n_{min}$, the exponent makes $|\varphi| \delta/2\pi$ oscillations on $[n_{min},n_{max}]$
and $\varphi$ should satisfy the condition $|\varphi|\gg 1/ \delta$. The problem arises whether this condition
is sufficient.

Note that since the Poincare analog of $n$ is $pR$ then, from the point of view of standard theory, 
the Poincare analog of $\varphi$ is $r/R$. Therefore in the Poincare limit the quantity 
$\varphi$ becomes infinitely small and, as noted in Chap. \ref{WPS}, infinitely small quantities cannot be
semiclassical. In standard theory the quantities $p$ and $r$ are semiclassical if $\Delta p \Delta r$ is of the
order of unity and this condition can be achieved, for example, by choosing the Gaussian WF in the
momentum or coordinate representation. However, we do not have an analog of this situation because
we consider only discrete dS momentum WFs with a finite support.
 
Our first example is such that $\psi(n)=exp(-i\varphi n)/\delta^{1/2}$ if $n\in [n_{min},n_{max}]$. 
Then a simple calculation gives
\begin{eqnarray}
&&{\bar G}=1-cos\varphi+\frac{1}{\delta}cos\varphi,\quad \Delta G=\frac{(\delta-1)^{1/2}cos\varphi}{\delta},
\quad {\bar {\cal A}}=(1-\frac{1}{\delta})sin\varphi\nonumber\\
&&{\bar n}=(n_{min}+n_{max})/2,\quad \Delta n= \delta(\frac{1-1/\delta^2}{12})^{1/2}
\label{Abar}
\end{eqnarray}
Since $\varphi$ is of the order of $r/R$, we will always assume that $\varphi\ll 1$. Therefore for the validity of 
the condition $\Delta G\ll {\bar G}$, $|\varphi|$ should be not
only much greater than $1/\delta$ but even much greater than $1/\delta^{1/4}$. Note also that 
$\Delta G \Delta n$ is of the order of $\delta^{1/2}$, i.e. much greater than ${\bar {\cal A}}$.  This result shows
that the state $\psi(\nu)$ is strongly non-semiclassical. The calculation shows that for ensuring the 
validity of semiclassical approximation, one should consider functions $\psi(\nu)$ which are small when $n$ is close to 
$n_{min}$ or $n_{max}$.

The second example is $\psi(\nu)=const\,C_{\delta}^{\nu}exp(-i\varphi \nu)$ where $\nu = n-n_{min}$ and
 $const$ can be defined from the 
normalization condition. Since $C_{\delta}^{\nu}=0$ when $\nu <0$ or $\nu>\delta$, this function is not zero only
when $\nu\in [0,\delta]$. The result of calculations is that $const^2=1/C_{2\delta}^{\delta}$ and
\begin{eqnarray}
&&{\bar G}=1-cos\varphi+\frac{cos\varphi}{\delta+1},\quad 
\Delta G=[\frac{sin^2\varphi}{\delta+1}+\frac{2}{\delta^2}+O(\frac{1}{\delta^3})]^{1/2},
\quad {\bar {\cal A}}=\frac{\delta sin\varphi}{\delta+1}\nonumber\\
&&{\bar n}=\frac{1}{2}(n_{min}+n_{max}),\quad \Delta n= \frac{\delta}{2(2\delta-1)^{1/2}}
\label{C}
\end{eqnarray}
Now for the validity of the condition $\Delta G\ll {\bar G}$, $|\varphi|$ should be much greater than 
$1/\delta^{1/2}$ and $\Delta G\Delta n$ is of the order of $|{\bar {\cal A}}|$ which shows that the function is
semiclassical. The matter is that $\psi(\nu)$ has a sharp peak at $\nu=\delta/2$ and by using Stirling's formula it
is easy to see that the width of the peak is of the order of $\delta^{1/2}$. It is also clear from the expression for
${\bar G}$ that this quantity equals the semiclassical value $1-cos\varphi$ with a high accuracy only when
$|\varphi|\gg 1/ \delta^{1/2}$. This example might be considered as an 
indication that a semiclassical WF such that the condition
$|\varphi|\gg 1/ \delta$ is sufficient, should satisfy the following properties. On one hand the width of the
maximum should be of the order of $\delta$ and on the other the function should be small when $n$ is close to 
$n_{min}$ or $n_{max}$. 

In view of this remark, the third 
example is $\psi(\nu)=const\, exp(-i\varphi \nu) \nu(\delta-\nu)$ if 
$n\in [n_{min},n_{max}]$. Then the normalization condition
is $const^2 = [\delta(\delta^4-1)/30]^{-1}$ and the result of calculations is
\begin{eqnarray}
&&{\bar G}=1-cos\varphi +\frac{5cos\varphi}{\delta^2}+O(\frac{1}{\delta^3}),\quad {\bar {\cal A}}=sin\varphi\, (1-\frac{5}{\delta^2}),\quad {\bar n}=(n_{min}+n_{max})/2\nonumber\\
&&\overline{G^2}=(1-cos\varphi)^2+\frac{10}{\delta^2}(cos\varphi - cos2\varphi)+\frac{15cos\varphi}{\delta^3}+
O(\frac{1}{\delta^4})\nonumber\\
&&\Delta G=\frac{1}{\delta}[10sin^2\varphi+\frac{15cos\varphi}{\delta}+O(\frac{1}{\delta^2})]^{1/2},\quad 
\Delta n= \frac{\delta}{2\sqrt{7}}
\label{sin}
\end{eqnarray}
Now ${\bar G}\approx 1-cos\varphi$ if $|\varphi|\gg 1/ \delta$ but $\Delta G\ll |{\bar G}|$ only if
$|\varphi|\gg 1/ \delta^{3/4}$ and $\Delta G\Delta n$ is of the order of $|{\bar {\cal A}}|$ only if 
$|\varphi|\gg 1/ \delta^{1/2}$. The reason why the condition $|\varphi|\gg 1/ \delta$ is not sufficient is that
$\overline{G^2}$ approximately equals its classical value $(1-cos\varphi)^2$ only when $|\varphi|\gg 1/ \delta^{3/4}$.
The term with $1/\delta^3$ in $\overline{G^2}$ arises because when $\nu$ is close to 0, $\psi(\nu)$ is proportional
only to the first degree of $\nu$ and when $\nu$ is close to $\delta$, it is proportional to $\delta - \nu$.

Our last example is $\psi(\nu)=const\, exp(-i\varphi \nu) [\nu(\delta-\nu)]^2$ if 
$n\in [n_{min},n_{max}]$. It will suffice to estimate sums $\sum_{\nu=1}^{\delta}\nu^k$ by $\delta^{k+1}/(k+1)+O(\delta^k)$.
In particular, the normalization condition is $const^2 = 35\cdot 18/\delta^9$ and the result of calculations is
\begin{eqnarray}
&&{\bar G}=1-cos\varphi +\frac{6cos\varphi}{\delta^2}+O(\frac{1}{\delta^4}),\quad {\bar {\cal A}}=sin\varphi\, (1-\frac{6}{\delta^2}),\quad {\bar n}=(n_{min}+n_{max})/2\nonumber\\
&&\overline{G^2}=(1-cos\varphi)^2+\frac{12}{\delta^2}(cos\varphi - cos2\varphi)+O(\frac{1}{\delta^4})\nonumber\\
&&\Delta G=\frac{1}{\delta}[12sin^2\varphi+O(\frac{1}{\delta^2})]^{1/2},\quad 
\Delta n= \frac{\delta}{2\sqrt{11}}
\label{nu2}
\end{eqnarray}
In this example the condition $|\varphi|\gg 1/ \delta$ is sufficient to ensure that $\Delta G\ll |{\bar G}|$ and
$\Delta G\Delta n$ is of the order of $|{\bar {\cal A}}|$. 

At the same time, the following question arises. If we wish
to perform mathematical operations with a physical quantity in classical theory, we should guarantee that not only
this quantity is semiclassical but a sufficient number of its powers is semiclassical too. Since the classical value
of $G$ is proportional to $\varphi^2$ and $\varphi$ is small, there is no guaranty that for the quantity $G$ this is
the case. Consider, for example, whether $G^2$ is semiclassical. It is clear from Eq. (\ref{nu2}) that 
$\overline{G^2}$ is close to its classical value $(1-cos\varphi)^2$ if $|\varphi|\gg 1/ \delta$. However, 
$\Delta (G^2)$ will be semiclassical only if $\overline{G^4}$ is close to its classical value $(1-cos\varphi)^4$.
A calculation with the WF from the last example gives
\begin{eqnarray}
&&\overline{G^4}=(1-cos\varphi)^4+\frac{24}{\delta^2}(1-cos\varphi)^3(3+4cos\varphi)+\nonumber\\
&&\frac{84}{\delta^4}(1-cos\varphi)^2(64cos^2\varphi+11cos\varphi-6)+\frac{35\cdot 9}{2\delta^5}+O(\frac{1}{\delta^6})
\label{G4B}
\end{eqnarray}
Therefore $\overline{G^4}$ will be close to its classical value $(1-cos\varphi)^4$ only if $|\varphi|\gg 1/ \delta^{5/8}$.
Analogously, if $\psi(\nu)=const[\nu(\delta-\nu)]^3$ then $G^2$ will be semiclassical but $G^3$ will not. This consideration 
shows that a sufficient number of powers of $G$ will be semiclassical only if $\psi(n)$ is sufficiently
small in vicinities of $n_{min}$ and $n_{max}$. On the other hand, in the example described by Eq. (\ref{C}), the
width of maximum is much less than $\delta$ and therefore the condition $|\varphi|\gg 1/ \delta$ is still insufficient.

The problem arises whether it is possible to find a WF such that the contributions of the values of $\nu$ close to
0 or $\delta$ is negligible while the effective width of the maximum is or order $\delta$. For example, it is known that 
for any segment $[a,b]$ and any $\epsilon < (b-a)/2$ it is possible to find an infinitely differentiable function $f(x)$ 
on $[a,b]$ such that $f(x)=0$ if $x\notin [a,b]$ and $f(x)=1$ if $x\in [a+\epsilon, b-\epsilon]$. However, we cannot use such
functions for several reasons. First of all, the values of $\nu$ can be only integers: $\nu=0,1,2,...\delta$. Another reason
is that for correspondence with FQT we can use only rational functions and even $exp(-i\nu\varphi)$ should be expressed in terms of rational functions (see Sec. \ref{consequences}). 

In view of this discussion, we accept that the functions similar to that described in the second example give the best approximation
for semiclassical approximation since in that case it is possible to prove that
the condition $|\varphi|\gg 1/ \delta^{1/2}$ guarantees that sufficiently many
quantities $G^k$ ($k=1,2, ...$) will be semiclassical. The first step of the proof is
to show by induction that
\begin{equation}
G^k\psi(\nu)=\frac{(-1)^k}{2^k}\sum_{l=0}^{2k} C_{2k}^l(-1)^l\psi(\nu+k-l)
\label{Gk}
\end{equation}
Then the calculation of the explicit expression for $\overline{G^k}$
involves hypergeometric functions
$$F(-\delta, -\delta+k;k+1;1)=\sum_{l=0}^{\infty}\frac{(-\delta)_l (-\delta+k)_l}{l!(k+1)_l}$$
where $(k)_l$ is the Pochhammer symbol. Such sums are finite and can be calculated by using the Saalschutz theorem
\cite{BE}: $F(-\delta, -\delta+k;k+1;1)=k!(2\delta+k)!/\delta!(\delta+k)!$. As a result,
\begin{equation}
\overline{G^k}=\frac{(-1)^k(\delta!)^2exp(-i\varphi k)}{2^k(\delta+k)!(\delta-k)!}
F(-2k,-\delta-k;\delta-k+1; exp[i(\varphi+\pi)])
\label{overlineGk}
\end{equation}
The hypergeometric function in this expression can be rewritten by using the formula \cite{BE}
$$F(a,b;1+a-b;z)=(1+z)^{-a}F[\frac{a}{2},\frac{a+1}{2};1+a-b;\frac{4z}{(1+z)^2}]$$
As a consequence 
\begin{equation}
\overline{G^k}=\frac{2^k(\delta!)^2}{(\delta+k)!(\delta-k)!}
\sum_{l=0}^k\frac{(-k)_l(-k+\frac{1}{2})_l}{l!(\delta+1-k)_l}(sin\frac{\varphi}{2})^{2(k-l)}
\label{overlineGkB}
\end{equation}
This result shows that $\overline{G^k}$ is given by a series in powers of 
$1/[\delta sin^2(\varphi/2)]$.
Hence if $\varphi\ll 1$ but $|\varphi|\gg 1/ \delta^{1/2}$ we get that the classical 
expression for $\overline{G^k}$ is $(\overline{G^k})_{class}=2^ksin^{2k}(\varphi /2)$
and the semiclassical approximation for $G^k$ is valid since if $k\ll \delta$ then
\begin{equation}
\frac{\Delta(G^k)}{\overline{G^k}}=\frac{(2k^2-k)^{1/2}}{\delta^{1/2}sin(\varphi/2)}
+O(\frac{1}{\delta sin^2(\varphi/2)}) 
\end{equation}

Since $\varphi$ is of the order of $r/R$, the condition $|\varphi|\gg 1/ \delta^{1/2}$ is definitely satisfied at
cosmological distances while the problem arises whether it is satisfied in the Solar System. Since $\delta$ can be 
treated as $2R\Delta q$ where $\Delta q$ is the width of the relative momentum distribution
in the internal two-body WF, $\varphi\delta$ is of the order of $r\Delta q$. For understanding what the order 
of magnitude of this quantity is, one should have estimations of $\Delta q$ for macroscopic WFs.
However, to the best of our knowledge, the existing theory does not make it possible to give reliable estimations 
of this quantity. 

Below we argue that $\Delta q$ is of the order of $1/r_g$ where $r_g$ is the gravitational (Schwarzschild)
radius of the component of the two-body system which has the greater mass. Then $\varphi\delta$ is of the order of
$r/r_g$. This is precisely the parameter defining when standard Newtonian gravity is a good approximation to GR. 
For example, the gravitational radius of the Earth is of the order of $0.01m$ while the radius of the Earth is 
$R_E=6.4\times 10^6m$. Therefore $R_E/r_g$ is of the order of $10^9$. The gravitational radius of the Sun is 
of the order of 
$3000m$, the distance from the Sun to the Earth is or order $150\times 10^9m$ and so $r/r_g$ is of the order of $10^8$.
At the same time, the above discussion shows that the condition $\varphi\delta\gg 1$ is not sufficient for ensuring
semiclassical approximation while the condition $|\varphi|\gg 1/ \delta^{1/2}$ is. Hence we should compare the
quantities $r/R$ and $(r_g/R)^{1/2}$. Then it is immediately clear that the requirement $|\varphi|\gg 1/ \delta^{1/2}$
will not be satisfied if $R$ is very large. For example, if $R$ is of the order of $10^{26}m$ then in the example with the Earth
$r/R$ is of the order of $10^{-19}$ and $(r_g/R)^{1/2}$ is of the order of $10^{-14}$ while in the example with the Sun
$r/R$ is of the order of $10^{-15}$ and $(r_g/R)^{1/2}$ is of the order of $10^{-10}$. Therefore in these examples the
requirement $|\varphi|\gg 1/ \delta^{1/2}$ is not satisfied. 

Our concusion follows. As shown in Chap. \ref{WPS}, even in standard Poincare invariant theory the position
operator should be defined not as $i\hbar\partial/\partial {\bf p}$ but by the operators
$({\cal R}_{||},{\bf {\cal R}}_{\bot})$. At the same time, the distance operator can be still defined in standard way,
i.e. by the
operator $-\hbar^2 (\partial/\partial{\bf p})^2$. However, explicit examples discussed in this section show that
for macroscopic bodies semiclassical approximation can be valid only if standard distance operator is modified too. 
In particular, as follows, for example, from Eq. (\ref{C}), the quantity $\Delta \varphi$ is of the order of
$1/\delta^{1/2}$ and therefore the quantity $\delta \Delta\varphi$ is not of the order of unity, as it should be in
standard theory, but of the order of $\delta^{1/2}$.

\section{Distance operator for macroscopic bodies}
\label{macroscopicdist}

As noted in Chap. \ref{WPS}, standard position operator in quantum theory is defined by the requirement that
the momentum and coordinate representations are related to each other by a Fourier transform and this requirement is
postulated by analogy with classical electrodynamics. However, as discussed in Chap. \ref{WPS}, the validity
of such a requirement is problematic and there exist situations when standard position operator does not work.
In addition, in Poincare invariant theories  there is no parameter $R$; in particular rapidly oscillating 
exponents do not contain this parameter. 

In the case of macroscopic bodies a new complication arises. 
It will be argued in the next chapters that in FQT the width $\delta$ of the $n$-distribution for a  
macroscopic body is inversely proportional to its mass. Therefore for nuclei and elementary particles 
the quantity $\delta$ is much greater than for macroscopic bodies and the requirement 
$|\varphi|\gg 1/ \delta^{1/2}$ can be satisfied in some situations. On the other hand, such a treatment 
of the distance operator for macroscopic bodies is incompatible with semiclassical approximation since, 
as discussed in the preceding section, if the distances are not cosmological then $\varphi$ is typically 
much less than $1/ \delta^{1/2}$. 
Hence the interpretation of the distance operator for macroscopic bodies should be modified.

As noted in Secs. \ref{classical} and \ref{NRWPS}, in standard theory the semiclassical WF in
momentum space has the form $exp(-i{\bf r}{\bf p})a({\bf p})$ where the amplitude $a({\bf p})$ has a sharp
maximum at the classical value of momentum ${\bf p}={\bf p}_0$ and ${\bf r}$ is the classical radius-vector.
This property is based on the fact that in standard theory the coordinate and momentum representations are related
to each other by the Fourier transform. However, as shown in Chap. \ref{WPS}, standard position operator should be
modified and hence the problem of the form of the semiclassical WF should be reconsidered.
In this section we discuss how the semiclassical WF in the $n$-representation should depend on
the classical value $\varphi$.

As noted in Sec. \ref{classical}, a necessary condition for semiclassical approximation is that the WF should make 
many oscillations in the region where its amplitude is not negligible. Hence if the 
rapidly oscillating exponent in the WF is $exp(-i\varphi n)$ then the number of oscillations is
of the order of $\varphi\delta$ and this number is large if $\varphi\gg 1/\delta$. As noted in the preceding section,
this condition is typically satisfied but for the validity of semiclassical approximation the value of
$\varphi$ should be not only much greater than $1/\delta$ but even much greater than $1/\delta^{1/2}$.
We assume that in the general case the rapidly oscillating exponent in the WF is not $exp(-i\varphi n)$
but $exp(-i\chi n)$ where $\chi$ is a function of $\varphi$ such that $\chi(\varphi)=\varphi$ when 
$\varphi\gg 1/\delta^{1/2}$ (in particular when $\varphi$ is of the order of cosmological distances) while for
macroscopic bodies in the Solar System (when $\varphi$ is very small), $\chi$ is a function of $\varphi=r/R$ to be 
determined. Note
that when we discussed the operator ${\bf D}^2$ compatible with standard interpretation of the distance
operator, we did not neglect ${\bf J}$ in this operator and treated $|\varphi|$ as $r_{||}/R$. However,
when we neglect all corrections of the order of $1/R$ and higher, we neglect ${\bf J}$ in ${\bf D}^2$ and
replace $\varphi$ by $\chi$ which does not vanish when $R\to\infty$. As shown in Sect. \ref{semiclass},
the operator ${\bf D}^2$ is rotationally invariant since the internal two-body momentum operator is a reduction
of the operator ${\bf J}^{"}$ on the two-body rest states, ${\bf D}$ satisfies Eq. (\ref{DJ}) and therefore
$[{\bf J}^{"},{\bf D}^2]=0$. Hence $\chi$ can be only a function of $r$ but not $r_{||}$.

Ideally, a physical interpretation of an operator of a physical quantity should be obtained from the
quantum theory of measurements which should describe the operator in terms of a measurement of the corresponding
physical quantity. However, although quantum theory is known for 90+ years, the quantum theory of measurements
has not been developed yet. Our judgment about operators of different physical quantities can be based only
on intuition and comparison of theory and experiment.  As noted in 
Sect. \ref{classical}, in view of our macroscopic experience, it seems unreasonable that if the
uncertainty $\Delta r$ of $r$ does not depend on $r$  then the relative accuracy 
$\Delta r/r$ in the measurement of $r$ is better when $r$ is greater. 

When $exp(-i\varphi n)$ is replaced by $exp(-i\chi n)$, the
results obtained in the preceding section remain valid but $\varphi$ should be
replaced by $\chi$. Suppose that when $\varphi$ is of the order of $1/\delta^{1/2}$ or less, $\chi=f(C(\varphi\delta)^{\alpha})$ where $C$ is a constant and $f(x)$ is a function such that $f(x)=x+o(x)$ where the correction 
$o(x)$ will be discussed later. Then if $\chi$ and $\varphi$ are treated not as classical but as quantum 
physical quantities we 
have that $\Delta\chi\approx C \varphi^{\alpha-1}\delta^{\alpha}\Delta\varphi$.
If $\varphi$ is replaced by $\chi$ then, as follows from the first expression in Eq. (\ref{C}), if 
$\chi\gg 1/\delta^{1/2}$ and $\chi\ll 1$,
the operator $G$ can be treated as the operator of the quantity $\chi^2/2$. Then it follows from the
second expression in Eq. (\ref{C}) that $\Delta(\chi^2)$ is of the order of $\chi/\delta^{1/2}$
and therefore $\Delta\chi$ is of the order of $1/\delta^{1/2}$. As a consequence,
$\Delta \varphi\approx const\cdot \varphi(\varphi\delta)^{-\alpha}/\delta^{1/2}$. Since
$(\varphi\gg 1/\delta)$, the accuracy of the 
measurement of $\varphi$ is better when $\alpha<0$. In that case the relative accuracy $\Delta\varphi/\varphi$ is
better for lesser values of $\varphi$. As noted in Sect. \ref{classical}, this is a desired behavior in view of
our macroscopic experience (in particular when the coordinates are defined by Eq. (\ref{coordmom})). Note also that the condition $\alpha<0$ is natural from the fact that $\chi\ll 1$ is
a necessary condition for the WF in momentum representation to be approximately continuous since the
standard momentum is of the order of $n/R$.

If $\alpha<0$ then $\Delta \varphi\approx const\cdot \varphi(\varphi\delta)^{|\alpha|}/\delta^{1/2}$.
In view of quantum mechanical experience, one might expect that the accuracy should be better if $\delta$ is greater. 
On the other hand, in our approach $\delta$ is inversely proportional to the masses of the bodies under consideration and our
macroscopic experience tells us that the accuracy of the measurement of relative distance does not depend on the mass.
Indeed, suppose that we measure a distance by sending a light signal. Then the accuracy of the measurement should not
depend on whether the signal is reflected by the mass $1kg$ or $1000kg$. Therefore at macroscopic level the accuracy should not depend on $\delta$. Hence the optimal choice is $\alpha=-1/2$. In that case 
$\Delta \varphi\approx const\cdot \varphi^{3/2}$ and $\chi=f(C/(\varphi\delta)^{1/2})$. Then, if $C$ is of the order of unity, the 
condition $\chi\gg 1/\delta^{1/2}$, which, as explained in the preceding section, guarantees that semiclassical
approximation is valid, is automatically satisfied since in the Solar System we always have $(R/r)^{1/2}\gg 1$. We will see
in the next section that such a dependence of $\chi$ on $\varphi$ and $\delta$ gives a natural explanation of the
Newton law of gravity.

\section{Newton's law of gravity}
\label{Newtonlaw}

As follows from Eqs. (\ref{C}), with $\varphi$ replaced by $\chi$, the mean value of the operator $G$
is $1-cos\chi$ with a high accuracy.
Consider two-body WFs having the form $\psi(N,n)=[\delta(N)/(\delta_1\delta_2)]^{1/2}\psi(n)$.
As follows from Eq. (\ref{states}), such functions are normalized to one. Then, as follows from Eq. (\ref{GW}),
the mean value of the operator $W$ can be written as
\begin{eqnarray}
&&\overline{W}=4R^2M_0^2+{\overline {\Delta W}},\quad 
{\overline {\Delta W}}=-2[(w_1+4n_1^2)(w_2+4n_2^2)]^{1/2}F(\delta_1,\delta_2,\varphi) \nonumber\\
&&F(\delta_1,\delta_2,\varphi)=\frac{1}{\delta_1\delta_2}\sum_{N=Nmin}^{Nmax}\delta(N)
\{1-cos[f(\frac{C}{(\varphi\delta(N))^{1/2}})]\}
\label{meanI2}
\end{eqnarray}
Strictly speaking, the semiclassical form of the WF $exp(-i\chi n)a(n)$ cannot be used if $\delta(N)$ is
very small; in particular, it cannot be used when $\delta(N)=0$. We assume that in these cases the internal wave
function can be modified such that the main contribution to the sum in Eq. (\ref{meanI2}) is given by those $N$ where
$\delta(N)$ is not small. 

If $\varphi$ is so large that the argument $\alpha$ of $cos$ in Eq. (\ref{meanI2}) is extremely small, 
then the correction to Poincare limit is negligible. The next approximation is that this argument is small such we can approximate $cos(\alpha)$ by
$1-\alpha^2/2$. Then, taking into account that $f(\alpha)=\alpha+o(\alpha)$ and that the number of values of $N$ is $\delta_1+\delta_2$ we get
\begin{equation}
{\overline {\Delta W}}=-C^2[(w_1+4n_1^2)(w_2+4n_2^2)]^{1/2}\frac{\delta_1+\delta_2}{\delta_1\delta_2|\varphi|}
\label{preNewton}
\end{equation}
Now, by analogy with the derivation of Eq. (\ref{33}), it follows that the 
classical nonrelativistic Hamiltonian is
\begin{equation}
H_{nr}({\bf r}, {\bf q}) =\frac{{\bf q}^2}{2m_{12}} - \frac{m_1m_2RC^2}{2(m_1+m_2)r}(\frac{1}{\delta_1}+\frac{1}{\delta_2})
\label{preNewton2}
\end{equation}
We see that the correction disappears if the width of the 
dS momentum distribution for each body becomes very large. In standard theory (over complex numbers)
there is no limitation on the width of distribution while, as noted in the preceding section, in semiclassical
approximation the only limitation is that the width of the dS momentum distribution should be much less
than the mean value of this momentum. In the next chapters we argue that in FQT it is natural that
the width of the momentum distribution for a macroscopic body is inversely proportional to its mass.
Then we recover the Newton gravitational law. Namely, if
\begin{equation}
\delta_j=\frac{R}{m_jG'}\quad (j=1,2), \quad C^2G'=2G 
\label{deltaj}
\end{equation}
then
\begin{equation}
H_{nr}({\bf r}, {\bf q}) =\frac{{\bf q}^2}{2m_{12}} - G\frac{m_1m_2}{r}
\label{Newton}
\end{equation}
We conclude that in our approach gravity is simply a dS the correction to standard nonrelativistic Hamiltonian. This
correction is spherically symmetric since, as noted in the beginning of this section, when all corrections of the 
order of $1/R$ 
are neglected, the dependence of ${\bf D}^2$ on ${\bf J}$ disappears.

\section{Special case: very large $m_2$}
\label{largem2}

Consider a special case when $m_2\gg m_1,|{\bf q}|$ and we do not assume that particle 1 is nonrelativistic.
As noted above, in the c.m. frame of the two-body system $n_1\approx n_2\approx n$. Since in this reference frame
the vectors ${\bf B}_1$ and ${\bf B}_2$ are approximately antiparallel and $|{\bf B}_1|\approx |{\bf B}_2|\approx 2n$,
it follows from Eq. (\ref{W0}) that
\begin{equation}
{\overline W_0}=[(w_1+4n^2)^{1/2}+(w_2+4n^2)^{1/2}]^2\approx [(w_1+4n^2)^{1/2}+w_2^{1/2}]^2\approx w_2+2w_2^{1/2}(w_1+4n^2)^{1/2}
\label{W0B}
\end{equation}
since $w_2\gg w_1,4n^2$.

Consider now the calculation of the quantity $F(\delta_1,\delta_2,\varphi)$ in Eq. (\ref{meanI2}). If the
quantities $\delta_i$ $(i=1,2)$ are inversely proportional to the corresponding masses then 
$\delta_1\gg \delta_2$. Now it is
clear from Fig. \ref{Fig.1} that in the sum for $F(\delta_1,\delta_2,\varphi)$ the number of terms approximately equals $\delta_1$ and in almost all of them $\delta(N)=\delta_2$. Hence $F(\delta_1,\delta_2,\varphi)\approx 1-cos\chi$ where
$\chi=f(C/(\varphi\delta_2)^{1/2})$. Then, as follows from Eqs. (\ref{meanI2}) and (\ref{W0B})
\begin{equation}
{\overline W^{1/2}}\approx w_2^{1/2}+(w_1+4n^2)^{1/2}cos\chi
\label{W12}
\end{equation}

Equation (\ref{W12}) is derived neglecting all the corrections of the order of $1/R$ and higher; 
in particular it is assumed that $k\ll n$. Hence the last term in Eq. (\ref{W12}) differs from the first term in 
Eq. (\ref{26}) only such that $\varphi$ is replaced by $\chi$. This is a consequence of the fact that the latter
has been derived by considering the single-particle WF and assuming that it contains
$exp(-i\varphi n)$ while the former has been derived by considering the WF in the c.m. frame and assuming 
that its dependence on the relative momentum variable $n$ contains $exp(-i\chi n)$. 

Since $W=4R^2M^2$ where $M$ is the standard two-body mass operator, it follows from Eq. (\ref{W12}) that if $m_2$ is very large then the mass operator of the two-body problem is fully defined by the energy of {\it free} particle 1 in the c.m. frame
of the two-body system. For example, when $f(x)\approx x$ then by analogy with the derivation of Eq. (\ref{Newton})
we get that the energy of particle 1 in the c.m. frame is
\begin{equation}
H_{rel}({\bf r}, {\bf q})=(m_1^2+{\bf q}^2)^{1/2}(1-\frac{Gm_2}{r})
\label{Hrel}
\end{equation}
and the nonrelativistic expression for this energy is 
\begin{equation}
H_{nr}({\bf r}, {\bf q})=\frac{{\bf q}^2}{2m_1}-\frac{Gm_1m_2}{r}
\label{Hnr}
\end{equation}

Let us stress that Eq. (\ref{Hnr}) is the nonrelativistic energy of {\it free} particle 1 in the c.m.
frame of the two-body system. In standard theory this expression is treated as a result of gravitational
interaction of particle 1 with the massive body having the mass $m_2$. {\it Hence in our approach gravity is simply
a kinematical consequence of dS symmetry}. 

By analogy with the single-body case, the internal two-body WF can be written as $\psi(n,k,\mu)$ where $n$
is the quantum number characterizing the magnitude of the relative dS momentum, $k$ is the quantum number 
characterizing the magnitude of the relative angular momentum and $\mu$ is the quantum number 
characterizing the $z$ projection of the relative angular momentum. If $m_1\ll m_2$ then in the c.m. frame the
radius-vector of particle 1 is much greater than the radius-vector of particle 2. As noted above, in the c.m. frame
$n_1\approx n_2\approx n$. Therefore the relative angular
momentum approximately equals the angular momentum of particle 1 in the two-body c.m. frame. As a consequence, 
$\psi(n,k,\mu)$ can be treated as a WF of particle 1 in the c.m. frame. The only difference between
this WF and the single-particle WF for the free particle 1 is that in the case $m_1\ll m_2$
the width of the $n$-distribution in the c.m. frame equals $\delta_2$, not $\delta_1$ as for the free particle 1.
As a consequence, the energy of particle 1 in the c.m. frame is described by Eqs. (\ref{Hrel}) and (\ref{Hnr}). 

In view of the analogy between the description of free particle 1 and particle 1 in the two-body c.m. frame, for
describing semiclassical values of the dS operators of particle 1 in the c.m. frame one can use the results of
Sec. \ref{S7} and Eq. (\ref{26xy}) where $\varphi$ is replaced by $\chi$. The classical motion of particle 1 in the
$xy$ plane such that $J_z>0$ corresponds to the case $\alpha =-\pi/2$ and $\mu=k$. Then, taking into account that
$k\ll n$, it follows from Eq. (\ref{26xy}) that
\begin{eqnarray}
&&B_x=-2nsin\beta,\quad B_y=2ncos\beta,\quad J_z=2k,\quad B_z=N_z={\bf J}_{\bot}=0\nonumber\\
&&{\cal E}=(w+4n^2)^{1/2}cos\chi,\quad N_x=(w+4n^2)^{1/2}(sin\chi sin\beta-\frac{k}{n}cos\chi cos\beta)\nonumber\\ 
&& N_y=(w+4n^2)^{1/2}(-sin\chi cos\beta-\frac{k}{n}cos\chi sin\beta) 
\label{xyplane}
\end{eqnarray}

For describing vectors in the $xy$ plane we will use the following notation. If the vector ${\bf A}$ has the
components $(A_x,A_y)$ then we will write ${\bf A}=(A_x,A_y)$. As in Sec. \ref{S7}, the relation between the
momentum ${\bf q}$ of particle 1 in the c.m. frame and the vector ${\bf B}$ is ${\bf q}={\bf B}/2R$, standard 
energy $E$ equals ${\cal E}/2R$ and the $\bot$ and $||$ components of the vector ${\bf N}$ are defined as in Sec. \ref{position}. Then, as follows from Eq. (\ref{xyplane}), ${\bf N}_{\bot}=-2ERk(cos\beta,sin\beta)/n$. Since ${\bf r}_{\bot}$ is defined such that ${\bf N}_{\bot}=-2E{\bf r}_{\bot}$
and $n=Rq$ where $q=|{\bf q}|$ we get that ${\bf r}_{\bot}=k(cos\beta,sin\beta)/q$ and hence the vector ${\bf r}$ can be written as
${\bf r}=r_{||}(sin\beta,-cos\beta)+|{\bf r}_{\bot}|(cos\beta,\sin\beta)$.

Since we work in units where $\hbar/2=1$ then $k=|{\bf r}_{\bot}|q$ and in standard units $J_z=L$ and 
$|{\bf r}_{\bot}|=L/q$. We now define the angles $\gamma_1$ and $\gamma_2$ such that $\beta=\pi/2+\gamma_1$,
$sin\gamma_2=L/qr$ and $cos\gamma_2=[1-(L/qr)^2]^{1/2}$ where $r=|{\bf r}|$. Then the final result for the vectors
${\bf q}$ and ${\bf r}$ can be written as
\begin{eqnarray}
{\bf q}=q(1-\frac{L^2}{q^2r^2})^{1/2}(cos\varphi,sin\varphi)+\frac{L}{r}(-sin\varphi,cos\varphi),\quad 
{\bf r}=r(cos\varphi,sin\varphi)
\label{pr}
\end{eqnarray}
where $\varphi=\gamma_1-\gamma_2$ and we assume that $sin\chi > 0$. Standard energy of particle 1 in the
c.m. frame is $E=(m^2+{\bf q}^2)^{1/2}cos\chi$ where $m=m_1$ and $\chi$ is a finction of $r$ discussed in the
preceding sections.

\section{Classical equations of motion}
\label{classeq}

Classical equations of motion should follow from quantum theory if the evolution operator is known. By analogy with
standard Schr\"{o}dinger equation one might think that the internal two-body evolution operator is $exp(-iMt)$ where
$M$ is the two-body mass operator. However, as discussed in Sec. \ref{ST}, the problem of time in quantum theory has
not been solved yet and such an evolution operator is problematic. Nevertheless, if the evolution operator is defined
by $M$ then on classical level the two-body mass and the quantities corresponding to operators commuting with $M$ are
conserved. In particular, if $L$ is the classical value of $J_3$ then $L$ is conserved.

In this section we show that classical equations of motion for all standard gravitational two-body problems can be
obtained according to the following scheme. We assume that classical values of the {\it free} two-body mass $M$ and $L$
are conserved. In the case when $m_1\ll m_2$ we assume that, according to Eq. (\ref{coordmom}), the coordinates are 
{\it defined} as
\begin{equation}
d{\bf r}=\frac{{\bf q}}{\epsilon(q)}dt
\label{rq}
\end{equation}
where $\epsilon(q)=(m^2+q^2)^{1/2}$. Then the results are generalized to the case when $m_1$ and $m_2$ are
comparable to each other. Note that the above conditions fully define the motion; in particular there is no need to
involve Lagrange equations, Hamilton equations, Hamilton-Jacobi equations etc.

Consider first the case when $m_1\ll m_2$. The three classical tests of GR --- precession of Mercury's perihelion,
gravitational red shift of light and deflection of light by the Sun --- can be discussed in this approximation.
If $\xi=sin^2\chi$ then, as discussed in the preceding sections,  
$\xi$ can be written as a series in powers of $(r_g/r)$ where $r_g$ is the gravitational radius of particle 2: 
$\xi=(r_g/r)+a(r_g/r)^2+...$.

The consideration of the gravitational red shift of light does not require Eq. (\ref{rq}) and equations of motion.
In that case it suffices to note that, according to Eq. (\ref{Hrel}), if particle 1 is the photon then 
in the approximation when $\xi=r_g/r$ its energy in standard units is 
\begin{equation}
E=qc(1-\frac{r_g}{2r})=qc(1-\frac{Gm_2}{c^2r})
\label{Ephoton}
\end{equation}
Consider the case when the photon travels in the radial direction from the Earth surface
to the height $h$. Let $R_E$ be the Earth radius, $q_1$ be the photon momentum on the Earth surface when $r=R_E$
and $q_2$ be the photon momentum when the photon is on the height $h$, i.e. when $r=R_E+h$. The corresponding photon
kinetic energies are $E_1=q_1c$ and $E_2=q_2c$, respectively. Since $E$ is the conserved quantity, it easily follows
from Eq. (\ref{Ephoton}) that if $h\ll R_E$ then $\Delta E_1=E_2-E_1\approx -E_1gh/c^2$ where $g$ is the free fall acceleration.
Therefore one can formally define the potential energy of the photon near the Earth surface by $U(h)=E_1gh/c^2$
and we have a full analogy with classical mechanics. From the formal point of view, the
result is in agreement with GR and the usual statement is that this effect has been 
measured in the famous Pound-Rebka experiment. We discuss this question in Sec. \ref{nonNewton}.

Consider now the derivation of equations of motions in the case when $m_1\ll m_2$. As follows from Eq. (\ref{pr})
\begin{equation}
\label{drB}
d{\bf r}=dr(cos\varphi,sin\varphi)+rd\varphi(-sin\varphi,cos\varphi)
\end{equation}
Therefore, as follows from Eqs. (\ref{pr}) and (\ref{rq}), the equations of motion have the form
\begin{equation}
\frac{dr}{dt}=\frac{1}{\epsilon(q)}(q^2-\frac{L^2}{r^2})^{1/2},\quad \frac{d\varphi}{dt}=\frac{L}{r^2\epsilon(q)}
\label{motion}
\end{equation}
where $q$ as a function of $r$ should be found from the condition that $E$ is a constant of motion. Since $E=\epsilon(q)cos\chi$, we have that 
\begin{equation}
q(r)^2=\frac{E^2}{1-\xi(r)}-m^2
\label{q(r)}
\end{equation}
In such problems as deflection of light by the Sun and precession of Mercury's perihelion it suffices to find only
the trajectory of particle 1. As follows from Eq. (\ref{motion}), the equation defining the trajectory is
\begin{equation}
\frac{d\varphi}{dr}=\frac{L}{r[r^2q(r)^2-L^2]^{1/2}}
\label{trajectory}
\end{equation}

Consider first deflection of light by the Sun. If $\rho$ is the minimal distance between the photon and the Sun
then when $r=\rho$ the radial component of the momentum is zero and hence, as follows from Eq. (\ref{Ephoton})
\begin{equation}
q(\rho)=\frac{L}{\rho},\quad E=\frac{L}{\rho}(1-\frac{r_g}{2\rho})
\label{Erho}
\end{equation}
Suppose that in Eq. (\ref{q(r)}) a good approximation is when only the terms linear in $r_g/r$ can be taken into
account. Then $q(r)^2\approx E^2(1+r_g/r)$ and, as follows from Eqs. (\ref{trajectory}) and (\ref{Erho}), in first order
in $r_g/r$
\begin{equation}
\frac{d\varphi}{dr}=\frac{\rho}{r}[r^2(1+\frac{r_g}{r}-\frac{r_g}{\rho})-\rho^2]^{-1/2}
\label{dphidr}
\end{equation}
Suppose that in the initial state the $y$ coordinate of the photon was $-\infty$, at the closest distance
to the Sun its coordinates are $(x=\rho,y=0)$ and in the final state the $y$ coordinate is $+\infty$. Then,
as follows from Eq. (\ref{dphidr}), the total change of the photon angle is
\begin{equation}
\Delta \varphi=2\int_{\rho}^{\infty}\frac{\rho}{r}[r^2(1+\frac{r_g}{r}-\frac{r_g}{\rho})-\rho^2]^{-1/2}dr 
\label{deltaphiB}
\end{equation}
The quantities $r_g/\rho$ and $r_g/r$ are very small and in the main approximation those quantities can be
neglected. Then $\Delta\varphi =\pi$ what corresponds to the non-deflected motion along a straight line.
In the next approximation in $r_g/\rho$
\begin{equation} 
\Delta\varphi=\pi+\frac{r_g}{\rho}
\label{deflB}
\end{equation}
This result is discussed in Sec. \ref{nonNewton}.

Consider now the trajectory of particle 1 if $m_1$ is arbitrary but such that $m_1\ll m_2$ and the
terms quadratic in $r_g/r$ should be taken into account. Then $E/(1-\xi)\approx E(1+\xi+\xi^2)$ and, as
follows from Eqs. (\ref{q(r)}) and (\ref{trajectory})
\begin{equation}
\frac{d\varphi}{dr}=\frac{L}{r}[(E^2-m^2)r^2+E^2r_gr+E^2r_g^2(1+a)-L^2]^{-1/2}
\label{Merc}
\end{equation}
If $E<m$ then it is clear from this expression that the quantity $r$ can be only in a finite range $[r_1,r_2]$.

For defining the trajectory one can use the fact that
$$\int \frac{dx}{x (-ax^2+bx-c)^{1/2} }=\frac{i}{\sqrt(c)}ln[A(x)+iB(x)]$$
where
$$A(x)=\frac{2}{x}(-ax^2+bx-c)^{1/2},\quad B(x)=\frac{i(bx-2c)}{xc^{1/2}}$$

Since $lnz=ln|z|+iarg(z)$, the result of integration of Eq. (\ref{Merc}) is
\begin{equation}
\label{MercB}
\varphi(r)=const+\frac{L}{[L^2-E^2r_g^2(1+a)]^{1/2}}arcsin[F(r)]
\end{equation}
where the explicit form of the function $F(r)$ is not important for our goal. It follows from this expression that 
the difference of the angles for consecutive perihelia is
\begin{equation}
\Delta\varphi=\frac{2\pi L}{[L^2-E^2r_g^2(1+a)]^{1/2}}
\label{MercC}
\end{equation}
If $E^2r_g^2(1+a) \ll L^2$ and particle 1 is nonrelativistic this expression  can be written as
\begin{equation}
\Delta\varphi=2\pi+ \frac{4\pi m^2m_2^2G^2(1+a)}{L^2}
\label{MercD}
\end{equation}
and the result of GR is recovered if $a=1/2$. This result is discussed in Sec. \ref{nonNewton}.

Note that in the three classical tests of GR we need only trajectories, i.e. the knowledge of the functions $r(t)$
and $\varphi(t)$ is not needed. Then it is clear that although for the derivation of Eq. (\ref{trajectory}) we
used Eq. (\ref{rq}), the only property of this equation needed for defining trajectories is that $d{\bf r}$ is
proportional to ${\bf q}$. However, for defining the functions $r(t)$ and $\varphi(t)$ it is important that
$d{\bf r}/dt$ is the velocity defined as ${\bf q}/\epsilon(q)$.

For example, as follows from Eqs. (\ref{motion}) and (\ref{q(r)}) the relation between $t$ and $r$ is
\begin{equation}
t(r)=E\int \frac{dr}{[1-\xi(r)]^{1/2}[(E^2-m^2)+E^2\xi(r)(1+\xi(r))-L^2/r^2]^{1/2}}
\label{t(r)}
\end{equation}
Taking into account corrections of the order of $r_g/r$ we get
\begin{equation}
t(r)=E\int \frac{(r+r_g/2)dr}{[(E^2-m^2)r^2+E^2r_gr+E^2(1+a)r_g^2-L^2]^{1/2}}
\label{t(r)B}
\end{equation}

Let $T$ be the period of rotations; for example it can defined as the time difference between two consecutive
perihelions. This quantity can be calculated by analogy with the above calculation of angular precession of
the perihelion and the result is
\begin{equation}
T=\frac{\pi Em^2r_g}{(m^2-E^2)^{3/2}}
\label{T}
\end{equation}
Suppose that particle 1 is nonrelativistic and define $E_{nr}=m-E$. Then 
\begin{equation}
T=T_{nr}(1-\frac{E_{nr}}{4m}),\quad T_{nr}=\frac{\pi m^3}{(2mE_{nr})^{3/2}}
\label{TB}
\end{equation} 
where $T_{nr}$ is the nonrelativistic expression for the period. It follows from this expression that the
relativistic correction to the period is $2.4\cdot 10^{-2}s$ for Mercury and $3.9\cdot 10^{-2}s$ for
Earth. In GR the period can be calculated by using the expression for $t(r)$ in this theory (see e.g. Ref. \cite{LLII}).
For Earth this gives an additional correction of $0.6s$. However, at present the comparison between theory and
experiment with such an accuracy seems to be impossible.

In standard nonrelativistic theory the acceleration $d^2{\bf r}/dt^2$ is directed toward the center and is proportional
to $1/r^2$. Let us check whether this property is satisfied in the above formalism. As follows from Eq. (\ref{pr})
\begin{eqnarray}
\frac{d^2{\bf r}}{dt^2}=[\frac{d^2r}{dt^2}-r(\frac{d\varphi }{dt})^2](cos\varphi ,sin\varphi )+
[2\frac{dr}{dt}\frac{d\varphi }{dt}+r\frac{d^2\varphi }{dt^2}](-sin\varphi ,cos\varphi )
\label{d2r}
\end{eqnarray}
Therefore $d^2{\bf r}/dt^2$ is directed toward ${\bf r}$ if the last term in the r.h.s. equals zero. A direct
calculation using Eqs. (\ref{motion}) and (\ref{q(r)}) gives
\begin{equation}
2\frac{dr}{dt}\frac{d\varphi }{dt}+r\frac{d^2\varphi }{dt^2}=-\frac{L}{2Er(1-\xi)^{1/2}}\frac{d\xi}{dt}
\label{perp}
\end{equation}
In the nonrelativistic approximation this quantity does equal zero but in the general case it does not.

An analogous calculation gives
\begin{equation}
\frac{d^2r}{dt^2}-r(\frac{d\varphi }{dt})^2=\frac{(m^2+L^2/r^2)E^2(1+2\xi)}{2(q^2-L^2/r^2)^{1/2}
\epsilon(q)^3}\frac{d\xi}{dt}
\label{parall}
\end{equation}
If $\xi=r_g/r$ then, as follows from Eq. (\ref{motion}), in the nonrelativistic approximation this quantity equals
$-Gm_2/r^2$. Hence in this approximation we indeed have the standard result $d^2{\bf r}/{dt^2}=-Gm_2{\bf r}/r^3$.

We now do not assume that $m_1\ll m_2$ but consider only nonrelativistic approximation. The relative angular
momentum ${\bf J}$ equals the total angular momentum in the c.m. frame. In this reference frame we have
${\bf r}_1\times {\bf p}_1+{\bf r}_2\times {\bf p}_2={\bf r}\times {\bf q}$ where ${\bf q}={\bf p}_1$ is the 
relative momentum and ${\bf r}={\bf r}_1-{\bf r}_2$ is the relative position. Therefore, by analogy with the derivation
of Eq. (\ref{pr}), one can derive the same relations where ${\bf q}$ is the relative momentum and ${\bf r}$ is the
relative position.

As follows from Eqs. (\ref{33}) and (\ref{Newton}), in the cases of dS antigravity and standard Newtonian gravity
the internal two-body nonrelativistic energy can be written as
\begin{equation}
E=\frac{{\bf q}^2}{2m_{12}}-\frac{1}{2}m_{12}\xi
\label{dsgrav}
\end{equation}
where $\xi=(r/R)^2$ for the dS antigravity and $\xi=r_g/r$ with $r_g=2G(m_1+m_2)$ for the Newtonian gravity.

By analogy with the above consideration, for deriving equations of notions one should define time by analogy
with Eq. (\ref{rq}). In our approach the effects of dS antigravity and standard Newtonian gravity are simply
kinematical manifestations of dS symmetry for systems of two free particles. The difference between those
cases is only that the quantity $\chi$ in the exponent $exp(-i\chi n)$ defining the behavior of the internal
two-body WF depends on $r$ differently in the cases when $r$ is of the order of cosmological distances
and much less than those distances. As follows from Eq. (\ref{rq}), in the nonrelativistic approximation
$$d{\bf r}=d{\bf r}_1-d{\bf r}_2=(\frac{{\bf p}_1}{m_1}-\frac{{\bf p}_2}{m_2})dt=\frac{{\bf q}}{m_{12}}dt$$
Then, by analogy with the derivation of Eq. (\ref{d2r}), we get
\begin{equation}
\frac{d^2{\bf r}}{dt^2}=\frac{{\bf r}}{R^2},\quad m_{12}\frac{d^2{\bf r}}{dt^2}=-\frac{Gm_1m_2}{r^3}{\bf r}
\label{dsgravaccel}
\end{equation}
for dS antigravity and Newtonian gravity, respectively.

As noted in Secs. \ref{antigravity} and \ref{LambdaDiscrete}, the first expression in Eq. (\ref{dsgravaccel}) is
a consequence of the Hamilton equations for the Hamiltonian (\ref{33}) and it is obvious that the second
expression is a consequence of the Hamilton equations for the Hamiltonian (\ref{Newton}). However, as shown above,
those expressions can be derived without involving the Hamilton equations but using only the relation (\ref{rq})
for each free particle.

\section{Discussion of classical effects of GR}
\label{nonNewton}

General Relativity is a pure classical theory and a common belief is that in the future quantum
theory of gravity the results of GR will be recovered in semiclassical approximation. Moreover, any quantum theory 
of gravity can be tested only on macroscopic level. Hence, the problem is not only to construct quantum theory
of gravity but also to understand a correct structure of the position operator on macroscopic level. However, in
the literature the latter problem is not discussed because it is tacitly assumed that the position operator on 
macroscopic level is the same as in standard quantum theory. This is a great extrapolation which should be substantiated.

As argued in Secs. \ref{validity} and \ref{macroscopicdist}, standard position operator is not semiclassical on
macroscopic level and therefore on this level it should be modified. In our approach gravity is simply a
manifestation of dS symmetry on quantum level for systems of free bodies. Then for calculating observable effects
one should know how the quantity $\chi$ in the exponent $exp(-i\chi n)$ for the internal two-body WF
depends on the distance between the bodies. As argued in Sec. \ref{macroscopicdist}, if $\xi=sin^2\chi$ then
the dependence $\xi=(r_g/r)+o(r_g/r)$ is reasonable and reproduces standard Newtonian gravity. In this
section we consider what our approach can say about the gravitational red shift of light, deflection of light 
by the Sun and precession of Mercury's perihelion which are treated as three classical tests of GR. 

As seen from Earth, the precession of Mercury's orbit is measured to be 5600" per century while the contribution of
GR is 43" per century. Hence the latter is less than 1\% of the total contribution. The main contribution to the total
precession arises as a consequence of the fact that Earth is not an inertial reference frame and when the precession
is recalculated with respect to the International Celestial Reference System the value of the precession becomes
($574.10\pm 0.65$)" per century. Celestial mechanics states that the gravitational tugs of the other planets contribute
($531.63\pm 0.69$)" while all other contributions are small. Hence there is a discrepancy of 43" per century and the
result of GR gives almost exactly the same value. Although there are different opinions on
whether, the contribution of GR fully explains the data or not, in the overwhelming majority of the literature
it is accepted that this is the case. However, a detailed analysis carried out in Ref. \cite{Vankov} shows that
both, experimental data and theoretical predictions of GR contain controversies and therefore the
problem of whether GR consistently describes the precession of Mercury perihelion remains open.

Our result (\ref{MercD}) is compatible with GR if $\xi=(r_g/r)+(r_g/r)^2/2 + o((r_g/r)^2)$. The result of GR is by
a factor of 3/2 greater than the results of several alternative theories of gravity which in our approach can
be reproduced if $\xi=(r_g/r)+o((r_g/r)^2)$. Hence the problem of the future quantum theory of gravity is to
understand the value of the quadratic correction to $\xi$.

The result for the gravitational red shift of
light given by Eq. (\ref{Ephoton}) is in agreement with GR and is treated such that it has been confirmed 
in the Pound-Rebka experiment. However,
the conventional interpretation of this effect has been criticized by L.B. Okun in Ref. \cite{okunphoton}.
In his opinion, "{\it a presumed analogy between a photon and a stone}" is wrong. The reason is that "{\it the energy of the
photon and hence its frequency $\omega=E/\hbar$ do not depend on the distance from the gravitational body, because
in the static case the gravitational potential does not depend on the time coordinate $t$. The reader who is not
satisfied with this argument may look at Maxwell's equations as given e.g. in section 5.2 of ref. \cite{weinbergcosm}.
These equations with time independent metric have solutions with frequencies equal to those of the emitter}".
In Ref. \cite{okunphoton} the result of the Pound-Rebka experiment is explained such that not the photon loses
its kinetic energy but the differences between the atom energy levels on the height $h$ are greater than on
the Earth surface and "{\it As a result of this increase the energy of a photon emitted in a transition of an atom 
downstairs is not enough to excite a reverse transition upstairs. For the observer upstairs this looks like
a redshift of the photon. Therefore for a competent observer the apparent redshift of the photon is a result of 
the blueshift of the clock.}".

As noted in Ref. \cite{okunphoton}, "{\it A naive (but obviously wrong!) way to derive the formula for the redshift
is to ascribe to the photon with energy $E$ a mass $m_{\gamma}= E/c^2$ and to apply to the photon a non-relativistic 
formula $\Delta E = -m_{\gamma}\Delta\phi$ treating it like a stone. Then the relative shift of photon energy is 
$\Delta E/E = -\Delta\phi/c^2$, which coincides with the correct result. But this coincidence cannot justify the 
absolutely thoughtless application of a nonrelativistic formula to an ultrarelativistic object.}" 

However, in our approach no nonrelativistic formulas for the photon have been used and the result 
$\Delta E_1/E_1=-gh/c^2$ has been obtained in a fully relativistic approach. As already noted, the only
problematic point in deriving this result is that the function $\xi (r)$ is not exactly known. 
In the framework of our approach a stone and a photon are simply particles with different
masses; that is why the stone is nonrelativistic and the photon is ultrarelativistic. Therefore there is no 
reason to think that in contrast to the stone, the photon will not lose its kinetic energy. At the same time,
we believe that Ref. \cite{okunphoton} gives strong arguments that energy levels on the Earth surface and
on the height $H$ are different.

We believe that the following point in the arguments of Ref. \cite{okunphoton} is not quite consistent. 
A stone, a photon and other particles can be characterized by their energies, momenta and 
other quantities for which there exist well defined operators. Those quantities might be measured in 
collisions of those particles with other particles. At the same time, as noted in Secs. \ref{ST} and
\ref{geom} the notions of "frequency
of a photon" or "frequency of a stone" have no physical meaning. If a particle WF 
(or, as noted in Sec. \ref{ST}, 
rather a state vector is a better name) contains $exp[i(px-Et)/\hbar]$ then by analogy with the theory of
classical waves one might say that the particle is a wave with the frequency $\omega=E/\hbar$ and the wave
length $\lambda=2\pi\hbar/p$. However, the fact that such defined quantities $\omega$ and $\lambda$ are 
the real frequencies and wave lengths measured e.g. in spectroscopic experiments needs to be substantiated.
Let $\omega$ and $\lambda$ be frequencies and wave lengths measured in experiments with classical waves.
Those quantities necessarily involve classical space and time.
Then the relation $E=\hbar\omega$ between the energies of particles in classical waves and frequencies of
those waves is only an assumption that those different quantities are related in such a way. This relation 
has been first proposed by Planck for the description of the blackbody radiation and
the experimental data indicate that it is valid with a high accuracy. As noted in Sec. \ref{geom}, this 
relation takes place in Poincare invariant electrodynamics. However, there is no guaranty that
this relation is always valid with the absolute accuracy, as the author of Ref. \cite{okunphoton} assumes.
In spectroscopic experiments not energies and momenta of emitted photons are measured but wave lengths 
of the radiation obtained as a result of transitions between different energy levels.
In particular, there is no experiment confirming that the relation $E=\hbar\omega$ is always exact, e.g.
on the Earth surface and on the height $h$. In summary, the Pound-Rebka experiment cannot be treated
as a model-independent confirmation of GR.

Consider now the deflection of light by the Sun. As shown in the preceding section, 
in the approximation $\xi=r_g/r$
the deflection is described by Eq. (\ref{deflB}). In the literature this result is usually represented such that if $\theta=\Delta\varphi -\pi$ is the deflection
angle then $\theta=(1+\gamma)r_g/\rho$ where $\gamma$ depends on the theory. Hence the result given by Eq. (\ref{deflB})
corresponds to $\gamma=0$. This result was first obtained by von Soldner in 1801 \cite{Soldner} and confirmed 
by Einstein in 1911. The known historical facts are that
in 1915 when Einstein created GR he obtained $\gamma=1$ and in 1919 this result was confirmed in observations of the
full Solar eclipse. Originally the accuracy of measurements was not high but now the quantity $\gamma$ is
measured with a high accuracy in experiments using the Very Long Base Interferometry (VLBI) technique and the
result $\gamma=1$ has been confirmed with the accuracy better than 1\%. The result $\gamma=1$ in GR is a consequence
of the fact that the post-Newtonian correction to the metric tensor in the vicinity of the Sun is not zero for both,
temporal and spatial components of this tensor. A question arises whether this result can be obtained in the framework
of a quantum approach. In the textbook \cite{Scadron}, 
the deflection is treated as a consequence of one-graviton exchange. The
author defines the vertices responsible for the interaction of a virtual graviton with a scalar nonrelativistic
particle and with a photon and in that case the cross-section of the process described by the one-graviton
exchange corresponds to the result with $\gamma=1$. The problem is that there is no other way of testing
the photon-graviton vertex and we believe that it is highly unrealistic that when the photon
travels in the $y$ direction from $-\infty$ to $+\infty$, it exchanges only by one virtual graviton with the Sun.
Therefore a problem of how to recover the result with $\gamma=1$ in quantum theory remains open. 

The confirmation of the result $\gamma=1$ in VLBI experiments is very difficult because corrections
to the simple geometric picture of deflection should be investigated. For example, the density of the
Solar atmosphere near the Solar surface is rather high and the assumption that the photon passes this
atmosphere practically without interaction with the particles of the atmosphere seems to be problematic.
In Ref. \cite{Lebach} the following corrections have been investigated at different radio-wave frequencies
$\omega$:
the brightness distribution of the observed source, the Solar plasma correction, the Earth's atmosphere, the
receiver instrumentation, and the difference in the atomic-clock readings at the two sites. All these corrections
are essentially model dependent. For example, the plasma delay $\tau_{plas}$ has been approximated by 
$I(t)/\omega^2$ where $I(t)$ depends on the electronic content
along the signal propagation paths to the two sites. 

The majority of authors investigating the deflection state that the experimental data confirm
the result $\gamma=1$ but several authors (see e.g. Ref. \cite{Marmet}) disagree with this conclusion.
It is also is not clear whether other effects might be important. For example, in Sec. \ref{experiment} we discussed
possible mechanisms which do not allow the photon WF to spread significantly. In particular, a
possible mechanism can be such that a photon is first absorbed by an atom and then is reemitted. 
Suppose that this mechanism plays an important role and photons encounter many atoms
on their way. In the period of time when the atom absorbs the photon but does not reemit it yet, the atom
acquires an additional acceleration as a result of its effective gravitational interaction with the Sun.
Then the absorbed and reemitted photons will have different accelerations and the reemitted photon is
expected to have a greater acceleration towards the Sun than the absorbed photon. This effect increases the
deflection angle and analogously other mechanisms of interaction of photons with the interstellar matter
are expected to increase the deflection angle since the matter moves with an acceleration towards the Sun.

\section{Discussion of the problem of gravitational radiation}

\label{Gravwaves}

Three classical effects of GR are treated as phenomena where the gravitational field is weak because
corrections to the Minkowskian metric are small. In recent
years considerable efforts have been made for investigating phenomena where the gravitational field is treated
as strong. 

One of the examples is the case of binary pulsars. 
In contrast to planets, conclusions about masses and radii of pulsars can be made only from models
describing their radiation. It is believed that typically pulsars are neutron stars with masses in the range
$(1.2-1.6)M_{\odot}$ and radii of the order of $10km$. In the case of binary pulsars, a typical situation
is that the second component of the binary system is not observable (at present the only known case where the
both components are pulsars is the binary pulsar J0737-3039).

The most famous case is the binary pulsar PSR B1913+16 discovered by Hulse and Taylor in 1974. A model 
with eighteen fitted parameters for this binary system has been described in Refs. \cite{Weisberg,Weisberg2} 
and references therein. In this model the masses of the pulsar and companion are approximately 
$1.4M_{\odot}$, the
period of rotation around the common center of mass is 7.75 hours, the values of periastron and apastron are 1.1 
and 4.8 $R_{\odot}$, respectively, and the orbital velocity of stars is 450 km/s and 110 km/s at periastron 
and apastron, respectively. Then relativistic effects are much stronger than in Solar System. For example,
the precession of periastron is 4.2 degrees per year. 

The most striking effect in the above model is that it predicts that the energy loss due to gravitational radiation
can be extracted from the data. As noted in Ref. \cite{Weisberg}, comparison of the measured and theoretical values 
requires a small correction for relative acceleration between the solar system and binary
pulsar system, projected onto the line of sight. The correction term depends on several rather poorly known 
quantities, including the distance and proper motion of the pulsar and the radius
of the Sun's galactic orbit. However, with the best currently available values the agreement between the data
and the Einstein quadrupole formula for the gravitational radiation is better than 1\%. The rate of decrease of orbital period is 76.5 microseconds per year (i.e. one second per 14000 years).

As noted by the authors of Ref. \cite{Weisberg}, "{\it Even with 30 years of observations, only a small portion of the North-South
extent of the emission beam has been observed. As a consequence, our model is
neither unique nor particularly robust. The North-South symmetry of the model
is assumed, not observed, since the line of sight has fallen on the same side of
the beam axis throughout these observations. Nevertheless, accumulating data
continue to support the principal features noted above.}"

The size of the invisible component is not known. The arguments that this component is a
compact object are as follows \cite{Will}: "{\it Because the orbit is so close ($ 1\, solar\, radius)$) and because there is no evidence of an eclipse of the pulsar signal or of mass transfer from the companion, it is generally agreed that 
the companion is compact. Evolutionary
arguments suggest that it is most likely a dead pulsar, while B1913+16 is a recycled pulsar.
Thus the orbital motion is very clean, free from tidal or other complicating effects. Furthermore,
the data acquisition is clean in the sense that by exploiting the intrinsic stability of the pulsar
clock combined with the ability to maintain and transfer atomic time accurately using GPS, the
observers can keep track of pulse time-of-arrival with an accuracy of $13\mu s$ , despite extended gaps
between observing sessions (including a several-year gap in the middle 1990s for an upgrade of the
Arecibo radio telescope). The pulsar has shown no evidence of glitches in its pulse period.}"
However, it is not clear whether or not there exist other reasons for substantial energy losses. 
For example, since the bodies have large velocities and are moving in
the interstellar medium, it is not clear whether their interaction with the medium can be neglected. 
In addition, a problem arises to what extent the effect of mass exchange in close binaries is important. 
The state-of-the-art review of the theory of close binaries can be found in Ref. \cite{Postnov} and references
therein. Nevertheless, the above results are usually treated as a strong indirect confirmation of the existence of gravitational waves (GWs). 

Those results have given a motivation for building powerful facilities aiming to 
detect GWs directly. After many years of observations 
no unambiguous detections of GWs have been reported \cite{Bizouard}. However,
recently the LIGO Collaboration has announced \cite{GW} the direct discovery of GWs and then
several similar phenomena have been found (see e.g. Ref. \cite{Abbott}).
On September 14, 2015 at 09:50:45 UTC the two LIGO detectors observed the event called GW150914 and treated 
as GWs for the following reasons.

The authors of Ref. \cite{GW} say that "{\it the most plausible explanation}" of the event is that the detected signals are
caused by {\it gravitational-wave emission in the coalescence of two black holes
--- i.e., their orbital inspiral and merger, and subsequent final black hole ringdown}. The motivation is that  
 the data are consistent with a system of parameters in numerical relativity models discussed in Ref. \cite{numericalA} 
and confirmed  to 99.9\% by an independent calculation based on Ref. \cite{numericalB}.
The data are consistent with the model where the initial black hole masses are 
$(36^{+5}_{-4})M_{\odot}$ and $(29\pm 4) M_{\odot}$ and the final black hole mass is $(62\pm 4) M_{\odot}$ 
with the energy $(3.0\pm 0.5) M_{\odot}c^2$ radiated in GWs during approximately $0.2s$. However, the
authors do not say explicitly how many initial parameters are needed in the model and do not display all the parameters.

The author of Ref. \cite{Cho} describes his interviews with well known gravitational scientists. In particular,
Professor Thorne, who is one of the founders of LIGO says: "{\it It is by far the most powerful explosion humans have ever detected 
except for the big bang}", and Professor Allen, who is the  director of the Max Planck Institute for Gravitational Physics and leader of 
the Einstein@Home project for the LIGO Scientific Collaboration says: "{\it For a tenth of a second the collision shines brighter 
than all of the stars in all the galaxies. But only in gravitational waves}".

From the particle physics point of view, the existence of neutron stars is not a problem because the process $p+e\to n+\nu$ is well
understood. As already noted, typical models say that the masses of neutron stars are in the range $(1.2-1.6)M_{\odot}$ and 
their radii are of the order of $10km$. It is believed that when the mass is greater then even such a dense neutron matter 
cannot prevent gravitational collapse. However, the existing particle theory does not know what happens to such
a matter under such extreme conditions. Therefore the theory does not know what type of matter black holes consist of. In the literature several models of black holes are discussed including those where
a black hole has a nonzero electric charge.

If gravity on quantum level is described in terms of gravitons then the following problem arises.
A black hole is a region of space that no real particles, including photons and gravitons can escape from inside it. However, at distances much greater than the size of the region the gravitational field 
of the black hole is the same as for the usual star with the same mass and spin. This implies  
that virtual gravitons can escape from the region without problems. The difference between real and virtual gravitons
is that the four-momenta squared of the latter do not equal $m_g^2$ where $m_g$ is the graviton mass. However, they can be very
close to $m_g^2$. Therefore it is not clear why the properties of real and virtual gravitons are so different. 

Another problem is whether or not it is natural that the only observed manifestation of the release of such a huge amount of energy 
during such a short period of time was that the 4km path of the laser beam in the LIGO interferometer was stretched by the 
value which is much less than the proton radius.
From the particle physics point of view, the merger of two black holes such that $3M_{\odot}c^2$ is released
in the form of GWs during $0.2s$ is a problem because even the type of matter black holes consist of is not
known. While from the point of view of GR gravitational waves are described as ripples in space-time, in particle theory 
any wave is treated as a collection of particles. In particular, GWs
are believed to consist of real (not virtual) gravitons. When two high energy particles smash in the accelerator, typically many
different particles are produced. By analogy, one might  think that in such a tremendous phenomenon, where the (unknown type of) matter
experiences extremely high accelerations, a considerable part of energy should be released not only in the form of
gravitons but also in the form of photons and other particles. For example, the electric charge of the neutron is zero but the neutron has a magnetic moment and consists of charged quarks. Therefore
the neutron moving with a large acceleration will emit photons. 
So the assumption that the energy is released only in
the form of GWs does not seem to be convincing and one might expect that the effect is accompanied by
extremely bright flashes in different parts of the electromagnetic spectrum. A problem of the electromagnetic energy
released in the black hole merger has been discussed by several authors (see e.g. Ref. \cite{BHEM} and references
therein) who state that the problem is extremely difficult and considerably model dependent. In addition, even if the energy
is released only in the form of GWs then a problem arises to what extent the orbits of Sun, Earth and
Moon will be affected by such strong GWs. This problem has not been discussed in the literature.

In experiments with two detectors the position of the source of GWs cannot be identified with a good
accuracy but in the experiment with three detectors \cite{Abbott} the area of the 
90\% credible region has been reduced from 1160 $deg^2$ to 60 $deg^2$. 
After the first LIGO  announcement the authors of Ref. \cite{Fermi} analyzed the data  of the Fermi Gamma-ray Burst Monitor 
obtained at the time of the LIGO event. The data reveal the presence of a weak
source above $50keV$, $0.4s$ after the LIGO event was detected. Its
localization is ill-constrained but consistent with the direction of the GW150914. However, in view of the weakness of the
signal it is highly questionable that it is related to the LIGO event. Since the energy released in the event is approximately known and the distance to the event also is
approximately known then it is easy to estimate that the energy received by Earth during 0.2$s$ is
by five orders of magnitude greater than the energy received from Sirius and by six orders of magnitude
less  than the energy received from Sun. 
In addition, it is usually assumed that photons
and gravitons are massless particles the speed of which can be only $c$. Therefore $0.4s$ corresponds to the 
distance $120000km$. This is incompatible with the fact that the monitor resides in 
a low-earth circular orbit at an altitude of $550km$.

We conclude that the problem of explaining strong gravitational 
effects is very complicated because GR, which is a pure classical
theory, should be reconciled with the present understanding of particle theory, and conclusions about GWs are based on models 
with many fitted parameters.
So the statements that those effects can be treated as strong confirmations of the existence of GWs are premature.  
In any case until the nature of gravity on classical and quantum level is well understood, different
approaches should be investigated.

\chapter{Why finite mathematics is the most general and FQT is more pertinent physical theory than standard one}
\label{Ch4}

As noted in Sec. \ref{crisis}, several strong arguments indicate that fundamental quantum theory should be 
based on finite mathematics. However, for the majority of physicists and mathematicians it is difficult
to accept those arguments because it is usually believed that classical mathematics (involving the notions
of infinitely small/large and continuity) is
fundamental while finite mathematics is something inferior which is used only in special applications.
In this section we argue that the situation is the opposite: classical mathematics is only a degenerate
case of finite one in the formal limit when the characteristic of the ring or field used in finite mathematics
goes to infinity. 

\section{Classical and finite mathematics in view of the philosophy of science}
\label{philosophy}

We first consider a problem whether classical mathematics can be substantiated as an abstract science. 
The investigation of this problem has a long history described in many textbooks and monographs
(see e.g. Ref. \cite{settheory}). Classical mathematics is based on the set theory proposed by Cantor.
One of the revolutionary features of this theory was that infinite sets are not only allowed but play a key role,
and there are different types of infinite sets. Before Cantor only potential infinity was acceptable,
i.e. infinity was treated only as a limit of something finite. However, in the Cantor set theory, actual infinity 
is extremely important. For example, here the set of all natural numbers {\it N} or the set of integers $Z$ are
not treated as limits of some finite sets. Even now many mathematicians advocate classical finitism, where infinity 
can be only potential, or strong finitism, where even potential infinity is not allowed. The examples are Peano arithmetic,
Robinson arithmetic and other theories. However, those theories are incomplete and all applications
to modern physics are based only on the Cantor set theory and its generalizations.

As shown by Russell and other mathematicians, the Cantor set theory contains several fundamental paradoxes.
To avoid them several axiomatic set theories have been proposed and the most known of them is the ZFC theory
developed by Zermelo and Fraenkel. The usual statements
in set theory are that the consistency of ZFC cannot be proven within ZFC itself
and the continuum hypothesis is independent of ZFC. Those statements have been
questioned by Woodin (see e.g. Ref. \cite{Woodin} and references therein) and the problem is
open. G\"{o}del's incompleteness theorems state that no system of axioms can ensure that all facts about natural 
numbers can be proven and the system of axioms in traditional approach to classical mathematics cannot demonstrate its own consistency.

In constructive approach to classical mathematics proposed by Brouwer there is no law of the excluded middle and it is required that any proof of existence be algorithmic. That is why constructive mathematics is treated such that, at least in principle, it 
can be implemented on a computer. Here "in principle" means that the number of steps might be not
finite. 

The majority of mathematicians prefer the traditional version. Physics is also based only on traditional mathematics. 
Hilbert was a strong opponent of constructive mathematics. He said: {\it "No one shall expel us from the paradise that Cantor has created for us"} and 
{\it "Taking the principle of excluded middle from the mathematician would be the same, say, as proscribing the telescope to the astronomer or to the boxer the use of his fists"}.

Some known results of classical mathematics are counterintuitive. For example, since the mapping $tgx$
from $(-\pi /2,\pi /2)$ to $(-\infty,\infty)$ is a bijection, those intervals have the same number of elements
although the former is a part of the latter. Another example is Hilbert's Grand Hotel paradox.
However, in classical mathematics those examples are not treated as contradictory.

Let us now consider classical mathematics from the point of view of philosophy of science.
In the 20s of the 20th century the Viennese circle of philosophers
under the leadership of Schlick developed an approach called logical positivism which contains
verification principle:  {\it A proposition is only cognitively meaningful if it can be definitively and 
conclusively 
determined to be either true or false} (see e.g. Refs. \cite{verificationism}). However, this principle does
not work in standard classical mathematics. For example, it cannot be determined whether the statement that 
$a+b=b+a$ for all natural numbers $a$ and $b$ is true or false. 

As noted by Grayling
\cite{Grayling}, {\it "The general laws of science are not, even in principle, verifiable, if verifying means 
furnishing conclusive proof of their truth. They can be strongly supported by repeated experiments and 
accumulated evidence but they cannot be verified completely"}. So, from the point of view of
classical mathematics and classical physics, verification principle is  
too strong. 

Popper proposed the 
concept of falsificationism \cite{Popper}: {\it If no cases where a claim is false can be found, then 
the hypothesis 
is accepted as provisionally true}. In particular, the statement that 
$a+b=b+a$ for all natural numbers $a$ and $b$ can be treated as provisionally true until one has found
some numbers $a$ and $b$ for which $a+b\neq b+a$.

According to the philosophy of quantum theory, there should be no statements
accepted without proof and based on belief in their correctness (i.e. axioms).
The theory should contain only those statements that can be verified, where by "verified" physicists mean 
an experiment involving only a finite number of steps. So the philosophy of quantum theory is similar to 
verificationism, not falsificationism. Note that Popper was a strong opponent of the philosophy of
quantum theory and supported Einstein in his dispute with Bohr.

From the point of view of verificationism and the philosophy of quantum theory, classical mathematics is not well 
defined
not only because it contains an infinite number of numbers. For example, let us pose a problem whether 
10+20 equals 30. Then we should describe an experiment
which should solve this problem. Any computing device can operate only with a finite 
amount of resources and can perform
calculations only modulo some number $p$. Say $p=40$, then the experiment will confirm that
10+20=30 while if $p=25$ then we will get that 10+20=5.

{\it So the statements that 10+20=30 and even that $2\cdot 2=4$
are ambiguous because they do not contain information on how they should be verified.} 
On the other hand, the statements
$$10+20=30\,(mod\, 40),\,\, 10+20=5\,(mod\, 25),$$
$$2\cdot 2=4\,(mod\, 5),\,\, 2\cdot 2=2\,(mod\, 2)$$
are well defined because they do contain such an information.
So only operations modulo some number are well defined.

We believe the following 
observation is very important: although classical 
mathematics (including its constructive version) is a part of our everyday life, people typically do not realize that {\it classical 
mathematics is implicitly 
based on the assumption that one can have any desired amount of resources}. 
So classical mathematics is based on the implicit assumption that we can consider an idealized case
when a computing device can operate with an infinite amount of resources. In other words, standard
operations with natural numbers are implicitly treated as limits of operations modulo $p$ 
when $p\to\infty$. As a rule, every limit in mathematics is thoroughly investigated but in the
case of standard operations with natural numbers it is not even mentioned that those
operations are limits of operations modulo $p$. In real life such limits even 
 might not exist if, for example, the Universe contains a finite number of elementary particles.

So classical mathematics has foundational problems which so far have not been solved in spite
of efforts of such great mathematicians as Cantor, Fraenkel, G\"{o}del, Hilbert, Kronecker, Russell, Zermelo and others, and, as noted above, classical mathematics is problematic from the point of view of verificationism and the philosophy of quantum theory. 
The philosophy of those great mathematicians was based on macroscopic experience in which the notions of infinitely small/large, continuity and standard division are natural. 
However, as noted above, those notions contradict the existence of elementary particles and are not natural
in quantum theory. The illusion of continuity arises when one neglects the discrete structure of matter.

G\"{o}del's works on the incompleteness theorems, saying that any mathematics
involving the set of all natural numbers has foundational problems, are written in highly technical terms
of mathematical logics. However, this fact is obvious from the philosophy of verificationism and
philosophy of quantum theory. On the other hand, since 
finite mathematics works only with a finite number of elements and all operations here are performed modulo some number, this mathematics satisfies
the principle of verifiability. As proved in Sec. \ref{finitemath},  as far as applications are concerned, classical mathematics is a special degenerate case of finite one. Hence foundational problems in classical mathematics are important only when it is treated as an abstract science. The technique of classical mathematics is very powerful and in many cases (but not all of them) describes reality with a high accuracy.

At the same time, finite mathematics does not contain infinities in principle (because, as described in the next
section, it starts
not from natural numbers but from a finite set) and therefore here G\"{o}del's incompleteness theorems do not apply.
Therefore finite mathematics does not have foundational problems
and here the validity of any statement can be directly verified (at least in principle).  

\section{Basic facts about finite mathematics}
\label{finmath}

Classical mathematics starts from the set of natural numbers, and the famous Kronecker's expression is: 
{\it "God made the natural numbers, all else is the work of man"}. By definition, finite mathematics can involve only a
finite number of elements. Therefore finite mathematics cannot start from the set of all natural numbers because it is infinite. 
It starts from a set $R_p$ of 
$p$ numbers 0, 1, 2, ... $p-1$ where addition and multiplication are defined
as usual but modulo $p$. In our opinion the notation $Z/p$ for $R_p$ is not quite adequate because it may
give a wrong impression that finite mathematics starts from the infinite set $Z$
and that $Z$ is more general than $R_p$. However, although $Z$ has more elements than $R_p$, $Z$ cannot 
be more general than $R_p$ because $Z$ does not contain operations modulo a number.

In the set $N$ only addition and multiplication are always possible. In order to make addition invertible we introduce negative integers. They do not have a direct physical meaning (e.g. the phrases "I have -2 apples" or "this computer
has -100 bits of memory" are meaningless) and their only goal is to get the ring of integers $Z$. In contrast to this
situation, the set $R_p$ is the ring without adding new elements and the number $p$ is called the characteristic
of this ring. For example, if $p=5$ then 3+1=4 as usual but 3+3=1, 
3$\cdot$2=1 and 3+2=0. Therefore -2=3 and -3=2. Moreover, if $p$ is prime then $R_p$ becomes the Galois field $F_p$
where all the four operations are possible. For example, 1/2=3, 1/4=4 etc.

One might say that those examples have nothing to do with physics and reality since 3+2 always equals 5 and not zero.
However, since operations in $R_p$ are modulo $p$, one can represent 
$R_p$ as a set $\{0,\pm 1,\pm 2,...,\pm(p-1)/2)\}$ if $p$ is odd and as a set
$\{0,\pm 1,\pm 2,...,\pm (p/2-1),p/2\}$ if $p$ is even. Let $f$ be a function from $R_p$ to $Z$ such that
$f(a)$ has the same notation in $Z$ as $a$ in $R_p$. Then for elements $a\in R_p$ such that $|f(a)|\ll p$, 
addition, subtraction and multiplication are the same as in $Z$. In other words, for
such elements we do not notice the existence of $p$. 

One might say that nevertheless the set $F_p$ cannot be used in physics since here $1/2=(p+1)/2$, i.e. a very large
number when $p$ is large. However, as explained in Sec. \ref{finitemath}, since quantum states are projective then, even in standard  
quantum theory, quantum states can be described with any desired accuracy by using only integers. 
Therefore even in standard quantum theory (to say nothing about quantum theory based on finite mathematics)
the notions of rational and real numbers play only an auxiliary role.

\begin{figure}[!ht]
\centerline{\scalebox{0.3}{\includegraphics{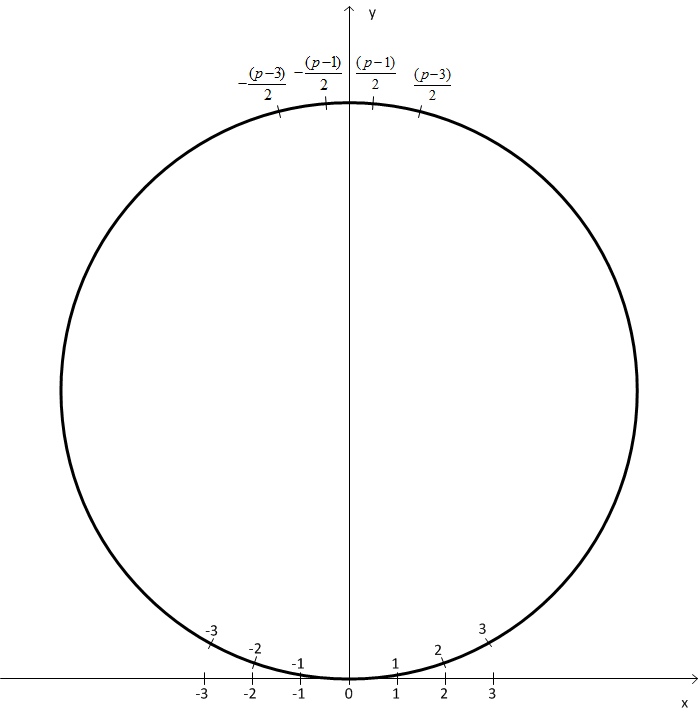}}}
\caption{
  Relation between $R_p$ and $Z$
}
\label{Fig.2}
\end{figure}

If elements of $Z$ are depicted as integer points on the $x$ axis of the $xy$ plane then, if $p$ is odd, the elements of $R_p$
can be depicted  as points of the circumference in Figure \ref{Fig.2}
and analogously if $p$ is even.
This picture is natural from the following considerations. As explained in textbooks, both $R_p$ and $Z$ 
are cyclic groups with respect to addition. However, 
$R_p$ has a higher symmetry because, in contrast to $Z$, $R_p$ has a property which we call {\it strong cyclicity}:
if we take any element $a\in R_p$ and sequentially add 1 then after $p$ steps we will exhaust
the whole set $R_p$ by analogy with the property that if we move along a circumference in the same direction
then sooner or later we will arrive at the initial point. At the same time, if we take an element $a\in Z$ then
the set $Z$ can be exhausted only if we first successively add +1 to $a$ and then -1 or {\it vice versa} and those
operations should be performed an infinite number of times.
As noted in Chap. \ref{AdS}, in quantum theory based on finite mathematics
strong cyclicity plays an important role. In particular, it explains why one IR of the symmetry algebra
describes a particle and its antiparticle simultaneously.

\section{Proof that the ring $Z$ is the limit of the ring $R_p$ when $p\to\infty$}
\label{Rp2Z}

In Sec. \ref{symmetry} we have formulated {\bf Definition 1.3} when theory A is more general than theory B and theory B is a special degenerate case of theory A. It has been proved that classical theory is
a special degenerate case of quantum one in the formal limit $\hbar\to 0$, NT is a special 
degenerate case of RT in the formal limit $c\to\infty$ and
Poincare invariant theory is a special degenerate case of dS/AdS theories in the
formal limit $R\to\infty$. In all those cases theory A contains a finite parameter, theory B is obtained from 
theory A in the formal limit when the parameter goes to zero or infinity and then theory  
B is a special degenerate case of theory A. The goal of the present section is to prove that, by analogy with the above examples, 

{\it Statement 1: The ring $R_p$ is more general than the ring $Z$
and the latter is a special degenerate case of the former in the formal limit $p\to\infty$}.

The proof of this statement is given following Ref. \cite{symm}. 

As noted in Sec. \ref{philosophy}, in the {\it technique} of classical mathematics infinity is understood only as a limit
but {\it the basis} of classical mathematics does involve actual infinity: classical mathematics starts from the infinite ring of
integers $Z$ and, even in standard textbooks on mathematics, it is not even posed a problem whether $Z$ can be treated as a limit of finite sets.
As noted in Sec. \ref{philosophy}, the problem of actual infinity is discussed in a vast literature, 
and in classical mathematics $Z$ is treated as actual and not potential infinity, i.e. there is 
no rigorous definition of $Z$ as a limit of finite sets. Moreover, classical set theory considers
infinite sets with different cardinalities.

As shown in the next section, {\it Statement 1} is the first stage in proving that finite mathematics
is more general than classical one. Therefore this statement should not be based on
results of classical mathematics. In particular, it should not be based on properties
of the ring $Z$ derived in classical mathematics. The statement should be proved
by analogy with standard proof that a sequence of natural numbers
$(a_n)$ goes to infinity if $\forall M>0$ $\exists n_0$ such that $a_n\geq M\,\, \forall n\geq n_0$. In particular, the proof should involve only potential infinity but
not actual one.

The meaning of {\it Statement 1} is
that for any $p_0>0$ there exists a set $S$ and a natural number $n$ such that for any 
$m\leq n$ the result of any $m$ operations of multiplication, summation or subtraction of elements from $S$ is the same for any $p\geq p_0$ and that cardinality
of $S$ and the number $n$ formally go to infinity when $p_0\to\infty$. This means 
that for the set $S$ and number $n$ there is no
manifestation of operations modulo $p$, i.e. the results of any $m\leq n$ operations
of elements from $S$ are formally the same in $R_p$ and $Z$. 

In practice this means that if experiments involve only such sets $S$ and numbers $n$
then it is not possible to conclude whether the experiments are described by a
theory involving $R_p$ with a large $p$ or by a theory involving $Z$.

As noted above, classical mathematics starts from the ring $Z$, and, even in standard textbooks on classical mathematics, it is not even
posed a problem whether $Z$ can be treated as a limit of finite sets. We did not
succeed in finding {\it a direct proof} of {\it Statement 1} in the literature.
However, the fact that $Z$ can be treated as a limit of $R_p$ when $p\to\infty$ follows from a sophisticated construction called ultraproducts. As shown e.g. in Refs. \cite{ultraproducts,ultraproducts1},
infinite fields of zero characteristic (and Z) can be embedded in ultraproducts of finite fields. This fact can also be proved by using only rings (see e.g. Theorem 3.1 in
Ref. \cite{Turner}). 
This is in the spirit of mentality of majority of mathematicians that sets with characteristic 0 are general, and for investigating those sets it is convenient to use properties of simpler sets of positive characteristics. 

The theory of ultraproducts (described in a wide literature --- see e.g. monographs
\cite{ultra,ultra1} and references therein) is essentially based on classical results on
infinite sets involving actual infinity. In particular, the theory is based on \L{}o\^s' theorem involving the
axiom of choice. Therefore theory of ultraproducts cannot be used in proving
that finite mathematics is more general than classical one.

Probably the fact that $Z$ can be treated as a limit of $R_p$ when $p\to\infty$, can also be proved in approaches not involving ultraproducts. For example, Theorem 1.1 in
Ref. \cite{WooWoo} states:

{\it Let $S$ be a finite subset of a characteristic zero integral domain $D$, and let 
$L$ be a finite set of non-zero elements in the subring $Z[S]$ of $D$. There exists an infinite sequence of primes with positive relative density such that for each prime $p$ in the sequence, there is a ring
homomorphism $\varphi_p$ : $Z[S] \to Z/pZ$ such that 0 is not in $\varphi_p(L)$.}

The theorem involves only primes, and the existence of homomorphism does not
guarantee that operations modulo $p$ are not manifested for a sufficient number
of operations. However, even if those problems can be resolved, the proof of
the theorem is based on the results of classical mathematics for characteristic zero
integral domains, and the proof involves real and complex numbers, i.e. the results involve
actual infinity. 

We conclude that the existing proofs that $Z$ can be treated as a limit of $R_p$ when 
$p\to\infty$ cannot be used in the proof that finite mathematics is more general than classical one. 

We now describe our proof of {\it Statement 1}. We define the function $h(p)$ such that $h(p)=(p-1)/2$ if $p$ is odd and $h(p)=p/2-1$ if $p$ is even.
Let $n$ be a natural number and $U(n)$ be a set of elements $a\in R_p$ such that 
$|f(a)|^n \leq h(p)$. Then $\forall m \leq n$ the result of any $m$ operations of addition,
subtraction or multiplication of elements $a\in U(n)$ is the same as for the corresponding elements $f(a)$ in $Z$, i.e. in this
case operations modulo $p$ are not explicitly manifested.

Let $n=g(p)$ be a function of $p$ and $G(p)$ be a function such that the set $U(g(p))$ contains
at least the elements $\{0,\pm 1,\pm 2,..., \pm G(p)\}$. In what follows $M>0$ is a natural number. 
If there is a sequence of natural numbers
$(a_n)$ then standard definition that $(a_n)\to \infty$ is that $\forall M\,\, \exists N$ 
such that $a_n\geq M\,\, \forall n\geq N$. By analogy with this
definition we will now prove 

{\it Proposition: There exist functions $g(p)$ and $G(p)$ such that $\forall M\,\, \exists p_0$ 
such that $g(p)\geq M$ and $G(p)\geq 2^M\,\, \forall p\geq p_0$.}
 
\begin{proof}
$\forall p>0$ there exists a unique natural $n$ such that $2^{n^2}\leq h(p)<2^{(n+1)^2}$. Define
$g(p)=n$ and $G(p)=2^n$. Then $\forall M \,\,\exists p_0$ such that $h(p_0)\geq 2^{M^2}$. Then
$\forall p\geq p_0$ the conditions of {\it Proposition} are satisfied.
\end{proof}

{\it Proposition} implies that  the ring $Z$ is the limit of the ring $R_p$ when $p\to\infty$, and the result of any finite combination of additions, subtractions and multiplications in
$Z$ can be reproduced in $R_p$ if $p$ is chosen to be sufficiently large. On the contrary, when the limit $p\to\infty$ is already taken then one cannot return back from $Z$ to
$R_p$, and in $Z$ it is not possible to 
reproduce all results in $R_p$ because in $Z$ there are no operations modulo a number. 
According to {\bf Definition 1.3} this means that {\it Statement 1} is valid, i.e. that {\it the ring $R_p$ is more general than $Z$, and $Z$ is the special degenerate case of $R_p$}.

When $p$ is very large then $U(g(p))$ is a relatively small part of $R_p$, and in general the results
in $Z$ and $R_p$ are the same only in $U(g(p))$. This is analogous to the fact mentioned in
Sec. \ref{symmetry} that the results of NT and RT are the same only in relatively small cases
when velocities are much less than $c$. However, 
 when the radius of the circumference in Figure \ref{Fig.2} becomes infinitely large then
a relatively small vicinity of zero in $R_p$ becomes the infinite set $Z$ when $p\to\infty$. 
{\it This example demonstrates that once we involve infinity and replace $R_p$ by $Z$ then we automatically 
obtain a degenerate theory because in $Z$ there are no operations modulo a number}.

\section{Proof that finite mathematics is more general than classical one and
quantum theory based on finite mathematics is more general than standard quantum theory}
\label{finitemath}

The goal of the present section is to prove

{\bf Main Statement: Even classical mathematics
itself is a special degenerate case of finite mathematics  in the formal limit when the characteristic of the 
field or ring in the latter goes to infinity}. 

Note that this statement is meaningful only if finite mathematics is more pertinent in
applications than classical one. Indeed if those theories are treated only as abstract ones
than the statement that one theory is more general than the other is meaningless.

In classical mathematics the ring $Z$ is the starting point for introducing the notions of rational and real numbers. Therefore those notions arise from a degenerate set. Then a question arises whether the fact that $R_p$ is more general than Z (proved in the preceding section) implies
that finite mathematics is more general than classical one. In particular, a problem arises whether finite 
mathematics can reproduce all results obtained by applications of classical mathematics.
For example, if $p$ is prime then $R_p$ becomes the Galois field $F_p$, and the results
in $F_p$ considerably differ from those in the set $Q$ of rational numbers even when $p$ is very large. In particular, 1/2 in 
$F_p$ is a very large number $(p + 1)/2$ if $p$ is very large.  Since quantum theory is the most
general physical theory, the answer to this question depends on whether standard quantum theory
based on classical mathematics is most general or is a special degenerate case of a more general 
quantum theory.

As shown in Secs. \ref{S6} and \ref{VS3},

{\it Statement 2: In standard dS and AdS quantum theories it is always possible to find a basis where the spectrum of all operators is purely discrete and the eigenvalues of those operators
are elements of $Z$}. Therefore, as follows from {\it Statement 1}, the remaining problem is whether or not quantum theory based
on finite mathematics can be a generalization of standard quantum theory where
states are described by elements of a separable complex Hilbert spaces $H$. 

Let $x$ be an element of $H$ and $(e_1,e_2,...)$ be
a basis of $H$ normalized such that the norm of each $e_j$ is an integer. Then {\it with any desired 
accuracy each element of $H$ can be approximated by a finite linear combination
\begin{equation}
x=\sum_{j=1}^n c_je_j 
\label{lincomb}
\end{equation}
where $c_j=a_j+ib_j$ and all the numbers $(a_j,b_j) \,\,(j=1,2,....n)$ are
rational}. 
This follows from the known fact that the set of such sums is dense in $H$.

The next observation is that spaces in quantum theory are projective, i.e. for any complex
number $c\neq 0$ the elements $x$ and $cx$ describe the same state. The meaning of this fact
is explained in Sec. \ref{classical}. In view of this property, both parts of Eq. (\ref{lincomb}) can be multiplied by a 
common denominator of all the numbers $a_j$ and $b_j$. As a result, we have

{\it Statement 3: Each element of $H$ can be approximated by a finite linear combination (\ref{lincomb})
where now all the numbers $a_j$ and $b_j$ are integers, i.e. belong to $Z$}.

We conclude that Hilbert spaces in standard quantum theory contain a big redundancy of
elements. Indeed, although formally the description of states in standard quantum theory
involves rational and real numbers, such numbers play only an auxiliary role because 
with any desirable accuracy each state can be described by using only integers. 
Therefore, as follows from {\bf Definition 1.3} and {\it Statements 1-3},
\begin{itemize}
\item Standard quantum theory based on classical mathematics is a special degenerate case
of quantum theory based on finite mathematics.
\item {\bf Main Statement} is valid.
\end{itemize}

{\bf Those results imply that mathematics describing nature at the most fundamental level involves only a finite number of numbers while the notions of limit and 
infinitely small/large and the notions constructed from them (e.g. continuity, derivative and integral) 
are needed only in calculations describing nature approximately}.

\section{Discussion}
\label{consequences}

The above construction has a well-known historical analogy. For many years people believed
that the Earth was flat and infinite, and only after a long period of time they realized that
it was finite and curved. It is difficult to notice the curvature when we deal only with
distances much less than the radius of the curvature. Analogously one might think that
the set of numbers describing physics in our Universe has a "curvature" defined by a very
 large number $p$ but we do not notice it when we deal only with numbers much less than $p$. 
As noted in Sec. \ref{philosophy}, finite mathematics is more natural than classical one 
from the point of view of the philosophy of quantum theory and the Viennese school of  logical positivism,
and, as proved in the preceding section,  finite mathematics is more general than classical one.

One might argue that introducing a new fundamental constant $p$ is not justified
and it is not clear why the choice of some $p$ is better than the choice of another $p$.
Let us note first that history of physics tells us that more general theories arise when a parameter,
which in the old theory was treated as infinitely small or infinitely large, becomes finite. 
For example, as noted in Sec. \ref{symmetry}, nonrelativistic physics is the degenerate case of 
relativistic one in the formal limit $c\to\infty$, 
classical physics is the degenerate case of quantum one
in the formal limit $\hbar\to 0$ and Poincare invariant theory is the degenerate case of dS or AdS invariant ones
in the formal limit $R\to\infty$. Therefore, it is natural to think that
in quantum physics the quantity $p$ should be not infinitely large but finite.

As far as the choice of a particular value of $p$ is concerned, let us note that 
any computer can be characterized by the number $p$ which shows that the computer can
perform operations only modulo this number. Analogously, our Universe can be treated as a computer
and the meaning of $p$ is that at present the state of the Universe is such that nature is described
by finite mathematics with this $p$. So $p$ is fundamental in the sense that it is defined by the present
state of the Universe. A problem arises whether $p$ is a constant or is different in various periods of time. 
In view of the problem of time in quantum theory, an extremely
interesting scenario is that the world time is defined by $p$ and this possibility is discussed in Chap. \ref{time}. 

The notions of infinitely small/large, continuity etc. were first proposed by Newton and Leibniz more than 300 years ago.
At that time people did not know about elementary particles, and those notions were in agreement with a 
belief based on everyday experience that any macroscopic object can be divided
into arbitrarily large number of arbitrarily small parts. However, as discussed in Sec. \ref{crisis}, the very 
existence of elementary particles indicates that those notions have only a limited meaning. 

I asked mathematicians whether in their opinion the fact that we cannot divide the electron by two or three means that
standard division is not universal. Some of them say that sooner or later the electron will be
divided. On the other hand, physicists typically are sure that the electron is indivisible. However, their mentality is
that since classical mathematics in many cases describes experiment then there is no need to philosophize
and to use mathematics which is not familiar to them. My observation is that
typically physicists not only have no idea about basics of finite mathematics but also believe that this
is exotics or pathology which has nothing to do with physics.

Note that in any computer the number of bits can be only a positive integer and such notions 
as 1/2 bit, 1/3 bit etc. are meaningless. So a bit is an analog of elementary particle.

Therefore mathematics involving the set of all rational numbers has only a limited 
applicability
and using classical mathematics in quantum physics and computer science is at least unnatural.
Rational numbers have arisen from macroscopic experience and it seems extremely unnatural that theories 
of quantum computing are based on complex numbers. 

The theory of quantum computing
is based on the notion of qubit which is a linear combination $c_0|0>+c_1|1>$  of states where a bit has the
value zero or one and $c_0$ and $c_1$ are complex numbers. From the formal point of view, if $c_0$ and $c_1$ are complex
numbers then the computer should have an infinite amount of resources what is unrealistic.

In finite mathematics the ring $R_p$ becomes the Galois field $F_p$ if $p$ is prime. 
In this case division is defined such that, by definition, $b=1/a$ if $ab=1\,\, (mod\,\, p)$.
For example, if $p=5$ then $1/2=3$, $1/4=4$ etc. For the transition from $R_p$ to $F_p$ no
new elements are needed. 

While at the level of rings classical and finite mathematics can be treated
as close to each other when $p$ is large, as explained above, at the level of fields they considerably differ
each other.  However, this does not mean that mathematics modulo $p$
cannot describe nature because, as explained in the preceding section and will be explained in more details
in the next sections, in view of the fact that spaces in quantum theory are projective, 
standard quantum theory based on classical mathematics is a special degenerate case
of quantum theory based on finite mathematics. 
Then classical mathematics describes many experiments
with a high accuracy as a consequence of the fact that in real life the number $p$ is very large.

As noted in Sec. \ref{crisis} , standard division has a limited applicability and has a physical meaning only in classical theory.
Therefore a problem arises whether for constructing FQT one should involve fields at all.
 In the spirit of Ref. \cite{Planat} (as Metod Saniga pointed out to me) one might think that the 
ultimate quantum theory will be based even on finite rings and not fields. This problem has several aspects which
are discussed below. 

\section{Quantum theory based on finite mathematics}
\label{FQT}

The official birth-year of quantum theory is 1925. The meaning of "quantum" is discrete and the presence of this word in
the name of the theory reflects the fact that some quantities have a discrete spectrum. 
The founders of the theory were highly educated physicists
but they used only classical mathematics and, as noted above, even now mathematical education at physics departments
does not involve discrete and finite mathematics. From the formal point of view, the existence of discrete
spectrum in classical mathematics is not a contradiction. On the other hand, discrete spectrum can be treated
as more general than continuous one: the latter can be treated as a formal degenerate special case of the former in a special case when distances between the levels of the discrete spectrum become (infinitely) small. 

In physics there are known examples in favor of this point of view. For example, the angular momentum operator has a pure discrete spectrum which becomes the continuous one in the formal limit $\hbar\to 0$. Another example is the following. As shown in Sec. \ref{symmetry}, Poincare symmetry is a special degenerate case of dS/AdS symmetries. The procedure when the latter becomes the former is performed as follows. Instead of angular momenta $M^{4\mu}$ we introduce standard Poincare 
four-momentum $P$ such that $P^{\mu}= M^{4\mu}/R$ where R is a parametr of contraction from dS/AdS algebras to Poincare algebra, and in general $R$ has nothing to do with the radius of the world. The spectrum of the operators $M^{4\mu}$ is discrete, the distances between the spectrum eigenvalues are of the order of $\hbar$ and therefore at this stage the Poincare four-momentum $P$ has the discrete spectrum such that the distances between the spectrum eigenvalues are of the order of $\hbar/R$. In the formal limit $R\to\infty$ the commutation relations for the dS/AdS algebras become the commutation relations for the Poincare algebra and instead of the discrete spectrum for the operators $M^{4\mu}$ we get the continuous spectrum for the operators $P$. 

In this work we consider an approach when quantum theory is based on a finite ring or field. 
One of the reasons why in many cases it is convenient to work with a Galois field rather than with a finite ring is mainly technical. 
The matter is that quantum theory is based on linear spaces, and such important notions as basis and invariance of dimension 
are well defined only if the linear space is over a field or body. 
In addition, the existence of division often simplifies calculations. However, history of physics tells us that for constructing a new theory
only those notions should be involved which are absolutely necessary. Below we discuss whether FQT can be constructed
without division.

Since the majority of physicists are not familiar with finite rings and fields, one of our goals is
to convince the reader that those notions are not only very simple and elegant, but
also are a natural basis for quantum physics. In Sec. \ref{finmath} we described basic facts about finite
mathematics. If a reader wishes to learn finite rings and fields on a more fundamental level,
he or she might start with standard textbooks (see e.g. Ref. \cite{VDW}).

In view of the present situation in modern quantum physics, a natural question arises
why, in spite of great efforts of thousands of highly qualified physicists for many years,
the problem of quantum gravity has not been solved yet. We believe that a possible answer
is that they did not use the most pertinent mathematics.

For example, the problem of infinities remains probably the most challenging 
one in standard formulation of
quantum theory. As noted by Weinberg \cite{Weinberg}, "{\it Disappointingly this problem 
appeared with even greater severity in the early days of quantum theory,
and although greatly ameliorated by subsequent improvements in the theory, it remains
with us to the present day}". The title of Weinberg's paper \cite{Wein3} is
"Living with infinities". A desire to have a theory without divergences is probably the
main motivation for developing modern theories extending QFT, e.g. loop quantum gravity, 
noncommutative quantum theory, string theory etc. On the other hand, in FQT infinities cannot exist in principle by construction.

As noted above, even for elements from $U(g(p))$ the result of division in $F_p$ differs generally speaking, from the corresponding result in the field of rational number $Q$. 
It is also clear that in general the meaning of square root in $R_p$ is not the same as in $Q$.
For example, even if $\sqrt{2}$ in $R_p$ exists, it is a very large number of the order of at least $p^{1/2}$. Another
obvious fact is that FQT cannot involve exponents and trigonometric functions since they are
represented by infinite sums. As follows from the fact that states in standard theory are projective, 
a direct correspondence between WF in FQT and standard theory can exist only for rational
functions.

Since standard quantum theory is based on complex numbers one might think that FQT should be based on
the ring $R_{p^2}$ or the field $F_{p^2}$ which contain $p^2$ elements such that any element can be represented as 
$a+bi$ where $a,b\in R_p$ or $a,b\in F_p$, respectively and $i$ is a formal element such
that $i^2=-1$. Then the definition of addition, subtraction and multiplication in $R_{p^2}$ and $F_{p^2}$ is obvious
and $R_{p^2}$ is obviously a ring regardless whether $p$ is prime or not. 
However, $F_{p^2}$ can be a field only if $p$ is prime but this condition is not sufficient.
By analogy with the field of complex
numbers, one could define division as $(a+bi)^{-1}=(a-ib)/(a^2+b^2)$.
This definition can be meaningful only if $a^2+b^2\neq 0$ in $F_p$
for any $a,b\in F_p$ i.e. $a^2+b^2$ is not divisible by $p$.
Therefore the definition is meaningful only if $p$ {\it cannot}
be represented as a sum of two squares and is meaningless otherwise. A simple example is that if $p=5$
then the field $F_{p^2}$ cannot be implemented in such a way because $2^2+1^2=5$.

We will not consider the case $p=2$ and therefore $p$ is necessarily odd.
Then we have two possibilities: the value of $p\,(mod \,4)$ 
is either 1 or 3. The known result of number theory
\cite{VDW} is that a prime number $p$ can be
represented as a sum of two squares only in the former case
and cannot in the latter one. Therefore the above
construction of the field $F_{p^2}$ is correct only if
$p\,(mod \,4)\,=\,3$. 

In Sec. \ref{Rp2Z} we discussed the correspondence  between $R_p$ and $Z$. It has been noted that if $n\leq g(p)$ then 
 for any $n$ operations of addition, subtraction and multiplication in the set $U(g(p))$ the results are the same as in $Z$.
By analogy, one can
define a set $U$ in $R_{p^2}$ such that $a+bi\in U$ if $a\in U(g(p))$ and $b\in U(g(p))$ and show that if 
$n\leq (g(p)-1)$ then 
 for any $n$ operations of addition, subtraction and multiplication in the set $U$ the results are the same as in $Z+iZ$.
Therefore if $f(a+bi)=f(a)+f(b)i$ then
$f$ is a local homomorphism between $R_{p^2}$ and $Z+Zi$.

In general, it is possible to consider linear spaces over any ring or field. Therefore a question arises what 
ring or field should be used in FQT. For example, it is known (see e.g. Ref. \cite{VDW}) that any Galois field 
can contain only $p^n$ elements where $p$
is prime and $n$ is natural. Moreover, the numbers $p$ and $n$ define the Galois field
up to isomorphism. It is natural to require that there should exist a correspondence
between any new theory and the old one, i.e. at some conditions the both theories should
give close predictions. In particular, there should exist a large number of quantum states
for which the probabilistic interpretation is valid. 

In view of the above discussion, the number $p$ should necessarily be very large and the problem
is to understand whether there exist deep reasons for choosing a particular
value of $p$, whether this is simply an accident that our Universe has been created with some value of $p$,
whether the number $p$ is dynamical, i.e. depends on the current state of the Universe etc.
For example, as noted above, the number $p$ defines the existing amount of resources. Several authors
considered models where our world is only a part of  the Universe and the amount of
resources in the world is not constant (see e.g. Ref. \cite{Netchitailo} and references therein). 

In any case, if we accept that $p$ is a universal parameter defining what ring or field describes nature
(at the present stage of the Universe or always) then the problem arises what the value of $n$ is.  If FQT is based on a field
then, since we treat FQT as a more general theory than standard one, it is desirable not to postulate
that it is based on $F_{p^2}$ (with $p=3\,\,(mod \,4)$) because standard theory is based on complex numbers
but vice versa, explain the fact that standard 
theory is based on complex numbers since FQT is based 
on $F_{p^2}$. Therefore we should find a motivation for
the choice of $F_{p^2}$ with $p=3\,\,(mod \,4)$. Arguments in favor of such a choice
are discussed in Refs. \cite{lev4,complex,symm1810} and in Secs. \ref{Hamiltonian} and \ref{VS7}.

\section{Correspondence between FQT and standard quantum theory}
\label{FQTvsStandard}

For any new theory there should exist a correspondence principle that at some conditions 
this theory and standard well tested one give close predictions. Known examples
are that classical nonrelativistic theory can be treated as a special case of relativistic
theory in the formal limit $c\to\infty$ and a special case of quantum mechanics in the
formal limit $\hbar\to 0$. Analogously, Poincare invariant theory is a special case of dS or
AdS invariant theories in the formal limit $R\to\infty$. We treat standard quantum theory
as a special case of FQT in the formal limit $p\to\infty$. Therefore a question arises which
formulation of standard theory is most suitable for its generalization to FQT. 

A known historical fact is that quantum mechanics has been originally proposed by Heisenberg and
Schr\"{o}dinger in two forms which seemed fully incompatible with each other. While in the Heisenberg
operator (matrix) formulation quantum states are described by infinite
columns and operators --- by infinite matrices, in the Schr\"{o}dinger wave formulations the states are described
by functions and operators --- by differential operators.
It has been shown later by Born, von Neumann, Dirac and others that
the both formulations are mathematically equivalent. In addition, the path integral approach has been
developed.

In the spirit of the wave or path integral approach one might
try to replace classical space-time by a finite lattice which
may even not be  a field. In that case the problem arises what
the natural quantum of space-time is and some of physical
quantities should necessarily have the field structure. A detailed discussion of this approach can be
found in Ref. \cite{Coish} and references therein. However, as argued in Sect. \ref{ST},
fundamental physical theory should not be based on space-time.

An approach for constructing a quantum theory over a Galois field similar to that proposed in our Refs. 
\cite{lev4,lev3} and subsequent publications has been discussed in Ref. \cite{Chang} and references therein. 

In the literature there have been discussed approaches where quantum theory is based on quaternions or 
 $p$-adic fields (see e.g. Ref. \cite{Adler} and references therein). In the cellular automation interpretation
of quantum mechanics proposed by 't Hooft (see ref. \cite{tHooft} and references therein) the observables can 
be only integers and the evolution of states with such observables is described by standard mathematics.
In those approaches  infinities still exist and so a problem remains whether or not it is possible to construct 
quantum theory without divergences. 

We treat FQT as a version of the matrix formulation when complex numbers
are replaced by elements of a finite ring or finite field. We will see below that in that case the columns and 
matrices are automatically truncated in a certain way, and the theory becomes not only
finite-dimensional but even finite.

In conventional quantum theory the state of a system is
described by a vector $\tilde x$ from a separable Hilbert
space $H$. We will now use a "tilde" to denote elements
of Hilbert spaces and complex numbers while elements
of linear spaces over a finite ring or field and elements of the corresponding ring or field will be denoted without a "tilde". Let $(\tilde e_1,\tilde e_2,...)$ be a basis in $H$. This
means that $\tilde x$ can be represented as
\begin{equation}
\tilde x =\tilde c_1 \tilde e_1+\tilde c_2 \tilde e_2+...
\label{G2}
\end{equation}
where $(\tilde c_1,\tilde c_2,...)$ are complex numbers.
It is assumed that there exists a complete set of commuting selfadjoint
operators $(\tilde A_1,\tilde A_2,...)$ in $H$ such that
each $\tilde e_i$ is the eigenvector of all these operators:
$\tilde A_j\tilde e_i ={\tilde \lambda}_{ji}\tilde e_i$. Then the
elements $(\tilde e_1,\tilde e_2,...)$ are mutually orthogonal:
$(\tilde e_i,\tilde e_j)=0$ if $i\neq j$ where (...,...) is
the scalar product in $H$. In that case the coefficients can
be calculated as
\begin{equation}
\tilde c_i = \frac{(\tilde e_i,\tilde x)}{(\tilde e_i,\tilde e_i)}
\label{G3}
\end{equation}
Their meaning is that
$|\tilde c_i|^2(\tilde e_i,\tilde e_i)/(\tilde x,\tilde x)$
represents the probability to find $\tilde x$ in the state
$\tilde e_i$. In particular, when $\tilde x$ and the basis
elements are normalized to one, the probability equals $|\tilde c_i|^2$.

As noted in Sec. \ref{finitemath}, Hilbert spaces in standard quantum theory contain a big redundancy of elements, and we do not need to know all of them. The reason is that, as explained in Sec.  \ref{finitemath}, with any desired accuracy 
each $\tilde x\in H$ can be approximated by a finite linear combination
\begin{equation}
\tilde x =\tilde c_1 \tilde e_1+\tilde c_2 \tilde e_2+...\tilde c_n\tilde e_n
\label{G4}
\end{equation}
where $\tilde c_j=\tilde a_j +i\tilde b_j$ and all the numbers $(\tilde a_j,\tilde b_j)\,\, (j=1,2,..n)$ are integers. 

This observation is very important from the following point of view. In classical mathematics the ring $Z$
is the starting point for introducing rational and real numbers. However, as follows from this observation, {\it even in standard
quantum theory all quantum states can be described with any desired accuracy by using only integers and therefore
rational and real numbers play only an auxiliary role}. 
This again poses a question whether the notions of infinitely small, continuity and division
are fundamental.

The meaning of the fact that Hilbert spaces in quantum theory are projective is very clear.
The matter is that not the probability itself but the relative probabilities of different measurement outcomes
have a physical meaning. We believe, the notion of probability is a
good illustration of the Kronecker expression about natural numbers (see Sec. \ref{finmath}). Indeed, this notion
arises as follows. Suppose that conducting
experiment $N$ times we have seen the first event $n_1$ times,
the second event $n_2$ times etc. such that $n_1 + n_2 + ... = N$.
We define the quantities $w_i(N) = n_i/N$ (these quantities
depend on $N$) and $w_i = lim\, w_i(N)$ when $N \rightarrow
\infty$. Then $w_i$ is called the probability of the
$ith$ event. We see that all the information about the experiment is given by a finite set of natural numbers, 
and all those numbers are finite. However, in order to define 
probabilities, people introduce additionally the notion of rational numbers and 
the notion of limit. Another example is the notion of mean value. Suppose we measure a physical
quantity such that in the first event its value is $q_1$, in the second event - $q_2$ etc.
Then the mean value of this quantity is defined as $(q_1n_1+q_2n_2+...)/N$ if $N$ is very large.
Therefore, even if all the $q_i$ are integers, the mean value might be not an integer. We again
see that rational numbers arise only as a consequence of our convention on how the results of
experiments should be interpreted. 

The Hilbert space is an example of a linear space over the field of complex numbers. Roughly speaking this means that
one can multiply the elements of the space by the elements
of the field and use the properties
$\tilde a(\tilde b\tilde x)=(\tilde a\tilde b)\tilde x$
and $\tilde a(\tilde b\tilde x+\tilde c\tilde y)=
\tilde a\tilde b\tilde x +\tilde a\tilde c\tilde y$ where
$\tilde a,\tilde b,\tilde c$ are complex numbers and
$\tilde x,\tilde y$ are elements of the space. 

Since complex conjugation is the automorphism of $R_{p^2}$ if $R_{p^2}=R_p+iR_p$ (and analogously in the
case of $F_{p^2}$) then, by analogy with conventional quantum theory, in FQT it is possible to consider situations when
linear spaces V over $R_{p^2}$ (or $F_{p^2}$) used for
describing physical states, are supplied by a scalar product (...,...) such that for any $x,y\in V$
and $a\in R_{p^2}$, $(x,y)$ is an element of $R_{p^2}$ and the following properties are satisfied:
\begin{equation}
(x,y) =\overline{(y,x)},\quad (ax,y)=\bar{a}(x,y),\quad
(x,ay)=a(x,y)
\label{G5}
\end{equation}

A correspondence between standard theory and FQT can be defined not only when $R_{p^2}=R_p+iR_p$ or
$F_{p^2}=F_p+iF_p$. For example, the field $F_{p^2}$ can be constructed by means of standard 
extension of 
$F_p$ as follows. Let the equation $x^2= -a_0$ ($a_0\in F_p$) has no solutions in $F_p$. 
Then $F_{p^2}$ can be formally described as the
set of elements $a+b\kappa$ where $a,b\in F_p$, $\kappa$ formally satisfies the condition $\kappa^2=-a_0$
but does not belong to $F_p$.
The actions in $F_{p^2}$ are defined in the natural way. The condition that the equation
$\kappa^2=-a_0$ has no solutions in $F_p$ is important in order to ensure that any nonzero
element from $F_{p^2}$ has an inverse. Indeed, the definition $(a+b\kappa)^{-1}=(a-b\kappa)/(a^2+a_0b^2)$ is
correct since the denominator cannot be equal to zero if at least $a$ or $b$ is
distinct from zero. The abovementioned correspondence is a special case when $a_0=1$, 
$p=3\,\, (mod\,\, 4)$ and $\kappa=i$. Another possible cases follow.

$a_0=2$: It is known \cite{VDW} that such a choice is possible if $p=5\,\, (mod\,\, 8)$.
The correspondence can be established as above, with the only difference that as the set
dense in H one can choose the set of such elements that ${\tilde c}_j={\tilde a}_j+\sqrt{2}{\tilde b}_j$, 
${\tilde a}_j,{\tilde b}_j\in Q$.

$a_0=3$: It is known \cite{VDW} that such a choice is possible if $p$ is the prime
number of the Fermata type, i.e. $p=2^n+ 1$. The correspondence can be established by
choosing ${\tilde c}_j={\tilde a}_j+\sqrt{3}{\tilde b}_j$, ${\tilde a}_j,{\tilde b}_j\in Q$.

Note that in cases $a_0=2$ and $a_0=3$ the equation $\kappa^2=-1$ does have solutions in $F_p$, and thus
in $F_{p^2}$ there is no element, which can be denoted as $i$. For this reason one might think that only the
case $a_0=1$ is natural. However, from the point
of view of correspondence between elements of projective spaces over $F_{p^2}$ and elements of
projective Hilbert spaces, all the three cases are on equal grounds. For simplicity we will mainly consider the case
$a_0=1$ when $R_{p^2}=R_p+iR_p$ or $F_{p^2}=F_p+iF_p$ but, as we will see below, the cases
$a_0=2$ and $a_0=3$ have their own advantages, since sometimes it is convenient that there exists such
an element $\epsilon\in F_p$ that $\epsilon^2=-1$. It is known \cite{VDW} that the field of $p^2$ elements has only one 
nontrivial automorphism. In all the considered cases it can be defined as $(a+b\kappa)\to \overline{a+b\kappa}=a-b\kappa$.

\section{Ring or field?}

We will always consider only finite dimensional spaces $V$ over
$R_{p^2}$ or $F_{p^2}$. Let $(e_1,e_2,...e_N)$ be a basis in such a space.
Consider subsets in $V$ of the form $x=c_1e_1+c_2e_2+...c_ne_n$
where for any $i,j$
\begin{equation}
c_i\in U(g(p)),\quad (e_i,e_j)\in U(g(p))
\label{G6}
\end{equation}
On the other hand, as noted above, in conventional quantum
theory we can describe quantum states by subsets of the form
Eq. (\ref{G4}). If $n\leq g(p)$ then
\begin{equation}
f(c_i)=\tilde c_i,\quad f((e_i,e_j))=(\tilde e_i,\tilde e_j)
\label{G7}
\end{equation}
and we have the correspondence between the description of
physical states in projective spaces over $R_{p^2}$ on 
one hand and projective Hilbert spaces on the other.
This means that if $p$ is very large then for a large number
of elements from $V$, linear combinations with the coefficients belonging to
$U(g(p))$ and scalar products look in the same way as for the elements from
a corresponding subset in the Hilbert space.

In the general case a scalar product in $V$ does not define
any positive definite metric and thus there is no probabilistic interpretation for all the elements from $V$.
In particular, $(e,e)=0$ does not necessarily imply that $e=0$.
However, the probabilistic interpretation exists for such a
subset in $V$ that the conditions (\ref{G7}) are satisfied.
Roughly speaking this means that for elements
$c_1e_1+...c_ne_n$ such that $(e_i,e_i),c_i{\bar c}_i\ll p$, $f((e_i,e_i))>0$
and $c_i{\bar c}_i>0$ for all $i=1,...n$, the probabilistic interpretation is valid.
Examples discussed below show that it is often possible to explicitly construct
a basis $(e_1,...e_N)$ such that $(e_j,e_k)=0$ for
$j\neq k$ and
$(e_j,e_j)\neq 0$ for all $j$ (see the subsequent section and Chap. \ref{AdS}).
Then $x=c_1e_1+...c_Ne_N$ where $c_j\in R_{p^2}$ or $c_j\in F_{p^2}$.

This discussion shows that for the correspondence between standard theory and FQT it is sufficient that Eqs. 
(\ref{G6},\ref{G7}) are satisfied and so it is sufficient that spaces in FQT are over $R_{p^2}$, not necessarily $F_{p^2}$.
As already noted, standard division does not have a fundamental meaning and so a problem arises whether it is
necessary to involve division for constructing FQT. However, division might be necessary from the following considerations.

Above we discussed the meaning of the fact that spaces in standard quantum theory are projective. Here $c\psi$ cannot be zero if
$\psi\neq 0$ and $c\neq 0$ because the set of complex numbers is a field. However, if we consider a space over a ring then
in general it is possible that $\psi\neq 0$, $c\neq 0$ but  $c\psi=0$. A simple example is that if $p$ is not prime, $a,b\in R_p$,
$a\neq 0$, $b\neq 0$ but $ab=0\,\, (mod\,\, p)$ then $ab\psi=0$. Another example is when $p$ is prime but $a^2+b^2=0$ in
$R_p$ (as noted in the preceding section, this can happen when $p=1\,\, (mod\,\, 4)$) then $(a^2+b^2)\psi =0$.

If a probabilistic interpretation is required then the situation when $\psi\neq 0$, $c\neq 0$ and $c\psi = 0$ should be obviously
excluded. However, as noted above, a probabilistic interpretation of quantum theory works only when 
a state is described by numbers which are much less than $p$. In this case such a situation cannot happen. Therefore our physical
intuition is insufficient on drawing a conclusion on whether or not such a situation should be excluded. It is automatically excluded
only if the space is over a field. 

Let us now suppose that there exist reasons why the above situation should be excluded when the space is over a ring.
This is the case when the ring $R$ is the integral domain, i.e. for $a,b\in R$, $ab=0$ is possible only when either $a=0$
 or $b=0$. A known theorem in number theory (see e.g. Ref. \cite{VDW}) is that every finite integral domain is a field.
Therefore a possible explanation of the requirement that spaces in quantum theory should be over a field is that there are
reasons that the above situation should be excluded. 

As shown in Sec. \ref{Matrix}, IRs describing Dirac singletons can be constructed over a ring but IRs describing massive
and massless particles can be constructed only over a field. In Sec. \ref{DirSingls} we
give strong arguments that only Dirac singletons can be true elementary particles. If this is the case then in
FQT division is not needed and the theory can be constructed over a ring. However, since now it is not clear
whether this possibility takes place then in what follows we assume that spaces in FQT are over a field
unless otherwise stated.

\section{Operators in FQT}
\label{operators}

In standard quantum theory operators act in a Hilbert space for the system under consideration. By analogy, we require that
in FQT operators act in a space over a finite ring or field.
As usual, if $A_1$ and $A_2$ are linear operators in $V$ such that
\begin{equation}
(A_1x,y)=(x,A_2y)\quad \forall x,y\in V
\label{G8}
\end{equation}
they are said to be conjugated: $A_2=A_1^*$.
It is easy to see
that $A_1^{**}=A_1$ and thus $A_2^*=A_1$.
If $A=A^*$ then the operator $A$ is said to be Hermitian.

If $(e,e)\neq 0$, $Ae=ae$, $a\in R_{p^2}$, and $A^*=A$, then it
is obvious that $a\in R_p$. In the subsequent section (see also Refs. \cite{lev4,complex}) 
we will see that there also exist situations when a Hermitian operator has eigenvectors $e$ such that
$(e,e)=0$ and the corresponding eigenvalue is pure imaginary. 

Let now $(A_1,...A_k)$ be a set of Hermitian commuting operators
in $V$, and $(e_1,...e_N)$ be a basis in $V$ with the properties
described above, such that $A_je_i=\lambda_{ji}e_i$.
Further, let $({\tilde A}_1,...{\tilde A}_k)$ be a set of
Hermitian  commuting
operators  in  a  Hilbert  space  $H$,   and
$(\tilde e_1,\tilde e_2,...)$ be a basis
in $H$ such that
$\tilde A_je_i={\tilde \lambda}_{ji}\tilde e_i$.
Consider a subset $c_1e_1+c_2e_2+...c_ne_n$
in $V$ such that,
in addition to the conditions (\ref{G7}), the
elements $e_i$
are the eigenvectors of the operators $A_j$ with
$\lambda_{ji}$
belonging to $U(g(p))$ and such that
$f(\lambda_{ji})={\tilde \lambda}_{ji}$. Then the
action of the operators on such
elements have the same form as the action of corresponding
operators on the subsets of elements in Hilbert
spaces discussed above.

Summarizing this discussion, we conclude that if $p$
is large then there exists a correspondence between the
description of physical states on the language of Hilbert spaces and selfadjoint operators
in them on one hand, and on the language of linear spaces over
$R_{p^2}$ and Hermitian operators in them on the other.

The field of complex numbers is algebraically closed (see
standard textbooks on modern algebra, e.g. Ref. \cite{VDW}).
This implies that any equation of the $nth$ order in this field
always has $n$ solutions. This is not, generally speaking, the
case for the field $F_{p^2}$. As a consequence, not every linear
operator in the finite-dimensional space over $F_{p^2}$ has an
eigenvector (because the characteristic equation may have no
solution in this field). One can define a field of
characteristic $p$ which is algebraically closed and
contains $F_{p^2}$. However such a field will necessarily be
infinite and we will not use it. We will see in this
chapter and Chap. \ref{AdS} that uncloseness of the field $F_{p^2}$ does not
prevent one from constructing physically meaningful representations
describing elementary particles in FQT.

In physics one usually considers Lie  algebras over real numbers $R$ and
their representations by Hermitian  operators in complex
Hilbert spaces. It is clear that analogs of such
representations in our case are representations of Lie
algebras over $R_p$  by Hermitian operators
in spaces over $R_{p^2}$ or $F_{p^2}$.
Representations in spaces over a field of nonzero
characteristics are called modular representations.
There exists a wide literature devoted to such
representations; detailed references can be found
for example in Ref.
\cite{FrPa} (see also Ref. \cite{lev4}).
In particular, it has been shown by Zassenhaus
\cite{Zass} that all modular IRs are finite-dimensional and
many papers have
dealt with the maximum dimension of such representations.
At the same time, mathematicians usually consider only representations
over an algebraically closed field.

From  the previous, it is natural to expect that the correspondence  between
ordinary and modular representations of two Lie algebras over $R$
and $F_p$, respectively, can be obtained if the structure constants of the Lie algebra
over $F_p$ - $c_{kl}^j$, and the structure constants of the Lie
algebra over $R$ - ${\tilde c}_{kl}^j$, are such that $f(c_{kl}^j)={\tilde c}_{kl}^j$
(the Chevalley basis \cite{Chev}), and
all the $c_{kl}^j$ belong to $U(g(p))$. In Refs.
\cite{lev4,lev3,levsusy} modular analogs of IRs
of su(2), sp(2), so(2,3), so(1,4) algebras and the osp(1,4)
superalgebra have been considered. Also modular
representations describing strings have been briefly mentioned. In all these cases the
quantities ${\tilde c}_{kl}^j$ take only the values
$0,\pm 1,\pm 2$ and the above correspondence does take place.

It is obvious that since all physical quantities in
FQT are discrete, this theory cannot involve any
dimensionful quantities and any operators having the
continuous spectrum. We have seen in the preceding chapters
than the so(1,4) invariant theory is dimensionless and 
it is possible to choose a basis such that all the operators have only discrete spectrum.
As shown in Chap. \ref{AdS}, the same is true for the so(2,3) invariant theories.
For this reason one might expect that those theories are
natural candidates for their generalization to FQT. This means that symmetry is defined
by the commutation relations (\ref{newCR}) which are now considered not in standard
Hilbert spaces but in spaces over $R_{p^2}$ or $F_{p^2}$.
We will see in this chapter that there exists a
correspondence in the above sense between modular IRs
of the finite field analog of the so(1,4) algebra and
IRs of the standard so(1,4) algebra and in Chap. \ref{AdS} the same will be shown for the so(2,3) algebra. 
At the same time,
there is no natural generalization of the Poincare invariant theory to FQT.

Since the main problems of QFT originate from
the fact that local fields interact at the same point,
the idea of all modern theories aiming to improve
QFT is to replace the interaction at a point by an
interaction in some small space-time region. From this
point of view, one could say that those theories involve
a fundamental length, explicitly or implicitly. Since
FQT is a fully discrete theory, one might wonder
whether it could be treated as a version of quantum
theory with a fundamental length. Although in FQT
all physical quantities are dimensionless and take
values in a finite ring or field, on a qualitative level
FQT might be thought to be a theory with the fundamental
length in the following sense. The maximum value
of the angular momentum in FQT cannot exceed $p$. Therefore
the Poincare momentum cannot exceed $p/R$. This
can be interpreted in such a way that
the fundamental length in FQT is of the order of $R/p$.
However, in view of the fact that standard uncertainty relations
are not well founded such a notion of fundamental length in FQT
does not have a physical meaning.

One might wonder how continuous transformations (e.g.
time evolution or rotations) can be described in the framework
of FQT. A general remark is that if theory $A$ is a
generalization of theory $B$ then the relation between
them is not always straightforward. For example,
quantum mechanics is a generalization of classical
mechanics, but in quantum mechanics the experiment outcome
cannot be predicted unambiguously, a particle cannot be
always localized etc. As
noted in Sec. \ref{ST}, even in the framework of standard
quantum theory, time evolution is well-defined only 
on macroscopic level. Suppose that this is the case and
the Hamiltonian $H_1$ in standard theory is a good approximation
for the Hamiltonian $H$ in FQT. Then one might think that
$exp(-iH_1t)$ is a good approximation for $exp(-iHt)$. However,
such a straightforward conclusion is problematic for the
following reasons. First, there can be no continuous parameters
in FQT. Second, even if $t$ is somehow discretized, 
it is not clear how the transformation $exp(-iHt)$ 
should be implemented in practice. On macroscopic level the
quantity $Ht$ is very large and therefore the Taylor series
for $exp(-iHt)$ contains a large number of terms
which should be known with a high accuracy. On the other hand,
one can notice that for computing $exp(-iHt)$ it is sufficient
to know $Ht$ only modulo $2\pi$ but in this case the question
about the accuracy for $\pi$ arises.
We see that a direct correspondence between 
standard quantum theory and FQT exists only on the level of
Lie algebras but not on the level of Lie groups.

\section{Modular IRs of dS algebra and spectrum of dS Hamiltonian}
\label{Hamiltonian}

Consider modular analogs of IRs constructed in Sec. \ref{S6}.
We noted that the basis elements of this IR are $e_{nkl}$ where at a fixed value of $n$, 
$k=0,1,...n$ and $l=0,1,...2k$. In standard case, IR is infinite-dimensional since
$n$ can be zero or any natural number. A modular analog of this IR can be only finite-dimensional.
The basis of the modular IR is again $e_{nkl}$ where at a fixed value of $n$ the numbers $k$
and $l$ are in the same range as above. The operators of such IR can be described by the same 
expressions as in Eqs. (\ref{20A}-\ref{22B}). Since they contain $i$ and nontrivial division,
in FQT one can consider those expressions only  in a space over $F_{p^2}$ with $p=3\,\, (mod\,\, 4)$. 
The quantity $n$ can now be only in the
range $0,1,...N$ where $N$ can be found from the condition that the algebra of
operators described by Eqs. (\ref{20A}) and (\ref{20B}) should be closed. This is the
case if $w+(2N+3)^2=0$ in $F_p$ and $N+k+2<p$. Therefore we have to show that
such $N$ does exist. 

In the modular case $w$ cannot be written as $w=\mu^2$ with $\mu\in F_p$ since the equality
$a^2+b^2=0$ in $F_p$ is not possible if $p=3\,\,(mod \,4)$. In terminology of number theory,
this means that $w$ is a quadratic nonresidue. Since $-1$ also is a quadratic nonresidue if
$p=3\,\,(mod \,4)$, $w$ can be written as $w=-{\tilde\mu}^2$ where ${\tilde\mu}\in F_p$ and
for ${\tilde\mu}$ obviously two solutions are possible. Then $N$ should satisfy one
of the conditions $N+3=\pm {\tilde\mu}$ and one should choose that with
the lesser value of $N$. Let us assume that both, ${\tilde\mu}$ and $-{\tilde\mu}$ are
represented by $0,1,...(p-1)$. Then if ${\tilde\mu}$ is odd, $-{\tilde\mu}=p-{\tilde\mu}$ is
even and {\it vice versa}. We choose the odd number as ${\tilde\mu}$. Then
the two solutions are $N_1=({\tilde\mu}-3)/2$ and $N_2=p-({\tilde\mu}+3)/2$. Since $N_1<N_2$, we choose
$N=({\tilde\mu}-3)/2$. In particular, this quantity satisfies the condition 
$N\leq (p-5)/2$. Since $k\leq N$, the condition $N+k+2<p$ is satisfied and
the existence of $N$ is proved. In any realistic scenario, $w$ is such that $w\ll p$ even for macroscopic bodies.
Therefore the quantity $N$ should be at least of the order of $p^{1/2}$. The dimension of IR is
\begin{equation}
Dim = \sum_{n=0}^{N}\sum_{k=0}^n(2k+1)=(N+1)(\frac{1}{3}N^2+\frac{7}{6}N+1)
\label{50}
\end{equation}
and therefore $Dim$ is at least of the order of $p^{3/2}$.

The relative probabilities are defined
by $||c(n,k,l)e_{nkl}||^2$. In standard theory the basis states and WFs can be normalized
to one such that the normalization condition is $\sum_{nkl}|{\tilde c}(n,k,l)|^2=1$. Since the values 
${\tilde c}(n,k,l)$ can be arbitrarily small, WFs can have an arbitrary support belonging to $[0,\infty)$.
However, in FQT the quantities $|c(n,k,l)|^2$ and $||e_{nkl}||^2$ belong to $F_p$.
Roughly speaking, this means that if they are not zero then they are greater or equal than one. Since
for probabilistic interpretation we should have that $\sum_{nkl}||c(n,k,l)e_{nkl}||^2\ll p$, 
the probabilistic interpretation may take place only if $c(n,k,l)=0$ for
$n>n_{max},\,\,n_{max}\ll N$. That is why in Chap. \ref{Ch3} we discussed only WFs having
the support in the range $[n_{min},n_{max}]$.  

As follows from the spectral theorem for selfadjoint operators in Hilbert spaces,  
any selfadjoint operator $A$ is fully decomposable, i.e. it is always possible to find a basis, such that
all the basis elements are eigenvectors (or generalized eigenvectors) of $A$. As noted in Sect.
\ref{operators}, in FQT this is not necessarily the case since the field $F_{p^2}$ is not 
algebraically closed. However, it can be shown \cite{VDW} that for any equation of the $Nth$ 
order, it is possible to extend the field such that the equation will have $N$ solutions.  
A question arises what is the minimum extension of $F_{p^2}$, which guarantees that all the
operators $({\cal E},{\bf N},{\bf B},{\bf J})$ are fully decomposable. 

The operators $({\bf B},{\bf J})$ describe a representation of the so(4) = su(2)$\times$su(2) subalgebra.
It is easy to show (see also Chap. \ref{AdS}) that the representation operators of the
su(2) algebra are fully decomposable in the field $F_{p^2}$. Therefore it is sufficient to investigate
the operators $({\cal E},{\bf N})$. They represent components of the so(4) vector operator $M^{0\nu}$
($\nu=1,2,3,4$) and therefore it is sufficient to investigate the dS 
energy operator ${\cal E}$, which with our choice of the basis has a rather simpler form (see Eqs. 
(\ref{20A}) and (\ref{22A})). This operator acts nontrivially only over the variable $n$ and its nonzero 
matrix elements are given by
\begin{equation}
{\cal E}_{n-1,n}=\frac{n+1+k}{2(n+1)}[w+(2n+1)^2],\quad {\cal E}_{n+1,n}=\frac{n+1-k}{2(n+1)}
\label{H1}
\end{equation}
Therefore, for a fixed value of $k$ it is possible to consider the action of ${\cal E}$ 
in the subspace with the basis elements $e_{nkl}$ ($n=k, k+1,...N)$.

Let $A(\lambda)$ be the matrix of the operator ${\cal E}-\lambda$ such that 
$A(\lambda)_{qr}={\cal E}_{q+k,r+k}-\lambda \delta_{qr}$. 
We use $\Delta_q^r(\lambda)$ to denote the determinant of the matrix obtained from
$A(\lambda)$ by taking into account only the rows and columns with the numbers $q,q+1,...r$. With our
definition of the matrix $A(\lambda)$, its first row and column have the number equal to 0 while the last ones 
have the number $K=N-k$. Therefore the characteristic equation can be written as
\begin{equation}
\Delta_0^K(\lambda)=0
\label{H2}
\end{equation}
In general, since the field $F_{p^2}$ is not algebraically closed, there is no guaranty that we will
succeed in finding even one eigenvalue. However, we will see below that in a special case of the operator
with the matrix elements (\ref{H1}), it is possible to find all $K+1$ eigenvalues.

The matrix $A(\lambda)$ is three-diagonal. It is easy to see that 
\begin{equation}
\Delta_0^{q+1}(\lambda)=-\lambda \Delta_0^q(\lambda)-A_{q,q+1}A_{q+1,q}\Delta_0^{q-1}(\lambda)
\label{H3}
\end{equation}
Let $\lambda_l$ be a solution of Eq. (\ref{H2}). We denote $e_q\equiv e_{q+k,kl}$. 
Then the element
\begin{equation}
\chi(\lambda_l)=\sum_{q=0}^K \{(-1)^q \Delta_0^{q-1}(\lambda_l)e_q/[\prod_{s=0}^{q-1}A_{s,s+1}]\}
\label{H4}
\end{equation}
is the eigenvector of the operator ${\cal E}$ with the eigenvalue
$\lambda_l$. This can be verified directly by using Eqs. (\ref{22A}) and (\ref{H1}-\ref{H4}).

To solve Eq. (\ref{H3}) we have to find the expressions for
$\Delta_0^q(\lambda)$ when $q=0,1,...K$. It is obvious that $\Delta_0^0(\lambda)=-\lambda$, and as follows from
Eqs. (\ref{H1}) and (\ref{H3}),
\begin{equation}
\Delta_0^1(\lambda)=\lambda^2-\frac{w+(2k+3)^2}{2(k+2)}
\label{H6}
\end{equation}
If $w=-{\tilde\mu}^2$ then it can be shown that $\Delta_0^q(\lambda)$ is given by the 
following expressions. If $q$ is odd then
\begin{eqnarray}
&&\Delta_0^q(\lambda)=\sum_{l=0}^{(q+1)/2}C_{(q+1)/2}^l
\prod_{s=1}^l[\lambda^2+({\tilde\mu}-2k-4s+1)^2](-1)^{(q+1)/2-l}\nonumber\\
&&\prod_{s=l+1}^{(q+1)/2}\frac{(2k+2s+1)({\tilde\mu}-2k-4s+1)({\tilde\mu}-2k-4s-1)}{2(k+(q+1)/2+s)}
\label{H7}
\end{eqnarray} 
and if $q$ is even then
\begin{eqnarray}
&&\Delta_0^q(\lambda)=(-\lambda)\sum_{l=0}^{q/2}C_{q/2}^l
\prod_{s=1}^l[\lambda^2+({\tilde\mu}-2k-4s+1)^2]
(-1)^{q/2-l}\nonumber\\
&&\prod_{s=l+1}^{(q+1)/2}\frac{(2k+2s+1)({\tilde\mu}-2k-4s-1)({\tilde\mu}-2k-4s-3)}{2(k+q/2+s+1)}
\label{H8}
\end{eqnarray}
Indeed, for $q=0$ Eq. (\ref{H8}) is compatible with
$\Delta_0^0(\lambda)=-\lambda$, and for $q=1$ Eq. (\ref{H7})
is compatible with Eq. (\ref{H6}). Then one can
directly verify that Eqs. (\ref{H7}) and (\ref{H8}) are compatible with Eq. (\ref{H3}).

With our definition of ${\tilde\mu}$, the only possibility for $K$ is such that 
\begin{equation}
{\tilde\mu} =2K+2k+3
\label{H10}
\end{equation}
Then, as follows
from Eqs. (\ref{H7}) and (\ref{H8}), when $K$ is odd or even, only the term with
$l=[(K+1)/2]$ (where $[(K+1)/2]$ is the integer part of $(K+1)/2$) contributes to 
$\Delta_0^K(\lambda)$ and, as a consequence
\begin{eqnarray}
&&\Delta_0^K(\lambda)=(-\lambda)^{r(K)}
\prod_{k=1}^{[(K+1)/2]}[\lambda^2+({\tilde\mu}-2j-4k+1)^2]
\label{H11}
\end{eqnarray}
where $r(K)=0$ if $K$ is odd and $r(K)=1$ if $K$ is even. 
If $p=3\,\,(mod\,\,4)$, this equation has solutions 
only if $F_p$ is extended, and the minimum extension
is $F_{p^2}$. Then the solutions are given by
\begin{equation}
\lambda =\pm i({\tilde\mu}-2k-4s+1)\quad (s=1,2...[(K+1)/2])
\label{H12}
\end{equation}
and when $K$ is even there also exists an additional
solution $\lambda=0$. When $K$ is odd, solutions can be represented as
\begin{equation}
\lambda =\pm 2i,\,\pm 6i,...\pm 2iK
\label{H13}
\end{equation}
while when $K$ is even, the solutions can be represented as
\begin{equation}
\lambda =0,\, \pm 4i,\,\pm 8i,...\pm 2iK
\label{H14}
\end{equation}
Therefore the spectrum is equidistant and the
distance between the neighboring elements is equal to $4i$. 
As follows from Eqs. (\ref{H10}), all the roots are simple and then, as follows from Eq. (\ref{H4}),
the operator ${\cal E}$ is fully decomposable. It can be shown by a direct calculation \cite{complex}
that the eigenvectors $e$ corresponding to pure imaginary eigenvalues are such that $(e,e)=0$
in $F_p$. Such a possibility has been mentioned in the preceding section.

Our conclusion is that if $p=3\,(mod \,4)$ then all the operators $({\cal E},{\bf N},{\bf B},{\bf J})$ 
are fully decomposable if $F_p$ is extended to $F_{p^2}$ but no further extension is necessary.
This might be an argument explaining why standard theory is based on complex numbers. On the
other hand, our conclusion is obtained by considering states where $n$ is not necessarily small
in comparison with $p^{1/2}$ and standard physical intuition does not work in this case. 
One might think that the solutions (\ref{H13}) and (\ref{H14}) for the eigenvalues of the dS
Hamiltonian indicate that FQT is unphysical since the Hamiltonian cannot have imaginary eigenvalues.
However, such a conclusion is premature since in standard quantum theory the
Hamiltonian of a free particle does not have normalized eigenstates (since the spectrum is pure
continuous) and therefore for any realistic state the width of the energy distribution cannot be zero.

If $A$ is an operator of a physical quantity in standard theory then the distribution of this quantity
in some state can be calculated in two ways. First, one can find eigenvectors of $A$, decompose the
state over those eigenvectors and then the coefficients of the decomposition describe the distribution.
Another possibility is to calculate all moments of $A$, i.e. the mean value, the mean square deviation
etc. Note that the moments do not depend on the choice of basis since they are fully defined by
the action of the operator on the given state. A standard result of the probability theory 
(see e.g. Ref. \cite{Billingsley}) is that the set of moments uniquely defines the moment distribution function,
which in turn uniquely defines the distribution. However in practice there is no need to know all
the moments since the number of experimental data is finite and knowing only several first moments
is typically quite sufficient. 

In FQT the first method does not necessarily defines the distribution. In particular, the above
results for the dS Hamiltonian show that its eigenvectors $\sum_{nkl}c(n,k,l)e_{nkl}$ 
are such that $c(n,k,l) \neq 0$ for all $n=k,...N$, where $N$ is at least of the order of $p^{1/2}$.
Since the $c(n,k,l)$ are elements of $F_{p^2}$, their formal modulus cannot be less than 1 and therefore
the formal norm of such eigenvectors cannot be much less than $p$ (the equality $(e,e)=0$ takes place
since the scalar product is calculated in $F_p$). Therefore eigenvectors of the dS Hamiltonian do not
have a probabilistic interpretation. On the other hand, as already noted, we can consider states
$\sum_{nkl}c(n,k,l)e_{nkl}$ such that $c(n,k,l)\neq 0$ only if $n_{min}\leq n\leq n_{max}$ where 
$n_{max}\ll N$. Then the probabilistic interpretation for such states might be a good approximation if 
at least several first moments give reasonable physical results 
(see the discussion of probabilities in Sect. \ref{FQTvsStandard}). In Chaps. \ref{Ch3} and \ref{twobody} we discussed semiclassical 
approximation taking into account only the first two moments: the mean value and mean square deviation.

\chapter{Semiclassical states in modular representations}
\label{Ch5}

\section{Semiclassical states in FQT}
\label{S11}

A possible approach for constructing semiclassical states in FQT is to use the basis defined by Eq. (\ref{19}) where the 
coefficients $c(n,k,\mu)$ should be elements of $F_{p^2}$. Such states should satisfy several criteria. First, as noted in
the preceding chapter, the probabilistic interpretation can be valid only if the quantities
$\rho_0(n,k,\mu)=(e_{nk\mu},e_{nk\mu})$ defined by Eq. (\ref{21}) are such that
$f(\rho_0(n,k,\mu))\geq 0$ and $f(\rho_0(n,k,\mu))\ll p$ where $f$ is the map from $F_p$ to $Z$ defined in Sec. \ref{finmath}.

By using the fact that spaces in quantum theory are projective one can replace the basis elements $e_{nk\mu}$ by
$Ce_{nk\mu}$ where $C\in F_{p^2}$ is any nonzero constant. Then the matrix elements of the operators in the new basis
are the same and the normalizations are defined by the quantities $\rho(n,k,\mu)=C{\bar C}\rho_0(n,k,\mu)$. As noted in
the preceding chapter, this reflects the fact that only ratios of probabilities have a physical meaning. Hence for
ensuring probabilistic interpretation one could try to find $C$ such that the quantities $f(\rho(n,k,\mu))$ have 
the least possible values.

As follows from Eq. (\ref{21}), 
\begin{equation}
\rho_0(n,k,\mu)=(2k+1)!C_{2k}^{k-\mu}C_n^kC_{n+k+1}^k\prod_{j=1}^n [w+(2j+1)^2]
\label{rho0}
\end{equation}
As noted in Chap. \ref{Ch4},
a probabilistic interpretation can be possible only if $c(n,k,\mu)\neq 0$ for $n\in [n_{min},n_{max}]$, 
$k\in [k_{min},k_{max}]$ and $\mu\in [\mu_{min},\mu_{max}]$. Hence our nearest goal is to find the
constant $C$ such that the quantities $\rho(n,k,\mu)$ have the least possible values when the quantum
numbers $(nk\mu)$ are in the above range.

We denote $\Delta n=n_{max}-n_{min}$, $\Delta k=k_{max}-k_{min}$ and $\Delta \mu=\mu_{max}-\mu_{min}$.
Since $R$ is very large, we expect that $\Delta n\gg \Delta k, \Delta \mu$ but since the exact value of $R$
is not known, we don't know whether a typical value of $k$ is much greater than $\Delta n$ or not. 
One can directly verify that $\rho_0(n,k,\mu)=C_1\rho(n,k,\mu)$ where
\begin{eqnarray}
&&\rho(n,k,\mu)=4^{k-k_{min}}\frac{(2k+1)!!(2k-1)!!(k_{max}-\mu_{min})!(k_{max}+\mu_{max})!}
{(2k_{min}+1)!!(2k_{min}-1)!!(k-\mu)!(k+\mu)!}\nonumber\\
&&\prod_{j=0}^{n+k-n_{min}-k_{min}-1}(n_{min}+k_{min}+2+j)\prod_{j=0}^{n_{max}-k_{min}-n+k-1}(n-k+1+j)\nonumber\\
&&\prod_{j=0}^{n-n_{min}-1}(n_{min}+1+j)\prod_{j=0}^{n_{max}-n-1}(n+2+j)\prod_{j=n_{min}+1}^n [w+(2j+1)^2]
\label{newrho}
\end{eqnarray}
\begin{eqnarray}
&&C_1=4^{k_{min}}\frac{(2k_{min}+1)!!(2k_{min}-1)!!}
{(k_{max}-\mu_{min})!(k_{max}+\mu_{max})!}\prod_{j=1}^{n_{min}} [w+(2j+1)^2]\nonumber\\
&&\prod_{j=0}^{k_{min}-\Delta n-1}(n_{max}+2+j)(n_{max}-k_{min}+1+j)
\label{C1}
\end{eqnarray}
if $k\gg \Delta n$ and
\begin{eqnarray}
&&\rho(n,k,\mu)=4^{k-k_{min}}\frac{(2k+1)!!(2k-1)!!(k_{max}-\mu_{min})!(k_{max}+\mu_{max})!}
{(2k_{min}+1)!!(2k_{min}-1)!!(k-\mu)!(k+\mu)!}\nonumber\\
&&\prod_{j=0}^{k-1}[(n+2+j)(n-k+1+j)]\prod_{j=n_{min}+1}^n [w+(2j+1)^2]
\label{newrhoB}
\end{eqnarray}
\begin{eqnarray}
&&C_1=4^{k_{min}}\frac{(2k_{min}+1)!!(2k_{min}-1)!!}
{(k_{max}-\mu_{min})!(k_{max}+\mu_{max})!}\prod_{j=1}^{n_{min}} [w+(2j+1)^2]
\label{C1B}
\end{eqnarray}
if $k$ is of the same order than $\Delta n$ or less.

The next step is to prove the existence of the constant $C$ such that $C{\bar C}=C_2$ where
$C_2=1/C_1$. For this purpose we note the following. It is known \cite{VDW} that any Galois field 
without its zero element is a cyclic multiplicative group. Let $r$ be a
primitive root in $F_p$, \emph{i.e.}, the element such that any nonzero
element of $F_p$ can be represented as $r^s$ $(s=1,2,...,p-1)$. Hence, if $C_2=r^s$ and $s$ is even then 
$C=r^{s/2}$ obviously satisfies the above requirement.

Suppose now that $s$ is odd. As noted in Chap. \ref{Ch4}, $-1$ is a quadratic residue in
$F_p$ if $p=1\,\, (mod\,\, 4)$ and a quadratic non-residue in $F_p$ if $p=3\,\, (mod\,\, 4)$. Therefore
in the case $p=3\,\, (mod\,\, 4)$ we have $-1=r^q$ where $q$ is odd. Hence $C_2=-C_3$ where $C_3=r^{s+q}$
is a quadratic residue in $F_p$. Now the quantity $C$ satisfying the above requirement exists if 
$C=\alpha r^{(s+q)2}$ and $\alpha$ satisfies the equation
\begin{equation}
\alpha {\bar \alpha}=-1
\label{V77}
\end{equation}

For proving that the solution of this equation exists we again use the property that any Galois field 
without its zero element is a cyclic multiplicative group but now this property is applied in the case of
$F_{p^2}$ with $p=3\,\, (mod\,\, 4)$. Let now $r$ be a primitive root in $F_{p^2}$. It is known \cite{VDW} that the only
nontrivial automorphism of $F_{p^2}$ is $\alpha\rightarrow {\bar \alpha}=\alpha^p$. Therefore if $\alpha =r^s$ then
$\alpha{\bar \alpha}= r^{(p+1)s}$. On the other hand, since $r^{(p^2-1)}=1$, $r^{(p^2-1)/2}=-1$. 
Therefore a solution of Eq. (\ref{V77}) exists at least with $s=(p-1)/2$.

The next step is to investigate conditions for the coefficients $c(n,k,\mu)$ such that the state
$\sum_{nk\mu}c(n,k,\mu)e(n,k,\mu)$ is semiclassical. As noted in Sect. \ref{S7}, in standard theory 
the quantities $c(n,k,\mu)$ contain the factor $exp[i(-n\varphi+k\alpha -\mu\beta)]$ and in the region
of maximum the quantities $|c(n,k,\mu)|^2$ are of the same order. To generalize these conditions to the
case of FQT we define a function $F$ from the set of complex numbers to $F_{p^2}$. If $a$ is a real number
then we define $l=Round(a)$ as an integer closest to $a$. This definition is ambiguous when $a=l\pm 0.5$ but
in the region of maximum the numbers in question are very large and the rounding errors $\pm 1$ are not important. Analogously, if $z=a+bi$ is a complex number then we define $Round(z)=Round(a)+Round(b)i$. Finally, we define $F(z)\in F_{p^2}$ as $f(Round(z))$.

As follows from Eqs. (\ref{newrho}) and (\ref{newrhoB}), the quantity $\rho(n,k,\mu)$ has the maximum at $n=n_{max}$, $k=k_{max}$, $\mu=\mu_{min}$. Consider the state $\sum_{nk\mu}c(n,k,\mu)e(n,k,\mu)$ such that
\begin{equation}
c(n,k,\mu)=a(n,k,\mu)F\{[\frac{\rho(n_{max},k_{max},\mu_{min}}{\rho(n,k,\mu)}]^{1/2}exp[i(-n\varphi+k\alpha -\mu\beta)]\}
\label{cnkmu}
\end{equation} 
where $a(n,k,\mu)$ is a slowly changing function in the region of maximum.

For the validity of semiclassical approximation the condition 
\begin{equation}
\rho(n_{max},k_{max},\mu_{min})\sum_{nk\mu} |a(n,k,\mu)|^2\ll p
\label{llp}
\end{equation}
should be satisfied. As follows from Eqs. (\ref{newrho}) and (\ref{newrhoB}), for a nonrelativistic particle it will be satisfied if
\begin{equation}
(4k_{max})^{\Delta k}[(k_{max}-\mu_{min})(k_{max}+\mu_{max})]^{(\Delta k+\Delta \mu)}n_{max}^{2(\Delta n+\Delta k)}w^{\Delta n}A
\Delta n \Delta k \Delta \mu \ll p
\label{llpB}
\end{equation} 
or 
\begin{equation}
(4k_{max})^{\Delta k}[(k_{max}-\mu_{min})(k_{max}+\mu_{max})]^{(\Delta k+\Delta \mu)}w^{\Delta n}A
\Delta n \Delta k \Delta \mu \ll p
\label{llpC}
\end{equation} 
respectively, where $A$ is the maximum value of $|a(n,k,\mu)|^2$.
If $A$ is not anomalously large then in the both cases those conditions can be 
approximately written as 
\begin{equation}
\Delta n lnw \ll lnp
\label{lnrho}
\end{equation}
Therefore not only the number $p$ should be very large, but even $lnp$ should be very large.

\begin{sloppypar}
\section{Many-body systems in FQT and gravitational constant}
\label{manybody}
\end{sloppypar}

In quantum theory, state vectors of a system of N bodies belong to the Hilbert space which is the 
tensor product of single-body Hilbert spaces. This means that state vectors of the $N$-body systems
are all possible linear combinations of functions
\begin{equation}
\psi(n_1,k_1,l_1,...n_N,k_N,l_N)=\psi_1(n_1,k_1,l_1)\cdots \psi_N(n_N,k_N,l_N)
\label{tprod}
\end{equation}
By definition, the bodies do not interact if all representation operators of the symmetry algebra for
the $N$-body systems are sums of the corresponding single-body operators. For example, the energy operator
${\cal E}$ for the $N$-body system is a sum ${\cal E}_1+{\cal E}_2+...+{\cal E}_N$ where the operator
${\cal E}_i$ ($i=1,2,...N$) acts nontrivially over its "own" variables $(n_i,k_i,l_i)$ while over other
variables it acts as the identity operator.  

If we have a system of noninteracting bodies in standard quantum theory, each $\psi_i(n_i,k_i,l_i)$ in
Eq. (\ref{tprod}) is fully independent of states of other bodies. However, in FQT the
situation is different. Here, as shown in the preceding section, a necessary condition for the WF to have a probabilistic 
interpretation is given by Eq. (\ref{lnrho}). Since we assume that $p$ is very large,
this is not a serious restriction. However, if a
system consists of $N$ components, a necessary condition that the WF of the system has a
probabilistic interpretation is
\begin{equation}
\sum_{i=1}^N \delta_i lnw_i \ll lnp
\label{70}
\end{equation} 
where $\delta_i=\Delta n_i$ and $w_i=4R^2m_i^2$ where $m_i$ is the mass of the subsystem $i$. 
This condition shows that in FQT the greater the number of components is,
the stronger is the restriction on the width of the dS momentum distribution for each component.
This  is a crucial difference between standard theory and FQT. A naive explanation is that if $p$ is
finite, the same set of numbers which was used for describing one body is now shared between $N$ bodies.
In other words, if in standard theory each body in the free $N$-body system does not feel the presence of
other bodies, in FQT this is not the case. This might be treated as an effective interaction in the free
$N$-body system.

In Chaps. \ref{Ch2} and \ref{twobody} we discussed a system of two free bodies such their relative motion can be described
in the framework of semiclassical approximation. We have shown that the mean value of the mass operator
for this system differs from the expression given by standard Poincare theory. The difference describes an
effective interaction which we treat as the dS antigravity at very large distances and gravity
when the distances are much less than cosmological ones. In the latter case the result depends on the total 
dS momentum distribution for each body (see Eq. (\ref{preNewton2})).
Since the interaction is proportional to the masses of the bodies, this effect is important only
in situations when at least one body is macroscopic. Indeed, if neither of the bodies is macroscopic, their
masses are small and their relative motion is not described in the framework of semiclassical approximation.
In particular, in this approach, gravity between two elementary particles has no physical meaning.

The existing quantum theory does not make it possible to reliably calculate the width of the total dS momentum
distribution for a macroscopic body and at best only a qualitative estimation of this quantity can be
given. The above discussion shows that the greater is the mass of the macroscopic body, the stronger is
the restriction on the dS momentum distribution for each subsystem of this body. Suppose that a
body with the mass $M$ can be treated as a composite system consisting of similar subsystems with the mass $m$.
Then the number of subsystems is $N=M/m$ and, as follows from Eq. (\ref{70}), the width $\delta$ of their dS momentum
distributions should satisfy the condition $N\delta lnw\ll lnp$ where $w=4R^2m^2$. Since the greater the value 
of $\delta$ is, the more accurate is the
semiclassical approximation, a reasonable scenario is that each subsystem tends to have the maximum possible
$\delta$ but the above restriction allows to have only such value of $\delta$ that it is of the order 
of magnitude not exceeding $lnp/(Nlnw)$. 

The next question is how to estimate the width of the total dS momentum distribution for a macroscopic body.
For solving this problem one has to change variables from individual dS momenta of subsystems to total and
relative dS momenta. Now the total dS momentum and relative dS momenta will have their own momentum distributions
which are subject to a restriction similar to that given by Eq. (\ref{70}). If we assume that all the variables
share this restriction equally then the width of the total momentum distribution also will be a quantity not
exceeding $lnp/(Nlnw)$. Suppose that $m=N_1m_0$ where $m_0$ is the nucleon mass. The value of $N_1$ should be 
such that our subsystem still can be described by semiclassical approximation. Then the estimation of $\delta$ is
\begin{equation}
\delta=N_1m_0lnp/[2Mln(2RN_1m_0)]
\label{estimation}
\end{equation}
Let $\mu_0$ be the dS nucleon mass and $M$ now means the dS mass of the macroscopic body.
Then this expression can be written as 
\begin{equation}
\delta=N_1\mu_0lnp/[2Mln(N_1\mu_0)]
\label{estimationB}
\end{equation}
Suppose that $N_1$ can be taken to be the same for all macroscopic bodies.
For example, it is reasonable to expect that when $N_1$ is of the order of $10^3$, the subsystems still can be
described by semiclassical approximation but probably this is the case even for smaller values of $N_1$.

In summary, although calculation of the width of the total dS momentum distribution for a macroscopic body is
a very difficult problem, FQT gives a reasonable qualitative explanation why this quantity is inversely
proportional to the mass of the body. With the estimation (\ref{estimation}), the result given by 
Eq. (\ref{preNewton2}) can be written in the form (\ref{Newton}) where 
\begin{equation}
G=\frac{2const\, Rln(2RN_1m_0)}{N_1m_0lnp}
\label{G}
\end{equation} 

In Chaps. \ref{Ch1} and \ref{Ch4} we argued that in theories based on dS invariance
neither the gravitational nor cosmological constant can be fundamental. In particular, in units $\hbar/2=c=1$, 
the dimension of $G$ is $length^2$ and its
numerical value is $l_P^2$ where $l_P$ is the Planck length ($l_P\approx 10^{-35}m$). Equation (\ref{G})
is an additional indication that this is the case since $G$ depends on $R$ and
there is no reason to think that it does not change with time. As noted in Sec. \ref{antigravity},
in our approach $\Lambda=3/R^2$ is only a formal parameter which has nothing to do with the curvature of dS space. However, since $G_{dS}=G\Lambda$ is dimensionless
in units $\hbar/2=c=1$, this quantity can be treated as the gravitational constant in dS theory. 
Let $\mu=2Rm_0$ be the dS nucleon mass. Then Eq. (\ref{G}) can be written as
\begin{equation}
G_{dS}=\frac{12const\, ln(N_1\mu )}{N_1\mu lnp}
\label{GLambda}
\end{equation} 
As noted in Sect. \ref{CC}, standard cosmological constant problem arises when one tries to explain the value of
$\Lambda$ from quantum theory of gravity assuming that this theory is QFT, $G$ is fundamental and dS symmetry is a manifestation of dark energy (or other fields) on flat Minkowski background. Such a theory contains strong divergences and the result depends on the
value of the cutoff momentum. With a reasonable assumption about this value, the quantity $\Lambda$ is of the order of
$1/G$ and this is reasonable since $G$ is the only parameter in this theory. Then $\Lambda$ is by more than 120 orders of
magnitude greater than its experimental value. However, in our approach we have an
additional fundamental parameter $p$. Equation (\ref{GLambda}) shows that $G\Lambda$ is
not of the order of unity but is very small since not only $p$ but even $ln p$ is very large. For a rough estimation, we assume that 
the values of $const$ and $N_1$ in this expression are of the order of unity. Then if, for example, $R$ is of the order of $10^{26}m$,
we have that $\mu$ is of the order of $10^{42}$ and $lnp$ is of the order of $10^{80}$. Therefore $p$ is a huge number of
the order of $exp(10^{80})$. In the preceding chapter we argued that standard theory can be treated as a special case of
FQT in the formal limit $p\to\infty$. The above discussion shows that restrictions on the width of the total dS
momentum arise because $p$ is not infinitely large. It is seen from Eq. (\ref{GLambda}) that gravity disappears in
the above formal limit. Therefore in our approach gravity is a consequence of the fact that dS symmetry is
considered over a finite field or ring rather than the field of complex numbers.

\chapter{Basic properties of AdS quantum theories}
\label{AdS}

As noted in Sec. \ref{dSvsAdS}, if one considers Poincare, AdS and dS symmetries in standard theory 
then only the latter symmetry does not contradict the possibility that gravity can be described in the
framework of a free theory. In addition, as shown in Secs. \ref{antigravity} and \ref{LambdaDiscrete}, 
the fact that $\Lambda > 0$
can be treated simply as an indication that among the three symmetries the dS one is the most pertinent
for describing nature. 

In standard theory the difference between IRs of the so(2,3) and so(1,4) algebras is that an IR of the
so(2,3) algebra where the operators $M^{\mu 4}$ ($\mu=0,1,2,3$)
are Hermitian can be treated as IRs of the so(1,4) algebra
where these operators are anti-Hermitian and vice versa. As
noted in Chap. \ref{Ch4}, in FQT a probabilistic interpretation is only approximate and
hence Hermiticy can be only a good approximation in some situations. 
Therefore one cannot exclude a possibility that elementary particles can be
described by modular analogs of IRs of the so(2,3) algebra while modular
representations describing symmetry of macroscopic bodies are modular analogs of standard
representations of the so(1,4) algebra.

In this chapter standard and modular IRs of the so(2,3) algebra are discussed in parallel in order
to demonstrate common features and differences between standard and modular cases.

\section{Modular IRs of the sp(2) and su(2) algebra}
\label{VS2}

The key role in constructing modular IRs of the so(2,3) algebra
is played by modular IRs of the sp(2) subalgebra. They are
described by a set of operators $(a',a'',h)$ satisfying the
commutation relations
\begin{equation}
[h,a']=-2a',\quad [h,a'']=2a'',\quad [a',a'']=h
\label{V2}
\end{equation}
The  Casimir operator of the second order for
the algebra (\ref{V2}) has the form
\begin{equation}
K=h^2-2h-4a''a'=h^2+2h-4a'a''
\label{V3}
\end{equation}
In general, representations of this algebra can be considered not only in spaces over $F_{p^k}$, where $k$ is a natural number,
but also in spaces over extensions of $R_p$.

We first consider representations with the
vector $e_0$ such that
\begin{equation}
a'e_0=0,\quad he_0=q_0e_0
\label{V4}
\end{equation}
where $q_0\in R_p$. Then it follows from Eqs.
(\ref{V3}) and (\ref{V4}), that \begin{equation}
he_n=(q_0+2n)e_n,\quad Ke_n=q_0(q_0-2)e_n,\quad  a'a''e_n=(n+1)(q_0+n)e_n
\label{V6}
\end{equation}

One can consider analogous representations in standard theory.
Then $q_0$ is a positive real number, $n=0,1,2,...$ and the
elements $e_n$ form a basis of the IR. In this case $e_0$ is a
vector with a minimum eigenvalue of the operator $h$ (minimum
weight) and there are no vectors with the maximum weight. The
operator $h$ is positive definite and bounded below by the
quantity $q_0$. For these reasons the above modular IRs can be
treated as modular analogs of such standard IRs that $h$ is
positive definite.

Analogously, one can construct modular IRs starting from the element $e_0'$ such that
\begin{equation}
a''e_0'=0,\quad he_0'=-q_0e_0'
\label{V7}
\end{equation}
and the elements $e_n'$ can be defined as $e_n'=(a')^ne_0'$.
Such modular IRs are analogs of standard IRs where $h$ is
negative definite. However, in the modular case Eqs.
(\ref{V4}) and (\ref{V7}) define the same IRs. This is clear from
the following consideration.

The set $(e_0,e_1,...e_N)$ will be a basis of IR if $a''e_i\neq 
0$ for $i<N$ and $a''e_N=0$. These conditions must be compatible with $a'a''e_N=0$. The case $q_0=0$ is of no interest
since, as follows from Eqs. (\ref{V4}-\ref{V7}), all the representation operators are null operators, the representation is 
one-dimensional and $e_0$ is the only basis vector in the representation space. If $q_0=1,...p-1$, it  
follows from Eq. (\ref{V6}) that $N$ is defined by the condition $q_0+N=0$. Hence $N=p-q_0$ and the dimension of IR equals 
\begin{equation}
Dim(q_0)=p-q_0+1
\label{Dim}
\end{equation}
This result is formally valid for all the values of $q_0$ if we treat $q_0$ as one of the numbers $1,...p-1, p$.
It is easy to see that $e_N$ satisfies Eq.
(\ref{V7}) and therefore it can be identified with $e_0'$. 

Let us forget for a moment that the eigenvalues of the operator
$h$ belong to $R_p$ and will treat them as integers. Then, as
follows from Eq. (\ref{V6}), the eigenvalues are
$$q_0,q_0+2,...,2p-2-q_0, 2p-q_0.$$
Therefore, if $f(q_0)>0$ and $f(q_0)\ll p$,
the maximum value of $q_0$ is $2p-q_0$, i.e. it is of the order of $2p$.

In standard theory, IRs are discussed in Hilbert spaces, i.e. the space of the IR is supplied by a positive definite
scalar product. It can be defined such that $(e_0,e_0)=1$, the operator $h$ is self-adjoint and
the operators $a'$ and $a''$ are adjoint to each other: $(a')^*=a''$. Then, as follows from Eq. (\ref{V6}),
\begin{equation}
(e_n,e_n)=n!(q_0)_n
\label{scalar}
\end{equation}
where we use the Pochhammer symbol $(q_0)_n=q_0(q_0+1)\cdots (q_0+n-1)$.
Usually the basis vectors are normalized to one. This is only a matter of convention but not a matter of
principle since not the probability itself but only ratios of probabilities have a physical meaning (see the
discussion in Chap. \ref{Ch4}). In FQT one can formally define the scalar product by the same formulas but
in that case this scalar product cannot be positive definite since in finite rings and fields the notions of positive and
negative numbers can be only approximate. Therefore (as noted in Chap. \ref{Ch4}) in FQT the probabilistic interpretation 
cannot be universal. 
However, if the quantities $q_0$ and $n$ are such that the r.h.s. of Eq. (\ref{scalar}) is much less than $p$
then the probabilistic interpretation is (approximately) valid if the IR is discussed in a space over
$R_{p^2}$ or $F_{p^2}$ (see Chap. \ref{Ch4} for a detailed discussion). Therefore if $p$ is very large, 
then for a large number of elements there is a correspondence between standard theory and FQT.

Representations of the su(2) algebra are defined by a set of operators $(L_+,L_-,L_3)$ satisfying the
commutations relations
\begin{equation}
[L_3,L_+]=2L_+,\quad [L_3,L_-]=-2L_-,\quad [L_+,L_-]=L_3
\label{su2CR}
\end{equation}
In the case of representations over the field of complex numbers, these relations can be formally
obtained from Eq. (\ref{V2}) by the replacements $h\to L_3$, $a'\to iL_-$ and $a"\to iL_+$. The difference between
the representations of the sp(2) and su(2) algebras in Hilbert spaces is that in the latter case the Hermiticity
conditions are $L_3^*=L_3$ and $L_+^*=L_-$. The Casimir operator for the algebra (\ref{su2CR}) is 
\begin{equation}
K=L_3^2-2L_3+4L_+L_-=L_3^2+2L_3+4L_-L_+
\label{su2Casimir}
\end{equation}

For constructing IRs, we assume that the representation space contains a vector $e_0$ such that 
\begin{equation}
L_3e_0=se_0\quad L_+e_0=0
\label{su2cyclic}
\end{equation}
where $s\geq 0$ for standard IRs and $s\in F_p$ for modular IRs. In the letter case we will denote $s$ by the
numbers $0,1,...p-1$. If $e_k=(L_-)^ke_0$ ($k=0,1,2,...$) then it is easy to see that
\begin{equation}
L_3e_k=(s-2k)e_k,\quad Ke_k=s(s+2)e_k,\quad L_+L_-e_k=(k+1)(s-k)e_k
\label{su2ek}
\end{equation}
The IR will be finite dimensional if there exists $k=k_{max}$ such that $L_+L_-e_k=0$ for this value of $k$.
As follows from the above expression, for modular IRs such a value of $k$ always exists, $k_{max}=s$ and
the dimension of the IR is $Dim(s)=s+1$. For standard IRs the same conclusion is valid if $s$ iz zero or a
natural number. In standard quantum theory, the representation operators of the su(2) algebra are associated with
the components of the angular momentum operator ${\bf L}=(L_x,L_y,L_z)$ such that $L_3=L_z$ and 
$L_{\pm}=(L_x\pm iL_y)/2$. The commutation relations for the components of ${\bf L}$ are usually written in
units where $\hbar=1$. Then $s$ can be only an integer or a half-integer and $Dim(s)=2s+1$. 

\section{Modular IRs of the so(2,3) Algebra}
\label{VS3}

Standard IRs of the so(2,3) algebra relevant for describing
elementary particles have been considered by several authors.
The description in this section is a combination of two elegant
ones given in Ref. \cite{Evans} for standard IRs and
Ref. \cite{Braden} for modular IRs. As already noted, in standard theory,
the commutation relations between the representation operators
are given by Eq. (\ref{newCR}) where $\eta^{44}=\pm 1$ for the AdS and dS
cases, respectively. As follows from the contraction procedure 
described in Sec. \ref{symmetry}, the operator $M^{04}$ can be treated as the AdS analog of
the energy operator.

If a modular IR is considered in a linear space over $F_{p^2}$
with $p=3\,\, (mod\,\, 4)$ then Eq. (\ref{newCR}) is also
valid. However, in the general case one can consider
modular IRs in linear spaces over any extension of $R_p$ or $F_p$. In this case it is convenient to work with another
set of ten operators. Let $(a_j',a_j'',h_j)$ $(j=1,2)$ be two
independent sets of operators satisfying the commutation
relations for the sp(2) algebra
\begin{equation}
[h_j,a_j']=-2a_j',\quad [h_j,a_j'']=2a_j'',\quad [a_j',a_j'']=h_j
\label{V9}
\end{equation}
The sets are independent in the sense that for different $j$
they mutually commute with each other. We denote additional
four operators as $b', b'',L_+,L_-$. The operators
$L_3=h_1-h_2,L_+,L_-$ satisfy the commutation relations (\ref{su2CR}) of the
su(2) algebra
while the other commutation relations are as follows
\begin{eqnarray}
&[a_1',b']=[a_2',b']=[a_1'',b'']=[a_2'',b'']=[a_1',L_-]=[a_1'',L_+]=\nonumber\\
&[a_2',L_+]=[a_2'',L_-]=0,\quad [h_j,b']=-b',\quad [h_j,b'']=b''\nonumber\\
&[h_1,L_{\pm}]=\pm L_{\pm},\quad [h_2,L_{\pm}]=\mp L_{\pm},\quad [b',b'']=h_1+h_2\nonumber\\
&[b',L_-]=2a_1',\quad [b',L_+]=2a_2',\quad [b'',L_-]=-2a_2'',\quad [b'',L_+]=-2a_1''\nonumber\\
&[a_1',b'']=[b',a_2'']=L_-,\quad [a_2',b'']=[b',a_1'']=L_+ \nonumber\\
&[a_1',L_+]=[a_2',L_-]=b',\quad [a_2'',L_+]=[a_1'',L_-]=-b''
\label{V11}
\end{eqnarray}
At first glance these relations might seem rather chaotic but
in fact they are very natural in the Weyl basis of the so(2,3)
algebra.

In spaces over $R_p+iR_p$ or $F_p+iF_p$ the
relation between the above sets of ten operators is
\begin{eqnarray}
&M_{10}=i(a_1''-a_1'-a_2''+a_2'),\quad M_{14}=a_2''+a_2'-a_1''-a_1'\nonumber\\
&M_{20}=a_1''+a_2''+a_1'+a_2', \quad M_{24}=i(a_1''+a_2''-a_1'-a_2') \nonumber\\
&M_{12}=L_3,\quad M_{23}=L_++L_-,\quad M_{31}=-i(L_+-L_-)\nonumber\\
&M_{04}=h_1+h_2,\quad M_{34}=b'+b'',\quad M_{30}=-i(b''-b')
\label{V12}
\end{eqnarray}
and therefore the sets are equivalent. However, the relations
(\ref{su2CR},\ref{V9},\ref{V11}) are more general since they can be used when
the representation space is a space over any extension of $R_p$ or $F_p$. 
It is also obvious that such a {\it definition} of the operators $M_{ab}$ is 
not unique. For example, any cyclic permutation of the indices
$(1,2,3)$ gives a new set of operators satisfying the same commutation relations.

In standard theory, the Casimir operator of the second order for the representation of the so(2,3) algebra
is given by 
\begin{equation}
I_2=\frac{1}{2}\sum_{ab} M_{ab}M^{ab}
\label{I2AdS}
\end{equation}
As follows from Eqs. (\ref{su2CR},\ref{V9}-\ref{V12}), $I_2$ can be written as
\begin{equation}
I_2=2(h_1^2+h_2^2-2h_1-4h_2-2b''b'+2L_-L_+-4a_1''a_1'-4a_2''a_2')
\label{I2B}
\end{equation}

We use the basis in which the operators $(h_j,K_j)$ $(j=1,2)$
are diagonal. Here $K_j$ is the Casimir operator (\ref{V3}) for the
algebra $(a_j',a_j",h_j)$. For constructing IRs we need
operators relating different representations of the
sp(2)$\times$sp(2) algebra. By analogy with Refs.
\cite{Evans,Braden}, one of the possible choices is as follows
\begin{eqnarray}
&A^{++}=b''(h_1-1)(h_2-1)-a_1''L_-(h_2-1)-a_2''L_+(h_1-1)
+a_1''a_2''b'\nonumber\\
&A^{+-}=L_+(h_1-1)-a_1''b',\quad
A^{-+}=L_-(h_2-1)-a_2''b',\quad A^{--}=b'
\label{V13}
\end{eqnarray}
We consider the action of these operators only on the space of
minimal sp(2)$\times$sp(2) vectors, i.e. such vectors
$x$ that $a_j'x=0$ for $j=1,2$, and $x$ is the eigenvector of
the operators $h_j$. If $x$ is a minimal vector such that
$h_jx=\alpha_jx$ then $A^{++}x$ is the minimal eigenvector of
the operators $h_j$ with the eigenvalues $\alpha_j+1$,
$A^{+-}x$ - with the eigenvalues $(\alpha_1+1,\alpha_2-1)$,
$A^{-+}x$ - with the eigenvalues $(\alpha_1-1,\alpha_2+1)$, and
$A^{--}x$ - with the eigenvalues $\alpha_j-1$.

By analogy with Refs. \cite{Evans,Braden}, we require
the existence of the vector $e_0$ satisfying the conditions
\begin{eqnarray}
&a_j'e_0=b'e_0=L_+e_0=0,\quad h_je_0=q_je_0\quad (j=1,2)
\label{V15}
\end{eqnarray}
where $q_j\in R_p$. As follows from Eq. (\ref{I2B}), in the IR characterized by the quantities $(q_1,q_2)$,
all the nonzero elements of the representation space are the eigenvectors of the operator $I_2$ with the 
eigenvalue
\begin{equation}
I_2=2(q_1^2+q_2^2-2q_1-4q_2)
\label{I2C}
\end{equation}

Since $L_3=h_1-h_2$ then, as follows from the results of Sec. \ref{VS2}, if $q_1$ and $q_2$ are characterized by the
numbers $0,1,...p-1$, $q_1\geq q_2$ and $q_1-q_2=s$ then the elements $(L_+)^ke_0$ ($k=0,1,...s$) form a basis of
the IR of the su(2) algebra with the spin $s$ such that the dimension of the IR is $s+1$. Therefore in finite theory the case when 
$q_1<q_2$ should be treated such that $s=p+q_1-q_2$. IRs with $q_1<q_2$ have no
analogs in standard theory and we will call them special IRs.

As follows from Eqs. (\ref{V9}) and (\ref{V11}), the operators
$(a_1',a_2',b')$ reduce the AdS energy $(h_1+h_2)$ by two units. Therefore
$e_0$ is an analog the state with the minimum energy which can
be called the rest state. For this reason we use $m_{AdS}$ to
denote $q_1+q_2$. In standard classification \cite{Evans}, the
massive case is characterized by the condition $q_2>1$ and the
massless one --- by the condition $q_2=1$. Hence in standard theory the 
quantity $m_{AdS}$ in the massive case is always greater than 2. 
There also exist two exceptional IRs discovered by Dirac \cite{DiracS} (Dirac
singletons). They are characterized by the conditions $(m_{AdS}=1,s=0)$ and 
$(m_{AdS}=2,s=1)$ or in terms of $(q_1,q_2)$, by the conditions $(q_1=1/2,q_2=1/2)$ and 
$(q_1=3/2,q_2=1/2)$, respectively.

In the theory over a Galois field or finite ring with odd $p$, $1/2$ should be treated as $(p+1)/2$ and 
$3/2$ --- as $(p+3)/2$. Hence the Dirac singletons are characterized by the conditions 
$(q_1=(p+1)/2,q_2=(p+1)/2)$ and $(q_1=(p+3)/2,q_2=(p+1)/2)$, respectively. In general, in this
theory it is possible that the quantities $(q_1,q_2)$ are given by the
numbers $2,3,...p-1$ but since $q_1+q_2$ is taken modulo $p$, it is possible that $m_{AdS}$ can take
one of the values (0,1,2). These cases also have no analogs in standard theory and we will 
call them special singleton IRs but will not treat the Dirac singletons as special.
In this section we will consider the massive case while the singleton, massless and special cases will be
considered in the next section.

As follows from the above remarks, the elements
\begin{equation}
e_{nk}=(A^{++})^n(A^{-+})^ke_0
\label{V16}
\end{equation}
represent the minimal sp(2)$\times$sp(2) vectors with the eigenvalues of the operators $h_1$ and $h_2$ equal to
$Q_1(n,k)=q_1+n-k$ and $Q_2(n,k)=q_2+n+k$, respectively. 

Consider the element $A^{--}A^{++}e_{nk}$. In view of the properties of the $A$ operators mentioned above,
this element is proportional to $e_{nk}$ and therefore one can write $A^{--}A^{++}e_{nk}=a(n,k)e_{nk}$.
One can directly verify that the actions of the operators $A^{++}$ and $A^{-+}$ on the space of minimal 
$sp(2)\times sp(2)$ vectors are commutative and therefore $a(n,k)$ does not depend on $k$.
A direct calculation gives
\begin{eqnarray}
&&(A^{--}A^{++}-A^{++}A^{--})e(n,k)=\{(Q_2-1)[Q_1-1)(Q_1+Q_2)-(Q_1-Q_2)]+\nonumber\\
&&(Q_1+Q_2-2)[n^2+k^2+n(q_1+q_2-3)-k(q_1-q_2+1)]\}e(n,k)
\label{A-A+}
\end{eqnarray}
where $Q_1\equiv Q_1(n,k)$ and $Q_2\equiv Q_2(n,k)$. As follows from this expression,
\begin{eqnarray}
&&a(n)-a(n-1)=q_1(q_2-1)(m_{AdS}-2)+2n(q_1^2+q_2^2+\nonumber\\
&&3q_1q_2-5q_1-4q_2+4)+6n^2(m_{AdS}-2)+4n^3
\label{an}
\end{eqnarray}
Since $b'e_0=0$ by construction, we have that $a(-1)=0$ and a direct calculation shows that, as a consequence of
Eq. (\ref{an})
\begin{equation}
\label{V17}
a(n)=(n+1)(m_{AdS}+n-2)(q_1+n)(q_2+n-1)
\end{equation}

Analogously, one can write $A^{+-}A^{-+}e_{nk}=b(k)e_{nk}$ and the result of a direct calculation is
\begin{equation}
b(k)=(k+1)(s-k)(q_1-k-2)(q_2+k-1)
\label{V18}
\end{equation}

As follows from these expressions, in the massive case $k$ can
assume only the values $0,1,...s$ and in standard theory
$n=0,1,...\infty$. However, in the modular case
$n=0,1,...n_{max}$ where $n_{max}$ is the first number for
which the r.h.s. of Eq. (\ref{V17}) becomes zero in $F_p$. Therefore $n_{max}=p+2-m_{AdS}$.

The full basis of the representation space can be chosen in the form
\begin{equation}
e(n_1n_2nk)=(a_1'')^{n_1}(a_2'')^{n_2}e_{nk}
\label{V19}
\end{equation}
In standard theory $n_1$ and $n_2$ can be any
natural numbers. However, as follows from the
results of the preceding section, Eq. (\ref{V9}) and
the properties of the $A$ operators,
\begin{eqnarray}
&n_1=0,1,...N_1(n,k),\quad n_2=0,1,...N_2(n,k)\nonumber\\
&N_1(n,k)=p-q_1-n+k,\quad N_2(n,k)=p-q_2-n-k
\label{V20}
\end{eqnarray}
As a consequence, the representation is finite dimensional in
agreement with the Zassenhaus theorem \cite{Zass} (moreover, it
is finite since any Galois field is finite).

Let us assume additionally that the representation space is
supplied by a scalar product (see Chap. \ref{Ch4}). The
element $e_0$ can always be chosen such that $(e_0,e_0)=1$.
Suppose that the representation operators satisfy the
Hermiticity conditions $L_+^*=L_-$, $a_j^{'*}=a_j''$,
$b^{'*}=b''$ and $h_j^*=h_j$. Then, as follows from Eq.
(\ref{V12}), in a special case when the representation space is
a space over $R_p+iR_p$ the
operators $M^{ab}$ are Hermitian as it should be. By using
Eqs. (\ref{V9}-\ref{V18}), one can show by a direct
calculation that the elements $e(n_1n_2nk)$ are mutually
orthogonal and the quantity
\begin{equation}
Norm(n_1n_2nk)=(e(n_1n_2nk),e(n_1n_2nk))
\label{V23}
\end{equation}
can be explicitly calculated. This quantity is an element of $R_p$ but below
it will be needed in the case when the representation space is over a field. For this
purpose it is convenient to represent this quantity as 
\begin{equation}
Norm(n_1n_2nk)=F(n_1n_2nk)G(nk)
\label{V24}
\end{equation}
where
\begin{eqnarray}
&F(n_1n_2nk)= n_1!(Q_1(n,k)+n_1-1)!n_2!(Q_2(n,k)+n_2-1)!\nonumber\\
&G(nk)=\{(q_2+k-2)!n!(m_{AdS}+n-3)!(q_1+n-1)!(q_2+n-2)!k!s!\}\nonumber\\
&\{(q_1-k-2)![(q_2-2)!]^3(q_1-1)!(m_{AdS}-3)!(s-k)!\nonumber\\
&[Q_1(n,k)-1][Q_2(n,k)-1]\}^{-1}
\label{V25}
\end{eqnarray}

In standard Poincare and AdS theories there also exist IRs with
negative energies. They can be constructed by analogy with
positive energy IRs. Instead of Eq. (\ref{V15}) one can
require the existence of the vector $e_0'$ such that
\begin{eqnarray}
&a_j''e_0'=b''e_0'=L_-e_0'=0,\quad h_je_0'=-q_je_0',\quad (e_0',e_0')\neq 0\quad (j=1,2)
\label{V26}
\end{eqnarray}
where the quantities $q_1,q_2$ are the same as for positive
energy IRs. It is obvious that positive and negative energy IRs
are fully independent since the spectrum of the operator
$M^{04}$ for such IRs is positive and negative, respectively.
However, {\it the modular analog of a positive energy IR
characterized by $q_1,q_2$ in Eq. (\ref{V15}), and the
modular analog of a negative energy IR characterized by the
same values of $q_1,q_2$ in Eq. (\ref{V26}) represent the
same modular IR.} This is the crucial difference between
standard quantum theory and FQT, and a proof is given below.

Let $e_0$ be a vector satisfying Eq. (\ref{V15}). Denote
$N_1=p-q_1$ and $N_2=p-q_2$. Our goal is to prove that the
vector $x=(a_1'')^{N_1}(a_2'')^{N_2}e_0$ satisfies the conditions
(\ref{V26}), \emph{i.e.} $x$ can be identified with $e_0'$.

As follows from the definition of $N_1,N_2$, the vector $x$ is
the eigenvector of the operators $h_1$ and $h_2$ with the
eigenvalues $-q_1$ and $-q_2$, respectively, and, in addition,
it satisfies the conditions $a_1''x=a_2''x=0$. Let us prove that
$b''x=0$. Since $b''$ commutes with the $a_j''$, we can write
$b''x$ in the form
\begin{equation}
b''x = (a_1'')^{N_1}(a_2'')^{N_2}b''e_0
\label{V27}
\end{equation}
As follows from Eqs. (\ref{V11}) and (\ref{V15}),
$a_2'b''e_0=L_+e_0=0$ and $b''e_0$ is the eigenvector of the
operator $h_2$ with the eigenvalue $q_2+1$. Therefore, $b''e_0$
is the minimal vector of the sp(2) IR which has the dimension
$p-q_2=N_2$. Hence $(a_2'')^{N_2}b''e_0=0$ and $b''x=0$.

The next stage of the proof is to show that $L_-x=0$.
As follows from Eq. (\ref{V11}) and the definition of $x$,
\begin{equation}
L_-x = (a_1'')^{N_1}(a_2'')^{N_2}L_-e_0-
N_1(a_1'')^{N_1-1}(a_2'')^{N_2}b''e_0
\label{V28}
\end{equation}
We have already shown that $(a_2'')^{N_2}b''e_0=0$, and therefore
it suffices to prove that the first term in the r.h.s. of
Eq. (\ref{V28}) is equal to zero. As follows from Eqs.
(\ref{V11}) and (\ref{V15}), $a_2'L_-e_0=b'e_0=0$, and $L_-e_0$
is the eigenvector of the operator $h_2$ with the eigenvalue
$q_2+1$. Therefore $(a_2'')^{N_2}L_-e_0=0$ and the proof is
completed.

Let us assume for a moment that the eigenvalues of the
operators $h_1$ and $h_2$ should be treated not as elements of
$R_p$ but as integers. Then, as follows from the consideration
in the preceding section, if $f(q_j)\ll p$ (j=1,2) then one
modular IR of the so(2,3) algebra corresponds to a standard
positive energy IR in the region where the energy is positive
and much less than $p$. At the same time, it corresponds to an
IR with the negative energy in the region where the AdS energy
is close to $4p$ but less than $4p$.

\section{Massless particles, Dirac singletons and special IRs}
\label{Singletons}

Those cases can be considered by analogy with the massive one. The case of Dirac singletons is especially simple. 
As follows from Eqs. (\ref{V17}) and (\ref{V18}), if $(m_{AdS}=1,s=0)$
then the only possible value of $k$ is $k=0$ and the only
possible values of $n$ are $n=0,1$ while if $(m_{AdS}=2,s=1)$ then
the only possible values of $k$ are $k=0,1$ and the only possible value of $n$ is $n=0$. 
This result does not depend on the value of $p$ and therefore it is valid in both, standard
theory and FQT. The only important difference between
standard and modular cases is that in the former $n_1,n_2=0,1,...\infty$ while in the latter the quantities
$n_1,n_2$ are in the range defined by Eq. (\ref{V20}). In the literature, the IR with $(m_{AdS}=2,s=1)$
is called Di and the IR with $(m_{AdS}=1,s=0)$ is called Rac.

The singleton IRs are indeed exceptional since the value of $n$ in them does not exceed 1 and therefore 
the impression is that singletons are two-dimensional objects, not three-dimensional
ones as usual particles. However, the singleton IRs have been obtained in the so(2,3) theory without reducing the algebra.
Dirac has titled his paper \cite{DiracS} "A Remarkable
Representation of the 3 + 2 de Sitter Group". Below we argue
that in FQT the singleton IRs are even more remarkable than in
standard theory.

First of all, as noted above, in standard theory there exist independent positive and negative IRs and the
latter are associated with antiparticles. In particular, in standard theory there exist four singleton IRs -
two IRs with positive energies and the corresponding IRs with negative energies, which can be called
antisingletons. However, at the end of the preceding section we have proved that in FQT one
IR contains positive and negative energy states simultaneously. This proof can be applied to the singleton
IRs without any changes. As a consequence, in the modular case there exist only two singleton IRs.

If $(m_{AdS}=1,s=0)$ then $q_1=q_2=1/2$ and, as noted in the preceding section, in FQT these relations
should be treated as $q_1=q_2=(p+1)/2$ where we assume that if the representation space is over a ring
then $p$ is odd. Analogously, if
$(m_{AdS}=2,s=1)$ then $(q_1=3/2,q_2=1/2)$ and in FQT
$(q_1=(p+3)/2,q_2=(p+1)/2)$. Therefore the values of $q_1$ and $q_2$ for the singleton
IRs are extremely large since they are of the order of $p/2$. As a consequence, the singleton IRs do not
contain states where all the quantum numbers are much less than $p$. Since some of the quantum numbers
are necessarily of the order of $p$, this is a natural explanation of the fact that singletons have not been
observed. In addition, as follows from the discussion in Chap. \ref{Ch4} and Secs. \ref{VS2} and \ref{VS3}, 
the fact that some quantum numbers are of the order of $p$ implies that 
the singletons cannot be described in terms of the probabilistic interpretation.

Note also that if we consider the singleton IRs as modular analogs of negative energy IRs then the
singleton IRs should be characterized either by $q_1=q_2=-1/2$ or by $q_1=-3/2,\,\, q_2=-1/2$. However,
since in FQT $-1/2=(p-1)/2$ and $-3/2=(p-3)/2$, those values are very close to ones characterizing
modular analogs of positive energy IRs. As a consequence, there is no approximation when singleton states
can be characterized as particles or antiparticles.

The Rac IR contains only minimal $sp(2)\times sp(2)$ vectors with $h_1=h_2=(p+1)/2$ and 
$h_1=h_2=(p+3)/2$ while the Di IR contains only minimal $sp(2)\times sp(2)$ vectors with 
$h_1=(p+3)/2,\,\, h_2=(p+1)/2$ and $h_1=(p+1)/2,\,\, h_2=(p+3)/2$. Hence it easily follows from Eq. (\ref{Dim})
that the dimensions of these IRs are equal to
\begin{equation}
Dim(Rac)=\frac{1}{2}(p^2+1)\quad Dim(Di)=\frac{1}{2}(p^2-1)
\label{DimDiRac}
\end{equation}
Additional arguments that in FQT singletons are exceptional are given in Sec. \ref{Matrix} and in
Chap.  \ref{DiracSingletons}.

Consider now the massless case when $q_2=1$. It follows from Eqs. (\ref{V17}) 
and (\ref{V18}) that $a(0)=0$ and $b(0)=0$. Therefore $A^{++}e_0=A^{-+}e_0=0$ and if the definition 
$e(n,k)=(A^{++})^n(A^{-+})^ke_0$ is used for $(n=0,1,...)$ and $(k=0,1,...)$ then
all the $e(n,k)$ will be the null elements. 

We first consider the case when $s\neq 0$ and $s\neq p-1$. In that case we define $e(1,0)$ not as $A^{++}e_0$ but as
$e(1,0)=[b"(h_1-1)-a_1"L_-]e_0$. A direct calculation using Eq. (\ref{V11}) shows that when $q_2=1$, this
definition is legitimate since $e(1,0)$ is the minimal $sp(2)\times sp(2)$ vector with the eigenvalues of the
operators $h_1$ and $h_2$ equal to $2+s$ and 2, respectively. With such a definition of $e(1,0)$, a
direct calculation using Eqs. (\ref{su2CR}) and (\ref{V11}) gives $A^{--}e(1,0)=b'e(1,0)=s(s+1)e_0$ and therefore
$e(1,0)\neq 0$. We now define $e(n,0)$ at $n\geq 1$ as $e(n,0)=(A^{++})^{n-1}e(1,0)$. Then Eq. (\ref{A-A+}) remains
valid when $n\geq 1$. Since $A^{++}b'e(1,0)=s(s+1)A^{++}e_0=0$, Eq. (\ref{an}) remains valid at $n=1,2,...$ and $a(0)=0$.
Hence we get
\begin{equation}
a(n)=n(n+1)(n+s+1)(n+s)\quad (n\geq 1)
\label{anmassless}
\end{equation}
As a consequence, the maximal value of $n$ in the modular case is $n_{max}=p-1-s$. This result has been obtained in
Ref. \cite{tmf}.

For analogous reasons, we now cannot define $e(0,k)$ as $(A^{-+})^ke_0$. However, if we define 
$e(0,k)=(L_-)^ke_0$ then, as follows from the discussion at the end of Sec. \ref{VS2}, the elements $e(0,k)$ ($k=0,1,...s$)
form a basis of the IR of the su(2) algebra with the spin $s$. Therefore the new definition of $e(0,k)$ is legitimate
since $e(0,k)$ is the minimal $sp(2)\times sp(2)$ vector with the eigenvalues of the
operators $h_1$ and $h_2$ equal to $1+s-k$ and $1+k$, respectively.

A direct calculation using Eqs. (\ref{su2CR}) and (\ref{V11}) gives that with the new definition of $e(0,k)$,
$A^{--}A^{++}e(0,k)=b'A^{++}e(0,k)=0$ and therefore $A^{++}e(0,k)=0$. When $1\leq k\leq s-1$, there is no way
to obtain nonzero minimal $sp(2)\times sp(2)$ vectors with the eigenvalues of the
operators $h_1$ and $h_2$ equal to $1+s-k+n$ and $1+k+n$, respectively, when $n>0$. However, when $k=s$, such vectors
can be obtained by analogy with the case $k=0$. We define $e(1,s)=[b''(h_2-1)-a_2''L_+]e(0,s)$. Then a direct calculation 
gives $b'e(1,s)=s(s+1)e(0,s)$ and therefore $e(1,s)\neq 0$. We now define $e(n,s)=(A^{++})^{n-1}e(1,s)$ for $n\geq 1$.
Then by analogy with the above discussion one can verify that if $A^{--}A^{++}e(n,s)=a(n)e(n,s)$ then $a(n)$ for 
$n\geq 1$ is again given by Eq. (\ref{anmassless}) and therefore in the modular case the maximal value of $n$ is
the same.

If $s=0$ then the only possible value of $k$ is $k=0$ and for the vectors $e(n,0)$ we have the same results
as above. In particular, Eq. (\ref{anmassless}) is valid with $s=0$. When $s=p-1$, we can define $e(n,0)$ and 
$e(n,s)$ as above but since $s+1=0\,\, (mod\,\, p)$, we get that $e(1,0)=e(1,s)=0$. This is in agreement with 
the above discussion since $n_{max}=0$ when $s=p-1$.
 
As shown in Sec. \ref{Weyl}, in dS theory there are no IRs which become Weil particles in Poincare limit. 
A problem arises whether the above results
can be treated as analogs of Weyl particles in standard and
modular versions of AdS invariant theory. In view of the relation $P^{\mu}=M^{4\mu}/2R$ (see Sec. \ref{symmetry}),
the AdS mass $m_{AdS}$ and the Poincare mass $m$ are related as $m=m_{AdS}/2R$. 
Since $m_{AdS}=2q_2+s$, the corresponding Poincare 
mass will be zero when $R\to\infty$ not only when $q_2=1$ but when $q_2$ is any
finite number. So a question arises why only the case $q_2=1$
is treated as massless. In Poincare invariant theory without spatial reflections massless particles
are characterized by the condition that they have a
definite helicity. In standard case the minimum value of the
AdS energy for massless IRs with positive energy is
$E_{min}=2+s$ when $n=0$. In contrast to the situation in
Poincare invariant theory, where massless particles cannot be
in the rest state, the massless particles in the AdS theory do
have rest states and, as shown above, the value of the $z$
projection of the spin in such states can be $-s,-s+2, ...s$ as
usual. However, we have shown that for any value of energy
greater than $E_{min}$, when $n\neq 0$, the spin state is
characterized only by helicity, which can take the values
either $s$ when $k=0$ or $-s$ when $k=s$, i.e. we have
the same result as in Poincare invariant theory. Note that in
contrast to IRs of the Poincare and dS algebras, standard IRs
describing particles in AdS invariant theory belong to the
discrete series of IRs and the energy spectrum in them is
discrete: $E=E_{min}, E_{min}+2, ...\infty$. Therefore,
strictly speaking, the rest states do not have measure zero as in
Poincare and dS invariant theories. 

Nevertheless, although the
probability that the energy is exactly $E_{min}$ is extremely
small, as a consequence of existence of rest states, one IR now contains states with both helicities,
$s$ and $-s$. Indeed, as noted above, the state $e(n,0)=(A^{++})^{n-1}e(1,0)$ for $n\geq 1$
has the energy $E=E_{min}+2n$ and helicity $s$. By acting by the operator $b'(A^{--})^{n-1}$ on $e(n,0)$
we obtain the element proportional to the rest state $e_0$. Then $(L_-)^se_0$ is the rest state $e(0,s)$.
Finally, as explained above, one can obtain $e(1,s)$ from $e(0,s)$ and $e(n,s)$ can be obtained as
$(A^{++})^{n-1}e(1,s)$. This state has the energy $E=E_{min}+2n$ and helicity $-s$. Therefore, as a
consequence of existence of rest states, states with the same energies but opposite helicities belong
to the same IR, and this is the case even if invariance under spatial reflection is not required. As a result,
in AdS theory there are no direct analogs of Weyl particles because massless IRs do not have only one
value of helicity.

A known case in Poincare theory is that if neutrino is massless then neutrino and antineutrino are
different particles with opposite helicities. However, in AdS theory, neutrino and antineutrino can be only
different states of the same particle. In experiment they manifest as different particles for the following reasons:
a) the probability to be in the rest state is extremely small; b) in Poincare limit the lepton number is strictly
conserved; c) in any weak reaction only states with definite helicities can take part because the interaction Hamiltonian 
contains projections on such states.  

Consider now dimensions of massless IRs. If $s=0$ then, as follows from the above results,
there exist only minimal $sp(2)\times sp(2)$ vectors with $h_1=h_2=1+n$, $n=0,1,...p-1$. Therefore, as
follows from Eq. (\ref{Dim}), the dimension of the massless IR with $s=0$ equals
\begin{equation}
Dim(s=0)=\sum_{n=0}^{p-1}(p-n)^2=\frac{1}{6}p(p+1)(2p+1)
\label{Dim0}
\end{equation}
If $s=1$, there exist only minimal $sp(2)\times sp(2)$ vectors with $(h_1=2+n,h_2=1+n)$ and 
$(h_1=1+n,h_2=2+n)$ where $n=0,1,...p-2$. Therefore
\begin{equation}
Dim(s=1)=2\sum_{n=0}^{p-2}(p-n)(p-n-1)=\frac{2}{3}p(p-1)(p+1)
\label{Dim1}
\end{equation}
If $s\geq 2$, there exist only minimal $sp(2)\times sp(2)$ vectors with $(h_1=1+s+n,h_2=1+n)$, 
$(h_1=1+n,h_2=1+s+n)$ where $n=0,1,...p-s$ and the minimal $sp(2)\times sp(2)$ vectors with 
$(h_1=1+s-k,h_2=1+k)$ where $k=1,...s-1$. Therefore, as
follows from Eq. (\ref{Dim})
\begin{eqnarray}
&&Dim(s\geq 2)=2\sum_{n=0}^{p-s}(p-n)(p-n-s)+\sum_{k=1}^{s-1}(p-k)(p-s+k)=\nonumber\\
&&\frac{p}{3}(2p^2-3s^2+1)+\frac{1}{2}s(s-1)(s+1)
\label{Dims}
\end{eqnarray}

As noted in Sec. \ref{VS3}, the cases of special IRs are such either $q_1$ and $q_2$
are represented by the numbers $0,1,..p-1$ and $q_1<q_2$ or in the case of special singletons,
$q_1,q_2=2,...p-1$ but $(q_1+q_2)\,\, (mod\,\, p)$ is one of the numbers (0,1,2). For example,
$(q_1=(p+1)/2,q_2=(p-1)/2)$ is a special singleton with $(m_{AdS}=0,s=1)$, 
$(q_1=(p+3)/2,q_2=(p-1)/2)$ is a special singleton with $(m_{AdS}=1,s=2)$ etc.
These cases can be investigated by analogy with massive IRs in Sec. \ref{VS3}. 
For reasons given in Sec. \ref{SS} and Chap. \ref{DiracSingletons}, among singleton IRs we will consider 
in detail only the Dirac singletons.
Then we will see that the only special IRs taking part in the decomposition
of the tensor product of the Dirac singletons are those with $q_1=0$. Then $s=p-q_2$. If $q_2=2,3,...p-1$ then, as follows
from Eq. (\ref{V17}), the quantum number $n$ can take only the value $n=0$. If $q_2=1$ then the special IR
can also be treated as the massless IR with $s=p-1$. As noted above, in this case the quantity $n$ also can take only
the value $n=0$. Let $Dim(q_1,q_2)$ be the dimension of the IR characterized by $q_1$ and $q_2$. Then, as follows
from Eq. (\ref{Dim})
\begin{equation}
Dim(0,q_2)=\sum_{k=0}^{p-q_2}(1+p-q_2-k)(1+k)=(1+p-q_2)^2+\frac{1}{2}(p-q_2)^2(1+p-q_2)
\label{dimspecial}
\end{equation}

\section{Matrix elements of representation operators}
\label{Matrix}

The matrix elements of the operator $A$ are defined as
\begin{equation}
Ae(n_1n_2nk)=\sum_{n_1'n_2'n'k'}
A(n_1'n_2'n'k';n_1n_2nk)e(n_1'n_2'n'k')
\label{V29}
\end{equation}
where the sum is taken over all possible values of $(n_1'n_2'n'k')$. The representation space of the
$so(2,3)$ algebra constructed above is obviously invariant under the action of the operators $(h_j,a_j',a_j'')$ ($j=1,2$)
by construction and the explicit result for the matrix elements of these operators is:
\begin{eqnarray}
&h_1e(n_1n_2nk)=[Q_1(n,k)+2n_1]e(n_1n_2nk)\nonumber\\
& h_2e(n_1n_2nk)=[Q_2(n,k)+2n_2]e(n_1n_2nk)
\label{V30}
\end{eqnarray}
\begin{eqnarray}
&a_1'e(n_1n_2nk)=n_1[Q_1(n,k)+n_1-1]e(n_1-1,n_2nk)\nonumber\\
&a_1''e(n_1n_2nk)=e(n_1+1,n_2nk)\nonumber\\
&a_2'e(n_1n_2nk)=n_2[Q_2(n,k)+n_2-1]e(n_1,n_2-1,nk)\nonumber\\
&a_2''e(n_1n_2nk)=e(n_1,n_2+1,nk)
\label{V31}
\end{eqnarray}

However, strictly speaking we have not proved yet that the representation space is invariant
under the action of the operators $(b',b'',L_+,L_-)$. In the massive case the explicit calculation
using Eqs. (\ref{su2CR},\ref{V9},\ref{V11},\ref{V16},\ref{V19}) 
gives
\begin{eqnarray}
&b''e(n_1n_2nk)=\{[Q_1(n,k)-1][Q_2(n,k)-1]\}^{-1}\}\nonumber\\
&[k(s+1-k)(q_1-k-1)(q_2+k-2)e(n_1,n_2+1,n,k-1)+\nonumber\\
&n(m_{AdS}+n-3)(q_1+n-1)(q_2+n-2)e(n_1+1,n_2+1,n-1,k)+\nonumber\\
&e(n_1,n_2,n+1,k)+e(n_1+1,n_2,n,k+1)]
\label{V32}
\end{eqnarray}
\begin{eqnarray}
&b'e(n_1n_2nk)=\{[Q_1(n,k)-1][Q_2(n,k)-1]\}^{-1}
[n(m_{AdS}+n-3)\nonumber\\
&(q_1+n-1)(q_2+n-2)(q_1+n-k+n_1-1)(q_2+n+k+n_2-1)\nonumber\\
&e(n_1n_2,n-1,k)+n_2(q_1+n-k+n_1-1)e(n_1,n_2-1,n,k+1)+\nonumber\\
&n_1(q_2+n+k+n_2-1)k(s+1-k)(q_1-k-1)(q_2+k-2)\nonumber\\
&e(n_1-1,n_2,n,k-1)+n_1n_2e(n_1-1,n_2-1,n+1,k)]
\label{V33}
\end{eqnarray}
\begin{eqnarray}
&L_+e(n_1n_2nk)=\{[Q_1(n,k)-1][Q_2(n,k)-1]\}^{-1}
\{(q_2+n+k+n_2-1)\nonumber\\
&[k(s+1-k)(q_1-k-1)(q_2+k-2)e(n_1n_2n,k-1)+\nonumber\\
&n(m_{AdS}+n-3)(q_1+n-1)(q_2+n-2)e(n_1+1,n_2,n-1,k)]+\nonumber\\
&n_2[e(n_1,n_2-1,n+1,k)+e(n_1+1,n_2-1,n,k+1)]\}
\label{V34}
\end{eqnarray}
\begin{eqnarray}
&L_-e(n_1n_2nk)=\{[Q_1(n,k)-1][Q_2(n,k)-1]\}^{-1}
\{n_1[k(s+1-k)\nonumber\\
&(q_1-k-1)(q_2+k-2)e(n_1-1,n_2n,k-1)+e(n_1-1,n_2,n+1,k)]\nonumber\\
&+(q_1+n-k+n_1-1)[e(n_1n_2n,k+1)+n(m_{AdS}+n-3)\nonumber\\
&(q_1+n-1)(q_2+n-2)e(n_1,n_2+1,n-1,k)]\}
\label{V35}
\end{eqnarray}
where we use a convention that $e(n_1n_2nk)$ is a null
vector if some of the numbers $(n_1n_2nk)$ are not in the range
described above. This result shows that in the massive case the representation can be selfconsistently constructed
only if it is over a field, not a ring. The analogous conclusion is valid in the massless case.

For the Rac singleton the only possible value of $k$ is $k=0$ and the only possible values of $n$ are $n=0,1$.
The basis consists of elements $e(n_1,n_2,0)=(a_1'')^{n_1}(a_2'')^{n_2}e_0$ and
$e(n_1,n_2,1)=(a_1'')^{n_1}(a_2'')^{n_2}b''e_0$. The result of explicit calculation is
\begin{eqnarray}
&&b'e(n_1,n_2,0)=n_1n_2e(n_1-1,n_2-1,1),\quad b''e(n_1,n_2,0)=e(n_1,n_2,1)\nonumber\\
&&L_+e(n_1,n_2,0)=n_2e(n_1,n_2-1,1),\quad L_-e(n_1,n_2,0)=n_1e(n_1-1,n_2,1)\nonumber\\
&&b'e(n_1,n_2,1)=(2n_1+1)(2n_2+1)e(n_1,n_2,0)\nonumber\\
&&b''e(n_1,n_2,1)=4e(n_1+1,n_2+1,0)\nonumber\\
&&L_+e(n_1,n_2,1)=2(2n_2+1)e(n_1+1,n_2,0)\nonumber\\
&&L_-e(n_1,n_2,1)=2(2n_1+1)e(n_1,n_2+1,0)
\label{MatrixRac}
\end{eqnarray}
For the Di singleton the only possible value of $n$ is $n=0$ and the only possible values of $k$ are $k=0,1$.
The basis consists of elements $e(n_1,n_2,0)=(a_1'')^{n_1}(a_2'')^{n_2}e_0$ and
$e(n_1,n_2,1)=(a_1'')^{n_1}(a_2'')^{n_2}L_-e_0$. The result of explicit calculation is
\begin{eqnarray}
&&b'e(n_1,n_2,0)=(1+2n_1)n_2e(n_1,n_2-1,1),\quad b''e(n_1,n_2,0)=2e(n_1+1,n_2,1)\nonumber\\
&&L_+e(n_1,n_2,0)=2n_2e(n_1+1,n_2-1,1),\quad L_-e(n_1,n_2,0)=(1+2n_1)e(n_1,n_2,1)\nonumber\\
&&b'e(n_1,n_2,1)=2n_1(n_2+1)e(n_1-1,n_2,0) ,\quad b''e(n_1,n_2,1)=2e(n_1,n_2+1,0)\nonumber\\
&&L_+e(n_1,n_2,1)=2(n_2+1)e(n_1,n_2,0)\nonumber\\
&&L_-e(n_1,n_2,1)=2n_1e(n_1-1,n_2+1,0)
\label{MatrixDi}
\end{eqnarray}
Therefore in FQT the additional exceptional feature of syngletons is that the representations for them can
be constructed over a ring, not necessarily over a field.

The important difference between standard and modular IRs is
that in the latter the trace of each representation operator is
equal to zero while in the former this is obviously not the
case (for example, the energy operator is positive definite for
IRs defined by Eq. (\ref{V15}) and negative definite for
IRs defined by Eq. (\ref{V26})). For the operators
$(a_j',a_j'',L_{\pm},b',b'')$ the validity of this statement is
clear immediately: since they necessarily change one of the
quantum numbers $(n_1n_2nk)$, they do not contain nonzero
diagonal elements at all. The proof for the diagonal operators
$h_1$ and $h_2$ follows. For each IR of the sp(2) algebra
with the "minimal weight" $q_0$ and the dimension $N+1$, the
eigenvalues of the operator $h$ are $(q_0,q_0+2,...q_0+2N)$.
The sum of these eigenvalues equals zero in $R_p$ since
$q_0+N=0$ in $R_p$ (see Sec. \ref{VS2}). Therefore we
conclude that for any representation operator $A$
\begin{equation}
\sum_{n_1n_2nk} A(n_1n_2nk,n_1n_2nk)=0
\label{V36}
\end{equation}
This property is very important for investigating a new
symmetry between particles and antiparticles in FQT
which is discussed in the subsequent section.

\section{Quantization and AB symmetry}
\label{VS4}

Let us first consider how the Fock space can be defined in standard
theory. As shown in Sec. \ref{VS3}, in the AdS case (in contrast to the situation in the
dS one) IRs with positive and negative energies are fully independent. Let $(n_1,n_2,n,k)$
be the set of all quantum numbers characterizing basis vectors of the IR and 
$a(n_1n_2nk)$ be the operator of particle
annihilation in the state described by the vector
$e(n_1n_2nk)$. Then the adjoint operator $a(n_1n_2nk)^*$ has
the meaning of particle creation in that state. Since we do not
normalize the states $e(n_1n_2nk)$ to one, we require that the
operators $a(n_1n_2nk)$ and $a(n_1n_2nk)^*$ should satisfy
either the anticommutation relations
\begin{eqnarray}
\{a(n_1n_2nk),a(n_1'n_2'n'k')^*\}=Norm(n_1n_2nk)
\delta_{n_1n_1'}\delta_{n_2n_2'}\delta_{nn'}\delta_{kk'}
\label{V37}
\end{eqnarray}
or the commutation relations
\begin{eqnarray}
[a(n_1n_2nk),a(n_1'n_2'n'k')^*]=Norm(n_1n_2nk)
\delta_{n_1n_1'}\delta_{n_2n_2'}\delta_{nn'}\delta_{kk'}
\label{V38}
\end{eqnarray}

A problem arises that in the case of negative energy IRs the operators
$a(n_1n_2nk)$ and $a(n_1n_2nk)^*$ have the meaning of the annihilation and creation operators,
respectively, for the states with negative energies and hence a question arises of whether such
operators are physical. An analogous problem for the dS case has been discussed in Sec.
\ref{InterpretationOfIRs}. One might think that since in the AdS case IRs with positive and negative
energies are fully independent, we can simply declare IRs with negative energies unphysical and
consider only IRs with positive energies. However, in QFT one cannot get rid of negative energy IRs
since here positive and negative energy IRs are combined together into a field satisfying a local
covariant equation. For example, the Dirac field combines together positive and negative energy IRs
into the Dirac field satisfying the Dirac equation.

For combining two IRs with positive and negative energies together, one can introduce a new quantum
number $\epsilon$ which will distinguish IRs with positive and negative energies; for example
$\epsilon=\pm 1$ for the positive and negative energy IRs, respectively. Then we have a set of
operators $a(n_1n_2nk,\epsilon)$ and $a(n_1n_2nk,\epsilon)^*$ such that by analogy with Eq.
(\ref{V37})
\begin{eqnarray}
\{a(n_1n_2nk,\epsilon),a(n_1'n_2'n'k',\epsilon')^*\}=Norm(n_1n_2nk)
\delta_{n_1n_1'}\delta_{n_2n_2'}\delta_{nn'}\delta_{kk'}\delta_{\epsilon\epsilon'}
\label{V37B}
\end{eqnarray}
and analogously in the case of commutators. The vacuum state ${\tilde \Phi}_{vac}$ can be
defined by the condition
\begin{equation}
a(n_1n_2nk,\epsilon){\tilde \Phi}_{vac}=0\quad \forall (n_1,n_2,n,k,\epsilon)
\label{Avac}
\end{equation}
As follows from Eqs. (\ref{V12}) and (\ref{V30}), the secondly quantized energy operator
has the form
\begin{equation}
M^{04}=\sum_{n_1n_2nk,\epsilon}\epsilon [m_{AdS}+2(n+n_1+n_2)]a(n_1n_2nk,\epsilon)^*a(n_1n_2nk,\epsilon)
\label{M04}
\end{equation}
and hence we have to solve the problem of the physical interpretation of the operators
$a(n_1n_2nk,-1)$ and $a(n_1n_2nk,-1)^*$. The two known ways of solving this problem follow.

In the spirit of Dirac's hole theory, one can define the new physical vacuum
\begin{equation}
\Phi_{vac}=\prod_{n_1n_2nk}a(n_1n_2nk,-1)^*{\tilde \Phi}_{vac}
\label{newphysvac}
\end{equation}
Then in the case of anticommutators  
each operator $a(n_1n_2nk,-1)$ creates a hole with a negative energy and 
the corresponding operator $a(n_1n_2nk,-1)^*$ annihilates this hole. Hence the operators
$a(n_1n_2nk,-1)^*$ can now be treated as the annihilation operators of states with positive
energies and the operators $a(n_1n_2nk,-1)$ --- as the creation operators of 
states with positive energies. A problem with such a treatment is that $\Phi_{vac}$ is the
eigenstate of the operator $M^{04}$ with the eigenvalue
\begin{equation}
{\cal E}_{vac}=-\sum_{n_1n_2nk}[m_{AdS}+2(n+n_1+n_2)]
\label{Ephysvac}
\end{equation} 
This is an infinite negative value and in quantum gravity a vacuum with an infinite energy is treated
as unacceptable. 

Another approach is that we consider only quantum numbers describing IRs with positive energies and,
in addition to the operators $a(n_1n_2nk)=a(n_1n_2nk,1)$ and $a(n_1n_2nk)^*=a(n_1n_2nk,1)^*$,
introduce new operators $b(n_1n_2nk)$ and $b(n_1n_2nk)^*$ instead of the 
operators $a(n_1n_2nk,-1)$ and $a(n_1n_2nk,-1)^*$ such that $b(n_1n_2nk)$ is proportional to
$a(n_1n_2nk,-1)^*$ and $b(n_1n_2nk)^*$ is proportional to $a(n_1n_2nk,-1)$. Then the $b$-operators
are treated as the annihilation operators of antiparticles with positive energies and the $b^*$
operators --- as the creation operators of antiparticles with positive energies. By
analogy with Eqs. (\ref{V37}) and (\ref{V38}), they should satisfy the relations
\begin{eqnarray}
\{b(n_1n_2nk),b(n_1'n_2'n'k')^*\}=Norm(n_1n_2nk)
\delta_{n_1n_1'}\delta_{n_2n_2'}\delta_{nn'}\delta_{kk'}
\label{V39}
\end{eqnarray}
\begin{eqnarray}
[b(n_1n_2nk),b(n_1'n_2'n'k')^*]=Norm(n_1n_2nk)
\delta_{n_1n_1'}\delta_{n_2n_2'}\delta_{nn'}\delta_{kk'}
\label{V40}
\end{eqnarray}
for anticommutation or commutation relations, respectively. In
this case it is assumed that in the case of anticommutation
relations all the operators $(a,a^*)$ anticommute with all the
operators $(b,b^*)$ while in the case of commutation relations
they commute with each other. It is also assumed that 
the vacuum vector $\Phi_0$ should satisfy the conditions
\begin{equation}
a(n_1n_2nk)\Phi_0=b(n_1n_2nk)\Phi_0=0\quad
\forall\,\, n_1,n_2,n,k
\label{V41}
\end{equation}
In QFT the second possibility is treated as more physical than that analogous
to Dirac's hole theory.

The Fock space in standard theory can now be defined as a
linear combination of all elements obtained by the action of
the operators $(a^*,b^*)$ on the vacuum vector, and the problem
of second quantization of representation operators can be
formulated as follows. Let $(A_1,A_2....A_n)$ be representation
operators describing IR of the AdS algebra. One should replace
them by operators acting in the Fock space such that the
commutation relations between their images in the Fock space
are the same as for original operators (in other words, we
should have a homomorphism of Lie algebras of operators acting
in the space of IR and in the Fock space). We can also require
that our map should be compatible with the Hermitian
conjugation in both spaces. It is easy to verify that a
possible solution satisfying all the requirements is as
follows. Taking into account the fact that the matrix elements
satisfy the proper commutation relations, the operators $A_i$
in the quantized form
\begin{eqnarray}
&A_i=\sum A_i(n_1'n_2'n'k',n_1n_2nk)
[a(n_1'n_2'n'k')^*a(n_1n_2nk)+\nonumber\\
&b(n_1'n_2'n'k')^*b(n_1n_2nk)]/Norm(n_1n_2nk)
\label{V42}
\end{eqnarray}
satisfy the commutation relations (\ref{su2CR},\ref{V9},\ref{V11}). Here the sum
is taken over all possible quantum numbers $(n_1',n_2',n',k',n_1,n_2,n,k)$. We will
not use special notations for operators in the Fock space since
in each case it will be clear whether the operator in question
acts in the space of IR or in the Fock space.

A known problem in standard theory is that the
quantization procedure does not define the order of the
annihilation and creation operators uniquely. For example,
another possible solution is
\begin{eqnarray}
&A_i=\mp \sum A_i(n_1'n_2'n'k',n_1n_2nk)
[a(n_1n_2nk)a(n_1'n_2'n'k')^*+\nonumber\\
&b(n_1n_2nk)b(n_1'n_2'n'k')^*]/Norm(n_1n_2nk)
\label{V43}
\end{eqnarray}
for anticommutation and commutation relations, respectively.
The solutions (\ref{V42}) and (\ref{V43}) are different since the
energy operators $M^{04}$ in these expressions differ by an
infinite constant. In standard theory the solution (\ref{V42})
is selected by imposing an additional requirement that all
operators should be written in the normal form where
annihilation operators precede creation ones. Then the vacuum
has zero energy and Eq. (\ref{V43}) should be rejected.
Such a requirement does not follow from the theory. Ideally
there should be a procedure which correctly defines the order
of operators from first principles.

In standard theory there also exist neutral particles. In that
case there is no need to have two independent sets of operators
$(a,a^*)$ and $(b,b^*)$, and Eq. (\ref{V42}) should be
written without the $(b,b^*)$ operators. The problem of neutral
particles in FQT is discussed in Sec. \ref{VS7}.

We now proceed to quantization in the modular case. The results
of Sec. \ref{VS3} show that one modular IR corresponds to two
standard IRs with positive and negative energies,
respectively. This indicates to a possibility that one modular
IR describes a particle and its antiparticle simultaneously.
However, we don't know yet what should be treated as a particle
and its antiparticle in the modular case. We have a description
of an object such that $(n_1n_2nk)$ is the full set of its
quantum numbers which take the values described in the
preceding sections.

We now assume that $a(n_1n_2nk)$ in FQT is the operator
describing annihilation of the object with the quantum numbers
$(n_1n_2nk)$ regardless of whether the numbers are physical or
nonphysical. Analogously $a(n_1n_2nk)^*$ describes creation of
the object with the quantum numbers $(n_1n_2nk)$. If these
operators anticommute then they satisfy Eq. (\ref{V37})
while if they commute then they satisfy Eq. (\ref{V38}).
Then, by analogy with standard case, the operators
\begin{eqnarray}
A_i=\sum A_i(n_1'n_2'n'k',n_1n_2nk)a(n_1'n_2'n'k')^*a(n_1n_2nk)/Norm(n_1n_2nk)
\label{V44}
\end{eqnarray}
satisfy the commutation relations (\ref{su2CR},\ref{V9},\ref{V11}). In this
expression the sum is taken over all possible values of the
quantum numbers in the modular case.

In the modular case the solution can be taken not only as in
Eq. (\ref{V44}) but also as
\begin{eqnarray}
A_i=\mp\sum A_i(n_1'n_2'n'k',n_1n_2nk)a(n_1n_2nk)a(n_1'n_2'n'k')^*/Norm(n_1n_2nk)
\label{V45}
\end{eqnarray}
for the cases of anticommutators and commutators, respectively.
However, as follows from Eqs. (\ref{V36}-\ref{V38}), the
solutions (\ref{V44}) and (\ref{V45}) are the same. Therefore in
the modular case there is no need to impose an artificial
requirement that all operators should be written in the normal form.

The problem with the treatment of the $(a,a^*)$ operators
follows.  When the values of $(n_1n_2n)$ are much less than
$p$, the modular IR corresponds to standard positive energy IR
and therefore the $(a,a^*)$ operator can be treated as those
describing the particle annihilation and creation,
respectively. However, when the AdS energy is negative, the
operators $a(n_1n_2nk)$ and $a(n_1n_2nk)^*$ become unphysical
since they describe annihilation and creation, respectively, in
the unphysical region of negative energies.

Let us recall that at any fixed values of $n$ and $k$, the
quantities $n_1$ and $n_2$ can take only the values described
in Eq. (\ref{V20}) and the eigenvalues of the operators
$h_1$ and $h_2$ are given by $Q_1(n,k)+2n_1$ and
$Q_2(n,k)+2n_2$, respectively. As follows from Eq. (\ref{Dim}) and the results of
Sec. \ref{VS3}, the first IR of the sp(2) algebra has the
dimension $N_1(n,k)+1$ and the second IR has the dimension
$N_2(n,k)+1$. If $n_1=N_1(n,k)$ then it follows from Eq.
(\ref{V20}) that the first eigenvalue is equal to $-Q_1(n,k)$ in
$F_p$, and if $n_2=N_2(n,k)$ then the second eigenvalue is
equal to $-Q_2(n,k)$ in $F_p$. We use ${\tilde n}_1$ to denote
$N_1(n,k)-n_1$ and ${\tilde n}_2$ to denote $N_2(n,k)-n_2$.
Then it follows from Eq. (\ref{V20}) that $e({\tilde
n}_1{\tilde n}_2nk)$ is the eigenvector of the operator $h_1$
with the eigenvalue $-(Q_1(n,k)+2n_1)$ and the eigenvector of
the operator $h_2$ with the eigenvalue $-(Q_2(n,k)+2n_2)$.

As noted above, standard theory involves the idea that creation of
the antiparticle with positive energy can be treated as
annihilation of the corresponding particle with negative energy
and annihilation of the antiparticle with positive energy can
be treated as creation of the corresponding particle with
negative energy. In FQT we also can define the operators $b(n_1n_2nk)$ and
$b(n_1n_2nk)^*$ in such a way that they will replace the
$(a,a^*)$ operators if the quantum numbers are unphysical. In
addition, if the values of $(n_1n_2n)$ are much less than $p$,
the operators $b(n_1n_2nk)$ and $b(n_1n_2nk)^*$ should be
interpreted as physical operators describing annihilation and
creation of antiparticles, respectively.

In FQT the $(b,b^*)$ operators cannot be independent of the
$(a,a^*)$ operators since the latter are defined for all
possible quantum numbers. Therefore the $(b,b^*)$ operators
should be expressed in terms of the $(a,a^*)$ ones. We can
implement the above idea if the operator $b(n_1n_2nk)$ is
defined in such a way that it is proportional to $a({\tilde
n}_1,{\tilde n}_2,n,k)^*$ and hence $b(n_1n_2nk)^*$ is
proportional to $a({\tilde n}_1,{\tilde n}_2,n,k)$.

Since we now consider massive and massless particles in FQT, and, as shown in Sec. \ref{Matrix}, representations for
them can be only over a field, Eq. (\ref{V25}) should now be considered in $F_p$.
Then  from the known Wilson theorem $(p-1)!=-1$ in $F_p$ \cite{VDW} it follows that
\begin{equation}
F(n_1n_2nk)F({\tilde n}_1{\tilde n}_2nk) = (-1)^s
\label{V46}
\end{equation}
We now define the $b$-operators as
\begin{equation}
a(n_1n_2nk)^*=\eta(n_1n_2nk) b({\tilde n}_1{\tilde n}_2nk)/
F({\tilde n}_1{\tilde n}_2nk)
\label{V47}
\end{equation}
where $\eta(n_1n_2nk)$ is some function. As a consequence,
\begin{eqnarray}
&a(n_1n_2nk)=\bar{\eta}(n_1n_2nk) b({\tilde n}_1{\tilde n}_2nk)^*/
F({\tilde n}_1{\tilde n}_2nk)\nonumber\\
&b(n_1n_2nk)^*=a({\tilde n}_1{\tilde n}_2nk)
F(n_1n_2nk)/{\bar \eta}({\tilde n}_1{\tilde n}_2nk)\nonumber\\
&b(n_1n_2nk)=a({\tilde n}_1{\tilde n}_2nk)^*
F(n_1n_2nk)/\eta({\tilde n}_1{\tilde n}_2nk)
\label{V48}
\end{eqnarray}

Equations (\ref{V47}) and (\ref{V48}) define a relation between
the sets $(a,a^*)$ and $(b,b^*)$. Although our motivation was
to replace the $(a,a^*)$ operators by the $(b,b^*)$ ones only
for the nonphysical values of the quantum numbers, we can
consider this definition for all the values of $(n_1n_2nk)$.
The transformation described by Eqs. ({\ref{V47}) and
(\ref{V48}) can also be treated as a special case of the
Bogolubov transformation discussed in a wide literature on
many-body theory (see e.g., Chap. 10 in Reference
\cite{Walecka} and references therein).

We have not discussed yet what exact definition of the physical
and nonphysical quantum numbers should be. This problem will be
discussed in Sec. \ref{VS5}. However, one might accept

{\it Physical-nonphysical states assumption: Each set of
quantum numbers $(n_1n_2nk)$ is either physical or unphysical.
If it is physical then the set $({\tilde n}_1{\tilde n}_2nk)$
is unphysical and vice versa.}

With this assumption we can conclude from Eqs. (\ref{V47})
and (\ref{V48}) that if some operator $a$ is physical then the
corresponding operator $b^*$ is unphysical and vice versa while
if some operator $a^*$ is physical then the corresponding
operator $b$ is unphysical and vice versa.

We have no ground to think that the set of the $(a,a^*)$
operators is more fundamental than the set of the $(b,b^*)$
operators and vice versa. Therefore the question arises whether
the $(b,b^*)$ operators satisfy the relations (\ref{V38}) or
(\ref{V39}) in the case of anticommutation or commutation
relations, respectively and whether the operators $A_i$ (see
Eq. (\ref{V44})) have the same form in terms of the
$(a,a^*)$ and $(b,b^*)$ operators. In other words, if the
$(a,a^*)$ operators in Eq. (\ref{V44}) are expressed in
terms of the $(b,b^*)$ ones then the problem arises whether
\begin{eqnarray}
A_i=\sum A_i(n_1'n_2'n'k',n_1n_2nk)b(n_1'n_2'n'k')^*b(n_1n_2nk)/Norm(n_1n_2nk)
\label{V49}
\end{eqnarray}
is valid. It is natural to accept the following

{\it Definition of the AB symmetry: If the $(b,b^*)$ operators
satisfy Eq. (\ref{V39}) in the case of anticommutators or
Eq. (\ref{V40}) in the case of commutators and all the
representation operators (\ref{V44}) in terms of the $(b,b^*)$
operators have the form (\ref{V49}) then it is said that the AB
symmetry is satisfied.}

To prove the AB symmetry we will first investigate whether
Eqs. (\ref{V39}) and (\ref{V40}) follow from Eqs.
(\ref{V37}) and (\ref{V38}), respectively. As follows from
Eqs. (\ref{V46}-\ref{V48}), Eq. (\ref{V39}) follows from
Eq. (\ref{V37}) if
\begin{equation}
\eta(n_1n_2nk) {\bar \eta}(n_1,n_2,nk)=(-1)^s
\label{V50}
\end{equation}
while Eq. (\ref{V40}) follows from Eq. (\ref{V38}) if
\begin{equation}
\eta(n_1n_2nk) {\bar \eta}(n_1,n_2,nk)=(-1)^{s+1}
\label{V51}
\end{equation}
We now represent $\eta(n_1n_2nk)$ in the form
\begin{equation}
\eta(n_1n_2nk)=\alpha f(n_1n_2nk)
\label{V52}
\end{equation}
where $f(n_1n_2nk)$ should satisfy the condition
\begin{equation}
f(n_1n_2nk) {\bar f}(n_1,n_2,nk)=1
\label{V53}
\end{equation}
Then $\alpha$ should be such that
\begin{equation}
\alpha {\bar \alpha}=\pm (-1)^s
\label{V54}
\end{equation}
where the plus sign refers to anticommutators and the minus
sign to commutators, respectively. If the normal
spin-statistics connection is valid, i.e. we have
anticommutators for odd values of $s$ and commutators for even
ones then the r.h.s. of Eq. (\ref{V54}) equals -1 while in
the opposite case it equals 1. In Sec. \ref{VS7}, Eq.
(\ref{V54}) is discussed in detail and for now we assume that
solutions of this relation exist.

A direct calculation using the explicit expressions
(\ref{V30}-\ref{V35}) for the matrix elements shows
that if $\eta(n_1n_2nk)$ is given by
Eq. (\ref{V52}) and
\begin{equation}
f(n_1n_2nk)=(-1)^{n_1+n_2+n}
\label{V55}
\end{equation}
then the AB symmetry is valid regardless of whether the normal
spin-statistics connection is valid or not.

\section{Physical and nonphysical states}
\label{VS5}

The operator $a(n_1n_2nk)$ can be the physical annihilation
operator only if it annihilates the vacuum vector $\Phi_0$.
Then if the operators $a(n_1n_2nk)$ and $a(n_1n_2nk)^*$ satisfy
the relations (\ref{V37}) or (\ref{V38}), the vector
$a(n_1n_2nk)^* \Phi_0$ has the meaning of the one-particle
state. The same can be said about the operators $b(n_1n_2nk)$
and $b(n_1n_2nk)^*$. For these reasons in standard theory it is
required that the vacuum vector should satisfy the conditions
(\ref{V41}). Then the elements
\begin{equation}
\Phi_+(n_1n_2nk)=a(n_1n_2nk)^*\Phi_0,\quad
\Phi_-(n_1n_2nk)=b(n_1n_2nk)^*\Phi_0
\label{V56}
\end{equation}
have the meaning of one-particle states for particles and
antiparticles, respectively.

However, if one requires the condition (\ref{V41}) in FQT, then
it is obvious from Eqs. (\ref{V47}) and (\ref{V48})
that the elements defined by Eq. (\ref{V56}) are null
vectors. Note that in standard approach the AdS energy is
always greater than $m_{AdS}$ while in FQT the AdS energy is not
positive definite. We can therefore try to modify Eq.
(\ref{V41}) as follows. Suppose that {\it Physical-nonphysical
states assumption} (see Sec. \ref{VS4}) can be substantiated.
Then we can break the set of elements $(n_1n_2nk)$ into two
nonintersecting parts with the same number of elements, $S_+$
and $S_-$, such that if $(n_1n_2nk)\in S_+$ then $({\tilde
n}_1{\tilde n}_2nk)\in S_-$ and vice versa. Then, instead of
the condition (\ref{V41}) we require
\begin{equation}
a(n_1n_2nk)\Phi_0=b(n_1n_2nk)\Phi_0=0\quad
\forall\,\, (n_1,n_2,n,k)\in S_+
\label{V57}
\end{equation}
In that case the elements defined by Eq. (\ref{V56}) will
indeed have the meaning of one-particle states for
$(n_1n_2nk)\in S_+$.

It is clear that if we wish to work with the full set of
elements $(n_1n_2nk)$ then, as follows from Eqs.
(\ref{V47}) and (\ref{V48}), the operators $(b,b^*)$ are
redundant and we can work only with the operators $(a,a^*)$.
However, if one works with the both sets, $(a,a^*)$ and
$(b,b^*)$ then such operators can be independent of each other
only for a half of the elements $(n_1n_2nk)$.

Regardless of how the sets $S_+$ and $S_-$ are defined, the
{\it Physical-nonphysical states assumption} cannot be
consistent if there exist quantum numbers $(n_1n_2nk)$ such
that $n_1={\tilde n}_1$ and $n_2={\tilde n}_2$. Indeed, in that
case the sets $(n_1n_2nk)$ and $({\tilde n}_1{\tilde n}_2nk)$
are the same what contradicts the assumption that each set
$(n_1n_2nk)$ belongs either to $S_+$ or $S_-$.

Since the replacements $n_1\rightarrow {\tilde n}_1$ and
$n_2\rightarrow {\tilde n}_2$ change the signs of the
eigenvalues of the $h_1$ and $h_2$ operators (see Sec.
\ref{VS4}), the condition that $n_1={\tilde n}_1$ and
$n_2={\tilde n}_2$ should be valid simultaneously implies that
the eigenvalues of the operators $h_1$ and $h_2$ should be
equal to zero simultaneously. Recall that (see Sec.
\ref{VS2}) if one considers IR of the sp(2) algebra and treats
the eigenvalues of the diagonal operator $h$ not as elements of
$R_p$ but as integers, then they take the values of
$q_0,q_0+2,...2p-q_0-2,2p-q_0$. Therefore the eigenvalue is
equal to zero in $R_p$ only if it is equal to $p$ when
considered as an integer. Since $m_{AdS}=q_1+q_2$ and the AdS energy
is $E=h_1+h_2$, the above situation can take place only if the
energy considered as an integer is equal to 2p. It now follows
from Eq. (\ref{V12}) that the energy can be equal to $2p$
only if $m_{AdS}$ is even. Since $s=q_1-q_2$, we conclude that $m_{AdS}$
can be even if and only if $s$ is even. In that case we will
necessarily have quantum numbers $(n_1n_2nk)$ such that the
sets $(n_1n_2nk)$ and $({\tilde n}_1{\tilde n}_2nk)$ are the
same and therefore the {\it Physical-nonphysical states
assumption} is not valid. On the other hand, if $s$ is odd
(\emph{i.e.} half-integer in the usual units) then there are no
quantum numbers $(n_1n_2nk)$ such that the sets $(n_1n_2nk)$
and $({\tilde n}_1{\tilde n}_2nk)$ are the same.

Our conclusion is as follows: {\it If the separation of states
should be valid for any quantum numbers then the spin $s$
should be necessarily odd.} In other words, if the notion of
particles and antiparticles is absolute then elementary
particles can have only a half-integer spin in the usual units.

In view of the above observations it seems natural to implement
the {\it Physical-nonphysical states assumption} as follows.
{\it If the quantum numbers $(n_1n_2nk)$ are such that
$m_{AdS}+2(n_1+n_2+n) < 2p$ then the corresponding state is physical
and belongs to $S_+$, otherwise the state is unphysical and
belongs to $S_-$.} However, one cannot guarantee that there are
no other reasonable implementations.

\section{AdS symmetry breaking}
\label{breaking}

In view of the above discussion, our next goal is the
following. We should take the operators in the form (\ref{V44})
and replace the $(a,a^*)$ operators by the $(b,b^*)$ ones only
if $(n_1n_2nk)\in S_-$. Then a question arises whether we will
obtain the standard result (\ref{V42}) where a sum is taken only
over values of $(n_1n_2nk)\in S_+$. The fact that we have
proved the AB symmetry does not guarantee that this is the case
since the AB symmetry implies that the replacement has been
made for all the quantum numbers, not only half of them.
However, the derivation of the AB symmetry shows that for the
contribution of such quantum numbers that $(n_1n_2nk)\in S_+$
and $(n_1'n_2'n'k')\in S_+$ we will indeed have the result
(\ref{V42}) up to some constants. This derivation also
guarantees that if we consider the action of the operators on
states described by physical quantum numbers and the result of
the action also is a state described by physical quantum
numbers then on such states the correct commutation relations
are satisfied. A problem arises whether they will be satisfied
for transitions between physical and nonphysical quantum
numbers.

Let $A(a_1^{'})$ be the secondly quantized operator
corresponding to $a_1^{'}$ and $A(a_1'')$ be the secondly
quantized operator corresponding to $a_1''$. Consider the
action of these operators on the state
$\Phi=a(n_1n_2nk)^*\Phi_0$ such that $(n_1n_2nk)\in S_+$ but
$(n_1+1,n_2nk)\in S_-$. As follows from Eqs. (\ref{V9}) and
(\ref{V30}), we should have
\begin{equation}
[A(a_1'),A(a_1'')]\Phi =[Q_1(n,k)+2n_1]\Phi
\label{correct}
\end{equation}
As follows from Eqs. (\ref{V31}) and (\ref{V47}),
$A(a_1'')\Phi=a(n_1+1,n_2nk)^*\Phi_0$. Since
$(n_1+1,n_2nk)\in S_-$, we should replace $a(n_1+1,n_2nk)^*$ by
an operator proportional to $b({\tilde n}_1-1,{\tilde n}_2nk)$
and then, as follows from Eq. (\ref{V41}),
$A(a_1'')\Phi=0$. Now, by using Eqs. (\ref{V31}) and
(\ref{V47}), we get
\begin{equation}
[A(a_1'),A(a_1'')]\Phi =n_1[Q_1(n,k)+n_1-1]\Phi
\label{incorrect}
\end{equation}
Equations (\ref{correct}) and (\ref{incorrect}) are
incompatible with each other and we conclude that our procedure
breaks the AdS symmetry for transitions between physical and
nonphysical states.

We conclude that if, by analogy with standard theory, one
wishes to interpret modular IRs of the dS algebra in terms of
particles and antiparticles then the commutation relations of
the dS algebra will be broken. This does not mean that such a
possibility contradicts the existing knowledge since they will
be broken only at extremely high dS energies of the order of $p$. At
the same time, a possible point of view is that since we
started from the symmetry algebra and treat the conditions
(\ref{newCR}) as a must, we should not sacrifice
symmetry because we don't know other ways of interpreting IRs.
So we have the following dilemma: {\it Either the notions of particles and
antiparticles are always valid and the commutation relations (\ref{newCR})
are broken at very large AdS energies of the order of p or the commutation relations (\ref{newCR})
are not broken and the notion of a particle and its antiparticle is only approximate.}
In the latter case such additive quantum numbers as the electric charge and the
baryon and lepton quantum numbers can be only approximately conserved.

\section{Dirac vacuum energy problem}
\label{VS6}

The Dirac vacuum energy problem is discussed in practically
every textbook on QFT. In its simplified form it can be
described as follows. Suppose that the energy spectrum is
discrete and $n$ is the quantum number enumerating the states.
Let $E(n)$ be the energy in the state $n$. Consider the
electron-positron field. As a result of quantization one gets
for the energy operator
\begin{equation}
E = \sum_n E(n)[a(n)^*a(n)-b(n)b(n)^*]
\label{V58}
\end{equation}
where $a(n)$ is the operator of electron annihilation in the
state $n$, $a(n)^*$ is the operator of electron creation in the
state $n$, $b(n)$ is the operator of positron annihilation in
the state $n$ and $b(n)^*$ is the operator of positron creation
in the state $n$. It follows from this expression that only
anticommutation relations are possible since otherwise the
energy of positrons will be negative. However, if
anticommutation relations are assumed, it follows from Eq.
(\ref{V58}) that
\begin{equation}
E = \{\sum_n E(n)[a(n)^*a(n)+b(n)^*b(n)]\}+E_0
\label{V59}
\end{equation}
where $E_0$ is some infinite negative constant. Its presence
was a motivation for developing Dirac's hole theory. In the
modern approach it is usually required that the vacuum energy
should be zero. This can be obtained by assuming that all
operators should be written in the normal form. However, this
requirement is not quite consistent since the result of
quantization is Eq. (\ref{V58}) where the positron
operators are not written in that form (see also the discussion
in Sec. \ref{VS4}).

Consider now the AdS energy operator $M^{04}=h_1+h_2$ in FQT.
As follows from Eqs. (\ref{V30}) and (\ref{V45})
\begin{eqnarray}
M^{04}=\sum [m_{AdS}+2(n_1+n_2+n)]a(n_1n_2nk)^*a(n_1n_2nk)/Norm(n_1n_2nk)
\label{V60}
\end{eqnarray}
where the sum is taken over all possible quantum numbers
$(n_1n_2nk)$. As noted in the preceding section, the two most well-known ways
of solving the problem of negative energies are either in the spirit of Dirac's
hole theory or by using the notion of antiparticles.

Consider first the second possibility. Then as follows from Eqs. (\ref{V46}-\ref{V48}) 
and (\ref{V52}-\ref{V54})
\begin{eqnarray}
&M^{04}=\{\sum_{S_+} [m+2(n_1+n_2+n)]
[a(n_1n_2nk)^*a(n_1n_2nk)+\nonumber\\
&b(n_1n_2nk)^*b(n_1n_2nk)]/Norm(n_1n_2nk)\}+{\cal E}_{vac}
\label{V61}
\end{eqnarray}
where the vacuum energy is given by
\begin{equation}
{\cal E}_{vac}=\mp \sum_{S_+}  [m_{AdS}+2(n_1+n_2+n)]
\label{V62}
\end{equation}
in the cases when the $(b,b^*)$ operators anticommute and
commute, respectively. For definiteness, we consider the case when 
the operators anticommute and therefore the sum in the r.h.s. of Eq. (\ref{V62}) is
taken with the minus sign.

In the approach similar to Dirac's hole theory one can define a new vacuum in FQT by analogy with Eq.
(\ref{newphysvac}):
\begin{equation}
\Phi_{vac}=\prod_{S_-}a(n_1n_2nk,-1)^*\Phi_0
\label{GFQTphysvac}
\end{equation}
where the product is taken over all the quantum numbers belonging to $S_-$. Then, as follows from the
definition of the sets $S_+$ and $S_-$, this vacuum will be the eigenstate of the operator $M^{04}$
with the the same eigenvalue ${\cal E}_{vac}$ as that given by Eq. (\ref{V62}) with the minus side in
the r.h.s. 

As noted in the dilemma at the end of the preceding section, in the approach involving the $b$ operators 
the commutation relations (\ref{newCR}) are necessarily broken at very large values of the AdS energy
while in the approach similar to Dirac's hole theory there is no need to introduce the $b$ operators.
In modern QFT the approach with the $b$ operators is treated as preferable since the
condition ${\cal E}_{vac}=0$ can be satisfied by imposing the (artificial) requirement
that all the operators should be written in the normal form while the in the approach similar to
Dirac's hole theory ${\cal E}_{vac}$ is necessarily an infinite negative constant. However, in FQT the
operators $a$ and $b$ are not independent and hence one cannot simply postulate that ${\cal E}_{vac}=0$.

Consider first the sum in Eq. (\ref{V62}) when the values
of $n$ and $k$ are fixed. It is convenient to distinguish the
cases $s > 2k$ and $s<2k$. If $s > 2k$ then, as follows from
Eq. (\ref{V20}), the maximum value of $n_1$ is such that
$m_{AdS}+2(n+n_1)$ is always less than $2p$. For this reason all the
values of $n_1$ contribute to the sum, which can be written as
\begin{eqnarray}
&S_1(n,k) =-\sum_{n_1=0}^{p-q_1-n+k}[(m_{AdS}+2n+2n_1)+\nonumber\\
&(m_{AdS}+2n+2n_1+2)+...+(2p-1)]
\label{V63}
\end{eqnarray}
A simple calculation shows that the result can be represented as
\begin{equation}
S_1(n,k)=\sum_{n_1=1}^{p-1}n_1^2-\sum_{n_1=1}^{n+(m_{AdS}-3)/2}n_1^2-
\sum_{n_1=1}^{(s-1)/2-k}n_1^2
\label{V65}
\end{equation}
where the last sum should be taken into account only if $(s-1)/2-k\geq 1$.

The first sum in this expression equals $(p-1)p(2p-1)/6$ and,
since we assume that $p\neq 2$ and $p\neq 3$, this quantity is
zero in $R_p$. As a result, $S_1(n,k)$ is represented as a sum
of two terms such that the first one depends only on $n$ and
the second --- only on $k$. Note also that the second term is
absent if $s=1$, i.e. for particles with the spin 1/2 in
the usual units.

Analogously, if $s < 2k$ the result is
\begin{equation}
S_2(n,k)=-\sum_{n_2=1}^{n+(m_{AdS}-3)/2}n_2^2-\sum_{n_2=1}^{k-(s+1)/2}n_2^2
\label{V67}
\end{equation}
where the second term should be taken into account only if
$k-(s+1)/2\geq 1$.

We now should calculate the sum
\begin{equation}
S(n)=\sum_{k=0}^{(s-1)/2}S_1(n,k) +\sum_{k=(s+1)/2}^s S_2(n,k)
\label{V68}
\end{equation}
and the result is
\begin{eqnarray}
&S(n)=-(s+1)(n+\frac{m_{AdS}-1}{2})[2(n+\frac{m_{AdS}-1}{2})^2-\nonumber\\
&3(n+\frac{m_{AdS}-1}{2})+1]/6-(s-1)(s+1)^2(s+3)/96
\label{V69}
\end{eqnarray}
Since the value of $n$ is in the range $[0,n_{max}]$, the final
result is
\begin{equation}
E_{vac}=\sum_{n=0}^{n_{max}}S(n)=(m_{AdS}-3)(s-1)(s+1)^2(s+3)/96
\label{V70}
\end{equation}
since in the massive case $n_{max}=p+2-m_{AdS}$.

Our final conclusion in this section is that {\it if $s$ is odd
and the separation of states into physical and nonphysical ones
is accomplished as in Sec. \ref{VS5} then $E_{vac}=0$ only if
$s=1$ (i.e. $s=1/2$ in the usual units)}. This result
shows that since the rules of arithmetic in Galois fields are
different from that for real numbers, it is possible that
quantities which are infinite in standard theory (e.g. the vacuum energy)
will be zero in FQT.

\section{Neutral particles and spin-statistics theorem}
\label{VS7}

In this section we discuss the relation between the
$(a,a^*)$ and $(b,b^*)$ operators only for all quantum numbers
(i.e. in the spirit of the AB-symmetry) and therefore
the results are valid regardless of whether the separation of
states into $S_+$ and $S_-$ can be justified or not (see the
discussion in Sec. \ref{breaking}). In other words, we treat the
set of the $(b,b^*)$ operators not necessarily as the one related to antiparticles
but simply as a set obtained from the $(a,a^*)$ operators by the transformation 
defined by Eqs. (\ref{V47}) and (\ref{V48}). 

The nonexistence of neutral elementary particles in FQT is one
of the most striking differences between FQT and standard
theory. One could give the following definition of neutral
particle:
\begin{itemize}
\item i) it is a particle coinciding with its
antiparticle
\item ii) it is a particle which does not coincide
with its antiparticle but they have the same properties
\end{itemize}
In standard theory only i) is meaningful since neutral
particles are described by real (not complex) fields and this
condition is required by Hermiticity. One might think that the
definition ii) is only academic since if a particle and its
antiparticle have the same properties then they are
indistinguishable and can be treated as the same. However, the
cases i) and ii) are essentially different from the operator
point of view. In the case i) only the $(a,a^*)$ operators are
sufficient for  describing the operators (\ref{V42}) in standard
theory. This is the reflection of the fact that the real field
has the number of degrees of freedom twice as less as the
complex field. On the other hand, in the case ii) both
$(a,a^*)$ and $(b,b^*)$ operators are required, i.e. in
standard theory such a situation is described by a complex
field. Nevertheless, the case ii) seems to be rather odd: it
implies that there exists a quantum number distinguishing a
particle from its antiparticle but this number is not
manifested experimentally. We now consider whether the
conditions i) or ii) can be implemented in FQT.

Since each operator $a$ is proportional to some operator $b^*$
and vice versa (see Eqs. (\ref{V47}) and (\ref{V48})), it is
clear that if the particles described by the operators
$(a,a^*)$ have a nonzero charge then the particles described by
the operators $(b,b^*)$ have the opposite charge and the number
of operators cannot be reduced. However, if all possible
charges are zero, one could try to implement i) by requiring
that each $b(n_1n_2nk)$ should be proportional to $a(n_1n_2nk)$
and then $a(n_1n_2nk)$ will be proportional to $a({\tilde
n}_1,{\tilde n}_2,nk)^*$. In this case the operators $(b,b^*)$
will not be needed at all.

Suppose, for example, that the operators $(a,a^*)$ satisfy the
commutation relations (\ref{V38}). In that case the operators
$a(n_1n_2nk)$ and $a(n_1'n_2'n'k')$ should commute if the sets
$(n_1n_2nk)$ and $(n_1'n_2'n'k')$ are not the same. In
particular, one should have $[a(n_1n_2nk), a({\tilde
n}_1{\tilde n}_2nk)]=0$ if either $n_1\neq {\tilde n}_1$ or
$n_2\neq {\tilde n}_2$. On the other hand, if $a({\tilde
n}_1{\tilde n}_2nk)$ is proportional to $a(n_1n_2nk)^*$, it
follows from Eq. (\ref{V38}) that the commutator cannot be
zero. Analogously one can consider the case of anticommutators.

The fact that the number of operators cannot be reduced is also
clear from the observation that the $(a,a^*)$ or $(b,b^*)$
operators describe an irreducible representation in which the
number of states (by definition) cannot be reduced. Our
conclusion is that in FQT the definition of neutral particle
according to i) is fully unacceptable.

Note that in standard theory there is a case of Majorana particles with spin 1/2. They are described by a real
equation and therefore are treated as neutral particles such that the equation describes a particle
and its antiparticle together and they are the same. However, from the point of view of IRs, Majorana
particles are simply neutral spin 1/2 particles. So the above discussion applies to such particles as well, and
they cannot be elementary.

Consider now whether it is possible to implement the definition
ii) in FQT. Recall that we started from the operators
$(a,a^*)$ and defined the operators $(b,b^*)$ by means of
Eq. (\ref{V47}). Then the latter satisfy the same
commutation or anticommutation relations as the former and the
AB symmetry is valid. Does it mean that the particles described
by the operators $(b,b^*)$ are the same as the ones described
by the operators $(a,a^*)$? If one starts from the operators
$(b,b^*)$ then, by analogy with Eq. (\ref{V47}), the
operators $(a,a^*)$ can be defined as
\begin{equation}
b(n_1n_2nk)^*=\eta'(n_1n_2nk) a({\tilde n}_1{\tilde n}_2nk)/
F({\tilde n}_1{\tilde n}_2nk)
\label{V72}
\end{equation}
where $\eta'(n_1n_2nk)$ is some function. By analogy with the
consideration in Sec. \ref{VS4} one can show that
\begin{equation}
\eta'(n_1n_2nk)=\beta (-1)^{n_1+n_2+n},\quad
\beta {\bar \beta}=\mp 1
\label{V73}
\end{equation}
where the minus sign refers to the normal spin-statistics
connection and the plus to the broken one.

As follows from Eqs. (\ref{V47}), (\ref{V50}-\ref{V53}),
(\ref{V72}), (\ref{V73}) and the definition of the quantities
${\tilde n}_1$ and ${\tilde n}_2$ in Sec. \ref{VS4}, the
relation between the quantities $\alpha$ and $\beta$ is $\alpha
{\bar \beta}=1$. Therefore, as follows from Eq.
(\ref{V73}), there exist only two possibilities, $\beta = \mp
\alpha$, depending on whether the normal spin-statistics
connection is valid or not. We conclude that the broken
spin-statistics connection implies that $\alpha{\bar
\alpha}=\beta{\bar\beta}=1$ and $\beta=\alpha$ while the normal
spin-statistics connection implies that $\alpha{\bar
\alpha}=\beta{\bar\beta}=-1$ and $\beta=-\alpha$. Since in the
first case there exist solutions such that $\alpha=\beta$ (e.g.
$\alpha = \beta = 1$), the particle and its antiparticle can be
treated as neutral in the sense of the definition ii). Since
such a situation is clearly unphysical, one might treat the Pauli
spin-statistics theorem \cite{Pauli2} as a requirement excluding neutral
particles in the sense ii).

We now consider another possible treatment of the
spin-statistics theorem, which seems to be much more
interesting. In the case of the normal spin-statistics connection $\alpha$ satisfies Eq. (\ref{V77}).
Such a relation is obviously impossible in standard theory.

As noted in Chap. \ref{Ch4}, $-1$ is a quadratic residue in
$F_p$ if $p=1\,\, (mod\,\, 4)$ and a quadratic non-residue in
$F_p$ if $p=3\,\, (mod\,\, 4)$. For example, $-1$ is a
quadratic residue in $F_5$ since $2^2=-1\,\, (mod\,\, 5)$ but
in $F_7$ there is no element $a$ such that $a^2=-1\,\, (mod\,\,
7)$. We conclude that if $p=1\,\, (mod\,\, 4)$ then Eq.
(\ref{V77}) has solutions in $F_p$ and in that case the theory
can be constructed without any extension of $F_p$.

Consider now the case $p=3\,\, (mod\,\, 4)$. Then Eq.
(\ref{V77}) has no solutions in $F_p$ and it is necessary to
consider this equation in an extension of $F_p$ (\emph{i.e.},
there is no "real" version of FQT). The minimum extension is
obviously $F_{p^2}$ and therefore the problem arises whether
Eq. (\ref{V77}) has solutions in $F_{p^2}$. As shown in Sec. \ref{S11}, this
equation does have solutions.

Our conclusion is that {\it if $p=3\,\, (mod\,\, 4 )$ then the
spin-statistics theorem implies that the field $F_p$ should
necessarily be  extended and the minimum possible extension is
$F_{p^2}$}. Therefore the spin-statistics theorem can be
treated as a requirement that if FQT is based on a field then it 
should be based on $F_{p^2}$
and standard theory should be based on complex numbers.

Let us now discuss a different approach to the AB symmetry. A
desire to have operators which can be interpreted as those
relating separately to particles and antiparticles is natural
in view of our experience in standard approach. However, one
might think that in the spirit of FQT there is no need to have
separate operators for particles and antiparticles since they
are different states of the same object. We can therefore
reformulate the AB symmetry in terms of only $(a,a^*)$
operators as follows. Instead of Eqs. (\ref{V47}) and
(\ref{V48}), we consider a {\it transformation} defined as
\begin{eqnarray}
&a(n_1n_2nk)^*\rightarrow \eta(n_1n_2nk)
a({\tilde n}_1{\tilde n}_2nk)/
F({\tilde n}_1{\tilde n}_2nk)\nonumber\\
&a(n_1n_2nk)\rightarrow \bar{\eta}(n_1n_2nk)
a({\tilde n}_1{\tilde n}_2nk)^*/
F({\tilde n}_1{\tilde n}_2nk)
\label{V78}
\end{eqnarray}
Then the AB symmetry can be formulated as a requirement that
physical results should be invariant under this transformation.

Let us now apply the AB transformation twice. Then we get
\begin{equation}
a(n_1n_2nk)^*\rightarrow \mp a(n_1n_2nk)^*,\quad
a(n_1n_2nk)\rightarrow \mp a(n_1n_2nk)
\label{V79}
\end{equation}
for the normal and broken spin-statistic connections,
respectively. Therefore, as a consequence of the
spin-statistics theorem, any particle (with the integer or
half-integer spin) has the (AB)$^2$ parity equal to $-1$.
Therefore in FQT any interaction can involve only an even
number of creation and annihilation operators. In particular,
this is additional demonstration of the fact that in FQT the
existence of neutral elementary particles is incompatible with
the spin-statistics theorem.

\section{Modular IRs of the osp(1,4) superalgebra}
\label{SS}

If one accepts supersymmetry then the results on modular IRs of
the so(2,3) algebra can be generalized by considering modular
IRs of the osp(1,4) superalgebra. Representations of the
osp(1,4) superalgebra have several interesting distinctions
from representations of the Poincare superalgebra. For this
reason we first briefly mention some known facts about the
latter representations (see e.g Ref. \cite{Wein-super} for
details).

Representations of the Poincare superalgebra are described by
14 operators. Ten of them are the representation
operators of the Poincare algebra---four momentum operators and
six representation operators of the Lorentz algebra, which
satisfy the commutation relations (\ref{PCR}). In addition,
there are four fermionic operators. The anticommutators
of the fermionic operators are linear combinations of the
momentum operators, and the commutators of the fermionic
operators with the Lorentz algebra operators are linear
combinations of the fermionic operators. In addition, the
fermionic operators commute with the momentum operators.

From the formal point of view, representations of the osp(1,4)
superalgebra are also described by 14 operators --- ten
representation operators of the so(2,3) algebra and four
fermionic operators. There are three types of relations: the
operators of the so(2,3) algebra commute with each other as
usual (see Sec. \ref{VS3}), anticommutators of the fermionic
operators are linear combinations of the so(2,3) operators and
commutators of the latter with the fermionic operators are
their linear combinations. However, in fact representations of
the osp(1,4) superalgebra can be described exclusively in terms
of the fermionic operators. The matter is as follows. In the
general case the anticommutators of four operators form ten
independent linear combinations. Therefore, ten bosonic
operators can be expressed in terms of fermionic ones. This is
not the case for the Poincare superalgebra since the Poincare
algebra operators are obtained from the so(2,3) one by
contraction. One can say that the representations of the
osp(1,4) superalgebra is an implementation of the idea that
supersymmetry is the extraction of the square root from the
usual symmetry (by analogy with the treatment of the
Dirac equation as a square root from the Klein-Gordon one).

We use $(d_1',d_2',d_1'',d_2'')$ to denote the fermionic operators of the osp(1,4) superalgebra. 
They should satisfy the following
relations. If $(A,B,C)$ are any fermionic operators, [...,...]
is used to denote a commutator and $\{...,...\}$ to denote an
anticommutator then
\begin{equation}
[A,\{ B,C\} ]=F(A,B)C + F(A,C)B
\label{S30}
\end{equation}
where the form $F(A,B)$ is skew symmetric, $F(d_j',d_j")=1$
$(j=1,2)$ and the other independent values of $F(A,B)$ are
equal to zero. The fact that the representation of the osp(1,4)
superalgebra is fully defined by Eq. (\ref{S30}) and the
properties of the form $F(.,.)$, shows that osp(1,4) is a
special case of the superalgebra.

We can now {\it define} the so(2,3) operators as
\begin{eqnarray}
&b'=\{d_1',d_2'\},\quad b''=\{d_1'',d_2''\},\quad
L_+=\{d_2',d_1''\},\quad L_-=\{d_1',d_2''\}\nonumber\\
&a_j'=(d_j')^2,\quad a_j''=(d_j'')^2,\quad
h_j=\{d_j',d_j''\} \quad (j=1,2)
\label{S31}
\end{eqnarray}
Then by using Eq. (\ref{S30}) and the properties of the
form $F(.,.)$, one can show by direct calculations that so
defined operators satisfy the commutation relations
(\ref{su2CR},\ref{V9},\ref{V11}). This result can be treated as a fact that
the operators of the so(2,3) algebra are not fundamental, only
the fermionic operators are.

By analogy with the construction of IRs of the osp(1,4)
superalgebra in standard theory \cite{Heidenreich}, we require
the existence of the generating vector $e_0$ satisfying the
conditions :
\begin{eqnarray}
d_j'e_0=d_2'd_1''e_0=0, \quad d_j'd_j''e_0=q_je_0\quad (j=1,2)
\label{S32A}
\end{eqnarray}
These conditions are written exclusively in terms of the $d$ operators. As follows from Eq. (\ref{S31}),
they can be rewritten as (compare with Eq. (\ref{V15}))
\begin{eqnarray}
d_j'e_0=L_+e_0=0, \quad h_je_0=q_je_0\quad (j=1,2)
\label{S32}
\end{eqnarray}
The full representation space can be obtained by successively
acting by the fermionic operators on $e_0$ and taking all
possible linear combinations of such vectors.

Let $E$ be an arbitrary linear combination of the
vectors $e_0$, $d_1''e_0$, $d_2''e_0$ and $d_2''d_1''e_0$. Our next goal
is to prove a statement analogous to that in Ref. \cite{Heidenreich}:

{\it Statement 1}: Any vector from the representation
space can be represented as a linear combination of the
elements $O_1O_2...O_nE$ where $n=0,1,...$ and $O_i$ is an operator
of the so(2,3) algebra.

The first step is to prove a simple

{\it Lemma:} If $D$ is any fermionic operator then DE is a
linear combination of elements $E$ and $OE$ where $O$ is an operator
of the so(2,3) algebra.

The proof is by a straightforward check using Eqs.
(\ref{S30}-\ref{S32}). For example,
$$d_1''(d_2''d_1''e_0)=\{d_1'',d_2''\}d_1''e_0-d_2''a_1''e_0=
b''d_1''e_0-a_1''d_2''e_0\,\, $$

To prove Statement 1 we define the height of a linear
combination of the elements $O_1O_2...O_nE$ as the maximum sum
of powers of the fermionic operator in this element. For
example, since each operator of the so(2,3) algebra is
composed of two fermionic operator, the height of the element
$O_1O_2...O_nE$ equals $2n+2$ if $E$ contains $d_2''d_1''e_0$,
equals $2n+1$ if $E$ does not contain $d_2''d_1''e_0$ but
contains either $d_1''e_0$ or $d_2''e_0$ and equals $2n$ if $E$
contains only $e_0$.

We can now prove Statement 1 by induction. The elements with
the heights 0, 1 and 2 obviously have the required form since,
as follows from Eq. (\ref{S31}),
$d_1''d_2''e_0=b''e_0-d_2''d_1''e_0$. Let us assume that
Statement 1 is correct for all elements with the heights $\leq
N$. Every element with the height $N+1$ can be represented as
$Dx$ where $x$ is an element with the height $N$. If
$x=O_1O_2...O_nE$ then by using Eq. (\ref{S30}) we can
represent $Dx$ as $Dx=O_1O_2...O_nDE+y$ where the height of the
element $y$ is $N-1$. As follows from the induction
assumption, $y$ has the required form, and, as follows from
Lemma, $DE$ is a linear combination of the elements $E$ and
$OE$. Therefore Statement 1 is proved.

As follows from Eqs. (\ref{S30}) and (\ref{S31}),
\begin{eqnarray}
[d_j',h_j]=d_j',\quad [d_j'',h_j]=-d_j'',\quad [d_j',h_l]=[d_j'',h_l]=0\quad (j,l=1,2\,\, j\neq l)
\label{S33}
\end{eqnarray}
It follows from these expressions that if $x$ is such that
$h_jx=\alpha_jx$ $(j=1,2)$ then $d_1''x$ is the eigenvector of
the operators $h_j$ with the eigenvalues
$(\alpha_1+1,\alpha_2)$, $d_2''x$ - with the eigenvalues
$(\alpha_1,\alpha_2+1)$, $d_1'x$ - with the eigenvalues
$(\alpha_1-1,\alpha_2)$, and $d_2'x$ - with the eigenvalues
$\alpha_1,\alpha_2-1$.

By analogy with the case of IRs of the so(2,3) algebra (see Sec. \ref{VS3}), 
we assume that $q_1$ and $q_2$
are represented by the numbers $0,1,...p-1$. We first consider the case when $q_2\geq 1$ and $q_1\geq q_2$.
We again use
$m_{AdS}$ to denote $q_1+q_2$ and $s$ to denote $q_1-q_2$. We first assume that $m_{AdS}\neq 2$ and $s\neq p-1$. 
Then Statement
1 obviously remains valid if we now assume that $E$ contains
linear combinations of $(e_0,e_1,e_2,e_3)$ where
\begin{eqnarray}
&&e_1=d_1''e_0,\quad e_2=[d_2''-\frac{1}{s+1}L_-d_1'']e_0\nonumber\\ 
&&e_3=(d_2''d_1''e_0-\frac{q_1-1}{m_{AdS}-2}b''+\frac{1}{m_{AdS}-2}a_1''L_-)e_0
\label{S34}
\end{eqnarray}

As follows from Eqs. (\ref{S30}-\ref{S33}), $e_0$
satisfies Eq. (\ref{V15}) and $e_1$ satisfies the same
condition with $q_1$ replaced by $q_1+1$. We see that the
representation of the osp(1,4) superalgebra defined by Eq.
(\ref{S32}) necessarily contains at least two IRs of the
so(2,3) algebra characterized by the values of the mass and
spin $(m_{AdS},s)$ and $(m_{AdS}+1,s+1)$ and the generating vectors $e_0$ and
$e_1$, respectively.

As follows from Eqs. (\ref{S30}-\ref{S33}), the vectors $e_2$ and $e_3$ satisfy the conditions
\begin{eqnarray}
&h_1e_2=q_1e_2,\quad h_2e_2=(q_2+1)e_2, \quad
h_1e_3=(q_1+1)e_3,\quad h_2e_3=(q_2+1)e_3\nonumber\\
&a_1'e_j=a_2'e_j=b'e_j=L_+e_j=0\quad (j=2,3)
\label{S36}
\end{eqnarray}
and therefore (see Eq. (\ref{V15})) they will be
generating vectors of IRs of the so(2,3) algebra if they are not equal to zero. 

If $s=0$ then, as follows from Eqs. (\ref{S30},\ref{S31},\ref{S34}), $e_2=0$. In the general case, as
follows from these expressions,
\begin{equation}
d_1'e_2=\frac{1-q_2}{s+1}L_-e_0,\quad d_2'e_2=\frac{s(q_2-1)}{s+1}e_0
\label{de2}
\end{equation}
Therefore $e_2$ is also a null vector if $e_0$ belongs to the massless IR (with $q_2=1$) while
$e_2\neq 0$ if $s\neq 0$ and $q_2\neq 1$. As follows from direct calculation using 
Eqs. (\ref{S30},\ref{S31},\ref{S34}) 
\begin{equation}
d_1'e_3=\frac{m_{AdS}-1}{m_{AdS}-2}[L_-d_1''-(2q_2+s-1)d_2'']e_0,\quad d_2'e_3=(q_2-\frac{q_1-1}{m_{AdS}-2})e_0
\label{de3}
\end{equation}
If $q_2=1$ then $d_1'e_3$ is proportional to $e_2$ (see Eq. (\ref{S34})) and hence $d_1'e_3=0$.
In this case $q_1-1=m_{AdS}-2$ and hence $d_2'e_3=0$. Therefore we conclude that $e_3=0$. It is also clear from Eq. (\ref{de3}) that $e_3=0$ if $m_{AdS}=1$. In all other cases $e_3\neq 0$.

Consider now the case $m_{AdS}=2$. If $s=0$ then $q_1=q_2=1$. The condition $e_2=0$ is still valid for the same
reasons as above but if $e_3$ is defined as $[d_2'',d_1'']e_0/2$ then $e_3$ is the minimal $sp(2)\times sp(2)$
vector with $h_1=h_2=2$ and, as a result of direct calculations using 
Eqs. (\ref{S30},\ref{S31},\ref{S34})
\begin{equation}
d_1'e_3=\frac{1}{2}(1-2q_1)d_2"e_0,\quad d_2'e_3=\frac{1}{2}(2q_2-1)e_0
\label{de3B}
\end{equation}
Hence in this case $e_3\neq 0$ and the IR of the osp(1,4) superalgebra corresponding to $(q_1,q_2)=(1,1)$
contains IRs of the so(2,3) algebra corresponding to $(1,1)$, $(2,1)$ and $(2,2)$. Therefore this 
IR of the osp(1,4) superalgebra should be treated as massive rather than massless.
 
At this point the condition that $q_1$ and $q_2$ are taken modulo $p$ has not been explicitly used and,
as already mentioned, our considerations are similar to those
in Ref. \cite{Heidenreich}. Therefore when $q_1\geq q_2$, modular IRs of the
osp(1,4) superalgebra can be characterized in the same way as conventional
IRs \cite{Heidenreich,F}:
\begin{itemize}
\item If $q_2>1$ and $s\neq 0$ (massive IRs), the osp(1,4)
supermultiplets contain four IRs of the so(2,3) algebra
characterized by the values of the mass and spin
$(m,s),(m+1,s+1),(m+1,s-1),(m+2,s)$.
\item If $q_2\geq 1$ and $s=0$ (collapsed massive IRs), the osp(1,4)
supermultiplets contain three IRs of the so(2,3) algebra
characterized by the values of the mass and spin
$(m,s),(m+1,s+1),(m+2,s)$.
\item If $q_2=1$ and $s=1,2,...p-2$ (massless IRs) the osp(1,4)
supermultiplets contains two IRs of the so(2,3) algebra
characterized by the values of the mass and spin
$(2+s,s),(3+s,s+1)$.
\item Dirac supermultiplet containing two Dirac
singletons (see Sec. \ref{Singletons}).
\end{itemize}

The first three cases have well-known analogs of IRs of the
super-Poincare algebra (see e.g., Ref. \cite{Wein-super})
while there is no super-Poincare analog of the Dirac
supermultiplet.

Since the space of IR of the superalgebra osp(1,4) is a direct
sum of spaces of IRs of the so(2,3) algebra, for modular IRs of
the osp(1,4) superalgebra one can prove results analogous to
those discussed in the preceding sections. In particular, one
modular IR of the osp(1,4) algebra is a modular analog of both
standard IRs of the osp(1,4) superalgebra with positive and
negative energies. This implies that one modular IR of the
osp(1,4) superalgebra contains both, a superparticle and its
anti-superparticle. 

At the same time, as noted in Sec. \ref{VS3}, there are special cases which have no analogs in standard theory.
The above results can be applied to those cases without any changes. For example, the special singleton
characterized by $(m_{AdS}=0,s)$, $s\neq 0$ generates a special supersingleton containing IRs of the so(2,3) algebra
with $(m_{AdS}=0,s)$, $(m_{AdS}=1,s+1)$, $(m_{AdS}=1,s-1)$ and $(m_{AdS}=2,s)$. In particular, when $s=1$ then
two of those IRs are the Di and Rac singletons. All other special
singletons also generate supersingletons containing more than two IRs of the so(2,3) algebra.
Hence the Dirac supersingleton can be treated as a more fundamental object than other special supersingletons.
For this reason, among supersingletons we will consider only the case of the Dirac supersingleton. Then
we will see below that the decomposition of the tensor product of the Dirac supersingletons  can contain only
special IRs of the osp(1,4) superalgebra with $q_1=0$. In this case we have that
$d_1'd_1''e_0=q_1e_0=0$, $d_2'd_1''e_0=L_+e_0=0$ and hence $d_1''e_0=0$.
Since $L_+d_2''e_0=d_1''e_0=0$ and $d_2'd_2''e_0=q_2e_0$, the vector $d_2''e_0$ is not zero and
if $e_0$ is the generating vector for the IR of the so(2,3) algebra with $(q_1=0,q_2)$ then
$d_2''e_0$ is the generating vector for the IR of the so(2,3) algebra with $(0,q_2+1)$. The IR
of the osp(1,4) superalgebra does not contain other IRs of the so(2,3) algebra since 
$d_2''d_1''e_0=0$ and $d_1''d_2''e_0=(d_1''d_2''+d_2''d_1'')e_0=b''e_0$.

By analogy with Sec. \ref{Singletons}, we use $SDim(s)$ to denote the dimension of the IR of the
osp(1,4) superalgebra in the massless case with the spin $s$ and $SDim(q_1,q_2)$ to denote the dimension 
of the IR of the osp(1,4) superalgebra characterized by the quantities $q_1$ and $q_2$. Then as follows
from the above discussion
\begin{eqnarray}
&&SDim(0,q_2)= Dim(0,q_2)+Dim(0,q_2+1)\quad (q_2=1,2,...p-1)\nonumber\\
&&SDim(s)=Dim(s)+Dim(s+1)\quad (s=1,2,...p-2)\nonumber\\
&&SDim(1,1)=Dim(1,1)+Dim(2,1)+Dim(2,2) 
\label{SDim}
\end{eqnarray}
and $Dim(p-1)=Dim(0,1)$.

\chapter{Dirac singletons as the only true elementary particles}
\label{DiracSingletons}

\section{Why Dirac singletons are indeed remarkable}
\label{elementary}

As already noted, Dirac singletons have been discovered by Dirac in his paper
\cite{DiracS} titled "A remarkable representation of the 3 + 2 de Sitter group".
In this section we argue that in FQT the Dirac singletons are even more remarkable than
in standard theory. As noted in Sec. \ref{VS3}, in the theory over a finite ring or field there
also exist special singleton-like IRs which have no analogs in standard theory.
As argued in Sec. \ref{SS}, from the point of view of supersymmetry they are less fundamental
than Dirac singletons. For this reason we will not consider such IRs and the term singleton will 
always mean the Dirac singleton.

As shown in Sec. \ref{VS3}, each IR of the so(2,3) algebra is characterized by the quantities
$(q_1,q_2)$. Consider a system of two particles such that the IR describing particle 1
is defined by the numbers $(q_1^{(1)},q_2^{(1)})$ and the IR describing particle 2 
is defined by the numbers $(q_1^{(2)},q_2^{(2)})$. The representation describing such a
system is the tensor product of the corresponding IRs defined as follows. Let
$\{e_i^{(1)}\}$ and $\{e_j^{(2)}\}$ be the sets of basis vectors for the IRs describing particle 1
and 2, respectively. Then the basis of the tensor product is formed by the 
elements $e_{ij}=e_i^{(1)}\times e_j^{(2)}$. Let $\{O_k^{(1)}\}$ and $\{O_l^{(2)}\}$ 
$(k,l=1,2,...10)$ be the sets of independent representation operators in the
corresponding IRs. Then the set of independent representation operators in the tensor
product is $\{O_k=O_k^{(1)}+O_k^{(2)}\}$. Here it is assumed that the
operator with the superscript $(j)$ acts on the elements $e_k^{(j)}$ in the same way as 
in the IR $j$ while on the elements $e_l^{(j')}$ where $j'\neq j$ it acts as the 
identity operator. For example,
$$h_1\sum_{ij}c_{ij}(e_i^{(1)}\times e_j^{(2)})=\sum_{ij}c_{ij}[(h_1^{(1)}e_i^{(1)})\times e_j^{(2)}+e_i^{(1)}\times (h_1^{(2)}e_j^{(2)})]$$
Then the operators $\{O_k\}$ satisfy the same commutation relations as in Eqs. 
(\ref{su2CR}), (\ref{V9}) and (\ref{V11}).

It is immediately clear from this definition that the tensor product of IRs
characterized by $(q_1^{(1)},q_2^{(1)})$ and $(q_1^{(2)},q_2^{(2)})$, respectively, 
contains at least the IR characterized by 
$(q_1=q_1^{(1)}+q_1^{(2)},q_2=q_2^{(1)}+q_2^{(2)})$.
Indeed, if $e_0^{(j)}$ $(j=1,2)$ is the generating vector for IR $j$ then the
vector $e_0=e_0^{(1)}\times e_0^{(2)}$ will satisfy Eq. (\ref{V15}). 

In Standard Model
(based on Poincare invariance) only massless particles are treated as elementary.
However, as shown in the seminal paper by Flato and Fronsdal \cite{FF}
(see also Ref. \cite{HeidenreichS}), in standard AdS theory each massless IR can be 
constructed from the tensor product of two singleton IRs and the authors of Ref. \cite{FF}
believe that this is a truly remarkable property. In general, in standard theory an IR characterized by $(q_1,q_2)$
can be constructed from tensor products of two IRs characterized by $(q_1^{(1)},q_2^{(1)})$ 
and $(q_1^{(2)},q_2^{(2)})$ if $q_1\geq (q_1^{(1)}+q_1^{(2)})$ and
$q_2\geq (q_2^{(1)}+q_2^{(2)})$. Since no interaction is assumed, a problem arises
whether a particle constructed from a tensor product of other two particles will be
stable. In standard theory a particle with the mass $m$ can be a stable composite
state of two particles with the masses $m_1$ and $m_2$ only if $m<(m_1+m_2)$ and
the quantity $(m_1+m_2-m)c^2$ is called the binding energy. The greater the binding
energy is the more stable is the composite state with respect to external interactions.

The authors of Ref. \cite{FF} and other works treat singletons as true elementary 
particles because their weight diagrams has only a single trajectory (that's why
the corresponding IRs are called singletons) and in AdS QFT 
singleton fields live on the boundary at infinity of the AdS bulk (boundary which
has one dimension less than the bulk). However, in that case one should
answer the following questions:
\begin{itemize}
\item a) Why singletons have not been observed yet.
\item b) Why such massless particles as photons and others are stable and their decays into singletons have not been observed.
\end{itemize}
There exists a wide literature 
(see e.g. Ref. \cite{FFS,Bekaert} and references therein) where this problem is investigated
from the point of view of standard AdS QFT. However, as noted in Sec. \ref{ST},
the physical meaning of field operators is not clear and products of local quantized
fields at the same points are not well defined.

In addition, the following question arises. Each massless boson (e.g. the photon) 
can be constructed from a tensor product of either two Dis or two Racs. Which of those 
possibilities (if any) is physically preferable? 
A natural answer is as follows. If the theory is supersymmetric then the AdS algebra should be extended 
to the superalgebra osp(1,4) which has only one positive energy IR combining Di and Rac into the Dirac supermultiplet.
For the first time, this possibility has been discussed probably in Refs. \cite{F,Heidenreich}. Therefore in standard theory 
there exists only one Dirac superparticle and its
antiparticle.

As shown in the preceding chapter, in FQT one IRs describes a particle and its
antiparticle simultaneously and hence in FQT there exists only one IR describing the
supersingleton. In addition, as shown in Sec. \ref{Singletons}, while dimensions of
massless IRs are of the order of $p^3$ (see Eqs. (\ref{Dim0}-\ref{Dims})), the dimensions
of the singleton IRs are of the order of $p^2$ (see Eq. (\ref{DimDiRac})) and, as follows
from Eq. (\ref{DimDiRac}), the dimension of the supersingleton IR is $p^2$. These
facts can be treated as arguments that in FQT the supersingleton can be the only
elementary particle. In Chap. \ref{Discussion} we argue that,
in contrast to standard theory, in FQT one can give natural explanations of a) and b). 
In addition, as shown in Sec. \ref{Matrix}, while massive and massless particles in FQT can
be described only over a field, the singletons can be also described over a ring.  

The chapter is organized as follows.  In Sec. \ref{semiclassical} we discuss in
detail how usual particles and singletons should be discussed in the Poincare and semiclassical limits of
standard theory. In Sec. \ref{singletontensorproduct} it is shown that, in contrast to
standard theory, the tensor products of singleton IRs in FQT contain not only massless 
IRs but also
special IRs, which have no analogs in standard theory. Beginning from 
Sec. \ref{supersingleton} we proceed to
the supersymmetric case, and the main result of the chapter is described in Sec. \ref{supertensorprod}. 
Here we explicitly find a complete list of IRs taking part in the decomposition of the tensor product
of two supersingletons. In standard theory the known results are recovered while in FQT this
list also contains special supersymmetric IRs which have no analogs in standard theory.

\section{Tensor product of modular IRs of the sp(2) algebra}
\label{tensorprod}

Consider two modular IRs of the sp(2) algebra in spaces $H_j$ ($j=1,2$). 
Each IR is defined by a set 
of operators $(h^{(j)},a^{(j)'},a^{(j)''})$ satisfying the commutation 
relations (\ref{V2}) and by 
a vector $e_0^{(j)}$ such that (see Eq. (\ref{V4}))
\begin{equation}
a^{(j)'}e_0^{(j)}=0,\quad h^{(j)}e_0=q_0^{(j)}e_0^{(j)}
\label{e0j}
\end{equation}
As follows from the results of the preceding section, the 
vectors $e_n^{(j)}=(a^{(j)''})^ne_0^{(j)}$ where
$k=0,1,...N^{(j)}$ and $N^{(j)}=p-q_0^{(j)}$ form a basis in $H_j$.

The tensor product of such IRs can be defined by analogy with the definition of
the tensor product of IRs of the so(2,3) algebra in the preceding section.
The basis of the representation space is formed by
the elements $e_{kl}=e_k^{(1)}\times e_l^{(2)}$ and the independent representation operators are $(h,a',a'')$ such
that $h=h^{(1)}+h^{(2)}$, $a'=a^{(1)'}+a^{(2)'}$ and $a''=a^{(1)''}+a^{(2)''}$. 
Then the operators $(h,a',a'')$ satisfy the same commutation relations as in Eq. 
(\ref{V2}) and hence they implement
a representation of the sp(2) algebra in the space $H_1\times H_2$. Our goal is to find a decomposition of this
representation into irreducible components. 

It is obvious that the cases when $q_0^{(1)}=0$ or $q_0^{(2)}=0$ are trivial and therefore we will assume
that $q_0^{(1)}\neq 0$ and $q_0^{(2)}\neq 0$. If $q_0^{(1)}$ and $q_0^{(2)}$ are
represented by the numbers $(1,2,...p-1)$ then we suppose that $q_0^{(1)}\geq q_0^{(2)}$ and consider the vector 
\begin{equation}
e(k)=\sum_{i=0}^kc(i,k)(e_i^{(1)}\times e_{k-i}^{(2)})
\label{ek}
\end{equation}
As follows from Eq. (\ref{V6}) and the definition of $h$, 
\begin{equation}
he(k)=(q_0^{(1)}+q_0^{(2)}+2k)e(k)
\label{hek}
\end{equation}
Therefore if $a'e(k)=0$ then the vector $e(k)$ generates a modular IR with the dimension 
$Dim(q_0^{(1)},q_0^{(2)},k)=p+1-(q_0^{(1)}-q_0^{(2)}-2k)$ where $q_0^{(1)}-q_0^{(2)}-2k$ is taken
modulo $p$. As follows from Eqs. (\ref{V6}) and (\ref{ek}),
\begin{equation}
a'e(k)=\sum_{i=0}^kc(i,k)[i(q_0^{(1)}+i-1)(e_{i-1}^{(1)}\times e_{k-i}^{(2)})+
(k-i)(q_0^{(2)}+k-i-1)(e_i^{(1)}\times e_{k-i-1}^{(2)})]
\label{hek2}
\end{equation}
This condition will be satisfied if
\begin{equation}
c(i,k)=C_k^i [\prod_{l=1}^i (q_0^{(2)}+k-i)][ \prod_{l=i}^k (q_0^{(1)}+l)]
\label{cik}
\end{equation}
It is clear from this expression that in standard case the possible values of $k$ are $0,1,...\infty$ while
in modular case $k=0,1,...k_{max}$ where $k_{max}=p-q_0^{(1)}$.

It is obvious that at different values of $k$, the IRs generated by $e(k)$ are linearly independent and
therefore the tensor product of the IRs generated by $e_0^{(1)}$ and $e_0^{(2)}$ contains all the IRs generated by $e(k)$.
A question arises whether the latter IRs give a full decomposition of the tensor product. This is the case
when the dimension of the tensor product equals the sum of dimensions of the IRs generated by $e(k)$. 
Below we will be interested in the tensor product of singleton IRs and, as shown in Sec. \ref{Singletons}, in that case
$q_0^{(1)}+q_0^{(2)}>p$. Therefore $q_0^{(1)}+q_0^{(2)}+2k\in [q_0^{(1)}+q_0^{(2)},2p-q_0^{(1)}+q_0^{(2)}]$
and for all values of $k$, $q_0^{(1)}+q_0^{(2)}+2k$ is in the range $(p,2p]$. Then, as follows from Eq. (\ref{Dim}),
the fact that the IRs 
generated by $e(k)$ give a full decomposition of the tensor product follows from the relation
\begin{equation}
\sum_{k=0}^{p-q_0^{(1)}}(2p+1-q_0^{(1)}-q_0^{(2)}-2k)=(p+1-q_0^{(1)})(p+1-q_0^{(2)})
\label{decomp}
\end{equation}

\section{Semiclassical approximation in Poincare limit}
\label{semiclassical}

The Flato-Fronsdal result \cite{FF} poses a fundamental 
question whether only singletons can be true elementary particles.
In the present work we consider singletons in the framework of FQT but the approach is
applicable in standard theory (over the complex numbers) as well. As already noted in 
Secs. \ref{Singletons} and \ref{Matrix}, the properties of singletons in standard theory and FQT are considerably different. 
In this chapter and Chap. 
\ref{Discussion} we argue that in FQT the singleton physics is even more interesting than in standard theory. 
However, since there exists a wide literature on singleton properties in standard theory, in the present section
we discuss what conclusions can be made about semiclassical approximation and Poincare limit for singletons in this theory.  

First we consider the case of massive and massless particles.
Since spin is a pure quantum phenomenon, one might expect that
in semiclassical approximation it suffices to consider the spinless case. 
Then, as shown in Sec. \ref{VS3}, the
quantum number $k$ can take only the value $k=0$, the basis vectors of the IR can be chosen as
$e(n_1n_2n)=(a_1'')^{n_1}(a_2'')^{n_2}e_n$ (compare with Eq. (\ref{V19})) where (see Eq. (\ref{V6}))
$e_n=(A^{++})^ne_0$. In the spinless case, $q_1=q_2=m/2$ and hence 
Eqs. (\ref{V30}-\ref{V36}) can be rewritten in the form:
\begin{eqnarray}
h_1e(n_1n_2n)=[Q+2n_1]e(n_1n_2n),\quad
h_2e(n_1n_2n)=[Q+2n_2]e(n_1n_2n)
\label{sem1}
\end{eqnarray}
\begin{eqnarray}
&a_1'e(n_1n_2n)=n_1[Q+n_1-1]e(n_1-1,n_2n),\,\, a_1''e(n_1n_2n)=e(n_1+1,n_2n)\nonumber\\
&a_2'e(n_1n_2n)=n_2[Q+n_2-1]e(n_1,n_2-1,n),\,\, a_2''e(n_1n_2n)=e(n_1,n_2+1,n)
\label{sem2}
\end{eqnarray}
\begin{eqnarray}
&b''e(n_1n_2n)=\frac{Q-2}{Q-1}n(m_{AdS}+n-3)e(n_1+1,n_2+1,n-1)+\nonumber\\
&\frac{1}{(Q-1)^2}e(n_1,n_2,n+1)
\label{sem3}
\end{eqnarray}
\begin{eqnarray}
&b'e(n_1n_2n)=\frac{Q-2}{Q-1}n(m_{AdS}+n-3)(Q+n_1-1)(Q+n_2-1)e(n_1,n_2,n-1)+\nonumber\\
&\frac{n_1n_2}{(Q-1)^2}e(n_1-1,n_2-1,n+1)
\label{sem4}
\end{eqnarray}
\begin{eqnarray}
&L_+e(n_1n_2n)=\frac{Q-2}{Q-1}n(m_{AdS}+n-3)(Q+n_2-1)e(n_1+1,n_2,n-1)+\nonumber\\
&\frac{n_2}{(Q-1)^2}e(n_1,n_2-1,n+1)
\label{sem5}
\end{eqnarray}
\begin{eqnarray}
&L_-e(n_1n_2n)=\frac{Q-2}{Q-1}n(m_{AdS}+n-3)(Q+n_1-1)e(n_1,n_2+1,n-1)+\nonumber\\
&\frac{n_1}{(Q-1)^2}e(n_1-1,n_2,n+1)
\label{sem6}
\end{eqnarray}
where $Q=Q(n)=m_{AdS}/2+n$.

The basis elements $e(n_1n_2n)$ are not normalized to one and in our special case
the results given by Eqs. (\ref{V23}-\ref{V25}) can be represented as
\begin{eqnarray}
&||e(n_1n_2n)||=F(n_1n_2n)=\{n!(m_{AdS}-2)_n[(\frac{m_{AdS}}{2})_n]^3
(\frac{m_{AdS}}{2}-1)_n\nonumber\\
&n_1!n_2!
(\frac{m_{AdS}}{2}+n)_{n_1}(\frac{m_{AdS}}{2}+n)_{n_2}\}^{1/2}
\label{fn1n2n}
\end{eqnarray}
By using this expression, Eqs. (\ref{sem1}-\ref{sem6}) can be rewritten in terms of the matrix
elements of representation operators with respect to the normalized basis 
${\tilde e}(n_1n_2n)=e(n_1n_2n)/F(n_1n_2n)^{1/2}$.

Each element of the representation space can be written as
$$x=\sum_{n_1n_2n}c(n_1n_2n){\tilde e}(n_1n_2n)$$
where $c(n_1n_2n)$ can be called the WF in the $(n_1n_2n)$ representation. 
It is normalized as 
$$\sum_{n_1n_2n}|c(n_1n_2n)|^2=1$$
In standard theory the quantum numbers $n_1$ and $n_2$ are in the range $[0,\infty )$ and for massive and
massless particles the quantum number $n$ also is in this range.  
By using Eqs. (\ref{sem1}-\ref{fn1n2n}), one can obtain the action of the representation operator on the
WF $c(n_1n_2n)$: 
\begin{eqnarray}
&h_1c(n_1n_2n)=[m_{AdS}/2+n+2n_1]c(n_1n_2n)\nonumber\\
&h_2c(n_1n_2n)=[m_{AdS}/2+n+2n_2]c(n_1n_2n)\nonumber\\
&a_1'c(n_1n_2n)=[(n_1+1)(m_{AdS}/2+n+n_1)]^{1/2}c(n_1+1,n_2n)\nonumber\\
&a_1"c(n_1n_2n)=[n_1(m_{AdS}/2+n+n_1-1)]^{1/2}c(n_1-1,n_2n)\nonumber\\
&a_2'c(n_1n_2n)=[(n_2+1)(m_{AdS}/2+n+n_2)]^{1/2}c(n_1,n_2+1,n)\nonumber\\
&a_2"c(n_1n_2n)=[n_2(m_{AdS}/2+n+n_2-1)]^{1/2}c(n_1,n_2-1,n)\nonumber\\
&b"c(n_1n_2n)=[\frac{n(m_{AdS}+n-3)(m_{AdS}/2+n+n_1-1)(m_{AdS}/2+n+n_2-1)}
{(m_{AdS}/2+n-1)(m_{AdS}/2+n-2)}]^{1/2}c(n_1,n_2,n-1)+\nonumber\\
&[\frac{n_1n_2(n+1)(m_{AdS}+n-2)}{(m_{AdS}/2+n)(m_{AdS}/2+n-1)}]^{1/2}c(n_1-1,n_2-1,n+1)\nonumber\\
&b'c(n_1n_2n)=[\frac{(n+1)(m_{AdS}+n-2)(m_{AdS}/2+n+n_1)(m_{AdS}/2+n+n_2)}{(m_{AdS}/2+n)(m_{AdS}/2+n-1)}]^{1/2}c(n_1,n_2,n+1)+\nonumber\\
&[\frac{(n_1+1)(n_2+1)n(m_{AdS}+n-3)}{(m_{AdS}/2+n-1)(m_{AdS}/2+n-2)}]^{1/2}c(n_1+1,n_2+1,n-1)\nonumber\\
&L_+c(n_1n_2n)=[\frac{(n+1)(m_{AdS}+n-2)n_1(m_{AdS}/2+n+n_2)}{(m_{AdS}/2+n)(m_{AdS}/2+n-1)}]^{1/2}c(n_1-1,n_2,n+1)+\nonumber\\
&[\frac{(n_2+1)n(m_{AdS}+n-3)(m_{AdS}/2+n+n_1-1)}{(m_{AdS}/2+n-1)(m_{AdS}/2+n-2)}]^{1/2}c(n_1,n_2+1,n-1)\nonumber\\
&L_-c(n_1n_2n)=[\frac{n(m_{AdS}+n-3)(n_1+1)(m_{AdS}/2+n+n_2-1)}{(m_{AdS}/2+n-1)(m_{AdS}/2+n-2)}]^{1/2}c(n_1+1,n_2,n-1)+\nonumber\\
&[\frac{n_2(n+1)(m_{AdS}+n-2)(m_{AdS}/2+n+n_1)}{(m_{AdS}/2+n)(m_{AdS}/2+n-1)}]^{1/2}c(n_1,n_2-1,n+1)
\label{sem12}
\end{eqnarray}

As noted in Sec. \ref{symmetry}, the contraction
to the Poincare invariant case can be performed as follows. If $R$ is a parameter with the dimension $length$
and the operators $P_{\mu}$ ($\mu=0,1,2,3$) are defined as $P_{\mu}=M_{\mu 4}/2R$ then in the formal limit
when $R\to\infty$, $M_{\mu 4}\to\infty$ but the ratio $M_{\mu 4}/R$ remains finite, one gets the commutation
relations of the Poincare algebra from the commutation relations of the so(2,3) algebra. Therefore in situations
where Poincare limit is valid with a high accuracy, the operators $M_{\mu 4}$ are much greater than the other
operators. The quantum numbers $(m_{AdS},n_1,n_2,n)$ should be very large since in the formal limit $R\to\infty$,
$m_{AdS}/2R$ should become standard Poincare mass and the quantities $(n_1/2R,n_2/2R,n/2R)$ should become
continuous momentum variables.

A typical form of the semiclassical WF is 
$$c(n_1,n_2,n)=a(n_1,n_2,n)exp[i(n_1\varphi_1+n_2\varphi_2+n\varphi)]$$
where the amplitude $a(n_1,n_2,n)$ has a sharp maximum at semiclassical values of $(n_1,n_2,n)$. Since the
numbers $(n_1,n_2,n)$ are very large, when some of them change by one, the major change of $c(n_1,n_2,n)$
comes from the rapidly oscillating exponent. As a consequence, in
semiclassical approximation each representation operator becomes the 
operator of multiplication by a function and, as follows from Eqs. (\ref{V12},\ref{sem12})
\begin{eqnarray}
&M_{04}=m_{AdS}+2(n_1+n_2+n)\quad M_{12}=2(n_1-n_2)\nonumber\\
&M_{10}=2[n_1(m_{AdS}/2+n+n_1)]^{1/2}sin\varphi_1-2[n_2(m_{AdS}/2+n+n_2)]^{1/2}sin\varphi_2\nonumber\\
&M_{20}=2[n_1(m_{AdS}/2+n+n_1)]^{1/2}cos\varphi_1+2[n_2(m_{AdS}/2+n+n_2)]^{1/2}cos\varphi_2\nonumber\\
&M_{14}=-2[n_1(m_{AdS}/2+n+n_1)]^{1/2}cos\varphi_1+2[n_2(m_{AdS}/2+n+n_2)]^{1/2}cos\varphi_2\nonumber\\
&M_{24}=2[n_1(m_{AdS}/2+n+n_1)]^{1/2}sin\varphi_1+2[n_2(m_{AdS}/2+n+n_2)]^{1/2}sin\varphi_2\nonumber\\
&M_{23}=2\frac{[n(m_{AdS}+n)]^{1/2}}{m_{AdS}/2+n}\{[n_1(m_{AdS}/2+n+n_2)]^{1/2}
cos(\varphi-\varphi_1)+\nonumber\\
&[n_2(m_{AdS}/2+n+n_1)]^{1/2}cos(\varphi-\varphi_2)\}\nonumber\\
&M_{31}=2\frac{[n(m_{AdS}+n)]^{1/2}}{m_{AdS}/2+n}\{[n_1(m_{AdS}/2+n+n_2)]^{1/2}sin(\varphi-\varphi_1)-\nonumber\\
&[n_2(m_{AdS}/2+n+n_1)]^{1/2}sin(\varphi-\varphi_2)\}\nonumber\\
&M_{34}=2\frac{[n(m_{AdS}+n)]^{1/2}}{m_{AdS}/2+n}\{[(m_{AdS}/2+n+n_1)(m_{AdS}/2+n+n_2)]^{1/2}cos\varphi+\nonumber\\
&(n_1n_2)^{1/2}cos(\varphi-\varphi_1-\varphi_2)\}\nonumber\\
&M_{30}=-2\frac{[n(m_{AdS}+n)]^{1/2}}{m_{AdS}/2+n}\{[(m_{AdS}/2+n+n_1)(m_{AdS}/2+n+n_2)]^{1/2}sin\varphi-\nonumber\\
&(n_1n_2)^{1/2}sin(\varphi-\varphi_1-\varphi_2)\}
\label{sem17}
\end{eqnarray}

We now consider what restrictions follow from the fact that in Poincare limit the operators $M_{\mu 4}$
($\mu=0,1,2,3$) should be much greater than the other operators. The first conclusion is that, as
follows from the first expression in Eq. (\ref{sem17}), the quantum numbers $n_1$ and $n_2$ should be such that $|n_1-n_2|\ll n_1,n_2$.
Therefore in the main approximation in $1/R$ we have that $n_1\approx n_2$. Then it follows from 
the last expression 
that $sin\varphi$ should be of the order of $1/R$ and hence $\varphi$ should be close either to zero or to $\pi$. Then
it follows from the last four expressions in Eq. (\ref{sem17}) that the operators $M_{\mu 4}$
will be indeed much greater than the other operators if $\varphi_2\approx \pi-\varphi_1$ and in the main 
approximation in $1/R$
\begin{eqnarray}
&M_{04}=m_{AdS}+2(2n_1+n),\quad M_{14}=-4[n_1(m_{AdS}/2+n+n_1)]^{1/2}cos\varphi_1 \nonumber\\ 
&M_{24}=4[n_1(m_{AdS}/2+n+n_1)]^{1/2}sin\varphi_1,\quad M_{34}=\pm 2[n(m_{AdS}+n)]^{1/2}
\label{sem22}
\end{eqnarray}
where $M_{34}$ is positive if $\varphi$ is close to zero and negative if $\varphi$ is close to $\pi$.
In this approximation we have that  $M_{04}^2-\sum_{i=1}^3M_{i4}^2=m_{AdS}^2$ which ensures that in Poincare
limit we have the correct relation between the energy and momentum.

Consider now the singleton case. Here the quantum numbers $(n,k)$ do not exceed 1 and in semiclassical approximation
the quantum numbers $(n_1,n_2)$ are very large. For calculating semiclassical approximation one can define the normalized basis 
${\tilde e}(n_1,n_2,n,k)$ by analogy with the above discussion. However, this is possible only in standard theory where 
1/2 and 3/2 are  understood as rational numbers. At the same time, in FQT they are understood as $(p+1)/2$ and 
$(p+3)/2$, respectively and here the notion of the normalized basis is meaningless. Therefore in FQT there is no 
semiclassical approximation for singletons and below we consider this approximation only in standard theory. 

Consider first the case of the Rac singleton. Here the basis of the
representation space is formed by the elements $e(n_1,n_2,n)$ where $n$ can take only the values 0 and 1. If $c(n_1,n_2,n)$
is the WF in the normalized basis and the dependence on $(n_1,n_2)$ is as above 
then a direct calculation using Eqs. (\ref{scalar},\ref{MatrixRac}) gives
\begin{eqnarray}
&b''c(n_1,n_2,n)=2(n_1n_2)^{1/2}\{c(n_1,n_2,0)\delta_{n1}+exp[-i(\varphi_1+\varphi_2)]
c(n_1,n_2,1)\delta_{n0}\}\nonumber\\
&b'c(n_1,n_2,n)=2(n_1n_2)^{1/2}\{c(n_1,n_2,1)\delta_{n0}+exp[i(\varphi_1+\varphi_2)]c(n_1,n_2,0)\delta_{n1}\}\nonumber\\
&L_+c(n_1,n_2,n)=2(n_1n_2)^{1/2}\{exp(-i\varphi_1)c(n_1,n_2,1)\delta_{n0}+exp(i\varphi_2)
c(n_1,n_2,0)\delta_{n1}\}\nonumber\\
&L_-c(n_1,n_2,n)=2(n_1n_2)^{1/2}\{exp(i\varphi_1)c(n_1,n_2,0)\delta_{n1}+\nonumber\\
&exp(-i\varphi_2)c(n_1,n_2,1)\delta_{n0}\}
\label{sem26}
\end{eqnarray}
where $\delta$ is the Kronecker symbol. Then the mean values of these operators can be written as
\begin{eqnarray}
&<b''>=A\{exp(i\varphi)+exp[-i(\varphi+\varphi_1+\varphi_2)]\},\quad <b'>=<b''>^*\nonumber\\
&<L_+>=A\{exp[-i(\varphi+\varphi_1)]+exp[i(\varphi+\varphi_2)]\},\quad <L_->=<L_+>^*
\label{sem28}
\end{eqnarray}
where 
$$\sum_{n_1,n_2}2(n_1n_2)^{1/2}c(n_1,n_2,1)^*c(n_1,n_2,0)=Aexp(i\varphi)$$
and we use $^*$ to denote the complex conjugation. By analogy with the above discussion, we conclude that 
the Poincare limit exists only if $\varphi_2\approx \pi-\varphi_1$ and $\varphi$ is close either to zero
or $\pi$. Then
\begin{equation}
M_{04}\approx 4n_1,\quad M_{14}\approx -4n_1cos(\varphi_1),\quad M_{24}\approx 4n_1sin(\varphi_1)
\label{sem29}
\end{equation}
and the mean value of the operator $M_{34}$ is much less than $M_{14}$ and $M_{24}$.

Consider now the case of the Di singleton. Here the quantum number $n$ takes only the value $n=0$ and the quantum
number $k$ can take only the values 0 and 1. We denote the WF $c(n_1,n_2,k)$ in the normalized basis as
a set $c_k(n_1,n_2)$ for $k=0,1$. Then an analogous direct calculation using Eqs. (\ref{scalar},\ref{MatrixDi}) gives
\begin{eqnarray}
&b'(c_0(n_1,n_2),c_1(n_1,n_2))\approx n_1(exp(-i\varphi_1)c_1(n_1,n_2),exp(-i\varphi_2)c_0(n_1,n_2))\nonumber\\
&b'(c_0(n_1,n_2),c_1(n_1,n_2))\approx n_1(exp(i\varphi_2)c_1(n_1,n_2),exp(i\varphi_1)c_0(n_1,n_2))\nonumber\\
&L_+(c_0(n_1,n_2),c_1(n_1,n_2))\approx n_1(exp[-i(\varphi_1-\varphi_2)]c_1(n_1,n_2),c_0(n_1,n_2))\nonumber\\
&L_-(c_0(n_1,n_2),c_1(n_1,n_2))\approx n_1(c_1(n_1,n_2),exp[i(\varphi_1-\varphi_2)]c_0(n_1,n_2))
\label{sem33}
\end{eqnarray}
Now by analogy with Eq. (\ref{V12}) it follows from Eq. (\ref{sem33}) that the mean values of the operators $M_{a3}$ are given by
\begin{eqnarray}
&<M_{34}>\approx 2A[cos(\varphi-\varphi_1)+cos(\varphi+\varphi_2)]\nonumber\\
&<M_{30}>\approx 2A[sin(\varphi-\varphi_1)-sin(\varphi+\varphi_2)]\nonumber\\
&<M_{23}>\approx 2A[cos(\varphi-\varphi_1+\varphi_2)+cos\varphi ]\nonumber\\
&<M_{31}>\approx 2A[sin(\varphi-\varphi_1+\varphi_2)-sin\varphi ]
\label{sem34}
\end{eqnarray}
where
$$\sum_{n_1n_2}n_1c_1(n_1,n_2)^*c_0(n_1,n_2)=Aexp(i\varphi)$$
If $\varphi_2\approx \pi-\varphi_1$ then it is easy to see that the Poincare limit for $<M_{23}>$ 
and $<M_{31}>$ exists if $\varphi\approx\varphi_1$ or $\varphi\approx\varphi_1 +\pi$. 
In that case the Poincare limit for $<M_{34}>$ 
and $<M_{30}>$ exists as well and $<M_{34}>$ disappears in the main approximation. 

We have shown that in Poincare limit the
$z$ component of the momentum is negligible for both, the Di and Rac singletons.  As noted in the remark after Eq. (\ref{V12}),
the definition (\ref{V12}) is not unique and, in particular, any definition obtained from Eq. (\ref{V12})
by cyclic permutation of the indices $(1,2,3)$ is valid as well. Therefore we conclude that in standard theory,
the Di and Rac singletons have the property that in the Poincare limit they are characterized by two
independent components of the momentum, not three as usual particles. This is a consequence of the fact
that for singletons only the quantum numbers $n_1$ and $n_2$ can be very large.

The properties of singletons in Poincare limit have been discussed by several authors, and their
conclusions are not in agreement with each other (a detailed list of references can be found e.g. in
Refs. \cite{FFS,Bekaert}). In particular, there are statements that the Poincare limit for singletons does
not exist or that in this limit all the components of the four-momentum become zero. The above consideration
shows that Poincare limit for singletons can be investigated in full analogy with Poincare limit
for usual particles. In particular, the statement that the singleton energy in Poincare limit becomes zero is not
in agreement with the fact that each massless particle (for which the energy in Poincare limit is not zero)
can be represented as a composite state of two singletons. The fact that standard singleton momentum can
have only two independent components does not contradict the fact that the momentum of a massless particle
has three independent components since, as noted above, the independent momentum components of two singletons can be
in different planes.

\section{Tensor products of singleton IRs}
\label{singletontensorproduct}

We now return to the presentation when the properties of singletons in standard and modular approaches
are discussed in parallel. The tensor products of singleton IRs have been defined 
in Sec. \ref{elementary}. If $e^{(j)}(n_1^{(j)},n_2^{(j)},n^{(j)},k^{(j)})$ ($j=1,2$) 
are the basis elements
of the IR for singleton $j$ then the basis elements in the representation space of the tensor product can be chosen as 
\begin{eqnarray}
&&e(n_1^{(1)},n_2^{(1)},n^{(1)},k^{(1)},n_1^{(2)},n_2^{(2)},n^{(2)},k^{(2)})=
e^{(1)}(n_1^{(1)},n_2^{(1)},n^{(1)},k^{(1)})\times\nonumber\\
&&e^{(2)}(n_1^{(2)},n_2^{(2)},n^{(2)},k^{(2)})
\label{e1e2}
\end{eqnarray}
In the case of the tensor product of singleton IRs of different types, we assume that singleton 1 is
Di and singleton 2 is Rac.

Consider a vector
\begin{equation}
e(q)=\sum_{i=0}^qc(i,q)e^{(1)}(i,0,0,0)\times e^{(2)}(q-i,0,0,0)
\label{e(q)}
\end{equation}
where the coefficients $c(i,q)$ are given by Eq. (\ref{cik}) such that the $q_0^{(j)}$ should be replaced
by $q_1^{(j)}$ ($j=1,2$). Since $h_2^{(j)}e^{(j)}(i,0,0,0)=((p+1)/2)e^{(j)}(i,0,0,0)$ ($j=1,2$) then the
vector $e(q)$ is the eigenvector of the operator $h_2=h_2^{(1)}+h_2^{(2)}$ with the eigenvalue $q_2=1$ and
satisfies the condition $a_2'e(q)=0$ where $a_2'=a_2^{(1)'}+a_2^{(2)'}$.
As follows from the results of Sec. \ref{tensorprod}, $e(q)$ is the eigenvector of the operator 
$h_1=h_1^{(1)}+h_1^{(2)}$ with the eigenvalue $q_1=q_1^{(1)}+q_1^{(2)}+2q$ and
satisfies the condition $a_1'e(q)=0$ where $a_1'=a_1^{(1)'}+a_1^{(2)'}$. It is obvious that the value of 
$q_1$ equals $3+2q$ for the tensor product $Di\times Di$, $2+2q$ for the tensor product $Di\times Rac$
and $1+2q$ for the tensor product $Rac\times Rac$. 

As follows from Eqs. (\ref{V11}) and (\ref{V19}), in the case of IRs
\begin{eqnarray}
\label{benk}
&&b'e(n_1n_2nk)=[(a_1'')^{n_1}(a_2'')^{n_2}b'+n_1(a_1'')^{n_1-1}(a_2'')^{n_2}L_++\nonumber\\
&&n_2(a_1")^{n_1}(a_2'')^{n_2-1}L_-+n_1n_2(a_1'')^{n_1-1}(a_2'')^{n_2-1}b'']e(0,0,n,k)\nonumber\\
&&L_+e(n_1n_2nk)=[(a_1'')^{n_1}(a_2'')^{n_2}L_++n_2(a_1'')^{n_1}(a_2'')^{n_2-1}b'']e(0,0,n,k)
\end{eqnarray}
Therefore, $e(q)$ satisfies the conditions $b'e(q)=L_+e(q)=0$ where $b'=b^{(1)'}+b^{(2)'}$ and 
$L_+=L_+^{(1)}+L_+^{(2)}$. Hence, $e(q)$ is an analog of the vector $e_0$ in Eq. (\ref{V15}) and 
generates an IR corresponding to the quantum numbers $(q_1,q_2=1)$. 

We conclude that the tensor product of singleton IRs contains massless IRs corresponding to
$q_1=q_1^{(1)}+q_1^{(2)}+2q$. As follows from the results of 
Sect. \ref{tensorprod} (see the remark after Eq. (\ref{cik})), $q$ can take the values 
$0,1,...,p-q_1^{(1)}$. Therefore $Rac\times Rac$ contains massless IRs with
$s=0,2,4,...,(p-1)$, $Di\times Rac$ contains massless IRs with
$s=1,3,5,...(p-2)$ and $Di\times Di$ contains massless IRs with $s=2,4,...(p-1)$. In addition,
as noted in Ref. \cite{FF}, $Di\times Di$ contains a spinless massive IR corresponding to
$q_1=q_2=2$. This question will be discussed in Sec. \ref{supertensorprod}

Our next goal is to investigate whether or not all those 
IRs give a complete decomposition of the corresponding tensor products. For example, as follows from
Eq. (\ref{DimDiRac}), for the product $Rac\times Rac$ this would be the case if the sum
$\sum_{k=0}^{(p-1)/2}Dim(2k)$ equals $(p^2+1)^2/4=p^4/4+O(p^2)$. However, as follows from Eqs. (\ref{Dim0}) and
(\ref{Dims}), this sum can be easily estimated as $11p^4/48+O(p^3)$ and hence, in contrast to the Flato-Fronsdal
result in standard theory, in the modular case the decomposition of 
$Rac\times Rac$ contains not only massles IRs. Analogously, the sum of dimensions of massless IRs entering 
into the decompositions of $Di\times Rac$ and $Di\times Di$ also can be easily estimated 
as $11p^4/48+O(p^3)$ what is less than $p^4/4+O(p^2)$. The reason is that in the modular case the 
decompositions of the tensor
products of singletons contain not only massles IRs but also special IRs. We will not investigate the
modular analog of the Flato-Fronsdal theorem \cite{FF} but concentrate our efforts on finding a full solution 
of the problem in the supersymmetric case.

\section{Supersingleton IR}
\label{supersingleton}

In this section we consider the supersingleton IR exclusively in terms of the fermionic
operators without decomposing the IR into the Di and Rac IRs.
As a preparatory step, we first consider IRs of a simple superalgebra
generated by two fermionic operators $(d',d")$ and one bosonic operator $h$ such that
\begin{equation}
h=\{d',d''\},\quad [h,d']=-d',\quad [h,d'']=d''
\label{osp12}
\end{equation} 
Here the first expression shows that, by analogy with the osp(1,4) superalgebra, the relations (\ref{osp12})
can be formulated only in terms of the fermionic operators. 

Consider an IR of the algebra (\ref{osp12}) generated by a vector $e_0$ such that
\begin{equation}
d'e_0=0,\quad d'd''e_0=q_0e_0
\label{e0}
\end{equation}
and define $e_n=(d'')^ne_0$. Then $d'e_n=a(n)e_{n-1}$ where, as follows from Eq. (\ref{e0}), $a(0)=0,\,\,a(1)=q_0$
and $a(n)=q_0+n-1-a(n-1)$. The solution of this equation in $R_p$ is 
\begin{equation}
a(n)=n\frac{p+1}{2}+\frac{p+1}{2}(q_0-\frac{p+1}{2})[1-(-1)^n]
\label{anB}
\end{equation}
When $p$ is prime, the equation can be considered in $F_p$ and the solution can be written as 
$a(n) =(q_0-1/2)\{[1-(-1)^n]+n\}/2$.

We will be interested in the special case of the supersingleton when $q_0=(p+1)/2$. 
The maximum possible value of n can be found from the condition that $a(n_{max})\neq 0,\,\,a(n_{max}+1)=0$.  
Therefore, as follows from Eq. (\ref{anB}), $n_{max}=p-1$ and the dimension of the IR is $p$. In the general case, if $q_0\neq 0$ then $a(n)=0$
if $n=2p+1-2q_0$ and the dimension of the IR is $D(q_0)=2p+1-2q_0$.

Consider now the supersingleton IR. Let $x=(d_1''d_2''-d_2''d_1'')e_0$. Then, as follows from Eq. (\ref{S30}),
$d_1'x=(2q_1-1)d_2''e_0$ and $d_2'x=(1-2q_2)d_1''e_0$. Since $q_1=q_2=(p+1)/2$ we have that $d_1'x=d_2'x=0$
and therefore $x=0$. Hence the actions of the operators $d_1''$ and $d_2''$ on $e_0$ commute with each other.
If $n$ is even then $d_1''(d_2'')^ne_0=(d_2'')^nd_1''e_0$ as a consequence of Eq. (\ref{S30}) and if $n$ is odd
then $d_1''(d_2'')^ne_0=(d_2'')^{n-1}d_1''d_2''e_0=(d_2'')^nd_1''e_0$ in view of the fact that $x=0$.
Analogously one can prove that $d_2''(d_1'')^ne_0=(d_1'')^nd_2''e_0$. We now can prove that
$d_1''(d_2'')^n(d_1'')^ke_0=(d_2'')^n(d_1'')^{k+1}e_0$. Indeed, if $n$ is even, this is obvious while if $n$ is odd then 
$$d_1''(d_2'')^n(d_1'')^ke_0=(d_2'')^{n-1}d_1''d_2''(d_1'')^ke_0=(d_2'')^{n-1}(d_1'')^{k+1}d_2''e_0
=(d_2'')^n(d_1'')^{k+1}e_0$$
and analogously $d_2''(d_1'')^n(d_2'')^ke_0=(d_1'')^n(d_2'')^{k+1}e_0$. 
Therefore the supersingleton IR is
distinguished among other IRs of the osp(1,4) superalgebra by the fact that the operators $d_1''$ and $d_2''$
commute in the representation space of this IR. Hence the basis of the representation space can be
chosen in the form $e(nk)=(d_1'')^n(d_2'')^ke_0$. As a consequence of the above consideration, $n,k=0,1,...p-1$
and the dimension of the IR is $p^2$ in agreement with Eq. (\ref{DimDiRac}).

The above results can be immediately generalized to the case of higher dimensions. Consider
a superalgebra defined by the set of operators $(d_j',d_j'')$ where $j=1,2,...J$ and, by analogy
with Eq. (\ref{S30}), any triplet of the operators $(A,B,C)$ satisfies the commutation-anticommutation relation
\begin{equation}
[A,\{ B,C\} ]=F(A,B)C + F(A,C)B
\label{triplet}
\end{equation}
where the form $F(A,B)$ is skew symmetric, $F(d_j',d_j'')=1$
$(j=1,2,...J)$ and the other independent values of $F(A,B)$ are
equal to zero. The higher-dimensional analog of the supersingleton IR can now be defined such
that the representation space contains a vector $e_0$ satisfying the conditions
\begin{equation}
d_j'e_0=0,\quad d_j'd_j''e_0=\frac{p+1}{2}e_0\quad (j=1,2,...J)
\end{equation}
The basis of the representation space can be chosen in the form 
$e(n_1,n_2,...n_J)=(d_1'')^{n_1}(d_2'')^{n_2}\cdots (d_J'')^{n_J}e_0$. 
In full analogy with the above consideration one can show that the operators
$(d_1'',...d_J'')$ mutually commute on the representation space. As a consequence,
in the modular case each of the numbers
$n_j\,\, (j=1,2,...J)$ can take the values $0,1,...p-1$
and the dimension of the IR is $p^J$. The fact that singleton physics can be directly
generalized to the case of higher dimensions has been indicated by several authors
(see e.g. Ref. \cite{FFS} and references therein).

\section{Tensor product of supersingleton IRs}
\label{supertensorprod}

We first consider the tensor product of IRs of the superalgebra (\ref{osp12}) with $q_0=(p+1)/2$.
The representation space of the tensor product consists of all linear combinations of elements 
$x^{(1)}\times x^{(2)}$ where $x^{(j)}$ is an element of the
representation space for the IR $j$ ($j=1,2$). The representation operators of the tensor product
are linear combinations of the operators $(d',d'')$ where $d'=d^{(1)'}+d^{(2)'}$ and
$d''=d^{(1)''}+d^{(2)''}$. Here $d^{(j)'}$ and $d^{(j)''}$ mean
the operators acting in the representation spaces of IRs 1 and 2, respectively. 
In contrast to the case of tensor products of IRs of the sp(2) and so(2,3) algebras,
we now require that if $d^{(j)}$ is some of the $d$-operators for the IR $j$ then the operators $d^{(1)}$ and $d^{(2)}$ anticommute rather than commute, i.e. $\{d^{(1)},d^{(2)}\}=0$
Then it is obvious that the independent operators defining the tensor product satisfy 
Eq. (\ref{osp12}). 

Let $e_0^{(j)}$ be the generating vector for IR $j$ and $e_i^{(j)}=(d^{(j)''})^ie_0^{(j)}$. Consider the 
following element of the representation space of the tensor product
\begin{equation}
e(k)=\sum_{i=0}^k c(i)(e_i^{(1)}\times e_{k-i}^{(2)})
\label{e1e2B}
\end{equation}
where $c(i)$ is some function. This element will be the generating vector of
the IR of the superalgebra (\ref{osp12}) if $d'e(k)=0$. As follows from the above
results and Eq. (\ref{e1e2B})
\begin{equation}
d'e(k)=\frac{1}{2}\sum_{i=1}^k ic(i)(e_{i-1}^{(1)}\times e_{k-i}^{(2)})+
\frac{1}{2}\sum_{i=0}^{k-1}(-1)^i(k-i)c(i)(e_i^{(1)}\times e_{k-i-1}^{(2)})
\label{d1eq}
\end{equation}
Therefore $d'e(k)=0$ is satisfied if $k=0$ or
\begin{equation}
(i+1)c(i+1)=(-1)^{i+1}(k-i)c(i),\quad i=0,1,...k-1
\label{ciB}
\end{equation}
when $k\neq 0$. As follows from this expression, if $c(0)=1$ then
\begin{equation}
c(i)=(-1)^{i(i+1)/2}C_k^i
\label{ci}
\end{equation}
where $C_k^i=k!/i!(k-i)!$ is the binomial coefficient.
As follows from Eq. (\ref{anB}), the possible values of $k$ are $0,1,...p-1$ and, as follows
from Eq. (\ref{e1e2B}), $he(k)=q_0e(k)$ where $q_0=1+k$. The fact that the tensor product is fully decomposable
into IRs with the different values of $k$ follows from the relation $\sum_{q_0=1}^p D(q_0)=p^2$.

The tensor product of the supersingleton IRs can be constructed as follows. 
The representation space of the tensor 
product consists of all linear combinations of elements $x^{(1)}\times x^{(2)}$ where $x^{(j)}$ is an element of the
representation space for the supersingleton $j$ ($j=1,2$). The fermionic operators of the
representation are linear combinations of the operators $(d_1',d_2',d_1'',d_2'')$ where $d_1'=d_1^{(1)'}+d_1^{(2)'}$
and analogously for the other operators. Here $d_k^{(j)'}$ and $d_k^{(j)''}$ ($k=1,2$) mean
the operators $d_k'$ and $d_k''$ acting in the representation spaces of supersingletons 1 and 2, respectively. 
We also assume that
if $d^{(j)}$ is some of the $d$-operators for supersingleton $j$ then $\{d^{(1)},d^{(2)}\}=0$.
Then all the $d$-operators of the tensor product satisfy Eq. (\ref{S30}) and 
the action of the bosonic operators in the tensor product can be defined by Eq. (\ref{S31}).

Let $e_0^{(j)}$ be the generating vector for supersingleton $j$ (see Eq. (\ref{S32})) and 
$e_0=e_0^{(1)}\times e_0^{(2)}$. Consider the following element of
the representation space of the tensor product:
\begin{eqnarray}
&&x(k_1,k_2)=\sum_{i=0}^{k_1}\sum_{j=0}^{k_2}(-1)^{[\frac{i(i+1)}{2}+\frac{j(j+1)}{2}+k_1j]}
C_{k_1}^iC_{k_2}^j\nonumber\\
&&(d_1^{(1)"})^i(d_1^{(2)''})^{k_1-i}(d_2^{(1)''})^j(d_2^{(1)''})^{k_2-j}e_0\quad (k_1,k_2=0,1,...p-1)
\label{xk1k2}
\end{eqnarray}
By using Eq. (\ref{S30}) and the results of this section, one can explicitly verify that all the
$x(k_1,k_2)$ are the nonzero vectors and
\begin{equation}
d_1'x(k_1,k_2)=d_2'x(k_1,k_2)=0,\quad d_2'd_1''x(k_1,k_2)=x(k_1+1,k_2-1)
\label{xk1k2B}
\end{equation} 

Since the $e_0^{(j)}$ ($j=1,2$) are the generating vectors of the IRs of the osp(1,4) superalgebra with
$(q_1,q_2)=((p+1)/2,(p+1)/2)$, it follows from Eq. (\ref{S32A}) that $x(k_1,k_2)$ is the generating vector
of the IRs of the osp(1,4) superalgebra with $(q_1,q_2)=(1+k_1,1+k_2)$ if $d_2'd_1''x(k_1,k_2)=0$.
Therefore, as follows from Eq. (\ref{xk1k2B}), this is the case if $k_2=0$. Hence the tensor product
of the supersingleton IRs contains IRs of the osp(1,4) algebra corresponding to $(q_1,q_2)=(1+k_1,1)$
($k_1=0,1,...p-1$).
As noted in Sect. \ref{SS}, the case $(0,1)$ can be treated either as the massless IR with $s=p-1$ or
as the special massive IR; the case $(1,1)$ can be treated as the massive IR of the osp(1,4) superalgebra
and the cases when $k_1=1,...p-2$ can be treated as massless IRs with $s=k_1$. 

The results of standard theory follow from the above results in the formal limit $p\to\infty$. Therefore
in standard theory the decomposition of tensor product of supersingletons contains the IRs of the osp(1,4)
superalgebra corresponding to $(q_1,q_2)=(1,1),\,(2,1),...(\infty,1)$ in agreement with  
the results obtained by Flato and Fronsdal \cite{FF} and Heidenreich \cite{HeidenreichS}.

As noted in Sect. \ref{singletontensorproduct}, the Flato-Fronsdal result for the tensor product $Di\times Di$
is that it also contains a massive IR corresponding to $q_1=q_2=2$. In terms of the fermionic operators
this result can be obtained as follows. If $y=(d_1^{(1)''}d_2^{(2)''}-d_2^{(1)''}d_1^{(2)''})e_0$ then, as follows
from Eqs. (\ref{S30}) and (\ref{S31}), 
\begin{eqnarray}
&&d_1^{(1)'}y=\frac{p+1}{2}d_2^{(2)''}e_0,\quad d_1^{(2)'}y=\frac{p+1}{2}d_2^{(1)''}e_0,\quad d_2^{(1)'}y=-\frac{p+1}{2}d_1^{(2)''}e_0\nonumber\\
&&d_2^{(2)'}y=-\frac{p+1}{2}d_1^{(1)''}e_0,\quad h_1y=h_2y=2y,\quad L_+y=L_-y=0
\end{eqnarray}
Since $a_j'=(d_j')^2$ for $j=1,2$ (see Eq. (\ref{S31})), it follows from these expressions that $a_1'y=a_2'y=0$, i.e.
$y$ indeed is the generating vector for the IR of the so(2,3) algebra characterized by $q_1 =q_2=2$. However, $y$ is not a
generating vector for any IR of the osp(1,4) superalgebra since it does not satisfy the condition $d_1'y=d_2'y=0$.

The vector $x(k_1,k_2)$ defined by Eq. (\ref{xk1k2}) becomes the null vector when $k_1=p$. Indeed, since 
$C_{k_1}^i=k_1!/[i!(k_1-i)!]$, the sum over $i$ in Eq. (\ref{xk1k2}) does not contain terms with $i\neq 0$ and $i\neq p$.
At the same time, if $i=0$ or $i=p$ the corresponding terms are also the null vectors since, as follows from the
results of the preceding section, $(d_1')^pe_0=(d_2')^pe_0=0$. It is obvious that this result is valid only in the
modular case and does not have an analog in standard theory. Therefore, as follows from Eq. (\ref{xk1k2B}), the 
decomposition of the tensor products of two supersingletons also contains IRs of the osp(1,4) superalgebra
characterized by $(q_1,q_2)=(0,0),\,(0,1),\,(0,2),...(0,p-1)$.

We have shown that the decomposition of the tensor products of two supersingletons contains IRs of the osp(1,4) superalgebra
characterized by the following values of $(q_1,q_2)$:
$$(0,0),\,(0,1),\,(0,2),...(0,p-1), (1,1),\,(2,1),...(p-1,1)$$
The question arises whether this set of IRs is complete, i.e. the decomposition of the tensor products of two supersingletons
does not contain other IRs of the osp(1,4) superalgebra. Since the dimension of the supersigleton IR is $p^2$ (see the
preceding section), this is the case if
\begin{equation}
\sum_{k=0}^{p-1}SDim(0,k)+\sum_{k=1}^{p-1}SDim(1,k)=p^4
\label{p4A}
\end{equation}
It is obvious that $SDim(0,0)=1$ since the IR characterized by $(q_1,q_2)=(0,0)$ is such that all the representation operators
acting on the generating vector give zero. Therefore, as follows from Eq. (\ref{SDim}), the condition (\ref{p4A}) can
be rewritten as
\begin{equation}
2+Dim(0)+Dim(2,2)+2\sum_{s=1}^{p-2}Dim(s)+2\sum_{q_2=1}^{p-1}Dim(0,q_2)=p^4
\label{p4B}
\end{equation}
since $Dim(1,1)=Dim(0)$. The expressions for $Dim(s)$ and $Dim(0,q_2)$ are given in Eqs.
(\ref{Dim0}-\ref{dimspecial}) and hence the only quantity which remains to be calculated is $Dim(2,2)$.

The IR of the so(2,3) algebra characterized by $(q_1,q_2)=(2,2)$ is the massive IR with $m_{AdS}=4$ and $s=0$.
Therefore, as follows from the results of Sect. \ref{VS3}, the quantity $k$ in Eq. (\ref{16}) can take only
the value $k=0$ and the quantity $n$ can take the values $0,1,...n_{max}$ where $n_{max}=p-2$. Hence, as follows from
Eqs. (\ref{Dim}) and (\ref{V20})
\begin{equation}
Dim(2,2)=\sum_{n=0}^{p-2}(p-1-n)^2=\frac{1}{6}p(p-1)(2p-1)
\label{Dim22}
\end{equation}
The validity of Eq. (\ref{p4B}) now follows from Eqs. (\ref{Dim0}-\ref{dimspecial},\ref{Dim22}).

The main result of this chapter can now be formulated as follows: 

{\it In FQT the tensor product of two Dirac supersingletons is fully decomposable into the following 
IRs of the osp(1,4) superalgebra:}
\begin{itemize}
\item {\it Massive IR characterized by $(q_1=1,q_2=1)$}
\item {\it Massless IRs characterized by $(q_1=2,...p-1,q_2=1)$}
\item {\it Special IRs characterized by $(q_1=0, q_2=0,1,...p-1)$}
\end{itemize}
{\it and the multiplicity of each IR in the decomposition equals one.}

\chapter{A conjecture on the nature of time}
\label{time}

In this work we discussed the effects of WPS, cosmological repulsion and gravity in the framework
of semiclassical approximation. We assumed that in this approximation the evolution can be described
by classical time $t$ and, as noted in Subsec. \ref{problemoftime}, in quantum theory the problem of 
time is very difficult. A problem arises whether the nature of classical time can be understood
proceeding from pure quantum notions. In this chapter we consider a conjecture that $t$ is a
manifestation of the fact that the parameter $p$ in FQT changes, i.e. the true evolution
parameter is $p$ and not $t$. 

At present Poincare approximation works with a high accuracy
because the quantities $p$ and $R$ are very large. In the spirit of cosmological models one might 
think that at early stages of the Universe the quantity $R$ was much less than now. Analogously,
one might think that the quantity $p$ was much less than now. Indeed, as explained in Chap. \ref{AdS},
the notion of particles and antiparticles can be only approximate when $p$ is very large. In view of
the problem of baryon asymmetry of the Universe one might think that the matter in the Universe
was created when such quantum numbers as electric charge and baryon and lepton quantum numbers were
strongly nonconserved and this could happen only if $p$ was much less than now. For this reason we
will consider a possibility that in our Universe the quantity $p$ is constantly increasing.

\section{One-dimensional model}
\label{1dim}

Consider a system of two particles with the masses $m_1$ and $m_2$ such that $m_2\gg m_1$. Then, as noted in Sec. \ref{largem2},
particle 1 can be considered in the framework of single-particle problem but the width of the $n_1$ distribution should be
replaced by the width of the $n$ distribution which equals $\delta=\delta_2$. For simplicity we will consider the case when on classical
level the particle is moving along the $z$-axes. The corresponding semiclassical WF is the eigenstate of the operator $J_z$ with the eigenvalue 
$\mu=0$ and such that the parameter $\alpha$ in Eq. (\ref{qclwf}) is zero or $\pi$. 
Our goal is to obtain classical results without using standard semiclassical approximation, position operators and time but proceeding only
from quantum states. However, the semiclassical results give a hint that if $k\ll n$ then a simple case which we can consider is the one-dimensional model
where the WF $c(n)$ depends only on $n$ and, as follows from the first expression in Eq. (\ref{22A}) 
\begin{eqnarray}
{\cal E}c(n)=\frac{1}{2}c(n-1)+\frac{1}{2}[w+(2n+3)^2]c(n+1)
\label{1E}
\end{eqnarray}
Although we work in FQT, it will be helpful to compare the results with those obtained in standard theory because our physical
intuition is based on that theory. Here, as follows from
Eq. (\ref{24A}), the dS energy operator acts on the normalized WF as
\begin{equation}
{\cal E}{\tilde c}(n)=\frac{1}{2}[(w+(2n+1)^2)]^{1/2}{\tilde c}(n-1)+
\frac{1}{2}[w+(2n+3)^2)]^{1/2}{\tilde c}(n+1)
\label{1Etilde}
\end{equation}

For the correspondence with standard theory, in FQT it is desirable to work with least possible
numbers in order to avoid comparisons modulo $p$ whenever possible. We now use $n_1$ and $n_2$
to define the minimum and maximum values of $n$ in the support of $c(n)$. Then by using the
fact that the space of states is projective, as follows from Eq. (\ref{normen}), the normalization of the
elements $e_n$ can be chosen as
\begin{equation}
(e_n,e_n)=\prod_{j=n_1+1}^n [w+(2j+1)^2]\quad (n\in[n_1, n_2])
\label{1norm}
\end{equation}
Then up to a normalization factor  the relation between the WFs in FQT and in standard theory can be written in the form
\begin{equation}
{\tilde c}(n_2-l)=c(n_2-l)\{\prod_{m=0}^{l-1}[w+(2n_2-2m+1)^2]\}^{-1/2}
\label{ctildec}
\end{equation}
where $l=n_2-n$.

Since $c(n)$ has a finite support it cannot be the eigenstate of the operator ${\cal E}$. For example,
$c(n_2+1)=0$ but, as follows from Eq. (\ref{1E}), ${\cal E}c(n_2+1)=c(n_2)/2\neq 0$. 
Analogously $c(n_1-1)=0$ but, as follows from Eq. (\ref{1E}), ${\cal E}c(n_1-1)=[w+(2n_1+1)^2]c(n_1)/2\neq 0$. 
We will see below
that the uncertainty of ${\cal E}$ is minimal when ${\cal E}c(n)=\lambda c(n)$ for $n\in [n_1,n_2]$.
This condition can be satisfied if the expression describing $c(n)$ at $n\in [n_1,n_2]$ is such that $c(n_1-1)=0$ and $c(n_2+1)=0$. 

Since the norm of $e_n$ is maximal when $n=n_2$, we want to work with least possible numbers, 
the states are projective, the minimum possible value of $c(n_2)$ in FQT is $c(n_2)=\pm 1$ then we
choose $c(n_2)=1$. Then, as follows from Eq. (\ref{1E}), for $n\in [n_1,n_2]$ all the values $c(n)$ can be found consecutively:
\begin{equation}
c(n-1)=2\lambda c(n) - [w+(2n+3)^2]c(n+1)
\label{c(n)}
\end{equation}
In particular, $c(n_2-1)=2\lambda$, $c(n_2-2)=4\lambda^2-W$ etc. However, it is
problematic to find an explicit expression for $c(n)$ if $n$ is arbitrary.

In the nonrelativistic case $w\gg n_2^2$ and for semiclassical WFs $\delta=(n_2-n_1)\ll n_2$. So one might think that a good approximation is to neglect the variations of $[w+(2n+1)^2]$ at $n\in [n_1,n_2]$ and
consider the following approximation of Eq. (\ref{c(n)}):
\begin{equation}
c(n-1)=2\lambda c(n) -Wc(n+1)
\label{cW}
\end{equation} 
 where $W=w+(2n_2+1)^2$. Then it is easy to prove by induction that
\begin{equation}
c(n_2-l)=\sum_{m=0}\frac{(-1)^m (l-m)!}{m!(l-2m)!}(2\lambda)^{l-2m}W^m
\label{cl}
\end{equation}
where the upper limit is defined by the condition that 
$1/(l-2m)!=0$ if $l<2m$. 
As follows from Eq. (\ref{ctildec}), in this approximation
\begin{equation}
{\tilde c}(n_2-l)=C(l)=C(l,x)=\sum_{m=0}\frac{(-1)^m (l-m)!}{m!(l-2m)!}(2x)^{l-2m}
\label{cl}
\end{equation}
where $x=\lambda/W^{1/2}$. This is the Gegenbauer polynomial which in the literature is
denoted as $C_l^1(x)$, and it is known that if $x=cos\theta$ then $C(l)=sin((l+1)\theta)/sin\theta$.
Since the notation $C_n^k$ is also used for binomial coefficients we will use for the
Gegenbauer polynomial $C_n^k(x)$ the notation $G_n^k(x)$.

Suppose that $sin((\delta +2)\theta)=0$. Then $(\delta+2)\theta=k\pi$ where $k$ is an integer,  
$sin((\delta+1)\theta )=(-1)^{k+1}sin\theta$ and 
\begin{equation}
Norm^2=\sum_{l=0}^{\delta}C(l)^2=\frac{1}{sin^2\theta}\sum_{l=0}^{\delta}sin^2((l+1)\theta)=
\frac{\delta+2}{2sin^2\theta}
\label{Norm}
\end{equation}
In this case ${\cal E}{\tilde c}(n)=\lambda {\tilde c}(n)$ for all 
$n\in [n_1,n_2]$, $\lambda$ is exactly the mean value of the operator ${\cal E}$: 
\begin{equation}
\bar{{\cal E}}=\frac{1}{Norm^2}({\tilde c},{\cal E}{\tilde c})=\frac{1}{Norm^2}\sum_{n=n_1}^{n_2} 
{\tilde c}(n){\cal E}{\tilde c}(n)=\lambda ,
\label{barE}
\end{equation}
and the uncertainty of ${\cal E}$ is
\begin{equation}
\Delta {\cal E}=\frac{1}{Norm}({\tilde c},({\cal E}-\bar{{\cal E}})^2{\tilde c})^{1/2}=
\frac{1}{Norm}||({\cal E}-\bar{{\cal E}}){\tilde c}||=(\frac{W}{\delta+2})^{1/2}|sin\theta|
\label{DeltaE}
\end{equation}

As follows from the first expression in Eq. (\ref{22A}), if $k\ll n$ then the dS energy of the particle which is far from other
particles approximately equals ${\cal E}\approx \pm W^{1/2}$ and, as follows from Eqs. (\ref{II70})
and (\ref{Hnr}), for nonrelativistic particles the effective interaction gives a small correction to ${\cal E}$.
Therefore $\lambda /W^{1/2}$ is close to 1 but is less than 1.
Hence one can choose $\theta$ such that $cos\theta=\lambda /W^{1/2}$, $\theta$ is small and $\theta>0$. 
Then, as follows from Eq. (\ref{DeltaE}), 
$\Delta {\cal E}/{\bar {\cal E}}\approx sin\theta/\delta^{1/2}$
is very small because $\delta$ is very large and $sin\theta$ is small. Indeed, a simple estimation shows
that if the kinetic and potential energies are of the same order then $\theta$ is of the order of $v/c$
and for the cosmological repulsion $\theta$ is of the order of $r/R$. As a consequence, the particle state
is strongly semiclassical.  

Another possible choice of the WF follows. We do not require that the condition $({\cal E}-\lambda)c(n)=0$ should
be satisfied at all $n\in [n_1,n_2]$,
choose an arbitrary value for $c(n_2-1)$ and find the values of $c(n)$ at $n=n_2-2,...,n_1$ from 
Eq. (\ref{cW}). Then in general the condition $({\cal E}-\lambda)c(n)=0$ will be satisfied only for $n\in [n_1+1,n_2-1]$. 
In particular, if $c(n_2-1)=\lambda$ then it follows from from Eqs. (\ref{ctildec}) and (\ref{cW}) that
${\tilde c}(n_2-l)=cos(l\theta)$. In that case the quantity ${\Delta\cal E}/\bar{{\cal E}}$ will be greater than
in the case of Eq. (\ref{DeltaE}) but will also be of the order not greater than $1/\delta^{1/2}$, i.e. very small.
We conclude that the requirement that the dS energy should be strongly semiclassical does not impose strong
restrictions on the WF.

A problem arises whether it is indeed a good approximation to neglect the variations of $[w+(2n+1)^2]$ at $n\in [n_1,n_2]$.
In what follows we describe two attempts to find the exact solution.

Consider this problem in standard theory and define
$$f(l)=\frac{w+(2(n_2-l)+1)^2}{w+(2n_2+1)^2}, \quad F(l)=[\prod_{m=0}^{l-1}f(m)]^{-1}C(l)$$
Then $F(l)=C(l)$ if $l=0,1$, $F(l) \neq C(l)$ at $l\geq 2$ and, as follows from Eq. (\ref{1Etilde})
\begin{equation}
F(l+1)=2cos\theta F(l) -f(l-1)F(l-1)
\label{F(l)}
\end{equation}
It is obvious that $F(l)\approx C(l)$ for $l\ll \delta$ but the problem is whether the approximate equality
takes place if $l$ is of the order of $\delta$.

We define $S(k,l)=\sum f(i_1)...f(i_l)$ where the sum is taken over all products of $l$ multipliers such that
$S(k,0)=S(0,1)=1$, $S(k,l)=0$ if $k>0$ and $k<l$, the indices $i_1,...i_l$ can take the values $0,1,...k$ in the ascending order and the difference between any
value and the previous one is greater or equal 2. 
Then it can be easily proved by induction that 
\begin{equation}
S(l,m)=S(l-1,m)+f(l)S(l-2,m-1)
\label{S(l,m)}
\end{equation}
We consider the first case discussed above, i.e. $c(n_2)=1$ and $c(n_2-1)=2\lambda$.  
Then it can be proved by induction that, as follows from Eq. (\ref{S(l,m)}), the solution of Eq. (\ref{F(l)}) is 
\begin{equation}
F(l)=\sum_{m=0}^{[l/2]}(-1)^m (2x)^{l-2m}S(l-2,m)
\label{F(l)B}
\end{equation}
where $[l/2]$ is the integer part of $l/2$.

Since we assume that $l\ll n_2$ then $f(l)\approx 1-ly$ where $y=4(2n_2+1)/W$. We assume that if
$l$ is of the order of $\delta$ then the approximate expression for $S(l,m)$ is
\begin{equation}
S(l,m)=\sum_{s=0}^m a(l,m,s), \quad a(l,m,s)=\frac{ (-y/2)^s(l+2-m)!l!}{(l+2-2m)!s!(m-s)!(l-s)!}
\label{S(l,mB)}
\end{equation}
It follows from this expression that only the values of $m\leq (l/2+1)$ contribute to the sum and 
$a(l,m,s+1)/a(l,m,s)=-y(m-s)(l-s)/[2(s+1)]$.

The value of $W$ is the Poincare analog of the energy squared: $W=4R^2(m^2+{\bf p}^2)$, $n_2$ is the Poincare analog of
$R|{\bf p}|$ and, as follows from Eq. (\ref{deltaj}), $\delta$ is of the order of $R/r_g$ where $r_g$ is the gravitational (Schwarzschild)
radius of the heavy body. Then if $l$ is of the order of $\delta$ and $R$ is of the order of $10^{26}meters$ then 
$yl^2 \ll 1$. However, as
noted above, the value of $R$ may be much greater than $10^{26}meters$, Poincare limit is defined as $R\to\infty$ and in the formal limit
$R\to\infty$, $y\delta^2\to\infty$. So if $m$ is of the order $l$ and $s\ll m$ then it is possible that 
$a(l,m,s+1)\gg a(l,m,s)$ but if $s$ if of the order of $m$ then $a(l,m,s+1)\ll a(l,m,s)$.

A direct calculation using Eq. (\ref{S(l,mB)}) gives
\begin{eqnarray}
&& S(l-1,m)+f(l)S(l-2,m-1)=\sum_{s=0}^m b(l,m,s), \nonumber\\
&&b(l,m,s)=a(l,m,s)[1+\frac{s(s-1)(l-s)}{l(l-1)(l+2-m)}]
\label{S(l,mC)}
\end{eqnarray}
Therefore Eq. (\ref{S(l,m)}) is satisfied with a high accuracy and Eq. (\ref{S(l,mB)}) is a good approximate
expression for $S(l,m)$.

As follows from Eqs. (\ref{F(l)}) and (\ref{S(l,mB)}), the expression for $F(l)$ can be represented as
\begin{equation}
F(l)=\sum_{s=0}^{[l/2]} (y/2)^s\frac{(l-2)!}{(l-2-s)!}\sum_{m=0}^{[\nu/2]}(-1)^m(2x)^{\nu-2m}\frac{(s+1)_{\nu-m}}{(\nu-2m)!m!} 
\label{F(l)C}
\end{equation} 
where $\nu=l-2s$ and $n_k=n(n-1)...(n-k+1)$ is the Pohhammer symbol. The last sum in this expression is the Gegenbauer polynomial 
$G_{\nu}^{s+1}(x)$ and therefore
\begin{equation}
F(l)=\sum_{s=0}^{[l/2]} (y/2)^s\frac{(l-2)!}{(l-2-s)!}G_{\nu}^{s+1}(x) 
\label{F(l)D}
\end{equation} 
Finally, by using the asymptotic expression for the Gegenbauer polynomial $G_{\nu}^{s+1}(x)$ when $\nu$ is large we
get
\begin{equation}
F(l)=\sum_{s=0}^{[l/2]}(y/4)^s\frac{(l-2)!(l-s)!}{(l-2-s)!s!(l-2s)!}\frac{cos[(l-s+1)\theta -(s+1)\pi/2]}{sin\theta^{s+1}}
\label{F(l)E}
\end{equation} 

If this expression is represented as $F(l)=\sum_s a(l,s)$ then for $l$ of the order of $\delta$ and $s\ll l$,
$a(l,s+1)/a(l,s)$ is of the order of $yl^2/sin\theta$. As noted above, the quantity $yl^2$ can be very large and
therefore the quantity $yl^2/sin\theta$ can be even larger, especially in cases when $\theta$ is of the order of $r/R$.
We see that even for the choice $c(n_2-1)=2\lambda$ understanding qualitative features of the
solution of Eq. (\ref{1E}) is very difficult. In addition, as noted above, the WF is strongly semiclassical for
other choices of $c(n_1-1)$. Therefore it is a great problem to understand what conditions govern the choice
of the semiclassical WF. 

The second attempt to find the exact solution follows. Consider the function 
\begin{equation}
{\tilde c}(n_2-l)=const\cdot  cos(\alpha(l)), \quad \alpha(l)=\sum_{m=1}^l arccos(\frac{\lambda}{[w+(2(n_2-m)+3)^2]^{1/2}})
\label{approxA}
\end{equation} 
where $const$ is a normalizing coefficient. When the variations of 
$[w+(2n+1)^2]$ at $n\in [n_1,n_2]$ are neglected this function becomes ${\tilde c}(n_2-l)=const\cdot cos(l\theta)$, i.e.
the approximate solution discussed above.
As follows from Eqs. (\ref{1Etilde}) and (\ref{approxA})
\begin{eqnarray}
&&{\cal E}{\tilde c}(n_2-l)=\lambda {\tilde c}(n_2-l) +\frac{1}{2}const\cdot sin((\alpha(l))\{[w+(2(n_2-l)+3)^2-\lambda^2]^{1/2}-\nonumber\\
&&[w+(2(n_2-l)+1)^2-\lambda^2]^{1/2}\}
\label{EA}
\end{eqnarray}
The presence of the second term in the r.h.s. shows that the function given by Eq. (\ref{approxA}) is not the exact solution. Typically this term is much less than the first one but this is not the case when $cos(\alpha(l))$ is small.

Analogously the function ${\tilde c}(n_2-l)=const\cdot  sin(\alpha(l))$ becomes ${\tilde c}(n_2-l)=const\cdot sin(l\theta)$
when the variations of $[w+(2n+1)^2]$ at $n\in [n_1,n_2]$ are neglected but it 
is not the exact solution because 
\begin{eqnarray}
&&{\cal E}{\tilde c}(n_2-l)=\lambda {\tilde c}(n_2-l) -\frac{1}{2}const\cdot cos((\alpha(l))\{[w+(2(n_2-l)+3)^2-\lambda^2]^{1/2}-\nonumber\\
&&[w+(2(n_2-l)+1)^2-\lambda^2]^{1/2}\}
\label{EB} 
\end{eqnarray} 

\section{Classical equations of motion}
\label{motion}

As already noted, Eq. (\ref{estimation}) gives the estimation of the width of the relative dS momentum if the mass of particle 2 is much greater
than the mass of particle 1. It also follows from Eq. (\ref{G}) that not only $p$ is a very large number but even $lnp$ is very large. Suppose
now that $p$ changes. We do not say that $p$ changes with time because time is a classical notion while we are considering a pure quantum problem.
Below we propose a scenario that classical time arises as a consequence of the fact that $p$ changes. As noted at the beginning
of this section, there are
reasons to think that at early stages of the Universe $p$ was much less than now i.e. $p$ is increasing.

If $p$ changes by $\Delta p$ then $\Delta p$ cannot be infinitely small because, roughly speaking, $p$ is an integer. Moreover, a possible scenario is that
at every step $p$ is multiplied by a number $k$ and if $k\gg 1$ then $\Delta p\gg p$. However, in that case $lnp$ changes by $\Delta lnp=lnk$. This
quantity also cannot be infinitely small but it is possible that $\Delta lnp/lnp$ is a very small real number.
As follows from Eq. (\ref{estimation}), 
$\Delta\delta /\delta = \Delta lnp /lnp$. Therefore $\Delta\delta /\delta$ does not depend on the heavy mass and depends only on the change of $p$. Since time is a dimensionful parameter, we {\it define} time such that its variation is given by $\Delta t=R \Delta lnp /lnp$. In that case $\Delta t$ also cannot be infinitely small but can be very small in
comparison with macroscopic times. 

In view of Eq. (\ref{G}) and the definition of time the following problem arises. If $p$ changes then does it mean that $G$ changes? In our approach the number $p$ is fundamental while $G$ is not. In view of the remarks in Secs. \ref{symmetry} and \ref{manybody}, a problem also arises whether dimensionful quantities can be fundamental. In particular,
as noted in Sec. \ref{manybody}, the quantity $G_{dS}$ given by Eq. (\ref{GLambda}) is more fundamental than $G$ because it is dimensionless. Equation (\ref{G}) shows
that $G$ depends not only on $p$ but also on $R$. This parameter has the dimension of meter because people want to deal with Poincare momenta
and not with dimensionless dS angular momenta. So it is not even clear whether $R$ expressed in meters 
changes or not. In any case, among the constants which
are treated as fundamental, $G$ is measured with the least accuracy and its value is known only for approximately 300 years. 
If $\Delta lnp\ll lnp$ then it is quite possible that the change of $G$ could not be noticed for such a short period of time. 
In view of these remarks we assume that relative variations of such quantities as $R$ and $\delta$ are much smaller than
relative variations of standard momenta and coordinates characterizing the particle under consideration. In what follows we use $p$ to denote the magnitude of standard momentum.

The problem arises how $n_2$ changes with the change of $\delta$. Understanding this problem is very difficult because, 
as discussed
in the preceding subsection, even understanding the behavior of the semiclassical WF is very difficult. 
For this reason we can only make
assumptions about the dependence of the variation of $n_2$ on the variation of $\delta$. Since the choice of
the WF is defined by the choice of $c(n_2-1)$, and $c(n_2-1)$ is a function of $\lambda$, we assume that 
$\lambda$ is the conserved quantity. For simplicity, in what follows we will write $n$ instead of $n_2$ and consider only nonrelativistic approximation.

Consider a situation in standard theory when a particle is moving along the $z$-axis
and is attracted or repulsed by a body in the origin. Consider first a possibility that
\begin{equation}
\Delta n=\pm (W-\lambda^2)^{1/2}\frac{\Delta\delta}{2\delta}
\label{1st}
\end{equation}
where the sign depends on whether the particle momentum and radius-vector are parallel or anti-parallel.
We treat Eq. (\ref{1st}) as an approximate consequence of FQT formulated in terms of real numbers and so we can use classical mathematics for
treating this expression with a good approximation.

If $\theta$ is {\it defined} such that $cos\theta=\lambda /W^{1/2}$ and $sin\theta$ is positive
then $\theta\approx sin\theta=(1-\lambda^2/W)^{1/2}$ and
\begin{equation}
\lambda \approx \pm W^{1/2}(1-\theta^2/2)\approx 2R(m+p^2/2m-m\theta^2/2),\quad p\Delta p=m^2 \theta\Delta \theta
\label{1stB}
\end{equation}
The last relation follows from the fact that $\lambda$ is a conserved quantity. Finally, we {\it define} $r$ such that 
$\theta=\varphi=r/R$. Then, as noted in Sec.  \ref{macroscopicdist}, this corresponds to standard position operator.
Note that $n=Rp$ and
then, as follows from the definition of time and Eqs. (\ref{1st}) and (\ref{1stB})
\begin{equation}
\Delta p=\pm \frac{mr}{R^2}\Delta t,\quad \Delta r=\pm \frac{p}{m}\Delta t
\label{1stC}
\end{equation}
In view of the remarks on Eq. (\ref{coordmom}), the second expression shows that the quantity $r$ defined above
indeed has the meaning of the coordinate. Since the quantities $p$ and $r$ are positive by construction, it is
clear that in our one-dimensional model the sign is $\pm$ when the momentum and radius-vector are collinear
and anticollinear, respectively.

In the approximation when $\Delta t$ in Eq. (\ref{1stC})  can be treated as infinitely small, we get 
$\dot{p}=\pm mr/R^2$, $\dot{r}=\pm p/m$, 
i.e. exactly the Hamilton equations obtained from the Hamiltonian $H=p^2/(2m) - mr^2/(2R^2)$. It
follows from these relations that $\ddot{r}=r/R^2$ in agreement with Eq. (\ref{accel}) (taking into account
that we work in units where $c=1$). Therefore we have repulsion as it should be in accordance with the
consideration in Sec. \ref{antigravity}. Here it has been noted that the result for dS antigravity is
compatible with the prescription of standard quantum theory that the coordinate and momentum representations
should be related to each other by the Fourier transform.

Consider now a possibility that
\begin{equation}
\Delta n=\pm \frac{(W-\lambda^2)^2}{4const^2W^{3/2}}\Delta \delta
\label{grav1}
\end{equation}
where $const$ is the same as in Eq. (\ref{deltaj}). We can define $\theta$, assume 
that $\theta\ll 1$  and use Eq. (\ref{1stB}) as above. 
Then $\Delta n=\pm W^{1/2}\theta^4\Delta \delta /(4const^2)$. However, if we define $r$ as above then this
quantity will not satisfy the second condition in Eq. (\ref{1stC}), i.e. it will not have the meaning of
coordinate. Therefore in the given case the momenta and coordinates cannot be related by the Fourier transform.
In accordance with Sec. \ref{macroscopicdist}, we now {\it define} $\theta=\chi=const /(\delta \varphi)^{1/2}$
where $\varphi=r/R$. Then as follows from the definition of time and Eqs. (\ref{deltaj}) and (\ref{1stB})
\begin{equation}
\Delta p=\pm \frac{MmG}{r^2}\Delta t,\quad \Delta r=\mp \frac{p}{m}\Delta t
\label{grav2}
\end{equation}
where $M$ is the mass of the heavy particle 2. As follows from the second expression, the quantity $r$ has now the
meaning of the coordinate in view of the remarks on Eq. (\ref{coordmom}). We conclude that the sign in 
Eq. (\ref{grav1}) should be opposite to that in Eq. (\ref{1st}): it is $\pm$ when the momentum and radius-vector are 
anticollinear and collinear, respectively. 
In the approximation when $\Delta t$ is infinitely small we get $\dot{p}=\pm MmG/r^2$, $\dot{r}=\mp p/m$ and
$\ddot{r}=-MG/r^2$. The last relation shows that in this case we have attraction as it should be for gravity.

We have considered two cases when $\Delta n$ is given by Eqs. (\ref{1st}) and (\ref{grav1}), respectively. The first case
reproduces standard dS antigravity and the second case --- standard gravity. The comparison of those expressions
shows that the first case takes place when $\delta\theta^3\ll 1$ and the second case --- in the opposite situation
when $\delta\theta^3\gg 1$. As follows from Eq. (\ref{deltaj}), $\delta$ is of the order $R/r_g$ where $r_g$ is the
gravitational radius of the heavy particle 2. As shown above, $\theta=r/R$ in the first case and 
$\theta=const (R/\delta r)^{1/2}\approx (r_g/r)^{1/2}$ in the second one. Therefore the above conditions are
indeed satisfied if $R$ is very large. 

Finally for illustration we consider a possibility to find the solution of the problem of time with the choice of the WF given by
Eq. (\ref{approxA}). Then if ${\tilde c}(n_2,\delta)={\tilde c}(n_2-\delta)$ and 
\begin{equation}
\alpha(n_2,\delta)=\sum_{m=1}^\delta arccos(\frac{\lambda}{[w+(2(n_2-m)+3)^2]^{1/2}})
\label{alpha}
\end{equation}
we have that 
\begin{eqnarray}
&&{\tilde c}(n_2,\delta)=const_1\cdot  cos[\alpha(n_2,\delta)], \nonumber\\
&&{\tilde c}(n_2+\Delta n,\delta+\Delta\delta)=const_2\cdot  cos[\alpha(n_2+\Delta n,\delta+\Delta\delta)]
\label{approxC}
\end{eqnarray} 

As noted in the preceding subsection, it is desirable that the solution ${\tilde c}(n)$ satisfies the condition
${\tilde c}(n_1-1)=0$. For this reason we assume that 
$$cos[\alpha(n_2,\delta)]=cos[\alpha(n_2+\Delta n,\delta+\Delta\delta)]=0.$$
This does not necessarily imply that $\alpha(n_2,\delta)=\alpha(n_2+\Delta n,\delta+\Delta\delta)$ but we
assume that for rather small values of $\Delta n$ and $\Delta\delta$ this is the case. Then
\begin{eqnarray}
&&\sum_{m=\delta+2}^{\delta+\Delta\delta+1} arccos(\frac{\lambda}{[w+(2(n_2+\Delta n-m)+3)^2]^{1/2}})=\nonumber\\
&&-\sum_{m=1}^{\delta+1} [arccos(\frac{\lambda}{[w+(2(n_2+\Delta n-m)+3)^2]^{1/2} })-\nonumber\\
&&arccos(\frac{\lambda}{[w+(2(n_2-m)+3)^2]^{1/2}})]
\label{Delta}
\end{eqnarray}
If $\Delta n\ll n_2$ then the l.h.s. approximately equals $\delta\theta$ and in the first order correction in $\Delta n$
we have the approximate relation
\begin{equation}
\Delta n=- (W-\lambda^2)\frac{\Delta\delta}{4\delta n_2}
\label{Delta2}
\end{equation} 
The r.h.s of this relation differs from the r.h.s. of Eq. (\ref{1st}) by the factor $(W-\lambda^2)^{1/2}/(2n_2)$.
This factor can be greater or less than unity but the solution (\ref{Delta2}) is unacceptable because in this case 
it is not possible to define $r$ satisfying Eq. (\ref{coordmom}). Nevertheless we believe that this example gives
hope that our conjecture on the problem of time can be substantiated with the exact solution of Eq. (\ref{1E}).

\section{Conclusion}

Although the number $p$ is a fundamental
parameter defining physical laws, this does not mean that this number is always the same in 
the history of Universe. We do not say that the number is the same at all times because time is
a pure classical notion and should not be present in quantum theory. Our conjecture is that {\it the existence of classical time 
is a consequence of the fact that $p$ changes} and in Sec. \ref{motion} we {\it define} time such that its 
variation $\Delta t$ is related to the variation of $p$ as
\begin{equation}
\Delta t=\frac{R}{c}\frac{\Delta lnp}{lnp}
\label{timedef}
\end{equation}
where $R$ is the parameter of contraction from the dS algebra to the Poincare one. Then, as shown in this section, there exist scenarios when classical equations of motions for cosmological acceleration and gravity 
can be obtained from pure quantum
notions without using space, time and standard semiclassical approximation.

In this scenario the goal of quantum theory is to determine how mean values of dS angular momenta change when 
the widths of their distribution change. As shown in Sec. \ref{1dim}, even in the one-dimensional model
discussed in this subsection the problem of finding exact solutions is very difficult. However, in Sec. \ref{motion}
we indicated two possibilities when classical equations of motion in standard dS antigravity and standard
gravity can be indeed obtained from pure quantum theory without involving any classical notions and standard semiclassical
approximation. 

\chapter{Discussion and conclusion}
\label{Discussion}

In Sec. \ref{crisis} we argue that the main reason of crisis in quantum physics is that nature, 
which is fundamentally discrete and even finite, is described by
continuous mathematics. Moreover, no ultimate physical theory can be based on continuous mathematics because
 that mathematics has its own foundational problems which cannot be resolved (for example, as a consequence of 
G\"{o}del's incompleteness theorems). In the first part of the work we discuss inconsistencies in standard approach 
to quantum theory and then we reformulate the theory such that it can be naturally generalized to a formulation 
based on finite mathematics. In this chapter we
discuss the main results of the present work in position operator, cosmological constant problem, gravity and
particle theory. 

\section{Position operator and wave packet spreading}

In standard physics education, the position operator is typically discussed only in nonrelativistic quantum mechanics.
Here it is postulated that coordinate and momentum representations are related
to each other by the Fourier transform and this leads to famous uncertainty relations. 
We argue that the postulate is based neither on strong 
theoretical arguments nor on experimental data.

In relativistic quantum theory local fields are discussed but typically in standard textbooks the argument $x$ of
those fields is not associated with any position operator (in spite of the principle of quantum theory that any physical
quantity can be discussed only in conjunction with the operator of this quantity). Probably one of the reasons is that
local quantum fields do not have a probabilistic interpretation and play only an auxiliary role for constructing
the $S$-matrix in momentum space. When this construction is accomplished the theory does not contain space-time
anymore in the spirit of the Heisenberg $S$-matrix program that in quantum theory one can describe only transitions
of states from the infinite past when $t\to -\infty$ to the distant future when $t\to +\infty$. As a consequence,
many physicists believe that the position operators is meaningful only in nonrelativistic theory. 

However, relativistic position operator is needed in several problems. For example, when we consider how photons 
from distant objects move to Earth we should know where those photons have been created (on Sun, Sirius or
other objects), what is the (approximate) trajectory of those photons etc. Meanwhile many quantum physicists are
not aware of the fact that relativistic position operator has been intensively discussed in papers by Newton and Wigner,
Hawton and other authors. By analogy with nonrelativistic quantum mechanics, in those papers the position and momentum operators are also related to each other by the Fourier transform.

Immediately after creation of quantum theory it has been realized that an inevitable consequence of the fact that
the position and momentum operators are related to each other by the Fourier transform is the effect
of wave packing spreading (WPS). Several well-known physicists (e.g. de Broglie) treated this fact as unacceptable
and proposed alternative approaches to quantum theory. At the same time,
it has not been shown that numerical results on WPS contradict experimental data. 
However, as shown in Chap. \ref{WPS}, in standard theory the results for WPS lead to
paradoxes. The most striking of them is that predictions of the theory contradict our experience in
observations of stars. 

We propose a consistent construction of the position operator where the position and momentum operators are not
related to each other by the Fourier transform. Then
the effect of WPS in directions perpendicular to the particle momentum is absent and the paradoxes are resolved.
Different components of the new position operator do not commute with each other and, as a consequence, there is
no WF in coordinate representation.

Our results give strong arguments that the notion of space-time is pure classical and does not exist on quantum
level. Hence fundamental quantum theory should not be based on Lagrangians and quantum fields in coordinate
representation.   

\section{Cosmological constant problem}

As noted in Sect. \ref{inter}, one of the main ideas of this work is that gravity might be 
not an interaction but simply a manifestation of
dS symmetry over a finite ring or field. This is obviously not in the spirit of mainstream approaches that gravity is a manifestation of the graviton exchange or holographic principle. 
Our approach does not involve GR, QFT, string theory, 
loop quantum gravity or other sophisticated 
theories. We consider only systems of {\it free} bodies in dS invariant quantum mechanics. 

Then the fact that we observe the cosmological repulsion is a strong argument that dS
symmetry is a more pertinent symmetry than Poincare or AdS ones. As shown in Refs. 
\cite{jpa1,symm1401} and in the present work, the phenomenon of the cosmological repulsion can be easily understood by considering semiclassical approximation in standard dS invariant quantum mechanics of two
free bodies. In the framework of this consideration it becomes immediately clear that the cosmological
constant problem does not exist and there is no need to involve empty space-time background,
dark energy or other artificial notions.  This phenomenon can be easily explained by using only standard 
quantum-mechanical notions  without involving dS space, metric, connections or
other notions of Riemannian geometry. 

One might wonder why such a simple explanation has not been widely discussed
in the literature. According to our observations, this is because even 
physicists working on dS QFT are not
familiar with basic facts about IRs of the dS algebra. It is difficult to imagine how 
standard Poincare invariant quantum theory can be constructed without involving known results on 
IRs of the Poincare algebra. Therefore it is reasonable to think that when Poincare invariance is replaced by dS one,
IRs of the Poincare algebra should be replaced by IRs of the dS algebra. However, physicists working
on QFT in curved space-time believe that fields are more fundamental than particles and therefore there is no need to involve IRs.

\section{Gravity}
\label{grav}

The mainstream approach to gravity is that gravity is the fourth (and probably the last)
interaction which should be unified with electromagnetic, weak and strong interactions.
While electromagnetic interaction is a manifestation of the photon exchange, weak
interaction is a manifestation of the W and Z boson exchange and strong interaction is a
manifestation of the gluon exchange, gravity is supposed to be a manifestation of the
graviton exchange. However, the notion of the exchange by virtual particles is taken from
particle theory while gravity is known only at macroscopic level. Hence thinking that gravity
can be explained by mechanisms analogous to those in particle theory is a great extrapolation.

There are several theoretical
arguments in favor of the graviton exchange. In particular, in the nonrelativistic approximation
Feynman diagrams for the graviton exchange can recover the Newton gravitational law by
analogy with how Feynman diagrams for the photon exchange can recover the Coulomb law.
However, the Newton gravitational law is known only on macroscopic level and, as noted in Sec. \ref{intropos},
the conclusion that the photon exchange reproduces the Coulomb law can be made only if one assumes
that coordinate and momentum representations are related to each other by the Fourier transform.
As discussed in Chaps. \ref{Ch1} and \ref{WPS}, on quantum level the coordinates are not needed and,
as shown in Chap. \ref{WPS}, standard position operator contradicts experiments. In addition, as
noted in Sec. \ref{intropos}, even on classical level the Coulomb law for pointlike electric charges has
not been verified with a high accuracy. So on macroscopic level the validity of the Newton gravitation law has been 
verified with a much greater accuracy than the Coulomb law. In view of these remarks, the argument that in
quantum theory the Newton gravitational law should be obtained by analogy with the Coulomb law is not
convincing.

The existence of gravitons can also be expected from the fact that GR (which is a classical
theory) predicts the existence of gravitational waves and that from the point of view of
quantum theory each classical wave consists of particles. However,
as discussed in Secs. \ref{nonNewton} and \ref{Gravwaves}, the statement that the data on 
binary pulsars and the recent LIGO data can be treated as a confirmation 
of the existence of gravitational waves is strongly model dependent and, as discussed in Sec. \ref{crisis}, 
the conclusion that the results \cite{BICEP2} of the BICEP2 collaboration can be treated as an indirect confirmation 
of the existence of gravitational waves is not based on strong theoretical arguments.

Any quantum theory 
of gravity can be tested only on macroscopic level. Hence, the problem is not only to construct quantum theory
of gravity but also to understand a correct structure of the position operator on macroscopic level. However, in
the literature the latter problem is not discussed because it is tacitly assumed that the position operator on 
macroscopic level is the same as in standard quantum theory. This is an additional great extrapolation which should be substantiated.

Efforts to construct quantum theory of gravity have not been successful yet.
Mainstream theories are based on the assumption that $G$ is a fundamental constant while,
as argued throughout this work, there are no solid reasons
to think so. The assumption that $G$ is a fundamental constant has been also adopted in GR. However, as
discussed in Secs. \ref{nonNewton} and \ref{Gravwaves}, the existing results on non-Newtonian gravitational experiments cannot 
be treated as an unambiguous confirmation of GR.

In recent years a number of works has appeared
where the authors treat gravity not as a fundamental interaction but as an emergent
phenomenon. We believe that until the nature of gravity has been unambiguously understood,
different approaches to gravity should be investigated. In the present work we consider
gravity as a pure kinematical manifestation of quantum dS symmetry in semiclassical 
approximation.

In contrast to IRs of the Poincare
and AdS algebras, in IRs of the dS algebra the particle mass {\it is not} the lowest eigenvalue of the dS Hamiltonian which
has the spectrum in the range $(-\infty,\infty)$. As a consequence, the free mass operator of the two-particle system
is not bounded below by $(m_1+m_2)$ where $m_1$ and $m_2$ are the particle masses. The discussion in Secs. \ref{antigravity}
and \ref{LambdaDiscrete} shows that this property by no means implies that the theory is unphysical.

Since in Poincare and AdS invariant theories the spectrum of the free mass operator is bounded below by $(m_1+m_2)$,
in these theories it is impossible to obtain the correction $-Gm_1m_2/r$ to the mean value of this operator.
However, in dS theory there is no law prohibiting such a correction. It is not a problem to indicate internal
two-body WFs for which the mean value of the mass operator contains $-Gm_1m_2/r$ with possible post-Newtonian
corrections. The problem is to show that such WFs are semiclassical with a high accuracy. As shown in Chaps. \ref{Ch2} and {\ref{twobody},
in semiclassical approximation any correction to the standard mean value of the mass operator is negative and
proportional to the energies of the particles. In particular, in the nonrelativistic approximation it is
proportional to $m_1m_2$. 

Our consideration in Chap. \ref{twobody} gives additional arguments (to those posed in Chap. \ref{WPS})
that standard distance operator should be modified since a problem arises whether it is physical at 
macroscopic distances. In Chap. \ref{twobody}
we argue that it is not and propose a modification of the distance operator which has correct properties
and gives for mean values of the free two-body mass operators the results compatible with   
Newton's gravity if the width of the de Sitter momentum distribution for a macroscopic body is inversely 
proportional to its mass. It has been also shown in Sec. \ref{classeq} that for all known gravitational
experiments, classical equations of motion can be obtained without involving the Lagrangian or Hamiltonian formalism
but assuming only that time is defined as in Eq. (\ref{coordmom}), i.e. that the relation between the spatial displacement
and the momentum is as in standard theory for free particles.

\section{Quantum theory over a finite ring or field}
 
In Chaps. \ref{Ch4} and \ref{Ch5} we argue that quantum theory based on a finite ring or field is more pertinent
than quantum theory  based on complex numbers. We tried to make the presentation as simple as possible without 
assuming that the reader is familiar with finite mathematics. Our version of a finite quantum theory (FQT) gives a natural qualitative
explanation why the width of the total dS momentum distribution of the macroscopic body is inversely proportional
to its mass. In this approach neither $G$ nor $\Lambda$ can be fundamental physical constants. 
We argue that only $G\Lambda$ might have physical meaning. The calculation of this quantity is a very difficult 
problem since it requires a detailed knowledge
of WFs of many-body systems. However, FQT gives clear indications that $G\Lambda$ contains a factor
$1/lnp$ where $p$ is the characteristic of the finite ring or field used in FQT. We treat standard theory as a special case of FQT in
the formal limit $p\to\infty$. Therefore gravity disappears in this limit. Hence in our approach gravity is a 
consequence of the fact that dS symmetry is considered over a finite ring or field rather than the field of complex numbers.

In our approach gravity is a phenomenon which has a physical meaning only in situations when at least one body is
macroscopic and can be described in the framework of semiclassical approximation. The result (\ref{meanI2})
shows that gravity depends on the width of the total dS momentum distributions for the bodies under consideration. 
However, when one mass is much greater than the other, the momentum distribution for the body with the lesser mass
is not important. In particular, this is the case when one body is macroscopic and the other is the photon. At the
same time, the phenomenon of gravity in systems consisting only of elementary particles has no physical meaning
since gravity is not an interaction but simply a kinematical manifestation of dS invariance in FQT
in semiclassical approximation. 
In this connection a problem arises what is the minimum mass when a body can be treated as macroscopic. This
problem requires understanding the structure of the many-body WF.

\section{Why finite mathematics is the most general}

The absolute majority of physicists believe that ultimate quantum theory will be based on classical mathematics 
involving the notions of infinitely small/large, continuity etc. Those notions were first proposed by Newton and
Leibniz more than 300 years ago when people did not know about the existence of atoms and elementary particles while in
quantum theory those notions are not natural. In addition, classical mathematics has foundational problems which,
according to G\"{o}del's incompleteness theorems, cannot be resolved.

The usual opinion is that finite mathematics is something inferior what is used only in special applications.
However, as proved in Sec.  \ref{finitemath}, the situation is the opposite: classical mathematics is a special degenerate case
of finite one in the formal limit when the characteristics of the field or ring in finite mathematics goes to
infinity. 

The first stage of the proof is the proof of {\it Statement 1} in Sec. \ref{Rp2Z} that the
ring $Z$ is the formal limit of the ring $R_p$ when $p\to\infty$. This fact poses a question on terminology of classical mathematics.
Here the phrase that $Z$ and the fields constructed from $Z$ (e.g. the fields of rational, real and complex numbers) are sets of characteristic 0 reflects the usual spirit that classical mathematics is more fundamental than finite one.  In our opinion it is natural to say that $Z$ is the ring of characteristic $\infty$ because it is a limit of rings of characteristics $p$ when
$p\to\infty$. The characteristic of the ring $p$ is understood such that all operations in the ring are modulo $p$ but operations modulo 0 are meaningless. Usually the characteristic of the ring is defined as the smallest positive number $n$ such that the sum of $n$ units $1+1+1...$ in the ring equals zero if such a number $n$ exists and 0 otherwise. However,  this sum can be written as $1\cdot n$ and the equality $1\cdot 0=0$ takes place in any ring.

Legitimacy of the limit of $R_p$ when $p\to\infty$ is problematic because when $R_p$ is replaced by $Z$
which is used as the starting point for constructing classical mathematics, we get classical
mathematics which has foundational problems, as discussed in Chap. \ref{Ch4}.

{\bf The fact that finite mathematics is more general than classical one implies that mathematics describing nature at the most fundamental level involves only a finite number of numbers while the notions of limit and 
infinitely small/large and the notions constructed from them (e.g. continuity, derivative and integral)
are needed only in calculations describing nature approximately}.

\section{Particle theory}

\subsection{Particle theory based on standard dS symmetry}

As noted above, in standard theory (based on complex numbers) the fact that $\Lambda > 0$
is a strong indication that dS symmetry is more pertinent than Poincare and AdS symmetries.
Hence it is reasonable to consider what happens when particle theory is considered from
the point of view of dS symmetry. Then  the key difference between IRs of the dS algebra 
on one hand and IRs of the Poincare and AdS algebras on the other
is that in the former case one IR can be implemented only on the upper and lower
Lorenz hyperboloids simultaneously. As a consequence, the number of states in IRs is always twice
as big as the number of states in the corresponding IRs of the AdS or Poincare algebra.
As explained in Sec. \ref{InterpretationOfIRs}, an immediate consequence of this
fact is that there are no neutral elementary particles in the theory.

Suppose that, by analogy with standard theory, one wishes
to interpret states with the support on the upper hyperboloid as particles and
states with the support on the lower hyperboloid as corresponding antiparticles. Then the first problem 
which arises is that the constant $C$ in Eq. (\ref{II60}) is infinite and one
cannot eliminate this constant by analogy with the AdS or Poincare theories.
Suppose, however, that this constant can be eliminated at least in Poincare approximation
where experiments show that the interpretation in terms of particles and antiparticles
is physical. Then, as shown in Sec. \ref{InterpretationOfIRs}, only fermions can be
elementary.

One might think that theories where only fermions can be elementary and the photon (and
also the graviton and the Higgs boson, if they exist) is not
elementary, cannot be physical. However, several authors
discussed models where the photon is composite; in particular,
in this work we discuss a possibility that the photon is a composite state of Dirac
singletons (see a discussion in the next section). An indirect confirmation of our
conclusions is that all known neutral elementary particles are bosons.

Another consequence of the fact that the IRs are implemented on the both hyperboloids is
that there is no superselection rule prohibiting states
which are superpositions of a particle and its antiparticle, and
transitions particle$\leftrightarrow$antiparticle are not prohibited. As a result, 
the electric charge and the baryon and lepton quantum numbers can be only approximately 
conserved. In particular, they are approximately conserved if Poincare approximation works
with a high accuracy.

This shows that dS invariant theory implies a considerably new
understanding of the notion of particles and antiparticles. In
contrast to Poincare or AdS theories, for combining a
particle and its antiparticle together, there is no need to
construct a local covariant field since they are already
combined at the level of IRs.

This is an important argument in favor of dS
symmetry. Indeed, the fact that in AdS and Poincare invariant
theories a particle and its antiparticle are described by
different IRs means that they are different objects. Then a
problem arises why they have the same masses and spins but
opposite charges. In QFT this follows from the CPT theorem
which is a consequence of locality since {\it we construct}
local covariant fields from a particle and its antiparticle
with equal masses. A question arises what happens if locality
is only an approximation: in that case the equality of masses,
spins {\it etc.}, is exact or approximate? Consider a simple
model when electromagnetic and weak interactions are absent.
Then the fact that the proton and the neutron have the same
masses and spins has nothing to do with locality; it is only a
consequence of the fact that the proton and the neutron belong
to the same isotopic multiplet. In other words, they are simply
different states of the same object---the nucleon. We see, that
in dS invariant theories the situation is analogous. The fact
that a particle and its antiparticle have the same masses and
spins but opposite charges (in the approximation when the
notions of particles, antiparticles and charges are valid) has
nothing to do with locality or non-locality and is simply a
consequence of the fact that they are different states of the
same object since they belong to the same IR. 

The non-conservation of the baryon and lepton quantum numbers
has been already considered in models of Grand Unification but
the electric charge has been always believed to be a strictly
conserved quantum number. In our approach all those quantum
numbers are not strictly conserved because in the case of dS
symmetry transitions between a particle and its
antiparticle are not prohibited. The experimental data that
these quantum numbers are conserved reflect the fact that at
present Poincare approximation works with a very high accuracy.
As noted in Sec. \ref{CC}, the cosmological constant is not
a fundamental physical quantity and if the quantity $R$ is very
large now, there is no reason to think that it was large
always. This completely changes the status of the problem known
as "baryon asymmetry of the Universe" since at early stages of
the Universe transitions between particles and antiparticles had a much
greater probability.

One might say that a possibility that only fermions can be elementary is not 
attractive since such a possibility would
imply that supersymmetry is not fundamental. There is no doubt that supersymmetry 
is a beautiful idea. On the other hand, one
might say that there is no reason for nature to have both, elementary fermions and 
elementary bosons since the latter can
be constructed from the former. A known historical analogy is that the simplest 
covariant equation is not the Klein-Gordon
equation for spinless fields but the Dirac and Weyl equations for the spin 1/2 fields 
since the former is the equation of the
second order while the latter are the equations of the first order.

In 2000, Clay Mathematics Institute announced seven Millennium Prize Problems. One of them is called
"Yang-Mills and Mass Gap" and the official description of this problem can be found in Ref. \cite{yangmills}.
In this description it is stated that the Yang-Mills theory should have three major properties where the
first one is as follows: 
"It must have a "mass gap;" namely there must be some constant $\Delta > 0$ such that every excitation of the
vacuum has energy at least $\Delta$." 
The problem statement assumes that quantum Yang-Mills theory should be constructed in the framework of Poincare
invariance. However, as follows from the above discussion, this invariance can be only approximate and dS
invariance is more general. Meanwhile, in dS theory the mass gap does not
exist. Therefore we believe that the problem has no solution. 

\subsection{Particle theory over a finite ring or field}

In standard theory a difference between representations of the so(2,3) and so(1,4) 
algebras is that 
IRs of the so(2,3) algebra where the operators $M^{\mu 4}$ ($\mu=0,1,2,3$)
are Hermitian can be treated as IRs of the so(1,4) algebra
where these operators are anti-Hermitian and vice versa. 
Suppose now that one accepts arguments of Chap. \ref{Ch4} that fundamental quantum theory
should be constructed over a finite ring or field rather than the field of complex numbers.
As noted in Chap. \ref{Ch4}, in FQT a probabilistic interpretation is only approximate and
hence Hermiticy can be only a good approximation in some situations. 
Therefore one cannot exclude a possibility that elementary particles can be
described by modular analogs of IRs of the so(2,3) algebra while modular
representations describing symmetry of macroscopic bodies are modular analogs of standard
representations of the so(1,4) algebra. In view of this observation, in Chap. \ref{AdS}
we consider standard and modular IRs of the so(2,3) algebra in parallel in order
to demonstrate common features and differences between standard and modular cases.

As noted in Chap. \ref{Ch4}, FQT does not contain infinities at all and all operators are
automatically well defined. In my discussions with physicists,
some of them commented this fact as follows. This is an
approach where a cutoff (the characteristic $p$ of the finite ring or 
field) is introduced from the beginning and for this reason
there is nothing strange in the fact that the theory does not
have infinities. It has a large number $p$ instead and this
number can be practically treated as infinite. The inconsistency of this argument is clear from the following analogy. 
It is not correct to say that relativistic theory is simply nonrelativistic one with the cutoff $c$ for
velocities. As a consequence of the fact that $c$ is finite, relativistic theory considerably differs from nonrelativistic one in several aspects.
The difference between finite rings or fields on one hand 
and usual complex numbers on the other
is not only that the former are finite and the latter are
infinite. If the set of usual numbers is visualized as a
straight line from $-\infty$ to $+\infty$ then the simplest
finite ring can be visualized not as a segment of this line
but as a circumference (see Figure \ref{Fig.2} in Sec. \ref{finmath}). This
reflects the fact that in finite mathematics the rules of arithmetic
are different and, as a result, FQT has many unusual features
which have no analogs in standard theory.

The Dirac vacuum energy problem discussed in Sec. \ref{VS6}
is a good illustration of this point. Indeed, in standard
theory the vacuum energy is infinite and, if FQT is treated
simply as a theory with a cutoff $p$, one would expect the
vacuum energy to be of the order of $p$. However, since the rules of
arithmetic in finite rings are different from standard ones,
the result of exact (i.e. non-perturbative) calculation
of the vacuum energy is precisely zero.

My original motivation for investigating FQT was as follows.
Let us take standard QED in dS or AdS space, write the
Hamiltonian and other operators in angular momentum basis and
replace standard IRs for the
electron, positron and photon by corresponding modular IRs. One
might treat this motivation as an attempt to substantiate
standard momentum regularizations (e.g., the Pauli-Villars
regularization) at momenta $p/R$ (where $R$ is the radius of
the Universe). In other terms this might be treated as
introducing fundamental length of the order of $R/p$. We now discuss
reasons explaining why this naive attempt fails.

One of the main results in Chap. \ref{AdS} is that (see Sec. \ref{VS3}) {\it in FQT the
existence of antiparticles follows from the fact that FQT is based on the finite
ring or field. Moreover, the very existence of
antiparticles is an indication that nature is described
by a finite field or ring rather than by complex numbers.} This result is not only very important but also
extremely simple and beautiful. A simple explanation follows.

In standard theory a particle is described by a positive energy IR where the energy has the spectrum in the range
$[mass,\infty)$. At the same time, the corresponding antiparticle is associated with a negative energy IR where the
energy has the spectrum in the range $(-\infty,-mass]$.
Consider now the construction of a modular IR for some particle.
We again start from the rest state (where energy=mass) and
gradually construct states with higher and higher energies.
However, in such a way we are moving not along a straight line
but along the circumference in Figure \ref{Fig.2}. Then sooner or later
we will arrive at the point where energy=-mass.
Therefore in FQT a particle and its antiparticle automatically belong to the same IR and have the same masses because 
the ring $R_p$ is finite and has the property of strong cyclicity. 

The fact that in FQT a particle and its antiparticle belong to the same IR
makes it possible to conclude that, in full analogy with the case of standard
dS theory (see the preceding section), there are no neutral particles in the theory,
the very notion of a particle and its antiparticle is only approximate and the
electric charge and the baryon and lepton quantum numbers can be only approximately
conserved. As shown in Sec. \ref{breaking}, if one tries to replace nonphysical annihilation
and creation operators $(a,a^*)$ by physical operators $(b,b^*)$ 
related to antiparticles then the symmetry on quantum level is inevitably broken. 
In FQT, by analogy with standard theory, it is possible not to introduce the notion
of antiparticles but work by analogy with Dirac's hole theory. Then the symmetry on
quantum level is preserved and, as shown in Sec. \ref{VS6}, in contrast to standard theory,
the vacuum can be chosen such that the vacuum energy is not infinite but zero.
This poses a problem whether there are physical reasons for such a choice of the
vacuum. 

As explained in Sec. \ref{VS7}, the spin-statistics theorem can be treated as a
requirement that standard quantum theory should be based on
complex numbers. This requirement also excludes the existence of
neutral elementary particles. 

Since FQT can be treated as the modular version of both, dS and AdS standard theories,
supersymmetry in FQT is not prohibited. In Sec. \ref{SS} we discuss common features and
differences between standard and modular IRs of the osp(1,4) algebra. One
of the most interesting feature of the modular case is how supersymmetry describes Dirac
singletons in FQT. This question is discussed in the next section.

\section{Dirac singletons}
\label{DirSingls}

One might think that since in
FQT the photon cannot be elementary, this theory cannot be
realistic and does not deserve attention. However,
the nonexistence of neutral elementary particles in FQT
shows that the photon (and the graviton and the Higgs boson if
they exist) should be considered on a deeper level. In Chap.
\ref{DiracSingletons} we argue that in FQT a possibility that
massless particles are composite states of Dirac singletons is
even more attractive than in standard theory.

As it has been noted in Chap. \ref{DiracSingletons}, the seminal result by Flato and Fronsdal \cite{FF} poses a
fundamental problem whether only Dirac singletons can be true elementary particles. In this case one has 
to answer the questions (see Sec. \ref{elementary}):
\begin{itemize}
\item a) Why singletons have not been observed yet.
\item b) Why such massless particles as photons and others are stable and their decays into singletons have not been observed.
\end{itemize}
In the literature, a typical explanations of a)
are that singletons are not observable because they cannot be considered in the Poincare limit or
because in this limit the singleton four-momentum becomes zero or because the singleton field lives
on the boundary of the AdS bulk or as a consequence of other reasons. As shown in Sec. \ref{semiclassical}, in standard theory,
semiclassical approximations for singletons in Poincare limit can be discussed in full analogy
with the case of massive and massless particles. As a result, in the general case the energy of singletons 
in Poincare limit is not zero but, in contrast to the case of usual particles, singletons can have
only two independent components of standard momentum, not three as usual particles. A problem arises
whether such objects can be detected by standard devices, whether they have a coordinate description etc.
At the same time, in standard theory there is no natural explanation of b). 

While in standard theory there
are four singleton IRs describing the Di and Rac singletons and their antiparticles, in FQT only two IRs
remain since standard Di and anti-Di now belong to the same IR and the same is true for standard Rac and anti-Rac.
We use Di and Rac to call the corresponding modular IRs, respectively. Nevertheless, 
since each massless boson can be represented as a composite state of two Dis or two Racs, a problem remains of what
representation (if any) is preferable. This problem has a natural solution if the theory is supersymmetric. 
Then the only IR is the (modular) Dirac supermultiplet combining (modular) Di and (modular) Rac into one IR.

The main result of Chap. \ref{DiracSingletons} is described in Sec. \ref{supertensorprod} where we explicitly describe a complete
set of supersymmetric modular IRs taking part in the decomposition of the tensor product of two modular 
Dirac supersingleton IRs. In particular, by analogy with the Flato-Fronsdal result, each massless superparticle
can be represented as a composite state of two Dirac supersingletons and one again can pose a question of whether
only Dirac (super)singletons can be true elementary (super)particles. 

This question is also natural in view of the following observation. As shown in Sec.
\ref{IRsdS}, the dS mass $m_{dS}$ and standard Poincare mass $m$
are related as $m_{dS}=Rm$ where $R$ is the radius of the Universe, and, as shown in Sec.
\ref{semiclassical}, the relation between the AdS and Poincare masses is analogous.
If for example one assumes that $R$ is of the order of $10^{26}m$ then the dS mass of the 
electron is of the order of $10^{39}$.
It is natural to think that a particle with such a dS mass cannot be elementary. Moreover, the present upper 
level for the photon mass is $10^{-16}ev$ which seems to be an extremely tiny quantity. However, the corresponding 
dS mass is of the order of $10^{17}$ and so even the mass which is treated as extremely small in Poincare
invariant theory might be very large in de Sitter theories. Nevertheless, assuming that only (super)singletons 
can be true elementary (super)particles, one still has to answer the questions a) and b). 

As explained in Sec. \ref{Singletons}, a crucial difference between Dirac singletons in standard theory and FQT follows.
In FQT 1/2 should be treated as $(p+1)/2$, the eigenvalues of the operators $h_1$ and $h_2$ for singletons 
in FQT are $(p+1)/2, (p+3)/2, (p+5)/2...$, i.e. huge numbers if $p$ is huge. Hence Poincare limit and semiclassical 
approximation for Dirac singletons in FQT have no physical meaning and they cannot be observable. In addition, 
as noted in Chap. \ref{Ch4}, the probabilistic interpretation for a particle
can be meaningful only if the eigenvalues of all the operators $M_{ab}$ are much less than $p$. Since for Dirac singletons
in FQT this is not the case, their state vectors do not have a probabilistic interpretation. These facts give a natural answer to the question a).

For answering question b) we note the following. In standard theory the notion of binding energy 
(or mass deficit) means that if a state with the mass $M$ is a bound state of two objects with the masses $m_1$ and $m_2$ then
$M<m_1+m_2$ and the quantity $|M-(m_1+m_2)|c^2$ is called the binding energy. The binding energy is a measure of
stability: the greater the binding energy is, the greater is the probability that the bound state will not decay
into its components under the influence of external forces. 

If a massless particle is a composite state of two Dirac singletons, and the eigenvalues of the operators 
$h_1$ and $h_2$ for the Dirac singletons in FQT are $(p+1)/2, (p+3)/2, (p+5)/2...$ then, since in FQT the eigenvalues 
of these operators should be taken modulo $p$, the corresponding eigenvalues for the massless particle are
$1, 2, 3...$. Hence an analog of the binding energy for the operators $h_1$ and $h_2$ is $p$, i.e. a huge number. 
This phenomenon can take place only in FQT: although, from the formal point of view, the Dirac singletons comprising
the massless state do not interact with each other, the analog of the binding energy for the operators $h_1$ and $h_2$
is huge. In other words, the fact that all the quantities in FQT are taken modulo $p$ implies a very strong
effective interactions between the singletons. It explains why the massless state does not decay into Dirac singletons
and why free Dirac singletons effectively interact pairwise for creating their bound state. 

As noted in the literature on singletons (see e.g. the review \cite{FFS} and references therein), the possibility that
only singletons are true elementary particles but they are not observable has some analogy with
quarks. However, the analogy is not full. According to Quantum Chromodynamics, forces between 
quarks at large distances prevent quarks from being observable in free states. In FQT Dirac
singletons cannot be in free states even if there is no interaction between them; the 
effective interaction between Dirac singletons arises as a consequence of the fact
that FQT is based on arithmetic modulo $p$. In addition, quarks and gluons are used
for describing only strongly interacting particles while in standard AdS theory and in FQT
quarks, gluons, leptons, photons, W and Z bosons can be constructed from Dirac singletons.

As noted at the end of Sec. \ref{supersingleton}, singleton physics can be directly generalized
to the case of higher dimensions, and this fact has been indicated in the literature on
singletons (see e.g. the review \cite{FFS} and references therein). 

Finally, in our opinion, an extremely important property of Dirac singletons in FQT is as follows.
As already noted, if the radius of the world $R$ is of the order of $10^{26}m$ or more then even the de 
Sitter mass of the electron is of the order of $10^{39}$ or more. It is unlikely that a particle with such
a large value of mass is elementary. It has been also noted several times that in quantum theory standard division cannot be fundamental. This poses
a problems whether it is necessary to have division in fundamental quantum theory, i.e. whether this theory 
should be based only on a ring and not a field. However, as noted in Sec. \ref{Matrix}, massive and massless
IRs in FQT can be constructed only over a field. At the same time, as shown in Sec. \ref{Matrix} and Chap.
\ref{DiracSingletons}, Dirac singletons in FQT can be discussed in a theory based only on a ring. In addition,
as shown in Chap. \ref{DiracSingletons}, massive and massless particles can be constructed from singletons.
Therefore the singleton physics in FQT is even more interesting than in standard theory, and the scenario that
only singletons are true elementary particles looks very appealing.

\section{Open problems} 

One of the main results of this work is that gravity can
be described as a pure kinematical manifestation of de Sitter symmetry over a finite ring or field. 
In this approach $G$ is not fundamental but a quantity which can be calculated.  
In Sec. \ref{inter} we argue that the very notion of interaction cannot be fundamental and interaction constants
can be treated only as phenomenological parameters. In particular, the Planck length has no fundamental meaning
and the notions of gravitational fields and gravitons are not needed.

In view of these results the following problems arise. Since gravity can be tested only on macroscopic level, any
quantum theory of gravity should solve the problem of constructing position operator on that level. As noted in Secs.
\ref{nonNewton} and \ref{grav}, in the literature this problem is not discussed because it is tacitly assumed that
the position operator on quantum level is the same as in standard quantum theory, but this is a great extrapolation.
In quantum theory it is postulated that any physical quantity is defined by an operator. However, quantum theory
does not define explicitly how the operator corresponding to a physical quantity is related to the measurement of
this quantity. As shown in Chap. \ref{twobody}, the mass operator for all known gravitational phenomena is fully
defined by a function describing the classical distance between the bodies in terms of their relative WF.
Therefore a fundamental problem is to understand the physical meaning of parameters characterizing WFs
of macroscopic bodies. 

In our approach quantum theory is based on a finite ring or field with the characteristic $p$. 
Although the number $p$ is a fundamental
parameter defining physical laws, this does not mean that this number is always the same in 
the history of Universe. In Chap. \ref{time} we discuss a  possibility that {\it the existence of classical time 
is a consequence of the fact that $p$ changes} and {\it define} time such that its 
variation $\Delta t$ is related to the variation of $p$ by Eq. (\ref{timedef}). 
In this scenario the goal of quantum theory is to determine how mean values of dS angular momenta change when 
the widths of their distribution change. As shown in Sec. \ref{1dim}, even in the one-dimensional model
discussed in this section the problem of finding exact solutions is very difficult. In Sec. \ref{motion}
we indicate two possibilities when classical equations of motion in dS antigravity and 
gravity can be indeed obtained from pure quantum theory without involving any classical notions and standard semiclassical
approximation. However, without exact solutions those possibilities can be treated only as arguments in favor
of the conjecture that the existence of classical time is a manifestation of the fact that $p$ changes.

Let us now discuss the following problem. Standard quantum theory is based on complex numbers for several reasons.
First, the theory involves momenta and coordinates which are related to each other by the Fourier transform. As noted in 
Chap. \ref{WPS}, this property is inherited from classical electrodynamics while adopting this property in quantum theory 
results in paradoxes. Another reason is that quantum theory involves selfadjoined operators in Hilbert spaces and, 
according to the spectral theorem, the spectral decomposition for them is always
valid only in complex Hilbert spaces. This property is related to the fact that the field of complex 
numbers {\cal C} is algebraically closed, i.e. any equation of the $n$th power in {\cal C} has exactly $n$ solutions.
However, quantum theory based on finite mathematics can involve only finite rings or fields which are not
algebraically closed; in particular here an equation of the $n$th power may have no solution at all. Nevertheless, in single-particle IRs
on finite rings or fields discussed in Chaps. \ref{Ch3} and \ref{AdS} the spectrum of all necessary physical operators in question is defined explicitly
by construction and therefore the fact that finite rings and fields are not algebraically closed is not important in this case.
Hence in FQT there are no strong arguments that the theory should be based on complex analogs of finite rings or fields,
and one can consider the possibility that complex extensions of those rings or fields are not necessary.

If ${\bf A}$ is the operator of a vector quantity then in quantum theory one can discuss the operators ($A_x,A_y,A_z$) representing projections of ${\bf A}$ on coordinate axes. However, the notion of coordinate axes is pure classical and it does not seem natural that this notion is present in
quantum theory. In classical approximation ${\bf A}$ becomes a vector and in this approximation quantum theory should determine projections
of the vector on coordinate axes but the notion of coordinate axis should be used only on classical level. 

As an example, consider the operators of the so(3) (or su(2)) algebra. They satisfy the commutation relations (\ref{J}) which necessarily
involve $i$ because the operators are selfadjoined. However, from the point of view of theory of Lie algebras, the most natural basis of 
operators in the representation space is not ($M_x,M_y,M_z$) but the Cartan-Weyl basis ($M_+,M_-,M_0$) where $M_0$ is the representation operator
of the basis element of the Cartan subalgebra and $M_{\pm}$ are the representation operators of the
root elements in the algebra. The commutation relations between these operators are given by
\begin{equation}
[M_0,M_-]=-2M_-,\quad [M_0,M_+]=2M_+,\quad [M_+,M_-]=M_0
\label{CW}
\end{equation}
and the Casimir operator is
\begin{equation}
K=M_0^2-2M_0+4M_+M_-=M_0^2+2M_0+4M_-M_+
\label{su2Casimir}
\end{equation}
These relations do not involve $i$ and if the basis of the representation space consists of eigenvectors of the operator 
$M_0$ then the matrix
elements of all representation operators are real.

If we now {\it define} $M_z=M_0/2$, $M_x=(M_++M_-)/2$, $M_y=-i(M_+-M_-)/2$ then the relations 
(\ref{CW}) become (\ref{J}) and $K=4{\bf M}^2$. 
We expect that in classical approximation the operators ($M_x,M_y,M_z$) 
become real values but it is clear that if the representation is considered
in the space over real numbers then it is possible to obtain real values only 
for ($M_x,M_z,{\bf M}^2$) but not for $M_y$. However, the real
value for the magnitude of $M_y$ can be found since $M_y^2={\bf M}^2-M_x^2-M_z^2$ 
and so only the sign of $M_y$ is not defined. It seems unnatural
that for defining only the sign of $M_y$ we must extend the representation space to the 
space over complex numbers. A problem arises whether
the real WF has a property which defines the direction of $M_y$ in classical limit. Analogously, a problem
arises whether for defining the direction of the momenta in the one-dimensional model of Sec. \ref{1dim}
it is necessary to involve complex functions. 

Consider now the following question. As already noted, in standard quantum theory the space of states is projective: $\psi$ and $const\cdot \psi$ are the same states because WFs have probabilistic interpretation
and only the ratio of probabilities has a physical meaning. On the other hand, in FQT probabilistic interpretation
takes place only when WFs are described by numbers which are much less than $p$. Therefore one might think
that, for example in semiclassical approximation, a system will try to go to a state described by least possible
numbers. But a question remains: do WFs $\psi$ and $-\psi$ describe the same state in FQT? Our conjecture is that this is not the case and the motivation follows.

From the point of view that each state is described by a computer, the states $\psi$ and $-\psi$ are
described by different numbers of bits. For example, if $a$ is a positive integer then $a$ and $-a$ are
described by different numbers of nonzero bits. One of the explanation is that in FQT -a=p-a and then it is
clear that $a$ and $-a$ are described by different numbers of nonzero bits. In computers the 
numbers $a$ and $-a$ differ by the bit defining the sign: usually this bit is zero for positive numbers
and 1 for negative ones. If $(c_1, c_2,...c_n)$ is the system WF then in semiclassical approximation 
the WF is approximately
the eigenfunction of all representation operators but in standard quantum theory the common
coefficient for all the numbers $(c_1, c_2,...c_n)$ is arbitrary. The above remarks make reasonable
the assumption that in FQT the system WF is such that the set $(c_1, c_2,...c_n)$ is described
by the minimum possible number of nonzero bits. If this assumption is correct then in FQT the WF
$\psi$ has a physical meaning itself in contrast to standard theory where only $|\psi|^2$ has a
physical meaning. This might solve the problem about the sign of $M_y$.

As shown in Chap. \ref{AdS}, in our approach  the notion of particle-antiparticle can be only approximate and 
the electric charge and other
additive quantum numbers (e.g. the baryon and lepton quantum numbers) can be only approximately conserved.
The extent of conservation depends on $p$: the greater is $p$, the greater is the extent of conservation.
One might think that at present the conservation laws work with a high accuracy because the present value of
$p$ is extremely large. However, if at early stages of the Universe the value of $p$ was much less than now then
the conservation laws were not so strict as now. In particular, this might be a reason of the baryonic 
asymmetry of the world.

By analogy with dS antigravity and gravity, one might think that electromagnetic, weak and strong interactions are not interactions
but manifestations of higher symmetries. Similar ideas have been already extensively discussed in the literature,
e.g. in view of compactification of extra dimensions. 

Our results indicate that fundamental quantum theory has a very long way ahead (in agreement with
Weinberg's opinion \cite{Wein2} that a new theory may be "centuries away").

\begin{center} {\bf Acknowledgements} \end{center}
I have greatly benefited from discussions with many physicists and mathematicians and 
it is difficult to mention all of them. A collaboration with Leonid Avksent'evich Kondratyuk and discussions with 
Skiff Nikolaevich Sokolov were very important for my understanding of basics of quantum theory.
They explained that the theory should not necessarily be based on a local Lagrangian and symmetry on quantum level 
means that proper commutation relations are satisfied. Also, Skiff Nikolaevich told me about an idea that
gravity might be a direct interaction. Eduard Mirmovich has proposed an idea that only angular momenta 
are fundamental physical quantities \cite{Mirmovich}. This idea and Dyson's paper \cite{Dyson} 
have encouraged me to study de Sitter invariant 
theories. At that stage the excellent book by Mensky \cite{Mensky} was very helpful. Metod Saniga has pointed out
that for constructing fundamental quantum theory division is not fundamental and it is reasonable to expect
that this theory will be constructed over a ring, not a field. I am also grateful to Bernard Bakker, Jos{\'e} Manuel Rodriguez Caballero,
Sergey Dolgobrodov, Boris Hikin, Anatoly Kamchatnov, Vladimir Karmanov, Gregory Keaton, Dmitry Logachev, Volodya Netchitailo, 
Mikhail Aronovich Olshanetsky, Michael Partenskii and Teodor Shtilkind for numerous useful discussions
and to Efim Zelmanov for telling me about ultraproducts and Refs. \cite{ultraproducts,ultraproducts1}.

\end{document}